\documentclass[a4paper,twoside,11pt]{scrbook}
\def\RPlus{\ensuremath{\mathbin{\rule[.13em]{.66em}{.22em}\hspace{-.44em}\rule[-.08em]{.22em}{.66em}\,}}} %Fat plus symbol
\usepackage[english]{babel}
\usepackage{graphicx}
\usepackage{dsfont}
\usepackage{bm}
\usepackage{amssymb}
\usepackage{amsthm}
\usepackage{amsmath}
\usepackage{amsfonts}
\usepackage{booktabs}
\usepackage{amsbsy}
\usepackage{fontenc}
\usepackage[utf8]{inputenc}
\usepackage{fancybox}
\usepackage[hypcap=false]{caption}
\usepackage{float}
\usepackage{subfigure} 
\usepackage{lmodern}
\usepackage[numbib,nottoc]{tocbibind}
\usepackage[pdfstartview={FitH},linkbordercolor={0 1 0}, colorlinks]{hyperref}
\usepackage{esint}
\usepackage{fancyhdr}
\usepackage{pdfpages}
\usepackage{trfsigns}
\usepackage{wasysym} %Smiley
\usepackage{siunitx} %SI Einheiten
\usepackage{mathtools} %für mathclap
\usepackage{relsize} %Integralgröße
\usepackage{scrhack} %Warnung bei book bezüglich toc
\usepackage{mleftright} %Distanz zu \left \right weg
\usepackage{tikz} %Diagramme
\usepackage[many]{tcolorbox} %Coloured boxes around theorems etc
\tcbuselibrary{theorems} % siehe oben

\usetikzlibrary{cd}

\makeatletter
\DeclareFontEncoding{LS1}{}{}
\DeclareFontSubstitution{LS1}{stix}{m}{n}
\DeclareMathAlphabet{\mathKel}{LS1}{stixscr}{m}{n}
\DeclareMathAlphabet{\mathcal}{LS1}{stixscr}{m}{n}
\makeatother
\DeclareMathOperator{\sAut}{\mathKel{A\mkern-5.5mu u\mkern-4mu t\mkern-1.5mu}}
\DeclareMathOperator{\saut}{\mathKel{a\mkern-4.5mu u\mkern-4mu t\mkern-1.5mu}}
\DeclareMathOperator{\sEnd}{\mathKel{E\mkern-4mu n\mkern-4.5mu d\mkern-1mu}}

\allowdisplaybreaks % mit \displaybreak einen Seitenumbruch in align-Umgebung erzeugen

\renewcommand{\theequation}{\arabic{chapter}.\arabic{equation}}

\def\be{\begin{equation}}
\def\ee{\end{equation}}
\def\bs{\begin{subequations}}
\def\es{\end{subequations}}
\def\ba#1\ea{\begin{align}#1\end{align}}
\def\bes{\begin{equation*}}
\def\ees{\end{equation*}}
\def\bas#1\eas{\begin{align*}#1\end{align*}}

\theoremstyle{plain}
\newtheorem{theorem}{Theorem}[section]
\newtcbtheorem
  [use counter*=theorem,number within=section]% init options
  {lemmata}% name
  {Lemma}% title
  {%
		fontupper=\itshape,
		breakable,
		enhanced,
    colback=gray!25,
    colframe=gray!0!black,
    fonttitle=\bfseries,
  }% options
  {lem}% prefix

\newtcbtheorem
  [use counter*=theorem,number within=section]% init options
  {propositions}% name
  {Proposition}% title
  {%
		fontupper=\itshape,
		breakable,
		enhanced,
    colback=gray!25,
    colframe=gray!0!black,
    fonttitle=\bfseries,
  }% options
  {prop}% prefix
\newtcbtheorem
  [use counter*=theorem,number within=section]% init options
  {theorems}% name
  {Theorem}% title
  {%
		fontupper=\itshape,
		breakable,
		enhanced,
    colback=gray!25,
    colframe=gray!0!black,
    fonttitle=\bfseries,
  }% options
  {thm}% prefix
\newtcbtheorem
  [use counter*=theorem,number within=section]% init options
  {corollaries}% name
  {Corollary}% title
  {%
		fontupper=\itshape,
		breakable,
		enhanced,
    colback=gray!25,
    colframe=gray!0!black,
    fonttitle=\bfseries,
  }% options
  {cor}% prefix

\theoremstyle{remark}
\newtcbtheorem
  [use counter*=theorem,number within=section]% init options
  {remarks}% name
  {Remark}% title
  {%
		breakable,
		enhanced,
    colback=gray!5,
    colframe=gray!50!black,
    fonttitle=\bfseries,
  }% options
  {rem}% prefix

\newtheorem{remark}[theorem]{Remarks}
\newtheorem*{remarkohne}{Remarks}

\newtcbtheorem
  [use counter*=theorem,number within=section]% init options
  {motivations}% name
  {Motivation}% title
  {%
		breakable,
		enhanced,
    colback=gray!5,
    colframe=gray!50!black,
    fonttitle=\bfseries,
  }% options
  {mot}% prefix

\theoremstyle{definition}

\newtcbtheorem
  [use counter*=theorem,number within=section]% init options
  {definitions}% name
  {Definition}% title
  {%
		breakable,
		enhanced,
    colback=gray!25,
    colframe=gray!0!black,
    fonttitle=\bfseries,
  }% options
  {def}% prefix
\newtcbtheorem
  [use counter*=theorem,number within=section]% init options
  {examples}% name
  {Example}% title
  {%colback=black!5,colframe=red!35!black
		breakable,
		enhanced,
    colback=gray!5,
    colframe=gray!50!black,
    fonttitle=\bfseries,
  }% options
  {ex}% prefix
\newtcbtheorem
  [use counter*=theorem,number within=section]% init options
  {situations}% name
  {Situation}% title
  {%colback=black!5,colframe=red!35!black
		breakable,
		enhanced,
    colback=gray!5,
    colframe=gray!50!black,
    fonttitle=\bfseries,
  }% options
  {sit}% prefix
\newtcbtheorem
  [use counter*=theorem,number within=section]% init options
  {conjectures}% name
  {Conjecture}% title
  {%colback=black!5,colframe=red!35!black
		breakable,
		enhanced,
    colback=gray!5,
    colframe=gray!50!black,
    fonttitle=\bfseries,
  }% options
  {conj}% prefix
\newtcbtheorem
  [use counter*=theorem,number within=section]% init options
  {fieldredefinitions}% name
  {Field redefinition}% title
  {%colback=black!5,colframe=red!35!black
		breakable,
		enhanced,
    colback=gray!5,
    colframe=gray!50!black,
    fonttitle=\bfseries,
  }% options
  {fieldredef}% prefix

\newcommand{\fullname}{Simon-Raphael Fischer}

\newcommand{\jahr}{2021}

\DeclareFontFamily{U}{mathx}{\hyphenchar\font45}
\DeclareFontShape{U}{mathx}{m}{n}{
      <5> <6> <7> <8> <9> <10>
      <10.95> <12> <14.4> <17.28> <20.74> <24.88>
      mathx10
      }{}
\DeclareSymbolFont{mathx}{U}{mathx}{m}{n}
\DeclareFontSubstitution{U}{mathx}{m}{n}
\DeclareMathAccent{\widecheck}{0}{mathx}{"71}
\DeclareMathAccent{\wideparen}{0}{mathx}{"75}

\usepackage[symbols,nogroupskip]{glossaries-extra}
\usepackage{glossary-mcols}  
\setglossarystyle{mcolindex}

\makenoidxglossaries

\glsxtrnewsymbol[description={Anchor of the space of derivations $\mathcal{D}(V)$ on a vector bundle $V$}]{a1}{\ensuremath{a}}
\glsxtrnewsymbol[description={Field of gauge bosons $A$}]{a0}{\ensuremath{A}}
\glsxtrnewsymbol[description={Field redefinition of the field of gauge bosons $A$}]{a0widetildelambda}{\ensuremath{\widetilde{A}^\lambda}}
\glsxtrnewsymbol[description={Field strength $F$}]{F}{\ensuremath{F}}
\glsxtrnewsymbol[description={(Generalized) field strength $G$}]{G}{\ensuremath{G}}
\glsxtrnewsymbol[description={Smooth manifolds}]{MN}{\ensuremath{M, N}}
\glsxtrnewsymbol[description={Exterior covariant derivative w.r.t. a connection $\nabla$}]{dnabla}{\ensuremath{\mathrm{d}^{\nabla}}}
\glsxtrnewsymbol[description={Exterior covariant derivative w.r.t. a $E$-connection ${}^E\nabla$}]{dEnabla}{\ensuremath{\mathrm{d}^{{}^E\nabla}}}
\glsxtrnewsymbol[description={Chevalley-Eilenberg differential}]{dCE}{\ensuremath{\mathrm{d}_{\mathrm{CE}}}}
\glsxtrnewsymbol[description={$E$-differential}]{dE}{\ensuremath{\mathrm{d}_{\mathrm{E}}}}
\glsxtrnewsymbol[description={Total differential of smooth maps, also viewed as functional}]{D}{\ensuremath{\mathrm{D}}}
\glsxtrnewsymbol[description={Space of derivations on a vector bundle $V$ at a base point $p \in N$}]{DApV}{\ensuremath{\mathcal{D}_p(V)}}
\glsxtrnewsymbol[description={Bundle of derivations on a vector bundle $V$}]{DAV}{\ensuremath{\mathcal{D}(V)}}
\glsxtrnewsymbol[description={Exterior covariant derivative w.r.t. a basic connection $\nabla^{\mathrm{bas}}$}]{dbasnabla}{\ensuremath{\mathrm{d}^{\nabla^{\mathrm{bas}}}}}
\glsxtrnewsymbol[description={de-Rham differential}]{dbas}{\ensuremath{\mathrm{d}}}
\glsxtrnewsymbol[description={Space of homomorphisms}]{Hom}{\ensuremath{\mathrm{Hom}}}
\glsxtrnewsymbol[description={Space of endomorphisms}]{End}{\ensuremath{\mathrm{End}}}
\glsxtrnewsymbol[description={Space of sections of endomorphisms}]{EndSection}{\ensuremath{\sEnd}}
\glsxtrnewsymbol[description={Space of sections of automorphisms}]{AutSection}{\ensuremath{\sAut}}
\glsxtrnewsymbol[description={Space of functionals for gauge theory as $k$-forms}]{Fk}{\ensuremath{\mathcal{F}_E^k}}
\glsxtrnewsymbol[description={Space of polynomial functionals for gauge theory as $k$-forms}]{Hk}{\ensuremath{\mathcal{H}_E^k}}
\glsxtrnewsymbol[description={Space of automorphisms}]{Aut}{\ensuremath{\mathrm{Aut}}}
\glsxtrnewsymbol[description={Adjoint representation of a Lie algebra}]{ad}{\ensuremath{\mathrm{ad}}}
\glsxtrnewsymbol[description={Space of diffeomorphisms}]{DAiff}{\ensuremath{\mathrm{Diff}}}
\glsxtrnewsymbol[description={Image of a function}]{Im}{\ensuremath{\mathrm{Im}}}
\glsxtrnewsymbol[description={Kernel of a function}]{Ker}{\ensuremath{\mathrm{Ker}}}
\glsxtrnewsymbol[description={Rank of a linear operator}]{Rzk}{\ensuremath{\mathrm{rk}}}
\glsxtrnewsymbol[description={Leaf of the (singular) orbit foliation of a Lie algebroid $E$}]{Leaf}{\ensuremath{L}}
\glsxtrnewsymbol[description={Tensors $\Gamma\left( \mathrm{T}^r_s(V) \right)$ ($r, s \in \mathds{N}_0$) for a vector bundle $V$}]{Trs(V)a}{\ensuremath{\mathcal{T}^r_s(V)}}
\glsxtrnewsymbol[description={$\mathrm{T}^r_s(V) := \bigotimes^s V^* \otimes \bigotimes^r V$ ($r, s \in \mathds{N}_0$) for a vector bundle $V$}]{Trs(V)}{\ensuremath{\mathrm{T}^r_s(V)}}
\glsxtrnewsymbol[description={Lie algebra}]{g1}{\ensuremath{\mathfrak{g}}}
\glsxtrnewsymbol[description={Lie group}]{g0}{\ensuremath{G}}
\glsxtrnewsymbol[description={Structure functions of an algebra (mainly of a Lie algebroid $E$)}]{Cbca}{\ensuremath{C_{bc}^a}}
\glsxtrnewsymbol[description={Jacobiator of an algebra}]{J}{\ensuremath{J}}
\glsxtrnewsymbol[description={Vector fields on a smooth manifold $N$}]{X(N)}{\ensuremath{\mathfrak{X}(N)}}
\glsxtrnewsymbol[description={Lie or $E$-Lie derivative (with Lie algebroid $E$})]{Lie}{\ensuremath{\mathcal{L}}}
\glsxtrnewsymbol[description={Vector bundle and mostly a smooth Lie algebroid}]{E}{\ensuremath{E}}
\glsxtrnewsymbol[description={Fiber of a vector bundle $E$ at a point $p$}]{Ep}{\ensuremath{E_p}}
\glsxtrnewsymbol[description={Smooth functions on a smooth manifold $N$}]{Cinfty(N)}{\ensuremath{C^\infty(N)}}
\glsxtrnewsymbol[description={Smooth functions on $N$ with values in $M$}]{Cinfty(N;M)}{\ensuremath{C^\infty(N; M)}}
\glsxtrnewsymbol[description={Tangent bundle of a smooth manifold $N$}]{TN}{\ensuremath{\mathrm{T}N}}
\glsxtrnewsymbol[description={Cotangent bundle of a smooth manifold $N$}]{T*N}{\ensuremath{\mathrm{T}^*N}}
\glsxtrnewsymbol[description={$E$-curvature of an $E$-connection ${}^E\nabla$}]{RnablaE}{\ensuremath{R_{{}^E\nabla}}}
\glsxtrnewsymbol[description={Curvature of infinitesimal gauge transformation $\delta$}]{Rdelta}{\ensuremath{R_{\delta}}}
\glsxtrnewsymbol[description={Basic curvature of a connection $\nabla$, also denoted by $S_\nabla$}]{Rnablabas}{\ensuremath{R^{\mathrm{bas}}_{\nabla}}}
\glsxtrnewsymbol[description={Curvature of a connection $\nabla$}]{Rnabla}{\ensuremath{R_{\nabla}}}
\glsxtrnewsymbol[description={Curvature of a vector bundle morphism $\xi$ of Lie algebroids}]{Rxi}{\ensuremath{R_{\xi}}}
\glsxtrnewsymbol[description={Basic curvature of a connection $\nabla$, also denoted by $R^{\mathrm{bas}}_{\nabla}$}]{Snabla}{\ensuremath{S_{\nabla}}}
\glsxtrnewsymbol[description={$E$-torsion of an $E$-connection ${}^E\nabla$}]{tEnabla}{\ensuremath{t_{{}^E\nabla}}}
\glsxtrnewsymbol[description={Projection onto $i$-th factor.}]{pri}{\ensuremath{\mathrm{pr}_i}}
\glsxtrnewsymbol[description={Abbreviation for Lie algebra bundle}]{LAB}{LAB}
\glsxtrnewsymbol[description={Abbreviation for curved Yang-Mills-Higgs gauge theory, also CYMH GT}]{CYMH}{CYMH}
\glsxtrnewsymbol[description={Yang-Mills-Higgs Lagrangian}]{LYMH}{\ensuremath{\mathfrak{L}_{\mathrm{YMH}}}}
\glsxtrnewsymbol[description={Curved Yang-Mills-Higgs Lagrangian}]{LZYMH}{\ensuremath{\mathfrak{L}_{\mathrm{CYMH}}}}
\glsxtrnewsymbol[description={Abbreviation for bundle of Lie algebras}]{BLA}{BLA}
\glsxtrnewsymbol[description={Derivation of sections along a path, denoted by $\mathrm{D}/\mathrm{d}t$}]{Ddt}{\ensuremath{\frac{\mathrm{D}}{\mathrm{d}t}}}
\glsxtrnewsymbol[description={Field redefinition of a Riemannian metric $g$}]{gwidetildelambda}{\ensuremath{\widetilde{g}^{\lambda}}}
\glsxtrnewsymbol[description={Space of fields, consisting of gauge bosons $A$ and Higgs-like fields $\Phi$}]{MSpaceOfFields}{\ensuremath{\mathfrak{M}_E}}
\glsxtrnewsymbol[description={Vertical bundle of a fibre bundle $F$}]{VF}{\ensuremath{\mathrm{V}F}}
\glsxtrnewsymbol[description={Subspace of $\mathfrak{X}\bigl(\mathfrak{M}_E(M;N)\bigr)$, vector fields along $B$-paths of a Lie algebroid $B$}]{XBM}{\ensuremath{\mathfrak{X}^B\bigl(\mathfrak{M}_E(M; N)\bigr)}}
\glsxtrnewsymbol[description={Centre of a Lie algebroid connection ${}^E\nabla$}]{ZENabla}{\ensuremath{Z\mleft( {}^E\nabla \mright)}}
\glsxtrnewsymbol[description={Centre of an LAB $K$}]{ZLAB}{\ensuremath{Z\mleft( K \mright)}}
\glsxtrnewsymbol[description={Lie bracket derivations of an LAB $K$}]{DAerK}{\ensuremath{\mathrm{Der}(K)}}
\glsxtrnewsymbol[description={Derivations of an LAB $K$ which are also Lie bracket derivations}]{DAVDerK}{\ensuremath{\mathcal{D}_{\mathrm{Der}}(K)}}
\glsxtrnewsymbol[description={Ideal of inner bracket derivations of an LAB $K$}]{adK}{\ensuremath{\mathrm{ad}(K)}}
\glsxtrnewsymbol[description={Outer bracket derivations of an LAB $K$}]{OutKA}{\ensuremath{\mathrm{Out}(K)}}
\glsxtrnewsymbol[description={Derivations of an LAB $K$ which are also outer bracket derivations}]{OutKDDerK}{\ensuremath{\mathrm{Out}\mleft(\mathcal{D}_{\mathrm{Der}}(K)\mright)}}
\glsxtrnewsymbol[description={Evaluation map with respect to the Higgs field}]{ev}{\ensuremath{\mathrm{ev}}}

\glsxtrnewsymbol[description={Space of sections of a vector bundle $V$}]{1Camma(V)}{\ensuremath{\Gamma(V)}}
\glsxtrnewsymbol[description={Dual bundle of a vector bundle $V$}]{V*}{\ensuremath{V^*}}
\glsxtrnewsymbol[description={Kronecker delta, the indices might shift their position to upper positions}]{1deltaijz}{\ensuremath{\delta_{ij}}}
\glsxtrnewsymbol[description={Infinitesimal gauge transformation parametrised by $\varepsilon$}]{1delta0epsilon}{\ensuremath{\delta_{\varepsilon}}}
\glsxtrnewsymbol[description={Variations of functionals along vector fields $\Psi_\varepsilon$ along Lie algebroid paths}]{1delta0Psiepsilon}{\ensuremath{\delta_{\Psi_\varepsilon}}}
\glsxtrnewsymbol[description={Infinitesimal gauge transformation on $E$-valued functionals using $\nabla_\rho$}]{1delta1varepsilon}{\ensuremath{\delta^{(1)}_\varepsilon}}
\glsxtrnewsymbol[description={Infinitesimal gauge transformation on $E$-valued functionals using $\nabla^{\mathrm{bas}}$, later just $\delta_\varepsilon$}]{1delta2varepsilon}{\ensuremath{\delta^{(2)}_\varepsilon}}
\glsxtrnewsymbol[description={Pre-bracket}]{1Delta}{\ensuremath{\Delta}}
\glsxtrnewsymbol[description={Components of the Levi-Civita tensor}]{1epsilonijk}{\ensuremath{\epsilon_{ijk}}}
\glsxtrnewsymbol[description={Contraction of tensors, or mainly "bookkeeping trick"}]{1jota}{\ensuremath{\iota}}
\glsxtrnewsymbol[description={Conormal bundle}]{1nu*}{\ensuremath{\nu^*}}
\glsxtrnewsymbol[description={Anchor of a Lie algebroid $E$, sometimes written as $\rho_E$}]{1rho}{\ensuremath{\rho}}
\glsxtrnewsymbol[description={An element of $\Omega^1(N;E)$ used for the field redefinition}]{1lambda}{\ensuremath{\lambda}}
\glsxtrnewsymbol[description={Physical field, major example is the Higgs field}]{1vhi}{\ensuremath{\Phi}}
\glsxtrnewsymbol[description={Minimal coupling $\mathfrak{D}$ of a physical field $\Phi$ with the field of gauge bosons $A$}]{DAPhi}{\ensuremath{\mathfrak{D}^A \Phi}}
\glsxtrnewsymbol[description={$\Lambda := \mathds{1}_E - \lambda \circ \rho \in \sAut(E)$, used for field redefinition}]{1Lambda}{\ensuremath{\Lambda}}
\glsxtrnewsymbol[description={Field redefinition of a fibre metric $\kappa$ of a vector bundle $E$}]{1kappawidetildelambda}{\ensuremath{\widetilde{\kappa}^{\lambda}}}
\glsxtrnewsymbol[description={$\widehat{\Lambda} := \mathds{1}_{\mathrm{T}N} - \rho \circ \lambda \in \sAut(\mathrm{T}N)$, used for field redefinition}]{1Lambdatilde}{\ensuremath{\widehat{\Lambda}}}
\glsxtrnewsymbol[description={Lie algebra action $\gamma: \mathfrak{g} \to \mathfrak{X}(N)$ of a Lie algebra $\mathfrak{g}$}]{1cammaz}{\ensuremath{\gamma}}
\glsxtrnewsymbol[description={Primitive of a connection in the context of curved gauge theory}]{1fZeta}{\ensuremath{\zeta}}
\glsxtrnewsymbol[description={Field redefinition of $\zeta$}]{1fZetaTilde}{\ensuremath{\widetilde{\zeta}^\lambda}}
\glsxtrnewsymbol[description={Field redefinition of $\zeta = 0$}]{1fZetaTilHat}{\ensuremath{\widehat{\zeta}^\lambda}}
\glsxtrnewsymbol[description={Space of $k$-forms of a smooth manifold $N$, $k \in \mathds{N}_0$}]{1ZOmegak(N)}{\ensuremath{\Omega^k(N)}}
\glsxtrnewsymbol[description={Space of $q$-forms of a vector bundle $E$, $q \in \mathds{N}_0$}]{1ZOmegas(E)}{\ensuremath{\Omega^s(E)}}
\glsxtrnewsymbol[description={Space of $(p,q)$-$E$-forms with values in $V$, $p, q \in \mathds{N}_0$}]{1ZOmegapq(NEV)}{\ensuremath{\Omega^{p, q}(N, E; V)}}
\glsxtrnewsymbol[description={Space of $p$-forms with values in a vector bundle $E$, $p \in \mathds{N}_0$}]{1ZOmegap(NV)}{\ensuremath{\Omega^{p}(N; V)}}
\glsxtrnewsymbol[description={Space of $E$-$q$-forms with values in a vector bundle $V$, $q \in \mathds{N}_0$}]{1ZOmegap(EV)}{\ensuremath{\Omega^{q}(E; V)}}
\glsxtrnewsymbol[description={Vector field describing the infinitesimal gauge transformation}]{1YPsiEpsilon}{\ensuremath{\Psi_{\varepsilon}}}
\glsxtrnewsymbol[description={Projection onto the field of gauge bosons as functional}]{1pivar}{\ensuremath{\varpi_2}}
\glsxtrnewsymbol[description={Field redefinition of $\varpi_2$}]{1pivarwidetildelambda}{\ensuremath{\widetilde{\varpi_2}^\lambda}}

\glsxtrnewsymbol[description={Lie bracket of a Lie algebroid $E$}]{0[]E}{\ensuremath{\mleft[\cdot, \cdot\mright]_E}}
\glsxtrnewsymbol[description={Lie bracket of the space of derivations $\mathcal{D}(V)$ on a vector bundle $V$}]{0[]D(V)}{\ensuremath{\mleft[\cdot, \cdot\mright]_{\mathcal{D}(V)}}}
\glsxtrnewsymbol[description={Lie bracket of a Lie algebra $\mathfrak{g}$}]{0[]g}{\ensuremath{[\cdot, \cdot]_{\mathfrak{g}}}}
\glsxtrnewsymbol[description={Lie bracket of vector fields $\mathfrak{X}(N)$} of a smooth manifold $N$]{0[]}{\ensuremath{[\cdot, \cdot]}}
\glsxtrnewsymbol[description={Coordinate vector fields $\frac{\partial}{\partial x^i}$}]{0partiali}{\ensuremath{\partial_i}}
\glsxtrnewsymbol[description={$E$-connection of a Lie algebroid $E$}]{0nablaE}{\ensuremath{{}^E\nabla}}
\glsxtrnewsymbol[description={$E$-connection defined by $(\mu, \nu) \mapsto \nabla_{\rho(\mu)} \nu$ w.r.t. a connection $\nabla$}]{0nablarho}{\ensuremath{\nabla_\rho}}
\glsxtrnewsymbol[description={Canonical basic connection of a Lie algebroid $E$}]{0nablabas}{\ensuremath{\nabla^{\text{bas}}}}
\glsxtrnewsymbol[description={Connection of a vector bundle $E$, the latter is often a Lie algebroid}]{0nabla}{\ensuremath{\nabla}}
\glsxtrnewsymbol[description={Field redefinition of a vector bundle connection}]{0nabla0widetildelambda}{\ensuremath{\widetilde{\nabla}^\lambda}}
\glsxtrnewsymbol[description={$E$-Bott connection of a Lie algebroid $E$}]{0nablaL}{\ensuremath{\nabla^L}}
\glsxtrnewsymbol[description={Exterior power of a vector bundle $V$}]{0bigwedgedotV}{\ensuremath{\bigwedge^\bullet V}}
\glsxtrnewsymbol[description={Pullback functional}]{0*omega}{\ensuremath{{}^*v}}
\glsxtrnewsymbol[description={Form-pullback functional}]{0!omega}{\ensuremath{{}^!\omega}}
\glsxtrnewsymbol[description={Pullback connection to functionals}]{0*nabla}{\ensuremath{{}^*\nabla}}

\begin{document}
%\delimitershortfall=-1pt

% Titelseite
\pagenumbering{Alph}
\begin{titlepage}

\begin{figure}
	\centering
	\includegraphics[width=1\textwidth]{Lyon.png}
	
	%\begin{flushleft}
	%\includegraphics[width=1\textwidth]{bilder/Logo-LABEX_MILYON.jpg}
	%\begin{flushright}
	%\includegraphics[width=0.4\textwidth]{bilder/SwissMAP-logo.png}
	%\end{flushright}
	%\end{flushleft}
%\end{center}
%\begin{figure}
	%\centering
		%\includegraphics[width=0.99\textwidth]{}
\end{figure}

\begin{minipage}[t]{0.5\textwidth}
%\hspace{-.6cm}
UNIVERSITY OF GENEVA\\
\vfill 
%\hspace{-.6cm}
Section of Mathematics
\end{minipage}
\begin{minipage}[t]{0.5\textwidth}
%\hspace{.8cm}
\hfill FACULTY OF SCIENCE\\
\vfill
\hfill Professor Anton Alekseev
\end{minipage}

\vspace{.3cm}
\hrule
\vspace{.3cm}

\begin{minipage}[t]{0.5\textwidth}
%\hspace{-.6cm}
CLAUDE BERNARD UNIVERSITY LYON 1\\
\vfill 
%\hspace{-.6cm}
Section of Mathematics
\end{minipage}
\begin{minipage}[t]{0.5\textwidth}
%\hspace{.8cm}
\hfill FACULTY OF SCIENCE\\
\vfill
\hfill Professor Thomas Strobl
\end{minipage}
\vspace{.3cm}
\hrule

\vfill\vfill
\begin{center}
{\Large \textbf{Geometry of curved Yang-Mills-Higgs gauge theories}}\\
\vspace{1.6cm}
{\large Ph.D.~Thesis}\\
\vspace{.3cm}
Presented at the Faculty of Science of the University of Geneva and Claude Bernard University Lyon 1\\
To obtain the dual Ph.D.~degree for mathematic\\
\vspace{1.4cm}
By\\
\vspace{.3cm}
{\large \textbf{Simon-Raphael Fischer}}\\
\vspace{.3cm}
from\\
\vspace{.3cm}
Straubing (Germany)
%\vfill\vfill
%{\large Th\`ese No.~5562}
\vfill\vfill
GENEVA and LYON\\
%\vspace{.3cm}
%Centre d'impression de l'UNIGE\\
%\vspace{.3cm}
\jahr
\end{center}
\vfill
\begin{figure}[H]
	\centering
	%\includegraphics[width=1\textwidth]{bilder/Lyon.png}
	
	%\begin{flushleft}
	\includegraphics[width=1\textwidth]{Logo-LABEX_MILYON.jpg}
	%\begin{flushright}
	%\includegraphics[width=0.4\textwidth]{bilder/SwissMAP-logo.png}
	%\end{flushright}
	%\end{flushleft}
%\end{center}
%\begin{figure}
	%\centering
		%\includegraphics[width=0.99\textwidth]{}
\end{figure}

\clearpage
\thispagestyle{empty}


N°d'ordre 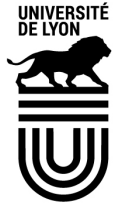 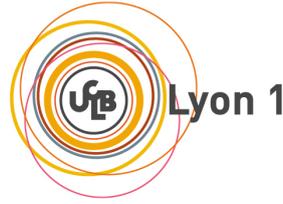 2021LYSE**1160**

# THESE de DOCTORAT DE L'UNIVERSITE DE LYON

opérée au sein de
**l'Université Claude Bernard Lyon 1**

**Ecole Doctorale** N° ED 512
**Informatique et Mathématiques de Lyon (InfoMaths)**

**Spécialité de doctorat**: Mathématiques

Soutenue publiquement le 31 mai 2021, par:
**Simon-Raphael Fischer**

# Geometry of curved Yang-Mills-Higgs gauge theories

Devant le jury composé de :

| | | |
|---|---|---|
| Kotov, Alexei | Chercheur, Université de Hradec Kralové | Rapporteur |
| Laurent-Gengoux, Camille | Professeur, Université de Lorraine | Rapporteur |
| Ilieva, Nevena | Professeure, Académie Bulgare des sciences | Examinatrice |
| Severa, Pavol | Chargé de cours, Université de Genève | Examinateur |
| Roger, Claude | Professeur Émerite, Université Lyon1 | Examinateur |
| Zhu, Chengchang | Professeure, Université de Göttingen | Examinatrice |
| Strobl, Thomas | Professeur, Université de Lyon | Directeur de thèse |
| Alekseev, Anton | Professeur, Université de Genève | Co-directeur de thèse |
| Frabetti, Alessandra | Maître de Conférences, Université Lyon 1 | Invitée |

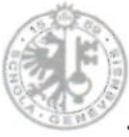

# UNIVERSITÉ DE GENÈVE
## FACULTÉ DES SCIENCES

DOCTORAT ÈS SCIENCES, MENTION MATHEMATIQUES

Thèse de Monsieur Simon-Raphaël FISCHER

intitulée :

«Geometry of Curved Yang-Mills-Higgs Gauge Theories»

La Faculté des sciences, sur le préavis de Monsieur A. ALEXEEV, professeur ordinaire et directeur de thèse (Section de mathématiques), Monsieur T. STROBL, professeur et codirecteur de thèse (Faculté des sciences, Université de Lyon 1, Lyon, France), Monsieur P. SEVERA, docteur (Section de mathématiques), Madame N. ILIEVA, professeure (Académie Bulgare des sciences, Bulgarie), Monsieur A. KOTOV, chercheur (Université de Hradec Kralové, Hradec Kralové, République Tchèque), Monsieur C. LAURENT-GENGOUX, professeur (Département UFR Mathématiques Informatique Mécanique, Université de Lorraine UMR 7502 - IECL, France), Monsieur C. ROGER, professeur émerite (Université de Lyon 1, Lyon, France), Madame C. ZHU, professeure (Universität Göttingen, Göttingen, Germany), autorise l'impression de la présente thèse, sans exprimer d'opinion sur les propositions qui y sont énoncées.

Genève, le 4 juin 2021

Thèse - 5562 -

Le Doyen

N.B. - La thèse doit porter la déclaration précédente et remplir les conditions énumérées dans les "Informations relatives aux thèses de doctorat à l'Université de Genève".



\end{titlepage}

%Doppelseite
%\newpage\thispagestyle{empty}\hspace{1em}
%\clearpage
%\ifodd\count0\else
%\thispagestyle{empty}
%\hbox{}\newpage
%\fi

%\vspace{0.2cm}
\newpage
\thispagestyle{empty}
\vspace*{\fill}
\begin{center}
\textit{Thanks to all my friends and family for their lasting support in the last years which were probably the most difficult of my life so far. Thanks to my mother, father, Dennis, Gregor, Marco, Nico, Jakob, Kathi, Konstantin, Lukas, Locki, Luciana, Gareth, Philipp, Dominik, Stefan, Ramona, Annerose, Michael, Maxim, and Anna. Also special thanks to their support also additionally in technical aspects of the thesis to Anton Alekseev, Mark Hamilton, and Alessandra Frabetti, and to Daniel for proof-reading my English, so, I finally have someone to blame for my bad English :) Without all your help this project would not have succeeded.}
\end{center}
\vspace*{\fill}
\normalsize

\newpage

\chapter*{Abstract}
%\begin{abstract}
This thesis is devoted to the study of the geometry of curved Yang-Mills-Higgs gauge theory (\textbf{CYMH GT}), a theory introduced by Alexei Kotov and Thomas Strobl. This theory reformulates classical gauge theory, in particular, the Lie algebra (and its action) is generalized to a Lie algebroid $E$, equipped with a connection $\nabla$, and the field strength has an extra term $\zeta$; there is a certain relationship between $\zeta$ and $\nabla$, for example, if $\zeta \equiv 0$, then $\nabla$ is flat. In the classical situation $E$ is an action Lie algebroid, a combination of a trivial Lie algebra bundle and a Lie algebra action, $\nabla$ is then the canonical flat connection with respect to such an $E$, and $\zeta\equiv 0$. The main results of this Ph.D.~thesis are the following:

\begin{itemize}
	\item Reformulating curved Yang-Mills-Higgs gauge theory, also including a thorough introduction and a coordinate-free formulation, while the original formulation was not completely coordinate-free. Especially the infinitesimal gauge transformation will be generalized to a derivation on vector bundle $V$-valued functionals. Those vector bundles $V$ will be the pullback of another bundle $W$, and the gauge transformation as derivation will be induced by a Lie algebroid connection on $W$, using a more general notion of pullbacks of connections. This also supports the usage of arbitrary types of connections on $W$ in the definition of the infinitesimal gauge transformation, not just canonical flat ones as in the classical formulation.
	%acting on the bundle in which a functional has values in. 
	\item Studying functionals as parameters of the infinitesimal gauge transformation, supporting a richer set of infinitesimal gauge transformations, especially the parameter itself can have a non-trivial gauge transformation. The discussion about the infinitesimal gauge transformation is also about what type of connection for the definition of the infinitesimal gauge transformation should be used, and this is argued by studying the commutator of two infinitesimal gauge transformations, viewed as derivations on $V$-valued functionals. We take the connection on $W$ then in such a way that the commutator is again an infinitesimal gauge transformation; for this flatness of the connection on $W$ is necessary and sufficient. For $W= E$ and $ W = \mathrm{T}N$ we use a Lie algebroid connection known as basic connection which is not the canonical flat connection in the classical non-abelian situation; this is not the connection normally used in the standard formulation, but it reflects the symmetries of gauge theory better than the usual connection, which is in general not even flat. For $W = \mathbb{R}$ the gauge transformation is uniquely given as the Lie derivative of a vector field on the space of fields given by the field of gauge bosons and the Higgs field, and the commutator is then just the Lie bracket of vector fields; in this case the bracket will also give again a vector field related to gauge transformations.
	%It will be the so-called basic connection, a generalization of Lie algebra representations.
	\item Defining an equivalence of CYMH GTs given by a field redefinition which is a transformation of structural data like the field of gauge bosons. In order to preserve the physics, this equivalence is constructed in such a way that the Lagrangian of the studied theory is invariant under this field redefinition. It is then natural to study whether there are equivalence classes admitting representatives with flat $\nabla$ and/or zero $\zeta$:
	\begin{enumerate}
	\item On the one hand, the equivalence class related to $E = \mathrm{T}\mathds{S}^7$, $\mathds{S}^7$ the seven-dimensional sphere, admits only representatives with non-flat $\nabla$, while locally the equivalence class of all tangent bundles admits a representative with flat $\nabla$.
	\item On the other hand, the equivalence class related to "$E=$ LAB" (Lie algebra bundle) has a relation with an obstruction class about extending Lie algebroids by LABs; this will imply that locally there is always a representative with flat $\nabla$ while globally this may not be the case, similar to the previous bullet point. Furthermore, a canonical construction for equivalence classes with no representative with zero $\zeta$ is given, which also works locally, and an interpretation of $\zeta$ as failure of the Bianchi identity of the field strength is provided. 
	\end{enumerate}
\end{itemize}
%\end{abstract}

\newpage

\chapter*{Résumé}
Cette thèse est consacrée à l'étude de la géométrie de la théorie de jauge Yang-Mills-Higgs courbe (\textbf{CYMH GT}), une théorie introduite par Alexei Kotov et Thomas Strobl. Cette théorie reformule la théorie de jauge classique, en particulier, l'algèbre de Lie (et son action) est généralisée à un algébroïde de Lie $E$, équipé d'une connexion $\nabla$, et l'intensité du champ a un terme supplémentaire $\zeta$; il existe une certaine relation entre $\zeta$ et $\nabla$, par exemple, si $\zeta \equiv 0$, alors $\nabla$ est plat. Dans la situation classique $E$ est un algébroïde de Lie d'action, une combinaison d'un fibré trivial d'algèbre de Lie et d'une action d'algèbre de Lie, $\nabla$ est alors la connexion plate canonique par rapport à un tel $E$, et $\zeta\equiv 0$. Les principaux résultats de cette thèse de doctorat sont les suivants:

\begin{itemize}
	\item Reformulation de la théorie de jauge courbée de Yang-Mills-Higgs, comprenant également une introduction approfondie et une formulation sans coordonnées, alors que la formulation originale n'était pas complètement sans coordonnées. En particulier, la transformation de jauge infinitésimale sera généralisée à une dérivation sur les fonctionnelle valuées des fibrés de vecteurs $V$. Ces fibrés de vecteurs $V$ seront le pullback d'un autre fibré $W$, et la transformation de jauge en tant que dérivation sera induite par une connexion algébroïde de Lie sur $W$, en utilisant une notion plus générale de pullbacks de connexions. Cela permet également d'utiliser des types arbitraires de connexions sur $W$ dans la définition de la transformation de jauge infinitésimale, et pas seulement des connexions plates canoniques comme dans la formulation classique.
	\item L'étude des fonctionnelles comme paramètres de la transformation de jauge infinitésimale permet d'obtenir un ensemble plus riche de transformations de jauge infinitésimales, en particulier le paramètre lui-même peut avoir une transformation de jauge non triviale. La discussion sur la transformation de jauge infinitésimale porte également sur le type de connexion à utiliser pour la définition de la transformation de jauge infinitésimale, ce que nous expliquons en étudiant le commutateur de deux transformations de jauge infinitésimales, considérées comme des dérivations sur des fonctionnelles valuées $V$. Nous prenons alors la connexion sur $W$ de telle sorte que le commutateur soit à nouveau une transformation de jauge infinitésimale; pour cela, la planéité de la connexion sur $W$ est nécessaire et suffisante. Pour $W= E$ et $W = \mathrm{T}N$, nous utilisons la connexion dite de base qui n'est pas la connexion plate canonique dans la situation non-abélienne classique; ce n'est pas la connexion normalement utilisée dans la formulation standard, mais elle reflète mieux les symétries de la théorie de jauge que la connexion habituelle, qui n'est en général même pas plate. Pour $W = \mathbb{R}$, la transformation de jauge est uniquement donnée comme la dérivée de Lie d'un champ vectoriel sur l'espace des champs donné par le champ des bosons de jauge et le champ de Higgs, et le commutateur est alors simplement le crochet de Lie des champs vectoriels; dans ce cas, le crochet donnera également à nouveau un champ vectoriel lié aux transformations de jauge.
	\item Définir une équivalence de GTs CYMH donnée par une redéfinition de champ qui est une transformation de données structurelles comme le champ des bosons de jauge. Afin de préserver la physique, cette équivalence est construite de telle manière que le Lagrangien de la théorie étudiée est invariant sous cette redéfinition de champ. Il est alors naturel d'étudier s'il existe des classes d'équivalence admettant des représentants avec des $\nabla$ plats et/ou des $\zeta$ nuls:
	\begin{enumerate}
	\item D'une part, la classe d'équivalence relative à $E = \mathrm{T}\mathds{S}^7$, $\mathds{S}^7$ la sphère à sept dimensions, n'admet que des représentants avec des $\nabla$ non plats, alors que localement la classe d'équivalence de tous les fibrés tangents admet un représentant avec des $\nabla$ plats.
	\item D'autre part, la classe d'équivalence liée à "$E=$ LAB" (Lie algebra bundle) a une relation avec une classe d'obstruction sur l'extension des algèbres de Lie par les LAB; cela impliquera que localement, il existe toujours un représentant avec $\nabla$ plat alors que globalement, cela peut ne pas être le cas, de manière similaire au point précédent. De plus, une construction canonique pour les classes d'équivalence sans représentant avec $\zeta$ nul est donnée, qui fonctionne également localement, et une interprétation de $\zeta$ comme échec de l'identité de Bianchi de l'intensité du champ est fournie. 
	\end{enumerate}
\end{itemize}

%\newpage
\setcounter{page}{10}
\thispagestyle{empty}\hspace{1em}\newpage
\pagenumbering{Roman}

%%%%%%%%%%%%%%%%%%%%%%%%%%%% Inhaltsverzeichnis %%%%%%%%%%%%%%%%%%%%%%%%%%%%

\tableofcontents

\setlength{\parindent}{12 pt}
%%%%%%%%%%%%%%%%%%%%%%%%%%%% Hier beginnt der Hauptteil %%%%%%%%%%%%%%%%%%%%%%%%%%%%

%\cleardoublepage
% für die leere(n) Seite(n)
\newpage
\thispagestyle{empty}\hspace{1em}
\newpage

\newpage
\thispagestyle{empty}\hspace{1em}
\newpage

\pagenumbering{arabic}
\chapter{Introduction}
%\hspace{12 pt}
This thesis concerns curved Yang-Mills-Higgs gauge theories (short: \textbf{CYMH GT}), introduced by Alexei Kotov and Thomas Strobl, a generalization of Yang-Mills-Higgs gauge theories, where we have essentially the following, as also summarized in \cite{CurvedYMH}:\footnote{Common conventions and notations are introduced at the end of the introduction; see Section \ref{StandardNotation}.}

\begin{itemize}
	\item $M$ a spacetime;
	\item $N$ a smooth manifold, serves as set for the values of the Higgs field $\Phi: M \to N$;
	\item $E \to N$ a Lie algebroid with anchor $\rho$, replacing the structural Lie algebra $\mathfrak{g}$ and its action $\gamma: \mathfrak{g} \to \mathfrak{X}(N)$ of the classical formulation;
	\item a vector bundle connection $\nabla$ on $E$;
	\item a fibre metric $\kappa$ on $E$, as a substitute of the ad-invariant scalar product on $\mathfrak{g}$;
	\item a Riemannian metric $g$ on $N$, replacing the scalar product on the vector space in which the Higgs field usually has values in and which is invariant under the action of $\gamma$, used for the kinetic term of $\Phi$ which is minimally coupled to the field of gauge bosons $A \in \Omega^1(M; \Phi^*E)$;
	\item a 2-form on $N$ with values in $E$, $\zeta \in \Omega^2(N;E)$, an additional contribution to the field strength of $A$.
\end{itemize}

A Lie algebroid is given by the following definition; especially, Lie algebroids can be thought as a generalization of both, tangent bundles and Lie algebras.

\begin{definitions*}{Lie algebroid, \cite[reduced definition of \S 16.1; page 113]{DaSilva}}
%\leavevmode\newline
Let $E \to N$ be a real vector bundle of finite rank. Then $E$ is a smooth Lie algebroid if there is a bundle map $\rho_E \coloneqq \rho: E \to \mathrm{T}N$, called the \textbf{anchor}, and a Lie algebra structure on $\Gamma(E)$ with Lie bracket $\mleft[ \cdot, \cdot \mright]_E$ satisfying
\bas
  \mleft[\mu, f \nu\mright]_E = f \mleft[\mu, \nu\mright]_E + \mathcal{L}_{\rho(\mu)}(f) ~ \nu
\eas
for all $f \in C^\infty(N)$ and $\mu, \nu \in \Gamma(E)$, where $\mathcal{L}_{\rho(\mu)}(f)$ is the action of the vector field $\rho(\mu)$ on the function $f$ by derivation. 
\end{definitions*}

Gauge invariance of the Yang-Mills-Higgs type functional leads to several \textbf{compatibility conditions} to be satisfied between those structures. If the connection $\nabla$ on $E$ is flat, the compatibilities imply that the Lie algebroid is locally what we call an action Lie algebroid. 

\begin{definitions*}{Action Lie algebroids, \cite[\S 16.2, Example 5; page 114]{DaSilva}}
Let $\mleft(\mathfrak{g}, \mleft[\cdot, \cdot \mright]_{\mathfrak{g}}\mright)$ be a Lie algebra equipped with a Lie algebra action $\gamma: \mathfrak{g} \to \mathfrak{X}(N)$ on a smooth manifold $N$. A \textbf{transformation Lie algebroid} or \textbf{action Lie algebroid} is defined as the bundle $E \coloneqq N \times \mathfrak{g}$ over $N$ with anchor
\bas
\rho(p, v) &\coloneqq \gamma(v)|_p
\eas
for $(p, v) \in E$, and Lie bracket
\bas
	\mleft.\mleft[\mu, \nu\mright]_E\mright|_p
	&\coloneqq 
	\mleft[\mu_p, \nu_p\mright]_{\mathfrak{g}}
		+ \mleft.\mleft(\mathcal{L}_{\gamma(\mu(p))}(\nu^a) - \mathcal{L}_{\gamma(\nu(p))}(\mu^a) \mright)\mright|_p ~ e_a
\eas
	for all $p \in N$ and $\mu, \nu \in \Gamma(E)$, where one views a section $\mu \in \Gamma(E)$ as a map $\mu: N \to \mathfrak{g}$ and $\mleft( e_a \mright)_a$ is some arbitrary frame of constant sections.
\end{definitions*}

Furthermore, $\nabla$ is then a canonical flat connection of the action Lie algebroid, and one arrives at the standard Yang-Mills-Higgs gauge theory if additionally $\zeta \equiv 0$. Thus, the theory represents a curved (with respect to $\nabla$) version of gauge theory equipped with an additional 2-form $\zeta$. If $\nabla$ is flat we say in general that we have a \textbf{pre-classical} gauge theory, and if additionally $\zeta \equiv 0$ we have a \textbf{classical} gauge theory. Every classical theory is also pre-classical, this is another implication of the compatibility conditions.

For a given $M, N$ and $E$ there is an equivalence of CYMH GTs given by a field redefinition, a transformation of the field of gauge bosons, but also of $\nabla$, $\kappa$, $g$ and $\zeta$. The Lagrangian is invariant under this transformation, hence, the physics is invariant. It is then natural to study whether it is possible that the equivalence class of a given CYMH GT has a (pre-)classical representative, and this is precisely the main motivation of this thesis. 
Along this study, CYMH GT is reintroduced in a coordinate-free way, especially providing a new coordinate-free formulation of the infinitesimal gauge transformations themselves. We proceed as follows:

In Chapter \ref{ClassicGaugeTheory} we recall the fundamental basics of classical gauge theory, mostly their infinitesimal information; that means that we always assume trivial principal bundles, thus, we do not need principal bundles altogether. In Section \ref{LieAlgebraActions} we introduce Lie algebras and their actions, comparing Lie algebra actions and representations; in Section \ref{IsotropyClassical} we discuss isotropies and their relation along orbits of a Lie group action. The classical Yang-Mills-Higgs gauge theory, especially the Yang-Mills-Higgs Lagrangian, is introduced in Section \ref{YMHGT}, and in Section \ref{InfGaugeTrafoClassical} we prove the infinitesimal gauge invariance of the Lagrangian. 
However, in Section \ref{NewInfGaugeTrafoTrafos} we are already reformulating infinitesimal gauge transformations, making the first step towards the generalized formulation of (infinitesimal) gauge theory. Even if the reader has a good knowledge about gauge theory, it is highly recommended to read Section \ref{NewInfGaugeTrafoTrafos} in order to understand later why CYMH GT is formulated as it is. The main result of this section is the reformulation of the infinitesimal gauge transformation as a derivation induced by what we call a Lie algebra connection; the key ingredients are the following, where the manifold $N$ is for simplicity a vector space, and $\mathfrak{g}$ is the structural Lie algebra with action $\gamma$:

\begin{itemize}
	\item The pair of infinitesimal gauge transformations, $\Psi_\varepsilon \coloneqq (\delta_\varepsilon \Phi, \delta_\varepsilon A)$, viewed as a \textbf{vector field on the space of fields $\mathfrak{M}_{\mathfrak{g}}$} whose elements are given as pairs $(\Phi, A)$, where $\Phi \in C^\infty(M;N)$ (Higgs field) and $A \in \Omega^1(M; \mathfrak{g})$ (field of gauge bosons); $\varepsilon$ is a functional with $(\Phi, A) \mapsto \varepsilon(\Phi, A) \in C^\infty(M; \mathfrak{g})$.
	\item The \textbf{evaluation map $\mathrm{ev}: M \times \mathfrak{M}_{\mathfrak{g}} \to N$} defined by
	\bas
	\mathrm{ev}(p, \Phi, A)
	&\coloneqq
	\Phi(p)
	\eas
	for all $(p, \Phi, A) \in M \times \mathfrak{M}_{\mathfrak{g}}$.
	\item The \textbf{"bookkeeping trick"} for functionals $L$, $(\Phi, A) \mapsto L(\Phi, A) \in \Omega^k(M; K)$ ($k \in \mathbb{N}_0$), where $K$ is a vector space. Let $\mleft( e_a \mright)_a$ be a basis of $K$, then locally $L = L^a \otimes e_a$, where $L^a \in \Omega^k(M)$. If viewing $\mleft( e_a \mright)_a$ as a constant frame of the trivial vector bundle $N \times K$ over $N$, then we can also write
	\bas
	L
	&=
	L^a \otimes \mathrm{ev}^*e_a
	\eas
	due to constancy of the frame. For bookkeeping reasons we formally denote this expression by $\iota(L)$; especially 
	\bas
	\iota(L)(Y_1, \dotsc, Y_k)
	&\in
	\Gamma\bigl( \mathrm{ev}^*(N \times K) \bigr)
	\eas
	for all $Y_1, \dotsc, Y_k \in  \mathfrak{X}(M)$, and
	\bas
	\iota(L)(\Phi, A)
	&\in
	\Omega^k(M; \Phi^*(N \times K))
	\eas
	for all $(\Phi, A) \in \mathfrak{M}_{\mathfrak{g}}$.
	\item A \textbf{$\mathfrak{g}$-connection ${}^{\mathfrak{g}}\nabla$} on $V \coloneqq N \times K \to N$, defined as an $\mathbb{R}$-bilinear map
	\bas
\mathfrak{g} \times \Gamma(V) &\to \Gamma(V), 
\\
(X, \nu) &\mapsto {}^\mathfrak{g}\nabla_X \nu,
\eas
satisfying
\bas
{}^\mathfrak{g}\nabla_X (f \nu)
&=
f ~ {}^\mathfrak{g}\nabla_X \nu
	+ \mathcal{L}_{\gamma(X)}(f) ~ \nu
\eas
for all $X \in \mathfrak{g}$, $\nu \in \Gamma(V)$ and $f \in C^\infty(N)$, where $\mathcal{L}_{\gamma(X)}(f)$ is the action of the vector field $\gamma(X)$ on the function $f$ by derivation.
\end{itemize}

The derived key statement is then the following theorem and definition, where we are going to use a generalized notion of pullbacks of connections.

\begin{theorems*}{}
There is a unique $\mathbb{R}$-linear operator $\delta_{\Psi_\varepsilon}: \Gamma\mleft(\mathrm{ev}^*(V)\mright) \to \Gamma\mleft(\mathrm{ev}^*(V)\mright)$ with
\bas
\delta_{\Psi_\varepsilon} (f s)
&=
\mathcal{L}_{\Psi_\varepsilon}(f) ~ s
	+ f ~ \delta_{\Psi_\varepsilon} s,
\\
\delta_{\Psi_\varepsilon} \mleft( \mathrm{ev}^*\vartheta \mright)
&=
- \mathrm{ev}^*\mleft( {}^{\mathfrak{g}}\nabla_\varepsilon \vartheta \mright)
\eas
for all $f \in C^\infty(M \times \mathfrak{M}_{\mathfrak{g}})$, $s \in \Gamma\mleft(\mathrm{ev}^*(V)\mright)$ and $\vartheta \in \Gamma(V)$, where we denote
\bas
\mleft.\mathrm{ev}^*\mleft( {}^{\mathfrak{g}}\nabla_\varepsilon \vartheta \mright)\mright|_{(p, \Phi_0, A_0)}
&=
\mleft.\mleft({}^{\mathfrak{g}}\nabla_{\varepsilon(\Phi, A)|_p} \vartheta\mright)\mright|_{\Phi(p)}
\eas
for all $(p, \Phi, A) \in M \times \mathfrak{M}_{\mathfrak{g}}$.
\end{theorems*}

\begin{definitions*}{Infinitesimal gauge transformation as derivation}
The \textbf{infinitesimal gauge transformation $\delta_\varepsilon L$ of a functional $L$}, $(\Phi, A) \mapsto L(\Phi, A) \in \Omega^k(M; K)$ ($k \in \mathbb{N}_0$), is then defined by
\bas
(\delta_\varepsilon L)(Y_1, \dotsc, Y_k)
&\coloneqq
\delta_{\Psi_\varepsilon}\bigl(
	\iota(L)(Y_1, \dotsc, Y_k)
\bigr)
\eas
for all $Y_1, \dotsc, Y_k$.
\end{definitions*}

Section \ref{NewInfGaugeTrafoTrafos} will then conclude that this definition of the infinitesimal gauge transformation recovers the typical definition by taking the canonical flat connection $\nabla$ of $V = N \times K$, \textit{i.e.}~given by $\nabla x = 0$ for all constant $x \in \Gamma(V)$, and then defining ${}^{\mathfrak{g}}\nabla \coloneqq \nabla_\gamma$, $(X, v) \mapsto \nabla_{\gamma(X)} v$ for all $X \in \mathfrak{g}$ and $v \in \Gamma(V)$.

Chapter \ref{MathematicalBasics} is mainly about introducing all the needed mathematical basics. Section \ref{LieAoids} starts with introducing Lie algebroids and related notions, especially introducing action Lie algebroids and Lie algebra bundles as a special example. Furthermore, small physical examples are provided, and isotropies are revisited to support a better understanding of the relationship to gauge theory. Section \ref{MorphsOfLieOids} discusses morphisms of Lie algebroids, but since we are mainly interested into base-preserving ones, this section is very short. An important basic notion are Lie algebroid connections, and we want to introduce them as certain morphisms of anchored vector bundles, similar to the introduction of Lie algebroid connections in \cite{mackenzieGeneralTheory}. In order to do so we first introduce the Lie algebroid of derivations of vector bundles in Section \ref{DerivationsOnvector}, and in Section \ref{SubsectionEDiffstuff} we finally introduce Lie algebroid connections as base- and anchor-preserving vector bundle morphisms; Lie algebroid connections on a vector bundle are essentially the same as typical vector bundle connections but the direction of differentiation is along sections of the Lie algebroid and the Leibniz rule is along the foliation of the anchor, similar to the Leibniz rule of the Lie bracket of a Lie algebroid. Section \ref{PullbacksAlsoGeneral} discusses pullbacks of Lie algebroid connections; first we follow a typical introduction using Lie algebroid paths, but concluding with a more general statement about pullbacks when one just differentiates along one direction:

\begin{corollaries*}{Pullbacks of connections just differentiating along one vector field}
Let $E_i \to N_i$ ($i \in\{1,2\}$) be two Lie algebroids over smooth manifolds $N_i$, $V \to N_2$ a vector bundle, and ${}^{E_2}\nabla$ an $E_2$-connection on $V$. Moreover, let $f \in C^\infty(N_1;N_2)$, $\nu_1 \in \Gamma(E_1)$ and $\nu_2 \in \Gamma(f^*E_2)$ such that
\bas
\mathrm{D}f\bigl(\rho_{E_1}(\nu_1)\bigr)
&=
\mleft(f^*\rho_{E_2}\mright)(\nu_2).
\eas

Then there is a unique $\mathbb{R}$-linear operator $\delta_{\nu_1}: \Gamma(f^*V) \to \Gamma(f^*V)$ with
\bas
\delta_{\nu_1}(h s)
&=
\mathcal{L}_{\rho(\nu_1)}(h) ~ s
	+ h ~ \delta_{\nu_1} s,
\\
\delta_{\nu_1} (f^*v)
&=
f^*\mleft(
	{}^{E_2}\nabla_{\nu_2} v
\mright)
\eas
for all $s \in \Gamma(f^*V)$, $v \in \Gamma(V)$ and $h \in C^\infty(N_1)$.
\end{corollaries*}

A major example of a Lie algebroid connection is the basic connection, induced by a vector bundle connection $\nabla$ on a Lie algebroid. The basic connection can be thought as a Lie algebra representation formulated as connection. Since the basic connection is related to conjugated connections, Section \ref{ConjugateConnections} introduces the notion of connections conjugate to each other, and Section \ref{SectionOfBasicConnStuff} then introduces the basic connection. Since Lie algebra representations are homomorphisms, one may want that the basic connection is flat. Hence, a tensor known as the basic curvature is also introduced and discussed; this tensor is in general not equivalent to the curvature of the basic connection, it encodes the curvature of the basic connection, but it also contains information about how $\nabla$ acts on the bracket of the Lie algebroid. We will see that the vanishing of the basic curvature is needed for the gauge invariance of the Yang-Mills-Higgs Lagrangian.

The remaining part of Chapter \ref{MathematicalBasics} is then again about very basic notions related to Lie algebroids. Section \ref{ExteriorCovariantDerivativesAoids} is about exterior covariant derivatives but generalized to Lie algebroid connections, and Section \ref{DirectProdsOfLieAlgoids} is about the natural Lie algebroid structure of the direct product of Lie algebroids. There is also the Splitting Theorem for Lie algebroids: The anchor of a Lie algebroid is a homomorphism of Lie brackets, thus, its image gives rise to a foliation on the base manifold by the Frobenius Theorem; the foliation is singular due to the fact that the anchor has not a constant rank in general. The Splitting Theorem is then about that Lie algebroids are locally a direct product of a Lie algebroid along a leaf of the foliation and along a submanifold transversal to the foliation. This is discussed in Section \ref{SectionAboutSplitting}, mostly in a simplified setting; however, references for more general statements will be provided. The last section, Section \ref{SectionOfLABStuff}, focuses on Lie algebra bundles, a trivial example of Lie algebroids with zero anchor. It starts with extending notions of Lie algebras like their centre to Lie algebra bundles and finishes with a discussion about Lie algebroids with surjective anchor and their quotients over ideals.

We then discuss the formulation of CYMH GT in Chapter \ref{GeneralizedGTfas}. This chapter reintroduces CYMH GT, using my own approach in many parts while the overall theory does not differ to the original one as \textit{e.g.}~presented in \cite{CurvedYMH}. It starts with the study of the \textbf{space of fields} in Section \ref{SpaceOfFieldsSection}, the infinite-dimensional manifolds of pairs of the Higgs field and the field of gauge bosons, similar to previously-mentioned $\mathfrak{M}_{\mathfrak{g}}$.

\begin{definitions*}{Space of fields}
Let $M, N$ be two smooth manifolds and $E\to N$ a Lie algebroid. Then we denote the \textbf{space of fields} by
\bas
\mathfrak{M}_E
&\coloneqq
\mathfrak{M}_E(M; N)
\coloneqq
\left\{ (\Phi, A)
~\middle|~
\Phi \in C^\infty(M;N) \text{ and } A \in \Omega^1(M; \Phi^*E)
\right\}.
\eas
%Thus for $\mathfrak{M}_E(M; N)$ we sometimes write
%\begin{center}
	%\begin{tikzcd}
		%\Omega^1(M;{}^*E) \arrow{d} \\
		%C^\infty(M;N)
	%\end{tikzcd}
%\end{center}

We will refer to $A \in \Omega^1(M; \Phi^*E)$ as the \textbf{field of gauge bosons} and $\Phi$ just as a \textbf{physical field} of this theory.
\end{definitions*}

The main idea is to define the infinitesimal gauge transformation as we did before in Section \ref{NewInfGaugeTrafoTrafos}, but especially generalized to Lie algebroids, their connections and to the setting described at the very beginning of this introduction; the Lie algebroid plays the role of the Lie algebra, and Lie algebroid connections will replace the Lie algebra connections, which we have suggested previously. One ingredient was to view the infinitesimal gauge transformation as a vector field $\Psi$ on $\mathfrak{M}_{\mathfrak{g}}$ which is now replaced by $\mathfrak{M}_E$. Thence, we will discuss the tangent space of the space of fields. Afterwards we discuss the definition and algebra of the functionals we are going to look at. Recall the "bookkeeping trick", the essential idea was that functionals have values in the $\mathrm{ev}$-pullback of a vector bundle over $N$, where the evaluation map is defined as before. Thus, we define functionals as certain forms on $M \times \mathfrak{M}_E$ with values in $\mathrm{ev}^*V$, where $V$ is a vector bundle over $N$; a similar argument will be applied to $A$ which explains why it has values in $\Phi^*E$ in the general setting. To avoid bloating formulas and definitions we will also introduce shortened notations which is why it is highly suggested to read Section \ref{SpaceOfFieldsSection}.

In Section \ref{NewPhysicQuants} we define physical quantities arising in gauge theory to the new generalised setting as in the beginning of this introduction but without $\zeta$, hence, without the extra term in the field strength. As a major example serves the following definition, where $t_{\nabla_\rho}$ is the torsion  of the $E$-connection $\nabla_\rho$ given by $\mleft( \nabla_\rho \mright)_\mu \nu = \nabla_{\rho(\mu)} \nu$.

\begin{definitions*}{Field of gauge bosons and their field strength, \newline \cite[especially Eq.~(11); $\Phi$ is denoted as $X$ there]{CurvedYMH}}
Let $M, N$ be smooth manifolds, and $E \to N$ a Lie algebroid equipped with a connection $\nabla$ on $E$. We define the \textbf{field strength $F$} by
\bas
F(\Phi, A)
\coloneqq
\mathrm{d}^{\Phi^*\nabla} A
	- \frac{1}{2} \mleft( \Phi^* t_{\nabla_\rho} \mright)\mleft( A \stackrel{\wedge}{,} A \mright)
\eas
for all $\Phi \in C^\infty(M;N)$ and $A \in \Omega^1(M; \Phi^*E)$.
\end{definitions*}

$\frac{1}{2} \mleft( \Phi^* t_{\nabla_\rho} \mright)\mleft( A \stackrel{\wedge}{,} A \mright)$ is an element of $\Omega^2(M; \Phi^*E)$ given by 
\bas
\mleft(\frac{1}{2} \mleft( \Phi^* t_{\nabla_\rho} \mright)\mleft( A \stackrel{\wedge}{,} A \mright)\mright)(X,Y)
&\coloneqq
\frac{1}{2} \mleft(
	\mleft( \Phi^* t_{\nabla_\rho} \mright)\mleft( A(X), A(Y) \mright)
	- \mleft( \Phi^* t_{\nabla_\rho} \mright)\mleft( A(Y), A(X) \mright)
\mright)
\\
&=
\mleft( \Phi^* t_{\nabla_\rho} \mright)\mleft( A(X), A(Y) \mright)
\eas
for all $X, Y \in \mathfrak{X}(M)$.

This section concludes that one has the classical definitions if $E$ is an action Lie algebroid and $\nabla$ its canonical flat connection. We then finally discuss infinitesimal gauge transformations in Section \ref{InfinitesimalGaugeTransformation}, defining them as in Section \ref{InfGaugeTrafoClassical} but extended to the generalized notions, and first omitting a definition of the infinitesimal gauge transformation of the field of gauge bosons; for this we also make use of the previously introduced corollary about pullbacks of connections if just differentiating along one direction. We will argue that the vector fields allowing such a pullback are precisely those vector fields $\Psi$ on the space of fields whose component along the "$\Phi$-direction" is given by the infinitesimal gauge transformation of the Higgs field.

That is, one milestone of this thesis is the formulation of infinitesimal gauge transformations of functionals as derivations induced by a generalized $\mathrm{ev}$-pullback of a Lie algebroid connection, while the infinitesimal gauge transformations of the fields are given by vector fields $\Psi$ on the space of fields; the classical formulation is recovered by using a canonical flat connection since functionals have values in a trivial vector bundle in the classical situation, such that a canonical flat connection is given. The parameters of the infinitesimal gauge transformations are functionals $\varepsilon$ such that $\varepsilon(\Phi, A) \in \Gamma(\Phi^*E)$; due to the fact that their values depends on $\Phi$ these parameters have in general also a non-trivial infinitesimal gauge transformation.

Afterwards the infinitesimal gauge transformation of the field of gauge bosons $A$ is formulated. We will see that its transformation $\delta_\varepsilon A$ does in general not live in the same space as $A$ itself due to horizontal components in the tangent space of the space of fields. Therefore we will apply a horizontal projection, however, to avoid loosing information about the "full" formula of $\delta_\varepsilon A$, this is done in such a way that the vector field $\Psi$ related to the given infinitesimal gauge transformation can uniquely be reconstructed. Essentially, the horizontal projection will only lead to a loss of information which is given by the infinitesimal gauge transformation of the Higgs field, and that information is already given, hence, one does not loose any real information. Technically, $\delta_\varepsilon A$ is given as the infinitesimal gauge transformation of the functional $\varpi_2$ given as the projection onto $A$, $\varpi_2(\Phi,A) \coloneqq A$. The vector field $\Psi = \Psi_\varepsilon$, parametrized by $\varepsilon$, is then uniquely encoded in the definition of the infinitesimal gauge transformation of $\Phi$ and in the condition
\bas
\mleft(\delta_\varepsilon \varpi_2\mright)(\Phi,A)
=
- (\Phi^*\nabla)\varepsilon,
\eas
where the Lie algebroid connection in the definition of $\delta_\varepsilon$ will be usually the basic connection in this thesis; this is also why there is not the typical Lie bracket term as usual in the definition of the infinitesimal gauge transformation of $A$, this information is saved in the basic connection itself. We will motivate that condition on $\varpi_2$ by how the minimal coupling between $\Phi$ and $A$ shall transform, similar to the typical motivation provided by physicists.

About the choice of using the basic connection: We will discuss what type of Lie algebroid connection should be used for the infinitesimal gauge transformation if the functional is not scalar-valued; the infinitesimal gauge transformation of scalar-valued functionals will uniquely be given as Lie derivative of the vector field behind the transformation. We do so by looking at the commutator of two infinitesimal gauge transformations; we expect that the commutator should be again an infinitesimal gauge transformation. This is the case for the vector fields behind the infinitesimal gauge transformations (the scalar-valued situation basically), denoted abstractly as $\Psi$ above, but now denoted as $\Psi_\varepsilon$ to account the parameter $\varepsilon$. We show that the relation is
\bas
[\Psi_\vartheta, \Psi_\varepsilon]
&=
- \Psi_{\Delta(\vartheta, \varepsilon)},
\eas
where $\vartheta$ is a second parameter and $\Delta$ is a Lie bracket for those parameters defined by
\bas
\Delta(\vartheta, \varepsilon)|_{(\Phi, A)}
&\coloneqq
\mleft( \delta_\varepsilon \vartheta - \delta_\vartheta \varepsilon \mright)|_{(\Phi, A)}
	+ (\Phi^*t_{\nabla_\rho})\bigl(\vartheta(\Phi,A), \varepsilon(\Phi, A)\bigr)
\eas
for all $(\Phi, A) \in \mathfrak{M}_E$; recall that the parameters themselves are functionals and have in general a non-trivial gauge transformation now. However, for vector-bundle functionals we use Lie algebroid connections as we motivated previously, the commutator of transformations is then essentially a lift of the bracket of the vector fields like $\Psi_\varepsilon$; we will see that then the relation of the commutator has essentially an extra term given by the $\mathrm{ev}$-pullback of the curvature of the used connection. Hence, if we want a similar behaviour as for the vector fields $\Psi_\varepsilon$, then we need to use a flat connection. We will see that the basic connection will be flat in the new formulation of gauge theory, hence, our choice, although we will argue that the gauge invariance of the Lagrangian is not affected by that choice since it is scalar-valued. 

Another canonical choice as connection would be $\nabla_\rho$. While the basic connection will not be the \emph{canonical} flat connection in the classical situation, $\nabla_\rho$ will be; thus, the condition for $\varpi_2$ would strongly resemble the typical formula of $\delta_\varepsilon A$ if using $\nabla_\rho$ instead. Therefore choosing the basic connection may be mainly an aesthetic choice, but we are going to see that the basic connection, as a generalization of Lie algebra representations, reflects the symmetries of gauge theory in a better way, simplifying calculations, while $\nabla_\rho$, among certain other difficulties, will not be flat in general such that its commutator of infinitesimal gauge transformations on vector bundle valued functionals would have an extra term.

In the discussion about the infinitesimal gauge invariance of the generalized gauge theory, starting in Section \ref{InfInvariance}, we will prove the gauge invariance of the Lagrangian in the more general setting (still without $\zeta$). However, after long calculations we will see that locally the new setting is the same as the classical setting, so, one may only have achieved a global formulation of gauge theory also allowing non-trivial bundles as values of functionals like the field strength; all of this is due to the fact that $\nabla$ has to be flat in order to have gauge invariance. Now $\zeta$ becomes important; in works like \cite{CurvedYMH} it is introduced as ansatz. However, we will introduce it by defining and studying a field redefinition in Section \ref{FieldRedefSection}. One can think of it as a coordinate-change as in classical mechanics, leaving an inertial frame, leading to extra terms in several physical relationships. As a next step one then reformulates classical mechanics such that it becomes coordinate-free and -independent; this is also denoted as \emph{covariantization} by physicists. Further steps are then generalizations of structures like assuming whether it is possible that those arising extra terms can always be mapped to zero by a coordinate change; if not, one may for example have a non-flat connection.

In our case the "coordinates" are structural data like the field of gauge bosons and $\nabla$. The study about the reformulation of the existing gauge theory in Section \ref{NastyCalculationsForTheseFieldRedefsBaeaeaeae}, such that it is "coordinate"-independent with respect to the field redefinition, will lead to a generalized gauge theory where the field strength has an extra term essentially given by the $\mathrm{ev}$-pullback of the previously-mentioned $\zeta \in \Omega^2(N;E)$. This will be then finalized in Section \ref{SectionAboutCYMHGTs}, and the field redefinition is then nothing else than an equivalence of such more general gauge theories, officially called curved Yang-Mills-Higgs gauge theories, abbreviated as CYMH GT. Finally, $\nabla$ is in general not required to be flat anymore in order to achieve gauge invariance, especially we have the relationship
\bas
R_\nabla
&=
- \mathrm{d}^{\nabla^{\mathrm{bas}}} \zeta
\eas
where $R_\nabla$ is the curvature of $\nabla$ and $\mathrm{d}^{\nabla^{\mathrm{bas}}}$ the exterior covariant derivative of the basic connection $\nabla^{\mathrm{bas}}$. This is also why $\zeta$ will be called \textbf{primitive of $\nabla$}. At this point we have finally recreated CYMH GTs, but in a coordinate-free way, while the original formulation is not completely coordinate-free, especially the infinitesimal gauge transformation was originally only formulated in a coordinate-dependent way, without using Lie algebroid connections as in this thesis. Chapter \ref{GeneralizedGTfas} will conclude with Section \ref{PropertiesOFNewTOlleGTs} which is about certain general properties of CYMH GTs needed for the following chapter.

Chapter \ref{ObstructionStuff} is then about whether or not there are CYMH GTs which are (pre-)classical, also after any field redefinition. It could be that a given $\zeta$ vanishes after the field redefinition; similar for $\nabla$ with respect to flatness. We first study Lie algebra bundles $E = K \to N$ (LABs) in Section \ref{ObstrLAB}: Subsection \ref{SumamryForLABSituation} shortly summarizes how a CYMH GT for LABs looks like, while in Subsection \ref{ConnectionIsALieDerivation} 
%we will see that the compatibility conditions imply that the connection will be equivalent to a so-called \textbf{Lie derivation law covering a pairing $\Xi$ of $\mathrm{T}N$ with the Lie algebra bundle $K$} as introduced in \cite[\S 7.2, Mackenzie writes coupling instead of pairing; page 271ff.]{mackenzieGeneralTheory}. $\Xi$ is essentially a Lie algebroid morphism of $\mathrm{T}N$ to the outer Lie bracket derivations of $K$, and Mackenzie has shown that $\Xi$ induces a differential with which we can define a cohomology class of $\mathrm{d}^\nabla \zeta$, denoted as $\mleft[ \mathrm{d}^\nabla \zeta \mright]_\Xi$, which is an invariant of the field redefinition. Using that terminology, we will see in 
and Subsection \ref{MackenzieZeugsUndExistenzvonPreclassical} we will see that the question, about whether we have a field redefinition transforming the gauge theory into a pre-classical one, has a strong relation to Mackenzie's study about extending Lie algebroids with Lie algebra bundles: $\nabla$ is by the compatibility conditions of a CYMH GT equivalent to a Lie derivation law covering what is called a pairing $\Xi$ which is a Lie algebroid morphism  $\mathrm{T}N \to \mathrm{Out}(\mathcal{D}_{\mathrm{Der}}(K)$, where $\mathrm{Out}(\mathcal{D}_{\mathrm{Der}}(K)$ is the Lie algebroid of outer bracket derivations of $K$, outer in the sense of the quotient of bracket derivations over inner bracket derivations. That is, $\nabla$ is also a bracket derivation and its equivalence class in the quotient space of the outer bracket derivations is equivalent to the pairing $\Xi$. We will see that the field redefinition is then just a transformation to any other Lie derivation law covering the same pairing. Furthermore, $\mathrm{d}^\nabla \zeta$ will be an invariant of the field redefinition, and the second Bianchi identity of $\nabla$ will imply that $\mathrm{d}^\nabla \zeta$ is a centre-valued form. By the compatibility conditions one can argue that $\nabla$ induces a differential $\mathrm{d}^\Xi$ on centre-valued forms, independent of the choice of $\nabla$. We will see that $\mathrm{d}^\nabla \zeta$ is closed with respect to $\mathrm{d}^\Xi$, such that it is natural to study the cohomology class of $\mathrm{d}^\nabla \zeta$ with respect to $\mathrm{d}^\Xi$; the invariance under the field redefinition will imply that this class only depends on $\Xi$. This class is precisely the obstruction class $\mathrm{Obs}(\Xi)$ developed by Mackenzie.

Therefore we will introduce and discuss Mackenzie's theory about extending Lie algebroids by LABs in Subsection \ref{MackenzieStuff}. On one hand, Mackenzie shows that the obstruction class is zero if and only if one can extend $\mathrm{T}N$ by $K$ in such a way that there is a transitive Lie algebroid for which the kernel of the anchor is given by $K$.\footnote{Actually, Mackenzie shows a general statement; in this thesis Mackenzie's statement is simplified to our setting.} On the other hand, Mackenzie also shows that, if $N$ is contractible, then there is always a \emph{flat} Lie derivation law $\nabla$ covering $\Xi$; for contractible $N$ the obstruction class is trivially zero. Due to these results of Mackenzie we derive in Subsection \ref{LABResultsWooooo} that a non-zero obstruction class implies that there is no field redefinition such that $\nabla$ becomes flat, and that for contractible $N$ there is always a field redefinition such that a given CYMH GT is pre-classical. 

\begin{theorems*}{Local existence of pre-classical gauge theory (simplified formulation)}
Let $(K, \Xi)$ be a pairing of $\mathrm{T}N$ over a contractible manifold $N$, and let $\nabla$ be a fixed Lie derivation law covering $\Xi$.

Then we have a field redefinition such that the redefinition of $\nabla$ is flat.
\end{theorems*}

\begin{theorems*}{Possible new and curved gauge theories on LABs}
Let $(K, \Xi)$ be a pairing of $\mathrm{T}N$ with $\mathrm{Obs}(\Xi) \neq 0$ and such that the fibre Lie algebra $\mathfrak{g}$ admits an $\mathrm{ad}$-invariant scalar product.

Then we can construct a CYMH GT for which there is no field redefinition with what it would become pre-classical.
\end{theorems*}

However, a zero obstruction class does not necessarily imply that a CYMH GT can be transformed to a pre-classical one, following an example of Mackenzie: The Hopf fibration $\mathds{S}^7 \to \mathds{S}^4$ has a zero obstruction class but no flat Lie derivation law covering its canonical pairing as an Atiyah sequence. 

Up to this point it was just about $\nabla$ and its field redefinition. In Subsection \ref{NonclassicalStuff} we quickly derive that for $\zeta$ it is easier to find an answer. If $\mathrm{d}^\nabla \zeta \neq 0$, then there is never a field redefinition making $\zeta$ vanish. We also provide a canonical construction of such $\zeta$ if starting with a certain classical gauge theory:

\begin{corollaries*}{Canonical construction of non-classical gauge theories (simplified formulation)}
Let $\mathfrak{g}$ be a Lie algebra with non-zero centre and admitting an $\mathrm{ad}$-invariant scalar product. Also let $(N, g)$ be any Riemannian manifold with at least three dimensions, and $K = N \times \mathfrak{g}$ be a trivial LAB over $N$, equipped with the canonical flat connection $\nabla$ and a metric $\kappa$ which restricts to an $\mathrm{ad}$-invariant scalar product on each fibre.

Then there is a $\zeta \in \Omega^2(N; Z(K))$, with $\mathrm{d}^\nabla \zeta \neq 0$, such that this set-up describes a non-classical CYMH GT with respect to an arbitrary spacetime $M$. Additionally, there is no field redefinition making $\zeta$ zero.
\end{corollaries*}

In Subsection \ref{BianchiStuff}, we turn shortly to the discussion about a possible physical meaning of $\mathrm{d}^\nabla \zeta \neq 0$ due to its influence to the obstruction of (pre-)classical CYMH GTs. We are going to see that it measures the failure of the Bianchi identity of the field strength, \textit{i.e.}~$\mathrm{d}^\nabla \zeta = 0$ if and only if the Bianchi identity is satisfied.

\begin{theorems*}{Bianchi identity of the field strength (simplified formulation)}
Let $M$ and $N$ be smooth manifolds, $K \to N$ an LAB, $\Phi \in C^\infty(M;N)$, and $\nabla$ and $\zeta \in \Omega^2(N; K)$ satisfying the compatibility conditions of a CYMH GT.

Then
\bas
\mathrm{d}^{\Phi^*\nabla}\bigl(G(\Phi,A)\bigr) + \mleft[ A \stackrel{\wedge}{,} G(\Phi,A) \mright]_{\Phi^*K}
&=
\Phi^! \mleft( \mathrm{d}^\nabla \zeta \mright),
\eas
where
\bas
G(\Phi,A)
&=
\mathrm{d}^{\Phi^*\nabla}A
	+ \frac{1}{2} \mleft[ A \stackrel{\wedge}{,} A \mright]_{\Phi^*K}
	+ \Phi^!\zeta
\eas
is the new field strength including the contribution of $\zeta$, and where $\mleft[ \cdot, \cdot \mright]_{\Phi^*K}$ is the $\Phi$-pullback of the field of Lie brackets of $K$.
\end{theorems*}

This concludes the discussion about LABs.

In Section \ref{TangentBundles} we turn to tangent bundles; again Subsection \ref{GeneralSituForTangent} will discuss the general situation for tangent bundles, and we will see that tangent bundles are locally always pre-classical in Subsection \ref{LocalTangentBundles}. 

\begin{theorems*}{Tangent bundles are locally pre-classical as CYMH GT (simplified version)}
Let $N = \mathbb{R}^n$ ($n \in \mathbb{N}_0$) be an Euclidean space as smooth manifold and $\nabla$ a connection on $E \coloneqq \mathrm{T}N$ which satisfies the compatibility conditions. Then there is a field redefinition such that $\nabla$ becomes flat.
\end{theorems*}

Globally however, we will see in Subsection \ref{UnitoctonionsasGT} that the seven-dimensional sphere $\mathds{S}^7$ admits a gauge theory in the sense of CYMH GT, related to a non-flat $\nabla$. A flat $\nabla$ would imply a Lie group structure on $\mathds{S}^7$ which does not exist as we know, and this will be the quintessence of its structure as CYMH GT for which there is no field redefinition towards a pre-classical theory.

\begin{theorems*}{Global example: Unit octonions (simplified version)}
$\mathbb{S}^7$ admits a CYMH GT such that the related connection $\nabla$ on $E \coloneqq \mathrm{T}\mathds{S}^7$ is not flat. Moreover, there is no field redefinition such that $\nabla$ becomes flat.
\end{theorems*}

The thesis concludes in Section \ref{GeneralObstrAoids} with a discussion about more general Lie algebroids; first stating a small general statement in Section \ref{GeneralGeneral}, but then turning to Lie algebroids given as the direct product of tangent bundles and Lie algebra bundles in Section \ref{GeneralSitDirectProducts}. We derive that the direct product of CYMH GTs has a natural structure as CYMH GT, and we can extend the existence of a redefinition towards a pre-classical theory by using previous results.

\begin{theorems*}{Direct products of CYMHG GTs around regular points are flat (simplified formulation)}
Let $N \coloneqq \mathbb{R}^n$ ($n \in \mathbb{N}_0$) be a smooth manifold such that its tangent bundle admits a CYMH GT, whose connection satisfying the compatibility conditions we denote by $\nabla^{N}$, and let $K \to S$ be an LAB over a smooth contractible manifold $S$ which also admits a CYMH GT, equipped with a connection $\nabla^K$ satisfying the compatibility conditions.

Then there is a field redefinition with respect to their direct product of CYMH GTs with connection $\nabla$ (satisfying the compatibility conditions) such that the field redefinition of $\nabla$ becomes flat, where $\nabla$ is canonically given as a product of $\nabla^N$ and $\nabla^K$.
\end{theorems*}

However, the discussion about general Lie algebroids will not go beyond this point, and the thesis will conclude with a possible conjecture, which may simplify further calculations related to direct products, especially allowing to extend other previous results.

\begin{conjectures*}{Existence of a splitted field redefinition (simplified formulation)}
Let $N$ be a smooth manifold such that its tangent bundle admits a CYMH GT, and let $K \to S$ be an LAB over a smooth manifold $S$ which also admits a CYMH GT.

If there is a field redefinition such that their direct product of CYMH GTs is pre-classical or classical, then there is also a field redefinition for each factor separately transforming each factor to a pre-classical or classical theory, respectively.
\end{conjectures*}

Subsection \ref{LastAnsatzes} just lists loose ansatzes and ideas for further calculations, not necessarily related to direct products; for the thesis itself it is not necessarily needed to read this subsection. Finally, Chapter \ref{ConclusionTheEnd} gives a short overview about possible future research plans.

\section{Notation and other conventions throughout this work}\label{StandardNotation}

In this thesis a lot of conventions are used, they are either in the following list or will be introduced later.

\begin{itemize}
	\item Throughout this work we always use Einstein's sum convention if suitable.
	\item Due to ambiguities about connectedness in the definition of \textbf{simply connected} manifolds, we emphasize that we will use the definition of simply connectedness which also requires that such a manifold is path-connected.
	\item A map $f: A \to B$ between two sets $A$ and $B$ we often also denote by $[A \ni a \mapsto f(a) \in B]$, or shortly $[a \mapsto f(a)]$, or also
\bas
A &\to B,\\
a &\mapsto f(a).
\eas
	\item Every time when we have a map with arguments from different sets, like a map $f$ defined on $A \times B$ with values in a set $C$, $(a,b) \mapsto f(a,b)$, where $A$ and $B$ are two sets, then we sometimes just insert one or a part of the arguments. Those we denote \textit{e.g.}~by $f(b)$ for $b \in B$, so, $f(b): A \to C, a\mapsto f(a,b)$. We may also write instead $f(\cdot, b)$. This only applies to situations where the arguments are not related by some condition like antisymmetry to avoid confusion when ordering of the arguments is important.
	\item $\gls{MN}$ will be smooth manifolds, although $M$ sometimes also denotes a spacetime; but the latter will be mentioned then.
	\item $\gls{TN}$ the tangent bundle of $N$.
	\item $\gls{X(N)}$ the space of vector fields of $N$ with Lie bracket $\gls{0[]}$.
	\item $\gls{DAiff}(N)$ will denote the space of diffeomorphisms of $N$ and $\gls{Cinfty(N)}$ the space of its smooth functions; when a smooth function has values in another smooth manifold $M$, then we denote that space by $ \gls{Cinfty(N;M)}$.
	\item With $\gls{0bigwedgedotV}$ we will denote the exterior power of a vector bundle $V$.
	\item $\gls{1Camma(V)}$ will be $V$'s vector space of sections.
	\item We will denote the bundle of automorphisms and endomorphisms of $V$ by $\gls{Aut}(V)$ and $\gls{End}(V)$, respectively. We also denote $\gls{AutSection}(V) \coloneqq \Gamma(\mathrm{Aut}(V))$ and $\gls{EndSection}(V) \coloneqq \Gamma(\mathrm{End}(V))$. With those we also always mean base-preserving ones, also called vertical automorphisms and vertical endomorphisms.
	\item We denote the space of \textbf{$(r,s)$-$V$-tensors} by $\gls{Trs(V)a} \coloneqq \Gamma\mleft(\gls{Trs(V)}\mright)$ for $r,s \in \mathds{N}_0$, where $\mathrm{T}^r_s(V) \coloneqq \bigotimes^s V^* \otimes \bigotimes^r V$ ($r, s \in \mathds{N}_0$).
	\item $\gls{V*}$ denotes the dual bundle of $V$, as a special example $\gls{T*N}$ denotes the cotangent bundle of $N$ and $\gls{1ZOmegak(N)} \coloneqq \Gamma\mleft(\bigwedge^k \mathrm{T}^*N\mright)$ the space of $k$-forms ($k \in \mathds{N}_0$).
	\item $\gls{0nabla}$ denotes a vector bundle connection on $V$ with $\gls{Rnabla}$ their curvature. Throughout this work we will also face a more general notion of connection, but when we just write \textbf{connection}, then we always mean a \textbf{vector bundle connection}. If some object is another type of connection, then it will be explicitly mentioned or clear by the context.
	\item As usual, one can extend a connection $\nabla$ to $\mathcal{T}^r_s(V)$ ($r, s\in \mathbb{N}_0$) by the Leibniz rule. We will denote such connections still with $\nabla$.
	\item In the following $\gls{D}$ is also the \textbf{total differential} or \textbf{tangent map} of smooth maps, \textit{i.e.}~for every smooth map $F: M \to N$ we have the canonical (total) differential $\mathrm{D}_pF: \mathrm{T}_pM \to \mathrm{T}_{F(p)}N$ for all $p\in M$. In the following we view $\mathrm{D}F$ as an element of $\Omega^1(M; F^*\mathrm{T}N)$ by $\mathfrak{X}(M) \ni Y \mapsto \mathrm{D}F(Y)$, where $\mathrm{D}F(Y) \in \Gamma(F^*\mathrm{T}N), M \ni p \mapsto \mathrm{D}_pF(Y_p)$.
	\item The de-Rham differential is denoted by $\gls{dbas}$.
	\item Coordinate vector fields on a smooth manifold we often denote by $\gls{0partiali}$.
	\item The Lie derivative of a vector field $X$ is denoted by $\gls{Lie}_X$, and with this we also denote the action of $X$ on smooth functions $f$ by derivation; the latter we may also denot with $X(f) = \mathcal{L}_X(f)$.
	\item With $\gls{1ZOmegap(NV)}$ ($p\in \mathbb{N}_0$) we denote the space of forms with values in $V$. There is a similar notation for vector spaces $W$, $\Omega^p(N; W)$; although $W$ is not defined as a bundle over $N$, with that we mean forms with values in the trivial bundle $N \times W \to N$; similar for all other type of tensors, and also for other vector spaces and their associated trivial vector bundles.
	\item When one has a connection $\nabla$ on a vector bundle $V \to N$, then one has the notion of the exterior covariant derivative on $\Omega^p(M;E)$, denoted by $\gls{dnabla}$. In the case of a trivial vector bundle $V=N \times W \to N$, where $W$ is some vector space, we will often use the \textbf{canonical flat connection} for $\nabla$, defined by $\nabla \nu = 0$, where $\nu$ is a constant section of $N \times W$, see \textit{e.g.}~\cite[Example 5.1.7; page 260f.]{hamilton} for a geometric interpretation as horizontal distribution. The canonical flat connection is clearly uniquely defined (if a trivialization is given) because constant sections generate all sections and due to the Leibniz rule and linearity of $\nabla$. That is, let $\nabla^\prime$ be another canonical flat connection with $\nabla^\prime \nu = 0$ for all constant sections $\nu$. Then every section of $N \times W$ is a sum of elements of the form $f \nu$, where $\nu$ is still a constant section and $f \in C^\infty(N)$, such that
\bas
\nabla(f \nu)
&=
\mathrm{d}f \otimes \nu
	+ f \underbrace{\nabla \nu}_{\mathclap{= 0 = \nabla^\prime \nu}}
=
\nabla^\prime(f \nu),
\eas
which proves the claim using the linearity of $\nabla$. Let $\mleft( e_a \mright)_a$ be a constant global frame of $N \times W$, thence,
\bas
\mathrm{d}^\nabla \omega
&=
\mathrm{d} \omega^a \otimes e_a
\eas
for all $\omega \in \Omega^p(M; W)$, where we write $\omega= \omega^a \otimes e_a$. Hence, we define
\ba
\mathrm{d}\omega
&\coloneqq
\mathrm{d}^\nabla \omega,
\ea
when $\nabla$ is the canonical flat connection. $\mathrm{d}$ is clearly a differential.
	\item With $\Phi^*V$ we denote the pullback/pull-back of the vector bundle $V$ under a smooth map $\Phi: M \to N$. We will also have sections $F$ as an element of $\Gamma\left( \left(\bigotimes_{m=1}^{l} E_m^*\right) \otimes E_{l+1} \right)$, where $E_1, \dots E_{l+1}$ ($l \in \mathbb{N}$) are real vector bundles of finite rank over $N$. Those pull-back as section, denoted by $\Phi^*F$, we will view as an element of $\Gamma\left( \mleft(\bigotimes_{m=1}^{l} \mleft(\Phi^*E_m\mright)^*\mright) \otimes \Phi^*E_{l+1} \right)$, and it is essentially given by
\bas
	(\Phi^*F)(\Phi^*\nu_1, \dotsc , \Phi^*\nu_l)
	&=
	\Phi^*\mleft( F\mleft( \nu_1, \dotsc, \nu_l \mright) \mright)
\eas
for all $\nu_1 \in \Gamma(E_1), \dotsc, \nu_l \in \Gamma(E_l)$, using that pullbacks of sections generate the sections of a pullback bundle. In general we also make use of that sections of $\Phi^*E$ can be viewed as sections of $E$ along $\Phi$, where $E \stackrel{\pi}{\to} N$ is any vector bundle over $N$. Let $\mu \in \Gamma(\Phi^*E)$, then it has the form $\mu_p = (p, v_p)$ for all $p \in M$, where $v_p \in E_{\Phi(p)}$, the fibre of $E$ at $\Phi(p)$; and a section $\nu$ of $E$ along $\Phi$ is a smooth map $M \to E$ such that $\pi(\nu) \coloneqq \pi \circ \nu = \Phi$. Then on one hand $\mathrm{pr}_2 \circ \mu$ is a section along $\Phi$, where $\mathrm{pr}_2$ is the projection onto the second component, and on the other hand $M \ni p \mapsto (p, \nu_p)$ defines an element of $\Gamma(\Phi^*E)$. With that one can show that there is a 1:1 correspondence of $\Gamma(\Phi^*E)$ with sections along $\Phi$. We do not necessarily mention it when we make use of that identification, it should be clear by the context.
	\item We will also often make use of that $\Gamma(\Phi^*E)$ is generated by pullbacks of $\Gamma(E)$. If we explicitly use this in calculations, then we take for example a local frame $\mleft( e_a \mright)_a$ of $E$, and then a frame of $\Phi^*E$ is given by $\mleft(\Phi^*e_a\mright)_a$. In such situations we implicitly assume that $\mleft(e_a\mright)_a$ is defined on a part of the image of $\Phi$. Similar for intersections of frames. 
	\item Furthermore, we will often need frames for bundles like $\Phi^*E$; we will then just write "Let $\mleft(e_a\mright)_a$ be a local frame of $E$" and implicitly mean that we take $\mleft(\Phi^*e_a\mright)_a$ as a frame for $\Phi^*E$.
	\item Do not confuse the previously discussed pull-back of sections with the pull-back of forms $F \in \Omega^l(N; V)$, here denoted by $\Phi^!F$, which is an element of $\Gamma\left( \mleft(\bigwedge_{m=1}^{l} \mathrm{T}^*M \mright) \otimes \Phi^*V \right) \cong \Omega^l(M; \Phi^*V)$, and not of $\Gamma\left( \mleft(\bigotimes_{m=1}^{l} \mleft(\Phi^*\mathrm{T}N\mright)^*\mright) \otimes \Phi^*E_{l+1} \right)$ like $\Phi^*F$. $\Phi^!F$ is defined by
\ba
\mleft.\mleft(\Phi^!F\mright)(Y_1, \dots, Y_l)\mright|_p
&\coloneqq
F_{\Phi(p)}\mleft(\mathrm{D}_p\Phi\mleft(\mleft.Y_1\mright|_p\mright), \dots, \mathrm{D}_p\Phi\mleft(\mleft.Y_l\mright|_p\mright)\mright)
\ea
for all $p \in M$ and $Y_1, \dots, Y_l \in \mathfrak{X}(M)$. 
	\item Unless otherwise stated, the considered manifolds and vector bundles are of finite dimension and rank, respectively, and smooth; arising fields are always real numbers, hence, we also view $\mathbb{C}^n$ ($n \in \mathbb{N}$) as $\mathbb{R}^{2n}$. 
	\item Morphisms of bundles over the same base are always base-preserving ones if not stated otherwise.
	\item In the case when we explicitly state that we now turn to infinite-dimensional manifolds, we always assume a convenient setting, for example that is, we assume that all the smooth structures \textit{etc.}~are given and well-defined such that we can treat those manifolds and objects as if they would be finite-dimensional for the constructions we are going to study. The tangent bundle of infinite-dimensional manifolds we will define by the approach of using equivalence classes of curves.
	\item As usual, there will be definitions of certain objects depending on other elements, and for keeping notations simple we will not always explicitly denote all dependencies. It will be clear by context on which it is based on, that is, when we define an object $A$ using the notion of Lie algebra actions $\gamma$ and we write "Let $A$ be [as defined before]", then it will be clear by context which Lie algebra action is going to be used, for example given in a previous sentence writing "Let $\gamma$ be a Lie algebra action".
	\item We have several identities shown in the Appendix \ref{CalculusIdentitiesNeeded}. We will use them throughout this work, but the thesis will be written in such a way that one only needs to know the appendix when starting to read Chapter \ref{GeneralizedGTfas}, and several notions arising in the appendix will be introduced before that chapter.
	\item At the very end is also a list of symbols. There we try to list all the needed symbols with page numbers where they got defined. When you read this thesis using its pdf, then all those symbols will be hyperlinked to that glossary. After clicking on such a link you may be able to get immediately back where you were using the return button on your mouse device if available, whether this works may also depend on your pdf reader; otherwise use the hyperlinks of the listed page numbers in the glossary for a quicker navigation.
	
	The list of symbols first lists generic symbols, then Greek letters, and afterwards Latin letters.
	\item References are not only given in the text, the references of referenced statements and definitions are especially given in the title of those statements. The title also mentions whether the statement as written in this thesis is a variation or generalization; when it is a strong generalization, then the reference will be mentioned in a remark after the statement or its proof.
\end{itemize}
%\newpage
\setcounter{equation}{0}

\chapter{Gauge theory}
\label{ClassicGaugeTheory}
\section{Lie algebras and their actions}\label{LieAlgebraActions}

In the following we will shortly introduce the basic setup of infinitesimal gauge theory where a trivial principal bundle is assumed and, thus, omitted. Equivalently, we assume a global gauge or we just look at some open neighbourhood of the spacetime admitting a local gauge. We will follow \cite{hamilton}.

Moreover, we will especially focus on the infinitesimal behaviour of gauge theory. That is, we will mainly concentrate on Lie algebras and not Lie groups. The following will also not be a deep discussion of the defined notions, just providing the very needed definitions, especially those which are going to be generalized later. Thus, it is in general recommended to have already knowledge about how gauge theory is mathematically formulated, especially Yang-Mills-Higgs gauge theory.

\begin{definitions}{Lie group, \cite[Definition 1.1.4; page 6]{hamilton}}{HamLieGroup}
A \textbf{Lie group} $G$ is a group which is also a smooth manifold such that 
\bas
G \times G &\to G, \\
(g,h) &\mapsto g \cdot h
\eas
is smooth, where $G \times G$ has the canonical smooth structure of a product manifold inherited by the smooth structure of $G$.
\end{definitions}

\begin{remark}
\leavevmode\newline
Usually, the definition of Lie groups contains also the condition about that the inverse map, $G\ni g \mapsto g^{-1}$, is smooth, which can be combined with the smoothness of the multiplication map to that 
\bas
G \times G &\to G, \\
(g,h) &\mapsto g \cdot h^{-1},
\eas
shall be smooth as a single condition for the definition of Lie groups. However, that is not needed as pointed out in \cite[Remark 1.1.8, page 7; see also Exercise 1.9.5, page 76f.]{hamilton}, which is why we just need to ask for smoothness of the product. 
%The idea for it is the following sketch of a proof: One first shows that $\mu: G \times G \to G, \mu(g,h) \coloneqq g \cdot h$, is a submersion, which is shown as usual by observing that
%\bas
%\mathrm{D}_{(g,h)}\mu(X, Y)
%&=
%\mathrm{D}_h L_g (Y)
	%+ \mathrm{D}_g R_h (X)
%\eas
%for all $(g,h) \in G \times G$ and $(X, Y) \in \mathrm{T}_{(g,h)}(G \times G) \cong \mathrm{T}_g G \oplus \mathrm{T}_h G$, where $L_g$ and $R_g$ denotes the left- and right-multiplication with $g \in G$, respectively. Trivially, by the smoothness of $\mu$ the left-multiplication $L_g$ and $L_{g^{-1}}$ are smooth and inverse to each other, hence, $L_g$ is as expected and known a diffeomorphism such that $\mathrm{D}_h L_g$ is an isomorphism, especially having a rank of $\mathrm{dim}(G)$. Thus, $\mathrm{D}_{(g,h)}\mu$ is a submersion.
%
%By the regular value theorem (the submersion theorem) one can conclude that for all $(g,h) \in G \times G$ there is an open neighbourhood $U$ of $(g,h)$ and $V$ of $g \cdot h$ in such a way that there is a section $s$ of $\mu$, that is $s: V \to U$ smooth, satisfying
%\bas
%s(g\cdot h)
%&=
%(g, h),
%\\
%\mu \circ s 
%&=
%\mathds{1}_V.
%\eas
%Also due to the fact that $\mu$ is a submersion, we know that $\mu^{-1}(e)$
\end{remark}

As known, the set of left invariant vector fields\footnote{This can be identified with the tangent space at the unit element as it is well-known.} on a Lie group form a Lie algebra.

\begin{definitions}{Lie algebra, \cite[Definition 1.4.1, page 36]{hamilton}}{HamLieAlgebra}
%\leavevmode\newline
Let $\gls{g1}$ be a vector space together with a map
\bas
\gls{0[]g}: \mathfrak{g} \times \mathfrak{g} &\to \mathfrak{g}, \\
(x, y) &\mapsto \mleft[ x, y \mright]_\mathfrak{g}.
\eas
This pair $\mleft( \mathfrak{g}, \mleft[ \cdot, \cdot \mright]_\mathfrak{g} \mright)$ is called a \textbf{Lie algebra} with \textbf{Lie bracket} $\mleft[ \cdot, \cdot \mright]_\mathfrak{g}$ when the following hold:
\begin{itemize}
	\item $\mleft[ \cdot, \cdot \mright]_\mathfrak{g}$ is bilinear.
	\item $\mleft[ \cdot, \cdot \mright]_\mathfrak{g}$ is antisymmetric.
	\item $\mleft[ \cdot, \cdot \mright]_\mathfrak{g}$ satisfies the \textbf{Jacobi identity}, \textit{i.e.}
	\bas
	\mleft[ x, \mleft[ y, z \mright]_\mathfrak{g} \mright]_\mathfrak{g}
		+ \mleft[ y, \mleft[ z, x \mright]_\mathfrak{g} \mright]_\mathfrak{g}
		+ \mleft[ z, \mleft[ x, y \mright]_\mathfrak{g} \mright]_\mathfrak{g}
	&=0
	\eas
	for all $x, y, z \in \mathfrak{g}$.
\end{itemize}
\end{definitions}

Such an algebra is characterized by the following constants.

\begin{definitions}{Structure constants, \cite[Definition 1.4.17; page 38]{hamilton}}{StructureConstants}
Let $\mleft( \mathfrak{g}, \mleft[ \cdot, \cdot \mright]_\mathfrak{g} \mright)$ be a Lie algebra. Then the \textbf{structure constants} $\gls{Cbca} \in C^\infty(\mathbb{R})$ are defined by
\ba
\mleft[ e_a, e_b \mright]_\mathfrak{g}
&=
C_{ab}^c e_c
\ea
for a given basis $\mleft( e_a \mright)_a$.
\end{definitions}

\begin{remark} \cite[Definition 1.4.17 \textit{et seq.}; page 38]{hamilton}
\leavevmode\newline
The antisymmetry and Jacobi identity of $\mleft[ \cdot, \cdot \mright]_\mathfrak{g}$ imply
\ba
C^a_{bc} &= - C^a_{cb}, \\
0
&=
C^d_{ae} C^e_{bc} + C^d_{be} C^e_{ca} + C^d_{ce} C^e_{ab}.
\ea
\end{remark}

For defining couplings we also need Lie group and Lie algebra representations.

\begin{definitions}{Lie group representation, \cite[Definition 2.1.1; page 84]{hamilton}}{LieGroupRepresentation}
Let $G$ be a Lie group and $W$ a vector space. Then a \textbf{representation} of $G$ on $W$ is a Lie group homomorphism
\bas
\Psi: G \to \mathrm{Aut}(W).
\eas
\end{definitions}

\begin{definitions}{Lie algebra representation \cite[Definition 2.1.5; page 85]{hamilton}}{LiealgebraRepresentation}
Let $\mathfrak{g}$ be a Lie algebra and $W$ a vector space. Then a \textbf{representation} of $\mathfrak{g}$ on $W$ is a Lie algebra homomorphism
\bas
\psi: \mathfrak{g} \to \mathrm{End}(W).
\eas
\end{definitions}

As known, these can be related as in the following lemma.

\begin{lemmata}{Every Lie group representation induces a Lie algebra representation \cite[Proposition 2.1.12; page 86]{hamilton}}{LieGroupRepInducesLieAlgRep}
Every representation $\Psi$ of a Lie group $G$ on $W$ defines a Lie algebra representation $\psi$ by $\psi \coloneqq \Psi_* \coloneqq \mathrm{D}_e \Psi$, where $e$ is the unit element of $G$.
\end{lemmata}

We will focus on the following examples of Lie algebra representations. The first example shows the homomorphism property directly, while the second one uses Lemma \ref{lem:LieGroupRepInducesLieAlgRep}.

\begin{examples}{$\mathrm{su}(2)$-action, \newline \cite[\S 6.2 \textit{et seq.}, page 586ff.; and \S 6.6 \textit{et seq.}; page 633ff.]{cohen2006quantum}}{sutwoliealgactionasLiealg}
Here we will view the Lie algebra $\mathfrak{g} = \mathrm{su}(2)$ as $\mathbb{R}^3$: Let $e_1, e_2, e_3$ denote the standard unit vectors corresponding to the coordinates $x^1, x^2, x^3$. Then the Lie bracket is given by the cross product, \textit{i.e.}~
\ba
\mleft[ e_i, e_j \mright]_{\mathrm{su}(2)} \coloneqq e_i \times e_j = \epsilon_{ijk} e_k,
\ea
where $\gls{1epsilonijk}$ is the Levi-Civita tensor. The representation on $W \coloneqq \mathbb{R}^3$ is given by
\ba
\psi(v)(w)
&\coloneqq
v \times w
=
\epsilon_{ijk} v^i w^j e_k
\ea
for all $v, w \in \mathbb{R}^3$. This is a homomorphism by 
\bas
\psi\mleft( \mleft[ u, v \mright]_{\mathrm{su}(2)} \mright)(w)
&=
u^i v^j w^k \underbrace{\epsilon_{ijl} \epsilon_{lkm}}_{\mathclap{= \delta_{ik} \delta_{jm} - \delta_{im} \delta_{jk}}} e_m
=
u^i w^i v^j e_j
	-  u^i w^j v^j e_i,
\eas
where $\gls{1deltaijz}$ is the Kronecker delta, and
\bas
\mleft( \mleft[ \psi(u), \psi(v) \mright]_{\mathrm{End}(\mathbb{R}^3)} \mright)(w)
&=
\mleft( u^i v^j \epsilon_{ilm} \epsilon_{jkl}
	- u^i v^j \epsilon_{jlm} \epsilon_{ikl} \mright) w^k e_m \\
&=
\mleft( - u^i v^i + u^i v^i \mright) w^m e_m
+ u^i v^j w^i e_j - u^i v^j w^j e_i \\
&=
\psi\mleft( \mleft[ u, v \mright]_{\mathrm{su}(2)} \mright)(w)
\eas
for all $u, v, w \in \mathbb{R}^3$.
\end{examples}

\begin{examples}{Electroweak interaction coupled to a Higgs field, \newline\cite[Example 8.1.9; page 449f.; and \S 8.3.1; page 465ff.]{hamilton}}{electroweakinteractionasLiealg}
The \textbf{electroweak interaction coupled to a Higgs field} is defined as $\mathfrak{g} \coloneqq \mathrm{su}(2) \oplus \mathrm{u}(1)$ acting on $W \coloneqq \mathbb{C}^2 (\cong \mathbb{R}^4)$. Let $\mathrm{i}$ be the imaginary number and $n_\gamma$ be a non-zero natural number (a normalization constant). The Lie algebra representation $\psi$ is then defined as the induced representation $\Psi_*$ of the Lie group representation $\Psi$ given by
	\bas
	(\mathrm{SU}(2) \times \mathrm{U}(1)) \times \mathbb{C}^2 &\to \mathbb{C}^2, \\
	\mleft(A, \e^{\mathrm{i} \alpha}, w\mright) &\mapsto \Psi\mleft( A, \e^{\mathrm{i} \alpha} \mright)(w) \coloneqq \mleft(A, \e^{\mathrm{i} \alpha}\mright) \cdot w
	\coloneqq
	\e^{\mathrm{i} n_\gamma \alpha} A w
	\eas
	for all $w \in \mathbb{C}^2$. This is clearly a Lie group representation.
\end{examples}

Another important examples are the adjoint representations.

\begin{examples}{Adjoint representations, \newline \cite[Theorem 2.1.45 and abstract before that; page 101]{hamilton} \& \cite[Theorem 2.1.52; page 105]{hamilton}}{AdjointReps}
We have the well-known \textbf{adjoint representation of a Lie group $G$}: For an element $g \in G$ we define the \textbf{conjugation $c_g$} as a map by
\bas
G &\to G,
\\
h
&\mapsto
c_g(h)
\coloneqq
ghg^{-1}.
\eas
It is easy to check that $c_g$ is a Lie group automorphism, \textit{i.e.}~a diffeomorphism and a homomorphism; moreover, the map $G \times G \to G, (g, h) \mapsto c_g(h),$ is a left action of $G$ on itself, especially we have $c_{gh} = c_g \circ c_h$ for all $g,h \in G$. All of those properties lead to the definition of the adjoint representation (of $G$) $\mathrm{Ad}: G \to \mathrm{Aut}(\mathfrak{g})$, a $G$-representation on $\mathfrak{g}$ defined as map by
\bas
G &\to \mathrm{Aut}(\mathfrak{g}),
\\
g
&\mapsto
\mathrm{Ad}(g)
\coloneqq
\mathrm{D}_e c_g,
\eas
where $e \in G$ is the neutral element; we defined Lie group representations with values in vector bundle automorphisms, but due to the properties of the conjugation one can also understand $\mathrm{Aut}(\mathfrak{g})$ here as the space of Lie algebra automorphisms, especially $\mathrm{Ad}(g)$ is additionally a homomorphism of the Lie bracket of $\mathfrak{g}$ for all $g \in G$.

The induced Lie algebra representation of $\mathrm{Ad}$ is given by $\gls{ad}: \mathfrak{g} \to \mathrm{End}(\mathfrak{g}), X \mapsto \mleft[ X, \cdot \mright]_{\mathfrak{g}}$, the \textbf{adjoint representation of $\mathfrak{g}$}.
\end{examples}

Representations can be generalized to actions on manifolds $N$.

\begin{definitions}{Left action on manifold, \cite[\S 3.2, Definition 3.2.1; page 130]{hamilton}}{LieGroupAction}
A \textbf{smooth left action} of a Lie group $G$ on a smooth manifold $N$ is a smooth map
\bas
G \times N &\to N, \\
(g, p) &\mapsto g \cdot p = gp,
\eas
where $G \times N$ is equipped with the canonical product structure, and we demand:
\begin{itemize}
	\item For all $g, h \in G$ and $p\in N$
		\bas
			(g \cdot h) \cdot p &= g \cdot (h \cdot p).
		\eas
	\item For all $p \in N$ and $e$ the neutral element of $G$
		\bas
			e \cdot p &= p.
		\eas
\end{itemize}
\end{definitions}

\begin{remark} \cite[\S 3.4; page 141ff.]{hamilton}\label{FundamentalVectorFields}
\leavevmode\newline
One may try to think about a left action as a generalization of Lie group representation when replacing the space of automorphisms of a vector space $W$ with the space of diffeomorphisms $N$, $\mathrm{Diff}(N)$, and then rewriting the left action as a map $G \to \mathrm{Diff}(N), g \mapsto \mleft[ p \mapsto gp \mright] \in \mathrm{Diff}(N)$. The definition of a left action then implies that this map would be a group homomorphism. 

Keep in mind that the definition of a representation of a Lie group demands smoothness of the representation such that we would need to define a smooth structure on (in general) infinite-dimensional sets like $\mathrm{Diff}(N)$ which we would like to avoid. Hence, when we also want to derive a Lie algebra action we just motivate it in the following way. Denote the action by $(g, p) \mapsto \Psi(g, p) \coloneqq g \cdot p$, then take any Lie algebra element $X \in \mathfrak{g}$ to conclude for $t, s \in \mathbb{R}$, by using Def.~\ref{def:LieGroupAction},
\bas
\mleft.\Psi\mleft( \e^{t X}, p \mright)\mright|_{t=0}
&=
e \cdot p
= p, \\
\Psi\mleft( \e^{(t+s) X}, p \mright)
&=
\Psi\mleft( \e^{t X} \cdot \e^{s X}, p \mright)
=
\Psi\mleft( \e^{t X}, \Psi\mleft( \e^{s X}, p \mright) \mright),
\eas
where $t \mapsto \e^{t X}$ denotes the 1-parameter subgroup through $X$. Thence, $\mathbb{R} \times N \to N, (t, p) \mapsto \Psi\mleft( \e^{tX}, p \mright)$ defines the flow of a (complete) vector field $\gamma(-X) \in \mathfrak{X}(N)$, defined at $p$ by $\gamma(-X)_p \coloneqq \mleft. \frac{\mathrm{d}}{\mathrm{d}t}\mright|_{t=0} \mleft[ t \mapsto \Psi\mleft( \e^{t X}, p \mright)\mright]$. This defines a map $\mathfrak{g} \to \mathfrak{X}(N), X \mapsto \gamma(X)$, which is known as the map to \textbf{fundamental vector fields}, and the change of the sign is needed to define $\gamma$ as a homomorphism of Lie algebras, see \textit{e.g.}~\cite[Proposition 3.4.4; page 144]{hamilton}. In fact, we are going to prove that in Prop.~\ref{prop:LieRepAndLieAct}, too, in the special situation of $N=W$ for some vector space $W$.
\end{remark}

Thence, we motivated the following definition.

\begin{definitions}{Lie algebra action, \cite[\S 16.2, Example 5; page 114]{DaSilva}}{LieAlgebraAction}
A \textbf{Lie algebra action} of a Lie algebra $\mathfrak{g}$ on a smooth manifold $N$ is a Lie algebra homomorphism 
\bas
\gls{1cammaz}: \mathfrak{g} \to \mathfrak{X}(N)
\eas
such that the map
\bas
N \times \mathfrak{g} &\to \mathrm{T}N, \\
(p, X) &\mapsto \gamma(X)_p
\eas
is smooth, equipping $N \times \mathfrak{g}$ with the canonical structure of product manifolds.
\end{definitions}

\begin{remark}
\leavevmode\newline
If $\gamma$ is induced by a (left) Lie group action as in Remark \ref{FundamentalVectorFields}, then we also call $\gamma$ the \textbf{induced Lie algebra action}.
\end{remark}

We can show that all Lie algebra representations define a Lie algebra action, not assuming any integrability to a Lie group representation.

\begin{propositions}{Lie algebra representation $\rightarrow$ Lie algebra action, \newline \cite[generalisation of parts of Example 3.4.2; page 143f.]{hamilton}}{LieRepAndLieAct}
Every Lie algebra representation $\psi$ on a vector space $W$ defines a Lie algebra action $\gamma$ by
\ba
\gamma(X)_v &\coloneqq - \psi(X)(v)
\ea
for all $X \in \mathfrak{g}$ and $v \in W$, where we view the right hand side as an element of $\mathrm{T}_vW$, making use of $\mathrm{T}_vW \cong W$.
\end{propositions}

\begin{remarks}{}{}
We then say that \textbf{$\gamma$ is induced by $\psi$}.
\end{remarks}

\begin{remark}\label{RemTVGleichV}
\leavevmode\newline
A few words about using $\mathrm{T}_vW \cong W$: In the following we will denote a basis of $W$ by $\mleft(e_a\mright)_a$, $v = v^a e_a$ for all $v \in W$, which we will also identify as a (constant) frame of $\mathrm{T}W$, \textit{i.e.}~$\partial_a \leftrightarrow e_a$ for some coordinate vector fields $\mleft( \partial_a \mright)_a$. Then the definition contained in Prop. \ref{prop:LieRepAndLieAct} reads
\bas
\gamma(X) &\coloneqq - \overline{\psi(X)},
\eas
where $\overline{T} \in \mathfrak{X}(W)$ for $T \in \mathrm{End}(W)$ is defined by
\bas
W &\to \mathrm{T}W, \\
v &\mapsto
\overline{T}(v)
\coloneqq
T^a_b v^b \mleft.\partial_a\mright|_v.
\eas
Normally, we will omit this notation most of the time and write $\overline{T} = T$ since the identification in $\mathrm{T}_vW \cong W$ is very natural. But until the proof of Prop. \ref{prop:LieRepAndLieAct} we are going to keep this notation.
\end{remark}

To prove Prop. \ref{prop:LieRepAndLieAct} we need to show the following Lemma and Corollary; these are basically the statements as for fundamental vector fields, \cite[\S 3.4; page 141ff.]{hamilton}, but just looking at $\mathfrak{g}= \mathrm{End}(W)$ with $\psi = \mathds{1}_{\mathrm{End}(W)}$ as representation on $W$, which is all one needs to prove Prop.~\ref{prop:LieRepAndLieAct}.

\begin{lemmata}{$\overline{\mathrm{End}(W)}$ a Lie subalgebra of $\mathfrak{X}(W)$, \newline \cite[\S 3.4; page 141ff.; especially second equation in Remark 3.4.5; page 145]{hamilton}}{LemmaEndGleichMinusVectorField}
Let $W$ be a vector space. Then $\overline{\mathrm{End}(W)}$ is a Lie subalgebra of $\mathfrak{X}(W)$, and we have
\ba
\overline{\mleft[ T, L \mright]}_{\mathrm{End}(W)}
&=
-\mleft[ \overline{T}, \overline{L} \mright]
\ea
for all $T, L \in \mathrm{End}(W)$. 
\end{lemmata}

\begin{proof}
\leavevmode\newline
That it is a subspace is clear due to $0 \in \overline{\mathrm{End}(W)}$ and
\bas
\overline{a T + b L}
&=
a \overline {T} + b \overline{L}
\eas
for all $T, L \in \mathrm{End}(W)$ and $a, b \in \mathbb{R}$. We also get for $v = v^a e_a \in W$
\bas
\mleft[ \overline{T}, \overline{L} \mright]_v
&=
\Big( 
\overline{T}^b ~ \underbrace{\partial_b \overline{L}^a}
_{\mathclap{= \partial_b \mleft[ v \mapsto L^a_c v^c \mright] = L^a_b}}
 - \overline{L}^b ~ \partial_b \overline{T}^a
\Big)\Big|_v ~ \mleft.\partial_a\mright|_v
=
- \mleft[T, L \mright]^a_{\mathrm{End}(W)}(v) ~ \mleft.\partial_a \mright|_v
=
- \mleft.\overline{\mleft[T, L \mright]}_{\mathrm{End}(W)} \mright|_v,
\eas
which also shows that it is a subalgebra.
\end{proof}

In fact, we can identify the endomorphisms of $W$ with this subalgebra.

\begin{corollaries}{Lie algebra isomorphism $\mathrm{End}(W) \cong \overline{\mathrm{End}(W)}$, \newline \cite[simplified Proposition 3.4.3; page 144]{hamilton}}{EndVGleichBarEndV}
Let $W$ be a vector space. Then there is a natural Lie algebra isomorphism 
\ba
\mathrm{End}(W) \cong \overline{\mathrm{End}(W)}. 
\ea
\end{corollaries}

\begin{proof}
\leavevmode\newline
Define $F: \mathrm{End}(W) \to \overline{\mathrm{End}(W)}$ by
\ba\label{defSuperEasyPeasyDefinitionvonhomomderactionsundReps}
F(L)
&\coloneqq
-\overline{L}
\ea
for all $L \in \mathrm{End}(W)$. Then observe for $T, L \in \mathrm{End}(W)$ that
\bas
\mleft[ F(T), F(L) \mright]
&=
\mleft[ \overline{T}, \overline{L} \mright]
\stackrel{\text{Lem. \ref{lem:LemmaEndGleichMinusVectorField}}}{=}
- \overline{\mleft[ T, L \mright]}_{\mathrm{End}(W)}
=
F\mleft(\mleft[ T, L \mright]_{\mathrm{End}(W)}\mright),
\eas
hence, $F$ is a homomorphism of Lie algebras, and it is clearly an isomorphism by definition \eqref{defSuperEasyPeasyDefinitionvonhomomderactionsundReps}.
\end{proof}

Using Lemma \ref{lem:LemmaEndGleichMinusVectorField} we can finally prove Prop. \ref{prop:LieRepAndLieAct}.

\begin{proof}[Proof of Prop. \ref{prop:LieRepAndLieAct}]
\leavevmode\newline
Smoothness is clearly inherited by the smoothness of $\psi$. We need to show that $\gamma$ defined by $\gamma(X) \coloneqq - \overline{\psi(X)}$ for all $X \in \mathfrak{g}$ is a homomorphism of Lie algebras. Then use the sign change of Lemma \ref{lem:LemmaEndGleichMinusVectorField} to show for $X, Y \in \mathfrak{g}$
\bas
\gamma\mleft(\mleft[ X, Y \mright]_{\mathfrak{g}}\mright)
&=
-\overline{\psi\mleft(\mleft[ X, Y \mright]_{\mathfrak{g}}\mright)}
\stackrel{\psi \text{ Homom.}}{=}
-\overline{\mleft[ \psi(X), \psi(Y) \mright]}_{\mathrm{End}(W)}
\stackrel{\ref{lem:LemmaEndGleichMinusVectorField}}{=}
\mleft[ \overline{\psi(X)}, \overline{\psi(Y)} \mright]
=
\mleft[ \gamma(X), \gamma(Y) \mright].
\eas
\end{proof}

Prop. \ref{prop:LieRepAndLieAct} immediately implies the following corollary.

\begin{corollaries}{Lie group representation defines actions, \newline \cite[Example 3.4.2, page 143f.]{hamilton}}{LieGroupRepsImplyActionStuff}
Every Lie group representation $\Psi$ on a vector space $W$ defines a Lie group and Lie algebra action on $W$.
\end{corollaries}

\begin{proof}
\leavevmode\newline
As it is well-known, every Lie group representation $\Psi$ defines a left action by
\bas
G \times W &\to W, \\
(g,v) &\mapsto g \cdot v \coloneqq \Psi(g)(v).
\eas
The Lie algebra action $\gamma$ is canonically given by the fundamental vector fields related to this action,
\bas
\gamma(X)_v 
&\coloneqq
\mleft. \frac{\mathrm{d}}{\mathrm{d}t}\mright|_{t=0}
\mleft[ t \mapsto \mleft( \e^{-t X} \cdot v \mright) \mright]
=
- \Psi_*(X)(v)
\eas
for $t \in \mathbb{R}$, for all $X \in \mathfrak{g}$ and $v \in W$. This is a Lie algebra action by Prop. \ref{prop:LieRepAndLieAct}.
\end{proof}

\section{Isotropy}\label{IsotropyClassical}

Of a special importance in this work will be the isotropy subalgebra of a Lie algebra $\mathfrak{g}$. We will define this without using group actions because we won't assume integrability in general throughout this work.

\begin{definitions}{The Isotropy Subalgebra, \newline \cite[infinitesimal version of Definition 3.2.4; page 132]{hamilton}}{IsotropySubalgebra}
Let $\mathfrak{g}$ be a Lie algebra, and $\gamma: \mathfrak{g} \to \mathfrak{X}(N)$ a Lie algebra action on a smooth manifold $N$. Then the \textbf{isotropy subalgebra $\mathfrak{g}_p$ at $p \in N$} is defined as
\ba
\mathfrak{g}_p
&\coloneqq
\left\{ X \in \mathfrak{g} ~\middle|~
\gamma(X)_p = 0
\right\}.
\ea
We also often call it just \textbf{isotropy (at $p$)}.

When we have a Lie algebra representation $\psi: \mathfrak{g} \to \mathrm{End}(W)$ on a vector space $W$, then its isotropy is related to its induced Lie algebra action as given in Prop.~\ref{prop:LieRepAndLieAct}.
\end{definitions}

\begin{remark}\label{ClassicalIsotropy}
\leavevmode\newline
Normally the isotropy subalgebra is defined by assuming a (left) Lie group action $\Psi: G \times N \to N, \Psi(g,p) = g\cdot p,$ of a Lie group $G$. Then the \textbf{isotropy group at $p \in N$}, \cite[Definition 3.2.4; page 132]{hamilton}, is defined as 
\ba
G_p
&\coloneqq
\left\{ g \in G ~ \middle|~
g \cdot p = p
\right\}.
\ea
By \cite[Proposition 3.2.9; page 134]{hamilton}, $G_p$ is an embedded Lie subgroup of $G$, and, by \cite[Proposition 3.2.10; page 134]{hamilton}, one can show that the Lie algebra of $G_p$ is the kernel of a map $\mathfrak{g} \to \mathrm{T}_pN$, defined by
\bas
X
&\mapsto
\mleft.\frac{\mathrm{d}}{\mathrm{d}t}\mright|_{t=0}\mleft[
t \mapsto \Psi\mleft( \e^{-tX}, p \mright)
\mright],
\eas
which is precisely the canonical action of fundamental vector fields defined by $\Psi$, evaluated at $p$. That is the motivation for Def.~\ref{def:IsotropySubalgebra}.
\end{remark}

In case of an integrable Lie algebra action we have the following relationship of isotropies.

\begin{corollaries}{Isotropy of integrable Lie algebra actions, \newline \cite[infinitesimal version of the abstract before Proposition 3.2.10; page 134]{hamilton}}{IsotropyVonLieAlgMitAdjoint}
Let $G$ be a Lie group with a (left) Lie group action $\Psi: G \times N \to N, (g,p) \mapsto \psi(g, p) = gp,$ on a smooth manifold $N$. Then
\ba
\mathrm{Ad}\mleft(g\mright)(\mathfrak{g}_p) = \mathfrak{g}_{gp}
\ea
for all $g \in G$ and $p \in N$, where $\mathfrak{g}_p$ and $\mathfrak{g}_{gp}$ are the corresponding isotropy subalgebras related to the Lie algebra action induced by $\Psi$. Especially, $\mathfrak{g}_p$ and $\mathfrak{g}_{gp}$ are isomorphic as Lie algebras.
\end{corollaries}

\begin{proof}
\leavevmode\newline
This corollary is the infinitesimal version of the other well-known relationship of isotropy groups, see \cite[abstract before Proposition 3.2.10; page 134]{hamilton},
\ba\label{isotropygrouprelation}
c_{g}(G_p)
&=
G_{gp}
\ea
for all $g \in G$ and $p \in N$, especially, $c_g: G_p \to G_{gp}$ is a Lie group isomorphism; this is easy to check. Because the isotropy algebras are here now induced by the Lie group action, we know that the induced Lie algebra action $\gamma$ is given by the fundamental vector fields, and, so, the isotropy subalgebras are the Lie algebras of the isotropy groups, recall Remark \ref{ClassicalIsotropy}.

First let us show that $\mathrm{Ad}\mleft(g\mright)(\mathfrak{g}_p) \subset \mathfrak{g}_{gp}$. Observe, making use of Eq.~\eqref{isotropygrouprelation},
\bas
c_g\mleft(
	\e^{tX}
\mright)
&\in
G_{gp}
%\mleft.\gamma\bigl(\mathrm{Ad}(g)(X)\bigr)\mright|_p
%&=
%\mleft.\frac{\mathrm{d}}{\mathrm{d}t}\mright|_{t=0}\mleft[
%t \mapsto \Psi\mleft( \e^{-t ~\mathrm{Ad}(g)(X)}, p \mright)
%\mright]
\eas
for all $g \in G$, $p\in N$, $X \in \mathfrak{g}_p$, and $t \in \mathbb{R}$. $\mleft[ \mathbb{R} \ni t \mapsto c_g\mleft( 	\e^{tX} \mright) \in G_{gp} \mright]$ is clearly a Lie group homomorphism as a composition of homomorphisms, especially a 1-parameter subgroup. Hence,
\bas
\mathfrak{g}_{gp}
&\ni
\mleft.\frac{\mathrm{d}}{\mathrm{d}t}\mright|_{t=0}
\mleft[
t \mapsto 
c_g\mleft(
	\e^{tX}
\mright)
\mright]
=
\mathrm{Ad}(g)(X),
\eas
and therefore $\mathrm{Ad}\mleft(g\mright)(\mathfrak{g}_p) \subset \mathfrak{g}_{gp}$.\footnote{Alternatively, use the well-known equation $c_g\mleft(\exp(tX) \mright) = \exp(t \mathrm{Ad}(g)(X))$, see \cite[Theorem 1.7.16; page 59]{hamilton}.}

That we have $\mathrm{Ad}\mleft(g\mright)(\mathfrak{g}_p) = \mathfrak{g}_{gp}$ simply comes from the fact that everything is finite-dimensional, so, $\mathrm{Ad}\mleft(g\mright)(\mathfrak{g}_p)$ is a finite-dimensional subspace of $\mathfrak{g}_{gp}$, and by the Lie group isomorphism in Eq.~\eqref{isotropygrouprelation} we have $\mathrm{dim}(\mathfrak{g}_p) = \mathrm{dim}(\mathfrak{g}_{gp})$. Thus, $\mathrm{Ad}\mleft(g\mright)(\mathfrak{g}_p) = \mathfrak{g}_{gp}$ follows, and that describes a Lie algebra automorphism $\mathfrak{g}_p \cong \mathfrak{g}_{gp}$ because $\mathrm{Ad}\mleft(g\mright)$ is a Lie algebra automorphism.
%
%Sketchy: By "previous statement" about Lie group actions, $g \in G$ and $p \in N$, $G_{gp} = \mathrm{Ad}(g)(G_p)$, where $G_p \coloneqq \left\{ q \in G ~ \middle| ~ qp=p \right\}$ is the isotropy group at $p$
%Just a sketch at the moment:
%We also have, for $g \in G$ and $p \in N$, $G_{gp} = \mathrm{Ad}(g)(G_p)$, where $G_p \coloneqq \left\{ q \in G ~ \middle| ~ qp=p \right\}$ is the isotropy group at $p$; (see ... add reference later)
%\bas
%\forall q \in G: ~
%K_p
%&=
%K_{qp},
%\eas
%\textit{i.e.}~infinitesimally for $v \in \mathfrak{g}$
%\bas
%&&&\forall t \in \mathbb{R}: ~
%K_p
%=
%\mathrm{Ad}(\exp(tv))(K_p) \\
%&\Rightarrow&
%&\forall w \in K_p: ~
%\mathrm{Ad}(\exp(tv))(w)
%\in K_p \\
%&\stackrel{\mathclap{K_p \text{ closed subalgebra of } \mathfrak{g}}}{\Rightarrow}&
%&\forall v \in \mathfrak{g}: ~ \forall w \in K_p: ~
%\mleft[ v, w \mright]_{\mathfrak{g}}
%\in K_p,
%\eas
\end{proof}

For the last statement we needed integrability. One may assume that isotropy subalgebras are in general ideals of the Lie algebra $\mathfrak{g}$ due to that result, by using that the induced Lie algebra representation of $\mathrm{Ad}$ is given by $\mathrm{ad}$. But the isotropy subalgebra is in general not an ideal, \textit{i.e.}~we have in general \textbf{not} $\mleft[ X, Y \mright]_{\mathfrak{g}} \in \mathfrak{g}_p$ for all $p \in N$, $X \in \mathfrak{g}_p$ and $Y \in \mathfrak{g}$. Given those, fix local coordinates $\mleft(\partial_i\mright)_i$ on $N$ around $p$ and a $\mathfrak{g}$-action $\gamma$ on $N$, then 
\bas
\gamma\mleft( \mleft[ X, Y \mright]_{\mathfrak{g}} \mright)_p
&=
\mleft.\mleft[ \gamma(X), \gamma(Y) \mright]\mright|_p
\\
&=
\biggl(
	\underbrace{\mleft.\mathcal{L}_{\gamma(X)}\mright|_p}_{=0}\mleft(\gamma^i(Y)\mright) 
	- \mleft.\mathcal{L}_{\gamma(Y)}\mright|_p\mleft(\gamma^i(X)\mright) 
\biggr) ~\partial_i
\\
&=
- \mleft.\mathcal{L}_{\gamma(Y)}\mright|_p\mleft(\gamma^i(X)\mright) ~\partial_i
\eas
for all $p \in N$, $X \in \mathfrak{g}_p$ and $Y \in \mathfrak{g}$, where we locally write $\gamma = \gamma^i ~ \partial_i$.
Therefore $\mathfrak{g}_p$ would be an ideal, if there is a coordinate system such that $\gamma^i(X)$ are constant along $\gamma$ around $p$; we will come back to this condition about constancy in another chapter. However, we will later see that the isotropy subalgebra is always an ideal of another Lie bracket, the bracket of a vector bundle which we will call a Lie algebroid. But let us now first shortly introduce the physical quantities.

\section{Yang-Mills-Higgs gauge theory}\label{YMHGT}

As introduced, we will only assume trivial principal bundles. Hence, the \textbf{field of gauge bosons} will be represented by an element $\gls{a0} \in \Omega^1(M; \mathfrak{g})$, where $\mathfrak{g}$ is a Lie algebra and $M$ is usually a spacetime (but often just a smooth manifold in the following).

We also need the following definition.

\begin{definitions}{Graded extension of the Lie bracket, \newline \cite[generalization of Definition 5.5.3; page 275]{hamilton}}{GradedExtensionOfBracket}
Let $M$ be a smooth manifold, $W$ and $W^\prime$ vector spaces and $F \in \bigwedge^2 W^* \otimes W^\prime$. Then for $\omega \in \Omega^k(M; W)$ and $\eta \in \Omega^l(M; W)$ ($k, l \in \mathbb{N}_0$) we define $F\mleft(\omega \stackrel{\wedge}{,} \eta\mright)$ as an element of $\Omega^{k+l}(M; W^\prime)$ by
\ba
&\bigl(F\mleft(\omega \stackrel{\wedge}{,} \eta\mright)\bigr)\mleft( X_1, \dotsc, X_{k+l} \mright)
\nonumber\\
&\coloneqq
\frac{1}{k!l!} \sum_{\sigma \in S_{k+l}}
	\mathrm{sgn}(\sigma) F\mleft( 
	\omega\mleft(X_{\sigma(1)}, \dotsc, X_{\sigma(k)}\mright),
	\eta\mleft(X_{\sigma(k+1)}, \dotsc, X_{\sigma(k+l)}\mright)
	\mright)
\ea
for all $X_1, \dotsc, X_{k+l} \in \mathfrak{X}(M)$, where $S_{k+l}$ is the group of permutations of $\{1, \dotsc, k+l\}$.

When either $\omega$ or $\eta$ is a zero-form, then we may also write $F(w, \eta)$ instead.
\end{definitions}

\begin{remark}
\leavevmode\newline
It is easy to check that $F\mleft(\omega \stackrel{\wedge}{,} \eta\mright)$ is well-defined, \textit{i.e.}~that it is an element of $\Omega^{k+l}(M; W^\prime)$ by construction. 

For $W = \mathfrak{g}$ and $F = \mleft[ \cdot, \cdot \mright]_{\mathfrak{g}}$ observe that we have for $A \in \Omega^1(M;\mathfrak{g})$
\bas
\mleft[ A \stackrel{\wedge}{,} A \mright]_{\mathfrak{g}}(X, Y)
&\coloneqq
F\mleft(A \stackrel{\wedge}{,} A\mright) (X, Y)
=
\mleft[ A(X), A(Y) \mright]_{\mathfrak{g}}
	- \mleft[ A(Y), A(X) \mright]_{\mathfrak{g}}
=
2 ~ \mleft[ A(X), A(Y) \mright]_{\mathfrak{g}}
\eas
for all $X, Y \in \mathfrak{X}(M)$. Making use of the structure constants $C^c_{ab}$ with respect to a given basis $\mleft( e_a \mright)_a$ of $\mathfrak{g}$, we can also write
\ba\label{LokaleFormVonAwedgeAClassical}
\mleft[ A \stackrel{\wedge}{,} A \mright]_{\mathfrak{g}}
&=
A^a \wedge A^b \otimes \mleft[ e_a, e_b \mright]_{\mathfrak{g}}
=
A^a \wedge A^b \otimes C^c_{ab} e_c.
\ea
\end{remark}

Let us now define the field strength.

\begin{definitions}{Field strength, \cite[Theorem 5.5.4; page 275]{hamilton}}{ClassicFieldStrength}
Let $\mathfrak{g}$ be a Lie algebra and $M$ a smooth manifold. The \textbf{field strength $\gls{F}(A)$ of $A \in \Omega^1(M; \mathfrak{g})$} is defined by
\ba
F(A)
&\coloneqq
\mathrm{d}A
	+ \frac{1}{2} \mleft[ A \stackrel{\wedge}{,} A \mright]_{\mathfrak{g}}.
\ea
We view the field strength also as a map $F: \Omega^1(M; \mathfrak{g}) \to \Omega^2(M; \mathfrak{g}), A \mapsto F(A)$.
\end{definitions}

The field strength satisfies the Bianchi Identity, encoding the homogeneous Maxwell equations in the case of electromagnetism.

\begin{theorems}{Bianchi identity of the field strength, \newline \cite[Theorem 5.14.2; page 311]{hamilton}}{ClassicBianchiIdenityOfFieldstrength}
Let $\mathfrak{g}$ be a Lie algebra and $M$ a smooth manifold. Then the field strength $F$ satisfies the \textbf{Bianchi Identity}
\ba
\mathrm{d}\bigl(F(A) \bigr)
	+ \mleft[ A \stackrel{\wedge}{,} F(A) \mright]_{\mathfrak{g}}
&=
0
\ea
for all $A \in \Omega^1(M; \mathfrak{g})$.
\end{theorems}

\begin{remark}
\leavevmode\newline
See the reference for a proof for now. We will later prove a more general Bianchi identity which will recover this statement; see Thm.~\ref{thm:BianchiIdentityOfFieldStrength}.
\end{remark}

Let us now define the needed Lagrangians; we are going to state later the typical conditions for gauge invariance, which is why we do not yet clarify any invariance of the used scalar products in the following.

\begin{definitions}{Yang-Mills Lagrangian, \cite[Definition 7.3.1; page 414]{hamilton}}{ClassicYMLagrangian}
Let $\mathfrak{g}$ be a Lie algebra, equipped with a scalar product $\kappa$, and $M$ a spacetime with spacetime metric $\eta$. Then we define the \textbf{Yang-Mills Lagrangian $\mathfrak{L}_{\mathrm{YM}}$} as a map $\Omega^1(M; \mathfrak{g}) \to \Omega^{\mathrm{dim}(M)}(M)$ by 
\ba
\mathfrak{L}_{\mathrm{YM}}(A)
&\coloneqq
-\frac{1}{2} ~ \kappa\bigl( F(A) \stackrel{\wedge}{,} * F(A) \bigr)
\ea
for all $A \in \Omega^1(M; \mathfrak{g})$, where $*$ is the Hodge star operator with respect to $\eta$.\footnote{As a reference, see for example \cite[Definition 7.2.4; page 408]{hamilton}.}
\end{definitions}

We also want to look at the Higgs field. The Higgs field is a map $\gls{1vhi} \in C^\infty(M;W)$, where $W$ is some vector space, and the field of gauge bosons $A$ are coupled to fields like the Higgs field via the minimal coupling.

\begin{definitions}{Minimal coupling, \newline \cite[Definition 5.9.3; page 292; Definition 7.5.5 \textit{et seq.}; page 426]{hamilton}}{ClassicMinimalCoupling}
Let $\mathfrak{g}$ be a Lie algebra, $M$ a smooth manifold, and $W$ a vector space. Furthermore, let $\psi: \mathfrak{g} \to \mathrm{End}(W)$ be a $\mathfrak{g}$-representation on $W$. Then we define the \textbf{minimal coupling $\mathfrak{D}$} as a map given by
\ba
C^\infty(M; W) \times \Omega^1(M; \mathfrak{g})
&\to
\Omega^1(M; W),
\nonumber \\
(\Phi, A)
&\mapsto
\mathfrak{D}(\Phi,A)
\coloneqq
\gls{DAPhi}
=
\mathrm{d}\Phi
	+ \psi(A)(\Phi),
\ea
where $\psi(A)(\Phi)$ is an element of $\Omega^1(M; W)$ given by
\bas
\bigl(\psi(A)(\Phi)\bigr)_p(Y)
=
\psi\bigl(A_p(Y)\bigr)\bigl(\Phi(p)\bigr)
\eas
for all $p \in M$ and $Y \in \mathrm{T}_pM$.
\end{definitions}

\begin{remark}
\leavevmode\newline
In \cite{hamilton} and other literature, minimal coupling also often just refers to the term $\psi(A)(\Psi)$.
\end{remark}

With that we can now define the Yang-Mills-Higgs Lagrangian.

\begin{definitions}{Yang-Mills-Higgs Lagrangian, \cite[Definition 8.1.1; page 446f.]{hamilton}}{ClassicYMHLagrangian}
Let $\mathfrak{g}$ be a Lie algebra, equipped with a scalar product $\kappa$, $M$ a spacetime with spacetime metric $\eta$, and $W$ a vector space, also equipped with a scalar product $g$. Furthermore, let $V \in C^\infty(W)$, the \textbf{potential of the Higgs field}, and $\psi: \mathfrak{g} \to \mathrm{End}(W)$ be a $\mathfrak{g}$-representation on $W$.
Then we define the \textbf{Yang-Mills-Higgs Lagrangian $\gls{LYMH}$} as a map $C^\infty(M; W) \times \Omega^1(M; \mathfrak{g}) \to \Omega^{\mathrm{dim}(M)}(M)$ by 
\ba
\mathfrak{L}_{\mathrm{YMH}}(\Phi, A)
&\coloneqq
-\frac{1}{2} ~ \kappa\bigl( F(A) \stackrel{\wedge}{,} * F(A) \bigr)
	+ g\mleft( \mathfrak{D}^A\Phi \stackrel{\wedge}{,} * \mathfrak{D}^A\Phi \mright)
	- *\bigl( V \circ \Phi \bigr)
\ea
for all $(\Phi, A) \in C^\infty(M; W) \times \Omega^1(M; \mathfrak{g})$, where $*$ is the Hodge star operator with respect to $\eta$.
\end{definitions}

The Higgs mechanism is needed for allowing masses of gauge bosons while keeping gauge invariance. We will not introduce and discuss this because it would exceed the scope of this thesis and it is already elaborated elsewhere, see for example \cite[\S 8; page 445ff.]{hamilton}. However, let us summarize the Higgs effect: The essential idea and result is that the components of $A$ along the isotropy subalgebras $\mathfrak{g}_p$ ($p \in W$) describe the massless gauge bosons, while the other components may describe the bosons with masses due to a non-trivial minimal coupling. That is, fix a point $p \in W$, take a basis $\mleft( f_\alpha \mright)_\alpha$ of $\mathfrak{g}_p$, and extend that basis to a basis of $\mathfrak{g}$, denoted by $\mleft( e_a \mright)_a$. Then write $A = A^a \otimes e_a$ and define $A_{\mathrm{iso}} \coloneqq A^\alpha \otimes f_\alpha$, and denote with $\gamma$ the Lie algebra action induced by $\psi$ as in Prop.~\ref{prop:LieRepAndLieAct}, such that
\bas
\gamma\bigl(A_{\mathrm{iso}}|_p(Y)\bigr)_p
&=
A_{\mathrm{iso}}^\alpha|_p(Y) \otimes \underbrace{\gamma(f_\alpha)_p}_{=0}
=
0
\eas
for all $p \in U$ and $Y \in \mathrm{T}_pM$. It is possible to extend that argument to certain open subsets of $W$, leading to that $A_{\mathrm{iso}}$ has a trivial (=0) coupling to any $\Phi$ such that $A_{\mathrm{iso}}$ is going to describe the massless gauge bosons like the photon and the gluons. While the remaining components of $A$ may be massive. Thus, in order to allow masses of gauge boson, one needs that the isotropy subalgebras are non-trivial subalgebras of $\mathfrak{g}$ at certain subsets of $W$ (especially around the minimum of the potential $V$). That is called \textbf{symmetry breaking}.

However, that is not the only factor needed, on one hand one needs a special form of the potential, and on the other hand there is also the known \textbf{unitary gauge} which essentially fixes the components of the Higgs field along the orbits of $\psi$ such that the gauge bosons only really couple to the components along the transversal structure. The components of the Higgs field along the orbits of $\psi$ generally describe the Nambu-Goldstone bosons, while the transversal components are the actual Higgs bosons. Therefore we would not have a Higgs effect without a transversal structure, and, thus, no masses of gauge bosons.

As mentioned, we will not prove or introduce anything of this in detail; see the given reference for an elaborated discussion. But after we will have introduced the generalized and new gauge theory, using Lie algebroids, we will very shortly revisit this behaviour, and it will be easier to formulate due to the fact that the new formulation supports Lie algebra bundles and vector bundles known as action Lie algebroids.

\section{Infinitesimal Gauge Invariance}\label{InfGaugeTrafoClassical}
Let us now turn to gauge invariance. We will only focus on its infinitesimal formulation because the generalized gauge theory we want to go to will not assume integrability in general. We will still follow \cite[especially \S 5; page 257ff.]{hamilton}, while we first give the observed space of fields in order to make following notations more compact.

\begin{definitions}{The space of fields}{ClassicSpaceofFieldsAgain}
Let $M$ be a smooth manifold, $W$ a vector space, and $\mathfrak{g}$ a Lie algebra. Then we define the \textbf{space of fields} by
\ba
\mathfrak{M}_{\mathfrak{g}}(M; W)
&\coloneqq
\left\{ (\Phi, A)
~\middle|~
\Phi \in C^\infty(M;W) \text{ and } A \in \Omega^1(M; \mathfrak{g})
\right\}.
\ea
\end{definitions}

\begin{definitions}{Infinitesimal gauge transformation of the Higgs field and the field of gauge bosons, \newline \cite[infinitesimal version of Theorem 5.3.9, see also comment afterwards; page 269f.]{hamilton} and \cite[infinitesimal version of Theorem 5.4.4; page 273]{hamilton}}{ClassicTrafos}
Let $M$ be a smooth manifold, $W$ a vector space, and $\mathfrak{g}$ a Lie algebra, equipped with a Lie algebra representation $\psi: \mathfrak{g} \to \mathrm{End}(W)$. Moreover, let $\varepsilon \in C^\infty(M; \mathfrak{g})$.

Then we define the \textbf{infinitesimal gauge transformation $\delta_\varepsilon \Phi$ of the Higgs field $\Phi \in C^\infty(M;W)$} also as an element of $C^\infty(M; W)$ by
\ba
\delta_\varepsilon \Phi
&\coloneqq
\psi(\varepsilon)(\Phi).
\ea
The \textbf{infinitesimal gauge transformation $\delta_\varepsilon A$ of the field of gauge bosons $A \in \Omega^1(M; \mathfrak{g})$} is defined as an element of $\Omega^1(M; \mathfrak{g})$ by
\ba
\delta_\varepsilon A
&\coloneqq
\mleft[ \varepsilon, A \mright]_{\mathfrak{g}}
	- \mathrm{d}\varepsilon.
\ea
\end{definitions}

With that one can define the infinitesimal gauge transformation of functionals.

\begin{definitions}{Infinitesimal gauge transformation of functionals, \newline \cite[motivated by statements like Theorem 7.3.2; page 414ff.]{hamilton}}{ClassFunctionalGaugeTrafoBlag}
Let $M$ be a smooth manifold, $W, K$ vector spaces, and $\mathfrak{g}$ a Lie algebra, equipped with a Lie algebra representation $\psi: \mathfrak{g} \to \mathrm{End}(W)$. Moreover, let $\varepsilon \in C^\infty(M; \mathfrak{g})$.

Then we define the \textbf{infinitesimal gauge transformation $\delta_\varepsilon L$ of $L: \mathfrak{M}_{\mathfrak{g}}(M; W) \to \Omega^k(M; K)$ ($k \in \mathbb{N}_0$)} as a map $\mathfrak{M}_{\mathfrak{g}}(M; W) \to \Omega^k(M;K)$ by
\ba
\mleft(\delta_\varepsilon L\mright)(\Phi, A)
&\coloneqq
\mleft.\frac{\mathrm{d}}{\mathrm{d}t}\mright|_{t=0}
\mleft[ t \mapsto
	L\mleft(
		\Phi + t \delta_\varepsilon \Phi,
		A + t \delta_\varepsilon A
	\mright)
\mright]
\ea
for $t \in \mathbb{R}$, where $\mathrm{d}/\mathrm{d}t$ is defined using the canonical flat connection on $M \times K \to M$.
\end{definitions}

\begin{remark}\label{RemabouttheddtOfClassicTrafos}
\leavevmode\newline
This definition leads to $(\delta_\varepsilon L)(\Phi,A) \in \Omega^k(M;K)$, because the vector space $W$ is viewed as a trivial vector bundle over $M$ such that one uses the canonical flat connection for the definition of $\mathrm{d}/\mathrm{d}t$, that is, one fixes a global trivialization, and then differentiates the components with respect to that trivialization separately. Thus, one actually uses a very trivial horizontal projection in that definition.

This definition is basically nothing else than a differential of functionals along the direction given by $(\delta_\varepsilon \Phi, \delta_\varepsilon A)$. But we want to keep it as presented in order to emphasize something later.
\end{remark}

One then calculates the typical formulas of the infinitesimal gauge transformations of the field strength and minimal coupling

\begin{propositions}{Infinitesimal gauge transformations of the field strength and minimal coupling, \newline \cite[infinitesimal version of Theorem 5.6.3; page 280]{hamilton} and \cite[infinitesimal version of Lemma 7.5.8; page 428]{hamilton}}{ClassicGaugeTrafoOfFieldStrengthAndMinimalCoupling}
Let $M$ be a smooth manifold, $W$ a vector space, and $\mathfrak{g}$ a Lie algebra, equipped with a Lie algebra representation $\psi: \mathfrak{g} \to \mathrm{End}(W)$. Moreover, let $\varepsilon \in C^\infty(M; \mathfrak{g})$.

Then we have
\ba
\mleft(\delta_\varepsilon F\mright)(\Phi, A)
&=
\mleft[ \varepsilon, F(A) \mright]_{\mathfrak{g}},
\\
\mleft(\delta_\varepsilon \mathfrak{D}\mright)(\Phi, A)
&=
\psi(\varepsilon)\mleft( \mathfrak{D}^A \Phi \mright)
\ea
for all $(\Phi, A) \in \mathfrak{M}_{\mathfrak{g}}(M; W)$.
\end{propositions}

\begin{remark}
\leavevmode\newline
The infinitesimal gauge transformation of $A$ can also motivated by conditioning that the gauge transformation of the minimal coupling has to look like as in this proposition. We will discuss this later in more detail in the general setting.
\end{remark}

\begin{proof}[Proof of Prop.~\ref{prop:ClassicGaugeTrafoOfFieldStrengthAndMinimalCoupling}]
\leavevmode\newline
We get\footnote{$F$ is independent of $\Phi$, so, one can omit it there.}
\bas
\mleft(\delta_\varepsilon F\mright)(A)
&=
\mleft.\frac{\mathrm{d}}{\mathrm{d}t}\mright|_{t=0}
\mleft[ t \mapsto	
	F\mleft(
		A + t \delta_\varepsilon A
	\mright)
\mright]
\\
&=
\mleft.\frac{\mathrm{d}}{\mathrm{d}t}\mright|_{t=0}
\mleft[ t \mapsto	
	\mathrm{d}\mleft(A + t \delta_\varepsilon A\mright)
	+ \frac{1}{2} \mleft[ A + t \delta_\varepsilon A \stackrel{\wedge}{,} A + t \delta_\varepsilon A \mright]_{\mathfrak{g}}
\mright]
\\
&=
\mathrm{d} \underbrace{\delta_\varepsilon A}
_{\mathclap{ = \mleft[ \varepsilon, A \mright]_{\mathfrak{g}} - \mathrm{d}\varepsilon }}
	+ \frac{1}{2} \mleft[ \delta_\varepsilon A \stackrel{\wedge}{,} A \mright]_{\mathfrak{g}}
	+ \frac{1}{2} \mleft[ A \stackrel{\wedge}{,} \delta_\varepsilon A \mright]_{\mathfrak{g}}
\\
&=
\mleft[ \mathrm{d}\varepsilon, A \mright]_{\mathfrak{g}}
	+ \mleft[ \varepsilon, \mathrm{d}A \mright]_{\mathfrak{g}}
	+ \mleft[ \mleft[ \varepsilon, A \mright]_{\mathfrak{g}} - \mathrm{d}\varepsilon \stackrel{\wedge}{,} A \mright]_{\mathfrak{g}}
\\
&=
	\mleft[ \varepsilon, \mathrm{d}A \mright]_{\mathfrak{g}}
	+ \mleft[ \mleft[ \varepsilon, A \mright]_{\mathfrak{g}} \stackrel{\wedge}{,} A \mright]_{\mathfrak{g}}
\eas
making use of Eq.~\eqref{LokaleFormVonAwedgeAClassical} which implies that we have a product rule with respect to the two arguments in $\mleft[ \cdot \stackrel{\wedge}{,} \cdot \mright]_{\mathfrak{g}}$ in sense of wedge products and the differential, and we clearly have $\mleft[ \omega \stackrel{\wedge}{,} \eta \mright]_{\mathfrak{g}} = \mleft[ \eta \stackrel{\wedge}{,} \omega \mright]_{\mathfrak{g}}$ for all $\omega, \eta \in \Omega^1(M; \mathfrak{g})$ due to the antisymmetry of the Lie bracket; see also Appendix \ref{CalculusIdentitiesNeeded} for their proof (as slightly generalized versions). Again using Eq.~\eqref{LokaleFormVonAwedgeAClassical}, the Jacobi identity of the Lie bracket and a basis $\mleft( e_a \mright)_a$ of $\mathfrak{g}$, we arrive
\bas
&&
\mleft[ \mleft[ \varepsilon, A \mright]_{\mathfrak{g}} \stackrel{\wedge}{,} A \mright]_{\mathfrak{g}}
&=
\varepsilon^a A^b \wedge A^c \otimes \mleft[ \mleft[ e_a, e_b \mright]_{\mathfrak{g}}, e_c \mright]_{\mathfrak{g}}
\\
&&
&=
\varepsilon^a A^b \wedge A^c \otimes \mleft(
	\mleft[ e_a, \mleft[ e_b, e_c \mright]_{\mathfrak{g}} \mright]_{\mathfrak{g}}
	+ \mleft[ \mleft[ e_a, e_c \mright]_{\mathfrak{g}}, e_b \mright]_{\mathfrak{g}}
\mright)
\\
&&
&=
\mleft[ \varepsilon, \mleft[ A \stackrel{\wedge}{,} A \mright]_{\mathfrak{g}} \mright]_{\mathfrak{g}}
	- \mleft[ \mleft[ \varepsilon, A \mright]_{\mathfrak{g}} \stackrel{\wedge}{,} A \mright]_{\mathfrak{g}}
\\
&\Leftrightarrow&
\mleft[ \mleft[ \varepsilon, A \mright]_{\mathfrak{g}} \stackrel{\wedge}{,} A \mright]_{\mathfrak{g}}
&=
\frac{1}{2} \mleft[ \varepsilon, \mleft[ A \stackrel{\wedge}{,} A \mright]_{\mathfrak{g}} \mright]_{\mathfrak{g}},
\eas
hence, 
\bas
\mleft(\delta_\varepsilon F\mright)(A)
&=
\mleft[ \varepsilon, \mathrm{d}A + \frac{1}{2} \mleft[ A \stackrel{\wedge}{,} A \mright]_{\mathfrak{g}} \mright]_{\mathfrak{g}}
=
\mleft[ \varepsilon, F(A) \mright]_{\mathfrak{g}}.
\eas

For the minimal coupling observe, also now using additionally a basis $\mleft( f_\alpha \mright)_\alpha$ of $W$,
\bas
\mathrm{d}\delta_\varepsilon \Phi
&=
\mathrm{d}\bigl( \psi(\varepsilon) (\Phi) \bigr)
\\
&=
\mathrm{d}\bigl( \varepsilon^a \Phi^\alpha \underbrace{\psi(e_a) (f_\alpha)}_{\in \mathfrak{g}} \bigr)
\\
&=
\mathrm{d}\varepsilon^a ~ \Phi^\alpha \psi(e_a) (f_\alpha)
	+ \varepsilon^a \mathrm{d}\Phi^\alpha \psi(e_a) (f_\alpha)
\\
&=
\psi(\mathrm{d}\varepsilon) (\Phi)
	+ \psi(\varepsilon) (\mathrm{d}\Phi),
\eas
and, thus,
\bas
\mleft(\delta_\varepsilon \mathfrak{D}\mright)(\Phi, A)
&=
\mleft.\frac{\mathrm{d}}{\mathrm{d}t}\mright|_{t=0}
\mleft[ t \mapsto	
	\mathrm{d}\mleft( \Phi + t \delta_\varepsilon \Phi \mright)
	+ \psi\mleft( A + t \delta_\varepsilon A \mright) \mleft( \Phi + t \delta_\varepsilon \Phi \mright)
\mright]
\\
&=
\mathrm{d}\delta_\varepsilon \Phi
	+ \psi\mleft( \delta_\varepsilon A \mright) \mleft( \Phi \mright)
	+ \psi\mleft( A \mright) \mleft( \delta_\varepsilon \Phi \mright)
\\
&=
\psi(\mathrm{d}\varepsilon) (\Phi)
	+ \psi(\varepsilon) (\mathrm{d}\Phi)
	+ \psi\mleft( \mleft[ \varepsilon, A \mright]_{\mathfrak{g}} - \mathrm{d}\varepsilon \mright) \mleft( \Phi \mright)
	+ \psi\mleft( A \mright) \mleft( \psi(\varepsilon) (\Phi) \mright)
\\
&=
\psi(\varepsilon) (\mathrm{d}\Phi)
	+ \underbrace{\mleft[ \psi(\varepsilon), \psi(A) \mright]_{\mathfrak{g}}
	+ \psi\mleft( A \mright) \mleft( \psi(\varepsilon) (\Phi) \mright)}
	_{= \psi\mleft( \varepsilon \mright) \mleft( \psi(A) (\Phi) \mright)}
\\
&=
\psi(\varepsilon)\mleft( \mathfrak{D}^A \Phi \mright),
\eas
where we used that $\psi$ is a homomorphism of Lie brackets.
\end{proof}

That leads to the typical well-known statement about the infinitesimal gauge invariance of the Yang-Mills-Higgs Lagrangian. For that we shortly recall what it means that a scalar product is invariant under a Lie algebra representation.

\begin{definitions}{Scalar products invariant under Lie algebra representations, \newline \cite[Definition 2.1.36; page 96]{hamilton}}{ClassicInvariance of metrics}
Let $\mathfrak{g}$ be a Lie algebra, $W$ a vector space and $\psi: \mathfrak{g} \to \mathrm{End}(W)$ a $\mathfrak{g}$-representation on $W$. Then we say that a scalar product $g$ on $W$ is \textbf{$\psi$-invariant}
\ba
g\mleft( \psi(X)(v), w \mright)
	+ g\mleft( v, \psi(X)(w) \mright)
&=
0
\ea
for all $X \in \mathfrak{g}$ and $v, w \in W$.
\end{definitions}

\begin{theorems}{Infinitesimal gauge invariance of the Yang-Mills-Higgs Lagrangian, \newline \cite[infinitesimal version of Theorem 7.3.2; page 414]{hamilton} and \cite[infinitesimal version of Theorem 7.5.10; page 429]{hamilton}}{ClassicGaugeInvarOfYMHLagrangians}
Let $\mathfrak{g}$ be a Lie algebra, equipped with a scalar product $\kappa$, $M$ a spacetime with spacetime metric $\eta$, and $W$ a vector space, also equipped with a scalar product $g$. Furthermore, let $V \in C^\infty(W)$ and $\psi: \mathfrak{g} \to \mathrm{End}(W)$ be a $\mathfrak{g}$-representation on $W$, whose induced Lie algebra action is denoted by $\gamma$. If we have
\ba
\kappa &\text{ is $\mathrm{ad}$-invariant},
\\
g &\text{ is $\psi$-invariant},
\\
0
&=
\mathcal{L}_{\gamma(\varepsilon)} V \circ \Phi\label{ClassicPotential}
\ea
for all $\varepsilon \in C^\infty(M; \mathfrak{g})$ and $\Phi \in C^\infty(M;W)$, then
\ba
\delta_\varepsilon \mathfrak{L}_{\mathrm{YMH}}
&=
0
\ea
for all $\varepsilon \in C^\infty(M; \mathfrak{g})$.
\end{theorems}

\begin{remark}
\leavevmode\newline
Condition \eqref{ClassicPotential} may be reduced to $\mathcal{L}_{\gamma(\varepsilon)}V = 0$; however, we will not discuss the potential, and that "weaker" formulation may be a good starting point if one wants to restrict the set of $\Phi$.
\end{remark}

\begin{proof}[Proof of Thm.~\ref{thm:ClassicGaugeInvarOfYMHLagrangians}]
\leavevmode\newline
We will prove the more general statement in more detail later, see Thm.~\ref{thm:GaugeInvariantStandardLagrangian}, but it is a trivial consequence of Prop.~\ref{prop:ClassicGaugeTrafoOfFieldStrengthAndMinimalCoupling}: We need to calculate
\bas
\mleft.\frac{\mathrm{d}}{\mathrm{d}t}\mright|_{t=0}
\mleft[
	\mathbb{R} \ni t \mapsto
	\mathfrak{L}_{\mathrm{YMH}}
	\mleft(
		\Phi + t \delta_\varepsilon \Phi,
		A + t \delta_\varepsilon A
	\mright)
\mright]
\eas
and we can do that on each summand in Def.~\ref{def:ClassicYMHLagrangian} separately. Applying the product rule when calculating $\frac{\mathrm{d}}{\mathrm{d}t}$ and using Prop.~\ref{prop:ClassicGaugeTrafoOfFieldStrengthAndMinimalCoupling}, it is clear that the first two summands, the Yang-Mills Lagrangian and the kinetic part of the Higgs field, vanish because of the imposed invariances on $\kappa$ and $g$. For the potential $V$ observe
\bas
\mleft.\mleft(\mleft.\frac{\mathrm{d}}{\mathrm{d}t}\mright|_{t=0}
\mleft[
	t \mapsto
	V
	\mleft(
		\Phi + t \delta_\varepsilon \Phi
	\mright)
\mright]\mright)\mright|_p
&=
\mleft(\mathrm{d}_{\Phi(p)}V\mright)\mleft( \psi\bigl(\varepsilon(p)\bigr)\bigl(\Phi(p)\bigr) \mright)
\stackrel{\text{ Prop.~\ref{prop:LieRepAndLieAct} } }{=}
- \mleft.\mathcal{L}_{\gamma\mleft(\epsilon(p)\mright)} V \mright|_{\Phi(p)},
\eas
which is also zero by the assumed condition on the potential. Hence, the infinitesimal gauge transformation of all three summands of the Yang-Mills-Higgs Lagrangian is zero.\footnote{The Hodge star operator can be ignored because the spacetime metric is independent of the fields $\Phi$ and $A$.}
\end{proof}

\begin{remark}
\leavevmode\newline
In \cite{hamilton} one assumes a function $\widetilde{V} \in C^\infty(\mathbb{R})$ instead of the general potential we took. There the potential is then given by $V (w) \coloneqq \widetilde{V} \bigl( g(w,w) \bigr)$ for all $w \in W$, \textit{e.g.}~$\widetilde{V}$ is a polynomial of the scalar product on $W$. Due to the $\psi$-invariance of $g$ we get
\bas
\mleft.\mleft(\mleft.\frac{\mathrm{d}}{\mathrm{d}t}\mright|_{t=0}
\mleft[
	t \mapsto
	V
	\mleft(
		\Phi + t \delta_\varepsilon \Phi
	\mright)
\mright]\mright)\mright|_p
&=
\mathrm{D}_{g(\Phi(p), \Phi(p))}\widetilde{V}\biggl(  
	g\Bigl( \psi(\varepsilon)(\Phi)|_p, \Phi(p) \Bigr)
	+ g\Bigl( \Phi(p), \psi(\varepsilon)(\Phi)|_p \Bigr)
\biggr)
\\
&=
0
\eas
for all $\Phi \in C^\infty(M;W)$, $\varepsilon \in C^\infty(M; \mathfrak{g})$ and $p \in M$. In the proof we also have seen
\bas
\mleft.\mleft(\mleft.\frac{\mathrm{d}}{\mathrm{d}t}\mright|_{t=0}
\mleft[
	t \mapsto
	V
	\mleft(
		\Phi + t \delta_\varepsilon \Phi
	\mright)
\mright]\mright)\mright|_p
&=
- \mleft.\mathcal{L}_{\gamma\mleft(\epsilon(p)\mright)} V \mright|_{\Phi(p)},
\eas
thus, Eq.~\eqref{ClassicPotential} is satisfied for such potentials. See \cite[\S 8; especially also the box at the top of page 450]{hamilton} for a thorough discussion about how the potential looks like for Yang-Mills-Higgs Lagrangians; in this work the potential will not play any important role, and besides conditions like Eq.~\eqref{ClassicPotential} it is not going to appear anywhere here.
\end{remark}

\section{Infinitesimal Gauge Invariance using connections}\label{NewInfGaugeTrafoTrafos}

We want to introduce and redefine infinitesimal gauge invariance in a different way now, already pointing out what the next sections will be about. Therefore this section also serves as a first step towards Lie algebroids and the new gauge theory. As we have seen, the common idea is to interpret infinitesimal gauge transformations as derivations of functionals, parametrised by Lie algebra valued functions $\varepsilon$. 

In this section we want to show that the infinitesimal gauge transformations can be viewed as a "connection-like" object on the infinite-dimensional spaces arising in the calculus of variations, but the connection will be inherited by a connection of a finite-dimensional vector bundle. Before we discuss this, let us introduce the connections we look at in the finite-dimensional situation; those will be a first step towards a generalization of typical vector bundle connections. In some sense, those are like Lie algebra actions, but as connections instead of a Lie derivative along a vector field.

\begin{definitions}{Lie algebra connection, \newline \cite[special situation of \S2, Definition 2.2]{basicconn}}{FirstStepLieDerivativeOfAnchors}
Let $\mathfrak{g}$ be a Lie algebra, and $\gamma: \mathfrak{g} \to \mathfrak{X}(N)$ be a Lie algebra action on a smooth manifold $N$. Then a \textbf{$\mathfrak{g}$-connection} on a vector bundle $E \to N$ is an $\mathbb{R}$-bilinear map ${}^\mathfrak{g}\nabla$
\bas
\mathfrak{g} \times \Gamma(E) &\to \Gamma(E), 
\\
(X, \nu) &\mapsto {}^\mathfrak{g}\nabla_X \nu,
\eas
satisfying
\ba\label{FirstStepToEDerivatives}
{}^\mathfrak{g}\nabla_X (f \nu)
&=
f ~ {}^\mathfrak{g}\nabla_X \nu
	+ \mathcal{L}_{\gamma(X)}(f) ~ \nu
\ea
for all $X \in \mathfrak{g}, \nu \in \Gamma(E)$ and $f \in C^\infty(N)$, where $\mathcal{L}_{\gamma(X)}(f)$ is the action of the vector field $\gamma(X)$ on the function $f$ by derivation.
\end{definitions}

\begin{remark}\label{DifferenceOfLieAlgConnections}
\leavevmode\newline
Similar to typical vector bundle connections, the Leibniz rule in the difference of two $\mathfrak{g}$-connections will cancel each other, resulting into an $\mathbb{R}$-linear map $\mathfrak{g} \to \sEnd(E)$; this is trivial to check. 

It is on purpose that there is no separate imposed $C^\infty(N)$-linearity in the $\mathfrak{g}$-argument, it is then in more alignment with the definition of $\mathfrak{g}$-actions. However, that is quickly recovered by defining 
\bas
\mleft.\mleft({}^{\mathfrak{g}}\nabla_{\varepsilon} \nu\mright)\mright|_p
&\coloneqq
\mleft.\mleft({}^{\mathfrak{g}}\nabla_{\varepsilon(p)} \nu\mright)\mright|_p
\eas
for all $\varepsilon \in C^\infty(N; \mathfrak{g})$, $\nu \in \Gamma(E)$ and $p \in N$. Furthermore, we will generalize this and the following concepts to Lie algebroid connections which will look more familiar again with the typical definition.
\end{remark}

\begin{examples}{Lie algebra action as a Lie algebra connection, \newline \cite[special situation of first example in Example 2.8]{ELeviCivita}}{LieAlgActionIsAConnection}
A major example is the Lie algebra action $\gamma$ itself: Let $E \to N$ be a trivial vector bundle over a smooth manifold $N$, whose global trivialization we denote by $\mleft( e_a \mright)_a$. As usual, also let $\mathfrak{g}$ be a Lie algebra, and $\gamma: \mathfrak{g} \to \mathfrak{X}(N)$ be a Lie algebra action on $N$. Then define ${}^\mathfrak{g}\nabla$ by
\bas
{}^\mathfrak{g}\nabla_X \nu
&\coloneqq
\mathcal{L}_{\gamma(X)}(\nu^a) ~ e_a
\eas
for all $X \in \mathfrak{g}$ and $\nu = \nu^a e_a \in \Gamma(E)$. 
%It is trivial to check that this satisfies the conditions in Def.~\ref{def:FirstStepLieDerivativeOfAnchors}.
Consider the canonical flat connection $\nabla$ of $E$ with respect to the chosen trivialization, \textit{i.e.}~defined by $\nabla e_a = 0$, then 
\bas
{}^\mathfrak{g}\nabla_X \nu
&=
\mathcal{L}_{\gamma(X)}(\nu^a) ~ e_a
=
\nabla_{\gamma(X)} \nu
\eas
for all $X \in \mathfrak{g}$ and $\nu \in \Gamma(E)$. This also proves that this defines a $\mathfrak{g}$-connection because it is trivial to check that all vector bundle connections $\nabla^\prime$ give rise to a $\mathfrak{g}$-connection defined by ${}^{\mathfrak{g}}\nabla^\prime_X = \nabla^\prime_{\gamma(X)}$ for all $X \in \mathfrak{g}$, regardless of triviality of $E$ or flatness of $\nabla^\prime$.

In general we therefore denote such connections by
\bas
{}^{\mathfrak{g}}\nabla^\prime
&=
\nabla^\prime_\gamma.
\eas
%
%Another canonical example is the so-called \textbf{basic connection $\nabla^{\mathrm{bas}}$}, which we will introduce with more details later and which is very important in this work. It is defined by
%\ba
%\mleft.\nabla^{\mathrm{bas}}_X \nu\mright|_p
%&\coloneqq
%\mleft[ X, \nu_p \mright]_{\mathfrak{g}}
	%+ \mleft.\nabla_{\gamma(X)} \nu\mright|_p
%\ea
\end{examples}

\begin{examples}{Basic connection, \newline \cite[special situation of \S2, Definition 2.9]{basicconn}}{ClassicAdRepIsAConnection}
Let $E = N \times W \to N$ be again a trivial bundle over $N$ with fibre type $W$, denote with $\mleft( e_a \mright)_a$ a global constant frame of $E$, and with $\nabla$ its canonical flat connection. Also now assume that the Lie algebra action $\gamma$ is induced by a Lie algebra representation $\psi: \mathfrak{g} \to \mathrm{End}(W)$. Then define a $\mathfrak{g}$-connection on $E$, denoted as $\nabla^{\mathrm{bas}}$, by
\ba
\mleft.\nabla^{\mathrm{bas}}_X \nu\mright|_p
&\coloneqq
\psi(X)(\nu_p)
	+ \mleft.\nabla_{\gamma(X)} \nu\mright|_p
\ea
for all $X \in \mathfrak{g}$, $\nu \in \Gamma(E)$ and $p \in N$. This defines clearly a $\mathfrak{g}$-connection, viewing $\psi(X)(\nu)$ as an element of $\Gamma(E)$ by $p \mapsto \psi(X)(\nu_p)$ such that we can view $\psi$ as an $\mathbb{R}$-linear map $\mathfrak{g} \to \sEnd(E)$; for this recall Rem.~\ref{DifferenceOfLieAlgConnections}.

Observe that for constant sections $\nu$ we get
\bas
\nabla^{\mathrm{bas}}_X \nu
&=
\psi(X)(\nu).
\eas
Of special importance is $W = \mathfrak{g}$ and $\psi = \mathrm{ad}$. 

Those $\mathfrak{g}$-connections are related to the notion of what is known as \textbf{basic connections}, which we will introduce with more details later and which will be very important throughout this work.
\end{examples}

Let us now assume that $N$ is a vector space $W$. Recall Def.~\ref{def:ClassFunctionalGaugeTrafoBlag} and Rem.~\ref{RemabouttheddtOfClassicTrafos}; the infinitesimal gauge transformation was essentially defined by expressing the differential as a derivative along a certain curve in $\mathfrak{M}_{\mathfrak{g}}(M; W)$, differentiating with $\mathrm{d}/\mathrm{d}t$ using a canonical flat connection of the involved finite-dimensional trivial vector bundles. However, especially because the aim of this work is also to present a covariantized formulation of gauge theory, one might want to reformulate this using general connections, not just the canonical flat connection, naturally supporting general vector bundles and manifolds as a result, while avoiding the problem of having horizontal components in some tangent bundle. The connections we want to use for that for now are the $\mathfrak{g}$-connections. But those are defined for vector bundles over $N=W$, not for a vector bundle over the spacetime $M$ (in which our functionals have values in); that is simply due to that the image of a Lie algebra action, used in the Leibniz rule, is a vector field on $N$. Therefore, in order to define a $\mathfrak{g}$-connection acting on forms of the spacetime $M$, we need to make a pullback to $M$, and the only map we have so far from $M$ to $N = W$ is $\Phi$. In other words, we want to define a "connection-like" object on functionals, which is inherited by a connection of some finite-dimensional vector bundle by making a pullback, and the differentiation of such a connection on functionals is along $\mathfrak{M}_{\mathfrak{g}}(M; W)$. Moreover, one could naively view functionals $L: \mathfrak{M}_{\mathfrak{g}}(M; W) \to \Omega^k(M; K)$ ($k \in \mathbb{N}_0$, $K$ a vector space) as sections of a bundle over $\mathfrak{M}_{\mathfrak{g}}(M; W)$ which has in general an infinite rank; more about that in a later chapter. Thus, we want to construct a "connection" on infinite-dimensional bundles coming from a finite-dimensional world.

Let us only focus on pullbacks along curves in this section for simplicity. By the Leibniz rule Eq.~\eqref{FirstStepToEDerivatives} the direction of the derivative is along the Lie algebra action $\gamma$, while the idea of a pullback of a connection is that it differentiates pullbacks of sections along the differential of the curve. Hence, one expects a technical obstacle when allowing every curve for the pullback, because the typical motivation is that the Leibniz rule is inherited by the pullbacked connection. So, we just allow certain curves, whose differential is in alignment with $\gamma$.

\begin{definitions}{Lie algebra paths, \newline \cite[\S 2, special situation of the Definition 2.4]{ELeviCivita}}{LieAlgebraPfadeKurvi}
Let $\mathfrak{g}$ be a Lie algebra, and $\gamma: \mathfrak{g} \to \mathfrak{X}(N)$ be a Lie algebra action on a smooth manifold $N$. Then a \textbf{$\mathfrak{g}$-path $\alpha$ with base path $\beta$} is a pair of smooth curves $(\alpha, \beta)$, where $\alpha: I \to \mathfrak{g}$ and $\beta: I \to N$, $I$ an open interval of $\mathbb{R}$, such that
\ba
\dot{\beta}(t)
&\coloneqq
\mleft.\frac{\mathrm{d}}{\mathrm{d}t} \beta\mright|_t
=
\mleft.\beta^*\Bigl(\gamma \bigl(\alpha(t)\bigr)\Bigr)\mright|_t
=
\mleft.\gamma \bigl(\alpha(t)\bigr)\mright|_{\beta(t)}.
\ea
We also say that \textbf{$\beta$ is lifted to $\alpha$}.
\end{definitions}

\begin{remark}\label{GpathBeiRep}
\leavevmode\newline
If $N = W$ is a vector space and $\gamma$ is induced by a Lie algebra representation $\psi: \mathfrak{g} \to \mathrm{End}(W)$, then, by Prop.~\ref{prop:LieRepAndLieAct}, we would also have
\ba
\mleft.\mleft(\frac{\mathrm{d}}{\mathrm{d}t} \beta \mright)\mright|_t
&=
-\psi \bigl(\alpha(t) \bigr)\mleft(\beta(t)\mright)
\ea
for all $w \in W$.
\end{remark}

\begin{propositions}{Pullbacks of $\mathfrak{g}$-connections along $\mathfrak{g}$-paths, \newline \cite[\S 2, special situation of the comment before Definition 2.4]{ELeviCivita}}{FirstEPullBACkConnectionFormula}
Let $\mathfrak{g}$ be a Lie algebra, $\gamma: \mathfrak{g} \to \mathfrak{X}(N)$ be a Lie algebra action on a smooth manifold $N$, and ${}^{\mathfrak{g}}\nabla$ a $\mathfrak{g}$-connection on a vector bundle $E \to N$. Also fix a $\mathfrak{g}$-path $\alpha: I \to \mathfrak{g}$ with base path $\beta: I \to N$, $I \subset \mathbb{R}$ an open interval. Then there is a unique vector bundle connection $\beta^*\mleft( {}^{\mathfrak{g}}\nabla \mright)$ on $\beta^*E \to I$ with
\ba\label{WieReagiertmanaufPullbacksbeigConnection}
\bigl(\beta^*\mleft( {}^{\mathfrak{g}}\nabla \mright)\bigr)_{c \frac{\mathrm{d}}{\mathrm{d}t}} (\beta^*\nu)
&=
\beta^*\mleft( {}^{\mathfrak{g}}\nabla_{c \alpha} \nu \mright)
\ea
for all $\nu \in \Gamma(E)$, $c \in \mathbb{R}$ and $t \in I$.
\end{propositions}

\begin{proof}
%[Sketch of proof of Prop.~\ref{prop:FirstEPullBACkConnectionFormula}]
\leavevmode\newline
The proof is basically the same as for pullbacks of vector bundle connections. The idea is the following: As usual, the idea is that the pullbacks of sections, $\beta^*\nu$ ($\nu \in \Gamma(E)$), generate $\Gamma(\beta^*E)$. Thus, Eq~\eqref{WieReagiertmanaufPullbacksbeigConnection} defines the connection uniquely, that is, sections $\mu$ of $\beta^*E$ are determined by sums of elements of the form $f \cdot\beta^*\nu$, $f \in C^\infty(I)$, and by the Leibniz rule any connection $\beta^*\mleft( {}^{\mathfrak{g}}\nabla \mright)$ satisfying Eq.~\eqref{WieReagiertmanaufPullbacksbeigConnection} also satisfies
\bas
\bigl(\beta^*\mleft( {}^{\mathfrak{g}}\nabla \mright)\bigr)_{c \frac{\mathrm{d}}{\mathrm{d}t}} (f~ \beta^*\nu)
&=
c ~ \frac{\mathrm{d}f}{\mathrm{d}t} ~ \beta^*\nu
	+ f ~ \beta^*\mleft( {}^{\mathfrak{g}}\nabla_{c \alpha} \nu \mright)
\eas
for all $c \in \mathbb{R}$ and $t \in I$, such that uniqueness follows by linearity, assuming existence is given, but for the existence one can simply take this equation as a possible definition for $\beta^*\mleft( {}^{\mathfrak{g}}\nabla \mright)$. Thus, let $\beta^*\mleft( {}^{\mathfrak{g}}\nabla \mright)$ locally be defined by
\ba\label{FullPulbackGConnection}
\bigl(\beta^*\mleft( {}^{\mathfrak{g}}\nabla \mright)\bigr)_{c \frac{\mathrm{d}}{\mathrm{d}t}} \mu
&\coloneqq
c~ \frac{\mathrm{d}\mu^a}{\mathrm{d}t} ~ \beta^*e_a
	+ \mu^a ~ \beta^*\mleft( {}^{\mathfrak{g}}\nabla_{c \alpha} e_a \mright)
\ea
for all $\mu = \mu^a ~ \beta^*e_a$,
where $\mleft( e_a \mright)_a$ is a local frame of $E$. Linearity in all arguments and the Leibniz rule follow by construction, also observe that for a function $h \in C^\infty(N)$ and $\nu \in \Gamma(E)$ we can calculate
\ba
%\mleft.\bigl(\beta^*\mleft( {}^{\mathfrak{g}}\nabla \mright)\bigr)_{c \frac{\mathrm{d}}{\mathrm{d}t}} \bigl(\beta^*(h\nu)\bigr)\mright|_t
%&=
\mleft.\beta^*\bigl( {}^{\mathfrak{g}}\nabla_{c \alpha} (h\nu) \bigr)\mright|_t
&=
	\underbrace{\mleft.\beta^*\bigl(\mathcal{L}_{c (\gamma \circ \alpha)}(h)\bigr)\mright|_t}_
	{\mathclap{ \stackrel{\text{Def.~\ref{def:LieAlgebraPfadeKurvi}}}{=} \mleft.c\mathcal{L}_{\dot{\beta}}(h)\mright|_t }} ~ \beta^*\nu
	+ \mleft.\beta^*\bigl( h ~ {}^{\mathfrak{g}}\nabla_{c \alpha} \nu \bigr)\mright|_t
\nonumber\\\label{ImportantEquationToCheckForPullbacks}
&=
\mleft.\mleft(c~\frac{\mathrm{d}(h \circ \beta)}{\mathrm{d}t} ~ \beta^*\nu
	+ (h \circ \beta) ~ \beta^*\bigl( {}^{\mathfrak{g}}\nabla_{c \alpha} \nu \bigr)\mright)\mright|_t
\ea
for all $t \in I$,
thus, 
\bas
\bigl(\beta^*\mleft( {}^{\mathfrak{g}}\nabla \mright)\bigr)_{c \frac{\mathrm{d}}{\mathrm{d}t}} \mleft( \beta^*\nu \mright)
&\stackrel{\eqref{FullPulbackGConnection}}{=}
c~\frac{\mathrm{d}(\nu^a \circ \beta)}{\mathrm{d}t} ~ \beta^*e_a
	+ (\nu^a \circ \beta) ~ \beta^*\bigl( {}^{\mathfrak{g}}\nabla_{c \alpha} e_a \bigr)
\stackrel{\eqref{ImportantEquationToCheckForPullbacks}}{=}
\beta^*\mleft( {}^{\mathfrak{g}}\nabla_{c \alpha} \nu \mright),
\eas
so, Eq.~\eqref{WieReagiertmanaufPullbacksbeigConnection} is satisfied.
Finally, by Eq.~\eqref{ImportantEquationToCheckForPullbacks} it also follows that \eqref{FullPulbackGConnection} is independent of the chosen frame and, thus, globally defined. To see this, observe that any other frame $\mleft( f_b \mright)_b$ of $E$, intersecting the neighbourhood of $\mleft( e_a \mright)_a$, is given by $e_a = M_a^b f_b$, where $M_a^b$ is a local invertible matrix function on $N$. Then
\bas
\mu
&=
\mu^a ~ \beta^*e_a
=
\mleft(M_a^b \circ \beta \mright) \mu^a ~ \beta^*f_b
\eqqcolon
\tilde{\mu}^b ~ \beta^*f_b,
\eas
such that $\mu^a = \mleft(\mleft( M^{-1} \mright)^a_b \circ \beta \mright) \tilde{\mu}^b$, and, thus, as a direct consequence of Eq.~\eqref{ImportantEquationToCheckForPullbacks},
\bas
\bigl(\beta^*\mleft( {}^{\mathfrak{g}}\nabla \mright)\bigr)_{c \frac{\mathrm{d}}{\mathrm{d}t}} \mu
&~~\stackrel{\mathclap{\eqref{FullPulbackGConnection}}}{=}~~
c ~ \frac{\mathrm{d}\mu^a}{\mathrm{d}t} ~ \beta^*e_a
	+ \mu^a ~ \beta^*\mleft( {}^{\mathfrak{g}}\nabla_{c \alpha} e_a \mright)
\\
&=
c ~ \frac{\mathrm{d}\mleft(\mleft(\mleft( M^{-1} \mright)^a_d \circ \beta \mright) \tilde{\mu}^d\mright)}{\mathrm{d}t} ~ \beta^*\mleft( M_a^b f_b \mright)
\\
&\hspace{1cm}
	+ \mleft(\mleft( M^{-1} \mright)^a_d \circ \beta \mright) \tilde{\mu}^d ~ \beta^*\mleft( {}^{\mathfrak{g}}\nabla_{c \alpha} \mleft( M_a^b f_b \mright) \mright)
%\\
%&=
%c ~ \mleft(
	%\frac{\mathrm{d}\mleft(\mleft( M^{-1} \mright)^a_d \circ \beta \mright)}{\mathrm{d}t} \tilde{\mu}^d
	%+ \mleft(\mleft( M^{-1} \mright)^a_d \circ \beta \mright) \frac{\mathrm{d}\tilde{\mu}^d}{\mathrm{d}t}
%\mright) ~ \beta^*\mleft( M_a^b f_b \mright)
%\\
%&\hspace{1cm}
	%+ \mleft(\mleft( M^{-1} \mright)^a_d \circ \beta \mright) \tilde{\mu}^d ~ \beta^*\mleft( 
		%M_a^b ~ {}^{\mathfrak{g}}\nabla_{c \alpha} f_b 
		%+ \mathcal{L}_{c (\gamma \circ \alpha)} M_a^b ~f_b 
	%\mright)
\\
&\stackrel{\mathclap{ \text{Eq.~\eqref{ImportantEquationToCheckForPullbacks}} }}{=}\quad~
c ~ \frac{\mathrm{d}\tilde{\mu}^b}{\mathrm{d}t} ~ \beta^*f_b
	+ \tilde{\mu}^b ~ \beta^*\mleft( {}^{\mathfrak{g}}\nabla_{c \alpha} f_b \mright)
\\
&\hspace{1cm}
	+ c\tilde{\mu}^d ~ \text{\Large$\Biggl($}
		- \frac{\mathrm{d}\mleft(M^b_f \circ \beta \mright)}{\mathrm{d}t} ~ \mleft( \mleft( M^{-1} \mright)^f_d \circ \beta  \mright)
\\
& \hphantom{+c\tilde{\mu}^d ~ \text{\Large$\Biggl($}} \hspace{2cm}
		+ \mleft(\mleft( M^{-1} \mright)^a_d \circ \beta \mright) ~ \frac{\mathrm{d} \mleft( M^b_a \circ \beta \mright)}{\mathrm{d}t}
	\text{\Large$\Biggr)$} ~ \beta^*f_b
\\
&=
c ~ \frac{\mathrm{d}\tilde{\mu}^b}{\mathrm{d}t} ~ \beta^*f_b
	+ \tilde{\mu}^b ~ \beta^*\mleft( {}^{\mathfrak{g}}\nabla_{c \alpha} f_b \mright),
\eas
using formulas of the differential of the inverse like $M ~ \mathrm{d}M^{-1} = - \mathrm{d}M ~ M^{-1}$ (similar for $\beta^*M = M \circ \beta$). Hence, Def.~\eqref{FullPulbackGConnection} is frame-independent, and this finishes the proof.
\end{proof}

\begin{remarks}{Essential condition for pullbacks of connections}{ImportantRemarkAboutPullbacks}
Observe that the essential part of the proof is Eq.~\eqref{ImportantEquationToCheckForPullbacks}, everything follows either by this equation or by the standard construction in \eqref{FullPulbackGConnection}. This will be important later because we are going to generalise such statements about the pullbacks of connections. To avoid doing the same all over again, we will just refer to this proof and remark, essentially one only needs to check something like Eq.~\eqref{ImportantEquationToCheckForPullbacks}. Eq.~\eqref{ImportantEquationToCheckForPullbacks} essentially proves that the Leibniz rule inherited by ${}^\mathfrak{g}\nabla$ is in alignment with the Leibniz rule of vector bundle connections on $\beta^*E \to I$.

Eq.~\eqref{ImportantEquationToCheckForPullbacks} also motivates why $\mathfrak{g}$-paths are precisely the objects one needs to provide a pullback of $\mathfrak{g}$-connections along curves.
\end{remarks}

Typically, this leads to the following construction.

\begin{propositions}{Derivations of sections along $\mathfrak{g}$-paths, \newline \cite[special situation of \S 2, beginning of subsection 2.3; there $\mathrm{D}/\mathrm{d}t$ is denoted as $\nabla^\alpha$]{ELeviCivita}}{DerivationAlonggLAlgPath}
Let $\mathfrak{g}$ be a Lie algebra, $\gamma: \mathfrak{g} \to \mathfrak{X}(N)$ be a Lie algebra action on a smooth manifold $N$, and ${}^{\mathfrak{g}}\nabla$ a $\mathfrak{g}$-connection on a vector bundle $E \to N$. Also fix a $\mathfrak{g}$-path $\alpha: I \to \mathfrak{g}$ with base path $\beta: I \to N$, $I \subset \mathbb{R}$ an open interval. Then there is a unique differential operator $\frac{\mathrm{D}}{\mathrm{d}t}: \Gamma\mleft(\beta^*V\mright) \to \Gamma\mleft(\beta^*V\mright)$ with
\ba
\frac{\mathrm{D}}{\mathrm{d}t} &\text{ is linear over } \mathbb{R}, \\
\frac{\mathrm{D}}{\mathrm{d}t}(f s)
&=
\frac{\mathrm{d}f}{\mathrm{d}t} ~ s
	+ f ~ \frac{\mathrm{D}}{\mathrm{d}t} s, \\
\mleft.\frac{\mathrm{D}}{\mathrm{d}t}\mright|_t \mleft( \beta^* v \mright)
&=
\mleft. \beta^*\mleft({}^{\mathfrak{g}}\nabla_{\alpha} v \mright)\mright|_t
\ea
for all $s \in \Gamma\mleft(\beta^*V\mright)$, $v \in \Gamma(V)$, $f \in C^\infty(I)$ and $t \in I$. 
\end{propositions}

\begin{proof}
\leavevmode\newline
Define
\ba
\frac{\mathrm{D}}{\mathrm{d}t}
&\coloneqq
\mleft(\beta^*\mleft({}^{\mathfrak{g}}\nabla \mright)\mright)_{\frac{\mathrm{d}}{\mathrm{d}t}},
\ea
where $\beta^*\mleft({}^{\mathfrak{g}}\nabla \mright)$ is given by Prop.~\ref{prop:FirstEPullBACkConnectionFormula}.
This operator satisfies the needed properties by Prop.~\ref{prop:FirstEPullBACkConnectionFormula}, and the uniqueness will follow by the uniqueness given in Prop.~\ref{prop:FirstEPullBACkConnectionFormula}.
\end{proof}

In the context of the previously introduced setting of gauge theory, we have $N = W$ a vector space, and $E$ will be a trivial vector bundle over $W$. Later, when we are going to introduce the generalized infinitesimal gauge transformation for the general theory, we will allow general manifolds and vector bundles. But to avoid certain difficulties, which we will face later, we keep it that simple most of the time in the following.

As argued earlier we want to make the pullback using $\Phi$, the Higgs field. But this is a field affected by the calculus of variations, and we want to show that a certain pullback of a $\mathfrak{g}$-connection describes infinitesimal gauge transformations, hence, $\Phi$ is a "coordinate" in that context. So, the map we make a pullback with is a different one, but strongly related to $\Phi$. Let us clarify with which map we actually make the pullback.

\begin{definitions}{The evaluation map}{FirstAttemptOfEvaluationMap}
Let $M$ be a smooth manifold, $W$ a vector space, and $\mathfrak{g}$ a Lie algebra. Then we define the \textbf{evaluation map} $\mathrm{ev}: M \times \mathfrak{M}_{\mathfrak{g}}(M; W) \to W$ by
\ba
\mathrm{ev}(p, \Phi, A)
&\coloneqq
\Phi(p)
\ea
for all $(p, \Phi, A) \in M \times \mathfrak{M}_{\mathfrak{g}}(M; W)$.
\end{definitions}

Given a $\mathfrak{g}$-connection ${}^{\mathfrak{g}}\nabla$, we may try $\mathrm{ev}^*\mleft( {}^{\mathfrak{g}}\nabla \mright)$ because the functionals we look at are of the form $L: \mathfrak{M}_{\mathfrak{g}}(M; W) \to \Omega^k(M; K)$ ($k \in \mathbb{N}_0$, $K$ a vector space), so, $L: M\times \mathfrak{M}_{\mathfrak{g}}(M; W) \to \bigwedge^k \mathrm{T}^*M \otimes K$. However, as we argued earlier, the pullback of a $\mathfrak{g}$-connection is not always given. Thus, the idea is to take a curve $\eta$ in $M \times \mathfrak{M}_{\mathfrak{g}}(M; W)$ such that $\mathrm{ev} \circ \eta$ can be lifted to a $\mathfrak{g}$-path. Then we can define $\mleft(\mathrm{ev} \circ \eta\mright)^*\mleft( {}^{\mathfrak{g}}\nabla \mright)$; in other words, we want to make the pullback with $\mathrm{ev}$ but the resulting pullback-connection just differentiates along certain directions.

Of course, we do not want to take any suitable curve. We want to identify this construction with the infinitesimal gauge transformations, which we denoted earlier by $(\delta \Phi, \delta A)$ (omitting the parameter $\varepsilon$ for now) for the fields $\Phi$ and $A$. Viewing $(\delta \Phi, \delta A)$ as a vector field on $\mathfrak{M}_{\mathfrak{g}}(M; W)$,\footnote{$(\delta \Phi, \delta A)$ is the value of that vector field at $(\Phi, A)$.} one wants to define $\eta$ as the (local) flow of that vector field. That is, we take a curve $\eta$ parallel to $\mathfrak{M}_{\mathfrak{g}}(M; W)$, so, the $M$-component is constant. 

\begin{remarks}{Tangent spaces of $\mathfrak{M}_{\mathfrak{g}}(M; W)$}{TangentSpaceOfMathfrakMg}
A note about the tangent bundle of $\mathfrak{M}_{\mathfrak{g}}(M; W)$: In the general setup, presented later, we need to study it, see Prop.~\ref{prop:TangentSpaceOfSpaceOfFields}. Due to that we assume vector spaces and trivial vector bundles for the values, it is trivial to check that we get 
\bas
\mathrm{T}_{(\Phi, A)}\mleft( \mathfrak{M}_{\mathfrak{g}}(M; W)\mright)
&\cong
\mathfrak{M}_{\mathfrak{g}}(M; W),
\eas
Hence, $\delta \Phi \in C^\infty(M;W)$ and $\delta A \in \Omega^1(M; \mathfrak{g})$ makes sense, even when interpreted as components of a vector field; still omitting the parameter $\varepsilon$.

Trivially, this comes from that one thinks of tangent vectors as velocities of curves in $\mathfrak{M}_{\mathfrak{g}}(M; W)$, which is basically just a pair of curves in $W$ and $\mathfrak{g}$ (after point evaluation, \textit{e.g.}~a curve in $C^\infty(M;W)$, $t \mapsto \Phi_t$, then viewed as $t \mapsto \Phi_t(p) \in W$). As usual, one uses then the canonical flat connections for $\mathrm{T}W \cong W \times W$ and $\mathrm{T}\mathfrak{g} \cong \mathfrak{g} \times \mathfrak{g}$ such that the velocities of the curves can be viewed as curves in the corresponding vector space. It is unusual to formulate it like this, or to even mention this, but with that we want to emphasize that one cannot expect that the vector field behind all of that has values $(\delta \Phi, \delta A) \in \mathfrak{M}_{\mathfrak{g}}(M; W)$ (globally) if canonical flat connections are not given. Especially, later in this work we will have $W= N$ an arbitrary smooth manifold such that $C^\infty(M;N) \ni \Phi$ will not carry a vector space structure in general, and, so, one could not even argue with an overall vector space structure of the infinite-dimensional space itself.
\end{remarks}

Fix now $(\Phi_0, A_0) \in \mathfrak{M}_{\mathfrak{g}}(M; W)$ and $p \in M$. Then take a curve $\eta = (p, \Phi, A): I \to M \times \mathfrak{M}_{\mathfrak{g}}(M; W)$ ($I \subset \mathbb{R}$ an open interval), $I \ni t \mapsto \eta_t = (p, \Phi_t, A_t)$, with
\bas
\eta_{t=0}
&=
(p, \Phi_0, A_0).
\eas
Observe then
\bas
\mathrm{ev}\circ \eta
&=
\Phi(p)
\coloneqq 
[t \mapsto \Phi_t(p)].
\eas
Given a Lie algebra action $\gamma: \mathfrak{g} \to \mathfrak{X}(W)$,\footnote{In general, the Lie algebra behind that action does not have to be related to the same Lie algebra as in the definition of $\mathfrak{M}_{\mathfrak{g}}(M; W)$ for the following definitions and constructions. But for simplicity we assume that.} $\mathrm{ev} \circ \eta$ can be lifted to a $\mathfrak{g}$-path, if there is a $\mathfrak{g}$-path $-\epsilon(p): I \to \mathfrak{g}, t \mapsto -\epsilon_t(p),$ such that
\bas
\mleft.\frac{\mathrm{d}}{\mathrm{d}t}\mright|_t \bigl( \Phi(p) \bigr)
&=
-\mleft.\gamma\bigl( \epsilon_t(p) \bigr)\mright|_{\Phi_t(p)}.
\eas
The sign is a convention, because if $\gamma$ is induced by a Lie algebra representation $\psi: \mathfrak{g} \to \mathrm{End}(W)$, then this equation can be written as, recall Rem.~\ref{GpathBeiRep},
\bas
\mleft.\frac{\mathrm{d}}{\mathrm{d}t}\mright|_t \bigl( \Phi(p) \bigr)
&=
\psi\bigl( \epsilon_t(p) \bigr)\bigl( \Phi_t(p) \bigr),
\eas
which resembles strongly the infinitesimal gauge transformation of the Higgs field (evaluated at $p$), here for the fixed $\Phi_0$ if $t=0$; recall Def.~\ref{def:ClassicTrafos}. Therefore we want to interpret the gauge transformation of the Higgs field as the "velocity" of those curves in $C^\infty(M;W)$ which can be lifted to a $\mathfrak{g}$-path, that is
\bas
\delta_{\epsilon_0} \Phi_0
&\coloneqq
\mleft.\frac{\mathrm{d}}{\mathrm{d}t}\mright|_{t=0} \bigl( \Phi(p) \bigr)
=
-\mleft.\gamma\bigl( \epsilon_{t=0}(p) \bigr)\mright|_{\Phi_0(p)}.
\eas
Since such lifts are in general not unique, we get naturally the parametrization of $\delta \Phi_0$ with respect to $\epsilon_0: M \to \mathfrak{g}, p \mapsto \epsilon_0(p) \coloneqq \epsilon_{t=0}(p)$.

\begin{definitions}{Infinitesimal gauge transformation of the Higgs field}{ClassicGaugeTrafoOfHiggs}
Let $M$ be a smooth manifold, $W$ a vector space, and $\mathfrak{g}$ a Lie algebra with Lie algebra action $\gamma$ on $W$, induced by a Lie algebra representation $\psi$. Then we define the subspace $\mathrm{T}^\psi_{(\Phi,A)}\bigl(\mathfrak{M}_{\mathfrak{g}}(M; W)\bigr)$ of $\mathrm{T}_{(\Phi,A)}\bigl(\mathfrak{M}_{\mathfrak{g}}(M; W)\bigr)$ for all $(\Phi, A) \in \mathfrak{M}_{\mathfrak{g}}(M; W)$ by
\ba
\mathrm{T}^\psi_{(\Phi,A)}\bigl(\mathfrak{M}_{\mathfrak{g}}(M; W)\bigr)
&\coloneqq
\left\{ (\delta \Phi, \delta A) \in \mathrm{T}_{(\Phi,A)}\mleft(\mathfrak{M}_{\mathfrak{g}}(M; W)\mright) ~ \middle| ~
	\exists \epsilon \in C^\infty(M; \mathfrak{g}): ~
	\delta \Phi
	= 
	\psi(\epsilon)(\Phi)
\right\}.
\ea
The set of sections with values in these subspaces is denoted by $\mathfrak{X}^\psi(\mathfrak{M}_{\mathfrak{g}}(M; W))$.

To emphasize the relation of the first component, $\delta \Phi$, with $\epsilon$, we also write
\ba
\delta_\epsilon \Phi
&\coloneqq
\psi(\epsilon)(\Phi)
\ea
instead of $\delta \Phi$. We call this the \textbf{infinitesimal gauge transformation of the Higgs field $\Phi$}.
\end{definitions}

\begin{remark}\label{PsiEpsilonDieErste}
\leavevmode\newline
For $\Psi \in \mathfrak{X}^\psi(\mathfrak{M}_{\mathfrak{g}}(M; W))$ observe that there is a smooth $\varepsilon: \mathfrak{M}_{\mathfrak{g}}(M; W) \to C^\infty(M; \mathfrak{g})$ with
\bas
\mleft.\Psi\mright|_{(\Phi, A)}
&=
\mleft( \delta_\epsilon \Phi, \delta A \mright) 
\eas
for all $(\Phi, A) \in \mathfrak{M}_{\mathfrak{g}}(M; W)$, where $\delta A \in \Omega^1(M; \mathfrak{g})$ and $\epsilon \coloneqq \varepsilon(\Phi, A) \in C^\infty(M; \mathfrak{g})$; and each such $\varepsilon$ defines a $\Psi\in \mathfrak{X}^\psi(\mathfrak{M}_{\mathfrak{g}}(M; W))$. With that one can easily see that $\mathfrak{X}^\psi(\mathfrak{M}_{\mathfrak{g}}(M; W))$ is a submodule of $\mathfrak{X}(\mathfrak{M}_{\mathfrak{g}}(M; W))$, respectively; but $\mathfrak{X}^\psi(\mathfrak{M}_{\mathfrak{g}}(M; W))$ is in general not a subalgebra, due to the fact that $\varepsilon$ itself depends on $\mathfrak{M}_{\mathfrak{g}}(M; W)$. To emphasize the relation between $\Psi$ and $\varepsilon$ we also often write $\Psi \eqqcolon \Psi_\varepsilon$. Keep in mind that $\Psi_\varepsilon$ is not unique for a given $\varepsilon$ because we did not fix $\delta A$ yet. Also observe the difference to the previous section: The parameter of the infinitesimal gauge transformation is going to be a functional $\mathfrak{M}_{\mathfrak{g}}(M; W) \to C^\infty(M; \mathfrak{g})$, while the typical formulation uses just $\epsilon \in C^\infty(M; \mathfrak{g})$ (basically a constant functional one could say). 
%One could already have guessed so because one can apply the previously discussed bookkeeping trick to the parameter of the infinitesimal gauge transformations.
 %$\varepsilon$ can also be viewed as a section of the trivial vector bundle over $M \times \mathfrak{M}_{\mathfrak{g}}(M; W)$ with fibre type $\mathfrak{g}$.
\end{remark}

To summarize, we have:

\begin{corollaries}{Flows of $\mathfrak{X}^\psi(\mathfrak{M}_{\mathfrak{g}}(M; W))$}{ClassicFLowsOfXgMg}
Let $M$ be a smooth manifold, $W$ a vector space, and $\mathfrak{g}$ a Lie algebra with Lie algebra action $\gamma$ on $W$, induced by a Lie algebra representation $\psi$. Also let $\Psi_\varepsilon \in \mathfrak{X}^\psi(\mathfrak{M}_{\mathfrak{g}}(M; W))$ for an $\varepsilon: \mathfrak{M}_{\mathfrak{g}}(M; W) \to C^\infty(M;\mathfrak{g})$ whose local flow through $(\Phi_0, A_0) \in \mathfrak{M}_{\mathfrak{g}}(M; W)$ we denote by $\mleft.\eta\mright|_{(\Phi_0, A_0)} \coloneqq (\Phi, A): I \to \mathfrak{M}_{\mathfrak{g}}(M; W), t \mapsto \mleft.\eta\mright|_{(\Phi_0, A_0)}(t)= (\Phi_t, A_t)$ ($I \subset \mathbb{R}$ an open interval).

Then there is a smooth curve $\epsilon: I \to C^\infty(M; \mathfrak{g}), t \mapsto \epsilon_t,$ with $\epsilon_{t=0} = \varepsilon(\Phi_0, A_0)$ and such that
\bas
-\epsilon(p)
&\coloneqq
[t \mapsto -\epsilon_t(p)]
\eas
is a $\mathfrak{g}$-path for all $p \in M$ with base path
\bas
\Phi(p)
&\coloneqq
[t \mapsto \Phi_t(p)],
\eas
that is
\ba\label{FlowStuffOfXgMg}
\mleft.\frac{\mathrm{d}}{\mathrm{d}t}\mright|_t \bigl( \Phi(p) \bigr)
&=
\psi\bigl( \epsilon_t(p) \bigr)\bigl( \Phi_t(p) \bigr)
=
\mleft(\delta_{\epsilon_t} \Phi_t \mright)(p).
\ea
\end{corollaries}

\begin{proof}
\leavevmode\newline
By construction and definition, \textit{i.e.}~there is an $\epsilon: I \to C^\infty(M; \mathfrak{g}), t \mapsto \epsilon_t,$ such that
\bas
\Psi^{(1)}_{\mleft.\eta\mright|_{(\Phi_0, A_0)}(t)}
&=
\psi(\epsilon_t)(\Phi_t),
\eas
where $\Psi^{(1)}$ is the first component of $\Psi$, the one along the "$\Phi$-direction"; thus, Eq.~\eqref{FlowStuffOfXgMg} follows by the definition of flows of vector fields, and one can take $\epsilon$ in such a way that $\epsilon_{t=0} = \varepsilon(\Phi_0, A_0)$ because we have at $t= 0$
\bas
\Psi^{(1)}_{\mleft.\eta\mright|_{(\Phi_0, A_0)}(0)}
&=
\Psi^{(1)}_{(\Phi_0, A_0)}
=
\psi\bigl(\varepsilon(\Phi_0, A_0)\bigr)(\Phi_0).
\eas
\end{proof}

Let us conclude this section with the definition of the infinitesimal gauge transformation of the studied functionals, making use of the previously-discussed relation between $\mathfrak{g}$-paths and the infinitesimal gauge transformation of the Higgs field. It is especially about pullbacks of $\mathfrak{g}$-connections, which were uniquely defined by their differentiation on pullbacks, but the definitions of the typical functionals like the field strength or the minimal coupling do not contain any visible pullback as if they do not live in a pullback bundle. But we will use a trivial bookkeeping trick: The bundle those functionals have values in is a trivial bundle over $M$, and trivial bundles are always trivially isomorphic to the pullback of another trivial bundle with the same fibre type, \textit{e.g.}~$M \times \mathfrak{g} \cong \Phi^*(W \times \mathfrak{g})$, $W \times \mathfrak{g}$ the trivial bundle over $N = W$. That is the following:

Let $K$ be a vector space, we viewed it as a trivial vector bundle over $M$, but we can do the same for $N=W$, so, $K$ can also be viewed as trivial vector bundle over $W$, and elements of $K$ are just constant sections of such a bundle. For bookkeeping, let us denote with $\iota_M$ and $\iota_W$ maps $K \hookrightarrow \Gamma(M \times K)$ and $K \hookrightarrow \Gamma(W \times K)$, respectively, which embed elements of $K$ canonically into the space of constant sections of the trivial bundles $M \times K$ and $W \times K$, respectively. Then take a smooth map $L: \mathfrak{M}_{\mathfrak{g}}(M; W) \to \Omega^k(M; K)$ ($k \in \mathbb{N}_0$) and a basis $\mleft( e_a \mright)_a$ of $K$. Previously we expressed $L$ then as, making use of $\iota_M$,
\bas
L
&=
L^a \otimes \iota_M(e_a),
\eas
where $L^a: \mathfrak{M}_{\mathfrak{g}}(M; W) \to \Omega^k(M)$. Fix $(\Phi, A) \in \mathfrak{M}_{\mathfrak{g}}(M; W)$, then we can trivially identify
\bas
\iota_M(e_a)
&=
\Phi^*\bigl( \iota_W (e_a) \bigr)
\eas
because $e_a$ is viewed as a constant section in both trivial vector bundles. Then observe
\bas
\mleft.\mathrm{ev}^*\bigl( \iota_W(e_a) \bigr)\mright|_{(p, \Phi, A)}
&=
\mleft.\iota_W (e_a)\mright|_{\Phi(p)}
=
\mleft.\Phi^*\bigl( \iota_W (e_a) \bigr)\mright|_p
=
\mleft.\iota_M(e_a)\mright|_p
\eas
for all $(p, \Phi, A) \in M \times \mathfrak{M}_{\mathfrak{g}}(M; W)$. Thus, we can also write
\bas
L
&=
L^a \otimes \mathrm{ev}^*\bigl( \iota_W(e_a) \bigr)
\eqqcolon
\gls{1jota}(L),
\eas
and that interpretation of $L$ we denote as $\iota(L)$ for bookkeeping reasons. Observe
\bas
\iota(L)(Y_1, \dotsc, Y_k)
&=
\underbrace{L^a(Y_1, \dotsc, Y_k)}_{\in C^\infty(M\times \mathfrak{M}_{\mathfrak{g}}(M;W))} ~ \mathrm{ev}^*\bigl( \iota_W(e_a) \bigr)
\in
\Gamma(\mathrm{ev}^*(W\times K))
\eas
for all $Y_1, \dotsc, Y_k \in \mathfrak{X}(M)$; therefore also $\iota(L)(\Phi, A) \in \Omega^k(M; \Phi^*K)$. With that we can now finally explicitly state the idea of describing infinitesimal gauge transformations as a certain pullback of a $\mathfrak{g}$-connection.

\begin{propositions}{Functional derivative along $\mathfrak{X}^\psi(\mathfrak{M}_{\mathfrak{g}}(M; W))$}{ClassicFunctionDerivativesAlongPsiEpsilon}
Let $M$ be a smooth manifold, $W, K$ vector spaces, and $\mathfrak{g}$ a Lie algebra with Lie algebra action $\gamma$ on $W$, induced by a Lie algebra representation $\psi$. Moreover, let $^{\mathfrak{g}}\nabla$ be a $\mathfrak{g}$-connection on the trivial vector bundle $W \times K$ over $W$, and $\Psi_\varepsilon \in \mathfrak{X}^\psi(\mathfrak{M}_{\mathfrak{g}}(M; W))$ for an $\varepsilon: \mathfrak{M}_{\mathfrak{g}}(M; W) \to C^\infty(M;\mathfrak{g})$.

Then there is a unique $\mathbb{R}$-linear operator $\delta_{\Psi_\varepsilon}: \Gamma\mleft(\mathrm{ev}^*( W \times K )\mright) \to \Gamma\mleft(\mathrm{ev}^*( W \times K )\mright)$ with
\ba
\delta_{\Psi_\varepsilon} (f s)
&=
\mathcal{L}_{\Psi_\varepsilon}(f) ~ s
	+ f ~ \delta_{\Psi_\varepsilon} s,
\label{ClassicGaugeTrafoLeibnizRule} \\
\delta_{\Psi_\varepsilon} \mleft( \mathrm{ev}^*\vartheta \mright)
&=
- \mathrm{ev}^*\mleft( {}^{\mathfrak{g}}\nabla_\varepsilon \vartheta \mright)\label{ClassicGaugeTrafoPullbackRelationtoEv}
\ea
for all $f \in C^\infty(M \times \mathfrak{M}_{\mathfrak{g}}(M; W))$, $s \in \Gamma\mleft(\mathrm{ev}^*(W \times K)\mright)$ and $\vartheta \in \Gamma(W \times K)$, where we denote
\bas
\mleft.\mathrm{ev}^*\mleft( {}^{\mathfrak{g}}\nabla_\varepsilon \vartheta \mright)\mright|_{(p, \Phi_0, A_0)}
&=
\mleft.\mleft({}^{\mathfrak{g}}\nabla_{\varepsilon(\Phi_0, A_0)|_p} \vartheta\mright)\mright|_{\Phi_0(p)}
\eas
for all $(p, \Phi_0, A_0) \in M \times \mathfrak{M}_{\mathfrak{g}}(M; W)$.
\end{propositions}

\begin{remark}\label{WecombineeverythingToAvoidStrictPullbacks}
\leavevmode\newline
This emphasizes that $\delta_{\Psi_\varepsilon}$ is the "$\mathrm{ev}$-pullback of ${}^{\mathfrak{g}}\nabla$ combined with a contraction along $\Psi_\varepsilon$" (up to a sign), and that combination leads to that we do not need an overall pullback with $\mathrm{ev}$. When we show this in the general setting, then we give a general condition about in which situations one can do such pullbacks, avoiding the ansatz using flows and curves, making the approach cleaner.
\end{remark}

\begin{proof}[Proof of Prop.~\ref{prop:ClassicFunctionDerivativesAlongPsiEpsilon}]
\leavevmode\newline
For $\Psi_\varepsilon$ let $\eta: I \times U \to \mathfrak{M}_{\mathfrak{g}}(M; W)$ be its local flow on an open subset $U \subset \mathfrak{M}_{\mathfrak{g}}(M; W)$, where $I \subset \mathbb{R}$ is an open interval containing 0, and we denote its flow through $(\Phi_0, A_0) \in U$ by $\eta|_{(\Phi_0, A_0)} = (\Phi, A), I \ni t \mapsto (\Phi_t, A_t)$. For the flow $\eta|_{(\Phi_0, A_0)}$ we can apply Cor.~\ref{cor:ClassicFLowsOfXgMg}, that is, there is an $\epsilon: I \to C^\infty(M; \mathfrak{g}), t \mapsto \epsilon_t,$ such that $\Phi(p)\coloneqq \mleft[ t \mapsto \Phi_t(p) \mright]$ is the base path of a $\mathfrak{g}$-path $-\epsilon(p) \coloneqq [t \mapsto -\epsilon_t(p)]$, and we have $\epsilon_{t=0} = \varepsilon(\Phi_0, A_0)$. Hence, fixing such a lift to a $\mathfrak{g}$-path, we can define by Prop.~\ref{prop:FirstEPullBACkConnectionFormula}
\ba\label{ClassicHiggsDerivation} 
\mleft.\delta_{\Psi_\varepsilon} s\mright|_{(p, \Phi_0, A_0)}
&\coloneqq
\biggl( 
	\underbrace{\mleft( \mathrm{ev} \circ \mleft(p, \eta|_{(\Phi_0, A_0)}\mright) \mright)^*}
	_{= (\Phi(p))^*}
	\mleft({}^{\mathfrak{g}}\nabla\mright)
\biggr)_{\mleft.\frac{\mathrm{d}}{\mathrm{d}t}\mright|_{t=0}}
\mleft(
	\mleft(p, \eta|_{(\Phi_0, A_0)}\mright)^*s
\mright)
\nonumber \\
&=
\Bigl( 
	\bigl(\Phi(p)\bigr)^*
	\mleft({}^{\mathfrak{g}}\nabla\mright)
\Bigr)_{\mleft.\frac{\mathrm{d}}{\mathrm{d}t}\mright|_{t=0}}
\mleft(
	\mleft(p, \eta|_{(\Phi_0, A_0)}\mright)^*s
\mright)
\ea
for all $s \in \Gamma\mleft(\mathrm{ev}^*(W \times K)\mright)$ and $p \in M$, where $\mleft(p, \eta|_{(\Phi_0, A_0)}\mright)^*s$ is by definition a section of $\mleft( \mathrm{ev} \circ \mleft(p, \eta|_{(\Phi_0, A_0)}\mright) \mright)^*(W \times K)$, especially,
\bas
\mleft.\mleft(p, \eta|_{(\Phi_0, A_0)}\mright)^*s\mright|_{t}
&=
\mleft.s\mright|_{(p, \Phi_t, A_t)}
\in
\{\Phi_t(p)\} \times K,
\eas
and, thus, it can also be seen as a section of $\bigl( \Phi(p) \bigr)^*(W \times K)$.
Then Def.~\ref{ClassicHiggsDerivation} is nothing else than the (restricted) definition of $\mleft.\mathrm{D}/\mathrm{d}t\mright|_{t=0}$ related to ${}^{\mathfrak{g}}\nabla$ and using the given $\mathfrak{g}$-path $-\epsilon(p)$ with base path $\Phi(p)$, see Prop.~\ref{prop:DerivationAlonggLAlgPath} and its proof. That is
\bas
\mleft(\delta_{\Psi_\varepsilon} s\mright)(p, \Phi_0, A_0)
&=
\mleft.\frac{\mathrm{D}}{\mathrm{d}t}\mright|_{t=0}
\mleft(
	\mleft(p, \eta|_{(\Phi_0, A_0)}\mright)^*s
\mright)
\eas
so, everything follows by Prop.~\ref{prop:DerivationAlonggLAlgPath}, \textit{i.e.}~$\mathbb{R}$-linearity is clearly implied, and
\bas
\mleft.\delta_{\Psi_\varepsilon} (f s)\mright|_{(p, \Phi_0, A_0)}
&=
\mleft.\frac{\mathrm{d}}{\mathrm{d}t}\mright|_{t=0}\mleft(
	f \circ \mleft(p, \eta|_{(\Phi_0, A_0)}\mright)
\mright)
~ \mleft.s\mright|_{(p, \Phi_0, A_0)}
+	f(p, \Phi_0, A_0) ~ \mleft.\frac{\mathrm{D}}{\mathrm{d}t}\mright|_{t=0}
\mleft(
	\mleft(p, \eta|_{(\Phi_0, A_0)}\mright)^*s
\mright)
\\
&=
\mleft.\mleft(\mathcal{L}_{\Psi_\varepsilon}(f) ~ s
	+ f ~ \delta_{\Psi_\varepsilon} s\mright)\mright|_{(p, \Phi_0, A_0)}
\eas
for all $f \in C^\infty(M \times \mathfrak{M}_{\mathfrak{g}}(M; W))$, and finally
\bas
\mleft.\delta_{\Psi_\varepsilon} \mleft( \mathrm{ev}^*\vartheta \mright)\mright|_{(p, \Phi_0, A_0)}
&=
\mleft.\frac{\mathrm{D}}{\mathrm{d}t}\mright|_{t=0}
\mleft(
	\mleft( \mathrm{ev} \circ \mleft(p, \eta|_{(\Phi_0, A_0)}\mright)\mright)^*\vartheta
\mright)
\\
&=
\mleft.\frac{\mathrm{D}}{\mathrm{d}t}\mright|_{t=0}
\bigl(
	\mleft( \Phi(p)\mright)^*\vartheta
\bigr)
\\
&=
- \mleft( \Phi(p)\mright)^*
\mleft(
	{}^{\mathfrak{g}}\nabla_{\epsilon_{t=0}} \vartheta
\mright)
\\
&\stackrel{\mathclap{ \epsilon_{t=0} = \varepsilon(\Phi_0, A_0) }}{=}\qquad
\mleft.
- \mathrm{ev}^*
\mleft(
	{}^{\mathfrak{g}}\nabla_{\varepsilon} \vartheta
\mright)
\mright|_{(p, \Phi_0, A_0)}
\eas
for all $\vartheta \in \Gamma(W \times K)$. Uniqueness also follows by Prop.~\ref{prop:DerivationAlonggLAlgPath}, although this $\mathrm{D}/\mathrm{d}t$ operator only differentiates sections of the form $\mleft(p, \eta|_{(\Phi_0, A_0)}\mright)^*s$; the vector space of such sections has $\bigl( \Phi(p) \bigr)^*\bigl(\Gamma(W \times K)\bigr)$ as a subset, the generators of sections of $\bigl( \Phi(p) \bigr)^*(W \times K)$, which was visible by having $s= \mathrm{ev}^*\vartheta$, that is
\bas
\mleft(p, \eta|_{(\Phi_0, A_0)}\mright)^*(\mathrm{ev}^*\vartheta)
&=
\mleft( \mathrm{ev} \circ \mleft(p, \eta|_{(\Phi_0, A_0)}\mright)\mright)^*\vartheta
=
\bigl(\Phi(p)\bigr)^*\vartheta.
\eas
Therefore the argument about uniqueness in the proof of Prop.~\ref{prop:DerivationAlonggLAlgPath} applies here, too.\footnote{Alternatively, one shows it directly in the same fashion, using again that $\mathrm{ev}$-pullbacks of sections generate $\Gamma(\mathrm{ev}^*(W\times K))$, such that Eq.~\eqref{ClassicGaugeTrafoPullbackRelationtoEv} uniquely defines the operator because Eq.~\ref{ClassicGaugeTrafoLeibnizRule} declares how the operator acts on the generated sections of pullbacks.}
\end{proof}

Now we extend it to functionals. We will now also recall the infinitesimal gauge transformation of the field of gauge bosons $A$ as in Def.~\ref{def:ClassicTrafos} and take that still as a definition; at this point there is nothing new to tell about that part of the infinitesimal gauge transformation, except that $\varepsilon: \mathfrak{M}_{\mathfrak{g}}(M; W) \to C^\infty(M; \mathfrak{g})$, and, thus, the derivation will be along a vector field $\Psi_\varepsilon$
\ba
\mleft.\Psi_\varepsilon\mright|_{(\Phi, A)}
&=
\mleft( \delta_\epsilon \Phi, \delta_\epsilon A \mright)
\ea
for all $(\Phi, A) \in \mathfrak{M}_{\mathfrak{g}}(M; W)$, where $\epsilon \coloneqq \varepsilon(\Phi, A)$ and $\delta_\epsilon A = \mleft[ \epsilon, A \mright]_{\mathfrak{g}} - \mathrm{d}\epsilon$. We shortly write for now $\Psi_\varepsilon = (\delta_\varepsilon \Phi, \delta_\varepsilon A)$. However, in the general setting later we need to discuss the gauge transformation of $A$ and how to define it, and therefore we will come back to this.

\begin{definitions}{Infinitesimal gauge transformation}{InfinitesimalGaugeTrafoClassicAsConnection}
Let $M$ be a smooth manifold, $W, K$ vector spaces, and $\mathfrak{g}$ a Lie algebra with Lie algebra action $\gamma$ on $W$, induced by a Lie algebra representation $\psi$. Moreover, let $^{\mathfrak{g}}\nabla$ be a $\mathfrak{g}$-connection on the trivial vector bundle $W \times K$ over $W$, and $\Psi_\varepsilon = (\delta_\varepsilon \Phi, \delta_\varepsilon A)$ for an $\varepsilon: \mathfrak{M}_{\mathfrak{g}}(M; W) \to C^\infty(M;\mathfrak{g})$.

Then we define the \textbf{infinitesimal gauge transformation $\delta_\varepsilon L$ for $L: \mathfrak{M}_{\mathfrak{g}}(M; W) \to \Omega^k(M; K)$} ($k \in \mathbb{N}_0$) as a map $\mathfrak{M}_{\mathfrak{g}}(M; W) \to \Omega^k(M; K)$ by
\ba
\mleft(\delta_\varepsilon L\mright)(Y_1, \dotsc, Y_k)
&\coloneqq
\delta_{\Psi_\varepsilon}\bigl(
	\iota(L)(Y_1, \dotsc, Y_k)
\bigr)
\ea
for all $Y_1, \dotsc, Y_k \in \mathfrak{X}(M)$, where $\delta_{\Psi_\varepsilon}$ is the unique operator given in Prop.~\ref{prop:ClassicFunctionDerivativesAlongPsiEpsilon} with respect to ${}^{\mathfrak{g}}\nabla$ and $\Psi_\varepsilon$.
 %as an $\mathbb{R}$-linear operator $\Omega^k(M \times \mathfrak{M}_{\mathfrak{g}}(M; W); \mathrm{ev}^*(W \times K)) \to \Omega^k(M \times \mathfrak{M}_{\mathfrak{g}}(M; W); \mathrm{ev}^*(W \times K))$
\end{definitions}

\begin{remark}
\leavevmode\newline
Recall that $\iota(L)$ was the bookkeeping trick, and, thus,
\bas
\iota(L)(Y_1, \dotsc, Y_k)
&\in
\Gamma(\mathrm{ev}^*(W\times K))
\eas
for all $Y_1, \dotsc, Y_k \in \mathfrak{X}(M)$. Hence, this definition is well-defined; that $\delta_\varepsilon L$ is a map $\mathfrak{M}_{\mathfrak{g}}(M; W) \to \Omega^k(M; K)$ also follows by construction. Especially observe that $C^\infty(M)$-multilinearity follows because $\mathcal{L}_{\Psi_\varepsilon} f = 0$ for all $f \in C^\infty(M)$ due to the fact that $\Psi_\varepsilon$ is a vector field on $\mathfrak{M}_{\mathfrak{g}}(M;W)$, viewed as a vector field in $M \times \mathfrak{M}_{\mathfrak{g}}(M;W)$. So, $C^\infty(M)$ is not affected by the Leibniz rule in $\delta_{\Psi_\varepsilon}$. The vector fields $Y_1, \dotsc, Y_k$ are similarly unaffected by the Lie derivative of $\mathcal{L}_{\Psi_\varepsilon}$; hence, this is a valid construction.
\end{remark}

We now compare it with the classic definition of the infinitesimal gauge transformation as in Def.~\ref{def:ClassFunctionalGaugeTrafoBlag}; for this also recall Ex.~\ref{ex:LieAlgActionIsAConnection}.

\begin{theorems}{Recover of classical definition of infinitesimal gauge transformation}{RecoverOfClassicInfgGaugeTrafo}
Let $M$ be a smooth manifold, $W, K$ vector spaces, and $\mathfrak{g}$ a Lie algebra with Lie algebra action $\gamma$ on $W$, induced by a Lie algebra representation $\psi$. Moreover, let $^{\mathfrak{g}}\nabla = \nabla_\gamma$ be the $\mathfrak{g}$-connection induced by the canonical flat connection $\nabla$ of the trivial vector bundle $W \times K \to W$ as in Ex.~\ref{ex:LieAlgActionIsAConnection}, and $\Psi_\varepsilon = (\delta_\varepsilon \Phi, \delta_\varepsilon A)$ for an $\varepsilon: \mathfrak{M}_{\mathfrak{g}}(M; W) \to C^\infty(M;\mathfrak{g})$.

Then we have
\ba
\mleft(\delta_\varepsilon L\mright)(\Phi, A)
&=
\mleft.\frac{\mathrm{d}}{\mathrm{d}t}\mright|_{t=0}
\mleft[ t \mapsto
	L\mleft(
		\Phi + t \delta_\epsilon \Phi,
		A + t \delta_\epsilon A
	\mright)
\mright]
\ea
for all $L: \mathfrak{M}_{\mathfrak{g}}(M;W) \to \Omega^k(M; K)$ $(k \in \mathbb{N}_0)$ and $(\Phi, A) \in \mathfrak{M}_{\mathfrak{g}}(M;W)$, where $\epsilon \coloneqq \varepsilon(\Phi,A)$, $t \in \mathbb{R}$, and $\delta_\varepsilon$ is as defined in Def.~\ref{def:InfinitesimalGaugeTrafoClassicAsConnection} with respect to $\nabla_\gamma$ and $\Psi_\varepsilon$.

In other words, we recover Def.~\ref{def:ClassFunctionalGaugeTrafoBlag}, especially when taking an $\varepsilon \in C^\infty(M;\mathfrak{g})$, \textit{i.e.}~a constant $\varepsilon$, "constant" in sense of
\bas
\varepsilon(\Phi,A)
&=
\varepsilon\mleft(\Phi^\prime,A^\prime\mright)
\eas
for all $(\Phi, A), \mleft(\Phi^\prime,A^\prime\mright) \in \mathfrak{M}_{\mathfrak{g}}(M;W)$.
\end{theorems}

\begin{remarks}{$\delta_\varepsilon A$ as transformation of a functional}{BosonsAsFunctionalies}
Recall that $\mathrm{d}/\mathrm{d}t$ is with respect to the canonical flat connection of $M \times K \to M$. Also observe that $\delta_\varepsilon A$ is here trivially given by $\delta_\varepsilon \varpi_2$, where $\varpi_2(\Phi, A) \coloneqq A$, the projection onto the second factor in $\mathfrak{M}_{\mathfrak{g}}$. Viewing the field of gauge bosons as the functional $\varpi_2$, one may want to define the infinitesimal gauge transformation of $A$ as the infinitesimal gauge transformation of $\varpi_2$; since $\varpi_2$ is $\mathfrak{g}$-valued, we would have
\bas
\iota(\varpi_2)(Y)
&\in
\Gamma(\mathrm{ev}^*(W \times \mathfrak{g}))
\eas
for all $Y \in \mathfrak{X}(M)$, and, thus, $\iota(A) \coloneqq \iota(\varpi_2)(\Phi, A) \in \Omega^1(M; \Phi^*(W \times \mathfrak{g}))$ for any fixed $\Phi$. For the infinitesimal gauge transformation of the field strength one also applies the bookkeeping trick such that it has values in $\mathrm{ev}^*(W \times \mathfrak{g})$, so, as we mentioned before, we want to view the Lie algebra as a bundle over $W$ instead of a bundle over $M$.
\end{remarks}

\begin{proof}[Proof of Thm.~\ref{thm:RecoverOfClassicInfgGaugeTrafo}]
\leavevmode\newline
Let $\mleft( e_a \mright)_a$ be a basis of $K$, that especially implies
\bas
\nabla \bigl( \iota_W(e_a) \bigr)
&=
0.
\eas
For $L: \mathfrak{M}_{\mathfrak{g}}(M;W) \to \Omega^k(M; K)$ we then write
\bas
\iota(L)
&=
L^a \otimes \mathrm{ev}^*\bigl(\iota_W(e_a)\bigr)
\eas
for $L^a: \mathfrak{M}_{\mathfrak{g}}(M;W) \to \Omega^k(M)$, so, $L^a \in \Omega^k(M\times \mathfrak{M}_{\mathfrak{g}}(M;W))$, and,
thus, by using Prop.~\ref{prop:ClassicFunctionDerivativesAlongPsiEpsilon},
\bas
\mleft.\mleft(\delta_\varepsilon L\mright)(Y_1, \dotsc, Y_k)\mright|_{(\Phi,A)}
&=
\mleft.
\delta_{\Psi_\varepsilon}\bigl(
	\iota(L)(Y_1, \dotsc, Y_k)
\bigr)
\mright|_{(\Phi,A)}
\\
&=
\mleft.\mathcal{L}_{\Psi_\varepsilon}\mleft(
	L^a(Y_1, \dotsc, Y_k)
\mright)\mright|_{(\Phi,A)} 
~ \underbrace{\mleft.\mathrm{ev}^*\bigl(\iota_W(e_a)\bigr)\mright|_{(\Phi,A)}}_
{= \Phi^*( \iota_W(e_a) ) = \iota_M(e_a)}
\\
&\hspace{1cm}
	- \underbrace{\mleft.\mleft(L^a(Y_1, \dotsc, Y_k) ~ \mathrm{ev}^*\mleft(\nabla_{\gamma(\varepsilon)}\bigl( \iota_W(e_a) \bigr) \mright)\mright)\mright|_{(\Phi,A)}}_{=0}
\\
&=
\mleft( 
	\mathcal{L}_{\mleft.\Psi_{\varepsilon}\mright|_{(\Phi,A)}}\mleft(L^a\mright)
	\otimes \iota_M(e_a)
\mright)(Y_1, \dotsc, Y_k)
\\
&=
\mleft( 
	\mleft.\frac{\mathrm{d}}{\mathrm{d}t}\mright|_{t=0}
\mleft[ t \mapsto
	L\mleft(
		\Phi + t \delta_\epsilon \Phi,
		A + t \delta_\epsilon A
	\mright)
\mright]
\mright)
(Y_1, \dotsc, Y_k)
\eas
for all $(\Phi, A) \in \mathfrak{M}_{\mathfrak{g}}(M;W)$ and $Y_1, \dotsc, Y_k \in \mathfrak{X}(M)$, using that $\mleft.\Psi_{\varepsilon}\mright|_{(\Phi,A)} = (\delta_\epsilon \Phi, \delta_\epsilon A)$.
\end{proof}

This concludes this section, we have shown how to write the infinitesimal gauge transformation using $\mathfrak{g}$-connections. One can even show that the gauge invariance of the Yang-Mills-Higgs Lagrangian can be shown with the same calculation of the previous section if $\varepsilon$ is allowed to depend on $\mathfrak{M}_{\mathfrak{g}}(M;W)$. Such a dependency starts to matter when applying the infinitesimal gauge transformation twice, which we will discuss later in full generality. Let us now shortly discuss what we have learned.

First of all, we needed to do the bookkeeping trick. That was due to the Lie algebra action $\gamma$, which acts on $N =W$ and not on $M$. Hence, the natural construction of $\mathfrak{g}$-connections using $\gamma$ is defined on bundles over $N$. This was why we needed to make a pullback and to think of functionals as having values in a pullback of a trivial bundle over $N$, especially using $\Phi \in C^\infty(M;N)$. For example, we thought of the Lie algebra $\mathfrak{g}$ as a trivial bundle over $M$ and $N$, $M \times \mathfrak{g}$ and $N \times \mathfrak{g}$, respectively, and it is more suitable to think of $M \times \mathfrak{g}$ as $\Phi^*(N \times \mathfrak{g})$. The aim of the presented generalised gauge theory is also to generalise the trivial Lie algebra bundle, especially getting rid of a global trivialisation by replacing it with some "suitable" bundle $E$. Hence, motivated by this section and as an ansatz, we are going to define $E$ in place of $N \times \mathfrak{g}$ later and $\Phi^*E$ will replace $M \times \mathfrak{g}$. In the same manner other vector spaces may be replaced like that, too.

Second, assume we have that non-trivial bundle $E$ now. Then we cannot impose the existence of a canonical flat connection anymore as we did in all the basic definitions before, like in Def.~\ref{def:ClassFunctionalGaugeTrafoBlag}; defining $\mathrm{d}/\mathrm{d}t$ using the tangent map would lead to arising horizontal components in the corresponding tangent bundle which may make further calculations more complicated when a functional is used in other functionals, like in contractions using scalar products and metrics, such that one may need to fix a horizontal distribution. Therefore the definition of infinitesimal gauge transformation as provided here is a first step towards a formulation using ($\mathfrak{g}$-)connections, \textit{e.g.}~taking a connection $\nabla$ and then defining ${}^{\mathfrak{g}}\nabla = \nabla_\gamma$.

Third, one could argue that one could just look at vector bundle connections $\nabla$ for which there is always a pullback, avoiding the problems discussed in this section. However, $\mathfrak{g}$-connections are more general, which we will see later, and we will then have an even more general notion. But, for example, allow infinite-dimensional Lie algebras, then take $\mathfrak{g} = \mathfrak{X}(N)$ and $\gamma = \mathds{1}$, the identity; then one clearly has the typical notion of a vector bundle connection. Especially when thinking about that the infinitesimal gauge transformations are just certain, not all, vector fields on $\mathfrak{M}_{\mathfrak{g}}$, one might argue why not using a different connection like a $\mathfrak{g}$-connection which is not directly related to $\nabla$. Recall Ex.~\ref{ex:ClassicAdRepIsAConnection}, we could also take $\nabla^{\mathrm{bas}}$, which is clearly different to $\nabla_\gamma$ as discussed there, even though $\nabla_\gamma$ contributes to its definition. We will later see that $\nabla^{\mathrm{bas}}$ does not necessarily have any notion of a parallel frame, even when it is assumed to be flat.\footnote{Flatness will be defined later for such connections, but the construction has the typical form.}
Actually, we are going to use the basic connection later, also for the infinitesimal gauge transformations. We will show that the gauge invariance of the Yang-Mills-Higgs Lagrangian can still be shown although we use $\nabla^{\mathrm{bas}}$, also in the context of the typical formulation of gauge theory. The advantage of the basic connection will be that it is always flat in the context of gauge theory, while $\nabla_\gamma$ might not be, which results into that we can generalize the well-known relation
\bas
\mleft[ \delta_\varepsilon, \delta_\varepsilon^\prime \mright]
&=
- \delta_{\mleft[ \varepsilon, \varepsilon^\prime \mright]_{\mathfrak{g}}},
\eas
where the sign comes from our sign conventions defined earlier. We will see that a possible curvature of $\nabla_\gamma$ will not result into a generalization of that equation, if we define the infinitesimal gauge transformations using $\nabla_\gamma$. Moreover, we have seen in Ex.~\ref{ex:ClassicAdRepIsAConnection} that $\nabla^{\mathrm{bas}}$ is a generalization of a Lie algebra representation; this will lead to that the basic connection supports the symmetries of gauge theories, leading to more convenient formulas of infinitesimal gauge transformations.

Last, the Lie algebra $\mathfrak{g}$ is not only important from an algebraic point of view, but also in sense of a connection besides the field of gauge bosons $A$, playing the role of a "direction of derivative" similar to the tangent bundle when defining typical vector bundle connections. Thus, let us now introduce an object generalizing both aspects, aspects of Lie algebras and tangent bundles: \textbf{Lie algebroids}.
\chapter{General theory of Lie algebroids}\label{MathematicalBasics}

\section{Lie algebroids}\label{LieAoids}

In the following we follow \cite[\S VII]{DaSilva}.

\begin{definitions}{Lie algebroid, \cite[reduced definition of \S 16.1, page 113]{DaSilva}}{test}
%\leavevmode\newline
Let $\gls{E} \to N$ be a real vector bundle of finite rank. Then $E$ is a smooth Lie algebroid if there is a bundle map $\gls{1rho}: E \to \mathrm{T}N$, called the \textbf{anchor}, and a Lie algebra structure on $\Gamma(E)$ with Lie bracket $\gls{0[]E}$ satisfying
\ba
  \mleft[\mu, f \nu\mright]_E = f \mleft[\mu, \nu\mright]_E + \mathcal{L}_{\rho(\mu)}(f) ~ \nu
\label{eq:E-Leibniz}
\ea
for all $f \in C^\infty(N)$ and $\mu, \nu \in \Gamma(E)$, where $\mathcal{L}_{\rho(\mu)}(f)$ is the action of the vector field $\rho(\mu)$ on the function $f$ by derivation. We will sometimes denote a Lie algebroid by $\mleft( E, \rho, \mleft[ \cdot, \cdot \mright]_E \mright)$.
\end{definitions}

\begin{remarks}{Transitive Lie algebroids, \cite[very beginning of \S 17; page 123]{DaSilva}}{TransitiveLieALgeoids}
If the anchor $\rho$ is surjective, then we say that \textbf{$E$ is transitive}.
\end{remarks}

\begin{remark}
\leavevmode\newline
We often will just write "Let $E$ be a Lie algebroid.", with that we canonically also denote the anchor by $\rho$ or $\rho_E$ and the Lie bracket by $\mleft[ \cdot, \cdot \mright]_E$ without further clarifying these notations. Furthermore, \cite[\S 16.1, page 113]{DaSilva} imposes that $\rho$ is a homomorphism of Lie brackets as a part of the definition of Lie algebroids, but we will see in the following that this is not needed, it will be already a consequence of this reduced definition as explained in \textit{e.g.}~\cite[page 68]{Homomrho}.
\end{remark}

\begin{examples}{\cite[\S 16.2, page 114]{DaSilva}}{LAoidsGeneralizeLAAndTN}
The two basic examples of Lie algebroids are the following.
\begin{enumerate}
	\item Each finite dimensional real Lie algebra is a Lie algebroid over a point set $\{*\}$ with zero anchor.
	\item The tangent bundle $\mathrm{T}N$ of any manifold $N$ where the anchor is the identity map and where the Lie bracket is the usual one of vector fields.
\end{enumerate}
\end{examples}

As shown by the basic examples above, the idea behind Lie algebroids is that they are a simultaneous generalization of tangent bundles and Lie algebras, this allows a generalization of specific terms of their calculus to Lie algebroids. We will also always view tangent bundles as Lie algebroids given by the structure presented in Ex.~\ref{ex:LAoidsGeneralizeLAAndTN}.

\begin{definitions}{Basic calculus on Lie algebroids $E$}{BasicCalculusOf LieAlgoide}
Let $E \to N$ be a Lie algebroid and $V\to N$ a vector bundle, then we define the following:
\begin{itemize}
	\item \textbf{Structure functions}, \cite[\S 16.5, page 119]{DaSilva}
	\newline Let $\mleft( e_a \mright)_a$ be some local frame over some open subset $U \subset N$. Then the \textbf{structure functions} $\gls{Cbca} \in C^\infty(U)$ are defined by 
	\ba
	[e_b, e_c]_E = C^a_{bc} e_a.
	\ea
	\item \textbf{$E$-Lie derivatives}, \cite[\S 16.1; page 113]{DaSilva}
	\newline One can define \textbf{$E$-Lie derivatives}, similar as in the situation of tangent bundles, by
	\ba
	\gls{Lie}_\mu(\nu) &\coloneqq [\mu, \nu]_E, \\
	\gls{Lie}_\mu(f) &\coloneqq \mathcal{L}_{\rho(\mu)}(f)
	\ea
	for all $f \in C^\infty(N)$ and $\mu, \nu \in \Gamma(E)$. The Leibniz rule \eqref{eq:E-Leibniz} then reads
	\ba
	\mathcal{L}_\mu (f\nu) &= f \mathcal{L}_\mu(\nu) + \mathcal{L}_\mu (f) ~ \nu
	\ea
	for all $f \in C^\infty(N)$ and $\mu, \nu \in \Gamma(E)$. We will use both notations, $\mathcal{L}_\mu$ and $\mathcal{L}_{\rho(\mu)}$; it is clear by context which is meant.
	\item \textbf{$E$-forms}, \cite[\S 18.1; page 131]{DaSilva}
	\newline The antisymmetric parts of $(0,s)$-$E$-tensors define the \textbf{$E$-forms}, \textit{i.e.}~$\gls{1ZOmegas(E)} \coloneqq \Gamma\mleft(\bigwedge^s E^*\mright)$ ($s \in \mathbb{N}_0$). The previously defined Lie derivative can be extended to those forms (and general $E$-tensors) with the typical definitions by imposing the Leibniz rule. As for typical forms, one can define \textbf{$E$-forms with values in $V$} by $\gls{1ZOmegap(EV)} \coloneqq \Gamma\mleft(\bigwedge^s E^* \otimes V \mright)$.
	\item \textbf{$E$-differential}, \cite[\S 18.1, page 131]{DaSilva}
	\newline The \textbf{$E$-differential} is defined as $\gls{dE}: \Omega^\bullet(E) \to \Omega^{\bullet+1}(E)$ by
	\ba
	\mleft(\mathrm{d}_E \omega\mright) \mleft( \nu_0, \dots, \nu_s \mright)
	&\coloneqq
	\sum_i (-1)^i ~ \mathcal{L}_{\nu_i} \mleft( \omega\mleft( \nu_0, \dots, \widehat{\nu}_i, \dots, \nu_s \mright) \mright) \nonumber \\
	&\hspace{1cm}
		+ \sum_{i < j} (-1)^{i+j} ~ \omega\mleft( \mleft[ \nu_i, \nu_j \mright]_E, \nu_0, \dots, \widehat{\nu}_i, \dots, \widehat{\nu}_j, \dots, \nu_s \mright)
	\ea
	for all $\omega \in \Omega^s(E)$ and $\nu_0, \dots, \nu_s \in \Gamma(E)$.
\end{itemize}
\end{definitions}

\begin{remark}
\leavevmode\newline
\indent $\bullet$ $\Gamma(E)$ is an infinite-dimensional Lie algebra w.r.t.~$[\cdot, \cdot]_E$ but it should be seen as a generalization of finite dimensional Lie algebras whose "finite dimension" is the finite rank of $E$: Choose a local frame $\mleft(e_a\mright)_a$ of $E$ over an open subset $U \subset N$. As introduced, one gets in general now \textit{structure functions} $C^a_{bc} \in C^\infty(U)$ instead of \textit{structure constants} and a base of the Lie algebra is replaced by such a (local) frame on the vector bundle; recall the last section about classical gauge theory where we viewed the basis of the Lie algebra as a global constant frame.

$\bullet$ In the following we will argue that the anchor of a Lie algebroid is a homomorphism of Lie brackets (if viewed as a tensor acting on sections). With that one can then show $\mathrm{d}_E^2=0$ by precisely the same calculation as one does with respect to the de-Rham differential. As argued in \cite[\S 18.1, page 131f.]{DaSilva}, there is a one-to-one correspondence between Lie algebroid structures and such differential operators squaring to zero and satisfying the graded Leibniz rule with respect to the wedge product. Moreover, there is also a correspondence to vector bundles admitting a cohomological vector field; but we won't use these relationships which is why we are not going to state or explain these relationships explicitly.
	%
	%Due to the Leibniz rule the structure functions do not transform as a tensor in general, \textit{i.e.} let $f_a \coloneqq M_a^b e_b$ be another local frame on some open subset $V \subset U$ for some invertible matrix function $M$, then with $[f_b, f_c]_E = \tilde{C}_{bc}^a f_a$ one can calculate in a straightforward manner that
	%\ba
	%\tilde{C}_{bc}^a
	%&= M_b^d M_c^g C_{dg}^h \left( M^{-1} \right)^a_h + M_b^d \left( M^{-1} \right)^a_g \mathcal{L}_{e_d}( M^g_c) - M_c^d \left( M^{-1} \right)^a_g \mathcal{L}_{e_d}( M^g_b).
	%\ea
	%
	%It is clear that the structure functions are exactly the structure constants in the case of a Lie algebra. In the situation of tangent bundles one could take \textit{e.g.} some coordinate vector fields $\mleft(\partial_i\mright)_i$ as a local frame, and their structure functions are zero.
	%
	%With other straightforward calculations one also shows
	%\ba
	%C^a_{bc} &= - C^a_{cb}, \\
	%0
	%&=
	%C^d_{ae} C^e_{bc} + C^d_{be} C^e_{ca} + C^d_{ce} C^e_{ab}
		%+ \mathcal{L}_{e_a}\mleft( C^d_{bc} \mright)
		%+ \mathcal{L}_{e_b}\mleft( C^d_{ca} \mright)
		%+ \mathcal{L}_{e_c}\mleft( C^d_{ab} \mright),
	%\ea
	%where the first equation comes from the antisymmetry and the second one from the Jacobi identity of $\mleft[ \cdot, \cdot \mright]_E$.
\end{remark}

In older works about Lie algebroids (also in \cite{DaSilva}) one often sees that the definition also contains the condition about that the induced map $\Gamma(\rho): \Gamma(E) \to \mathfrak{X}(N)$ (which we will still denote as $\rho$) is a homomorphism of Lie algebras w.r.t. $\mleft[ \cdot, \cdot \mright]_E$ and $\mleft[ \cdot, \cdot \mright]$, the Lie bracket of vector fields $\mathfrak{X}(N)$. But that is not needed, see \textit{e.g.}~\cite[page 68]{Homomrho}. To show this we want to introduce some measures for the homomorphism property and the Jacobi identity. Let us start with the former.

\begin{definitions}{Curvature of morphisms, \newline \cite[variant of Definition 5.2.9; page 187]{mackenzieGeneralTheory}}{GeneralDefOfCurvMorphisms}
Let $E_1, E_2$ be two Lie algebroids over the same base manifold $N$. Then the \textbf{curvature of a vector bundle morphism $\xi: E_1 \to E_2$} is a map $\gls{Rxi}: \Gamma(E_1) \times \Gamma(E_1) \to \Gamma(E_2)$ defined by
\ba
R_\xi(\mu, \nu)
&\coloneqq
\mleft[ \xi(\mu), \xi(\nu) \mright]_{E_2}
	- \xi \mleft(
		\mleft[ \mu, \nu \mright]_{E_1}
	\mright)
\ea
for all $\mu, \nu \in \Gamma(E_1)$.
\end{definitions}

\begin{remark}
\leavevmode\newline
$R_\xi$ is clearly anti-symmetric.

For an anchor $\rho$ of a Lie algebroid we therefore expect $R_\rho = 0$ in case it is a homomorphism of Lie brackets.
\end{remark}

Later, in the sections about connections, we will see that it makes sense to call $R_\xi$ curvature, though one may already see why by its definition. What we want to show is that $R_\rho = 0$ for an anchor $\rho$ of a Lie algebroid. Hence, let us first show that those curvature are tensors if $\xi$ is an anchor preserving vector bundle morphism, which basically describes a morphism related to the structure given by the anchor:\footnote{In fact, one can also define vector bundles known as anchored vector bundles which are just vector bundles with a bundle map like the anchor; see \textit{e.g.}~\cite[\S 3, first part of Definition 3.1]{meinrenkensplitting}. Then the following definition is the definition of morphisms of anchored vector bundles.}

\begin{definitions}{Anchor-preserving vector bundle morphism, \newline \cite[\S 4.3, Equation (22); page 157]{mackenzieGeneralTheory}}{DefOfAnchorPreservingStuff}
Let $E_i\stackrel{\pi_i}{\to} N_i$ ($i \in \{1,2\}$) be two Lie algebroids over smooth manifolds $N_i$. Then we say that a vector bundle morphism $\xi: E_1 \to E_2$ over a smooth map $f: N_1 \to N_2$\footnote{That means $\pi_2 \circ \xi = f \circ \pi_1$.} is \textbf{anchor-preserving} if it satisfies
\ba\label{EqFuerAnchorBundleMorphisms}
\mathrm{D}f \circ \rho_{E_1}
&=
\rho_{E_2} \circ \xi.
\ea
\end{definitions}

\begin{remarks}{Notations and base-preserving morphisms}{SomeExtraNotationForAnchorBundleMorphs}
$\bullet$ As it is well-known, $\xi$ does not necessarily induce a map $\Gamma(E_1) \to \Gamma(E_2)$ on sections, that depends on how $f$ is structured. However, we have 
\bas
\pi_2 \bigl( \xi(\nu) \bigr)
&=
f\bigl( \underbrace{\pi_1 (\nu)}_{= \mathds{1}_{N_1}} \bigr)
=
f
\eas
for all $\nu \in \Gamma(E_1)$, such that $\xi$ induces a tensor on $\Gamma(E_1) \to \Gamma(f^*E_2)$ (the $C^\infty(N_1)$-linearity follows trivially); see \textit{e.g.}~\cite[paragraph after Propositon 7.10]{meinrenkenlie}, too. Recall, that we introduced that already for maps like $\mathrm{D}f$ at the end of the introduction, that is, $\mathrm{D}f \in \Omega^1(N_1; f^*\mathrm{T}N_2)$, which is also trivially an anchor-preserving vector bundle morphism over $f$. This is why we write equations like Eq.~\eqref{EqFuerAnchorBundleMorphisms} often as
\ba
\mathrm{D}f \circ \rho_{E_1}
&=
(f^*\rho_{E_2}) \circ \xi
\ea
when we view that condition as an equation for sections, in order to emphasize the relationship with the pullback;
recall that $f^*\rho_{E_2}: \Gamma(f^*E_2) \to \Gamma(f^*\mathrm{T}N_2)$. However, sometimes we also omit the notation of that pullback in that case.
\newline
\newline
$\bullet$ If $E_1, E_2$ are two Lie algebroids over the same base manifold $N$, then a vector bundle morphism $\xi: E_1 \to E_2$ is anchor-preserving if it satisfies
\ba
\rho_{E_1}
&=
\rho_{E_2} \circ \xi.
\ea
For this recall, that in this case we always mean base-preserving morphisms if not mentioning otherwise, that is, $f = \mathds{1}_N$. The anchor is therefore a trivial example for an anchor-preserving morphism.
\end{remarks}

\begin{remark}
\leavevmode\newline
As in \cite[Definition 5.2.5; page 186]{mackenzieGeneralTheory} one may also call such anchor-preserving morphisms ($E_1$-) connections; also here it will be clearer later why, but to avoid confusion with typical connections carrying a Leibniz rule (also called Koszul connection in \cite{mackenzieGeneralTheory}), we will not denote those as such.
\end{remark}

\begin{lemmata}{Curvatures are tensorial in case of anchor-preservation, \newline \cite[variant of Lemma 5.2.8; page 187]{mackenzieGeneralTheory}}{KruemmungenSindTensorenMitAnkerErhaltung}
Let $E_1, E_2$ be two Lie algebroids over the same base manifold $N$, and $\xi:E_1\to E_2$ an anchor-preserving vector bundle morphism. Then $R_\xi$ is an anti-symmetric tensor, \textit{i.e.}~it is $C^\infty(N)$-bilinear.
\end{lemmata}

\begin{remark}
\leavevmode\newline
This also shows that one could test the homomorphism property of anchors in just one frame around each point locally, because anchors are trivially anchor-preserving morphisms.
\end{remark}

\begin{proof}[Proof of Lemma \ref{lem:KruemmungenSindTensorenMitAnkerErhaltung}]
\leavevmode\newline
$R_\xi$ is clearly antisymmetric and, thus, we only need to show the $C^\infty(N)$-linearity with respect to one argument. That is, applying the Leibniz rule on both summands,
\bas
R_\rho(\mu, f \nu)
&=
\mleft[ \xi(\mu),f \xi(\nu) \mright]_{E_2}
	- \xi \mleft(
		\mleft[ \mu, f \nu \mright]_{E_1}
	\mright)
\\
&=
f R_\xi(\mu, \nu)
	+  \underbrace{\mathcal{L}_{(\rho_{E_2} \circ \xi)(\mu)}(f)}
	_{\mathcal{L}_{\rho_{E_1}(\mu)}(f)}
	~ \xi(\nu)
	- \xi\mleft( \mathcal{L}_{\rho_{E_1}(\mu)}(f) ~ \nu \mright)
\\
&=
f R_\xi(\mu, \nu)
\eas
for all $\mu, \nu \in \Gamma(E_1)$ and $f \in C^\infty(N)$.
\end{proof}

\begin{remark}
\leavevmode\newline
By using what we discussed in Remark \ref{rem:SomeExtraNotationForAnchorBundleMorphs}, one can define a curvature also for vector bundle morphisms of Lie algebroids over different bases, and that notion should still be a tensor in case of anchor-preserving morphisms, too.
\end{remark}

There is a certain relationship between the curvature of an anchor $\rho$ using the Jacobiator which will help us to show that anchors are also Lie bracket homomorphisms.

\begin{definitions}{Jacobiator, \cite[Remark 6.12; page 35]{meinrenkenlie}}{JacobiatorOfLieAlgebras}
Let $W$ be a vector space, not necessarily finite-dimensional, equipped with an antisymmetric bilinear bracket $\mleft[ \cdot, \cdot \mright]_W: W\times W \to W, (v, w) \mapsto \mleft[ v, w \mright]_W$. Then we define the \textbf{Jacobiator $\gls{J}: W \times W \times W \to W$} by 
\ba
J(\mu, \nu, \eta)
&\coloneqq 
\mleft[\mu, \mleft[\nu, \eta\mright]_W\mright]_W
	+ \mleft[\nu, \mleft[\eta, \mu\mright]_W\mright]_W 
	+ \mleft[\eta, \mleft[\mu, \nu\mright]_W\mright]_W
\ea
for all $\mu, \nu \in W$.
\end{definitions}

\begin{remark}
\leavevmode\newline
It is clear that $J = 0$ if $W = \Gamma(E)$ as Lie algebra, for $E$ a Lie algebroid. It is also trivial to see that $J$ is $\mathbb{R}$-trilinear and antisymmetric.
\end{remark}

%\begin{definitions}{Jacobiator and curvature of a bundle map}{DefMeasureofJacobiandHomom}
%Let $E \to N$ be a real vector bundle of finite rank, equipped with a bundle map $\rho: E \to \mathrm{T}N$ and an antisymmetric bilinear bracket $\mleft[ \cdot, \cdot \mright]_E$ on the space of sections $\Gamma(E)$ satisfying the Leibniz rule \eqref{eq:E-Leibniz}. Then we define the following objects.
%\begin{itemize}
	%\item \textbf{Curvature of $\rho$}, \cite[variant of Definition 5.2.9, page 187]{mackenzieGeneralTheory}
	%\newline We define the curvature of $\rho$ by $R_\rho: \Gamma(E) \times \Gamma(E) \to \mathfrak{X}(N)$, \textbf{COMMENT: Change sign here}
	%\ba
	%R_\rho(\mu, \nu)
	%&\coloneqq
	%\rho \mleft( \mleft[ \mu, \nu \mright]_E \mright)
		%- \mleft[ \rho(\mu), \rho(\nu) \mright]
	%\ea
	%for all $\mu, \nu \in \Gamma(E)$.
	%\item \textbf{Jacobiator}, \cite[Remark 6.12, page 35]{meinrenkenlie}
	%\newline We define the Jacobiator $\gls{J}: \Gamma(E) \times \Gamma(E) \times \Gamma(E) \to \Gamma(E)$,
	%\ba
	%J(\mu, \nu, \eta)
	%&\coloneqq [\mu, [\nu, \eta]_E]_E + [\nu, [\eta, \mu]_E]_E + [\eta, [\mu, \nu]_E]_E
	%\ea
	%for all $\mu, \nu, \eta \in \Gamma(E)$.
%\end{itemize}
%\end{definitions}
%

\begin{propositions}{Relation of Jacobiator and anchor, \cite[page 68]{Homomrho}}{MeasureofJacobiandHomom}
Let $E \to N$ be a real vector bundle of finite rank, equipped with a bundle map $\rho: E \to \mathrm{T}N$ and an antisymmetric bi-linear bracket $\mleft[ \cdot, \cdot \mright]_E$ on the space of sections $\Gamma(E)$ satisfying the Leibniz rule \eqref{eq:E-Leibniz} with respect to $\rho$. Then the following are equivalent:
\begin{itemize}
	\item $J$ is a tensor, where $J$ is the Jacobiator related to $\Gamma(E)$ with bracket $\mleft[ \cdot, \cdot\mright]_E$.
	\item $R_\rho = 0$.
\end{itemize}
\end{propositions}

\begin{remarks}{Anchor is a Homomorphism}{AnchorAHomom}
This implies that the anchor of a Lie algebroid is a homomorphism of Lie algebras because the definition of Lie algebroids assumes the Jacobi identity on $\mleft[ \cdot, \cdot \mright]_E$, so, $J = 0$, the zero-tensor. Vice versa, when we know that $R_\rho = 0$, then we only need to check the Jacobi identity in one frame around each point because $J$ behaves like a tensor.
\end{remarks}

\begin{proof}[Proof of Prop.~\ref{prop:MeasureofJacobiandHomom}]
\leavevmode\newline
We have
\bas
J(\mu, \nu, f \eta)
&= 
\mleft[\mu, \mleft[\nu, f \eta\mright]_E\mright]_E 
	+ \mleft[\nu, \mleft[f \eta, \mu\mright]_E\mright]_E 
	+ \mleft[f \eta, \mleft[\mu, \nu\mright]_E\mright]_E 
\\
&= \left[\mu, f \left[\nu, \eta\right]_E + \mathcal{L}_{\rho(\nu)}(f) ~ \eta\right]_E + \left[\nu, f \left[\eta, \mu\right]_E - \mathcal{L}_{\rho(\mu)}(f) ~ \eta\right]_E 
\\
&\hspace{1cm} 
+ f [\eta, [\mu, \nu]_E]_E - \mathcal{L}_{\rho([\mu, \nu]_E)}(f) ~ \eta 
\\ 
&= 
f \underbrace{\left( [\mu, [\nu, \eta]_E]_E + [\nu, [\eta, \mu]_E]_E + [\eta, [\mu, \nu]_E]_E \right)}_{= J(\mu, \nu, \eta)}
\\
&\hspace{1cm}
				+ \mathcal{L}_{\rho(\mu)}(f) ~ [\nu, \eta]_E + \mathcal{L}_{\rho(\nu)}(f) ~ [\eta, \mu]_E  - \mathcal{L}_{\rho(\mu)}(f) ~ [\nu, \eta]_E + \mathcal{L}_{\rho(\nu)}(f) ~ [\mu, \eta]_E 
\\
&\hspace{1cm}
		 + \mathcal{L}_{\rho(\mu)}\left( \mathcal{L}_{\rho(\nu)}(f) \right) \eta - \mathcal{L}_{\rho(\nu)}\left( \mathcal{L}_{\rho(\mu)}(f) \right) \eta
		- \mathcal{L}_{\rho([\mu, \nu]_E)}(f) ~ \eta 
\\
&=
f J(\mu,\nu,\eta) + \left[ \mathcal{L}_{\rho(\mu)}, \mathcal{L}_{\rho(\nu)} \right](f) ~ \eta - \mathcal{L}_{\rho([\mu, \nu]_E)}(f) ~ \eta 
\\
&= 
f J(\mu,\nu,\eta) + \mathcal{L}_{[\rho(\mu), \rho(\nu)]}(f) ~ \eta - \mathcal{L}_{\rho([\mu, \nu]_E)}(f) ~ \eta 
\\
&= 
f J(\mu,\nu,\eta) - \mathcal{L}_{R_\rho(\mu, \nu)}(f) ~ \eta
\eas
for all $\mu, \nu, \eta \in \Gamma(E)$ and $f \in C^\infty(N)$. Thus, we have
\bas
J(\mu, \nu, f \eta)
&= 
f J(\mu, \nu, f \eta)
\eas
if and only if
\bas
R_\rho(\mu,\nu)
&=
0,
\eas
	%\bas
	%J \text{ is a tensor}
	%&\Leftrightarrow
	%\mathcal{L}_{R_\rho(\mu, \nu)}(f) = 0
	%\Leftrightarrow
	%R_\rho(\mu, \nu) = 0,
	%\eas
where we use that a vector field of $N$ is zero when it always acts as zero derivation. The same argument holds for all arguments due to the antisymmetry of $J$. Hence, we get the desired equivalence of statements.
\end{proof}

In the following we introduce other important examples of Lie algebroids which we need later, see \cite[\S 16.2]{DaSilva}.

\begin{examples}{Bundle of Lie algebras, \newline \cite[\S 16.2, Example 2; page 114]{DaSilva} and \cite[\S 16.3; page 116f.]{DaSilva}}{BLA}
A \textbf{bundle of Lie algebras}, or \textbf{\gls{BLA}}, is a bundle whose fibers consist of Lie algebras, necessarily of the same dimension, giving rise to structure functions on the base manifold which should be smooth.
	
Such a bundle is a Lie algebroid with the anchor $\rho \equiv 0$.
	
The converse is also true, every Lie algebroid with zero anchor is a bundle of Lie algebras because then $\mleft[ \cdot, \cdot  \mright]_E$ behaves as a tensor due to the lack of a real Leibniz rule and is thence a field of Lie algebra brackets. This is why BLAs may be just defined as Lie algebras with zero anchor.
\end{examples}

As argued in \cite[Theorem 6.4.5; page 238f.]{mackenzieGeneralTheory}, when the Lie algebras of each fibre of a bundle of Lie algebras are isomorphic to each as Lie algebras, then we denote that as \textbf{Lie algebra bundle} (in short \textbf{LAB}).

\begin{definitions}{Lie algebra bundle (LAB), \cite[Definition 3.3.8; page 104]{mackenzieGeneralTheory}}{LAB}
Let $\mathfrak{g}$ be a Lie algebra. A \textbf{Lie algebra bundle}, or \textbf{\gls{LAB}}, is a vector bundle $K \to N$ equipped with a field of Lie algebra brackets $\mleft[ \cdot, \cdot \mright]_{\mathfrak{g}}: \Gamma(K) \times \Gamma(K) \to \Gamma(K)$, \textit{i.e.} $\mleft[ \cdot, \cdot \mright]_{\mathfrak{g}} \in \Gamma\mleft(\bigwedge^2 K^* \otimes K \mright)$ such that it restricts to a Lie algebra bracket on each fibre, and such that $K$ admits an \textbf{LAB atlas $\{ \psi_i: K|_{U_i} \to U_i \times \mathfrak{g} \}$ of LAB charts} subordinate to some open covering $\mleft( U_i \mright)_i$ of $N$, that is, an atlas such that each induced map $\psi_{i, p}: K_p \to \mathfrak{g}$ is a Lie algebra isomorphism, where $p \in U_i$, $K_p$ the fiber at $p$, $\psi_{i, p} \coloneqq \mathrm{pr}_2 \circ \mleft.\psi_i\mright|_{K_p}$ and $\mathrm{pr}_2$ is the projection onto the second factor.
\end{definitions}

We are going to discuss those later in more detail. For gauge theory the following example is of special importance, and this example emphasizes why we are interested into Lie algebroids.

\begin{definitions}{Action Lie algebroids, \cite[\S 16.2, Example 5; page 114]{DaSilva}}{ActionLieAlgebroids}
Let $\mleft(\mathfrak{g}, \mleft[\cdot, \cdot \mright]_{\mathfrak{g}}\mright)$ be a Lie algebra equipped with a Lie algebra action $\gamma: \mathfrak{g} \to \mathfrak{X}(N)$ on a smooth manifold $N$. A \textbf{transformation Lie algebroid} or \textbf{action Lie algebroid} is defined as the bundle $E \coloneqq N \times \mathfrak{g}$ over $N$ with anchor
\ba
\rho(p, v) &\coloneqq \gamma(v)|_p
\ea
for $(p, v) \in E$, and Lie bracket
\ba\label{LieBracketActionLieAlg}
	\mleft.\mleft[\mu, \nu\mright]_E\mright|_p
	&\coloneqq 
	\mleft[\mu_p, \nu_p\mright]_{\mathfrak{g}}
		+ \mleft.\mleft(\mathcal{L}_{\gamma(\mu(p))}(\nu^a) - \mathcal{L}_{\gamma(\nu(p))}(\mu^a) \mright)\mright|_p ~ e_a
\ea
	for all $p \in N$ and $\mu, \nu \in \Gamma(E)$, where one views a section $\mu \in \Gamma(E)$ as a map $\mu: N \to \mathfrak{g}$ and $\mleft( e_a \mright)_a$ is some arbitrary frame of constant sections.
\end{definitions}

\begin{remark}
\leavevmode\newline
$\mleft[ \cdot,\cdot, \mright]_E$ is here clearly well-defined since one just allows global constant frames. That is, another global and constant frame is just given by $f_b = M_b^a e_a$, where $M_b^a$ are constants (and invertible as matrix). Due to this constancy, $\mleft.\mleft(\mathcal{L}_{\gamma(\mu(p))}(\nu^a) - \mathcal{L}_{\gamma(\nu(p))}(\mu^a) \mright)\mright|_p ~ e_a$ is clearly independent of the chosen global constant frame. 

Observe also that we have
\bas
\rho(\nu)
&=
\gamma(\nu),
\\
\mleft[\mu, \nu\mright]_E
&=
\mleft[\mu, \nu\mright]_{\mathfrak{g}}
\eas
for all constant sections $\mu, \nu \in \Gamma(E)$. We can trivially view constant sections of $E$ as elements of $\mathfrak{g}$ as we did in Chapter \ref{ClassicGaugeTheory}; doing so implies that action Lie algebroids encode the Lie algebra and its action.
\end{remark}

\begin{propositions}{Action Lie algebroids are Lie algebroids, \newline \cite[\S 16.2, Example 5; page 114]{DaSilva}}{ActionLieoidsAreOids}
Let $\mleft(\mathfrak{g}, \mleft[\cdot, \cdot \mright]_{\mathfrak{g}}\mright)$ be some Lie algebra equipped with a Lie algebra action $\gamma: \mathfrak{g} \to \mathfrak{X}(N)$ on a smooth manifold $N$. Then the action Lie algebroid as defined in Def.~\ref{def:ActionLieAlgebroids} is a Lie algebroid structure on $E = N \times \mathfrak{g}$. Moreover, it is the unique Lie algebroid structure on $E$ with
\ba
\rho(\nu)
&=
\gamma(\nu),
\\
\mleft[\mu, \nu\mright]_E
&=
\mleft[\mu, \nu\mright]_{\mathfrak{g}}
\ea
for all constant sections $\mu, \nu \in \Gamma(E)$.
\end{propositions}

\begin{remark}
\leavevmode\newline
The statement about uniqueness is equivalent to say that the action Lie algebroid is the unique Lie algebroid structure on $E= N \times \mathfrak{g}$ such that the map $h$, defined by
\bas
\mathfrak{g} &\to \Gamma(E),
\\
X &\mapsto h(X) = X,
\eas
is a Lie algebra homomorphism with $\rho \circ h = \gamma$,\footnote{Observe the similarity to the definition of anchor-preserving morphisms.} where we mean with $h(X) = X$ that $h(X)$ is $X$ as constant section in $E$. That emphasizes why we are interested into Lie algebroids when we want to generalize gauge theory. Together with the uniqueness this also implies that action Lie algebroids are the unique Lie algebroid structure related to classical gauge theory; which is why we want to use those later to recover the classical theory.
\end{remark}

\begin{proof}[Proof of Prop.~\ref{prop:ActionLieoidsAreOids}]
\leavevmode\newline
First, let us show that we have a Lie algebroid structure.
By construction it is clear that $\rho$ is a bundle map, $\mleft[ \cdot, \cdot \mright]_E$ is antisymmetric and satisfies the Leibniz rule w.r.t.~$\rho$. Using a global frame of constant sections $\mleft( e_a \mright)_a$, the curvature $R_\rho$ of $\rho$ (see Def.~\ref{def:GeneralDefOfCurvMorphisms}) is zero, in fact, for any $p \in N$ we have 
\bas
R_\rho(e_a, e_b)|_p
&=
[ \rho(e_a), \underbrace{\rho(e_b)}_{\mathclap{\stackrel{\text{const.}}{=} ~\gamma(e_b|_p) = \gamma(e_b) }} ]|_p 
	- \rho_p \mleft( \mleft.\mleft[ e_a, e_b \mright]_E\mright|_p \mright)
\\
&\stackrel{\mathclap{\text{const.}}}{=} ~~
\mleft.\mleft[ \gamma(e_a), \gamma(e_b) \mright]\mright|_p 
	- \mleft.\gamma\mleft( \mleft[ e_a, e_b \mright]_{\mathfrak{g}} \mright)\mright|_p
\\
&= 
0,
\eas
where we used that $\gamma$ is a homomorphism for the last equality. Thence, $\rho$ is a homomorphism.
	
Then by using Prop.~\ref{prop:MeasureofJacobiandHomom} one can finally show that the Jacobi identity is satisfied. By using again a global constant frame $\mleft( e_a \mright)_a$ and $\mleft[ e_a, e_b \mright]_E = \mleft[ e_a, e_b \mright]_{\mathfrak{g}}$, we get
\bas
J(e_a, e_b, e_c)
&=
\mleft[e_a, \mleft[e_b, e_c\mright]_E\mright]_E 
	+ \mleft[e_b, \mleft[e_c, e_a\mright]_E\mright]_E
	+ \mleft[e_c, \mleft[e_a, e_b\mright]_E\mright]_E(p) 
\\
&\stackrel{\mathclap{\text{const.}}}{=} ~~
\mleft[e_a, \mleft[e_b, e_c\mright]_E\mright]_{\mathfrak{g}} 
	+ \mleft[e_b, \mleft[e_c, e_a\mright]_E\mright]_{\mathfrak{g}} 
	+ \mleft[e_c, \mleft[e_a, e_b\mright]_E\mright]_{\mathfrak{g}} 
\\
&\stackrel{\mathclap{\text{const.}}}{=} ~~
\mleft[e_a, \mleft[e_b, e_c\mright]_{\mathfrak{g}}\mright]_{\mathfrak{g}} 
	+ \mleft[e_b, \mleft[e_c, e_a\mright]_{\mathfrak{g}}\mright]_{\mathfrak{g}}
	+ \mleft[e_c, \mleft[e_a, e_b\mright]_{\mathfrak{g}}\mright]_{\mathfrak{g}} 
\\
&= 0.
\eas
Therefore we can conclude that this defines a Lie algebroid. Uniqueness comes by construction because constant sections describe a global frame and since we require that the anchor is a bundle morphism, and that the Lie bracket on $\Gamma(E)$ needs to satisfy the Leibniz rule; in other words the definition of the action Lie algebroid comes precisely from the motivation to impose those conditions. That is, assume that we have another bundle map $\rho^\prime: E \to \mathrm{T}N$ with
\bas
\rho^\prime(\nu)
&=
\gamma(\nu)
=
\rho(\nu)
\eas
for all constant sections $\nu \in \Gamma(E)$. Then for all sections $\eta= \eta^a e_a \in \Gamma(E)$ we have
\bas
\rho^\prime(\eta)
&=
\eta^a \rho^\prime(e_a)
=
\eta^a \rho(e_a)
=
\rho(\eta),
\eas
hence, $\rho^\prime = \rho$ follows trivially, and, so, we can assume the same anchor for any other Lie algebroid structure. For the Lie bracket assume that there is another Lie bracket $\mleft[ \cdot, \cdot \mright]_E^\prime$ on $\Gamma(E)$, satisfying the Leibniz rule with respect to $\rho^\prime=\rho$, with
\bas
\mleft[ \mu, \nu \mright]_E^\prime
&=
\mleft[\mu, \nu\mright]_{\mathfrak{g}}
=
\mleft[\mu, \nu\mright]_E
\eas
for all constant sections $\mu, \nu$. Therefore we can show for all sections $\eta = \eta^a e_a, \xi = \xi^b e_b \in \Gamma(E)$ that
\bas
\mleft[ \eta, \xi \mright]_E^\prime
&=
\eta^a \xi^b ~ \underbrace{\mleft[e_a, e_b\mright]_E^\prime}_{=\mleft[e_a, e_b\mright]_E}
	+ \mleft(\mathcal{L}_{\rho(\eta)}(\xi^a)
	- \mathcal{L}_{\rho(\xi)}(\eta^a) \mright) ~ e_a
\\
&=
\mleft[ \eta, \xi \mright]_E
\eas
for all $p \in N$, using the Leibniz rule of both brackets with respect to $\rho^\prime = \rho$. This proves the uniqueness.
\end{proof}

Recall Prop.~\ref{prop:LieRepAndLieAct}, with that we can use previous examples of Lie algebra actions to construct action Lie algebroids.

\begin{examples}{$\mathrm{su}(2)$-action Lie algebroid, recall Ex. \ref{ex:sutwoliealgactionasLiealg} and its references}{sutwoliealgactionasLiealgoid}
Let $E \coloneqq \mathbb{R}^3 \times \mathbb{R}^3 \to \mathbb{R}^3$; $e_x, e_y, e_z$ are the standard unit vectors (which we will also denote by $e_1, e_2, e_3$, corresponding to $x^1=x, x^2=y, x^3=z$), the anchor is given by $\rho(e_j) = - \epsilon_{jkl} x^k ~ \partial/\partial x^l$, where $\epsilon_{jkl}$ is the Levi-Civita tensor. The Lie bracket is given by the cross product w.r.t. $\mleft(e_i\mright)_i$, \textit{i.e.} $\mleft[ e_i, e_j \mright]_E \coloneqq e_i \times e_j$.
	
That this is an action Lie algebroid simply follows by that its Lie algebra action is induced by the Lie algebra representation introduced in Ex.~\ref{ex:sutwoliealgactionasLiealg}.
\end{examples}

\begin{examples}{Electroweak interaction coupled to a Higgs field, \newline recall Ex. \ref{ex:electroweakinteractionasLiealg} and its references}{electroweakinteractionasLiealgoid}
The action Lie algebroid corresponding to the \textbf{electroweak interaction coupled to a Higgs field} is defined as action Lie algebroid for $\mathfrak{g} \coloneqq \mathrm{su}(2) \times \mathrm{u}(1)$ over $N \coloneqq \mathbb{C}^2 (\cong \mathbb{R}^4)$. Let $\mathrm{i}$ be the imaginary number, $g_w$ and $g^\prime$ be positive real numbers (the \textbf{coupling constants}), $n_\gamma$ be a non-zero natural number (a normalization constant) and
\bas
\beta_l
&\coloneqq
g_w \frac{\mathrm{i} \sigma_l}{2}
\in \mathrm{su}(2), \quad l \in \{1,2,3\}, \\
\beta_4
&\coloneqq
g^\prime \frac{\mathrm{i}}{2 n_\gamma}
\in \mathrm{u}(1),
\eas
where the $\sigma_l$ are the Pauli matrices
\bas
\sigma_1
&\coloneqq
\begin{pmatrix}
0 & 1 \\
1 & 0
\end{pmatrix}, &
\sigma_2
&\coloneqq
\begin{pmatrix}
0 & -\mathrm{i} \\
\mathrm{i} & 0
\end{pmatrix}, &
\sigma_3
&\coloneqq
\begin{pmatrix}
1 & 0 \\
0 & -1
\end{pmatrix}.
\eas
Writing $\mathbb{C}^2 \ni \omega \coloneqq \begin{pmatrix} \omega^1 \\ \omega^2 \end{pmatrix} = \begin{pmatrix} x^1 + \mathrm{i} x^2 \\ x^3 + \mathrm{i} x^4 \end{pmatrix} \cong \begin{pmatrix} x^1 \\ x^2 \\ x^3 \\ x^4 \end{pmatrix}$ and denoting the coordinate vector fields for the $\mleft(x^i\mright)_i$ by $\partial_i$, the Lie algebra action $\gamma$ is then defined by 
\bas
\gamma\mleft(\beta_1\mright)_\omega
&\coloneqq
\frac{g_w}{2} ~ \mleft.\mleft( x^4 \partial_1 - x^3 \partial_2 + x^2 \partial_3 - x^1 \partial_4 \mright)\mright|_\omega, 
\\
\gamma\mleft(\beta_2\mright)_\omega
&\coloneqq
\frac{g_w}{2} ~ \mleft.\mleft( -x^3 \partial_1 - x^4 \partial_2 + x^1 \partial_3 + x^2 \partial_4 \mright)\mright|_\omega, 
\\
\gamma\mleft(\beta_3\mright)_\omega
&\coloneqq
\frac{g_w}{2} ~ \mleft.\mleft( x^2 \partial_1 - x^1 \partial_2 - x^4 \partial_3 + x^3 \partial_4 \mright)\mright|_\omega, 
\\
\gamma\mleft(\beta_4\mright)_\omega
&\coloneqq
\frac{g^\prime}{2} ~ \mleft.\mleft( x^1 \partial_1 + x^2 \partial_2 + x^3 \partial_3 + x^4 \partial_4 \mright)\mright|_\omega,
\eas
which is induced by the Lie algebra representation introduced in Ex.~\ref{ex:electroweakinteractionasLiealg}, hence, it defines an action Lie algebroid.
\end{examples}

Let us conclude this section by revisiting the isotropy introduced in Section \ref{IsotropyClassical}. In order to do so it is useful to start with action Lie algebroids $E = N \times \mathfrak{g} \to N$ related to a Lie algebra $\mathfrak{g}$ action $\gamma$ on a smooth manifold $N$. By Def.~\ref{def:IsotropySubalgebra} the isotropy at $p \in N$ is given by the kernel of $\gamma$ with point evaluation at $p$. However, as we have seen, this is precisely the kernel of the anchor then at point $p$. Hence, we can immediately generalize the definition of isotropies.

\begin{definitions}{Isotropies of Lie algebroids, \newline \cite[\S 16.1, comment after the remark on page 113]{DaSilva}}{IsotropyForLieAlgeoids}
Let $E \to N$ be a Lie algebroid over a smooth manifold $N$. Then the \textbf{isotropy of $E$} is defined as the kernel of the anchor $\rho$, $\mathrm{Ker}(\rho)$.
\end{definitions}

Recall the discussion after Cor.~\ref{cor:IsotropyVonLieAlgMitAdjoint}, the isotropy at a point is in general not an ideal of $\mathfrak{g}$, however, the isotropy as a kernel of the anchor is an ideal of $E$ in the sense of
\bas
\rho\bigl(\mathrm{ad}(\nu)\bigr)
&=
\rho\mleft(
	\mleft[ \nu, \cdot \mright]_E
\mright)
=
0
\eas
for all $\nu \in \Gamma(E)$ with $\rho(\nu) = 0$, using that $\rho$ is a homomorphism of Lie brackets; one can generalize this of course to open subsets of $N$. The Leibniz rule in $\mleft[ \cdot, \cdot \mright]_E$ is basically canceling the failure of being an ideal as it happened in the discussion after Cor.~\ref{cor:IsotropyVonLieAlgMitAdjoint}. Also observe that
\bas
\mleft.\mleft[ \nu, f\mu \mright]_E\mright|_p
&=
f(p) ~ \mleft.\mleft[ \nu, \mu \mright]_E\mright|_p
	+ \underbrace{\mathcal{L}_{\rho(\nu)_p}(f)}_{=0} ~ \mu_p
\eas
for all $f \in C^\infty(N)$ and $\nu, \mu \in \Gamma(E)$ such that $\rho(\nu)_p = 0$ at a fixed point $p \in N$. Hence, the Lie bracket becomes tensorial if restricted onto sections with values in the isotropy (at a point), therefore it is then a typical Lie bracket and it restricts onto each fibre such that $\mathrm{Ker}(\rho_p)$ is a Lie algebra at each point $p \in N$, as also argued in \cite[\S 16.1, comment after the remark on page 113]{DaSilva}. However, the dimension of the isotropy is in general not constant which is why the isotropy is in general not a bundle of Lie algebras; simply take an action Lie algebroid as in Ex.~\ref{ex:electroweakinteractionasLiealgoid}, especially the action is induced by a Lie algebra representation on a vector space $N = W$. The isotropy at $0 \in W$ is then always the full Lie algebra while aside that this is in general of course not the case; we called this symmetry breaking, recall the discussion after Def.~\ref{def:ClassicYMHLagrangian}. 

If the anchor is always zero, then the rank of the isotropy is constant and equals the ranks of $E$. Hence, a Lie algebroid with zero anchor is a bundle of Lie algebras, as also argued in \cite[second example in \S 16.2; page 114]{DaSilva}.

In general, the anchor gives rise to a singular foliation on $N$ due to that it is a homomorphism of Lie brackets; we will discuss this later. Let us first turn very shortly to morphisms and then to Lie algebroid connections.

\section{Morphism of Lie algebroids}\label{MorphsOfLieOids}

It is of course a natural question what a morphism of Lie algebroids is; we will only need the easier definition of morphisms for Lie algebroids over the same base, which is straightforward to formulate.

\begin{definitions}{Base-preserving morphism of Lie algebroids, \newline \cite[\S 3.3, second part of Definition 3.3.1; page 100]{mackenzieGeneralTheory}}{BasePreservingMorphismOfLieAlgebroids}
Let $\mleft( E_1, \rho_{E_1}, \mleft[ \cdot, \cdot \mright]_{E_1} \mright)$ and $\mleft( E_2, \rho_{E_2}, \mleft[ \cdot, \cdot \mright]_{E_2} \mright)$ be two Lie algebroids over the same base manifold $N$. Then a \textbf{morphism of Lie algebroids $\phi: E_1 \to E_2$ over $N$}, or a \textbf{base-preserving morphism of Lie algebroids}, is a vector bundle morphism with
\bas
\rho_{E_2} \circ \phi &= \rho_{E_1}, \\
\phi\mleft( \mleft[ \mu, \nu \mright]_{E_1} \mright) &= \mleft[ \phi(\mu), \phi(\nu) \mright]_{E_2}
\eas
for all $\mu, \nu \in \Gamma(E_1)$.

When $\phi$ is additionally an isomorphism of vector bundles then we call it an \textbf{isomorphism of Lie algebroids over $N$}, or a \textbf{base-preserving isomorphism of Lie algebroids}.
\end{definitions}

\begin{remark}
\leavevmode\newline
For a Lie algebroid $E \to N$ over a smooth manifold $N$ its anchor $\rho$ is therefore also a Lie algebroid morphism $E \to \mathrm{T}N$; recall Remark \ref{rem:AnchorAHomom}.

The first condition is actually the same as for anchor-preservation for morphisms over the same base; recall the second point in Remark \ref{rem:SomeExtraNotationForAnchorBundleMorphs}.
\end{remark}

There is also a definition of morphisms for Lie algebroids over different bases, but we will not need it which is why we are going to omit its definition; see \textit{e.g.}~\cite[\S 7]{meinrenkenlie}.

We want to introduce connections as anchor-preserving morphisms; flatness is then equivalent to say that connections are morphisms of Lie algebroids. In order to define connections like that we need to introduce the derivations on vector bundles.
 %For two Lie algebroids $E_i \to N_i$ ($i \in \{1,2\}$) over smooth manifolds $N_i$, the idea is to use that a vector bundle morphism $\phi$ over a smooth map $f$ induces a map $\Gamma(E_1) \to \Gamma(f^*E_2)$ as we have discussed before. $f^*$Then replace the previously-used 

\section{\texorpdfstring{Derivations on vector bundles $V$}{Derivations on a vector bundle}}\label{DerivationsOnvector}

In Chapter \ref{ClassicGaugeTheory} we defined Lie algebra connections to define infinitesimal gauge transformations. Let us now start to reintroduce that concept for Lie algebroids, going towards Lie algebroid connections, generalizing typical vector bundle connections.

Moreover, we want to view connections slightly different, as a certain morphism of Lie algebroids. Before we can do this we need to introduce the Lie algebroid of derivations now, which have a relationship to certain vector fields known as \textbf{linear vector fields} on a vector bundle. The following constructions are motivated by \cite[Example 3.3.4; page 102f.; and \S 3.4; page 110ff.]{mackenzieGeneralTheory}.

\begin{definitions}{Derivations on a vector bundle at a fixed point, \newline \cite[variation of Example 3.3.4, page 102f.]{mackenzieGeneralTheory}}{DifferentialOperatorsOfLieAlgebroids}
Let $V \to N$ be a vector bundle over a smooth manifold $N$ and $p \in N$; the fibre of $V$ at $p$ we denote with $V_p$. Then a \textbf{derivation on $V$ at $p$} is an $\mathbb{R}$-linear map $L: \Gamma(V) \to V_p$ for which there exists a tangent vector $a_p(L) \in \mathrm{T}_pN$ such that
\ba
L(fv)
&=
f(p) ~ L(v) + \mathcal{L}_{a_p(L)}(f) ~ v_p
\ea
for all $f \in C^\infty(N)$ and $v \in \Gamma(V)$. We say that $L$ lifts $a_p(L)$.

We define the space of all derivations on $V$ at $p$ by
\ba
\gls{DApV}
&\coloneqq
\left\{ L: \Gamma(V) \to V_p ~ \middle| ~ L \text{ a derivation on $V$ at $p$} \right\}.
\ea
%
%Similarly, a \textbf{derivation on $V$} is defined as an $\mathbb{R}$-linear map $D: \Gamma(V) \to \Gamma(V)$ for which there exists a vector field $a(D) \in \mathfrak{X}(N)$ such that
%\bas
%D(fv)
%&=
%f D(v) + \mathcal{L}_{a(D)}(f) ~ v
%\eas
%for all $f \in C^\infty(N)$ and $v \in \Gamma(V)$.
\end{definitions}

\begin{remark} \label{RemDerivationsatapointarelocally}
\leavevmode\newline
It is clear that $\mathcal{D}_p(V)$ is a vector space, where the zero element is just the zero map with $a_p(0) = 0$, and all $L \in \mathcal{D}_p(V)$ can be restricted to open subsets $U$ around $p$ with the typical arguments.
\end{remark}

Our aim is to show that the disjoint union $\mathcal{D}(V)$ of $\mathcal{D}_p(V)$ admits a vector bundle structure and even forms a Lie algebroid. Its sections have then the following form, formally already denoted by $\Gamma(\mathcal{D}(V))$.

\begin{definitions}{Derivations on a vector bundle $V$, \newline \cite[Example 3.3.4; page 102f.]{mackenzieGeneralTheory}}{DerivationsOnV}
Let $V \to N$ be a vector bundle over a smooth manifold $N$. Then a \textbf{derivation on $V$} is an $\mathbb{R}$-linear map $\mathcal{T}: \Gamma(V) \to \Gamma(V)$ such that there is a smooth vector field $a\mleft( \mathcal{T} \mright) \in \mathfrak{X}(N)$ with
\ba\label{eqDerivationsLiftASuperDuperVectorField}
\mathcal{T}(fv)
&=
f ~ \mathcal{T}(v)
	+ \mathcal{L}_{a\mleft( \mathcal{T} \mright)}(f) ~ v
\ea
for all $f \in C^\infty(N)$ and $v \in \Gamma(V)$. We say that $\mathcal{T}$ lifts $a(\mathcal{T})$.

We define the space of all derivations on $V$ by
\ba
\Gamma(\mathcal{D}(V))
&\coloneqq
\left\{
\mathcal{T}: \Gamma(V) \to \Gamma(V) ~ \middle| ~
\mathcal{T} \text{ a derivation on } V
\right\}.
\ea
\end{definitions}

\begin{remark}
\leavevmode\newline
%This definition of course also applies to $V|_U$ where $U$ is some open subset of $N$. One could also write $\Gamma(\mathcal{D}(U)) \coloneqq \Gamma(\mathcal{D}(V|_U))$.
%
It is clear that $\sEnd(V) \subset \Gamma(\mathcal{D}(V))$ with $a(A) \equiv 0$ for all $A \in \sEnd(V)$, and that $\Gamma(\mathcal{D}(V))$ is a $C^\infty(N)$-module.
\end{remark}

The following result can be seen as a generalization of the section around Remark \ref{RemTVGleichV}.
 %recall also the definition of $\overline{\mathrm{End}}$ mentioned there. 

\begin{propositions}{Isomorphisms of the space of derivations of $V$ at $p$, \newline \cite[Example 3.10]{basicconn}}{IsomorphismofDerivationonVectorbundleatabasepoint}
Let $V \to N$ be a real vector bundle with non-zero finite rank and $p \in N$ whose fiber we denote with $V_p$. Then each vector bundle connection $\nabla$ on $V$ induces a vector space isomorphism
\ba
\mathcal{D}_p(V)
&\cong
\mathrm{T}_pN \oplus \mathrm{End}(V_p)
\ea
%and
%\ba
%\mathcal{D}_p(V)
%&\cong
%\mathrm{T}_pN \oplus \overline{\mathrm{End}(V_p)}
%\subset
%\mathrm{T}_pN \oplus \mathfrak{X}(V_p).
%\ea
Under such isomorphisms $a_p: \mathcal{D}_p(V) \to \mathrm{T}_pN$, $L \mapsto a_p(L)$ is the projection onto the first factor.
\end{propositions}

\begin{remark}
\leavevmode\newline
The last statement shows why we say that $a_p(L)$ is lifted by $L \in \mathcal{D}_p(V)$.
\end{remark}

\begin{proof}
\leavevmode\newline
Define $T: \mathrm{T}_pN \oplus \mathrm{End}(V_p) \to \mathcal{D}_p(V)$ by
\ba 
(X, A) &\mapsto T(X, A), 
\nonumber\\ \label{EqFibrewisesupderduperisomorphismofderivations}
\mleft( T(X, A) \mright) (v) 
&\coloneqq T(X, A)(v) \coloneqq 
\nabla_X v|_p + A(v_p)
\ea
for all $v \in \Gamma(V)$. $T$ is clearly bilinear, and $T(X,A)$ clearly defines a derivation at $p$. For injectivity, observe
\bas
\nabla_X
&=
-A,
\eas
for all $(X, A)$ in the kernel of $T$, which is clearly a contradiction to the Leibniz rule in $\nabla_X$ when $X \neq 0$ due to the fact that $V$ has a non-zero rank. Thus, for such $(X,A)$, $X=0$ and then clearly also $A=0$; so, injectivity is given.

For surjectivity observe for all $L \in \mathcal{D}_p(V)$, 
%using a local frame $\mleft( e_a \mright)_a$ of $V$,
\bas
L(v)
&=
\mleft.\nabla_{a_p(L)} v\mright|_p
	+ L(v) - \mleft.\nabla_{a_p(L)} v\mright|_p
\eas
hence, use $X \coloneqq a_p(L) \in \mathrm{T}_pN$ and define $A \coloneqq L - \nabla_{a_p(L)}$, which is clearly an element of $\mathrm{End}(V_p)$. Hence, $T$ is surjective, too.

That $a_p$ is under such an isomorphism the projection onto the first factor is clear by construction.
%
%The last isomorphism comes by using Cor. \ref{CorEndVGleichBarEndV}, but just using the underlying vector space isomorphism because we have not defined any Lie bracket on $\mathcal{D}_p(V)$. That $a_p$ is the projection onto the first factor is clear by construction, as argued, $X = a_p(L)$, and that is independent of the choice of frame in the definition of $T$.
\end{proof}

%\begin{remark}
%\leavevmode\newline
%Recall the isomorphism mentioned in Cor. \ref{CorEndVGleichBarEndV} such that we can now combine it with the isomorphism of the proof above to explicitly construct $\mathcal{D}_p(V) \cong \mathrm{T}_pN \oplus \overline{\mathrm{End}(V_p)}$, so, we are now going to use exactly the same notation as in the previous proof and as in Cor. \ref{CorEndVGleichBarEndV}. Take for all $x \in V_p$ for a fixed $p \in N$ a (local) section $v \in \Gamma(V)$ with $v_p = x$, and then observe that
%\bas
%T(v)
%&\stackrel{\text{Prop. \ref{PropIsomorphismofDerivationonVectorbundleatabasepoint}}}{\mapsto}
%\mleft(a_p(T), \overline{A}_{v_p} \mright)
%\in
%\mathrm{T}_pN \oplus \mathrm{T}_{x}V_p
%\eas
%for all $T \in \mathcal{D}_p(V)$
%\end{remark}

Trivially extending that isomorphism to all $p \in N$, leads to a canonical vector bundle structure inherited by the Whitney sum $\mathrm{T}N \oplus \mathrm{End}(V)$.

\begin{lemmata}{Vector bundle of derivations, \newline \cite[variation of the introduction in Example 3.3.4, page 102f.]{mackenzieGeneralTheory} and \cite[Example 3.10]{basicconn}}{LemmaVectorbundlestructureofDV}
Let $V \to N$ be a real vector bundle with non-zero rank. Then there is a unique vector bundle structure on $\mathcal{D}(V) \coloneqq \coprod_{p \in N} \mathcal{D}_p(V)$ such that $\Gamma(\mathcal{D}(V))$ of Def.~\ref{def:DerivationsOnV} is its space of smooth sections, where $\coprod$ is the disjoint union of sets.

Moreover, each connection $\nabla$ on $V$ defines a vector bundle isomorphism
\ba
\mathcal{D}(V) &\cong \mathrm{T}N \oplus \mathrm{End}(V), 
\ea
where $\mathrm{T}N \oplus \mathrm{End}(V)$ is the Whitney sum of vector bundles.
\end{lemmata}

\begin{proof}
\leavevmode\newline
This follows by Prop.~\ref{prop:IsomorphismofDerivationonVectorbundleatabasepoint}: Given a connection $\nabla$, we can define an isomorphism $T: \mathfrak{X}(N) \oplus \sEnd(V) \to \Gamma(\mathcal{D}(V))$ of $C^\infty(N)$-modules 
\ba
T(X, A)
&\coloneqq
\nabla_X + A
\ea
for all $(X, A) \in \mathfrak{X}(N) \oplus \sEnd(V)$. This shows that $\Gamma(\mathcal{D}(V))$ is a locally free sheaf of modules of constant rank, and it restricts to $\mathcal{D}_p(V)$ at $p \in N$ because $T$ restricts to the isomorphism of Prop.~\ref{prop:IsomorphismofDerivationonVectorbundleatabasepoint}. Then we make use of the 1:1 correspondence of vector bundles and locally free sheaf of modules of constant rank (over a sheaf of rings coming from a ringed space), which implies a unique vector bundle structure on $\mathcal{D}(V) \coloneqq \coprod_{p \in N} \mathcal{D}_p(V)$ such that $\Gamma(\mathcal{D}(V))$ is its space of smooth sections. Since $T$ is clearly $C^\infty(N)$-linear, we also have an isomorphism of vector bundles $\mathcal{D}(V) \cong \mathrm{T}N \oplus \mathrm{End}(V)$ by $T$.
\end{proof}

This leads to the following definitions.

\begin{definitions}{The bundle of derivations, \newline\cite[variation of Example 3.3.4, page 102f.]{mackenzieGeneralTheory}}{LieAlgebroidOfDerivations}
Let $V \to N$ be a real vector bundle with finite rank. Then we define the \textbf{bundle of derivations on $V$} as the vector bundle $\gls{DAV}$ equipped with the vector bundle structure of Lemma \ref{lem:LemmaVectorbundlestructureofDV}, assuming that the rank of $V$ is non-zero; if the rank is zero, then we define $\mathcal{D}(V) \coloneqq N \times \{0\}$.
\end{definitions}

%\begin{remark}
%\leavevmode\newline
%Sections of $\mathcal{D}(V)$
%\end{remark}

\begin{propositions}{Lie algebroid structure on $\mathcal{D}(V)$,\newline \cite[Example 3.3.4, page 102f.]{mackenzieGeneralTheory}}{LieAlgebroidOfDerivationOnV}
%\leavevmode\newline
Let $V \to N$ be a real vector bundle. $\mathcal{D}(V)$ together with $\gls{a1}$ defined by
\ba
\mathcal{D}(V) &\to \mathrm{T}N, \\
\mathcal{D}_p(V) \ni D &\mapsto a(D) \coloneqq a_p(D),
\ea
and $\mleft[ \cdot, \cdot \mright]_{\mathcal{D}(V)}$, defined by
\ba
\Gamma(\mathcal{D}(V)) \times \Gamma(\mathcal{D}(V)) &\to \Gamma(\mathcal{D}(V)), \\
(\mathcal{T}_1, \mathcal{T}_2) &\mapsto \mleft[ \mathcal{T}_1, \mathcal{T}_2 \mright]_{\mathcal{D}(V)}
\coloneqq
\mathcal{T}_1 \circ \mathcal{T}_2 - \mathcal{T}_2 \circ \mathcal{T}_1,
\ea
is a Lie algebroid with anchor $a$ and Lie bracket $\mleft[ \cdot, \cdot \mright]_{\mathcal{D}(V)}$. The anchor extended on sections is exactly the same $a$ as in Def. \ref{def:DerivationsOnV}.
\end{propositions}

\begin{remark}
\leavevmode\newline
By Prop.~\ref{prop:IsomorphismofDerivationonVectorbundleatabasepoint}, $\mathcal{D}(V)$ is also transitive.
\end{remark}

\begin{proof}
\leavevmode\newline
For $p \in N$ and for all $f \in C^\infty(N)$, $v \in \Gamma(V)$, $\alpha, \beta \in \mathbb{R}$ and $D_1, D_2 \in \mathcal{D}_p(V)$ we have
\bas
\mleft( \alpha D_1 + \beta D_2 \mright)(fv)
&=
f(p) ~ \mleft( \alpha D_1 + \beta D_2 \mright)(v) + \mathcal{L}_{a_p(\alpha D_1 + \beta D_2)}(f) ~ v_p
\eas
and
\bas
\mleft( \alpha D_1 + \beta D_2 \mright)(fv)
&=
\alpha D_1 (fv)
	+ \beta D_2 (fv) \\
&=
f(p) ~ \mleft( \alpha D_1 + \beta D_2 \mright)(v)
	+ \mleft( \alpha \mathcal{L}_{a_p(D_1)}(f) + \beta \mathcal{L}_{a_p(D_2)}(f) \mright) ~ v_p \\
&=
f(p) ~ \mleft( \alpha D_1 + \beta D_2 \mright)(v)
	+ \mathcal{L}_{\alpha a_p(D_1) + \beta a_p(D_2)}(f) ~ v_p
\eas
and, hence,
\bas
\mathcal{L}_{\alpha a_p(D_1) + \beta a_p(D_2)}(f) ~ v_p 
&=
\mathcal{L}_{a_p(\alpha D_1 + \beta D_2)}(f) ~ v_p.
\eas
For a non-zero rank we can therefore conclude
\bas
\alpha a_p(D_1) + \beta a_p(D_2)
&=
a_p(\alpha D_1 + \beta D_2).
\eas
That means that $a$ extends on sections, which gives the $a$ given in Def. \ref{def:DerivationsOnV} on sections by Lemma \ref{lem:LemmaVectorbundlestructureofDV} ($\Rightarrow$ Def. \ref{def:DerivationsOnV} gives the sections of $\mathcal{D}(V)$ $\Rightarrow$ point evaluation at $p$ of $\mathcal{T} \in \Gamma(\mathcal{D}(V))$ gives a derivation of $V$ at $p$ lifting the tangent vector $a(\mathcal{T})|_p$ which we therefore identify as $a_p(\mathcal{T}_p)$). While all of that is trivial for zero rank since then $a \equiv 0$.
%
%Now take two sections $\mathcal{T}_1, \mathcal{T}_2 \in \Gamma(\mathcal{D}(V))$, at each point $p \in N$ they describe a derivation on $V$ at $p$ which are projected onto some tangent vector at $p$, $a_p\mleft(\mleft.\mathcal{T}_1\mright|_p\mright)$ and $a_p\mleft(\mleft.\mathcal{T}_2\mright|_p\mright)$. Since $a$ is a smooth vector bundle morphism we can conclude that $\mathfrak{X}(N) \ni a\mleft(\mathcal{T}_i\mright), p \mapsto a_p\mleft( \mathcal{T}_i \mright)$, and
%\ba
%\mathcal{T}_i(fv)
%&=
%f ~ \mathcal{T}_i(v)
	%+ \mathcal{L}_{a\mleft(\mathcal{T}_i\mright)}(f) ~ v\label{eqDerivationsLiftASuperDuperVectorField}
%\ea
%for all $f \in C^\infty(N)$, $v \in \Gamma(V)$ and $i \in \{1,2\}$.

That $\mleft[ \cdot, \cdot \mright]_{\mathcal{D}(V)}$ is a Lie bracket is clear since it is just the typical commutator of linear operators on a (infinite-dimensional) vector space. Thence, the only thing left is to show the Leibniz rule, which simply follows by
\bas
\mleft[ \mathcal{T}_1, f \mathcal{T}_2 \mright]_{\mathcal{D}(V)}(v)
&=
\mathcal{T}_1\mleft( f \mathcal{T}_2(v) \mright)
	- f ~ \mathcal{T}_2\mleft( \mathcal{T}_1(v) \mright)
\stackrel{\text{Eq. \eqref{eqDerivationsLiftASuperDuperVectorField}}}{=}
f ~ \mleft[ \mathcal{T}_1, \mathcal{T}_2 \mright]_{\mathcal{D}(V)}(v)
	+ \mathcal{L}_{a\mleft( \mathcal{T}_1 \mright)}(f) ~ \mathcal{T}_2(v)
\eas
for all $f \in C^\infty(N)$, $v \in \Gamma(V)$.
\end{proof}

As usual for differential operators, we will identify those derivations as certain vector fields, following \cite[beginning of \S 2; $\Gamma(\mathcal{D}(V))$ is there denoted as $\saut(E)$]{meinrenkensplitting} and \cite[\S 3.4 \textit{et seq.}; page 110ff.]{mackenzieGeneralTheory}. For the following recall that for each vector bundle $V \stackrel{\pi}{\to} N$ there is also a vector bundle structure for $\mathrm{T}V \stackrel{\mathrm{D}\pi}{\to} \mathrm{T}N$, and the following diagram describes a double vector bundle
\begin{center}
	\begin{tikzcd}
		 \mathrm{T}V \arrow{r}{\mathrm{D}\pi} \arrow{d}{\pi_{\mathrm{T}V}} & \mathrm{T}N \arrow{d}{\pi_{\mathrm{T}N}} \\
		V \arrow[r, "\pi"]& N
	\end{tikzcd}
\end{center}
that is, each horizontal and vertical line is a vector bundle, and the horizontal and vertical scalar multiplications on $\mathrm{T}V$ commute, see \textit{e.g.}~\cite[\S 3ff.]{Highervectorbundles}. Let us shortly recap the vector bundle structure of $\mathrm{T}V \stackrel{\mathrm{D}\pi}{\to} \mathrm{T}N$, following \cite[discussion at the beginning of \S 3.4; page 110ff.]{mackenzieGeneralTheory}: The linear structure at $v \in \mathrm{T}_p N$ ($p \in N$) is basically given by the vertical structure of $V$ prolonged along the fibre $V_p$, but as an affine space whose offset is given by $v$. That is, let $\xi, \eta \in \mathrm{T}V$ with 
\bas
\mathrm{D}_{\pi_{\mathrm{T}V}(\xi)}\pi(\xi)
&=
\mathrm{D}_{\pi_{\mathrm{T}V}(\eta)}\pi(\eta)
\eqqcolon
v,
\eas
and, hence, due to $\pi_{\mathrm{T}N}(v) = p$,
\bas
p
&= 
(\pi \circ \pi_{\mathrm{T}V})(\xi)
=
(\pi \circ \pi_{\mathrm{T}V})(\eta).
\eas
Thus, one can take curves $f,h: I \to V$ ($I \in \mathbb{R}$ an open interval around 0) with
\bas
f(0)
&=
\pi_{\mathrm{T}V}(\xi),
&
\mleft.\frac{\mathrm{d}}{\mathrm{d}t}\mright|_{t=0} f
&=
\xi,
\\
h(0)
&=
\pi_{\mathrm{T}V}(\eta),
&
\mleft.\frac{\mathrm{d}}{\mathrm{d}t}\mright|_{t=0} h
&=
\eta,
\eas
such that
\bas
\pi \circ f = \pi \circ h,
\eas
because the condition on $\xi$ and $\eta$ imply on the base paths $\pi \circ f, \pi \circ h: I \to N$ that
\bas
(\pi\circ f)(0)
&=
p
=
(\pi \circ h)(0),
\\
\mleft.\frac{\mathrm{d}}{\mathrm{d}t}\mright|_{t=0} \bigl( \pi \circ f \bigr)
&=
\mathrm{D}_{\pi_{\mathrm{T}V}(\xi)}(\xi)
=
\mathrm{D}_{\pi_{\mathrm{T}V}(\eta)}(\eta)
=
\mleft.\frac{\mathrm{d}}{\mathrm{d}t}\mright|_{t=0} \bigl( \pi \circ h \bigr).
\eas
%So, one takes one base path satisfying those, and then $f$ and $h$ are the lifts this base path, which is possible because $\xi$ and $\eta$ just differ along the vertical structures (base points in same fibre, and sharing the same projection under $\mathrm{D}\pi$).
Then the addition and scalar multiplication with $\lambda \in \mathbb{R}$ for $\mathrm{T}V \stackrel{\mathrm{D}\pi}{\to} \mathrm{T}N$ is defined by
\bas
\xi \RPlus \eta
&\coloneqq
\mleft.\frac{\mathrm{d}}{\mathrm{d}t}\mright|_{t=0} (f + h),
\\
\lambda \boldsymbol{\cdot} \xi
&\coloneqq
\mleft.\frac{\mathrm{d}}{\mathrm{d}t}\mright|_{t=0} (\lambda h),
\eas
where the addition of curves is well-defined because of $\pi \circ f = \pi \circ h$ which implies $\pi(f+h)= \pi(f) = \pi(h)$; so, one can take the sum of the curves and
\bas
\mathrm{D}\pi(\xi \RPlus \eta)
&=
\mleft.\frac{\mathrm{d}}{\mathrm{d}t}\mright|_{t=0} \bigl( \underbrace{\pi (f+h)}_{= \pi(f)} \bigr)
=
\mathrm{D}\pi(\xi)
=
v.
\eas
In other words, those operations come from interpreting tangent vectors as equivalence classes of curves, assuming there are representatives of the classes sharing the same base path ($\pi \circ f = \pi \circ h$) with which one can do those operations.
It is trivial to show that we have a double vector bundle. The operations of the linear structure in $\mathrm{T}V \stackrel{\pi_{\mathrm{T}V}}{\to} V$ is still denoted in the same manner as usual, and by definition one also gets
\bas
\pi_{\mathrm{T}V}(\xi \RPlus \eta)
&=
\pi_{\mathrm{T}V}(\xi)
	+ \pi_{\mathrm{T}V}(\eta),
\\
\pi_{\mathrm{T}V}(\lambda \boldsymbol{\cdot} \xi)
&=
\lambda ~ \pi_{\mathrm{T}V}(\xi).
\eas

\begin{definitions}{Linear vector fields, \cite[Definition 3.4.1; page 113]{mackenzieGeneralTheory}}{LinearVectorFieldsOnVectorBundles}
Let $V \stackrel{\pi}{\to} N$ be a vector bundle over a smooth manifold $N$. Then a \textbf{linear vector field on $V$} is a vector field $\xi \in \mathfrak{X}(V)$ which is also a vector bundle morphism $V \to \mathrm{T}V$ over a vector field $X \in \mathfrak{X}(N)$, \textit{i.e.}~on one hand the following diagram commutes
\begin{center}
	\begin{tikzcd}
		 V \arrow{r}{\xi} \arrow{d}{\pi} & \mathrm{T}V \arrow{d}{\mathrm{D}\pi} \\
		N \arrow[r, "X"]& \mathrm{T}N
	\end{tikzcd}
\end{center}
that is
\ba\label{LiftingVectorFieldsByLinearOnes}
\mathrm{D}\pi \circ \xi &= X \circ \pi = \pi^*X,
\ea
and on the other hand we have additionally
\ba\label{LinearityOfLinearVectorFields}
\xi_{\alpha x + \beta y}
&=
\alpha \boldsymbol{\cdot} \xi_x
	\RPlus \beta \boldsymbol{\cdot} \xi_y
\ea
for all $x, y \in V$ with $\pi(x) = \pi(y)$ and $\alpha, \beta \in \mathbb{R}$.

We say that \textbf{$\xi$ lifts $X$}.
\end{definitions}

\begin{remarks}{Coordinates on $\mathrm{T}V$}{CoordinateOnTangentStuffFOrLinearVectorFields}
As usual, vector fields are locally determined by their action on coordinate functions, that is, denote with $x^i$ coordinates on $N$, then coordinates on $V$ are given by $\pi^*x^i$ and $y^j$, where the latter are the fibre coordinates, given by a local trivialization, especially $y^j$ are (local) smooth and fibre-linear functions on $V$, elements of $\Gamma(V^*)$, whose set we denote by $C^\infty_{\mathrm{lin}}(V) \coloneqq \Gamma(V^*)$ as in \cite{mackenzieGeneralTheory}. That means that (linear) vector fields on $V$ are uniquely given by their action on $\pi^*C^\infty(N)$ and $C^\infty_{\mathrm{lin}}(V) \coloneqq \Gamma(V^*)$, we will emphasize this in the following proposition.
%
%Using a trivialisation, it is trivial to see that the coordinate vector fields of the defined coordinates are linear vector fields:
%\ba
%\frac{\partial}{\partial \mleft(\pi^*x^i\mright)}
%&\text{ linear and lifting }
%\frac{\partial}{\partial x^i},
%\\
%\frac{\partial}{\partial y^j}
%&\text{ linear and lifting }
%0.
%\ea
%For the former use curves parallel to the base manifold, and for the latter curves perpendicular to the base, using the trivialization induced by those coordinates.
\end{remarks}

The following proposition shows the idea behind the linear vector fields.

\begin{propositions}{Action of linear vector fields, \newline \cite[first two statements of Proposition 3.4.2; page 113f.]{mackenzieGeneralTheory}}{ActionOfLinearVecFields}
Let $V \stackrel{\pi}{\to} N$ be a vector bundle over a smooth manifold $N$, and $\xi \in \mathfrak{X}(V)$. Then $\xi$ is a linear vector field on $V$ if and only if $\xi\mleft( \pi^*C^\infty(N) \mright) \subset \pi^*C^\infty(N)$ and $\xi \mleft( C^\infty_{\mathrm{lin}}(V) \mright) \subset C^\infty_{\mathrm{lin}}(V)$.
\end{propositions}

\begin{proof}
\leavevmode\newline
\indent $\bullet$ We prove that by first showing that Eq.~\eqref{LiftingVectorFieldsByLinearOnes} is equivalent to $\xi\mleft( \pi^*C^\infty(N) \mright) \subset \pi^*C^\infty(N)$ for $\xi \in \mathfrak{X}(V)$. Let $f \in C^\infty(N)$, then 
\bas
\xi\mleft(\pi^*f\mright)
&=
\mathrm{d}\mleft(\pi^*f\mright)(\xi)
=
\mleft(\pi^*\mathrm{d}f \mright)\bigl( \mathrm{D}\pi(\xi) \bigr).
\eas
If $\mathrm{D}\pi(\xi) = \pi^*X$ for an $X \in \mathfrak{X}(N)$, then clearly
\bas
\xi\mleft(\pi^*f\mright)
&=
\pi^*\mleft( \mathrm{d}f(X) \mright)
\in \pi^*(C^\infty(N)).
\eas
Therefore let us now show the other direction.
 %by assuming the contrary, \textit{i.e.}~assume now $\xi\mleft(\pi^*f\mright) \notin \pi^*(C^\infty(N))$.
We know that $\mathrm{D}\pi(\xi) \in \Gamma(\pi^*\mathrm{T}N)$. Let $\mleft( \partial_i = \partial/\partial x^i \mright)_i$ local coordinate vector fields on $N$, then we can write
\bas
\mathrm{D}\pi(\xi)
&=
\mathrm{d}\pi^i(\xi) ~ \pi^*\partial_i,
\eas
and, so, we get the well-known formula $\mleft(\text{for } f = x^j\mright)$
\bas
\xi\mleft(\pi^*x^j\mright)
&=
\mathrm{d}\pi^i(\xi) ~ \pi^*\mleft(\partial_i x^j\mright)
=
\mathrm{d}\pi^j(\xi).
\eas
Hence, when there is for all $f$ an $h_f \in C^\infty(N)$ with $\xi(\pi^*f) = \pi^* h_f$,\footnote{That restricts trivially to local subsets, that is, it will work for $f=x^j$, too.} then
\bas
\mathrm{D}\pi(\xi)
&=
\underbrace{\mathrm{d}\pi^i(\xi)}_{= \xi\mleft(\pi^*x^i\mright)} ~ \pi^*\partial_i
=
\pi^*\mleft(
	\sum_i h_{x^i} ~ \partial_i
\mright).
\eas
Since the coordinates $x^j$ were arbitrary, we can conclude that there is a vector field $X \in \mathfrak{X}(N)$ such that $\mathrm{D}\pi(\xi) =\pi^*X$; that is, define $X \coloneqq \sum_i h_{x^i} ~ \partial_i$, and then show it is independent of coordinates, that is, take another coordinate system $\mleft( \partial_\alpha^\prime= \partial/\partial z^\alpha \mright)_\alpha$ of $N$. Then denote with $M$ the (local) invertible Jacobian with $\partial^\prime_\alpha = M_\alpha^i \partial_i$. Since terms like $\xi\mleft(\pi^*x^i\mright)$ describe the components of $\xi$ along the coordinates $\pi^*x^i$, we can immediately conclude
\bas
\pi^*h_{z^\alpha}
&=
\xi\mleft(\pi^*z^\alpha\mright)
=
\pi^*\mleft(\mleft(M^{-1}\mright)^\alpha_i\mright) ~ \xi\mleft(\pi^*x^i\mright)
=
\pi^*\mleft(\mleft(M^{-1}\mright)^\alpha_i ~ h_{x^i}\mright).
\eas
Therefore
\bas
\sum_\alpha h_{z^\alpha} ~ \partial^\prime_\alpha
&=
\sum_i h_{x^i} ~ \partial_i,
\eas
thence, $X$ is well-defined. Thus, Eq.~\eqref{LiftingVectorFieldsByLinearOnes} is equivalent to $\xi\mleft( \pi^*C^\infty(N) \mright) \subset \pi^*C^\infty(N)$.

$\bullet$ Now let $\xi \in \mathfrak{X}(V)$ satisfying Eq.~\eqref{LiftingVectorFieldsByLinearOnes} and lifting a vector field $X \in \mathfrak{X}(N)$, $x, y \in V$ with $\pi(x)=\pi(y)$ (such that $\mathrm{D}_x\pi(\xi_x) = \mathrm{D}_y\pi(\xi_y)$ by Eq.~\eqref{LiftingVectorFieldsByLinearOnes}), and let $f_x, f_y:I \to V,$ ($I \subset \mathbb{R}$ an open interval around 0) be curves with $f_x(0)=x, f_y(0)=y$, $\pi(f_x) = \pi(f_y)$ and
\bas
\mleft. \frac{\mathrm{d}}{\mathrm{d}t} \mright|_{t=0} f_x
&=
\xi_x,
&
\mleft. \frac{\mathrm{d}}{\mathrm{d}t} \mright|_{t=0} f_y
&=
\xi_y,
\eas
then observe for all $\lambda \in C^\infty_{\mathrm{lin}}(V)$ that
\bas
\mleft(
	\alpha \boldsymbol{\cdot} \xi_x
	\RPlus \beta \boldsymbol{\cdot} \xi_y
\mright)(\lambda)
&=
\mleft(
	\mleft. \frac{\mathrm{d}}{\mathrm{d}t} \mright|_{t=0}\mleft(
		\alpha f_x + \beta f_y
	\mright)
\mright)
(\lambda)
\\
&=
\mleft. \frac{\mathrm{d}}{\mathrm{d}t} \mright|_{t=0}\bigl(
	\underbrace{\lambda \circ \mleft(
		\alpha f_x + \beta f_y
	\mright)}_
	{\mathclap{ \stackrel{\lambda \text{ linear}}{=} \alpha (\lambda \circ f_x) + \beta (\lambda \circ f_y) }}
\bigr)
\\
&=
\alpha ~ \xi_x(\lambda)
	+ \beta ~ \xi_y(\lambda)
\eas
for all $\alpha, \beta \in \mathbb{R}$. 

If $\xi$ satisfies Eq.~\eqref{LinearityOfLinearVectorFields}, then by those results
\bas
\xi_{\alpha x +\beta y}(\lambda)
&=
\alpha ~ \xi_x(\lambda)
	+ \beta ~ \xi_y(\lambda),
\eas
therefore $\xi(\lambda) \in C^\infty_{\mathrm{lin}}(V)$ and the proof is finished (due to the previous bullet point).

If, on the other hand, $\xi(\lambda) \in C^\infty_{\mathrm{lin}}(V)$, then also
\bas
\xi_{\alpha x +\beta y}(\lambda)
&=
\alpha ~ \xi_x(\lambda)
	+ \beta ~ \xi_y(\lambda)
=
\mleft(
	\alpha \boldsymbol{\cdot} \xi_x
	\RPlus \beta \boldsymbol{\cdot} \xi_y
\mright)(\lambda).
\eas
For an $h \in C^\infty(N)$ observe
\bas
\mleft(
	\alpha \boldsymbol{\cdot} \xi_x
	\RPlus \beta \boldsymbol{\cdot} \xi_y
\mright)(\pi^*h)
&=
\mleft(
	\mleft. \frac{\mathrm{d}}{\mathrm{d}t} \mright|_{t=0}\mleft(
		\alpha f_x + \beta f_y
	\mright)
\mright)
(\pi^*h)
\\
&=
\mleft. \frac{\mathrm{d}}{\mathrm{d}t} \mright|_{t=0}\bigl(
	h \circ
	\underbrace{\pi \circ \mleft(
		\alpha f_x + \beta f_y
	\mright)}_{= \pi \circ f_x}
\bigr)
\\
&=
\mathrm{d}_p h \underbrace{\mleft( \mathrm{D}_x\pi (\xi_x) \mright)}
_{ \mathclap{ \stackrel{\text{Eq.~\eqref{LiftingVectorFieldsByLinearOnes}}}{=} \mathrm{D}_{\alpha x + \beta y} \pi (\xi_{\alpha x + \beta y}) } }
\\
&=
\xi_{\alpha x + \beta y} (\pi^*h).
\eas
This proves the claim by Remark \ref{rem:CoordinateOnTangentStuffFOrLinearVectorFields}; that is, fix additionally to the coordinates $\pi^*x^i$ fibre coordinates $y^j \in C^\infty_{\mathrm{lin}}(V)$, then express $\xi$ in those coordinates by
\bas
\xi_{\alpha x + \beta y}
&=
\xi_{\alpha x + \beta y}\mleft( \pi^*x^i \mright) ~ \mleft.\pi^*\mleft(\frac{\partial}{\partial x^i}\mright)\mright|_{\alpha x + \beta y}
	+ \xi_{\alpha x + \beta y}\mleft( y^j \mright) \mleft.\frac{\partial}{\partial y^j}\mright|_{\alpha x + \beta y}
\\
&=
\mleft(
	\alpha \boldsymbol{\cdot} \xi_x
	\RPlus \beta \boldsymbol{\cdot} \xi_y
\mright)\mleft(\pi^*x^i\mright) 
~ \mleft.\pi^*\mleft(\frac{\partial}{\partial x^i}\mright)\mright|_{\alpha x + \beta y}
	+ \mleft(
	\alpha \boldsymbol{\cdot} \xi_x
	\RPlus \beta \boldsymbol{\cdot} \xi_y
\mright)\mleft(y^j\mright)
 \mleft.\frac{\partial}{\partial y^j}\mright|_{\alpha x + \beta y}
\\
&=
\alpha \boldsymbol{\cdot} \xi_x
	\RPlus \beta \boldsymbol{\cdot} \xi_y.
\eas
\end{proof}

As vector fields the linear vector fields carry a natural Lie algebroid structure when they are a closed algebra, and this is trivial to check.
%; with Prop.~\ref{prop:ActionOfLinearVecFields} we have shown that linear vector fields form a subspace of $\mathfrak{X}(V)$, giving rise to a subbundle of $\mathrm{T}V\to V$.\footnote{Observe that for this the double vector bundle structure plays a role to achieve the needed compatibility with Eq.~\eqref{LiftingVectorFieldsByLinearOnes}.}

\begin{corollaries}{Linear vector fields are a subalgebra, \newline \cite[Corollary 3.4.3; page 114]{mackenzieGeneralTheory}}{LinFieldsAsClosedSubalge}
Let $V \stackrel{\pi}{\to} N$ be a vector bundle over a smooth manifold $N$, and $\xi, \varsigma \in \mathfrak{X}(V)$ linear vector fields on $V$ lifting vector fields $X, Y \in \mathfrak{X}(N)$, respectively. Then $[\xi, \varsigma]$ is a linear vector field lifting $[X, Y]$.
\end{corollaries}

\begin{proof}
\leavevmode\newline
That $[\xi, \varsigma]$ is a linear vector field trivially follows by Prop.~\ref{prop:ActionOfLinearVecFields}, that is, compositions of linear vector fields like $\xi \circ \varsigma$ are clearly also lineary vector fields by Prop.~\ref{prop:ActionOfLinearVecFields}, thus, also $[\xi, \varsigma] = \xi \circ \varsigma - \varsigma \circ \xi$.

We also have $\mathrm{D}\pi (\xi) = \pi^*X$ and $\mathrm{D}\pi (\varsigma) = \pi^*Y$. That immediately implies
\bas
\mathrm{D}\pi\mleft( [\xi, \varsigma] \mright)
&=
\pi^*\mleft( [X, Y] \mright),
\eas
which is a well-known fact, as also given in \cite[Proposition A.1.49; page 615]{hamilton}. 

In case this is unknown for the reader:
It can be quickly shown by first observing that 
\bas
\mathcal{L}_\xi(\pi^*f)
&=
\mathcal{L}_\xi \mleft( f \circ \pi \mright)
=
\pi^*\mleft(
	\mathrm{d}f\mleft( \mathrm{D}\pi (\xi) \mright)
\mright)
=
\pi^*\mleft(
	\mathcal{L}_X(f)
\mright)
\eas
for all $f \in C^\infty(N)$,
as also given in \cite[Lemma A.1.48; page 615]{hamilton}; basically the same as for pullback connections. By definition we also clearly have $\mathrm{D}\pi(\xi)(f) = \mathcal{L}_\xi(\pi^*f)$. Therefore altogether
\bas
\pi^*\mleft( \mleft( \mathcal{L}_{X} \circ \mathcal{L}_Y \mright) (f) \mright)
&=
\mathcal{L}_{\xi}\mleft( \pi^* \mleft( \mathcal{L}_Y(f) \mright) \mright)
=
\mleft( \mathcal{L}_\xi \circ \mathcal{L}_\varsigma \mright)(\pi^*f),
\eas
thus,
\bas
\pi^*\mleft( [X, Y](f) \mright)
&=
\pi^*\bigl( \mleft(
	\mathcal{L}_{X} \circ \mathcal{L}_Y 
	- \mathcal{L}_{Y} \circ \mathcal{L}_X 
\mright) (f) \bigr)
=
\mathcal{L}_{[\xi, \varsigma]}(\pi^*f)
=
\mathrm{D}\pi\bigl([\xi, \varsigma]\bigr)(f),
\eas
which finishes the proof.
\end{proof}

Finally we can relate it to the derivations of $V$, denoting the Lie algebra of linear vector fields by $\saut(V)$; the notation comes from that one can motivate that linear vector fields are the Lie algebra of $\sAut(V)$, but we are neither going to prove nor use this, see \textit{e.g.}~the beginning of \cite{meinrenkensplitting} for a short motivation.

\begin{theorems}{Derivations as linear vector fields, \newline \cite[Theorem 3.4.5; page 116]{mackenzieGeneralTheory}}{DerivationsSindEigentlichLineareVektorfelderKrass}
Let $V \stackrel{\pi}{\to} N$ be a vector bundle over a smooth manifold $N$, and let $D$ be a map defined by
\ba
\saut(V) &\to \Gamma(\mathcal{D}(V)),
\nonumber\\
\xi &\mapsto D_\xi,
\ea
where $D_\xi \in \Gamma(\mathcal{D}(V))$ is given by
\ba
\lambda\mleft( D_\xi v \mright)
&\coloneqq
X \bigl( \lambda(v) \bigr)
	- \xi_v(\lambda)
\ea
for all $v \in \Gamma(V)$ and $\lambda \in \Gamma(V^*) = C^\infty_{\mathrm{lin}}(V)$, and where $X \in \mathfrak{X}(N)$ is the vector field lifted by $\xi$.

Then $D$ is a bracket-preserving isomorphism of $C^\infty(N)$-modules.
\end{theorems}

\begin{remark}
\leavevmode\newline
Let us show that $D$ is well-defined. Observe
\bas
\lambda\bigl(
	D_\xi (\alpha v + \beta w)
\bigr)
&=
X \bigl( \lambda(\alpha v + \beta w) \bigr)
	- \xi_{\alpha v + \beta w}(\lambda)
\\
&=
\alpha \Bigl( X \bigl( \lambda(v) \bigr) - \xi_v(\lambda) \Bigr)
	+ \beta \Bigl( X \bigl( \lambda(w) \bigr) - \xi_w(\lambda) \Bigr)
\\
&=
\alpha ~ \lambda(D_\xi v)
	+ \beta ~ \lambda(D_\xi w)
\\
&=
\lambda\mleft( \alpha D_\xi v + \beta D_\xi w \mright)
\eas
for all $v, w \in \Gamma(V)$, $\lambda \in \Gamma(V^*)$, $\xi \in \saut(V)$ (lifting $X \in \mathfrak{X}(N)$) and $\alpha, \beta \in \mathbb{R}$, using $\pi(v) = \mathds{1}_N = \pi(w)$ and Prop.~\ref{prop:ActionOfLinearVecFields}, that is, $\xi(\lambda)$ is linear. Similarly one shows for all $f \in C^\infty(N)$ that
\bas
\lambda \mleft( D_\xi (fv) \mright)
&=
X \underbrace{\bigl( \lambda(fv) \bigr)}
	_{=f \lambda(v)}
	- \xi_{fv}(\lambda)
\\
&=
f ~ \mleft( X \bigl( \lambda(v) \bigr) - \xi_v(\lambda) \mright)
	+ \mathcal{L}_X(f) ~ \lambda(v)
\\
&=
f ~ \lambda(D_\xi v)
	+ \mathcal{L}_X(f) ~ \lambda(v)
\\
&=
\lambda \mleft( f D_\xi v + \mathcal{L}_X(f) ~ v \mright).
\eas
Hence, $D_\xi \in \Gamma(\mathcal{D}(V))$.
\end{remark}

\begin{proof}[Very short sketch for the proof of Thm.~\ref{thm:DerivationsSindEigentlichLineareVektorfelderKrass}]
\leavevmode\newline
We are not going to show this because we will not need this statement, please see the reference; the proof is relatively straightforward, but using several tricks. One first shows that $\saut(V)$ are sections of a certain Lie algebroid isomorphic to $\mathcal{D}(V^*)$ such that one essentially needs to show that $\mathcal{D}(V) \cong \mathcal{D}(V^*)$. For all $L \in \Gamma(\mathcal{D}(V))$ one can define a $T \in \Gamma(\mathcal{D}(V^*))$ as usual by forcing the Leibniz rule as in
\bas
\bigl( T(\lambda) \bigr)(v)
&\coloneqq
a(L)\bigl( \lambda(v) \bigr)
	- \lambda\bigl(L(v)\bigr)
\eas
for all $\lambda \in \Gamma(V^*)$ and $v \in \Gamma(V)$. This defines also an isomorphism of Lie algebroids $\mathcal{D}(V) \cong \mathcal{D}(V^*)$; see more in \cite[discussion after Corollary 3.4.3; page 114ff.]{mackenzieGeneralTheory}.
\end{proof}

%With this we have an immediate natural Lie algebroid structure; for this recall that the base manifold $N$ is an embedded submanifold of the vector bundle $V$, embedded by the zero section of $V$. By the previous corollary and the Frobenius theorem. The set of linear vector fields restricted to $N$ is denoted by 
%\ba
%\mathfrak{X}_{\mathrm{lin}}
%&\coloneqq
%\left\{
%~\middle|~
%\right\}
%\ea
%
%\begin{propositions}{Lie algebroid of linear vector fields, \newline \cite[part of Theorem 3.4.5; page 116]{mackenzieGeneralTheory}}{LinVecFieldsAreSuperDuperAlgoids}
%Let $V \to N$ be a vector bundle over a smooth manifold $N$. Then the set of all linear vector fields on $V$ carry a Lie algebroid structure over $N$ whose anchor $\rho_{\mathrm{lin}}$ is given by
%\ba
%\rho_{\mathrm{lin}}(\xi)
%&\coloneqq
%X
%\ea
%for all linear vector fields $\xi$ lifting $X \in \mathfrak{X}(N)$, and its Lie bracket is $\mleft[ \cdot, \cdot \mright]_{\mathrm{lin}}$ is given by
%\ba
%\mleft[ \xi, \varsigma \mright]_{\mathrm{lin}}
%&=
%\mleft.\mleft[ \xi, \varsigma \mright]\mright|_{N}
%\ea
%\end{propositions}
%
\section{Lie algebroid connections}\label{SubsectionEDiffstuff}

In the following we will introduce the notion of $E$-connections, following partially \cite[\S 2]{basicconn}. See also \cite[\S 2.5]{ELeviCivita} \textit{e.g.} for a discussion about an $E$-Levi-Civita connection and other similar terms similar to Riemannian geometry. However, we want to introduce connections using the previous section, as in \cite{mackenzieGeneralTheory}.

\begin{definitions}{$E$-connection, $E$-curvature and $E$-torsion, \newline \cite[variation of Definition 5.2.5; page 186]{mackenzieGeneralTheory} \newline \cite[variation of Definition 5.2.9; page 187]{mackenzieGeneralTheory} \newline \cite[\S 4.1, trivial generalization of Equation (14); page 154]{mackenzieGeneralTheory}}{Econnection}
Let $E \to N$ be a Lie algebroid over a smooth manifold $N$ and $V \to N$ be a vector bundle over $N$. 
\begin{enumerate}
\item An $E$-connection on the vector bundle $V$ is a base- and anchor-preserving vector bundle morphism $\gls{0nablaE}: E \to \mathcal{D}(V)$, $\nu \mapsto {}^E\nabla_\nu$.
\item The $E$-curvature $\gls{RnablaE}$ of $^E\nabla$ is defined as in Def.~\ref{def:GeneralDefOfCurvMorphisms} by
\ba
R_{{}^E\nabla}(\mu, \nu) 
&\coloneqq 
\mleft[{}^E\nabla_\mu, {}^E\nabla_\nu\mright]_{\mathcal{D}(V)}
	- {}^E\nabla_{[\mu, \nu]_E}
\ea
for all $\mu, \nu \in \Gamma(E)$. ${}^E\nabla$ is called \textbf{flat} if its curvature vanishes.
\item In the special case of $V = E$ we can define also the $E$-torsion $\gls{tEnabla}$ as an element of $\mathcal{T}^1_2(E)$ given by
\ba
t_{{}^E\nabla}(\mu, \nu) \coloneqq {}^E\nabla_\mu \nu - {}^E\nabla_\nu \mu - [\mu, \nu]_E
\ea
for all $\mu, \nu \in \Gamma(E)$.
\end{enumerate}
\end{definitions}

\begin{remark}
\leavevmode \newline
\indent $\bullet$ The base- and anchor-preservation in the definition of an $E$-connection especially means
\bas
a \circ {}^E\nabla
&=
\rho,
\eas
so, for all $\mu \in E$ we have that ${}^E\nabla_\mu$ is $\mathbb{R}$-linear and
\bas
^E\nabla_\mu (fv) &= f ~{}^E\nabla_\mu v + \mathcal{L}_{\rho(\mu)}(f)~ v,
\eas
for all $f \in C^\infty(N)$ and $v \in \Gamma(V)$.
That it is a base-preserving vector bundle morphism, implies that one can extend ${}^E\nabla$ to sections, giving rise to an $\mathbb{R}$-linear map $\Gamma(E) \to \Gamma\bigl( \mathcal{D}(V) \bigr)$, with
\bas
^E\nabla_{f\nu} (v) &= f ~{}^E\nabla_\nu v
\eas
for all $\nu \in \Gamma(E)$, $f \in C^\infty(N)$ and $v \in \Gamma(V)$. This is precisely the typical definition of a connection, besides that the Leibniz rule is along a more general anchor. In the case of $E = \mathrm{T}N$, especially $\rho_E = \mathds{1}_{\mathrm{T}N}$, we have a typical vector bundle connection, and it is trivial to see that both definitions are equivalent in that situation.

$\bullet$ As noted at the end of the introduction, when write "connection" or "vector bundle connection", then we always mean typical $\mathrm{T}N$-connections.

$\bullet$ This clearly generalizes the concept of Lie algebra connections as in Def.~\ref{def:FirstStepLieDerivativeOfAnchors}, for example look at an action Lie algebroid, but now with the tensorial behaviour again due to the bundle structure.

$\bullet$ As for vector bundle connections, one can view the curvature as a map 
\bas
R_{{}^E\nabla}: \Gamma(E) \times \Gamma(E) \times \Gamma(V) &\to \Gamma(V),
\\
(\mu,\nu,v) &\mapsto R_{{}^E\nabla}(\mu,\nu)v
=
{}^E\nabla_\mu {}^E\nabla_\nu v
	- {}^E\nabla_\nu {}^E\nabla_\mu v
	- {}^E\nabla_{\mleft[ \mu, \nu \mright]_E} v.
\eas
In Lemma \ref{lem:KruemmungenSindTensorenMitAnkerErhaltung} we have that it is tensorial the first two arguments. For the third it is as for vector bundle connections,
\bas
R_{{}^E\nabla}(\mu,\nu) (fv)
&=
f ~ R_{{}^E\nabla}(\mu,\nu)v
	+ \underbrace{\mleft(
		\mathcal{L}_{\rho(\mu)}\mleft(\mathcal{L}_{\rho(\nu)} (f) \mright)
		- \mathcal{L}_{\rho(\nu)}\mleft(\mathcal{L}_{\rho(\mu)} (f) \mright)
		- \mathcal{L}_{\mleft[ \rho(\mu), \rho(\nu) \mright]_E (f)}
	\mright)}_{=0} ~ v
\\
&=
f ~ R_{{}^E\nabla}(\mu,\nu)v
\eas
for all $f \in C^\infty(N)$, $\mu, \nu \in \Gamma(E)$ and $v \in \Gamma(V)$, using that $\rho$ is a homomorphism of Lie brackets. To summarize, $a \circ R_{{}^E\nabla} = 0$, and $R_{{}^E\nabla}$ can be viewed as an element of $\mathcal{T}^1_3(E)$.

$\bullet$ As in the situation of vector bundle connections it is trivial and straightforward to check that $t_{{}^E\nabla}$ is an anti-symmetric tensor because of the fact the Leibniz rules in the connections and the Lie bracket cancel each other.
\end{remark}

In Ex.~\ref{ex:LieAlgActionIsAConnection} we had a canonical Lie algebra connection, induced by a Lie algebra action and vector bundle connection. We can generalize this connection.

\begin{examples}{Canonically induced $E$-connection, \newline \cite[first example in Example 2.8]{ELeviCivita}}{NablaRhoConnection}
Let $E \to N$ be a Lie algebroid over a smooth manifold $N$ and $V \to N$ be a vector bundle over $N$, equipped with a vector bundle connection $\nabla$. Then define ${}^E\nabla$ on $V$ by 
\ba
{}^E\nabla_\mu
&\coloneqq 
\nabla_{\rho(\mu)}
\ea
for all $\mu \in \Gamma(E)$. This is a canonical example of an $E$-connection which we will denote as $\gls{0nablarho}$.
\end{examples}

As for vector bundle connections, we can extend a given $E$-connection to $\mathcal{T}^r_s(V)$ ($r, s \in \mathbb{N}_0$).

\begin{examples}{Dual Lie algebroid connections,\newline very typical construction forcing the Leibniz rule as in \cite[Definition 2.1.36, but using connections; page 96]{hamilton}}{DualEConns}
Let $E \to N$ be a Lie algebroid over a smooth manifold $N$ and $V \to N$ be a vector bundle over $N$, equipped with an $E$-connection ${}^E\nabla$. Then we define its \textbf{dual $E$-connection} on $V^*$, still denoted as ${}^E\nabla$, by
\ba
\mleft({}^E\nabla_\nu \omega\mright)(v)
&\coloneqq
\mathcal{L}_\nu\bigl( \omega(v) \bigr)
	- \omega \mleft( {}^E\nabla_\nu v \mright)
\ea
for all $\nu \in \Gamma(E)$, $\omega \in \Gamma(V^*)$ and $v \in \Gamma(V)$. It is trivial to prove that ${}^E\nabla_\nu \omega \in \Gamma(V^*)$ and that this ${}^E\nabla$ is an $E$-connection on $V^*$. Similarly, as for vector bundle connections, one extends ${}^E\nabla$ to $\mathcal{T}^r_s(V)$ for all $r, s \in \mathbb{N}_0$, always denoted by ${}^E\nabla$.
\end{examples}

Flatness just means trivially the following by definition.

\begin{corollaries}{Flat connections, \cite[\S 5.2, Definition 5.2.9; page 187]{mackenzieGeneralTheory}}{FlatConnectionsAreLieAlgebroidMorphisms}
Let $E\to N$ be a Lie algebroid over a smooth manifold $N$ and $V \to N$ a vector bundle. Then an $E$-connection ${}^E\nabla: E \to \mathcal{D}(V)$ on $V$ is flat if and only if it is a (base-preserving) morphism of Lie algebroids.
\end{corollaries}

\begin{proof}
\leavevmode\newline
This simply follows by definition.
\end{proof}

Of special importance regarding curvatures are of course the Bianchi identities.

\begin{theorems}{Bianchi identities, \newline \cite[Satz 8.3, generalization of second statement there; page 90]{LangeIstEsHerMitDerRaumzeit} \newline \cite[reformulation of Proposition 7.1.9; page 265]{mackenzieGeneralTheory}}{1stBianchi}
%\leavevmode\newline
Let $E\to N$ be a Lie algebroid over a smooth manifold $N$, and ${}^E\nabla$ be an $E$-connection on $E$. Then the curvature $R_{{}^E\nabla}$ satisfies both Bianchi identities, \textit{i.e.}~for all $\mu, \nu, \eta \in \Gamma(E)$ we have the \textbf{first Bianchi identity}
\ba\label{eq:firstBianchi}
&R_{{}^E\nabla}(\mu, \nu) \eta + R_{{}^E\nabla}(\nu, \eta) \mu + R_{{}^E\nabla}(\eta, \mu) \nu 
\nonumber \\
&=
t_{{}^E\nabla}\mleft(t_{{}^E\nabla}(\mu, \nu), \eta\mright) + t_{{}^E\nabla}(t_{{}^E\nabla}(\nu, \eta), \mu) + t_{{}^E\nabla}(t_{{}^E\nabla}(\eta, \mu), \nu)
\nonumber \\
&\hspace{1cm}
+ \left({}^E\nabla_\mu t_{{}^E\nabla}\right)(\nu, \eta) 
+ \left({}^E\nabla_\nu t_{{}^E\nabla}\right)(\eta, \mu) + \left({}^E\nabla_\eta t_{{}^E\nabla}\right)(\mu, \nu),
\ea
and we also get the \textbf{second Bianchi identity}
\ba
0&=
\left( {}^E\nabla_\mu R_{{}^E\nabla}\right)(\nu, \eta) + \left( {}^E\nabla_\nu R_{{}^E\nabla}\right)(\eta, \mu) + \left( {}^E\nabla_\eta R_{{}^E\nabla}\right)(\mu, \nu)
\nonumber\\
&\hspace{1cm}
+ R_{{}^E\nabla}\left( t_{{}^E\nabla}(\mu, \nu), \eta \right) 
	+R_{{}^E\nabla}\left( t_{{}^E\nabla}(\nu, \eta), \mu \right)
	+ R_{{}^E\nabla}\left( t_{{}^E\nabla}(\eta, \mu), \nu \right).
\ea
\end{theorems}

\begin{remark}
\leavevmode\newline
Eq. \eqref{eq:firstBianchi} implies that $t_{{}^E\nabla}$ satisfies the Jacobi identity if ${}^E\nabla$ is flat and $t_{{}^E\nabla}$ is covariantly constant with respect to ${}^E\nabla$. Thence, it would define another Lie bracket on $\Gamma(E)$ which is $C^\infty$-bilinear. Moreover, this Lie bracket then also defines a Lie bracket on each fibre $E_p$.
\end{remark}

\begin{proof}[Proof of the first Bianchi identity]
\leavevmode\newline
The second Bianchi identity we will prove later by its generalization (see Thm.~\ref{thm:2ndBianchi} and Remark \ref{rem:FinallyTheOtherBianchiStuff}). The former statement we can prove now by showing that it is equivalent to the Jacobi identity for $[\cdot, \cdot]_E$. First observe for $\mu, \nu, \eta \in \Gamma(E)$ that
\bas
\left[\mu, \left[\nu, \eta\right]_E\right]_E
&=
\left[ \mu, -t_{{}^E\nabla}(\nu, \eta) + {}^E\nabla_\nu \eta - {}^E\nabla_\eta \nu \right]_E \\
&=
t_{{}^E\nabla}( \mu, t_{{}^E\nabla}(\nu, \eta) )
- {}^E\nabla_\mu \left( t_{{}^E\nabla} (\nu,\eta) \right)
+ {}^E\nabla_{t_{{}^E\nabla} (\nu,\eta)} \mu \\
&\hspace{1cm}-
t_{{}^E\nabla}\left( \mu, {}^E\nabla_\nu \eta \right)
+ {}^E\nabla_\mu {}^E\nabla_\nu \eta
- {}^E\nabla_{{}^E\nabla_\nu \eta} \mu \\
&\hspace{1cm}+
t_{{}^E\nabla}\left( \mu, {}^E\nabla_\eta \nu \right)
- {}^E\nabla_\mu {}^E\nabla_\eta \nu
+ {}^E\nabla_{{}^E\nabla_\eta \nu} \mu \\
&=
-t_{{}^E\nabla}( t_{{}^E\nabla}(\nu, \eta), \mu )
- {}^E\nabla_\mu \left( t_{{}^E\nabla} (\nu,\eta) \right)
+ t_{{}^E\nabla}\left( {}^E\nabla_\nu \eta, \mu \right)
+ t_{{}^E\nabla}\left( \mu, {}^E\nabla_\eta \nu \right) \\
&\hspace{1cm}+
{}^E\nabla_\mu {}^E\nabla_\nu \eta
- {}^E\nabla_\mu {}^E\nabla_\eta \nu
- {}^E\nabla_{[\nu, \eta]_E} \mu.
\eas
With $\sigma$ we will denote the cyclic sum and thence by the Jacobi identity (and the cyclic property of the total sum)
\bas
&&0&=
\sigma\left( \left[\mu, \left[ \nu, \eta \right]_E\right]_E \right) \\
&&&=
\sigma\big( -t_{{}^E\nabla}( t_{{}^E\nabla}(\nu, \eta), \mu )
- {}^E\nabla_\mu \left( t_{{}^E\nabla} (\nu,\eta) \right)
+ t_{{}^E\nabla}\left( {}^E\nabla_\nu \eta, \mu \right)\\
&&&\qquad~+
t_{{}^E\nabla}\left( \mu, {}^E\nabla_\eta \nu \right)
+ {}^E\nabla_\mu {}^E\nabla_\nu \eta
- {}^E\nabla_\mu {}^E\nabla_\eta \nu
- {}^E\nabla_{[\nu, \eta]_E} \mu \big) \\
&&&=
\sigma\big( -t_{{}^E\nabla}( t_{{}^E\nabla}(\mu, \nu), \eta )
- {}^E\nabla_\mu \left( t_{{}^E\nabla} (\nu,\eta) \right)
+ t_{{}^E\nabla}\left( {}^E\nabla_\mu \nu, \eta \right)\\
&&&\qquad~+
t_{{}^E\nabla}\left( \nu, {}^E\nabla_\mu \eta \right)
+ {}^E\nabla_\mu {}^E\nabla_\nu \eta
- {}^E\nabla_\nu {}^E\nabla_\mu \eta
- {}^E\nabla_{[\mu, \nu]_E} \eta \big) \\
&\Leftrightarrow&
\sigma\big( R_{{}^E\nabla}(\mu, \nu)\eta \big)
&=
\sigma\left( t_{{}^E\nabla}( t_{{}^E\nabla}(\mu, \nu), \eta)
+ \left( {}^E\nabla_\mu t_{{}^E\nabla} \right)(\nu, \eta) \right).
\eas
\end{proof}

In Section \ref{NewInfGaugeTrafoTrafos} we have seen that pullbacks of Lie algebra connections were important to define the infinitesimal gauge transformation. Hence, let us turn to pullbacks of Lie algebroid connections.

\section{Pullbacks of Lie algebroid connections}\label{PullbacksAlsoGeneral}

As in the discussion around Def.~\ref{def:LieAlgebraPfadeKurvi} we need to be careful about how and when we can make a pullback of Lie algebroid connections. We want to generalize Prop.~\ref{prop:FirstEPullBACkConnectionFormula}, especially recall its proof and Remark \ref{rem:ImportantRemarkAboutPullbacks}. For simplicity let us first look again at curves.

\begin{definitions}{$E$-paths, \cite[\S 2, Definition 2.4]{ELeviCivita}}{EPaths}
Let $\mleft(E, \rho, \mleft[ \cdot, \cdot \mright]_E\mright) \stackrel{\pi}{\to} N$ be a Lie algebroid and $I \subset \mathbb{R}$ an open interval. Then an \textbf{$E$-path} is a smooth map $\alpha: I \to E$ with
\ba
(\gamma^*\rho) ( \alpha)
&=
\frac{\mathrm{d}}{\mathrm{d}t} \gamma,
\ea
where the curve $\gamma: I \to N$, $t \mapsto \pi(\alpha(t))$, is the \textbf{base path of $\alpha$}. We also say that \textbf{$\gamma$ is lifted by $\alpha$}.
\end{definitions}

\begin{remark}\label{SectionsAlongCurvesAreCurvePullbacksections}
\leavevmode\newline
Recall that for a vector bundle $V \stackrel{\mathrm{pr}}{\to} N$ we say that a section of $V$ along $\gamma$ is a smooth map $v: I \to V$ with $\mathrm{pr} \circ v = \gamma$, and that we identify sections of $\gamma^*V$ with sections of $V$ along $\gamma$. That means that an $E$-path $\alpha$ can be viewed as a section of $\gamma^*E$.
\end{remark}

Using this we can define a pullback $E$-connection and a derivation along an $E$-path.

\begin{propositions}{Pull-back of an $E$-connection along an $E$-path, \newline \cite[\S 2, comment before Definition 2.4]{ELeviCivita}}{PullBackEconnAlongEPaths}
Let $E \to N$ be a Lie algebroid, $V \to N$ a vector bundle and ${}^E\nabla$ an $E$-connection on $V$. Fix an $E$-path $\alpha$, $I \ni t \mapsto \alpha(t) \in E$, with base path $\gamma$. Then there is a unique vector bundle connection $\gamma^*\mleft( {}^E\nabla \mright)$ on $\gamma^*V \to I$ with
\ba\label{eqPullbackEconnectioncondition}
\gamma^*\mleft( {}^E\nabla \mright)_{c  \frac{\mathrm{d}}{\mathrm{d}t}} \mleft( \gamma^* v \mright)
&=
\gamma^*\mleft( {}^E\nabla_{c  \alpha} v \mright)
\ea
for all $v \in \Gamma(V)$ and $c \in \mathbb{R}$.
\end{propositions}

\begin{remark}\label{RemarkNotationvonPullbackConnection}
\leavevmode\newline
As introduced, we will view ($E$-)connections as base- and anchor-preserving morphisms, and, when acting on sections, as 1-forms. In the latter case, ${}^E\nabla v \in \Omega^1(E; V)$, and the pull-back as a section gives then $\gamma^*\mleft( {}^E\nabla v \mright) \in \Gamma\mleft( \mleft(\gamma^*E\mright)^* \otimes \gamma^*V \mright)$, therefore we define $\mleft(\gamma^*\mleft( {}^E\nabla v \mright)\mright)(c\alpha) \eqqcolon \gamma^*\mleft( {}^E\nabla_{c\alpha} v \mright)$ when viewing $\alpha$ as a section of $\gamma^*E$. One could also just write ${}^E\nabla_{c\alpha} v$ when using the interpretation of connections as morphisms, because ${}^E\nabla_{c\alpha(t)}$ is then a derivation of $V$ at $\gamma(t)$ such that it is immediate that we have a section along $\gamma$ and, hence, of $\gamma^*V$. However, most of the time we prefer to write the pull-back as an accentuation.

When $\alpha = \gamma^* \nu$ for $\nu \in \Gamma(V)$, then we write $\gamma^*\mleft( {}^E\nabla_{c\nu} v \mright)$, although it looks ambiguous with the notation just discussed previously,
\bas
\gamma^*\mleft( {}^E\nabla_{c~ \gamma^*\nu} v \mright)
&=
\mleft(\gamma^*\mleft( {}^E\nabla v \mright)\mright)\mleft(c~ \gamma^*\nu\mright)
=
\gamma^*\mleft(\mleft( {}^E\nabla v \mright)(c\nu)\mright)
=
\gamma^*\mleft( {}^E\nabla_{c \nu} v \mright),
\eas
but the notation should be clear by the context.
\end{remark}

\begin{proof}[Proof of Prop.~\ref{prop:PullBackEconnAlongEPaths}]
\leavevmode\newline
As usual, the condition \eqref{eqPullbackEconnectioncondition} uniquely defines $\gamma^*\mleft( {}^E\nabla \mright)$ by using that $\gamma^*(\Gamma(V))$ generates $\Gamma(\gamma^*V)$ and extending Eq.~\eqref{eqPullbackEconnectioncondition} by forcing the Leibniz rule, \textit{i.e.}~we define
\bas
\gamma^*\mleft( {}^E\nabla \mright)_{c  \mleft.\frac{\mathrm{d}}{\mathrm{d}t}\mright|_t} \mleft( f^i ~ \gamma^* v_i \mright)
\coloneqq
c  \mleft.\frac{\mathrm{d} f^i}{\mathrm{d}t}\mright|_t ~ \mleft.\gamma^*\mleft( v_i \mright)\mright|_t
	+ f^i(t) ~ \mleft.\gamma^*\mleft({}^E\nabla_{c  \alpha} v_i\mright)\mright|_t
\eas
for all $v_i \in \Gamma(V)$, $f^i \in C^\infty(I)$, $t \in I$ and $c \in \mathbb{R}$, where the index $i$ runs over an arbitrary range; recall Def.~\eqref{FullPulbackGConnection} in the proof of Prop.~\ref{prop:FirstEPullBACkConnectionFormula}. Every other connection satisfying Eq.~\eqref{eqPullbackEconnectioncondition} has the same form by the Leibniz rule, and, so, uniqueness follows if existence is given. Hence, it is only left to prove that this gives a well-defined connection, that is, we need to prove that it is independent of the choice of generators $v_i$ as in the proof of Prop.~\ref{prop:FirstEPullBACkConnectionFormula} and that it is a connection satisfying Eq.~\eqref{eqPullbackEconnectioncondition}. Recall Remark \ref{rem:ImportantRemarkAboutPullbacks}, we especially need to check whether the Leibniz rule inherited by ${}^E\nabla$ is compatible with the Leibniz rule of connections of $\gamma^*V \to I$, for this we need to calculate
\bas
\mleft. \gamma^*\mleft({}^E\nabla_{c  \alpha} (h v)\mright)\mright|_t
&=
\underbrace{\mathcal{L}_{c\rho(\alpha(t))}}_{=~ \mathcal{L}_{c\dot{\gamma}(t)}}(h) ~ v|_{\gamma(t)}
	+ h(\gamma(t)) ~ \mleft. \gamma^*\mleft({}^E\nabla_{c  \alpha}  v\mright)\mright|_t
\\
%&=
%\gamma^*\mleft( h {}^E\nabla_{c  \alpha(t)}  v \mright)
	%+ \gamma^*\mleft(c ~ \mleft.\frac{\mathrm{d}}{\mathrm{d}t}\mright|_t(h \circ \gamma) ~ v|_{\gamma(t)} \mright) \\
&=
\mleft.\mleft(
	c ~ \frac{\mathrm{d} (h \circ \gamma)}{\mathrm{d}t} ~ \gamma^*v 
	+ ( h \circ \gamma ) ~ \gamma^*\mleft( {}^E\nabla_{c \alpha} v \mright)
\mright)\mright|_{t}
\eas
for all $v \in \Gamma(V)$ and $h \in C^\infty(N)$. Thus, the proof is then the same as for Prop.~\ref{prop:FirstEPullBACkConnectionFormula}; linearity and the Leibniz rule follow by construction, and Eq.~\eqref{eqPullbackEconnectioncondition} and the independence of the taken generators follows by the previous calculation.
\end{proof}

As usual, one can use this to define parameter derivatives.

\begin{propositions}{Derivations of sections along $E$-paths, \newline \cite[\S 2, beginning of subsection 2.3; there $\mathrm{D}/\mathrm{d}t$ is denoted as $\nabla^\alpha$]{ELeviCivita}}{DerivationAlongEPath}
Let $E \to N$ be a Lie algebroid, $V \to N$ a vector bundle and ${}^E\nabla$ an $E$-connection on $V$. Fix an $E$-path $\alpha$, $I \ni t \mapsto \alpha(t) \in E$, with base path $\gamma$. Then there is a unique differential operator $\gls{Ddt}: \Gamma\mleft(\gamma^*V\mright) \to \Gamma\mleft(\gamma^*V\mright)$ with
\ba
\frac{\mathrm{D}}{\mathrm{d}t} &\text{ is linear over } \mathbb{R}, \\
\frac{\mathrm{D}}{\mathrm{d}t}(f s)
&=
\frac{\mathrm{d}f}{\mathrm{d}t} ~ s
	+ f ~ \frac{\mathrm{D}}{\mathrm{d}t} s, \\
\mleft.\frac{\mathrm{D}}{\mathrm{d}t}\mright|_t \mleft( \gamma^* v \mright)
&=
\mleft. \gamma^*\mleft({}^E\nabla_{\alpha} v \mright)\mright|_t \label{ParameterderivativeonCurvePullbackSections}
\ea
for all $s \in \Gamma\mleft(\gamma^*V\mright)$, $v \in \Gamma(V)$, $f \in C^\infty(I)$ and $t \in I$. 
\end{propositions}

\begin{proof}
\leavevmode\newline
Uniqueness will follow again by using that $\gamma^*(\Gamma(V))$ generates $\Gamma(\gamma^*V)$ and extending Eq.~\eqref{ParameterderivativeonCurvePullbackSections} by forcing the Leibniz rule, this is given by choosing
\bas
\frac{\mathrm{D}}{\mathrm{d}t}
&\coloneqq
\gamma^*\mleft( {}^E\nabla \mright)_{\frac{\mathrm{d}}{\mathrm{d}t}}
\eas
and then everything follows by Prop.~\ref{prop:PullBackEconnAlongEPaths}.
\end{proof}

\begin{remark}\label{DdtGleichddt}
\leavevmode\newline
%\indent $\bullet$ We often write
%\bas
%\frac{\mathrm{D}_{\mleft(\alpha, ^E\nabla\mright)}}{\mathrm{d}t}
%\eas
%instead of $\frac{\mathrm{D}}{\mathrm{d}t}$. 
When $V = N \times \mathbb{R}$, then we clearly have $\mathrm{D}/\mathrm{d}t = \mathrm{d}/\mathrm{d}t$, for this use the uniqueness and define ${}^E\nabla \coloneqq \nabla^0_\rho$, where $\nabla^0 = \mathrm{d}$ is the canonical flat connection, and
\bas
\mleft.\frac{\mathrm{d}}{\mathrm{d}t}\mright|_t\underbrace{(\gamma^*v)}_{\mathclap{= v \circ \gamma: I \to \mathbb{R}}}
&=
\mathrm{d}_{\gamma(t)}v \underbrace{\mleft( \mleft.\frac{\mathrm{d}}{\mathrm{d}t}\mright|_t \gamma \mright)}
_{\mathclap{= (\gamma^*\rho)(\alpha(t))}}
=
\mleft.\gamma^*\Bigl(
	\mathrm{d}v\bigl( (\gamma^*\rho)(\alpha) \bigr)
\Bigr)\mright|_t
=
\mleft.\gamma^*\mleft( {}^E\nabla_\alpha v \mright)\mright|_t.
\eas
\end{remark}

Prop.~\ref{prop:PullBackEconnAlongEPaths} can be generalized, using the notion defined in Def.~\ref{def:DefOfAnchorPreservingStuff}.

\begin{corollaries}{Pullbacks of Lie algebroid connections by anchor-preserving morphisms}{GeneralPullbackAnchorPreserving}
Let $E_i \to N_i$ ($i \in\{1,2\}$) be two Lie algebroids over smooth manifolds $N_i$, $V \to N_2$ a vector bundle, and ${}^{E_2}\nabla$ an $E_2$-connection on $V$. Also fix an anchor-preserving vector bundle morphism $\xi: E_1 \to E_2$ over a smooth map $f: N_1 \to N_2$. Then there is a unique $E_1$-connection $f^*\mleft( {}^{E_2}\nabla \mright)$ on $f^*V$ with
\ba\label{GeneralPullbackDef}
\mleft(f^*\mleft( {}^{E_2}\nabla \mright)\mright)_\nu (f^*v)
&=
f^*\mleft(
	{}^{E_2}\nabla_{\xi(\nu)} v
\mright)
\ea
for all $v \in \Gamma(V)$ and $\nu \in \Gamma(E_1)$.
\end{corollaries}

\begin{remark}
\leavevmode\newline
This result is motivated by \cite[Example 7.7]{meinrenkenlie} where it is shown that there is a 1:1 correspondence of Lie algebroid paths and anchor-preserving morphisms. That is, let $E_1 = \mathrm{T}I$, where $I \subset \mathbb{R}$ is an open interval. Then define 
\ba\label{1to1AnchorPresAndEpAth}
\alpha
&\coloneqq
\xi\mleft( \frac{\mathrm{d}}{\mathrm{d}t} \mright),
\ea
which is a map $I \to E_2, t \mapsto \xi\mleft( \mathrm{d}/\mathrm{d}t|_t \mright)$,
such that the anchor-preservation implies
\bas
\mleft(f^*\rho_{E_2}\mright)(\alpha)
&=
\mathrm{D}f\mleft( \frac{\mathrm{d}}{\mathrm{d}t} \mright)
=
\frac{\mathrm{d}}{\mathrm{d}t} f.
\eas
Hence, $\alpha$ is an $E_2$-path lifting $f$. Vice versa one can define $\xi$ by Eq.~\eqref{1to1AnchorPresAndEpAth} if $\alpha$ is given, and then extending $\xi$ canonically to a tensor.

Furthermore, as one can see, the presented definitions of connections and their pullbacks can also be extended to vector bundles with just an anchor, without the need of a Lie bracket ($\Rightarrow$ anchored vector bundle). But as we have seen before, for example recall Remark \ref{WecombineeverythingToAvoidStrictPullbacks}, one can even generalize it further which we will do in the next statement.
\end{remark}

\begin{proof}[Proof of Cor.~\ref{cor:GeneralPullbackAnchorPreserving}]
\leavevmode\newline
We only give a sketch because the proof is exactly as in Prop.~\ref{prop:PullBackEconnAlongEPaths}, and all other similar statements as in Section \ref{NewInfGaugeTrafoTrafos}; instead of $\mathrm{d}/\mathrm{d}t$ one has essentially $\mathcal{L}_{\rho_{E_1}(\nu)}$ for $\nu \in \Gamma(E_1)$ which does neither change the structure nor the arguments of the proof. Making use of Def.~\ref{def:DefOfAnchorPreservingStuff} we get
\bas
f^*\mleft({}^{E_2}\nabla_{\xi(\nu)} (h v)\mright)
&=
(h\circ f) ~ f^*\mleft({}^{E_2}\nabla_{\xi(\nu)}  v\mright)
	+ f^*\bigl( \underbrace{\mathcal{L}_{(\rho_{E_2}\circ\xi)(\nu)}}
	_{\mathclap{ =~ \mathcal{L}_{ \mleft(\mathrm{D}f \circ \rho_{E_1}\mright)(\nu)}  }}
	(h)\bigr) ~ f^*v 
\\
%&=
%\gamma^*\mleft( h {}^E\nabla_{c  \alpha(t)}  v \mright)
	%+ \gamma^*\mleft(c ~ \mleft.\frac{\mathrm{d}}{\mathrm{d}t}\mright|_t(h \circ \gamma) ~ v|_{\gamma(t)} \mright) \\
&=
(h\circ f) ~ f^*\mleft({}^{E_2}\nabla_{\xi(\nu)}  v\mright)
	+ \mathcal{L}_{\rho_{E_1}(\nu)}(h \circ f) ~ f^*\mleft( v \mright)
\eas
for all $h \in C^\infty(N_2)$, $v \in \Gamma(V)$ and $\nu \in \Gamma(E_1)$, using
\bas
f^*\Bigl(
	\mathcal{L}_{ \mleft(\mathrm{D}f \circ \rho_{E_1}\mright)(\nu)} (h)
\Bigr)
&=
\mleft(f^*\mathrm{d}h\mright)\bigl(\mleft(\mathrm{D}f \circ \rho_{E_1}\mright)(\nu)\bigr)
=
\underbrace{\mleft(f^!\mathrm{d}h\mright)}_{\mathclap{ = \mathrm{d}f^! h }}\bigl( \rho_{E_1}(\nu) \bigr)
=
\mathcal{L}_{\rho_{E_1}(\nu)}(h \circ f).
\eas
As mentioned in the proof of Prop.~\ref{prop:PullBackEconnAlongEPaths} and Remark \ref{rem:ImportantRemarkAboutPullbacks}, this proves that the inherited Leibniz rule of ${}^{E_2}\nabla$ is compatible with the Leibniz rule of $E_1$-connections on $f^*V$. Hence, the remaining proof is then precisely as in Prop.~\ref{prop:PullBackEconnAlongEPaths} and \ref{prop:FirstEPullBACkConnectionFormula}; locally, $f^*\mleft( {}^{E_2}\nabla \mright)$ is defined by
\bas
\mleft(f^*\mleft( {}^{E_2}\nabla \mright)\mright)_\nu \mu
&\coloneqq
\mathcal{L}_{\rho_{E_1}(\nu)}\mleft(\mu^a\mright) ~ f^*e_a
	+ \mu^a ~ f^*\mleft( {}^{E_2}\nabla_{\xi(\nu)} e_a \mright)
\eas
for all $\mu = \mu^a ~ f^*e_a$, where $\mleft( e_a \mright)_a$ is a local frame of $V$. Linearity and the Leibniz rule follow by construction, and the well-definedness and Eq.~\eqref{GeneralPullbackDef} additionally by the first calculation about the compatibility of Leibniz rules.
\end{proof}

What we need is an even more general statement as in Section \ref{NewInfGaugeTrafoTrafos}, with still precisely the same proof as before; recall Prop.~\ref{prop:ClassicFunctionDerivativesAlongPsiEpsilon}.

\begin{corollaries}{Pullbacks of connections just differentiating along one vector field}{VeryGeneralPullbackConnection}
Let $E_i \to N_i$ ($i \in\{1,2\}$) be two Lie algebroids over smooth manifolds $N_i$, $V \to N_2$ a vector bundle, and ${}^{E_2}\nabla$ an $E_2$-connection on $V$. Moreover, let $f \in C^\infty(N_1;N_2)$, $\nu_1 \in \Gamma(E_1)$ and $\nu_2 \in \Gamma(f^*E_2)$ such that
\ba\label{WeakAnchorPreserv}
\mathrm{D}f\bigl(\rho_{E_1}(\nu_1)\bigr)
&=
\mleft(f^*\rho_{E_2}\mright)(\nu_2).
\ea

Then there is a unique $\mathbb{R}$-linear operator $\delta_{\nu_1}: \Gamma(f^*V) \to \Gamma(f^*V)$ with
\ba
\delta_{\nu_1}(h s)
&=
\mathcal{L}_{\nu_1}(h) ~ s
	+ h ~ \delta_{\nu_1} s,
\\
\delta_{\nu_1} (f^*v)
&=
f^*\mleft(
	{}^{E_2}\nabla_{\nu_2} v
\mright)
\ea
for all $s \in \Gamma(f^*V)$, $v \in \Gamma(V)$ and $h \in C^\infty(N_1)$.
\end{corollaries}

\begin{remarks}{Commutating diagram behind pullbacks}{CommutingDiagramOfPullbacks}
Recall Remark \ref{rem:SomeExtraNotationForAnchorBundleMorphs}, the pullback in $\mleft(f^*\rho_{E_2}\mright)(\nu_2)$ in Eq.~\eqref{WeakAnchorPreserv} is just for emphasizing that $\nu_2$ is a section along $f$; one can omit this in the notation, especially if one views sections like $\nu_2$ as a map $N_1 \to E_2$. Then we can equivalently write
\ba
\mathrm{D}f \circ \rho_{E_1}(\nu_1)
&=
\rho_{E_2} \circ \nu_2,
\ea
that is equivalent to that the following diagram commutes
\begin{center}
	\begin{tikzcd}
		N_1 \arrow{r}{\nu_2} \arrow[d, "\rho_{E_1}(\nu_1)", swap]	& E_2 \arrow{d}{\rho_{E_2}} 
		\\
		\mathrm{T}N_1 \arrow{r}{\mathrm{D}f} & \mathrm{T}N_2
	\end{tikzcd}
\end{center}
\end{remarks}

\begin{remark}\label{JustLieDerivativeForGeneralPullbackAndlineBundle}
\leavevmode\newline
\indent $\bullet$ In general one may want to write $\delta_{\nu_1} = \mleft(f^*\mleft( {}^{E_2}\nabla \mright)\mright)_{\nu_1}$, because it is precisely this by uniqueness if a general pullback is possible. But to avoid confusion about the existence of a general pullback we will stick with $\delta_{\nu_1}$, and it will be clear by context which connection and $\nu_2$ is used for the definition of $\delta_{\nu_1}$.

$\bullet$ As in Remark \ref{DdtGleichddt}, in the case of $V = \mathbb{R} \times N_2$, the trivial line bundle over $N_2$, we canonically use ${}^{E_2}\nabla \coloneqq \nabla^0_{\rho_{E_2}}$, where $\nabla^0 \coloneqq \mathrm{d}$. Then one can similarly show as before that 
\bas
\delta_{\nu_1}
&=
\mathcal{L}_{\nu_1}.
\eas
\end{remark}

\begin{proof}[Proof of Cor.~\ref{cor:VeryGeneralPullbackConnection}]
\leavevmode\newline
That is precisely the same proof as in the previous statements and as in Section \ref{ClassicGaugeTheory}; the only difference is just the meaning, $\nu_i$ are fixed sections, but that does not matter in the calculations. Eq.~\eqref{WeakAnchorPreserv} is just the condition about anchor-preservation in the case of a fixed pair of sections, and one uses this equation in the same fashion to how we used an anchor-preserving morphism in the previous proofs. Essentially replace $\nu$ with $\nu_1$ and $\xi(\nu)$ with $\nu_2$ in the proof of Cor.~\ref{cor:GeneralPullbackAnchorPreserving}.
\end{proof}

The advantage of this weak formulation is that we do not need to know whether or not $f$ can be lifted to any morphism with certain properties like anchor-preservation. Eq.~\eqref{WeakAnchorPreserv} states what one needs to make a pullback of a Lie algebroid connection to just differentiate along one direction. That was precisely the idea in the discussion around Prop.~\ref{prop:ClassicFunctionDerivativesAlongPsiEpsilon}, but now more compactly written down, not using flows of the involved vector fields.

\section{\texorpdfstring{Conjugated $E$-connections}{Conjugated Lie algebroid connections}}\label{ConjugateConnections}

Later we will introduce a Lie algebroid connection known as basic connection, and it has a special form which we want to study in a more general sense of conjugated $E$-connections; the name is motivated by \cite[paragraph after Proposition 2.12]{basicconn}, while we especially refer to \cite{blaomTangentBundleAsLieGroup} where the conjugate connections are called dual connections.

\begin{definitions}{Conjugated $E$-connections, \newline \cite[beginning of \S 4.6]{blaomTangentBundleAsLieGroup}}{ConjugationOfConnections}
Let $E \to N$ be a Lie algebroid over a smooth manifold $N$, and $\overline{\nabla}$ be an $E$-connection on $E$. We define its \textbf{conjugated $E$-connection} $\widehat{\nabla}$ by
\ba
\widehat{\nabla}_\mu \nu &\coloneqq \mleft[ \mu, \nu \mright]_E + \overline{\nabla}_\nu \mu
\ea
for all $\mu, \nu \in \Gamma(E)$. We also say that \textbf{$\widehat{\nabla}$ and $\overline{\nabla}$ are conjugate to each other}.
\end{definitions}

\begin{remark}
\leavevmode\newline
It is straightforward to check that the conjugate is an $E$-connection on $E$, linearity over $\mathbb{R}$ is clear, and we have
\bas
\widehat{\nabla}_\mu \mleft( f \nu \mright)
&=
\mleft[ \mu, f \nu \mright]_E + \overline{\nabla}_{f\nu} \mu
=
f ~\widehat{\nabla}_\mu \nu
	+ \mathcal{L}_{\rho(\mu)}(f) ~ \nu, \\
\widehat{\nabla}_{f\mu} \nu
&=
\mleft[ f \mu, \nu \mright]_E + \overline{\nabla}_\nu \mleft( f \mu \mright)
=
f ~\widehat{\nabla}_\mu \nu
	- \mathcal{L}_{\rho(\nu)}(f) ~ \mu
	+ \mathcal{L}_{\rho(\nu)}(f) ~ \mu
=
f ~\widehat{\nabla}_\mu \nu
\eas
for all $\mu, \nu \in \Gamma(E)$ and $f \in C^\infty(N)$,
using the Leibniz rule of the Lie bracket, and that $\overline{\nabla}$ is an $E$-connection. It also makes sense to say that both $E$-connections are conjugate to each other because $\overline{\nabla}$ is also the conjugate to $\widehat{\nabla}$ by definition, that is,
\bas
\mleft[ \mu, \nu \mright]_E + \widehat{\nabla}_\nu \mu
&=
\overline{\nabla}_\mu \nu,
\eas
and the conjugate of a connection is unique, that follows trivially by definition.
\end{remark}

We need several relations between their curvatures and torsions throughout this work.

\begin{corollaries}{Torsion of conjugated $E$-connections \newline \cite[first statement in the first proposition of \S 4.6]{blaomTangentBundleAsLieGroup}}{TorsionOfDualTorsions}
Let $\widehat{\nabla}$ and $\overline{\nabla}$ be two $E$-connections, conjugate to each other, on a Lie algebroid $E \to N$ over a smooth manifold $N$. Then we get for their torsions
\ba
t_{\widehat{\nabla}}(\mu, \nu)
&=
- t_{\overline{\nabla}}(\mu, \nu)
\ea
for all $\mu, \nu \in \Gamma(E)$.
\end{corollaries}

\begin{proof}
\leavevmode\newline
We have
\bas
t_{\widehat{\nabla}}(\mu, \nu)
&=
\widehat{\nabla}_\mu \nu
	- \widehat{\nabla}_\mu \nu
	- \mleft[ \mu, \nu \mright]_E
\\
&=
\mleft[ \mu, \nu\mright]_E
	+ \overline{\nabla}_\nu \mu
	- \mleft[ \nu, \mu\mright]_E
	- \overline{\nabla}_\mu \nu
	- \mleft[ \mu, \nu \mright]_E
\\
&=
\mleft[ \mu, \nu\mright]_E
	+ \overline{\nabla}_\nu \mu
	- \overline{\nabla}_\mu \nu
\\
&=
- t_{\overline{\nabla}}(\mu, \nu)
\eas
for all $\mu, \nu \in \Gamma(E)$.
\end{proof}

\begin{lemmata}{Curvature of conjugated $E$-connections, \newline the first identity comes from \cite[second statement of the first proposition in \S 4.6]{blaomTangentBundleAsLieGroup}}{CurvatureOfDualConnectionsGeneral}
Let $\widehat{\nabla}$ and $\overline{\nabla}$ be two $E$-connections, conjugate to each other, on a Lie algebroid $E \to N$ over a smooth manifold $N$. Then we have for their curvatures
\ba
R_{\overline{\nabla}}(\mu, \nu) \eta
&=
\left(\widehat{\nabla}_\eta t_{\widehat{\nabla}}\right)(\mu, \nu)
	+ R_{\widehat{\nabla}}(\mu, \eta) \nu
	- R_{\widehat{\nabla}}(\nu, \eta) \mu \label{DualCurvaRemainin1} \\
&=
- \mleft(
\widehat{\nabla}_{\eta} \left( \left[ \mu, \nu \right]_E \right)
	- \left[ \widehat{\nabla}_{\eta} \mu, \nu \right]_E
	- \left[ \mu, \widehat{\nabla}_{\eta} \nu \right]_E
	- \widehat{\nabla}_{\overline{\nabla}_\nu \eta} \mu
	+ \widehat{\nabla}_{\overline{\nabla}_\mu \eta} \nu \mright) \label{DualCurvaRemainin2} 
	%\\
%R_{\widehat{\nabla}}(\mu, \nu) \eta
%&=
%\left(\overline{\nabla}_\eta t_{\overline{\nabla}}\right)(\mu, \nu)
	%+ R_{\overline{\nabla}}(\mu, \eta) \nu
	%- R_{\overline{\nabla}}(\nu, \eta) \mu \label{DualCurvaRemainin1} \\
%&=
%- \mleft(
%\overline{\nabla}_{\eta} \left( \left[ \mu, \nu \right]_E \right)
	%- \left[ \overline{\nabla}_{\eta} \mu, \nu \right]_E
	%- \left[ \mu, \overline{\nabla}_{\eta} \nu \right]_E
	%- \overline{\nabla}_{\widehat{\nabla}_\nu \eta} \mu
	%+ \overline{\nabla}_{\widehat{\nabla}_\mu \eta} \nu \mright)
\ea
for all $\mu, \nu, \eta \in \Gamma(E)$.
\end{lemmata}

\begin{remark}
\leavevmode\newline
The second statement is a generalization of what is shown for a special type of connection in \cite[Proposition 2.12]{basicconn}.
\end{remark}

\begin{proof}[Proof of Lemma \ref{lem:CurvatureOfDualConnectionsGeneral}]
\leavevmode\newline	
We will show Eq.~\eqref{DualCurvaRemainin1} by first showing Eq.~\eqref{DualCurvaRemainin2}, but the latter for $R_{\widehat{\nabla}}$ instead of $R_{\overline{\nabla}}$; this does not matter of course, because when we know the formula for one connection, then also for the conjugated connection. Just by the definition of duality and the Jacobi identity we have
\bas
&\overline{\nabla}_{\mu} \left( \left[ \eta, \nu \right]_E \right)
	- \left[ \overline{\nabla}_{\mu} \eta, \nu \right]_E
	- \left[ \eta, \overline{\nabla}_{\mu} \nu \right]_E
	- \overline{\nabla}_{\widehat{\nabla}_\nu \mu} \eta
	+ \overline{\nabla}_{\widehat{\nabla}_\eta \mu} \nu \\
&=
\mleft[ \mu, \mleft[ \eta, \nu \mright]_E \mright]_E
	+ \mleft[ \nu, \mleft[ \mu, \eta \mright]_E \mright]_E 
	+ \mleft[ \eta, \mleft[ \nu, \mu \mright]_E \mright]_E 
\\
&\hspace{1cm}
	- \mleft[ \widehat{\nabla}_{\eta} \mu , \nu \mright]_E
	- \mleft[ \eta, \widehat{\nabla}_{\nu} \mu \mright]_E
	- \mleft[ \widehat{\nabla}_\nu \mu, \eta  \mright]_E
	+ \mleft[ \widehat{\nabla}_\eta \mu, \nu \mright]_E 
\\
&\hspace{1cm}
	+\widehat{\nabla}_\nu \widehat{\nabla}_\eta \mu
	- \widehat{\nabla}_\eta \widehat{\nabla}_\nu \mu
	+ \widehat{\nabla}_{\mleft[ \eta, \nu \mright]_E} \mu
\\
&=
R_{\widehat{\nabla}}(\nu, \eta) \mu
\\
&=
- R_{\widehat{\nabla}}(\eta, \nu) \mu
\eas
for all $\mu, \nu, \eta \in \Gamma(E)$. Eq.~\eqref{DualCurvaRemainin2} is therefore shown, and using this and Cor.~\ref{cor:TorsionOfDualTorsions} we also have
\bas
\left(\widehat{\nabla}_\eta t_{\widehat{\nabla}}\right)(\mu, \nu)
&=
-\left(\widehat{\nabla}_\eta t_{\overline{\nabla}}\right)(\mu, \nu)\\
&=
-\widehat{\nabla}_\eta \left( t_{\overline{\nabla}}(\mu, \nu) \right)
	+ t_{\overline{\nabla}}\left( \widehat{\nabla}_\eta \mu, \nu \right)
	+ t_{\overline{\nabla}}\left( \mu, \widehat{\nabla}_\eta \nu \right) \\
&=
\left[ \eta, \left[\mu, \nu\right]_E - \overline{\nabla}_{\mu} \nu + \overline{\nabla}_{\nu} \mu \right]_E
+ \overline{\nabla}_{\underbrace{ \left[\mu, \nu\right]_E - \overline{\nabla}_{\mu} \nu + \overline{\nabla}_{\nu} \mu }_{=\overline{\nabla}_{\nu} \mu - \widehat{\nabla}_\nu \mu}} \eta \\
&\hspace{1cm}
	+\overline{\nabla}_{\widehat{\nabla}_\eta \mu} \nu
	-\overline{\nabla}_{\nu} \left( \left[ \eta, \mu \right]_E + \overline{\nabla}_{\mu} \eta \right)
	- \left[ \left[ \eta, \mu \right]_E + \overline{\nabla}_{\mu} \eta, \nu \right]_E \\
&\hspace{1cm}+
\overline{\nabla}_{\mu} \left( \left[ \eta, \nu \right]_E + \overline{\nabla}_{\nu} \eta \right)
	- \overline{\nabla}_{\widehat{\nabla}_\eta \nu} \mu
	- \left[ \mu, \left[ \eta, \nu \right]_E + \overline{\nabla}_{\nu} \eta \right]_E \\
&=
\underbrace{\mleft[ \eta, \mleft[\mu, \nu \mright]_E \mright]_E
	+ \left[ \mu, \left[\nu, \eta \right]_E \right]_E
	+ \left[ \nu, \left[\eta, \mu \right]_E \right]_E}_{=0} \\
&\hspace{1cm}+
\underbrace{\overline{\nabla}_{\mu} \left( \left[ \eta, \nu \right]_E \right)
	- \left[ \overline{\nabla}_{\mu} \eta, \nu \right]_E
	- \left[ \eta, \overline{\nabla}_{\mu} \nu \right]_E
	- \overline{\nabla}_{\widehat{\nabla}_\nu \mu} \eta
	+ \overline{\nabla}_{\widehat{\nabla}_\eta \mu} \nu}_{= R_{\widehat{\nabla}}(\nu, \eta) \mu} \\
&\hspace{1cm}
\underbrace{-\overline{\nabla}_{\nu} \left( \left[ \eta, \mu \right]_E \right)
	+ \left[ \overline{\nabla}_{\nu} \eta, \mu \right]_E
	+ \left[ \eta, \overline{\nabla}_{\nu} \mu \right]_E
	+ \overline{\nabla}_{\widehat{\nabla}_\mu \nu} \eta
	- \overline{\nabla}_{\widehat{\nabla}_\eta \nu} \mu}_{= -R_{\widehat{\nabla}}(\mu, \eta) \nu} \\
&\hspace{1cm}
\underbrace{-\overline{\nabla}_{\widehat{\nabla}_\mu \nu} \eta
	+ \overline{\nabla}_{\overline{\nabla}_{\nu} \mu} \eta}_{= - \overline{\nabla}_{\left[ \mu, \nu \right]_E} \eta}
	+ \overline{\nabla}_{\mu} \overline{\nabla}_{\nu} \eta
	- \overline{\nabla}_{\nu} \overline{\nabla}_{\mu} \eta \\
&=
R_{\overline{\nabla}}(\mu, \nu) \eta
	+ R_{\widehat{\nabla}}(\nu, \eta) \mu
	- R_{\widehat{\nabla}}(\mu, \eta) \nu.
\eas
This gives Eq.~\eqref{DualCurvaRemainin1}.
\end{proof}

We are especially interested into the curvature if the conjugated $E$-connection is flat.

\begin{corollaries}{Curvature of conjugated $E$-connections where one connection is flat, \newline \cite[second and third statement of the first proposition in \S 4.6]{blaomTangentBundleAsLieGroup}}{LemmaCurvatureOfDualConnections}
Let $\widehat{\nabla}$ and $\overline{\nabla}$ be two $E$-connections, conjugate to each other, on a Lie algebroid $E \to N$ over a smooth manifold $N$. If $\widehat{\nabla}$ is flat, then
\ba
R_{\overline{\nabla}}(\mu, \nu) \eta
&=
\left(\widehat{\nabla}_\eta t_{\widehat{\nabla}}\right)(\mu, \nu),
\ea
also written as
\ba
R_{\overline{\nabla}}
&=
\widehat{\nabla} t_{\widehat{\nabla}}.
\ea
\end{corollaries}

\begin{proof}
\leavevmode\newline
This simply follows by Lemma \ref{lem:CurvatureOfDualConnectionsGeneral}.
\end{proof}

If both connections conjugate to each other are flat, then we have another Lie bracket by the first Bianchi identity.

\begin{corollaries}{Torsion as Lie bracket}{TOrsionCanBeLieBracketIfFlat}
Let $\widehat{\nabla}$ and $\overline{\nabla}$ be two flat $E$-connections, conjugate to each other, on a Lie algebroid $E \to N$ over a smooth manifold $N$. Then their torsions are Lie brackets for $\Gamma(E)$ which restrict to Lie brackets on the fibres, giving rise to a BLA structure on $E$.
\end{corollaries}

\begin{proof}
\leavevmode\newline
This follows by the flatness of both connections first Bianchi identity in Thm.~\ref{thm:1stBianchi} and Cor.~\ref{cor:LemmaCurvatureOfDualConnections}, the latter implies 
\bas
\widehat{\nabla} t_{\widehat{\nabla}}
&=
0,
\eas
and the former, the first Bianchi identity, then gives
\bas
t_{\widehat{\nabla}}\mleft(t_{\widehat{\nabla}}(\mu, \nu), \eta\mright) 
	+ t_{\widehat{\nabla}}\mleft(t_{\widehat{\nabla}}(\nu, \eta), \mu\mright) 
	+ t_{\widehat{\nabla}}\mleft(t_{\widehat{\nabla}}(\eta, \mu), \nu\mright)
&=
0
\eas
for all $\mu, \nu, \eta \in \Gamma(E)$. Bilinearity and antisymmetry is given, thus, $t_{\widehat{\nabla}}$ is a Lie bracket for $\Gamma(E)$, therefore also $t_{\overline{\nabla}}$ by Cor.~\ref{cor:TorsionOfDualTorsions}. Since torsions are tensors we can conclude that the torsion describes a Lie bracket on each fibre, too.
\end{proof}

\section{Basic connection and the basic curvature}\label{SectionOfBasicConnStuff}

As mentioned and already introduced in a simplified form in Ex.~\ref{ex:ClassicAdRepIsAConnection}, there is also another canonical example of $E$-connection, the \textbf{basic connection} $\nabla^{\text{bas}}$. We follow mainly \cite[\S 2.3]{basicconn}; however, in \cite[\S 3.4]{fernandes} the basic connection is introduced as a certain Bott connection along certain leaves given by the anchor, but we will neither use nor introduce that notion. The basic connection is actually the conjugate connection of $\nabla_\rho$.

\begin{definitions}{Basic connection, \cite[Definition 2.9]{basicconn}}{CanonicalBasicConnection}
%\leavevmode\newline
Let $E \to N$ be a Lie algebroid over a smooth manifold $N$, and let $\nabla$ be a vector bundle connection on $E$. We then define the \textbf{basic connection (induced by $\nabla$)} as a pair of $E$-connections, one on $E$ itself and the other one on $\mathrm{T}N$, both denoted by $\gls{0nablabas}$.
\begin{enumerate}
\item \textbf{(Basic $E$-connection on $E$)}
\newline The basic connection on $E$ is defined as the conjugate of $\nabla_\rho$, that is,
\ba
\nabla^{\mathrm{bas}}_\mu \nu \coloneqq [\mu, \nu]_E + \nabla_{\rho(\nu)} \mu
\ea
for all $\mu, \nu \in \Gamma(E)$
\item \textbf{(Basic $E$-connection on $\mathrm{T}N$)}
\newline The basic connection on $\mathrm{T}N$ is defined by
\ba
\nabla^{\mathrm{bas}}_\mu X \coloneqq [\rho(\mu), X] + \rho\left( \nabla_X \mu \right)
\ea
for all $\mu \in \Gamma(E)$ and $X \in \mathfrak{X}(N)$
\end{enumerate}
\end{definitions}

\begin{remark}
\leavevmode\newline
It is trivial to see that these are $E$-connections.

In the physics' part, Chapter \ref{GeneralizedGTfas}, we will discuss the use of this connection in physics, as also arising in \cite[discussion around Equation (17)]{CurvedYMH}. Nevertheless one can see here already that one gets the adjoint representation for bundle of Lie algebras, \textit{i.e.}~$\rho \equiv 0$, because then the basic connection on $E$ is just the field of Lie brackets.

In the following we often just write of the "basic connection" or $\nabla^{\mathrm{bas}}$, while we then always mean both connections. It should be clear by context which of both connections we mean then. Similar for its curvature $R_{\nabla^\mathrm{bas}}$; but the torsion $t_{\nabla^{\mathrm{bas}}}$ will only denote the torsion for the basic connection on $E$ since only on $E$ the torsion is formulated.
\end{remark}
We will use the following essential property of the basic connection very often.

\begin{corollaries}{Compatibility of the basic connection with the anchor, \newline \cite[comment after Definition 2.9]{basicconn}}{ENablaMitRhoVertauschung}
Let $E \to N$ be a Lie algebroid over a smooth manifold $N$, and let $\nabla$ be a vector bundle connection on $E$. Then
\ba
\rho\circ\nabla^{\mathrm{bas}}
&=
\nabla^{\mathrm{bas}} \circ \rho.
\ea
\end{corollaries}

\begin{proof}
\leavevmode\newline
We have
\bas
\rho \mleft(
	\nabla^{\mathrm{bas}}_\mu \nu
\mright)
&=
\rho \mleft(
	\mleft[ \mu, \nu \mright]_E
	+ \nabla_{\rho(\nu)} \mu
\mright)
=
\mleft[ \rho(\mu), \rho(\nu) \mright]_E
	+ \rho\mleft( \nabla_{\rho(\nu)} \mu \mright)
=
\nabla^{\mathrm{bas}}_\mu \bigl( \rho(\nu) \bigr)
\eas
for all $\mu,\nu \in \Gamma(E)$, using that the anchor is a homomorphism of Lie brackets.
\end{proof}

As in \cite{CurvedYMH}, we will later see that $\nabla^{\mathrm{bas}}$ should be flat for a given $\nabla$ in order to formulate a gauge theory (among other conditions). Thence, it is important to study the curvature of $\nabla^{\mathrm{bas}}$. Its curvature is encoded in another tensor, the \textbf{basic curvature}.
%With $\gls{Hom}$ we denote the space of Homomorphisms (here of vector bundles).

\begin{definitions}{Basic curvature, \cite[Definition 2.10]{basicconn}}{basiccurvature}
Let $E \to N$ be a Lie algebroid over a smooth manifold $N$, and let $\nabla$ be a connection on $E$. The \textbf{basic curvature $\gls{Rnablabas}$} is then defined as an element of $\Gamma\left(\bigwedge^2E^* \otimes \mathrm{T}^*N \otimes E \right)$ by
\ba
R^{\mathrm{bas}}_\nabla(\mu, \nu) X
&\coloneqq
\nabla_X\mleft(\mleft[\mu, \nu\mright]_E\mright) 
	- \mleft[ \nabla_X \mu, \nu \mright]_E 
	- \mleft[ \mu, \nabla_X \nu \mright]_E 
	- \nabla_{\nabla^{\mathrm{bas}}_\nu X} \mu 
	+ \nabla_{\nabla^{\mathrm{bas}}_\mu X} \nu,
\ea
where $\mu, \nu \in \Gamma(E)$ and $X \in \mathfrak{X}(N)$.
\end{definitions}

\begin{remark}
\leavevmode\newline
\indent $\bullet$ As stated in \cite{basicconn} one may think of this as $\nabla_X([\mu, \nu]_E) - [ \nabla_X \mu, \nu ]_E - [ \mu, \nabla_X \nu ]_E$ which is a measure of the derivation property of $\nabla$ w.r.t. $[\cdot, \cdot]_E$, but corrected in such a way that it is tensoriel in all arguments. For a zero anchor the basic curvature would be equivalent to $\nabla_X([\mu, \nu]_E) - [ \nabla_X \mu, \nu ]_E - [ \mu, \nabla_X \nu ]_E$ since then the basic connection on $\mathrm{T}N$ is identically zero.

$\bullet$ Compare the form of the basic curvature also with Lemma \ref{lem:CurvatureOfDualConnectionsGeneral}.

$\bullet$ It is trivial to see that the basic curvature is antisymmetric in the Lie algebroid arguments and that it is trilinear. Also let $f \in C^\infty(N)$ and observe
\bas
R^{\mathrm{bas}}_\nabla(\mu, \nu) (fX)
&=
\nabla_{fX}\mleft(\mleft[\mu, \nu\mright]_E\mright) 
	- \underbrace{\mleft[ \nabla_{fX} \mu, \nu \mright]_E}
		_{\mathclap{ = f \mleft[ \nabla_{X} \mu, \nu \mright]_E - \mathcal{L}_\nu(f)~ \nabla_X \mu }}
	- \mleft[ \mu, \nabla_{fX} \nu \mright]_E 
	- \underbrace{\nabla_{\nabla^{\mathrm{bas}}_\nu (fX)} \mu}
		_{\mathclap{ = f \nabla_{\nabla^{\mathrm{bas}}_\nu X} \mu + \mathcal{L}_\nu(f) ~ \nabla_X \mu }}
	+ \nabla_{\nabla^{\mathrm{bas}}_\mu (fX)} \nu
\\
&=
f ~ R^{\mathrm{bas}}_\nabla(\mu, \nu) X
\eas
for all $\mu, \nu \in \Gamma(E)$ and $X \in \mathfrak{X}(N)$, and
\bas
R^{\mathrm{bas}}_\nabla(\mu, f \nu) X
&=
\nabla_X\mleft(\mleft[\mu, f\nu\mright]_E\mright) 
	- \mleft[ \nabla_X \mu, f\nu \mright]_E 
	- \mleft[ \mu, \nabla_X (f\nu) \mright]_E 
	- \nabla_{\nabla^{\mathrm{bas}}_{f\nu} X} \mu 
	+ \nabla_{\nabla^{\mathrm{bas}}_\mu X} (f\nu)
\\
&=
f R^{\mathrm{bas}}_\nabla(\mu, \nu) X
\\
&\hspace{1cm}
	+ \mathcal{L}_X(f) ~ \mleft[ \mu, \nu \mright]_E
	+ \mathcal{L}_{\rho(\mu)}(f) ~ \nabla_X (\nu )
	+ \mathcal{L}_X \mathcal{L}_{\rho(\mu)}(f) ~ \nu
	- \mathcal{L}_{\rho(\nabla_X \mu)}(f) ~ \nu
\\
&\hspace{1cm}
	- \mathcal{L}_{\rho(\mu)}(f) ~ \nabla_X \nu 
	- \mathcal{L}_X(f) ~ \mleft[ \mu, \nu \mright]_E
	- \mathcal{L}_{\rho(\mu)} \mathcal{L}_X(f) ~ \nu
	+ \underbrace{\mathcal{L}_{\nabla^{\mathrm{bas}}_\mu X}(f)}
		_{\mathclap{ = \mathcal{L}_{[\rho(\mu), X] + \rho(\nabla_X \mu)}(f) }} 
		~ \nu
\\
&=
f R^{\mathrm{bas}}_\nabla(\mu, \nu) X
	+ \underbrace{\mathcal{L}_X \mathcal{L}_{\rho(\mu)}(f) ~ \nu
	- \mathcal{L}_{\rho(\mu)} \mathcal{L}_X(f) ~ \nu
	- \mathcal{L}_{[X, \rho(\mu)]}(f) ~ \nu}_{= ~ 0}
\\
&=
f R^{\mathrm{bas}}_\nabla(\mu, \nu) X,
\eas
that the basic curvature is also tensorial in $\mu$ follows by the antisymmetry.
\end{remark}

Do not confuse this tensor with $R_{\nabla^{\mathrm{bas}}}$, the curvature of the basic connection, either on $E$ or $\mathrm{T}N$. However, the curvatures are encoded in the basic curvature.

\begin{propositions}{Relations between the curvatures, \newline \cite[Proposition 2.11]{basicconn}, \cite[Equation (9)]{CurvedYMH}, \cite[generalization of second statement of the first proposition in \S 4.6]{blaomTangentBundleAsLieGroup}}{SnablamitREnabla}
Let $E \to N$ be a Lie algebroid over a smooth manifold $N$, and let $\nabla$ be a connection on $E$. Then one has:
\begin{enumerate}
\item The curvature of $\nabla^{\mathrm{bas}}$ on $E$ is equal to $- R_\nabla^{\mathrm{bas}}(\cdot, \cdot) \circ \rho$.
\item The curvature of $\nabla^{\mathrm{bas}}$ on $\mathrm{T}N$ is equal to $- \rho \circ R^{\mathrm{bas}}_\nabla$.
\end{enumerate}
We also have an important relation to the curvature $R_\nabla$ of $\nabla$,
\ba\label{eq:compcondfast}
R_\nabla^{\mathrm{bas}}(\mu, \nu)X 
&= \left( \nabla_X t_{\nabla^{\mathrm{bas}}} \right)(\mu, \nu) 
- R_\nabla(\rho(\mu), X) \nu + R_\nabla(\rho(\nu), X) \mu
\ea
for all $\mu, \nu \in \Gamma(E)$ and $X \in \mathfrak{X}(N)$,
where $t_{\nabla^{\mathrm{bas}}}$ is the $E$-torsion of the basic connection on $E$. 
\end{propositions}

\begin{remark}\label{rem:vanishingbasicconn}
\leavevmode\newline
This implies that both $\nabla^{\mathrm{bas}}$ are flat if $R_\nabla^{\mathrm{bas}} \equiv 0$. The converse is in general not true. But for invertible $\rho$ the converse would hold.
For $R_\nabla^{\mathrm{bas}} \equiv 0$ one also gets 
\ba
( \nabla_X t_{\nabla^{\mathrm{bas}}})(\mu, \nu)
&= R_\nabla(\rho(\mu), X) \nu - R_\nabla(\rho(\nu), X) \mu,
\ea
and by Cor.~\ref{cor:TorsionOfDualTorsions} we also have $t_{\nabla^{\mathrm{bas}}} = - t_{\nabla_\rho}$ such that one can rewrite this with the torsion of $\nabla_\rho$.
\end{remark}

\begin{proof}[Proof of Prop.~\ref{prop:SnablamitREnabla}]
\leavevmode\newline
For the curvature of $\nabla^{\mathrm{bas}}$ on $E$ observe, using Cor.~\ref{cor:ENablaMitRhoVertauschung},
\bas
-R_\nabla^{\mathrm{bas}}(\mu, \nu) \bigl( \rho(\eta)\bigr)
&=
-\mleft(
	\nabla_{\rho(\eta)}([\mu, \nu]_E) 
	- [ \nabla_{\rho(\eta)} \mu, \nu ]_E 
	- [ \mu, \nabla_{\rho(\eta)} \nu ]_E 
	- \nabla_{\nabla^{\mathrm{bas}}_\nu \rho(\eta)} \mu 
	+ \nabla_{\nabla^{\mathrm{bas}}_\mu \rho(\eta)} \nu
\mright)
\\
&=
-\mleft(
	\nabla_{\rho(\eta)}\mleft(\mleft[\mu, \nu\mright]_E\mright) 
	- \mleft[ \nabla_{\rho(\eta)} \mu, \nu \mright]_E 
	- \mleft[ \mu, \nabla_{\rho(\eta)} \nu \mright]_E 
	- \nabla_{\rho\mleft(\nabla^{\mathrm{bas}}_\nu \eta\mright)} \mu 
	+ \nabla_{\rho\mleft(\nabla^{\mathrm{bas}}_\mu \eta\mright)} \nu
\mright)
\\
&\stackrel{\mathclap{ \text{Lem.~\ref{lem:CurvatureOfDualConnectionsGeneral}} }}{=}\quad~
R_{\nabla^{\mathrm{bas}}}(\mu,\nu)\eta
\eas
for all $\mu, \nu, \eta \in \Gamma(E)$.
%\bas
%S_\nabla(\mu, \nu) \rho(\eta)
%&= \nabla_{\rho(\eta)}([\mu, \nu]_E) - [ \nabla_{\rho(\eta)} \mu, \nu ]_E - [ \mu, \nabla_{\rho(\eta)} \nu ]_E - \nabla_{\nabla^{\mathrm{bas}}_\nu \rho(\eta)} \mu + \nabla_{\nabla^{\mathrm{bas}}_\mu \rho(\eta)} \nu \\
%&= \nabla_{\rho(\eta)}([\mu, \nu]_E) - [ \nabla_{\rho(\eta)} \mu, \nu ]_E - [ \mu, \nabla_{\rho(\eta)} \nu ]_E - \nabla_{\rho\left(\nabla^{\mathrm{bas}}_\nu \eta\right)} \mu + \nabla_{\rho\left(\nabla^{\mathrm{bas}}_\mu \eta\right)} \nu \\ 
%&\quad\underbrace{- [\mu, [\nu, \eta]_E]_E - [\nu, [\eta, \mu]_E]_E - [\eta, [\mu, \nu]_E]_E}_{= 0 \text{ (Jacobi)}} \\
%&= \underbrace{\left[ [\mu, \nu]_E, \eta \right]_E + \nabla_{\rho(\eta)}([\mu, \nu]_E)}_{= \nabla^{\mathrm{bas}}_{[\mu, \nu]_E} \eta} - [ \underbrace{[\mu, \eta]_E + \nabla_{\rho(\eta)} \mu}_{= \nabla^{\mathrm{bas}}_\mu \eta}, \nu ]_E + \nabla_{\rho\left(\nabla^{\mathrm{bas}}_\mu \eta\right)} \nu \\
%&\quad - [ \mu, \underbrace{[\nu, \eta]_E + \nabla_{\rho(\eta)} \nu}_{= \nabla^{\mathrm{bas}}_\nu \eta} ]_E - \nabla_{\rho\left(\nabla^{\mathrm{bas}}_\nu \eta\right)} \mu \\
%&= - \left( \nabla^{\mathrm{bas}}_\mu \nabla^{\mathrm{bas}}_\nu \eta - \nabla^{\mathrm{bas}}_\nu \nabla^{\mathrm{bas}}_\mu \eta - \nabla^{\mathrm{bas}}_{[\mu, \nu]_E} \eta \right)
%= - R_{\nabla^{\mathrm{bas}}}(\mu, \nu) \eta
%\eas
In the same fashion as in the proof of Lemma \ref{lem:CurvatureOfDualConnectionsGeneral}, using the Jacobi identity and that $\rho$ is a homomorphism, we also have
\bas
\rho\left( R_\nabla^{\mathrm{bas}}(\mu, \nu) X \right)
&= \rho\left( \nabla_X([\mu, \nu]_E) - [ \nabla_X \mu, \nu ]_E - [ \mu, \nabla_X \nu ]_E - \nabla_{\nabla^{\mathrm{bas}}_\nu X} \mu + \nabla_{\nabla^{\mathrm{bas}}_\mu X} \nu \right) \\
&\hspace{1cm}
	+ [[\rho(\mu), \rho(\nu)], X] + [[\rho(\nu), X], \rho(\mu)] + [[X, \rho(\mu)], \rho(\nu)] \\
&= \underbrace{[\rho([\mu, \nu]_E), X] + \rho\left( \nabla_X([\mu, \nu]_E)\right)}_{= \nabla^{\mathrm{bas}}_{[\mu, \nu]_E}X}
\\
&\hspace{1cm}
+ [\rho(\nu), \underbrace{[\rho(\mu), X] + \rho(\nabla_X\mu)}_{= \nabla^{\mathrm{bas}}_\mu X}]
%&\hspace{1cm} 
	+ \rho\left( \nabla_{\nabla^{\mathrm{bas}}_\mu X} \nu \right)
\\
&\hspace{1cm}
	- [\rho(\mu), \underbrace{[\rho(\nu), X] + \rho(\nabla_X \nu)}_{= \nabla^{\mathrm{bas}}_\nu X}]
	- \rho\left( \nabla_{\nabla^{\mathrm{bas}}_\nu X} \mu \right) 
\\
&=
\nabla^{\mathrm{bas}}_{[\mu, \nu]_E}X
	+ \nabla^{\mathrm{bas}}_\nu \nabla^{\mathrm{bas}}_\mu X
	- \nabla^{\mathrm{bas}}_\mu \nabla^{\mathrm{bas}}_\nu X
\\
&= - R_{\nabla^{\mathrm{bas}}}(\mu, \nu)X
\eas
for all $X \in \mathfrak{X}(N)$.
By Cor.~\ref{cor:TorsionOfDualTorsions} we know that that $t_{\nabla^{\mathrm{bas}}} = - t_{\nabla_\rho}$, thus,
\bas
\left( \nabla_X t_{\nabla^{\mathrm{bas}}} \right)(\mu, \nu) 
&=
-(\nabla_X t_{\nabla_\rho})(\mu, \nu)
\\
&= - \nabla_X \left(t_{\nabla_\rho}(\mu, \nu)\right)
+ t_{\nabla_\rho}(\nabla_X \mu, \nu)
+ t_{\nabla_\rho}(\mu,\nabla_X \nu) \\
&= \nabla_X \left([\mu, \nu]_E - \nabla_{\rho(\mu)} \nu + \nabla_{\rho(\nu)} \mu\right)
\\
&\hspace{1cm} 
+ \nabla_{\rho\left( \nabla_X \mu \right)} \nu - \nabla_{\rho(\nu)} \nabla_X \mu - \left[ \nabla_X \mu, \nu \right]_E \\
&\hspace{1cm}
	+ \nabla_{\rho(\mu)} \nabla_X \nu - \nabla_{\rho\left( \nabla_X \nu \right)} \mu - [\mu, \nabla_X \nu]_E 
\\
&= 
\nabla_X ([\mu, \nu]_E) 
	- \left[ \nabla_X \mu, \nu \right]_E 
	- [\mu, \nabla_X \nu]_E 
	+ \nabla_{\rho\left( \nabla_X \mu \right)} \nu 
	- \nabla_{\rho\left( \nabla_X \nu \right)} \mu \\
&\hspace{1cm}
	+ R_\nabla(\rho(\mu), X) \nu + \nabla_{[\rho(\mu), X]} \nu - R_\nabla(\rho(\nu), X) \mu - \nabla_{[\rho(\nu), X]} \mu \\
&= R^{\mathrm{bas}}_\nabla(\mu, \nu)X + R_\nabla(\rho(\mu), X) \nu - R_\nabla(\rho(\nu), X) \mu.
\eas
\end{proof}

The basic connection on $E$ is conjugate to $\nabla_\rho$ by definition, and it will be later very important that the basic connection is flat for gauge theory as we will see. By our discussion about conjugate Lie algebroid connections we can immediately derive the following by Cor.~\ref{cor:LemmaCurvatureOfDualConnections}.

\begin{theorems}{Curvature of $\nabla_\rho$ for a vanishing basic curvature}{modBianchithm}
Assume $R_\nabla^{\mathrm{bas}} (\cdot, \cdot) \circ \rho = 0$, then we have 
\ba
R_{\nabla_\rho} 
&= \nabla^{\mathrm{bas}} t_{\nabla^{\mathrm{bas}}},\label{eq:BaufOrbit}
\ea
\textit{i.e.}
\bas
R_{\nabla_\rho}(\mu, \nu) \eta = R_\nabla(\rho(\mu), \rho(\nu)) \eta 
&= \left(\nabla^{\mathrm{bas}}_\eta t_{\nabla^{\mathrm{bas}}}\right)(\mu, \nu)
\eas
for all $\mu, \nu, \eta \in \Gamma(E)$.
\end{theorems}

\begin{proof}
\leavevmode\newline
By Prop.~\ref{prop:SnablamitREnabla} we know that the assumption implies that $\nabla^{\mathrm{bas}}$ on $E$ is flat. Thence, we can use Cor.~\ref{cor:LemmaCurvatureOfDualConnections} because of that $\nabla^{\mathrm{bas}}$ on $E$ and $\nabla_\rho$ are conjugate to each other. This concludes the proof.
\end{proof}

\section{Exterior covariant derivatives}\label{ExteriorCovariantDerivativesAoids}

As for standard connections one can now define exterior covariant derivatives related to Lie algebroid connections. 

\begin{definitions}{Exterior covariant derivatives using Lie algebroid connections, \newline \cite[the discussion after Def. 2.2]{basicconn}}{AllgemeineExteriorCovariantDerivativeSch}
Let $E \to N$ be a Lie algebroid over a smooth manifold $N$, ${}^E\nabla$ an $E$-connection on a vector bundle $V \to N$. Then we define the \textbf{exterior covariant derivative $\gls{dEnabla}$} as an operator $\Omega^q(E;V) \to \Omega^{q+1}(E;V)$ ($q \in \mathbb{N}_0$) by
\ba
\left(\mathrm{d}^{{}^E\nabla} \omega \right) (\nu_0, \dots, \nu_q)
&\coloneqq 
\sum_{i = 0}^{q} (-1)^i ~ {}^E\nabla_{\nu_i}\left(\omega\left( \nu_0, \dots, \widehat{\nu}_i, \dots, \nu_q \right) \right) \nonumber \\
&\hspace{1cm}
	+ \sum_{0 \leq i < j \leq q} (-1)^{i+j} \omega( [\nu_i, \nu_j]_E, \nu_0, \dots, \widehat{\nu}_i, \dots, \widehat{\nu}_j, \dots, \nu_q)
\ea
for all $\omega \in \Omega^q(E;V)$ and $\nu_0, \dots, \nu_q \in \Gamma(E)$.
\end{definitions}

\begin{remark}
\leavevmode\newline
That this is a well-defined operator can be shown as in the case of vector bundle connections.
\end{remark}

Moreover, in the case of a connection $\nabla$ on $E$ one has also the previously discussed basic connection $\nabla^{\mathrm{bas}}$ as $E$-connection on $E$ and $\mathrm{T}N$. $\nabla$ is typical vector bundle connection and $\nabla^{\mathrm{bas}}$ a pair of $E$-connections. Hence, it may make sense to look at forms with two degrees, one for $\mathrm{T}N$ and the other one with respect to $E$. 

The following space is also developed and studied by Alexei Kotov, communicated to me in private communication, his studies are planned to be published in 2021.

\begin{definitions}{$(p,q)$-$E$-forms}{ExteriorCovariantDerivatives}
Let $E \to N$ be a Lie algebroid over a smooth manifold $N$, and $V \to N$ a vector bundle.
Then the \textbf{space of $(p,q)$-$E$-forms with values in $V$} ($p, q \in \mathds{N}_0$), will is defined by
\ba
\gls{1ZOmegapq(NEV)} \coloneqq \Gamma\left(\bigwedge^p \mathrm{T}^*N \otimes \bigwedge^q E^* \otimes V\right).
\ea
\end{definitions}

Let us study possible exterior covariant derivatives on this space in the case of $E =V$.

\begin{remarks}{Exterior covariant derivatives induced by $\nabla$}{ExteriorStuffRemark}
Let $E \to N$ be a Lie algebroid over a smooth manifold $N$ and $\nabla$ a connection on $E$.
\newline\newline
$\bullet$ For $q = 0$ one gets the space of $p$-forms with values in $E$, $\Omega^p(N;E)$, or more general, those are forms on $N$ with values in $\bigwedge^q E^* \otimes E$, \textit{i.e.}~
\ba\label{FirstInterpretationOfDoubleDegree}
\Omega^{p, q}(N, E; E) \cong \Omega^p\left(N; \bigwedge^q E^* \otimes E\right).
\ea
\newline\newline
$\bullet$ Analogously 
\ba\label{SecondInterpretationOfDoubleDegree}
\Omega^{p, q}(N, E; E) \cong \Omega^q\left(E; \bigwedge^p \mathrm{T}N^* \otimes E\right).
\ea
\newline\newline
$\bullet$ Using Eq.~\eqref{FirstInterpretationOfDoubleDegree}, denote with $\nabla$ also the canonically induced connection on $\bigwedge^q E^* \otimes E$; then we have a canonical definition of $\mathrm{d}^\nabla$ on $\Omega^{p,q}(N,E;E)$. Since the canonically induced connection on $\bigwedge^q E^* \otimes E$ is defined by using the Leibniz rule, one can rewrite the exterior covariant derivative $\mathrm{d}^\nabla$ of $\omega \in \Omega^{p, q}(N, E; E)$ as an element of $\Omega^{p+1, q}(N, E; E)$ by
\ba
&\mleft(\mathrm{d}^\nabla \omega\mright)\mleft( X_0, \dots, X_p, \nu_1, \dots, \nu_q \mright) 
\nonumber \\
&= 
\sum_{i=0}^p (-1)^i \biggl( \nabla_{X_i} \Bigl( \omega\left(X_0, \dots, \widehat{X}_i, \dots, X_p, \nu_1, \dots, \nu_q\right) \Bigr) 
\nonumber \\
&\hphantom{\sum_{i=0}^p (-1)^i \biggl(} \hspace{1cm}
	- \sum_{j=1}^q \omega\mleft( X_0, \dots, \widehat{X}_i, \dots, X_p, \nu_1, \dots, \nabla_{X_i} \nu_j, \dots, \nu_q \mright) \biggr) 
\nonumber \\
&\hspace{1cm}
	+ \sum_{0 \leq i < j \leq p} (-1)^{i+j} \omega\mleft( [X_i, X_j], X_0, \dots, \widehat{X}_i, \dots, \widehat{X}_j, \dots, X_p, \nu_1, \dots, \nu_q \mright),
\ea
where $X_0, \dots, X_p \in \mathfrak{X}(N)$ and $\nu_1, \dots, \nu_q \in \Gamma(E)$.
\newline\newline
$\bullet$ Similarly one proceeds with $\nabla^{\mathrm{bas}}$, using that the basic connection acts on both, $E$ and $\mathrm{T}N$, such that there is a canonically induced notion of $\nabla^{\mathrm{bas}}$ on $\bigwedge^p \mathrm{T}N^* \otimes E$. By Eq.~\eqref{SecondInterpretationOfDoubleDegree} we have $\mathrm{d}^{\nabla^{\mathrm{bas}}}: \Omega^{p, q}(N, E; E) \to \Omega^{p, q+1}(N, E; E)$ given by
\ba
&\mleft(\mathrm{d}^{\nabla^{\mathrm{bas}}} \omega\mright)\mleft( X_1, \dots, X_p, \nu_0, \dots, \nu_q \mright) 
\nonumber \\
&=
\sum_{i=0}^q (-1)^i \biggl( \nabla^{\mathrm{bas}}_{\nu_i} \bigl( \omega\left(X_1, \dots, X_p, \nu_0, \dots, \widehat{\nu}_i, \dots \nu_q\right) \bigr) 
\nonumber \\
&\hphantom{\sum_{i=0}^q (-1)^i \biggl(} \hspace{1cm}
	- \sum_{j=1}^p \omega\mleft( X_1, \dots, \nabla^{\mathrm{bas}}_{\nu_i} X_j, \dots, X_p, \nu_0, \dots, \widehat{\nu}_i, \dots, \nu_q \mright) \biggr) 
\nonumber \\
&\hspace{1cm}
	+ \sum_{0 \leq i < j \leq q} (-1)^{i+j} \omega\mleft( X_1, \dots, X_p, [\nu_i, \nu_j]_E, \nu_0, \dots, \widehat{\nu}_i, \dots, \widehat{\nu}_j, \dots, \nu_q \mright),
\ea
where $\omega \in \Omega^{p, q}(N, E; E)$, $X_1, \dots, X_p \in \mathfrak{X}(N)$ and $\nu_0, \dots, \nu_q \in \Gamma(E)$.
\newline\newline
$\bullet$ For LABs one can see that $\mathrm{d}^{\nabla^{\mathrm{bas}}}$ acts as the Chevalley-Eilenberg differential $\mathrm{d}_{\mathrm{CE}}$ because the basic connection on $\mathrm{T}N$ is then identically to zero and the one on $E$ is just the adjoint.
\end{remarks}

The commutation of the basic curvature with the anchor carries over to the differential.

\begin{lemmata}{Differential of basic curvature commutes with anchor}{commutationanchordifferential}
Let $E \to N$ be a Lie algebroid over a smooth manifold $N$ and $\nabla$ a connection on $E$.
Then
\ba
\Big( \nabla^{\mathrm{bas}}_\mu \big( \omega \circ \underbrace{(\rho, \dots, \rho)}_{p \text{ times}} \big) \Big) (\nu_1, \dots, \nu_p)
&=
\left(\mathrm{d}^{\nabla^{\mathrm{bas}}} \omega \right)( \rho(\nu_1), \dots, \rho(\nu_p), \mu),
\ea
for all $\omega \in \Omega^{p}(N;E)$ ($p \in \mathbb{N}_0$) and $\mu, \nu_1, \dots \nu_p \in \Gamma(E)$;
in short
\ba
\nabla^{\mathrm{bas}} \left( \omega \circ (\rho, \dots, \rho) \right)
&=
\left(\mathrm{d}^{\nabla^{\mathrm{bas}}} \omega \right) \circ (\rho, \dots, \rho, \mathds{1}_E).
\ea
\end{lemmata}

\begin{proof}
\leavevmode\newline
Recall $\rho \circ \nabla^{\mathrm{bas}} = \nabla^{\mathrm{bas}} \circ \rho$ by Cor.~\ref{cor:ENablaMitRhoVertauschung}, then
\bas
\left( \nabla^{\mathrm{bas}}_\mu \left( \omega \circ (\rho, \dotsc, \rho) \right) \right)(\nu_1, \dots, \nu_p)
&=
\nabla^{\mathrm{bas}}_\mu \bigl( \omega\mleft( \rho(\nu_1), \dotsc, \rho(\nu_p) \mright) \bigr) \\
&\hspace{1cm}
	- \sum_{j=1}^p \omega\Bigl( \rho(\nu_1), \dotsc ,
	\underbrace{\rho\mleft(\nabla^{\mathrm{bas}}_\mu \nu_j\mright)}
		_{= \nabla^{\mathrm{bas}}_\mu (\rho( \nu_j ))},
		\dotsc, \rho(\nu_p) \Bigr) 
\\
&= 
\left( \nabla^{\mathrm{bas}}_\mu \omega \right) (\rho(\nu_1), \dotsc, \rho(\nu_p))
\\
&= 
\left(\mathrm{d}^{\nabla^{\mathrm{bas}}} \omega \right)( \rho(\nu_1), \dotsc, \rho(\nu_p), \mu).
\eas
\end{proof}

Recall that we did not prove the second Bianchi identity in Thm.~\ref{thm:1stBianchi}. We are going to prove the second Bianchi identity using the following theorem.

\begin{theorems}{Second Bianchi identity, \newline \cite[reformulation of Proposition 7.1.9; page 265]{mackenzieGeneralTheory}}{2ndBianchi}
Let $E \to N$ be a Lie algebroid over a smooth manifold $N$, $V \to N$ a vector bundle, and let ${}^E\nabla$ be an $E$-connection on $V$, while we denote its naturally induced definition on $\mathrm{End}(V)$ also ${}^E\nabla$. Viewing its curvature $R_{{}^E\nabla}$ as an element of $\Omega^2(E; \mathrm{End}(V))$ we then have
\ba
\mathrm{d}^{{}^E\nabla} R_{{}^E\nabla} = 0.
\ea
\end{theorems}

\begin{proof}[Proof of Thm.~\ref{thm:2ndBianchi}]
\leavevmode\newline
Let $\mu, \nu, \eta \in \Gamma(E)$ and $v \in \Gamma(V)$, then
\bas
\left(\left(\mathrm{d}^{{}^E\nabla} R_{{}^E\nabla}\right)(\mu, \nu, \eta)\right) (v)
&=
\Bigl( {}^E\nabla_\mu \left( R_{{}^E\nabla}(\nu, \eta) \right) - {}^E\nabla_\nu \left( R_{{}^E\nabla}(\mu, \eta) \right) + {}^E\nabla_\eta \left( R_{{}^E\nabla}(\mu, \nu) \right) \\
&\hphantom{\Bigl(}\hspace{1cm}
	- R_{{}^E\nabla}([\mu, \nu]_E, \eta) + R_{{}^E\nabla}([\mu, \eta]_E, \nu) - R_{{}^E\nabla}([\nu, \eta]_E, \mu) \Bigr)(v)
\\
&=
{}^E\nabla_\mu\left( R_{{}^E\nabla}(\nu, \eta)v \right) - R_{{}^E\nabla}(\nu, \eta) \left( {}^E\nabla_\mu v\right)
\\
&\hspace{1cm}
- {}^E\nabla_\nu\left( R_{{}^E\nabla}(\mu, \eta)v \right) + R_{{}^E\nabla}(\mu, \eta) \left( {}^E\nabla_\nu v\right) 
\\
&\hspace{1cm}
	+ {}^E\nabla_\eta\left( R_{{}^E\nabla}(\mu, \nu)v \right) - R_{{}^E\nabla}(\mu, \nu) \left( {}^E\nabla_\eta v\right) 
\\
&\hspace{1cm}
	- R_{{}^E\nabla}([\mu, \nu]_E, \eta)v + R_{{}^E\nabla}([\mu, \eta]_E, \nu)v - R_{{}^E\nabla}([\nu, \eta]_E, \mu)v 
\\
&=
{}^E\nabla_\mu {}^E\nabla_\nu {}^E\nabla_\eta v - {}^E\nabla_\mu {}^E\nabla_\eta {}^E\nabla_\nu v - {}^E\nabla_\mu {}^E\nabla_{[\nu, \eta]_E} v 
\\
&\hspace{1cm}
	- {}^E\nabla_\nu {}^E\nabla_\eta {}^E\nabla_\mu v + {}^E\nabla_\eta {}^E\nabla_\nu {}^E\nabla_\mu v + {}^E\nabla_{[\nu, \eta]_E} {}^E\nabla_\mu v 
\\
&\hspace{1cm}
	- {}^E\nabla_\nu {}^E\nabla_\mu {}^E\nabla_\eta v + {}^E\nabla_\nu {}^E\nabla_\eta {}^E\nabla_\mu v + {}^E\nabla_\nu {}^E\nabla_{[\mu, \eta]_E} v 
\\
&\hspace{1cm}
	+ {}^E\nabla_\mu {}^E\nabla_\eta {}^E\nabla_\nu v - {}^E\nabla_\eta {}^E\nabla_\mu {}^E\nabla_\nu v - {}^E\nabla_{[\mu, \eta]_E} {}^E\nabla_\nu v 
\\
&\hspace{1cm}
	+ {}^E\nabla_\eta {}^E\nabla_\mu {}^E\nabla_\nu v - {}^E\nabla_\eta {}^E\nabla_\nu {}^E\nabla_\mu v - {}^E\nabla_\eta {}^E\nabla_{[\mu, \nu]_E} v 
\\
&\hspace{1cm}
	- {}^E\nabla_\mu {}^E\nabla_\nu {}^E\nabla_\eta v + {}^E\nabla_\nu {}^E\nabla_\mu {}^E\nabla_\eta v + {}^E\nabla_{[\mu, \nu]_E} {}^E\nabla_\eta v 
\\
&\hspace{1cm}
	- {}^E\nabla_{[\mu, \nu]_E} {}^E\nabla_\eta v + {}^E\nabla_\eta {}^E\nabla_{[\mu, \nu]_E} v + {}^E\nabla_{\left[ [\mu, \nu]_E, \eta \right]_E}v 
\\
&\hspace{1cm}
	+ {}^E\nabla_{[\mu, \eta]_E} {}^E\nabla_\nu v - {}^E\nabla_\nu {}^E\nabla_{[\mu, \eta]_E} v - {}^E\nabla_{\left[ [\mu, \eta]_E, \nu \right]_E}v 
\\
&\hspace{1cm}
	- {}^E\nabla_{[\nu, \eta]_E} {}^E\nabla_\mu v + {}^E\nabla_\mu {}^E\nabla_{[\nu, \eta]_E} v + {}^E\nabla_{\left[ [\nu, \eta]_E, \mu \right]_E}v
\\
&= 
0,
\eas
where we also used the Jacobi identity.
\end{proof}

\begin{remarks}{Proof of the second Bianchi identity of Thm.~\ref{thm:1stBianchi}}{FinallyTheOtherBianchiStuff}
We can now finally prove the second statement of Thm.~\ref{thm:1stBianchi} by showing that it is equivalent to Thm.~\ref{thm:2ndBianchi} if $V=E$; for $\mu, \nu, \eta \in \Gamma(E)$ we have
\bas
&\mleft( {}^E\nabla_\mu R_{{}^E\nabla}\mright)(\nu, \eta) + \mleft( {}^E\nabla_\nu R_{{}^E\nabla}\mright)(\eta, \mu) + \mleft( {}^E\nabla_\eta R_{{}^E\nabla}\mright)(\mu, \nu)
\\
&\hspace{1cm}
	+ R_{{}^E\nabla}\left( t_{{}^E\nabla}(\mu, \nu), \eta \right) 
	+ R_{{}^E\nabla}\left( t_{{}^E\nabla}(\nu, \eta), \mu \right)
	+ R_{{}^E\nabla}\left( t_{{}^E\nabla}(\eta, \mu), \nu \right) \\
&=
{}^E\nabla_\mu \left( R_{{}^E\nabla} (\nu, \eta) \right) - R_{{}^E\nabla} \left( {}^E\nabla_\mu \nu, \eta \right) - R_{{}^E\nabla}\left(\nu, {}^E\nabla_\mu \eta\right) 
\\
&\hspace{1cm}
	+ {}^E\nabla_\nu \left( R_{{}^E\nabla} (\eta, \mu) \right) - R_{{}^E\nabla} \left( {}^E\nabla_\nu \eta, \mu \right) - R_{{}^E\nabla}\left(\eta, {}^E\nabla_\nu \mu\right) \\
&\hspace{1cm}
	+ {}^E\nabla_\eta \left( R_{{}^E\nabla} (\mu, \nu) \right) - R_{{}^E\nabla} \left( {}^E\nabla_\eta \mu, \nu \right) - R_{{}^E\nabla}\left(\mu, {}^E\nabla_\eta \nu\right) \\
&\hspace{1cm}
	+ R_{{}^E\nabla}\left( {}^E\nabla_\mu \nu - {}^E\nabla_\nu \mu - [\mu, \nu]_E, \eta \right)
+ R_{{}^E\nabla}\left( {}^E\nabla_\nu \eta - {}^E\nabla_\eta \nu - [\nu, \eta]_E, \mu \right) \\
&\hspace{1cm}
	+ R_{{}^E\nabla}\left( {}^E\nabla_\eta \mu - {}^E\nabla_\mu \eta - [\eta, \mu]_E, \nu \right) \\
&=
{}^E\nabla_\mu \left( R_{{}^E\nabla} (\nu, \eta) \right) - {}^E\nabla_\nu \left( R_{{}^E\nabla} (\mu, \eta) \right) + {}^E\nabla_\eta \left( R_{{}^E\nabla} (\mu, \nu) \right)
\\
&\hspace{1cm}
	- R_{{}^E\nabla}([\mu, \nu]_E, \eta)
	+ R_{{}^E\nabla}([\mu, \eta]_E, \nu) - R_{{}^E\nabla}([\nu, \eta]_E, \mu) \\
&=
\mleft( \mathrm{d}^{{}^E\nabla} R_{{}^E\nabla} \mright)(\mu, \nu, \eta) 
\\
&\stackrel{\mathclap{ \text{Thm.~\ref{thm:2ndBianchi}} }}{=}\quad~~
0.
\eas
So, both formulations are equivalent for $V=E$, but Thm.~\ref{thm:2ndBianchi} is valid for any vector bundle $V$ and, thus, more general.
\end{remarks}

\begin{remark}
\leavevmode\newline
With a similar calculation as in Remark \ref{rem:FinallyTheOtherBianchiStuff} one can also rewrite the first Bianchi identity of Thm.~\ref{thm:1stBianchi} to
\bas
R_{{}^E\nabla}(\mu, \nu) \eta + R_{{}^E\nabla}(\nu, \eta) \mu + R_{{}^E\nabla}(\eta, \mu) \nu 
&=
\mleft( \mathrm{d}^{{}^E\nabla} t_{{}^E\nabla} \mright)(\mu, \nu, \eta)
\eas
for all $\mu, \nu, \eta \in \Gamma(E)$. Be careful, the right hand side is not the same as \textit{e.g.}~in Thm.~\ref{thm:modBianchithm}, \textit{i.e.}~not the same as ${}^E\nabla t_{{}^E\nabla}$ because the torsion is an element of $\Omega^{0,2}(N,E;E)$ such that ${}^E\nabla$ and $\mathrm{d}^{{}^E\nabla}$ do act differently.
\end{remark}

It is now natural to ask whether there is some usable commutation relation between both differentials, $\mathrm{d}^\nabla$ and $\mathrm{d}^{\nabla^{\mathrm{bas}}}$ for a fixed connection $\nabla$.

\begin{propositions}{Commutation relation}{commutationrelation}
Let $E \to N$ be a Lie algebroid over a smooth manifold $N$ and $\nabla$ a connection on $E$. Then
\ba
&\mleft(\mathrm{d}^\nabla \mathrm{d}^{\nabla^{\mathrm{bas}}} \omega\mright)\mleft( X_0, \dots, X_p, \nu_0, \dots, \nu_q \mright) \nonumber \\
&= \mleft(\mathrm{d}^{\nabla^{\mathrm{bas}}} \mathrm{d}^\nabla \omega\mright) \mleft( X_0, \dots, X_p, \nu_0, \dots, \nu_q \mright) \nonumber \\
&\hspace{0.8cm}
	+ \sum_{i=0}^p \sum_{k=0}^q (-1)^{i+k} R_\nabla^{\mathrm{bas}}\mleft(\nu_k, \omega\mleft( X_0, \dots, \widehat{X}_i, \dots, X_p, \nu_0, \dots, \widehat{\nu}_k, \dots, \nu_q \mright)\mright)X_i \nonumber \\
&\hspace{0.8cm}
	+ \sum_{i=0}^p \sum_{k=0}^q (-1)^{i+k} R_\nabla\Big( X_i, \rho\mleft( \omega \mleft( X_0, \dots, \widehat{X}_i, \dots, X_p, \nu_0, \dots, \widehat{\nu}_k, \dots, \nu_q \mright) \mright)\Big) \nu_k \nonumber \\
&\hspace{0.8cm}
	+ \sum_{\substack{ i,j=0 \\ i < j }}^p \sum_{k=0}^q (-1)^{i+j+k} \omega\mleft( \rho\mleft( R_\nabla\mleft( X_i, X_j \mright)\nu_k \mright), X_0, \dots, \widehat{X}_i, \dots, \widehat{X}_j, \dots, X_p, \nu_0, \dots, \widehat{\nu}_k, \dots, \nu_q \mright) \nonumber \\
&\hspace{0.8cm}
	+ \sum_{i=0}^p \sum_{\substack{ k,l=0 \\ k < l  }}^q (-1)^{i+k+l} \omega\mleft( X_0, \dots, \widehat{X}_i, \dots, X_p, R_\nabla^{\mathrm{bas}}(\nu_k, \nu_l)X_i, \nu_0, \dots, \widehat{\nu}_k, \dots, \widehat{\nu}_l, \dots, \nu_q \mright)
\ea
for all $\omega \in \Omega^{p, q}(N, E; E)$ ($p, q \in\mathbb{N}_0$), $X_0, \dots, X_p \in \mathfrak{X}(N)$ and $\nu_0, \dots, \nu_q \in \Gamma(E)$.
\end{propositions}

\begin{remark}
\leavevmode\newline
If $\nabla$ is flat and if $R_\nabla^{\mathrm{bas}} = 0$, then one has simply
\ba\label{eq:flatcommutation}
\mathrm{d}^\nabla \mathrm{d}^{\nabla^{\mathrm{bas}}} \omega
&= \mathrm{d}^{\nabla^{\mathrm{bas}}} \mathrm{d}^\nabla \omega.
\ea
Both differentials, $\mathrm{d}^\nabla$ and $\mathrm{d}^{\nabla^{\mathrm{bas}}}$, square to zero (recall Prop. \ref{prop:SnablamitREnabla})\footnote{As for vector bundle connections, one can also show for general Lie algebroid connections that the square of their exterior covariant derivatives is directly related to their curvature. We will not need this and the statements about $\mathrm{d}_1$ and $\mathrm{d}_2$, hence, we do not show this. But the calculation is precisely the same.} and, so, also the differentials
\ba
\omega &\mapsto 
\mathrm{d}_1 \omega
\coloneqq
\left( \mathrm{d}^\nabla + (-1)^p \mathrm{d}^{\nabla^{\mathrm{bas}}} \right) \omega, 
\label{def:differential1} \\
\omega &\mapsto 
\mathrm{d}_2 \omega
\coloneqq
\left( (-1)^q \mathrm{d}^\nabla + \mathrm{d}^{\nabla^{\mathrm{bas}}} \right) \omega 
\label{def:differential2}
\ea
for all $\omega \in \Omega^{p,q}(N,E;E)$, that can be seen by
\bas
\mathrm{d}_1^2 \omega
&= \mathrm{d}_1 \left( \mathrm{d}^\nabla + (-1)^p \mathrm{d}^{\nabla^{\mathrm{bas}}} \right) \omega \\
&= \left( \mathrm{d}^\nabla + (-1)^{p+1} \mathrm{d}^{\nabla^{\mathrm{bas}}} \right) \mathrm{d}^\nabla \omega
+ \left( \mathrm{d}^\nabla + (-1)^p \mathrm{d}^{\nabla^{\mathrm{bas}}} \right) (-1)^p \mathrm{d}^{\nabla^{\mathrm{bas}}} \omega \\
&= \underbrace{\left(\mathrm{d}^\nabla\right)^2}_{=0} \omega + \underbrace{\left( \mathrm{d}^{\nabla^{\mathrm{bas}}} \right)^2}_{=0} \omega + (-1)^{p+1} \mathrm{d}^{\nabla^{\mathrm{bas}}} \mathrm{d}^\nabla \omega + (-1)^p \mathrm{d}^\nabla \mathrm{d}^{\nabla^{\mathrm{bas}}} \omega \\
&= (-1)^p \left( \mathrm{d}^\nabla \mathrm{d}^{\nabla^{\mathrm{bas}}} - \mathrm{d}^{\nabla^{\mathrm{bas}}} \mathrm{d}^\nabla \right) \omega \\
&\stackrel{\mathclap{\text{Eq. } \eqref{eq:flatcommutation}}}{=} \quad~ 
0,
\eas
similarly with $\mathrm{d}_2$.

For $\nu \in \Gamma(E)$ one gets
\ba
\mleft[ \mathrm{d}^{\nabla^{\mathrm{bas}}}, \mathrm{d}^\nabla \mright] \nu
&=
\iota_\nu R_\nabla^{\mathrm{bas}} + \iota_{\rho(\nu)} R_\nabla,
\ea
here, $\gls{1jota}$ denotes the contraction. Especially for flat $\nabla$, $R_\nabla^{\mathrm{bas}}$ describes the commutation relation of both exterior covariant derivatives.
\end{remark}

\begin{proof}[Proof of Prop.~\ref{prop:commutationrelation}]
\leavevmode\newline
That is an extremely long and tedious but completely straightforward calculation. There is no trick to use, "just" insert the definitions of all tensors and exterior covariant derivatives on both sides of the equation and compare.
\end{proof}

We can immediately conclude the following.

\begin{corollaries}{Commutation for vanishing basic curvature}{commutationS=0}
Let $E \to N$ be a Lie algebroid over a smooth manifold $N$ and $\nabla$ a connection on $E$.
Then $R_\nabla^{\mathrm{bas}} = 0$ if and only if
\ba\label{EasyDifferentialCommutationSGleichNuuull}
&\mleft(\mathrm{d}^\nabla \mathrm{d}^{\nabla^{\mathrm{bas}}} \omega\mright)\mleft( X_0, \dots, X_p, \nu_0, \dots, \nu_q \mright) \nonumber \\
&= \mleft(\mathrm{d}^{\nabla^{\mathrm{bas}}} \mathrm{d}^\nabla \omega\mright) \mleft( X_0, \dots, X_p, \nu_0, \dots, \nu_q \mright) \nonumber \\
&\hspace{0.8cm}
	+ \sum_{i=0}^p \sum_{k=0}^q (-1)^{i+k} R_\nabla\Big( X_i, \rho\mleft( \omega \mleft( X_0, \dots, \widehat{X}_i, \dots, X_p, \nu_0, \dots, \widehat{\nu}_k, \dots, \nu_q \mright) \mright)\Big) \nu_k \nonumber \\
&\hspace{0.8cm}
	+ \sum_{\substack{ i,j=0 \\ i<j }}^p \sum_{k=0}^q (-1)^{i+j+k} \omega\mleft( \rho\mleft( R_\nabla\mleft( X_i, X_j \mright)\nu_k \mright), X_0, \dots, \widehat{X}_i, \dots, \widehat{X}_j, \dots, X_p, \nu_0, \dots, \widehat{\nu}_k, \dots, \nu_q \mright)
\ea
for all $\omega \in \Omega^{p, q}(N, E; E)$ ($p,q \in \mathbb{N}_0$), $X_0, \dots, X_p \in \mathfrak{X}(N)$ and $\nu_0, \dots, \nu_q \in \Gamma(E)$.
\end{corollaries}

\begin{remark}
\leavevmode\newline
The "$\Rightarrow$"-direction was also found by Alexei Kotov. While I have derived it with the more general previous proposition, Alexei Kotov has directly shown it from the point of view of differentialgraded manifolds. This was communicated in a personal communication but there is a paper planned about that by Alexei Kotov and Thomas Strobl, planned for 2021.
\end{remark}

\begin{proof}[Proof of Cor.~\ref{cor:commutationS=0}]
\leavevmode\newline
The "$\Rightarrow$" direction, \textit{i.e.}~we assuming a vanishing basic curvature, is clear by Prop.~\ref{prop:commutationrelation}. For the "$\Leftarrow$" direction we want to use Eq.~\eqref{eq:compcondfast} in Prop.~\ref{prop:SnablamitREnabla}. Observe that
\bas
\mleft(\mathrm{d}^{\nabla^{\mathrm{bas}}} \mathds{1}_E\mright) (\mu, \nu)
&=
\nabla^{\mathrm{bas}}_\mu \nu - \nabla^{\mathrm{bas}}_\nu \mu - [\mu, \nu]_E
=
t_{\nabla^{\mathrm{bas}}}(\mu, \nu)
\eas
for all $\mu, \nu \in \Gamma(E)$,
and
\bas
\mleft( \mathrm{d}^\nabla \mathds{1}_E \mright)(X, \mu)
&= \mleft( \nabla_X \mathds{1}_E\mright) (\mu)
= \nabla_X \mu - \nabla_X \mu
= 0
\eas
for all $X \in \mathfrak{X}(N)$ and $\mu \in \Gamma(E)$.
Using these and by choosing $\omega = \mathds{1}_E \in \Omega^{0,1}(M,E;E)$ we have by Eq.~\eqref{EasyDifferentialCommutationSGleichNuuull}
\bas
&&\mleft(\mathrm{d}^\nabla t_{\nabla^{\mathrm{bas}}} \mright)(X, \mu, \nu)
&=
\mleft(\mathrm{d}^\nabla \mathrm{d}^{\nabla^{\mathrm{bas}}} \mathds{1}_E \mright)(X, \mu, \nu)
= 
R_\nabla(X, \rho(\nu))\mu - R_\nabla(X, \rho(\mu))\nu \\
&\stackrel{\text{Eq. } \eqref{eq:compcondfast}}{\Leftrightarrow}&
R_\nabla^{\mathrm{bas}}(\mu, \nu)X 
&= \underbrace{\mleft( \nabla_X t_{\nabla^{\mathrm{bas}}} \mright)(\mu, \nu)}_{= \mleft(\mathrm{d}^\nabla t_{\nabla^{\mathrm{bas}}} \mright)(X, \mu, \nu)}
- R_\nabla(\rho(\mu), X) \nu + R_\nabla(\rho(\nu), X) \mu
= 0.
\eas
\end{proof}
\section{Direct product of Lie algebroids}\label{DirectProdsOfLieAlgoids}

We will also need to know how to define the direct products of Lie algebroids where we especially refer to \cite[Lemma 6.25]{meinrenkenlie} or \cite[beginning of \S 4.2; page 155]{mackenzieGeneralTheory}.

In the following we will have two Lie algebroids $(E_1, \mleft[ \cdot, \cdot \mright]_{E_1}, \rho_1) \to N_1$ and $(E_2, \mleft[ \cdot, \cdot \mright]_{E_2}, \rho_2) \to N_2$ over two smooth manifolds $N_1$ and $N_2$. With $\gls{pri}: N_1 \times N_2 \to N_i$ we will denote in the following part of this section the projection onto the $i$-th factor ($i \in \{1,2\}$), and $\mathrm{T}\mleft( N_1 \times N_2 \mright)$ can be regarded as the Whitney sum of vector bundles $\mathrm{pr}_1^*\mleft( \mathrm{T}N_1 \mright) \oplus \mathrm{pr}_2^*\mleft( \mathrm{T}N_2 \mright)$, as usual and as mentioned in \cite{mackenzieGeneralTheory}. We want to define a Lie algebroid structure on $\mathrm{pr}_1^*\mleft( E_1 \mright) \oplus \mathrm{pr}_2^*\mleft( E_2 \mright) \to N_1 \times N_2$ (Whitney sum of $\mathrm{pr}_i^*\mleft( E_i \mright)$), and, thus, a canonical candidate of the anchor is immediately given by $\mathrm{pr}_1^*\rho_{E_1} \oplus \mathrm{pr}_2^*\rho_{E_2}$.

Sections of $\mathrm{pr}_i^*\mleft( E_i \mright)$ can be viewed as compositions of the form $\mu^a ~ \mathrm{pr}_i^*\mleft( V^i_a \mright)$, where $V^i_a \in \Gamma(E_i)$ and $\mu^a \in C^\infty\mleft( N_1 \times N_2 \mright)$, simply using that pullbacks of sections generate all sections. Using such decompositions has the advantage that the frames are given by (pullbacks of) frames of $E_i$, especially, $\mathrm{pr}_i^*\mleft( V^i_a \mright)$ (no sum over $i$) is constant along $N_j$, $j \neq i$. We then say that we take a \textbf{frame induced by $E_1$ and $E_2$}.

\begin{lemmata}{Uniqueness of the Lie algebroid structure on $E_1 \times E_2$, \newline \cite[Lemma 6.25]{meinrenkenlie} \newline \cite[beginning of \S 4.2; page 155]{mackenzieGeneralTheory}}{LemmaUniquenessOfDirectProductStructure}
Let $(E_1, \mleft[ \cdot, \cdot \mright]_{E_1}, \rho_1) \to N_1$ and $(E_2, \mleft[ \cdot, \cdot \mright]_{E_2}, \rho_2) \to N_2$ be two Lie algebroids over two smooth manifolds $N_1$ and $N_2$, and let $E_1 \times E_2 \coloneqq \mathrm{pr}_1^*\mleft( E_1 \mright) \oplus \mathrm{pr}_2^*\mleft( E_2 \mright) \to N_1 \times N_2$ be the Whitney sum of vector bundles, equipped with the direct product of anchors. Then there is a unique Lie algebroid structure on $E_1 \times E_2$ such that
\ba\label{defGenerationOfDirectProductSections}
\Gamma(E_1) \oplus \Gamma(E_2) &\to \Gamma\mleft(E_1 \times E_2\mright),
\nonumber \\
(\mu, \nu) &\mapsto \mathrm{pr}_1^*\mu \oplus \mathrm{pr}_2^*\nu
=
\mleft( \mathrm{pr}_1^*\mu, \mathrm{pr}_2^*\nu \mright)
\ea
is a Lie algebra homomorphism, where $\Gamma(E_i)$ are viewed as (infinite-dimensional) Lie algebras.
\end{lemmata}

\begin{remark}
\leavevmode\newline
With the direct product of anchors we mean here
\bas
\rho_{E_1 \times E_2}
\coloneqq
\rho_{E_1} \times \rho_{E_2}
&\coloneqq
\mathrm{pr}_1^*\rho_{E_1} \oplus \mathrm{pr}_2^*\rho_{E_2}.
\eas
\end{remark}

\begin{proof}[Sketch of the proof of Lemma \ref{lem:LemmaUniquenessOfDirectProductStructure}]
\leavevmode\newline
We just give a sketch of the proof since the calculations are all very straightforward, but tedious to write down explicitly; the construction is as usual, making use of that some certain subset of sections generate all sections and that one knows how to define structures on that subset given by the map in \eqref{defGenerationOfDirectProductSections}. The full structure then uniquely follows by forcing the Leibniz rule on the Lie bracket.

In the following we will also omit all the pullback notations, so, when we write for example that we take a section of $\Gamma(E_1)$, then we actually mean a pullback of that section along $\mathrm{pr}_1$. Especially, we understand $\Gamma(E_1) \oplus \Gamma(E_2)$ as embedded in the sense of \eqref{defGenerationOfDirectProductSections}.

$\bullet$ For the existence we define the Lie bracket $\mleft[ \cdot, \cdot \mright]_{E_1 \times E_2}$ as in the following: Let $\mleft( f_a^{(i)} \mright)_a$ be a frame of $E_i$ ($i \in \{1,2\}$) and their pullbacks give combined a frame of $E_1 \times E_2$ which we denote by $\mleft( e_a \mright)_a$; note that $e_a \in \Gamma(E_1) \oplus \Gamma(E_2)$. The bracket $\mleft[ e_a, e_b \mright]_{E_1 \times E_2}$ of this frame is then canonically defined as direct product of the brackets $\mleft[ \cdot, \cdot \mright]_{E_1}$ and $\mleft[ \cdot, \cdot \mright]_{E_2}$ given by the direct product of Lie algebras $\Gamma(E_1) \oplus \Gamma(E_2)$. Making use of that $\Gamma(E_1) \oplus \Gamma(E_2)$ generates $\Gamma(E_1 \times E_2)$, we then write for two sections $\mu = \mu^a e_a, \nu = \nu^a e_a \in \Gamma(E_1 \times E_2)$, and we then apply the typical construction to force the Leibniz rule on the full set of sections,
\ba\label{1defLieBracketOfDirectProductAlgebroids}
\mleft[ \mu, \nu \mright]_{E_1 \times E_2}
&\coloneqq
\mu^a \nu^b ~ \mleft[ e_a, e_b \mright]_{E_1 \times E_2}
	+ \mu^a \mathcal{L}_{\rho_{E_1 \times E_2}(e_a)}\mleft(\nu^b\mright) ~ e_b
	- \nu^b \mathcal{L}_{\rho_{E_1 \times E_2}(e_b)}\mleft(\mu^a\mright) ~ e_a,
\ea
where $\rho_{E_1 \times E_2} = \rho_{E_1} \times \rho_{E_2}$ is the direct product of anchors. This is well-defined, because any other frames $\mleft( f_a^{(i)} \mright)_a$ are locally related by a matrix on $N_i$, so, a change constant along $N_j$ ($j \in \{1,2\}$, $i \neq j$). Hence, $E_1$-$E_2$-mixed terms of $\mleft[ e_a, e_b \mright]_{E_1 \times E_2}$ are unaffected by a change of such frames, and, so, it is still a direct product of Lie brackets for another frame. Especially, it follows that the bracket is the direct product of the brackets on $\Gamma(E_1) \oplus \Gamma(E_2)$. That the whole bracket is independent of the chosen frame is also trivial and straightforward to check; that essentially follows by construction since the Lie derivatives $\mathcal{L}_{\rho_{E_1 \times E_2}(e_a)}$ will cancel the Leibniz rule of $\mleft[ e_a, e_b \mright]_{E_1 \times E_2}$ when changing the frame.

The calculations that this gives a Lie algebroid structure is now straightforward, similar to the proof of Prop.~\ref{prop:ActionLieoidsAreOids}. That is, the curvature of $\rho_{E_1 \times E_2}$ is trivially the direct product of the curvature of $\rho_{E_1}$ and $\rho_{E_2}$
\bas
R_{\rho_{E_1 \times E_2}}
&=
R_{\rho_{E_1}} \times R_{\rho_{E_2}}
\eas
recall Def.~\ref{def:GeneralDefOfCurvMorphisms}. That simply follows by the fact that the anchor is a direct product and that the Lie bracket is a direct product on $\Gamma(E_1) \oplus \Gamma(E_2)$, so, the curvature is a direct product in the frame $(e_a)$, and therefore always because the curvature is a tensor (Lemma \ref{lem:KruemmungenSindTensorenMitAnkerErhaltung}) and $\Gamma(E_1) \oplus \Gamma(E_2)$ generates $\Gamma(E_1 \times E_2)$. Since $E_i$ are Lie algebroids, the curvature is zero.

The Lie bracket clearly satisfies the Leibniz rule with respect to $\rho_{E_1 \times E_2}$, and hence by Prop.~\ref{prop:MeasureofJacobiandHomom}, we can test the Jacobi identity in a given frame; by construction, with respect to the frame $(e_a)$ the bracket is a direct product of Lie brackets given by the direct product of Lie algebras $\Gamma(E_1) \oplus \Gamma(E_2)$. So, Jacobi identity immediately follows.

$\bullet$ That the map defined in \eqref{defGenerationOfDirectProductSections} is a Lie algebra homomorphism follows by construction since the anchor and the Lie bracket are defined as direct products on $\Gamma(E_1) \oplus \Gamma(E_2)$.

$\bullet$ Uniqueness will follow by using that $\Gamma(E_1 \times E_2)$ is generated by $\Gamma(E_1) \oplus \Gamma(E_2)$ as a module over $C^\infty(N_1 \times N_2)$ using the map defined in \eqref{defGenerationOfDirectProductSections}, now denoted by $\Phi$. Since $\Phi$ shall be a homomorphism, the bracket on $\Gamma(E_1) \oplus \Gamma(E_2)$ embedded into $\Gamma(E_1 \times E_2)$ is given by the direct product of $\mleft[ \cdot, \cdot \mright]_{E_1}$ and $\mleft[ \cdot, \cdot \mright]_{E_2}$ in sense of Lie algebras; similarly as for $\mathfrak{X}(N_1) \oplus \mathfrak{X}(N_2)$. Then take any Lie algebroid bracket on $E_1 \times E_2$ such that $\Phi$ is a homomorphism and express sections with respect to $(e_a)_a$. Using the Leibniz rule, every other possible Lie bracket has then the form of \eqref{1defLieBracketOfDirectProductAlgebroids}, therefore uniqueness is given.
\end{proof}

Hence, we define:

\begin{definitions}{Direct product of Lie algebroids}{DirecProductOfLieAlgebroids}
Let $(E_1, \mleft[ \cdot, \cdot \mright]_{E_1}, \rho_1) \to N_1$ and $(E_2, \mleft[ \cdot, \cdot \mright]_{E_2}, \rho_2) \to N_2$ be two Lie algebroids over two smooth manifolds $N_1$ and $N_2$, and let $E_1 \times E_2 \coloneqq \mathrm{pr}_1^*\mleft( E_1 \mright) \oplus \mathrm{pr}_2^*\mleft( E_2 \mright) \to N_1 \times N_2$ be the Whitney sum of vector bundles.

Then we call the Lie algebroid structure as given in Lemma \ref{lem:LemmaUniquenessOfDirectProductStructure} the \textbf{direct product of Lie algebroids}.
\end{definitions}

There are some examples of direct products, especially also the Higgs mechanism of the standard model.

\begin{examples}{Examples of direct products of Lie algebroids}{ExamplesOfDirectProductsLieAoids}
We provide two canonical examples; the first one directly comes by the construction for which we viewed $\mathrm{T}\mleft( N_1 \times N_2 \mright)$ as the Whitney sum $\mathrm{pr}_1^*\mleft( \mathrm{T}N_1 \mright) \oplus \mathrm{pr}_2^*\mleft( \mathrm{T}N_2 \mright)$.
\begin{enumerate}
	\item The first example is the direct product of two tangent bundles, $E_i \coloneqq \mathrm{T}N_i$ where the Lie brackets are the ones from the tangent bundles and $\rho_i \coloneqq \mathds{1}_{\mathrm{T}N_i}$. Then $E_1 \times E_2 = \mathrm{T}\mleft( N_1 \times N_2 \mright)$.
	\item Let $E_1$ be the action Lie algebroid of the electroweak interaction, see Ex. \ref{ex:electroweakinteractionasLiealgoid}, and $E_2$ be the Lie algebra $\mathrm{su}(3) \to \{*\}$ over a point set $\{*\}$ (with zero anchor). Then $E_1 \times E_2$ is called the \textbf{Higgs mechanism of the standard model}.
\end{enumerate}
\end{examples}

As usual, if we have several structures given on both factors, then we can often take their product to define a similar structure on the whole product of Lie algebroids. For tensors and connections this is straightforward, however, we also have Lie algebroid connections and we have seen that pullbacks of those may not always been given; especially recall Cor.~\ref{cor:GeneralPullbackAnchorPreserving}, that is, anchor-preserving vector bundle morphisms are needed.

\begin{lemmata}{Projections have lifts to anchor-preserving morphisms}{LiftsOfProjections}
Let $(E_1, \mleft[ \cdot, \cdot \mright]_{E_1}, \rho_1) \to N_1$ and $(E_2, \mleft[ \cdot, \cdot \mright]_{E_2}, \rho_2) \to N_2$ be two Lie algebroids over two smooth manifolds $N_1$ and $N_2$, and let $E_1 \times E_2$ be the direct product of Lie algebroids.

Then the projections $\pi_i: E_1 \times E_2 \to E_i$ ($i \in \{1,2\}$) are anchor preserving vector bundle morphisms over $\mathrm{pr}_i: N_1 \times N_2 \to N_i$.
\end{lemmata}

\begin{remark}
\leavevmode\newline
To clarify: $\pi_i$ project to $E_i \to N_i$ as Lie algebroid, not onto $\mathrm{pr}_i^*E_i \to N_1 \times N_2$. However, extended to sections, $\pi_i$ maps to $\Gamma(\mathrm{pr}_i^*E_i)$; recall Remark \ref{rem:SomeExtraNotationForAnchorBundleMorphs}.
\end{remark}

\begin{proof}[Proof of Lemma \ref{lem:LiftsOfProjections}]
\leavevmode\newline
$\pi_i$ are clearly vector bundle morphisms by definition. Denote with $p_i$ the projection of the bundle $E_i \stackrel{p_i}{\to} N_i$, similarly $p$ the projection of $E_1 \times E_2 \stackrel{p}{\to} N_1 \times N_2$, then
\bas
p_i \circ \pi_i
&=
\mathrm{pr}_i \circ p
\eas
by definition, \textit{i.e.}~using that $E_1 \times E_2 = \mathrm{pr}_1^*E_1 \oplus \mathrm{pr}_2^*E_2$. Hence, $\pi_i$ are vector bundle morphisms over $\mathrm{pr}_i$. Therefore we only need to check the anchor-preservation, that is, observe that with precisely the same arguments
\bas
\mathrm{Dpr}_1: \mathrm{pr}_1^*\mathrm{T}N_1 \oplus \mathrm{pr}_2^*\mathrm{T}N_2
&\to
\mathrm{T}N_1,
\\
(X, Y)
&\mapsto
X
\eas
is a vector bundle morphism over $\mathrm{pr}_1$ as it is also well-known, similarly for $\mathrm{Dpr}_2$.\footnote{Essentially, the $\mathrm{Dpr}_i$ are the "$\pi_i$ for $E_i = \mathrm{T}N_i$".} Then
\bas
\mleft(\mathrm{Dpr}_i \circ
	\rho_{E_1 \times E_2}
\mright)(\mu_1, \mu_2)
&=
\mathrm{Dpr}_i \mleft(
	\rho_{E_1 \times E_2}(\mu_1, \mu_2)
\mright)
\\
&=
\mathrm{Dpr}_i \Bigl(
	\bigl( (\mathrm{pr}_1^*\rho_{E_1})(\mu_1), (\mathrm{pr}_2^*\rho_{E_2})(\mu_2) \bigr)
\Bigr)
\\
&=
(\mathrm{pr}_i^*\rho_{E_i})(\mu_i)
\\
&=
(\mathrm{pr}_i^*\rho_{E_i}\circ\pi_i)(\mu_1, \mu_2)
\eas
for all $(\mu_1, \mu_2) \in \Gamma(E_1 \times E_2)$. Thus, $\pi_i$ is anchor-preserving; also recall Remark \ref{rem:SomeExtraNotationForAnchorBundleMorphs}.
\end{proof}

By Cor.~\ref{cor:GeneralPullbackAnchorPreserving} we can therefore also make pullbacks of Lie algebroid connections along those projections. As a conclusion of this section, let us summarize and introduce the following.

\begin{remarks}{Products of inherited structures}{NotationAboutProductStructures}
Let $(E_1, \mleft[ \cdot, \cdot \mright]_{E_1}, \rho_1) \to N_1$ and $(E_2, \mleft[ \cdot, \cdot \mright]_{E_2}, \rho_2) \to N_2$ be two Lie algebroids over two smooth manifolds $N_1$ and $N_2$, and let $E_1 \times E_2$ be the direct product of Lie algebroids. 
Furthermore, let $\pi_i$ ($i \in \{1,2\}$) be the projections $E_1 \times E_2 \to E_i$ as in Lemma \ref{lem:LiftsOfProjections}.

Then, roughly in general, if we have some object $B_i$ on $E_i$, then we define their product by
\ba\label{ProductOfObjects}
B_1 \times B_2
&\coloneqq
\mathrm{pr}_1^*B_1 \oplus \mathrm{pr}_2^*B_2,
\ea
in case there is a well-defined notion for $\mathrm{pr}_i^*B_i$. This is of course well-defined for tensors, \textit{i.e.}~$B_i \in \mathcal{T}^r_s(E_i)$ ($r,s \in \mathbb{N}_0$).

Another examples are vector bundle connections $B_i \coloneqq \nabla^i$ on $E_i$, or $E_i$-connections $B_i \coloneqq {}^{E_i}\nabla$ on vector bundles $V_i \to N_i$ by using Cor.~\ref{cor:GeneralPullbackAnchorPreserving}. Especially the latter means that we always canonically use $\pi_i$ for the pullbacks of $E_i$-connections, and observe
\bas
\mleft(\mathrm{pr}_i^*\mleft( {}^{E_i}\nabla \mright)\mright)_{(\mu_1, \mu_2)} (\mathrm{pr}_i^*v)
&=
\mathrm{pr}_i^*\mleft(
	{}^{E_i}\nabla_{\mu_i} v
\mright)
\eas
for all $v \in \Gamma(V_i)$ and $(\mu_1, \mu_2) \in \Gamma(E_1 \times E_2)$. Thence, exactly what one naturally expects, for example "mixed terms are zero", that is, for example 
\bas
\mleft(\mathrm{pr}_1^*\mleft( {}^{E_1}\nabla \mright)\mright)_{(0, \mu_2)} (\mathrm{pr}_1^*v)
&=
0.
\eas

That is of special usage if one uses that $\Gamma(E_1) \oplus \Gamma(E_2)$ generates $\Gamma(E_1 \times E_2)$ and that the mentioned structures are uniquely given by how they act on $\Gamma(E_1) \oplus \Gamma(E_2)$; also recall Lemma \ref{lem:LemmaUniquenessOfDirectProductStructure}. So, one just needs to take a frame induced by frames of $E_i$, and if a given structure restricts in that frame to a structure on $E_i$, if just using the part of the frame induced by $E_i$, and has no "mixed terms", then one knows that this object can be written as direct product. 

All of that above similarly for structures given by $\mathrm{T}N_i$, and structures involving the tangent bundles and the $E_i$ as in the case of the anchors.

For example, let us have vector bundle connections $\nabla^i$ on $E_i$, then we have the induced basic connections $\nabla^{i,\mathrm{bas}}$. We have a vector bundle connection on $E_1 \times E_2$ by
\bas
\nabla
&\coloneqq
\nabla^1 \times \nabla^2,
\eas
whose curvature also splits as it is well-known (trivial to check with a frame induced by frames of $E_1$ and $E_2$). With $\nabla^{1,\mathrm{bas}} \times \nabla^{2,\mathrm{bas}}$ one has a pair of $E_1 \times E_2$-connections on $E_1 \times E_2$ and $\mathrm{T}N_1 \times \mathrm{T}N_2$. Taking a frame induced by frames of $E_1$ and $E_2$ and $\mathrm{T}N_1$ and $\mathrm{T}N_2$, all of those connections and Lie algebroid connections restrict to the factors in $E_1 \times E_2$ by definition. Using Lemma \ref{lem:LemmaUniquenessOfDirectProductStructure}, also the Lie bracket and anchor are a direct product on such a frame, for both $E_1 \times E_2$ and $\mathrm{T}N_1 \times \mathrm{T}N_2$, hence, 
\bas
\mleft( \nabla^1 \times \nabla^2 \mright)^{\mathrm{bas}}
&=
\nabla^{1,\mathrm{bas}} \times \nabla^{2,\mathrm{bas}}
\eas
and
\bas
R_{\nabla^1 \times \nabla^2}^{\mathrm{bas}}
&=
R_{\nabla^1}^{\mathrm{bas}}
\times
R_{\nabla^2}^{\mathrm{bas}}.
\eas
Similarly the exterior covariant derivatives of $\nabla^1 \times \nabla^2$ and $\mleft( \nabla^1 \times \nabla^2 \mright)^{\mathrm{bas}}$ split on products of forms $\omega_i \in \Omega^{p_i, q_i}(N, E;E)$ ($p_i, q_i \in \mathbb{N}_0$) given by
\ba
\omega_1 \times \omega_2
&\coloneqq
\mathrm{pr}_1^!\omega_1 \oplus \mathrm{pr}_2^!\omega_2.
\ea
The differentials of $\mathrm{Dpr}_i$ are projections $\mathrm{T}(N_1 \times N_2) \to \mathrm{T}N_i$ such that there is not really a significant distinction between $\mathrm{pr}_i^*$ and $\mathrm{pr}_i^!$. This is why we are not going to clarify in such situations whether the product is using pullbacks in sense of sections or forms. It will be clear by context.
\end{remarks}

\section{Splitting theorem for Lie algebroids}\label{SectionAboutSplitting}

Using the last section, one can locally formulate Lie algebroids as direct products of certain Lie algebroids. Let us study that, but first we need some basic notions; we are mainly following \cite{DaSilva} now.

\begin{definitions}{Singular and regular points of vector bundle morphisms, \newline \cite[\S 4; generalization of third remark after Theorem 4.1; page 17]{DaSilva}}{RegularPointsOfVectorBundleMorphisms}
Let $V_1 \stackrel{\pi_1}{\to} N_1$ and $V_2 \stackrel{\pi_2}{\to} N_2$ be vector bundles over smooth manifolds $N_1$ and $N_2$, respectively. Also let $P: V_1 \to V_2$ be a continuous vector bundle morphism over some continuous map $f: N_1 \to N_2$, \textit{i.e.}~$\pi_2 \circ P = f \circ \pi_1$.

We call a point $p \in N_1$ a \textbf{regular point} if there is an open neighbourhood around $p$ onto which $\gls{Rzk}(P)$, the rank of $P$, is constant. \textbf{Singular points} are points $p \in N_1$ which are not regular.
\end{definitions}

In our case $P$ will be the anchor $\rho$, and since $\rho$ is a homomorphism we know that the image of $\rho$, $\gls{Im}(\rho)$, is closed under the Lie bracket of the tangent bundle such that we expect a foliation related to the image of $\rho$ by the Frobenius Theorem; however, since the rank of an anchor is not constant as we pointed out earlier, the foliation induced by the image of the anchor is a singular foliation. Formally, this is proven as a more general Frobenius theorem as also discussed in \cite[discussion after the definition in \S 16.1; page 113]{DaSilva}; also see \cite[beginning of \S 3.1]{meinrenkensplitting}. Essentially, one gets still a foliation if a subset of the tangent bundle is closed under the Lie bracket, but the foliation is singular (non-constant dimension of the leaves). We are interested into those leaves of the anchor, also called orbits, such that we need to study the rank of $\rho$. There is a statement about that the amount of singular points is "small".

\begin{propositions}{Amount of singular and regular points, \newline \cite[generalization of second remark after Theorem 4.1; page 17]{DaSilva}}{RegularPointsAreDense}
Let the situation be as in Def.~\ref{def:RegularPointsOfVectorBundleMorphisms}. Then the set of all regular points is dense in $N_1$.
\end{propositions}

\begin{proof}
\leavevmode\newline
Let $S_{\text{reg}}$ and $S_{\text{locmax}}$ be the sets of regular points and of local maxima of $\text{rk}(P)$ in $N_1$, respectively. It is clear that $S_{\text{reg}} \subset S_{\text{locmax}}$ but we can also show $S_{\text{locmax}} \subset S_{\text{reg}}$: Let $p \in N_1$ be a local maximum of $\text{rk}(P)$ with value $k \in \mathds{N}_0$ and let $k \geq 1$ w.l.o.g. (since for $k=0$ it is clear that then $p \in S_{\text{reg}}$). Then there is a minor $m$ of order $k$ of $P$ such that $m(p) \neq 0$. By continuity of $P$ there is an open neighbourhood $U \subset N_1$ containing $p$ such that $\left. m\middle|_{ U} \right. \neq 0$ and, thus, $\left.rk(P)\middle|_U\right. \geq k$. Therefore also $\left.rk(P)\middle|_U\right. = k$ due to $p \in S_{\text{locmax}}$. Thence, $p \in S_{\text{reg}}$ and so $S_{\text{reg}} = S_{\text{locmax}} \eqqcolon S$.

Now let $x_0 \in N_1 \setminus S$ and $U$ an open neighbourhood of $x_0$. $\mathrm{rk}(P)$ reaches its upper bound on $U$, \textit{i.e.}
\bas
\exists y \in U: ~ \forall x \in U: ~ (\text{rk}(P))(x) \leq (\text{rk}(P))(y).
\eas
This follows by the fact that $\sup_{x \in U} (\mathrm{rk}(P))(x) \eqqcolon l < \infty$ by the boundedness of $\mathrm{rk}(P)$ and w.l.o.g.~we can say that $l \in \mathds{N}_0$ by the $\mathds{N}_0 \text{-valuedness}$ of $\mathrm{rk}(P)$; there must be a $y \in U$ such that $l = (\mathrm{rk}(P))(y)$ since for any other upper bound $l' \in \mathds{N}_0$ of the rank on $U$, for which there is no $y \in U$ with $l' = (\mathrm{rk}(P))(y)$, one can lower $l'$ by 1 such that $l' - 1$ is still an upper bound (follows again by the $\mathds{N}_0\text{-valuedness}$). This procedure is repeated until one gets an upper bound which is the value of some element in $U$. Thus, the supremum is also a maximum. Thence
\bas
&\forall x_0 \in N_1 \setminus S: ~ \forall \text{ open neighbourhoods } U \text{ of } x_0: \exists y \in U: ~ y \in S_{\text{locmax}} = S_{\text{reg}} \\
&\Rightarrow x_0 \text{ is an accumulation point of } S_{\text{reg}}
\\
&\Rightarrow N_1 \setminus S \subset \overline{S}
\\
&\Rightarrow \overline{S_{\text{reg}}} = N_1,
\eas
where $\overline{S}$ denotes the closure of $S = S_{\mathrm{reg}}$.
\end{proof}

\begin{remark}
\leavevmode\newline
This means, assuming $N_1$ is connected, one has "walls of measure zero" of singular points between the connected components of the set of all regular points, \textit{i.e.}~between zones of different rank of $P$. By the previous proof one can also see that the rank of $P$ is locally not maximal at a singular point.
\end{remark}

Around regular points of $\rho$, its distribution is also an integrable foliation since the rank is constant. In general the natural question arises if one can split the Lie algebroid structure locally along this distribution, in sense of "orbital plus transversal structure". Indeed, there are several statements about such \textbf{splitting theorems}, starting with the important splitting theorem of Poisson manifolds by Weinstein as in \cite[Theorem 4.2; page 19]{DaSilva}, another splitting theorem for Lie algebroids can be found in \cite[Theorem 1.1]{fernandes}. If you are interested into a more general approach and theorem then see \cite{meinrenkensplitting}; in this paper the locality is just along the foliation while it can be "arbitrary big" along the transversal direction.

To discuss the splitting theorem for Lie algebroids would certainly exceed the work of this thesis. Hence, we will just state the most simplified statement around regular points without further proof; see the listed references for a thorough discussion. Recall the discussion after Def.~\ref{def:IsotropyForLieAlgeoids}, the kernel of the anchor at a point is a Lie algebra. Around regular points this means that the kernel is a bundle of Lie algebras, $\mathrm{Ker}(\rho) \to N$, one makes use of that in the following statement. For the following statement also recall that two submanifolds $M_1, M_2$ of $N$ are transversal if
\bas
\mathrm{T}_pM_1 + \mathrm{T}_p M_2
&=
\mathrm{T}_p N
\eas
for all $p \in M_1 \cap M_2$. We speak of a \textbf{direct transversal} if the sum is a direct sum/product.

\begin{theorems}{Splitting theorem around regular points, \cite[Corollary 4.2]{meinrenkensplitting}}{DirectSplitting}
Let $E \to N$ be a Lie algebroid over a connected manifold $N$ such that $N$ only consists of regular points of the anchor $\rho$. Fix a point $p \in N$, and denote with $L$ the leaf through $p$, given by the foliation of $\rho$. Furthermore, take a submanifold $S$ with $p \in S$ and which is transversal to the foliation of the anchor and which is a direct transversal of $L$. Then
\ba
E
&\stackrel{\text{locally around } p}{\cong}
\mathrm{T}L \times \mathrm{Ker}(\rho)|_S,
\ea
where $\mathrm{T}L \times \mathrm{Ker}(\rho)|_S$ is the direct product of Lie algebroids $\mathrm{T}L \to L$ and $\mathrm{Ker}(\rho)|_S \to S$ (the bundle of Lie algebras given by the $\mathrm{Ker}(\rho)$ restricted to $S$).
\end{theorems}

\begin{remarks}{Local frame of the splitting theorem}{LocalSplittingFrame}
This theorem implies that around regular points $p \in N$ are coordinate vector field $\mleft( \partial_i \mright)_i$ of $L$, and a frame $\mleft( e_a \mright)_a$ of $\mathrm{Ker}(\rho)|_S$ such that
\bas
\rho(\partial_i)
&=
\partial_i,
\\
\rho(e_a)
&=
0,
\\
\mleft[ \partial_i, e_a \mright]_E
&=
0,
\eas
using Lemma \ref{lem:LemmaUniquenessOfDirectProductStructure}. We will later define the field of gauge bosons $A$ as a form on the spacetime with values in (the pullback of) a Lie algebroid; the components of $A$ along $e_a$ are then the massless gauge bosons, while the other ones may get mass. The Higgs field will be a smooth map of the spacetime to $N$, and its components along $L$ are then the Nambu-Goldstone bosons, while the transversal components are the Higgs bosons; for this recall the discussion about the Higgs mechanism after Def.~\ref{def:ClassicYMHLagrangian} and the isotropy around Def.~\ref{def:IsotropyForLieAlgeoids}.
\end{remarks}

Using such a frame we conclude this section with a short statement about the existence of parallel frames of Lie algebroid connections.

\begin{lemmata}{Parallel frames of flat Lie algebroid connections around regular points, \newline \cite[Lemma 2.9]{parallelFrameEconn}}{ParallelFramesForEConnections}
Let $E\to N$ be a Lie algebroid over a smooth manifold $N$, and ${}^E\nabla$ be an $E$-connection on a vector bundle $V \to N$. Moreover, assume that ${}^E\nabla_\nu = 0$ for all $\nu \in E$ with $\rho(\nu) = 0$. Then there is locally around each regular point a frame $\mleft( e_a \mright)_a$ of $E$ such that
\bas
{}^E\nabla e_a 
&=
0.
\eas
\end{lemmata}

\begin{proof}[Sketch of the proof]
\leavevmode\newline
Fix a regular point $p \in N$. We just give a short sketch of the proof, using a frame around $p$ as given in Remark \ref{rem:LocalSplittingFrame}, denoted by $\mleft( f_a \mright)_a$, such that a subset of the frame, denoted as $\mleft(g_i\mright)_i$, satisfies $\rho(g_i) = \partial_i$ for some local coordinate vector fields $\mleft( \partial_i \mright)_i$ of the leaf through $p$. The remaining part of the frame, denoted as $\mleft( h_\alpha \mright)_\alpha$, spans the kernel of the anchor, that is, $\rho(h_\alpha) = 0$. Then 
\bas
{}^E\nabla_{f_b} v
&=
\mathcal{L}_{\rho(f_a)}(v^a) ~ f_a
	+ v^a ~ {}^E\nabla_{f_b} f_a
=
\mathcal{L}_{\rho(f_a)}(v^a) ~ f_a
	+ v^a \omega_{ab}^c f_c
\eas
for all $v = v^a f_a \in \Gamma(V)$ and $\mu \in \Gamma(E)$, where $\omega_{ab}^c$ are smooth functions locally on $N$ given by $\omega_{ab}^c f_c = {}^E\nabla_{f_b} f_a$. Let us study the equation ${}^E\nabla v = 0$. If $f_b = g_i$, then
\bas
0
&=
\partial_i v^a ~ f_a
	+ v^a \omega_{ai}^c f_c,
\eas
that is just the standard well-known PDEs, which we can solve. However, if $f_b = h_\alpha$, then
\bas
0
&=
v^a \omega_{a\alpha}^c f_c,
\eas
and that is an algebraic equation, which may or may not be solvable. By the condition ${}^E\nabla_\nu = 0$ for all $\nu \in E$ with $\rho(\nu) = 0$ we know that ${}^E\nabla_{h_\alpha} = 0$ and, so, $\omega_{a\alpha}^c=0$. This resolves the problem of the algebraic equations which are now trivially satisfied. Hence, the remaining proof of the existence of the parallel frame is then similar to flat vector bundle connections, making use of the vanishing mixed components of the Lie bracket as given in the third equation in Remark \ref{rem:LocalSplittingFrame} when studying the curvature with respect to such statements, in order to allow similar arguments about parallel transport as for vector bundle connections; see the reference for the remaining proof.
\end{proof}

Especially the proof emphasizes why one cannot expect in general to have a parallel frame for flat Lie algebroid connections. For example take an action Lie algebroid $E = N \times \mathfrak{g}$ over a smooth manifold $N$, related to a Lie algebra $\mathfrak{g}$, and denote with $\nabla$ its canonical flat connection. Then the basic connection on $E$ gives
\bas
\nabla^{\mathrm{bas}}_{\mu} \nu
&=
\mleft[ \mu, \nu \mright]_{\mathfrak{g}}
\eas
for all constant sections $\mu, \nu \in \Gamma(N \times \mathfrak{g})$. Therefore the basic connection is also flat because it is just the Lie bracket (by the Jacobi identity); but it is a canonical flat connection if and only if $\mathfrak{g}$ is abelian. If the basic connection on $E$ has a parallel frame $\mleft( e_a \mright)_a$, then
\bas
\nabla_{\rho(e_a)} e_b
&=
\mleft[ e_a, e_b \mright]_E,
\eas
which may not necessarily hold for any frame. Since the left hand side is tensorial in $e_a$ we could then derive for all sections $\nu$ with (in that neighbourhood) $\rho(\nu)=0$ that
\bas
0
&=
\nu^a ~ \mleft[ e_a, e_b \mright]_E.
\eas
 However, the important piece of information in this work is to know that the basic connection is in general not the canonical flat connection for action Lie algebroids if $\nabla$ is already the canonical flat connection.

\section{Lie algebra bundles}\label{SectionOfLABStuff}

Of special importance are the Lie algebra bundles (LABs), defined in Def.~\ref{def:LAB}. As Lie algebroids they are rather easy since the anchor is zero. But they will still play an important role later; also the kernel of each anchor is a bundle of Lie algebras around regular points, which is why it is important to study those. LABs are a special case of bundle of Lie algebras, but we will see later why we are mainly interested into those.

We will summarise the most important results of this section in Ex.~\ref{ex:BigCoolDiagramOfMackenzieAboutLABsStuff}.

\subsection{Notions similar to Lie algebras}

Many constructions related to Lie algebras carry over to LABs. We will explain why.

\begin{propositions}{sub-LABs, \cite[Proposition 3.3.9; page 105]{mackenzieGeneralTheory}}{SubLABS}
Let $K\to N$ be an LAB over a smooth manifold $N$ with fibre type $\mathfrak{g}$ as Lie algebra. Moreover, let $\mathfrak{h}$ be a Lie \textbf{characteristic subalgebra of $\mathfrak{g}$}, that is, a subalgebra of $\mathfrak{g}$ such that $\varphi(\mathfrak{h}) = \mathfrak{h}$ for all Lie algebra automorphism $\varphi: \mathfrak{g} \to \mathfrak{g}$. 

Then there is a well-defined \textbf{sub-LAB $L$ of $K$}, that is, a subbundle $L$ of $K$ which is also an LAB such that each LAB chart $\psi: K|_U \to U \times \mathfrak{g}$ restricts to an LAB chart $L|_U \to U \times \mathfrak{h}$, where $U$ is an open subset of $N$ on which an LAB chart is defined.
\end{propositions}

\begin{remark}
\leavevmode\newline
It is an immediate consequence that the field of Lie brackets of $L$ is given by the field of Lie brackets of $K$ restricted to $L$.
\end{remark}

\begin{proof}[Proof of Prop.~\ref{prop:SubLABS}]
\leavevmode\newline
That is trivial. The essential thing to note is that we need $\varphi(\mathfrak{h}) = \mathfrak{h}$ for all Lie algebra automorphisms $\varphi: \mathfrak{g} \to \mathfrak{g}$ as a condition for gluing the canonical construction of a sub-LAB in given a trivialization, \textit{i.e.}~it is trivial to construct a sub-LAB for a trivial LAB, and for gluing those constructions it is important that each LAB chart can restrict to a Lie algebra isomorphism $L|_U \to U \times \mathfrak{h}$ corresponding to the same subalgebra $\mathfrak{h}$. To make this possible, the local images/restrictions must be stable under transition maps in case two LAB charts of $K$ overlap in some open neighbourhood. The transition maps are Lie algebra automorphisms, and, so, if two overlapping LAB charts of $K$ restrict as stated, then their transition map is in alignment with this due to $\varphi(\mathfrak{h}) = \mathfrak{h}$ for all Lie algebra automorphisms $\varphi: \mathfrak{g} \to \mathfrak{g}$.

Hence, restricting the inverse of each LAB chart of $K$ to $U \times \mathfrak{h}$ defines a subbundle $L$ of $K$, such that each fibre is essentially the subalgebra $\mathfrak{h}$ and its bracket is canonically the restriction of the field of Lie brackets of $K$; all of that is well-defined by the previous paragraph, and that gives an LAB structure on $L$.
\end{proof}

\begin{examples}{Centres of LABs, \newline \cite[first parapgraph after Proposition 3.3.9; page 105]{mackenzieGeneralTheory}}{CentreOfLABK}
With this proposition we can quickly generalize certain constructions of Lie algebras to the level of LABs. For example, possible subalgebras $\mathfrak{h}$ of a Lie algebra $\mathfrak{g}$ with $\varphi(\mathfrak{h}) = \mathfrak{h}$ for all Lie algebra automorphisms $\varphi: \mathfrak{g} \to \mathfrak{g}$ are trivially, due to that $\varphi$ is a homomorphism of brackets, the centre $Z(\mathfrak{g})$ of $\mathfrak{g}$ and $\mleft[ \mathfrak{g}, \mathfrak{g} \mright]_{\mathfrak{g}}$, the corresponding sub-LABs are denoted by $\gls{ZLAB}$ and $\mleft[ K, K \mright]_K$, respectively; we especially need the former. Moreover, the sections of $Z(K)$ are also the centre of the Lie algebra $\Gamma(K)$.
\begin{center}
	\begin{tikzcd}
	Z(\mathfrak{g}) \arrow{r} & Z(K) \arrow{d} \\
			& N
	\end{tikzcd}
\end{center}
\end{examples}

\begin{examples}{Derivations of LABs, \newline \cite[second and third parapgraph after Proposition 3.3.9, and discussion around Proposition 3.3.10; page 105]{mackenzieGeneralTheory}}{DerivationsOFLABSK}
Another important LABs will be related to Lie bracket derivations $\mathrm{Der}(\mathfrak{g})$ of a Lie algebra $\mathfrak{g}$; those are as usual defined as those endomorphisms $T \in \mathrm{End}(\mathfrak{g})$ of $\mathfrak{g}$ such that
\bas
T\mleft( \mleft[ x, y \mright]_{\mathfrak{g}} \mright)
&=
\mleft[ T(x), y \mright]_{\mathfrak{g}}
	+ \mleft[ x, T(y) \mright]_{\mathfrak{g}}
\eas
for all $x, y \in \mathfrak{g}$. Recall, that we derived the derivations of a vector bundle $V \to N$, denoted by $\mathcal{D}(V)$, whose anchor was denoted by $a$ and its kernel is trivially given by $\mathrm{End}(V)$. Since the rank of $\mathrm{End}(V)$ is constant, so, $a$ has constant rank, and the kernel of anchors around regular points is a bundle of Lie algebras, we can conclude that $\mathrm{End}(V)$ is an LAB, also because of that the Lie algebra fibre type is trivially given by $\mathrm{End}(W)$ where $W$ is the fibre type of of $V$.

In case of $V=K$ an LAB over $N$, we have an LAB with fibre type $\mathrm{End}(\mathfrak{g})$, and $\mathrm{Der}(\mathfrak{g})$ is a subalgebra as it is well-known and trivial to check. Now let $\varphi \in \mathrm{Aut}(\mathfrak{g})$, then take $T \in \mathrm{Der}(\mathfrak{g})$, and observe for $\varphi \circ T \circ \varphi^{-1}$ that
\bas
\mleft(\varphi \circ T \circ \varphi^{-1}\mright)\mleft( \mleft[ x, y \mright]_{\mathfrak{g}} \mright)
&=
(\varphi \circ T)\mleft(
	\mleft[ \varphi^{-1}(x),\varphi^{-1}(y) \mright]_{\mathfrak{g}}
\mright)
\\
&=
\varphi\mleft(
	\mleft[ T\mleft(\varphi^{-1}(x)\mright),\varphi^{-1}(y) \mright]_{\mathfrak{g}}
	+ \mleft[ \varphi^{-1}(x),T\mleft(\varphi^{-1}(y)\mright) \mright]_{\mathfrak{g}}
\mright)
\\
&=
\mleft[ \mleft(\varphi \circ T \circ \varphi^{-1}\mright)(x), y \mright]_{\mathfrak{g}}
	+ \mleft[ x, \mleft(\varphi \circ T \circ \varphi^{-1}\mright)(y) \mright]_{\mathfrak{g}}
\eas
for all $x, y \in \mathfrak{g}$. Thus, $\varphi \circ T \circ \varphi^{-1} \in \mathrm{Der}(\mathfrak{g})$; similar for the inverse of $\varphi$ such that $\varphi \circ \mathrm{Der}(\mathfrak{g}) \circ \varphi^{-1} = \mathrm{Der}(\mathfrak{g})$. The conjugation with $\varphi$ is just a certain type of elements in $\mathrm{Aut}\bigl(\mathrm{End}(\mathfrak{g})\bigr)$ such that it looks like that we cannot yet use Prop.~\ref{prop:SubLABS}. However, the proof of Prop.~\ref{prop:SubLABS} was just about transition maps and in case of $\mathrm{End}$-bundles the typical atlas\footnote{This is also clearly its LAB atlas.} has such transition maps as we know in general, which is why we can conclude similarly as in the proof of Prop.~\ref{prop:SubLABS} that there is a well-defined sub-LAB $\gls{DAerK}$ of $\mathrm{End}(K)$ with fibre type $\mathrm{Der}(\mathfrak{g})$.
\begin{center}
	\begin{tikzcd}
	\mathrm{Der}(\mathfrak{g}) \arrow{r} & \mathrm{Der}(K) \arrow{d} \\
			& N
	\end{tikzcd}
\end{center}
There is a special set of derivations, the \textbf{ideal of inner derivations $\mathrm{ad}(\mathfrak{g})$ of $\mathfrak{g}$}; that is, an inner derivation is of the form $\mathrm{ad}(x)$ for an $x \in \mathfrak{g}$. It is trivially a derivation by the Jacobi identity, and an ideal of $\mathrm{Der}(\mathfrak{g})$ by
\bas
\mleft( \mleft[ \mathrm{ad}(x), T \mright]_{\mathrm{Der}(K)} \mright)(y)
&=
\mleft[ x, T(y) \mright]_{\mathfrak{g}}
	- \underbrace{T\mleft( \mleft[ x, y \mright]_{\mathfrak{g}} \mright)}
	_{\mathclap{ = \mleft[ T(x), y \mright]_{\mathfrak{g}} + \mleft[ x, T(y) \mright]_{\mathfrak{g}} }}
\\
&=
- \mleft(\mathrm{ad}\bigl(T(x)\bigr)\mright)(y)
\eas
for all $x, y \in \mathfrak{g}$ and $T \in \mathrm{Der}(\mathfrak{g})$. As above, observe that for all $\varphi \in \mathrm{Aut}(\mathfrak{g})$ we have
\bas
\mleft(\varphi \circ \mathrm{ad}(x) \circ \varphi^{-1}\mright)(y)
&=
\varphi\mleft(
	\mleft[ x, \varphi^{-1}(y) \mright]_{\mathfrak{g}}
\mright)
=
\mleft(\mathrm{ad}\bigl( \varphi(x) \bigr)\mright)(y),
\eas
hence, the discussed conjugation above restricts to inner derivations. Therefore we can apply the same argument as above to derive that $\mathrm{ad}(\mathfrak{g})$ gives rise to a sub-LAB of $\mathrm{Der}(K)$ and of $\mathrm{End}(K)$, denoted by $\gls{adK}$, the \textbf{ideal of inner derivations of $K$}.
\end{examples}

\begin{remark}
\leavevmode\newline
As shown in \cite[discussion around Proposition 3.3.10; page 105]{mackenzieGeneralTheory}, one can quickly derive that $\mathrm{ad}(K)$ is the image of $\mathrm{ad}: K \to \mathrm{Der}(K)$, which is just defined as the fibre-wise extended adjoint map of $\mathrm{ad}$ on $\mathfrak{g}$. Since it is a tensor, the adjoint extends to sections.
\end{remark}

$\mathrm{ad}(K)$ is trivially an ideal in the following sense.

\begin{definitions}{Ideals of LABs, \cite[Definition 3.3.11; page 106]{mackenzieGeneralTheory}}{IdealsOfLABSK}
Let $K \to N$ be an LAB over a smooth manifold $N$ and $L$ a sub-LAB of $K$. Then $L$ is an \textbf{ideal of $K$} if each fibre of $L_p$ is an ideal of $K_p$ for all $p \in N$.
\end{definitions}

One can construct a quotient of $\mathrm{Der}(K)$ over $\mathrm{ad}(K)$ in the usual way, but we need such quotients a bit more general. For this we need to discuss extensions of tangent bundles where LABs play an important role. Those are best described as certain short exact sequences.

\subsection{Extensions of tangent bundles with Lie algebra bundles}

\begin{definitions}{Extension of tangent bundles by LABs and transversals, \newline \cite[\S 7.1, Definition 7.1.11; page 266; and Definition 7.3.1; page 277]{mackenzieGeneralTheory}}{ExtensionOfTNByLABs}
Let $K \to N$ be an LAB. Then an \textbf{extension of $\mathrm{T}N$ by $K$} is a short exact sequence of Lie algebroids over $N$
\begin{center}
	\begin{tikzcd}
		0 \arrow{r} & K \arrow{r}{\iota} & E \arrow{r}{\pi} & \mathrm{T}N \arrow{r} & 0,
	\end{tikzcd}
\end{center}
where $E \to N$ is a Lie algebroid and 
the sequence is exact as a sequence of vector bundles but each arrow represents a Lie algebroid morphism,
equivalently denoted as\footnote{The hooked arrow emphasizes the inclusion, and the two-headed arrow the surjectivity.}
\be\label{defShortExactSeqExtensionOfTNByK}
	\begin{tikzcd}
		K \arrow[hook]{r}{\iota} & E \arrow[two heads]{r}{\pi} & \mathrm{T}N.
	\end{tikzcd}
\ee
A transversal of \eqref{defShortExactSeqExtensionOfTNByK} is a vector bundle morphism $\chi: \mathrm{T}N \to E$ such that $\pi \circ \chi = \mathds{1}_{\mathrm{T}N}$.
\end{definitions}

\begin{remark}
\leavevmode\newline
\indent $\bullet$ As in this definition, we will use those sequences also to define the corresponding notation of the Lie algebroid morphisms, in order to avoid separately writing "[$\dotsc$] where $\iota: K \to E$ is a Lie algebroid morphism [$\dotsc$]". We also only give the sequence, implicitly meaning that $K$ will be an LAB and $E$ a Lie algebroid over $N$ without mentioning it further. 

$\bullet$ Furthermore, $\iota$ is an injective Lie algebroid morphism, especially an embedding since it is also vector bundle morphism. Hence, $\iota$ is up to Lie algebroid isomorphisms the inclusion in this work and can be thought as such, which is why we often omit it. These notations normally emphasize that a change of the explicit description of $K$ is possible, in that case the inclusion would be replaced by a composition of the corresponding inclusion with a Lie algebroid isomorphism; however, we will not need this.

$\bullet$ We will, as usual, denote the Lie bracket of $E$ by $\mleft[ \cdot, \cdot \mright]_E$, and $\pi$ is its anchor $\rho$ due to that $\pi$ is anchor-preserving and that the anchor of $\mathrm{T}N$ is the identity. Therefore we will use the typical notation of anchors in the following instead of $\pi$; we also clearly have $\iota(K) = \mathrm{Ker}(\rho)$ by the exactness of the sequence.

$\bullet$ $E$ is a transitive Lie algebroid because $\rho = \pi$ is surjective in that case; in fact, by \cite[Theorem 6.5.1; page 248]{mackenzieGeneralTheory} each transitive Lie algebroid $E$ is such a short exact sequence. The rank of the anchor is constant for transitive Lie algebroids such that there are only regular points and, so, the kernel of the anchor, $\mathrm{Ker}(\rho)$, is a bundle of Lie algebras. One can show that $\mathrm{Ker}(\rho)$ is also a Lie algebra bundle by Thm.~\ref{thm:BLALAB}; the essential trick is to take a vector bundle morphism $\chi: \mathrm{T}N \to E$ with $\rho \circ \chi = \mathds{1}_{\mathrm{T}N}$, and then to define a connection $\nabla$ on $\mathrm{Ker}(\rho)$ by $\mathrm{ad} \circ \chi$, \textit{i.e.}~$\nabla_X \nu \coloneqq \mleft[ \chi(X), \nu \mright]_E$ for all $X \in \mathfrak{X}(N)$ and $\nu \in \mathrm{Ker}(\rho)$. This connection will be a Lie bracket derivation of $\mathrm{Ker}(\rho)$ such that Thm.~\ref{thm:BLALAB} can be used. We will not prove this, since we are not going to need it, hence, see the reference; however, the essential calculations will be done later in Section \ref{ObstrLAB}. Moreover, it is useful for the following constructions to keep this information in mind, in order to understand why it is a useful simplification to assume transitive Lie algebroids.

$\bullet$ So, in our case, extensions are equivalent to transitive Lie algebroids, such that one may wonder about the different name. Often, especially in Section \ref{ObstrLAB}, we will have a given $K$ and $N$, then there is the question whether there is an $E$ in the sense of an extension involving $K$ and $\mathrm{T}N$. Thence, the idea is that $E$ \textbf{extends} $\mathrm{T}N$ by $K$ in sense of Lie algebroids. The different name here is especially to emphasize a different context. Moreover, the idea of extensions can be generalized in the sense of replacing $\mathrm{T}N$ by an arbitrary Lie algebroid as in \cite[Definition 3.3.19; page 109]{mackenzieGeneralTheory}.
\end{remark}

\begin{examples}{Derivations as extension and connections as transversal, \newline \cite[second statement of Corollary 3.6.11; page 140]{mackenzieGeneralTheory}}{DerivationsAreExtensions}
Let $V \to N$ be a vector bundle over a smooth manifold $N$. Then $\mathcal{D}(V)$ with anchor $a$ describes an extension as a transitive Lie algebroid as we have seen,
\be\label{ExtensioNofDerivations}
	\begin{tikzcd}
		\mathrm{End}(V) \arrow[hook]{r} & \mathcal{D}(V) \arrow[two heads]{r}{a} & \mathrm{T}N.
	\end{tikzcd}
\ee
By definition, a vector bundle connection $\nabla$ of $V$ is then a transversal of \eqref{ExtensioNofDerivations}, and each transversal a connection.

In the case of $V = K$ an LAB, we can define $\gls{DAVDerK}$ as the subset of those derivations generated by sections $T \in \Gamma(\mathcal{D}(K))$ with
\bas
T\mleft( \mleft[ \mu, \nu \mright]_{K} \mright)
&=
\mleft[ T(\mu), \nu \mright]_{K}
	+ \mleft[ \mu, T(\nu) \mright]_{K}
\eas
for all $\mu, \nu \in \Gamma(K)$. Since $\mleft[ \cdot, \cdot \mright]_{\mathcal{D}(K)}$ is just defined as a commutator, it follows as trivial as for $\mathrm{Der}(\mathfrak{g})$ of a Lie algebra $\mathfrak{g}$ that $\Gamma\mleft(\mathcal{D}_{\mathrm{Der}}(K)\mright)$ is a subalgebra of $\Gamma(\mathcal{D}(K))$; and at each point $p\in N$ we have that $\mathcal{D}_{\mathrm{Der}}(K)$ is a subspace of $\mathcal{D}(K)$. It is also a Lie algebroid, whose structure is inherited by $\mathcal{D}(K)$; for this take a connection $\nabla$ on $K$ which is a Lie bracket derivation, see Thm.~\ref{thm:BLALAB} for its existence later. Then define a map 
\bas
\mathrm{T}N \times \mathrm{Der}(K) &\to \mathcal{D}_{\mathrm{Der}}(K),
\\
(X, A) &\mapsto \nabla_X + A,
\eas
which is clearly well-defined because of the fact that the difference of two connections is always an element $L$ of $\Omega^1(N; \mathrm{End}(K))$; if then both of these connections are Lie bracket derivations, then so also $L$ such that $L \in \Omega^1(N; \mathrm{Der}(K))$. Hence, $\nabla_X + A \in \mathcal{D}_{\mathrm{Der}}(K)$. As in the proof of Prop.~\ref{prop:IsomorphismofDerivationonVectorbundleatabasepoint}, see also Lemma \ref{lem:LemmaVectorbundlestructureofDV}, this defines an isomorphism of vector spaces at each point, and as for $\mathcal{D}(K)$ this leads to that $\mathcal{D}_{\mathrm{Der}}(K)$ has constant rank and it admits a transitive Lie algebroid structure with precisely the same arguments as for general derivations; since this structure is inherited by $\mathcal{D}(K)$, we may say that $\mathcal{D}_{\mathrm{Der}}(K)$ is a transitive Lie subalgebroid. The kernel of its anchor, $a|_{\mathcal{D}_{\mathrm{Der}}(K)}$, consists by definition of those elements of $\mathrm{End}(K)$ which are also Lie bracket derivations, so, the kernel is $\mathrm{Der}(K)$. Therefore we arrive at another extension, basically the restriction of \eqref{ExtensioNofDerivations} onto $\mathcal{D}_{\mathrm{Der}}(K)$,
\be\label{SequenceForBracketDerivations}
	\begin{tikzcd}
		\mathrm{Der}(K) \arrow[hook]{r} & \mathcal{D}_{\mathrm{Der}}(K) \arrow[two heads]{r}{a} & \mathrm{T}N,
	\end{tikzcd}
\ee
and also here, a vector bundle connection of $K$ which is also a Lie bracket derivation is equivalent to a transversal for \eqref{SequenceForBracketDerivations}.
\end{examples}

As for Lie algebras, we want to take the quotient of $\mathrm{Der}(K)$ and $\mathcal{D}_{\mathrm{Der}}(K)$ over $\mathrm{ad}(K)$. That is, as usual, done over ideals of Lie algebroids, which shall be subsets of the kernel of the anchor; the reason behind this is to avoid problems in quotients with respect to the anchor. The typical constructions for quotients will then apply because the anchor of an equivalence class is going to be independent of the chosen representative.

\begin{definitions}{Ideals of transitive Lie algebroids, \newline \cite[Definition 6.5.6; page 250]{mackenzieGeneralTheory}}{IdealsOfTransitiveLieAlgebroids}
Let
\begin{center}
	\begin{tikzcd}
		K \arrow[hook]{r}{\iota} & E \arrow[two heads]{r}{\rho} & \mathrm{T}N.
	\end{tikzcd}
\end{center}
be an extension. Then an \textbf{ideal $L$ of $E$} is a sub-LAB of $K$ with
\ba
\mleft[ \nu, \mu \mright]_E &\in \Gamma(L)
\ea
for all $\nu \in \Gamma(E)$ and $\mu \in \Gamma(L)$.
\end{definitions}

\begin{remark}
\leavevmode\newline
As we know, the kernel of $\rho$, $K$, is a canonical example of an ideal.
\end{remark}

\begin{propositions}{Quotient Lie algebroids of transitive Lie algebroids, \newline \cite[Proposition 6.5.8]{mackenzieGeneralTheory}}{QuotientsOfTransitiveLAOids}
Let
\begin{center}
	\begin{tikzcd}
		K \arrow[hook]{r}{\iota} & E \arrow[two heads]{r}{\rho} & \mathrm{T}N.
	\end{tikzcd}
\end{center}
be an extension and $L$ an ideal of $E$. Furthermore, we denote with $E\Big/\iota(L)$ and $K \Big/ L$ the quotient bundle as vector bundles, whose natural projections we denote by $\sharp: E \to E\Big/\iota(L), \mu \mapsto \mu + \iota(L)$, and $\sharp|_K$, respectively. Then naturally define
\ba
K \Big/ L &\stackrel{\overline{\iota}}{\to} E \Big/ \iota(L),
\\
\sharp|_K(\mu) &\mapsto \overline{\iota}\bigl(\sharp|_K(\mu)\bigr) \coloneqq \sharp\bigl(\iota(\mu)\bigr),\label{SharpRestricted}
\ea
and
\ba
E \Big/ \iota(L) &\stackrel{\overline{\rho}}{\to} \mathrm{T}N,
\\
\sharp(\nu) &\mapsto \overline{\rho}\bigl( \sharp(\nu) \bigr) \coloneqq \rho(\nu),
\ea
and finally equip $E \Big/ \iota(L)$ with the bracket $\mleft[ \cdot, \cdot \mright]_{E \big/ \iota(L)}$
\ba
\mleft[ \sharp(\nu), \sharp(\eta) \mright]_{E \big/ \iota(L)}
&\coloneqq
\sharp\mleft(\mleft[ \nu, \eta \mright]_E\mright)
\ea
for all $\nu, \eta \in \Gamma(E)$. Then
\begin{center}
	\begin{tikzcd}
		K \Big/ L \arrow[hook]{r}{\overline{\iota}} & E \Big/ \iota(L) \arrow[two heads]{r}{\overline{\rho}} & \mathrm{T}N
	\end{tikzcd}
\end{center}
is an extension such that $\sharp$ is a surjective submersion with kernel $\iota(L)$.
\end{propositions}

\begin{remarks}{}{LabelOfQuotient}
We call $E \Big/ \iota(L)$ the \textbf{quotient (transitive) Lie algebroid of $E$ over $L$}. By definition $\sharp$ is a Lie algebroid morphism, as is $\sharp|_K$ by Eq.~\eqref{SharpRestricted} since $\overline{\iota}$ and $\iota$ are injective Lie algebroid morphisms and embeddings.
\end{remarks}

\begin{proof}[Sketch of the proof of Prop.~\ref{prop:QuotientsOfTransitiveLAOids}]
\leavevmode\newline
The proof is straightforward because the constructions are the typical ones for such structures. We just give a sketch, one essentially needs to check that everything is well-defined, that we have a Lie bracket in combination with an anchor and that the sequence of the quotients is exact. First of all, everything has constant rank such that the taken quotients as vector bundles are valid. Moreover, $\overline{\iota}$ is well-defined because $\iota$ is injective by the exactness of the sequence, hence, let $\mu, \mu^\prime \in K$ with $\sharp|_K(\mu) = \sharp|_K(\mu^\prime)$
\bas
\sharp\bigl(  \iota(\mu) \bigr)
&=
\sharp\bigl( \underbrace{\iota(\mu - \mu^\prime)}_{\in \iota(L)} + \iota(\mu^\prime) \bigr)
=
\sharp\bigl( \iota(\mu^\prime) \bigr),
\eas
such that $\overline{\iota}\bigl( \sharp|_K(\mu) \bigr) =\overline{\iota}\bigl( \sharp|_K(\mu^\prime) \bigr)$;
similarly for $\hat{\nu}, \hat{\nu}^\prime \in E$ with $\sharp(\hat{\nu}) = \sharp(\hat{\nu}^\prime)$
\bas
\rho(\hat{\nu})
&=
\rho(\underbrace{\hat{\nu} - \hat{\nu}^\prime}_{\mathclap{ \in ~ \iota(L)~ \subset ~ \iota(K) }} + \hat{\nu}^\prime)
=
\rho(\hat{\nu}^\prime),
\eas
thus, $\overline{\rho}\bigl( \sharp(\hat{\nu}) \bigr) = \overline{\rho}\bigl( \sharp(\hat{\nu}^\prime) \bigr)$, and, finally for $\nu, \nu^\prime, \eta, \eta^\prime \in \Gamma(E)$ with $\sharp(\nu) = \sharp(\nu^\prime)$ and $\sharp(\eta) = \sharp(\eta^\prime)$,
\bas
\sharp\bigl(
	\mleft[ \nu, \eta \mright]_E
\bigr)
&=
\sharp\bigl(
	[ \underbrace{\nu - \nu^\prime}_{\mathclap{ \in \iota(L) \subset \iota(K) }} + \nu^\prime,
	\underbrace{\eta - \eta^\prime}_{\mathclap{ \in \iota(L) \subset \iota(K) }} + \eta^\prime ]_E
\bigr)
=
\sharp\bigl(
	\mleft[ \nu^\prime, \eta^\prime \mright]_E
\bigr),
\eas
using that the kernel of the anchor is an ideal of the Lie bracket, therefore also $\mleft[ \sharp(\nu), \sharp(\eta) \mright]_{E \big/ \iota(L)} = \mleft[ \sharp(\nu^\prime), \sharp(\eta^\prime) \mright]_{E \big/ \iota(L)}$. The (bi-)linearity of all those maps follows trivially, the bracket is also clearly anti-symmetric, and 
\bas
[ \sharp(\nu), \underbrace{f ~ \sharp(\eta)}_{\mathclap{ =\sharp(f \eta) }} ]_{E \big/ \iota(L)}
&=
\sharp\bigl(
	\mleft[ \nu, f \eta \mright]_E
\bigr)
\\
&=
\sharp\bigl(
	f \mleft[ \nu, \eta \mright]_E
	+	\underbrace{\mathcal{L}_{\rho(\nu)}}_{\mathclap{ = \mathcal{L}_{\overline{\rho}( \sharp(\nu) )} }}(f) ~ \eta
\bigr)
\\
&=
f ~ \sharp\bigl(
	\mleft[ \nu, \eta \mright]_E
\bigr)
	+ \mathcal{L}_{\overline{\rho}( \sharp(\nu) )}(f) ~ \sharp(\eta)
\\
&=
f \mleft[ \sharp(\nu), \sharp(\eta) \mright]_{E \big/ \iota(L)}
	+ \mathcal{L}_{\overline{\rho}( \sharp(\nu) )}(f) ~ \sharp(\eta)
\eas
for all $f\in C^\infty(N)$. The Jacobi identity is clearly inherited by $\mleft[ \cdot, \cdot \mright]_E$, so, it is a Lie bracket and $\overline{\rho}$ is the anchor by Prop.~\ref{prop:MeasureofJacobiandHomom}.
By construction, $\overline{\iota}$ is still injective, that is, assume
\bas
\overline{\iota}\bigl(\sharp|_K(\mu)\bigr)
&=
\overline{\iota}\bigl(\sharp|_K(\mu^\prime)\bigr)
\eas
for two fixed $\mu, \mu^\prime \in K$, then
\bas
0
&=
\sharp\bigl(
	\iota(\mu - \mu^\prime)
\bigr),
\eas
thus, $\mu - \mu^\prime \in L$ such that $\sharp|_K(\mu) = \sharp|_K(\mu^\prime)$, which proves the injectivity of $\overline{\iota}$.
Moreover,
\bas
\overline{\rho}\mleft(
	\overline{\iota}\bigl(\sharp|_K(\mu)\bigr)
\mright)
&=
\overline{\rho}\mleft(
	\sharp\bigl( \iota(\mu) \bigr)
\mright)
=
\rho\bigl(
	\iota(\mu)
\bigr)
=
0
\eas
for all $\mu \in K$; the anchor $\overline{\rho}$ is clearly surjective by $\overline{\rho} \circ \sharp= \rho$ and because the quotient is just over a subbundle of $K = \mathrm{Ker}(\rho)$, that is, for all $X \in \mathfrak{X}(N)$ let $\nu \in \Gamma(E)$ such that $X = \rho(\nu)$, then
\bas
\overline{\rho}\bigl( \sharp(\nu)\bigr)
&=
\rho(\nu)
=
X.
\eas
Thence, the sequence of the quotients is exact. That $\sharp$ is a surjective submersion with kernel $\iota(L)$ follows trivially by construction as natural projection of quotient spaces.
\end{proof}

\begin{examples}{Outer bracket derivations of K, \newline \cite[Definition 7.2.1 and Equation (7); page 271]{mackenzieGeneralTheory}}{OuterDerivationsOfK}
Let $K \to N$ be an LAB over a smooth manifold $N$. Then we have the following quotient
\be
	\begin{tikzcd}
		\mathrm{Der}(K)\Big/\mathrm{ad}(K) \arrow[r, hook] & \mathcal{D}_{\mathrm{Der}}(K)\Big/ \mathrm{ad}(K) \arrow[two heads]{r}{\overline{a}} & \mathrm{T}N,
	\end{tikzcd}
\ee
which we denote by
\be
	\begin{tikzcd}
		\gls{OutKA} \arrow[r, hook] & \gls{OutKDDerK} \arrow[two heads]{r}{\overline{a}} & \mathrm{T}N,
	\end{tikzcd}
\ee
where $\mathrm{Out}(K) \coloneqq \mathrm{Der}(K)\Big/\mathrm{ad}(K)$ are the \textbf{outer bracket derivations of $K$}, and $\mathrm{Out}\mleft(\mathcal{D}_{\mathrm{Der}}(K)\mright) \coloneqq \mathcal{D}_{\mathrm{Der}}(K)\Big/ \mathrm{ad}(K)$ are those derivations in $\mathcal{D}(K)$ which are also outer bracket derivations. This quotient is possible because exactly as in Ex.~\ref{ex:DerivationsOFLABSK} one can show that $\mathrm{ad}(K)$ is also an ideal of $\mathcal{D}_{\mathrm{Der}}(K)$ and not just of $\mathrm{Der}(K)$, that is, we get again as in Ex.~\ref{ex:DerivationsOFLABSK}
\ba
\mleft[ T, \mathrm{ad}(\nu) \mright]_{\mathcal{D}_{\mathrm{Der}(K)}}
&=
\mathrm{ad}\bigl( T(\nu) \bigr)
\ea
for all $\nu \in \Gamma(K)$ and $T \in \Gamma\mleft(\mathcal{D}_{\mathrm{Der}(K)}\mright)$.
\end{examples}

Let us finish this chapter with a summary of this section, also recall Remark \ref{rem:LabelOfQuotient}.

\begin{examples}{Summary of Section \ref{SectionOfLABStuff}, \newline \cite[\S 7.2, Figure 7.1; page 272; we omit the labels of the inclusion arrows]{mackenzieGeneralTheory}}{BigCoolDiagramOfMackenzieAboutLABsStuff}
Let $K \to N$ be an LAB over a smooth manifold $N$. Then the main results of Section \ref{SectionOfLABStuff} can be summarized in the following commuting diagram
\be
	\begin{tikzcd}
		Z(K) \arrow[hook]{d} \arrow[equal]{r} & Z(K) \arrow[hook]{d} \\
		K \arrow{d}{\mathrm{ad}} \arrow[equal]{r} & K \arrow{d} \\
		\mathrm{Der}(K) \arrow[two heads]{d}{\sharp^+} \arrow[hook]{r} & \mathcal{D}_{\mathrm{Der}}(K) \arrow[two heads]{d}{\sharp} \arrow[two heads]{r}{a} & \mathrm{T}N \arrow[equal]{d} \\
		\mathrm{Out}(K) \arrow[hook]{r} & \mathrm{Out}\mleft(\mathcal{D}_{\mathrm{Der}}(K)\mright) \arrow[two heads]{r}{\overline{a}} & \mathrm{T}N
	\end{tikzcd}
\ee
where both rows and columns are short exact sequences of Lie algebroid morphisms, especially the last two rows are extensions, and the diagram serves as a definition of the notation of the new Lie algebroid morphisms, for example $\sharp^{(+)}$ denotes the projection of derivations into the space of outer derivations.
\end{examples}
\chapter{Generalized gauge theory}\label{GeneralizedGTfas}

The purpose of the following sections is now to introduce a new and more general formulation of gauge theory which we have introduced in Chapter \ref{ClassicGaugeTheory}. Especially recall the section about the infinitesimal gauge transformation using Lie algebra connections, Section \ref{NewInfGaugeTrafoTrafos}. Again, we do not want to assume integrability, and so we only compare the new theory with a classical gauge theory whose principal bundle is trivial and can thus be avoided completely by fixing a global gauge.\footnote{We will use Lie algebroids; their integration is more complicated than the integrability of Lie algebras, see \textit{e.g.}~\cite[\S 16.4; page 117]{DaSilva}.}

In that chapter we have used a "bookkeeping trick", denoted by $\iota$;\footnote{Recall the discussion about $\iota$ after Cor.~\ref{cor:ClassicFLowsOfXgMg}.} that is, generalized, that we had a spacetime $M$ and the Higgs field $\Phi$ is a smooth map $M \to N$. The physical quantities like the field strength then had values in $\mathrm{ev}^*K$ and hence in $\Phi^*K$ after point evaluation at $\Phi$, where $\mathrm{ev}$ was the evaluation map of Def.~\ref{def:FirstAttemptOfEvaluationMap} and $K$ was some vector bundle over $N$ (like the Lie algebra); also recall Remark \ref{rem:BosonsAsFunctionalies} where we argued that one can do something similar for the field of gauge bosons and its infinitesimal gauge transformation, we are going to do so, thus, viewing the field of gauge bosons of the classical formulation as forms with values in a $\Phi$-pullback of a trivial Lie algebra bundle. Moreover, we used $\mathfrak{g}$-connections, where $\mathfrak{g}$ is a Lie algebra acting on $N$ via a Lie algebra action $\gamma$. By Prop.~\ref{prop:ActionLieoidsAreOids} action Lie algebroids as bundle over $N$ are a good candidate describing that notion, or more general, Lie algebroids and the notion of Lie algebroid connections.

This is why we are going to define the following physical quantities as having values in some pullback using the evaluation map and $\Phi$ as for the field of gauge bosons, why we are going to use a Lie algebroid $E$ over $N$ instead of a Lie algebra $\mathfrak{g}$, and why we will compare the following definitions with action Lie algebroids in order to allow a comparison with Chapter \ref{ClassicGaugeTheory}. We will see that action Lie algebroids with their canonical flat connection will be the standard formulation of gauge theory.

Although we speak of $\Phi$ as the Higgs field it can be of course any other field with a similar Lagrangian, since we never really discuss the potential term. The Higgs field is just a main example.

If you are interested into the calculations of this and the following chapter, then read Appendix \ref{CalculusIdentitiesNeeded} first and the proofs listed there; certain steps of calculations are explained there which will be simply used in the following without further explanation. We also need a similar notation as in Def.~\ref{def:GradedExtensionOfBracket}, but extended to more than two arguments.

\begin{definitions}{Graded extension of products, \newline \cite[generalization of Definition 5.5.3; page 275]{hamilton}}{GradingOfProducts}
Let $l \in \mathbb{N}$ and $E_1, \dots E_{l+1} \to N$ be vector bundles over a smooth manifold $N$, and $F \in \Gamma\left( \left(\bigotimes_{m=1}^{l} E_m^*\right) \otimes E_{l+1} \right)$. Then we define the \textbf{graded extension of $F$} as
	\bas
\Omega^{k_1}(N; E_1) \times \dots \times \Omega^{k_l}(N; E_l)
&\to \Omega^{k}(N; E_{l+1}), \\
(A_1, \dots, A_l)
&\mapsto
F\mleft(A_1\stackrel{\wedge}{,} \dotsc \stackrel{\wedge}{,} A_l\mright),
\eas
where $k := k_1+\dots k_l$ and $k_i \in \mathbb{N}_0$ for all $i\in \{1, \dots, l\}$. $F\mleft(A_1\stackrel{\wedge}{,} \dotsc \stackrel{\wedge}{,} A_l\mright)$ is defined as an element of $\Omega^{k}(N; E_{l+1})$ by
\bas
&F\mleft(A_1\stackrel{\wedge}{,} \dotsc \stackrel{\wedge}{,} A_l\mright)\mleft(Y_1, \dots, Y_{k}\mright)
\coloneqq \\
&\frac{1}{k_1! \cdot \dots \cdot k_l!} \sum_{\sigma \in S_{k}} \mathrm{sgn}(\sigma) ~ F\left( A_1\left( Y_{\sigma(1)}, \dots, Y_{\sigma(k_1)} \right), \dots, A_l\left( Y_{\sigma(k-k_l+1)}, \dots, Y_{\sigma(k)} \right) \right)
\eas
for all $Y_1, \dots, Y_{k} \in \mathfrak{X}(N)$, where $S_{k}$ is the group of permutations of $\{1, \dots, k\}$ and $\mathrm{sgn}(\sigma)$ the signature of a given permutation $\sigma$. 

$\stackrel{\wedge}{,}$ may be written just as a comma when a zero-form is involved.

Locally, with respect to given frames $\mleft( e^{(i)}_{a_i} \mright)_{a_i}$ of $E_i$, this definition has the form
\ba\label{CoordExprOfGradedExtension}
F\mleft(A_1\stackrel{\wedge}{,} \dotsc \stackrel{\wedge}{,} A_l\mright)
&=
F\mleft(e^{(1)}_{a_1}, \dotsc, e^{(l)}_{a_l}\mright) \otimes A_1^{a_1} \wedge \dotsc \wedge A_l^{a_l}
\ea
for all $A_i = A_i^{a_i} \otimes e^{(i)}_{a_i}$, where $A_i^{a_i}$ are $k_i$-forms on $N$.
\end{definitions}

\begin{remark}
\leavevmode\newline
Using this notation, one has a useful way to compare pullbacks of forms, denoted by an exclamation mark, and pullbacks of sections, denoted by a star. That is, let $\Phi \in C^\infty(M;N)$ and $F \in \Omega^l(N;W)$ for $W \to N$ a vector bundle, then
\ba\label{EqPullBackFormelFuerVerschiedeneDefinitionen}
\Phi^!F 
&=
\frac{1}{l!}~
\mleft(\Phi^*F\mright) ( \underbrace{\mathrm{D}\Phi \stackrel{\wedge}{,} \dotsc \stackrel{\wedge}{,} \mathrm{D}\Phi}_{l \text{ times}} )
\ea
by using the anti-symmetry of $F$ and Def. \ref{def:GradingOfProducts}, \textit{i.e.}
\bas
&\mleft.\frac{1}{l!}~
\Big(\mleft(\Phi^*F\mright) ( \mathrm{D}\Phi \stackrel{\wedge}{,} \dotsc \stackrel{\wedge}{,} \mathrm{D}\Phi ) \Big) (Y_1, \dots, Y_l)\mright|_p \\
&\hspace{1cm}
=
\frac{1}{l!}~
\sum_{\sigma \in S_{l}} \mathrm{sgn}(\sigma) ~ \underbrace{(\Phi^*F)\mleft(\mathrm{D}\Phi\mleft(Y_{\sigma(1)}\mright), \dots, \mathrm{D}\Phi\mleft(Y_{\sigma(l)}\mright)\mright)}_{\mathclap{= \mathrm{sgn}(\sigma) ~ (\Phi^*F)\mleft(\mathrm{D}\Phi\mleft(Y_{1}\mright), \dots, \mathrm{D}\Phi\mleft(Y_{l}\mright)\mright)}}\Big|_p \\
&\hspace{1cm}
=
\frac{1}{l!}~ \underbrace{\mleft( \sum_{\sigma \in S_{l}} 1 \mright)}_{= l!} ~
F_{\Phi(p)}\mleft(\mathrm{D}_p\Phi\mleft(\mleft.Y_{1}\mright|_p\mright), \dots, \mathrm{D}_p\Phi\mleft(\mleft.Y_{l}\mright|_{p}\mright)\mright) \\
&\hspace{1cm}
= \mleft.\mleft(\Phi^!F\mright)(Y_1, \dots, Y_l)\mright|_p
\eas
for all $p \in M$ and $Y_1, \dots, Y_l \in \mathfrak{X}(M)$.
\end{remark}

In case of antisymmetric tensors we of course preserve that.

\begin{propositions}{Graded extensions of antisymmetric tensors}{GradedExtensionPlusAntiSymm}
Let $E_1, E_2 \to N$ be real vector bundles of finite rank over a smooth manifold $N$, $F \in \Omega^2(E_1; E_2)$. Then
\ba
F \mleft( A \stackrel{\wedge}{,} B \mright)
&=
-\mleft( -1 \mright)^{km}
F \mleft( B \stackrel{\wedge}{,} A \mright)
\ea 
for all $A \in \Omega^k(N; E_1)$ and $B \in \Omega^m(N; E_2)$ ($k,m \in \mathbb{N}_0$). Similarly extended to all $F \in \Omega^l(E_1; E_2)$.
\end{propositions}

\begin{remark}
\leavevmode\newline
This is a generalization of similar relations just using the Lie algebra bracket $\mleft[ \cdot, \cdot\mright]_{\mathfrak{g}}$ of a Lie algebra $\mathfrak{g}$, see \cite[\S 5, first statement of Exercise 5.15.14; page 316]{hamilton}.
\end{remark}

\begin{proof}
\leavevmode\newline
Trivial by using Eq.~\eqref{CoordExprOfGradedExtension}.
\end{proof}

\section{Space of fields}\label{SpaceOfFieldsSection}

Before we can define quantities like the field strength, we need to define and study the infinite-dimensional manifold of the arising fields as we did in the classical situation; recall Def.~\ref{def:ClassicSpaceofFieldsAgain}. Because of the non-triviality of the following bundles we need to take a closer look at this space. Recall that we assume convenient settings when treating infinite-dimensional objects.

\begin{definitions}{Space of fields}{SpaceOfFields}
Let $M, N$ be two smooth manifolds and $E\to N$ a Lie algebroid. Then we denote the \textbf{space of fields} by
\ba
\gls{MSpaceOfFields}
&\coloneqq
\mathfrak{M}_E(M; N)
\coloneqq
\left\{ (\Phi, A)
~\middle|~
\Phi \in C^\infty(M;N) \text{ and } A \in \Omega^1(M; \Phi^*E)
\right\}
\ea
which we sometimes view as a fibration over $C^\infty(M;N)$
\begin{center}
	\begin{tikzcd}
		\mathfrak{M}_E(M; N) \arrow{d} \\
		C^\infty(M;N)
	\end{tikzcd}
\end{center}
where the projection is given by $\mathfrak{M}_E(M; N) \ni (\Phi, A) \mapsto \Phi$.
%Thus for $\mathfrak{M}_E(M; N)$ we sometimes write
%\begin{center}
	%\begin{tikzcd}
		%\Omega^1(M;{}^*E) \arrow{d} \\
		%C^\infty(M;N)
	%\end{tikzcd}
%\end{center}

We will refer to $\gls{a0} \in \Omega^1(M; \Phi^*E)$ as the \textbf{field of gauge bosons} and $\gls{1vhi}$ just as a \textbf{physical field} of this theory.
\end{definitions}

Let us look at the tangent space of $\mathfrak{M}_E(M; N)$; we are interested into that because of the identification of infinitesimal gauge transformations as tangent vectors. Also recall the discussion about the double vector bundle structure before Def.~\ref{def:LinearVectorFieldsOnVectorBundles} which we need now again.

\begin{propositions}{Tangent space of $\mathfrak{M}_E(M; N)$}{TangentSpaceOfSpaceOfFields}
Let $M, N$ be two smooth manifolds and $E \stackrel{\pi}{\to} N$ a Lie algebroid. Then the tangent space $\mathrm{T}_{(\Phi_0,A_0)} \bigl(\mathfrak{M}_E(M; N)\bigr)$ of $\mathfrak{M}_E(M; N)$ at $(\Phi_0, A_0)$ consists of pairs $(\mathcal{v}, \mathcal{a})$ with $\mathcal{v} \in \Gamma\mleft(\Phi_0^*\mathrm{T}N\mright)$ and $\mathcal{a} \in \Omega^1\mleft(M; \mathcal{v}^*\mathrm{T}E\mright)$, where $\mathcal{v}^*\mathrm{T}E$ is the pullback of $\mathrm{T}E \stackrel{\mathrm{D}\pi}{\to}\mathrm{T}N$ as a vector bundle, viewing $\mathcal{v}$ as a map $M \to \mathrm{T}N$. This pair also satisfies
\ba
\pi_{\mathrm{T}E}(\mathcal{a})
&=
A_0,
\ea
where $\pi_{\mathrm{T}E}$ denotes the projection of the vector bundle $\mathrm{T}E \to E$.
\end{propositions}

\begin{remarks}{Total situation as commuting diagram}{TangentCommutingDiagram}
This implies that we have in total\footnote{Recall that we view sections of pullback bundles also as sections along maps; see Section \ref{StandardNotation}.}
\begin{center}
	\begin{tikzcd}
		 \mathrm{T}E \arrow{rr}{\mathrm{D}\pi} \arrow[dd, "\pi_{\mathrm{T}E}", swap] && \mathrm{T}N \arrow{dd}{\pi_{\mathrm{T}N}} \\
		&M \arrow[ld, "A_0(Y)", pos=0.3] \arrow{rd}{\Phi_0} \arrow{ru}{\mathcal{v}} \arrow[lu, "\mathcal{a}(Y)", pos=0.2]  \\
		E \arrow[rr, "\pi", swap]&& N
	\end{tikzcd}
\end{center}
for all $(\Phi_0,A_0) \in \mathfrak{M}_E(M;N)$, $(\mathcal{v}, \mathcal{a}) \in \mathrm{T}_{(\Phi_0, A_0)}\bigl( \mathfrak{M}_E(M;N) \bigr)$ and $Y \in \mathfrak{X}(M)$, that is,
\ba
\pi\bigl(A_0(Y)\bigr)
&=
\Phi_0,
\\
\pi_{\mathrm{T}N} (\mathcal{v})
&=
\Phi_0,
\\
\pi_{\mathrm{T}E}(\mathcal{a})
&=
A_0,\label{AGaugeTrafoIsOverA}
\\
\mathrm{D}\pi \bigl( \mathcal{a}(Y) \bigr)
&=
\mathcal{v}\label{HorizontalCompOfDeltaA}
\ea
for all $Y \in \mathfrak{X}(M)$, where the projections of the vector bundles $\mathrm{T}E \to E$ and $\mathrm{T}N \to N$ are denoted by $\pi_{\mathrm{T}E}$ and $\pi_{\mathrm{T}N}$, respectively.
\end{remarks}
\newpage
\begin{remark}\label{RemarkAboutThatWeStillHaveLinearStructureinDeltaA}
\leavevmode\newline
Especially for Eq.~\eqref{HorizontalCompOfDeltaA} recall the discussion about the double vector bundle structure before Def.~\ref{def:LinearVectorFieldsOnVectorBundles}. That is,
\bas
\mathcal{a}(f Y + h Z)
&=
f \boldsymbol{\cdot} \mathcal{a}(Y)
	\RPlus h \boldsymbol{\cdot} \mathcal{a}(Z)
\eas
for all $Y, Z \in \mathfrak{X}(M)$ and $f, h \in C^\infty(M)$, because $\mathcal{a}$ has values in $\mathrm{T}E$ viewed as a vector bundle over $\mathrm{T}N$. Therefore also
\bas
\mathrm{D}\pi\bigl( \mathcal{a}(f Y + h Z) \bigr)
&=
\mathrm{D}\pi\bigl( \mathcal{a}(Y) \bigr).
\eas
This is also in alignment with Eq.~\eqref{AGaugeTrafoIsOverA} although it is about the vector bundle $\mathrm{T}E \to E$, so,
\bas
\pi_{\mathrm{T}E}\bigl( \mathcal{a}(f Y + h Z) \bigr)
&=
\pi_{\mathrm{T}E}\bigl( f \boldsymbol{\cdot} \mathcal{a}(Y)
	\RPlus h \boldsymbol{\cdot} \mathcal{a}(Z) \bigr)
\\
&=
f ~ \pi_{\mathrm{T}E}\bigl( \mathcal{a}(Y) \bigr)
	+ h ~ \pi_{\mathrm{T}E}\bigl( \mathcal{a}(Z) \bigr)
\\
&=
A_0\bigl( f Y + h Z \bigr).
\eas
\end{remark}

\begin{proof}[Proof of Prop.~\ref{prop:TangentSpaceOfSpaceOfFields}]
\leavevmode\newline
We identify the tangent spaces of $(\Phi_0, A_0) \in \mathfrak{M}_E(M; N)$ with the set consisting of elements of the form
\bas
\mleft.\frac{\mathrm{d}}{\mathrm{d}t}\mright|_{t=0} \gamma,
\eas
where $\gamma: I \to \mathfrak{M}_E(M; N)$ is a curve with $\gamma(0) = (\Phi_0, A_0)$ and $I$ an open interval of $\mathbb{R}$ around 0. Since we do not have any conditions on $\mathfrak{M}_E(M; N)$ besides that $A_0$ has values in $\Phi_0^*E$, we will see that we just need to describe where the "velocity" of the curves live, and surjectivity will then just follow by that we always can find curves with arbitrary initial conditions on position and velocity. Let us write $\gamma = (\Phi, A)$, $t \mapsto \gamma(t)=(\Phi_t, A_t)$, with
\bas
\Phi_t &\in C^\infty(M;N), 
&
\Phi_{t=0} &= \Phi_0,
\\
A_t &\in \Omega^1(M; \Phi^*_t E),
&
A_{t=0} &= A_0
\eas
for all $t \in I$. As usual, the tangent space consists of elements of the form
\bas
\mleft( \mleft.\frac{\mathrm{d}}{\mathrm{d}t}\mright|_{t=0} \mleft[t\mapsto\Phi_t\mright], 
	\mleft.\frac{\mathrm{d}}{\mathrm{d}t}\mright|_{t=0} \mleft[t\mapsto A_t\mright]
\mright).
\eas
Hence, for all $p\in M$ we have a curve $\Phi(p) \coloneqq \mleft[ t \mapsto \Phi_t(p) \mright]$ in $N$ with
\bas
\mleft.\frac{\mathrm{d}}{\mathrm{d}t}\mright|_{t=0} (\Phi(p)) &\in \mathrm{T}_{\Phi_0(p)}N,
\eas
such that for all curves $\Phi$
\bas
\mleft.\frac{\mathrm{d}}{\mathrm{d}t}\mright|_{t=0} [t\mapsto \Phi_t]
&\in
\Gamma(\Phi_0^*\mathrm{T}N),
\eas
and besides $\Phi_{t=0}(p) = \Phi_0(p)$ there is no other condition on $\Phi(p)$, thus, for all $v \in \mathrm{T}_{\Phi_0(p)}N$ there is a curve $\Phi(p)$ such that its "initial velocity" is $v$, \textit{i.e.}~
\bas
v &= \mleft.\frac{\mathrm{d}}{\mathrm{d}t}\mright|_{t=0} \bigl(\Phi(p)\bigr),
\eas
and extending this argument we can achieve that for all $\mathcal{v} \in \Gamma(\Phi_0^*\mathrm{T}N)$ there is a curve $\Phi$ such that
\bas
\mathcal{v}
&=
\mleft.\frac{\mathrm{d}}{\mathrm{d}t}\mright|_{t=0} [t\mapsto\Phi_t],
\eas

Now we fix such a curve $\Phi$ for a fixed $\mathcal{v}$. Let us look at the curve $A(Y) \coloneqq \mleft[ t \mapsto A_t(Y) \mright]$ for all $Y \in \mathfrak{X}(M)$, that is $A(Y): I \times M \to E$, $(t,p) \mapsto A_{t,p}(Y_p)$ with $\pi \circ A(Y) = \Phi$, where $\pi$ is the projection of $E$ onto N. So,
\bas
\mathrm{T}_{\Phi_0(p)}N \ni
\mathcal{v}_p
&=
\mleft.\frac{\mathrm{d}}{\mathrm{d}t}\mright|_{t=0} \bigl(\pi ( A_p(Y_p))\bigr)
=
\mathrm{D}_{A_{0}(Y)|_p} \pi \mleft( \mleft.\frac{\mathrm{d}}{\mathrm{d}t}\mright|_{t=0} A_p(Y_p) \mright)
=
\mathrm{D}_{A_{0}(Y)|_p} \pi (\mathcal{a}_p(Y_p)),
\eas
where 
\bas
\mathcal{a}_p(Y_p)
&\coloneqq
\mleft.\frac{\mathrm{d}}{\mathrm{d}t} \mright|_{t=0} [ t\mapsto A_{t,p}(Y_p) ]
\in \mathrm{T}_{A_{0}(Y)|_p} E
\eas
for all $p \in M$. Hence, we can also see $\mathcal{a}$ equivalently as a form on $M$ with values in $\mathrm{T}E$ such that
\bas
\pi_{\mathrm{T}E}(\mathcal{a})
&=
A,
\\
\mathrm{D}\pi\bigl( \mathcal{a}(Y) \bigr)
&=
\mathcal{v}
\eas
for all $Y\in\mathfrak{X}(M)$, and we view $\mathcal{a}$ as an element of $\Omega^1(M; \mathcal{v}^*\mathrm{T}E)$, too, where we view $\mathrm{T}E$ as the vector bundle $\mathrm{T}E \stackrel{\mathrm{D}\pi}{\to} \mathrm{T}N$; that is because of the following: Let $Z \in \mathfrak{X}(M)$ be another vector field and $f,h \in C^\infty(M)$, then
\bas
\mathcal{a}_p\mleft(f(p) ~ Y_p + h(p) ~ Z_p\mright)
&=
\mleft.\frac{\mathrm{d}}{\mathrm{d}t} \mright|_{t=0} \mleft[ t\mapsto 
	A_{t,p}\mleft(f(p) ~ Y_p + h(p) ~ Z_p\mright) 
\mright]
\\
&=
\mleft.\frac{\mathrm{d}}{\mathrm{d}t} \mright|_{t=0} \mleft[ t\mapsto 
	f(p) ~ A_{t,p}\mleft( Y_p \mright) 
	+ h(p) ~ A_{t,p}\mleft( Z_p \mright) 
\mright]
\\
&=
f(p) \boldsymbol{\cdot} \mathcal{a}_p(Y_p)
	\RPlus h(p) \boldsymbol{\cdot} \mathcal{a}_p(Z_p),
\eas
because of
\bas
\mathrm{D}_{A_p(Y_p)}\pi\mleft(\mathcal{a}_p(Y_p)\mright)
&=
\mathcal{v}_p
=
\mathrm{D}_{A_p(Z_p)}\pi\mleft(\mathcal{a}_p(Z_p)\mright)
\eas
and since $[ t\mapsto A_{t,p}(Y_p) ]$ and $[ t\mapsto A_{t,p}(Z_p) ]$ are the representing curves of $\mathcal{a}_p(Y_p)$ and $\mathcal{a}_p(Z_p)$ as tangent vectors, respectively, satisfying
\bas
\pi\mleft(A_{t,p}(Y_p)\mright)
&=
\Phi_t(p)
=
\pi\mleft(A_{t,p}(Z_p)\mright),
\eas
such that we precisely get the definitions of $\boldsymbol{\cdot}$ and $\RPlus$.

As before, we can conclude that we can find a curve $A$ for all $\mathcal{a} \in \Omega^1(M; \mathcal{v}^*\mathrm{T}E)$ such that
\bas
\mathcal{a}
&=
\mleft.\frac{\mathrm{d}}{\mathrm{d}t} \mright|_{t=0} A.
\eas
(In this proof, we make use of the homotopy lifting property of fibrations such that we can find an $A(Y): I \times M \to E$ for each $\Phi: I \times M \to N$ for all $(\Phi_0, A_0) \in \mathfrak{M}_E(M; N)$ with the suitable properties.)
\end{proof}

Think of $(\mathcal{v}, \mathcal{a})$ again as candidates for the infinitesimal gauge transformations, for which we wrote $(\delta \Phi, \delta A)$ in Chapter \ref{ClassicGaugeTheory}; also recall Remark \ref{rem:TangentSpaceOfMathfrakMg}. But other than in Remark \ref{rem:TangentSpaceOfMathfrakMg} we cannot assume canonical flat connections now which is why the last result shows that we cannot view $(\mathcal{v}, \mathcal{a})$ as an element of $\mathfrak{M}_E(M;N)$ in general, thus, we changed the notation to $(\mathcal{v}, \mathcal{a})$ for now. So, we do not have any canonical horizontal distribution given, and therefore let us study the vertical structure first.

Recall that there is the notion of a \textbf{vertical bundle} for fibre bundles $F \stackrel{\pi}{\to} N$ (as \textit{e.g.}~introduced in \cite[\S 5.1.1, for principal bundles, but it is straightforward to extend the definitions]{hamilton}), which is defined as a subbundle $\gls{VF}$ of the tangent bundle $\mathrm{T}F \to F$ given as the kernel of $\mathrm{D}\pi : \mathrm{T}F \to \mathrm{T}N$. The fibres $\mathrm{V}_eF$ of $F$ at $e \in F$ are then given by 
\bas
\mathrm{V}_e F
&=
\mathrm{T}_e F_p,
\eas
where $p \coloneqq \pi(e) \in N$ and $F_p$ is the fibre of $F$ at $p$.
Now consider a vector bundle $E \stackrel{\pi}{\to} N$, then $\mathrm{V}_e E = \mathrm{T}_e E_p \cong E_p$ because the fibres are vector spaces.
%This can be emphasized by looking at the pullback vector bundle $\pi^*E \to E$ (\textbf{COMMENT: Reference?})
%\begin{center}
	%\begin{tikzcd}
		 %\pi^*E \arrow{d} & E \arrow{d}{\pi} \\
		%E \arrow{r}{\pi} & N
	%\end{tikzcd}
%\end{center}
%Elements of $\pi^*E$ are pairs $(e, e^\prime)$ defined by $e, e^\prime \in E$ and $\pi(e)= \pi(e^\prime)$, hence, $e^\prime \in E_{p}$ for $p = \pi(e)$. Thus, each fibre $(\pi^*E)_e$ of $\pi^*E$ is given by $E_p$, and therefore there is a very natural vector bundle isomorphism $\pi^*E \cong \mathrm{V}E$ because the tangent bundle $\mathrm{T}E$ also consists of pairs $(e, v)$ with $e \in E$ and $v \in \mathrm{T}_e E$, for the latter we then make use of the natural isomorphism $\mathrm{T}_e E_p \cong E_p$ when looking at $\mathrm{V}E$.

\begin{propositions}{Vertical bundle of $\mathfrak{M}_E(M; N)$}{VerticalBundleOfFracM}
Let $M, N$ be two smooth manifolds and $E \stackrel{\pi}{\to} N$ a Lie algebroid. Then the vertical bundle of $\mathfrak{M}_E(M; N)$, viewed as a fibration over $C^\infty(M;N)$, is given by
\ba
\mathrm{V}_{(\Phi,A)}\bigl(\mathfrak{M}_E(M; N)\bigr)
&\cong
\left\{
	(\mathcal{v}, \mathcal{a})
~\middle|~
	\mathcal{v}= 0 \in \Gamma(\Phi^*\mathrm{T}N), ~
	\mathcal{a} \in \Omega^1(M; \Phi^*E)
\right\}
\cong
\Omega^1(M; \Phi^*E).
\ea
%In total, $\mathrm{V}\mathfrak{M}_E(M; N)$ is isomorphic to
%\begin{center}
	%\begin{tikzcd}
		%\Omega^1(M; {}^*E) \arrow{d} \\
		%\mathfrak{M}_E(M; N)
	%\end{tikzcd}
%\end{center}
\end{propositions}

%\begin{remark}
%\leavevmode\newline
%We can quickly motivate the result of the proposition by using what we discussed before this statement, \textit{i.e.}~we identify $\mathrm{V}\mathfrak{M}_E(M; N)$ with $\varpi^*\mathfrak{M}_E(M; N)$, where $\varpi$ is the projection of $\mathfrak{M}_E(M; N) \to C^\infty(M;N)$,
%\begin{center}
	%\begin{tikzcd}
		 %\varpi^*\mathfrak{M}_E(M; N) \arrow{d} & \mathfrak{M}_E(M; N) \arrow{d}{\varpi} \\
		 %\mathfrak{M}_E(M; N) \arrow{r}{\varpi} & C^\infty(M; N)
	%\end{tikzcd}
%\end{center}
%Therefore the fibre of $\mathrm{V}\mathfrak{M}_E(M; N) = \varpi^*\mathfrak{M}_E(M; N)$ at $(\Phi,A)\in \mathfrak{M}_E(M; N)$ is given by the fibre of $\mathfrak{M}_E(M; N)$ at
%\bas
%\varpi(\Phi, A) &=\Phi,
%\eas
%which is $\Omega^1(M; \Phi^*E)$.
%\end{remark}

\begin{proof}[Proof of Prop.~\ref{prop:VerticalBundleOfFracM}]
\leavevmode\newline
We have the fibration $\mathfrak{M}_E(M; N) \stackrel{\varpi}{\to} C^\infty(M;N)$, where $\varpi(\Phi, A) \coloneqq \Phi$ for all $(\Phi, A) \in \mathfrak{M}_E(M; N)$. Hence,
\bas
\mathrm{D}_{(\Phi, A)}\varpi(\mathcal{v}, \mathcal{a})
=
\mathcal{v}
\eas
for all $(\mathcal{v}, \mathcal{a}) \in \mathrm{T}_{(\Phi,A)}\mathfrak{M}_E(M; N)$. The kernel of $\mathrm{D}\varpi$ at $(\Phi,A) \in \mathfrak{M}_E(M; N)$ is then given by 
\bas
\mathrm{Ker}\mleft( \mathrm{D}_{(\Phi,A)} \varpi \mright)
&=
\left\{
(\mathcal{v}, \mathcal{a}) \in \mathrm{T}_{(\Phi,A)} \mathfrak{M}_E(M; N)
~\middle|~
\mathcal{v}=0
\right\}.
\eas
By Prop.~\ref{prop:TangentSpaceOfSpaceOfFields}, we then know that $\mathcal{a}$ has values in the vertical bundle $\mathrm{V}E$, that is, for $\mathcal{a}_p(Y_p) \in \mathrm{T}_{A_p(Y_p)} E$ ($p \in M$, $Y \in \mathfrak{X}(M)$) we have
\bas
&&
\mathrm{D}_{A_p(Y_p)}\pi\mleft(\mathcal{a}_p(Y_p)\mright)
&=
0
\\
&\Leftrightarrow&
\mathcal{a}_p(Y_p)
&\in
\mathrm{V}_{A_p(Y_p)} E
\cong
E_{\Phi(p)}.
\eas
Thus, we can view $\mathcal{a}$ equivalently as an element of $\Omega^1(M; \Phi^*E)$, so,
\bas
\mathrm{V}_{(\Phi,A)}\mathfrak{M}_E(M; N)
&\cong
\left\{
	(\mathcal{v}, \mathcal{a})
~\middle|~
	\mathcal{v}= 0 \in \Gamma(\Phi^*\mathrm{T}N), ~
	\mathcal{a} \in \Omega^1(M; \Phi^*E)
\right\}
\cong
\Omega^1(M; \Phi^*E).
\eas
\end{proof}

That is, we can in general only expect to have $(\mathcal{v}, \mathcal{a}) \in \mathfrak{M}_E(M;N)$ if at least $\mathcal{v}=0$. Recall that we identified this component with the infinitesimal gauge transformation of the Higgs field which was proportional to the Lie algebra representation, see Def.~\ref{def:ClassicTrafos}. Even when we do not have yet the general definition of that infinitesimal gauge transformation, it is natural to assume that this transformation is therefore only zero when there is no coupling of the gauge bosons to the Higgs field (= zero action), but in general there will be of course a coupling. As already mentioned, we circumvented that problem in Chapter \ref{ClassicGaugeTheory} by choosing canonical flat connections; moreover, observe that this condition about $\mathcal{v}=0$ comes from that the field of gauge bosons $A$ has values in $\Phi^*E$, as if we would have applied the "bookkeeping trick" to $A$ in Section \ref{NewInfGaugeTrafoTrafos}, too. Thus, we are going to treat the infinitesimal gauge transformation of $A$ similar to how we defined the infinitesimal gauge transformation for functionals in Section \ref{NewInfGaugeTrafoTrafos}, then we also achieve that its transformation can be viewed again as an element of $\Omega^1(M; \Phi^*E)$, simplifying further calculations, without really loosing information about the transformation of $A$; we will explain this later. That the infinitesimal gauge transformation of the Higgs field is in general not a smooth map $M \to N$ will be on the other hand actually less of a problem.

But before we can make that mathematical precise, we need to define at what type of functionals we are going to look at. One key step is to look at $M \times \mathfrak{M}_E(M;N)$ as we did in Def.~\ref{def:FirstAttemptOfEvaluationMap} and afterwards.

\begin{definitions}{Evaluation map of $M \times \mathfrak{M}_E$}{EvaluationMap}
Let $M, N$ be manifolds, and $E \to N$ a Lie algebroid over $N$.
Then we define the \textbf{evaluation map} $\mathrm{ev}$ by
\ba
M \times \mathfrak{M}_E(M;N) &\to N
\nonumber\\
(\Phi,A)
&\mapsto
\mathrm{ev}(p, \Phi, A)
\coloneqq
\Phi(p)
\ea
for all $p\in M$ and $(\Phi, A) \in \mathfrak{M}_E$.
\end{definitions}

\begin{remarks}{Bigrading of forms on $M \times \mathfrak{M}_E$}{Bigrading}
Let $\pi_i$ ($i \in \{1,2\}$) be the projection onto the $i$-th factor in $M \times \mathfrak{M}_E$, then
\ba
\mathrm{T}\mleft( M \times \mathfrak{M}_E \mright)
\cong
\pi_1^*\mathrm{T}M \oplus \pi_2^*\mathrm{T}\mathfrak{M}_E.
\ea
Gives rise to a bigrading of $\bigwedge^k \mathrm{T}^*\mleft( M \times \mathfrak{M}_E \mright)$ ($k \in \mathbb{N}_0$),
\ba
\bigwedge^k \mathrm{T}^*\mleft( M \times \mathfrak{M}_E \mright)
\cong
\bigoplus_{\substack{p,q \in \mathbb{N}_0 \\ p+q = k}} \mleft(
	\bigwedge^{p,q} \mathrm{T}^*\mleft( M \times \mathfrak{M}_E \mright)
\mright),
\ea
where
\ba
\bigwedge^{p,q} \mathrm{T}^*\mleft( M \times \mathfrak{M}_E \mright)
&\coloneqq
\pi_1^*\mleft(\bigwedge^p \mathrm{T}^*M\mright) \otimes \pi_2^*\mleft(\bigwedge^q \mathrm{T}^*\mathfrak{M}_E\mright).
\ea
Similarly, for $V$ a vector bundle over $M \times \mathfrak{M}_E$,
\ba
\Omega^k\mleft( M \times \mathfrak{M}_E; V \mright)
\cong
\bigoplus_{\substack{p,q \in \mathbb{N}_0 \\ p+q = k}} \bigl(
	\Omega^{p,q} \mleft( M \times \mathfrak{M}_E; V \mright)
\bigr),
\ea
with 
\ba
\Omega^{p,q} \mleft( M \times \mathfrak{M}_E; V \mright)
\coloneqq
\Gamma\mleft(
	\pi_1^*\mleft(\bigwedge^p \mathrm{T}^*M\mright) \otimes \pi_2^*\mleft(\bigwedge^q \mathrm{T}^*\mathfrak{M}_E\mright) \otimes V
\mright).
\ea
When $V$ is the trivial line bundle, then we just write $\Omega^{p,q}(M \times \mathfrak{M}_E)$.

If $V$ is instead a vector bundle over $N$, then we have $\mathrm{ev}^*V$ naturally as bundle over $M \times \mathfrak{M}_E$. Then, when taking a slice through $(\Phi, A) \in \mathfrak{M}_E$, \textit{i.e.}~evaluating a form at points $M \times \{\Phi, A\}$ while $(\Phi, A) \in \mathfrak{M}_E$ is fixed,
\ba\label{SliceOfBIiiigManifold}
\mleft.L\mright|_{M \times \{\Phi, A\}}
&\in
\Omega^p(M; \Phi^*V)
\ea
for all $L \in \Omega^{p,0} \mleft( M \times \mathfrak{M}_E; \mathrm{ev}^*V \mright)$.
Similarly, the de-Rham differential splits on $\Omega^k(M \times \mathfrak{M}_E)$ as a differential along $M$ and $\mathfrak{M}_E$, $\mathrm{d}_{\text{total}} = \mathrm{d}_M + \mathrm{d}_{\mathfrak{M}_E}$. When using exterior derivatives, then we focus on directions along $M$, and we will denote that de-Rham differential by $\mathrm{d}$, \textit{i.e.}~$\mathrm{d}= \mathrm{d}_M$.
\end{remarks}

\begin{remark}
\leavevmode\newline
Do not confuse notations like $\Omega^{p,q} \mleft( M \times \mathfrak{M}_E; V \mright)$ with the notation given in Def.~\ref{def:ExteriorCovariantDerivatives}; it will be clear by the context which we mean, and, besides the next paragraphs, we actually will not really use $\Omega^{p,q} \mleft( M \times \mathfrak{M}_E; V \mright)$ as notation anymore because we only want to motivate the next and some following definitions with this notation.
\end{remark}

Eq.~\eqref{SliceOfBIiiigManifold} is precisely the space our functionals should take values in when evaluated at $(\Phi, A) \in \mathfrak{M}_E$. This leads to the following definition.

\begin{definitions}{Space of functionals in gauge theory}{FunctionalsAsForms}
Let $M, N$ be two smooth manifolds, $E\to N$ a Lie algebroid, and $V \to N$ a vector bundle. Then the \textbf{space of functionals $\gls{Fk}(M; {}^*V)$} ($k \in \mathbb{N}_0$) is defined as
\ba
\mathcal{F}^k_E(M; {}^*V)
&\coloneqq
\Omega^{k,0}\bigl(M \times \mathfrak{M}_E(M;N); \mathrm{ev}^*V\bigr).
\ea

If $V = N \times \mathbb{R}$ is the trivial line bundle over $N$, then we just write $\mathcal{F}_E^k(M)$ instead of $\mathcal{F}^k_E(M;{}^*V)$.
\end{definitions}

\begin{remark}
\leavevmode\newline
We often write for $L \in \mathcal{F}^k_E(M; {}^*V)$
\bas
\mathfrak{M}_E \ni (\Phi, A)
&\mapsto
L(\Phi, A) 
\coloneqq 
\mleft.L\mright|_{M \times \{\Phi, A\}}
\in \Omega^k(M; \Phi^*V)
\eas
especially when we do not evaluate at $p \in M$; recall Eq.~\eqref{SliceOfBIiiigManifold}. Observe that $L$ acts non-trivially only on $\mathrm{T}M$.
\end{remark}

\begin{examples}{Projection onto the field of gauge bosons}{ProjectionOntoGaugeBosonies}
Besides the physical quantities which we will define later, we have an important and trivial functional $\gls{1pivar} \in \mathcal{F}^1_E(M; {}^*E)$ given as the projection onto the field of gauge bosons, that is
\ba
\varpi_2(\Phi,A)
&\coloneqq
A
\ea
for all $(\Phi, A) \in \mathfrak{M}_E$.
We will especially need this functional to define the infinitesimal gauge transformation of $A$ and in several combinations with other functionals.
\end{examples}

\begin{examples}{Tangent map, total differential as functional}{DAsFunctional}
Also the total differential $\gls{D}$ can be viewed as a functional. That is $\mathrm{D} \in \mathcal{F}^1_E(M; {}^*\mathrm{T}N)$ by
\ba
\mathrm{D}(\Phi, A)
&\coloneqq
\mathrm{D}\Phi
\in
\Omega^1(M; \Phi^*\mathrm{T}N).
\ea
Hence, when we just write $\mathrm{D}$, then we mean precisely that. 
\end{examples}

For the following discussion and definitions we use a similar convention of notation as in Section \ref{DirectProdsOfLieAlgoids}. That is, we have $\mathrm{T}(M \times \mathfrak{M}_E) \cong \pi_1^*\mathrm{T}M \oplus \pi_2^*\mathrm{T}\mathfrak{M}_E$ as in Remark \ref{rem:Bigrading}. If we speak for example about $\mathrm{T}M$, especially sections thereof, $\mathfrak{X}(M)$, then we mean their canonical embedding as a subalgebra of $\mathfrak{X}(M \times \mathfrak{M}_E)$; so, $X \in \mathfrak{X}(M)$ is also viewed as an element of $\mathfrak{X}(M\times \mathfrak{M}_E)$ but constant along $\mathfrak{M}_E$. For vector bundle morphisms defined on $\mathrm{T}(M \times \mathfrak{M}_E)$ we then also mean that forms restricted onto $\mathrm{T}M$ extend to maps acting on $\mathfrak{X}(M)$.

\begin{remarks}{Notions on $\mathcal{F}^k_E$ and further pullbacks with $\mathrm{ev}$}{NotionsOnFunctionals}
By Def.~\ref{def:FunctionalsAsForms}, we recover typical notions on the space of functionals, notions like wedge products, Def.~\ref{def:GradingOfProducts} and contractions \textit{etc.}~by restricting notions on $\Omega^\bullet(M \times \mathfrak{M}_E)$ and $\Omega^\bullet(M \times \mathfrak{M}_E; \mathrm{ev}^*V)$ to $\Omega^{\bullet,0}(M \times \mathfrak{M}_E)$ and $\Omega^{\bullet,0}(M \times \mathfrak{M}_E; \mathrm{ev}^*V)$ ($\bullet$ as placeholder for the degree), respectively. Hence, we will not need to define all those notions in that setting, and, especially, $\Gamma(\mathrm{ev}^*V)$ is therefore generated by elements of the form $\mathrm{ev}^*v$, where $v \in \Gamma(V)$. 

\hspace{0.3cm} Now assume we have a vector bundle connection $\nabla$ on $V$, then $\mathrm{ev}^*\nabla$ is a connection on $\mathrm{ev}^*V$. We want to restrict the exterior covariant derivative related to that connection just to vector fields on $M$. Observe for all $X\in \mathfrak{X}(M) \subset \mathfrak{X}(M \times \mathfrak{M}_E)$, with flow $\gamma$ in $M$ through a $p \in M$, $(t, p) \mapsto \gamma_t(p)$ ($t \in I$ for some open interval in $\mathbb{R}$ containing 0),
\ba\label{DGleichDev}
\mathrm{D}_{(p, \Phi, A)} \mathrm{ev} (X)
&=
\mleft.\frac{\mathrm{d}}{\mathrm{d}t}\mright|_{t=0}
\bigl(
	\mathrm{ev} \circ (\gamma(p), \Phi, A)
\bigr)
=
\mleft.\frac{\mathrm{d}}{\mathrm{d}t}\mright|_{t=0}
\bigl(
	(\Phi \circ \gamma )(p) 
\bigr)
=
\mathrm{D}_p \Phi (X)
\ea
for all $(p, \Phi, A) \in M \times \mathfrak{M}_E$, where $(\gamma(p), \Phi, A)$ is the flow of $X \in \mathfrak{X}(M)$ at $(p, \Phi, A)$, viewed as an element of $\mathfrak{X}(M \times \mathfrak{M}_E)$. So, the pushforward of $X$ with $\mathrm{ev}$ at $(\Phi,A)$ is the same as the pushforward of $X$ with $\Phi$, thus
\bas
\mleft(\mathrm{ev}^* \nabla\mright)_{X_{(p, \Phi, A)}}
&=
\mleft( \Phi^*\nabla \mright)_{X_p}
\eas
for all $(p, \Phi, A)$,
viewing $X$ as an element of $\mathfrak{X}(M \times \mathfrak{M}_E)$ on the left hand side and as an element of $\mathfrak{X}(M)$ on the right hand side. Hence, we then also have
\bas
\mleft.\bigl(\mleft(\mathrm{ev}^* \nabla\mright)_{X} v\bigr)\mright|_{(p, \Phi, A)}
=
\mleft.\mleft(\mleft( \Phi^*\nabla \mright)_{X_p} v|_{(\Phi, A)} \mright) \mright|_p
\eas
for all $v \in \Gamma(\mathrm{ev}^*V)$,
since $X$ does not differentiate along $\mathfrak{M}_E$, and viewing $v|_{(\Phi, A)} \coloneqq [p \mapsto v|_{(p, \Phi, A)}]$ as an element of $\Gamma(\Phi^*V)$ on the right hand side.
Therefore this naturally leads on one hand to an exterior covariant derivative on the space of functionals by restricting $\mathrm{ev}^*\nabla$ to $\mathrm{T}M$ because then the exterior covariant derivative of $\mleft.\mleft(\mathrm{ev}^*\nabla\mright)\mright|_{\mathrm{T}M}$ clearly restricts to $\mathcal{F}^\bullet_E(M; {}^*V)$, and on the other hand
\bas
\mleft.\mleft(\mathrm{d}^{\mleft.\mleft(\mathrm{ev}^*\nabla\mright)\mright|_{\mathrm{T}M}} L\mright)\mright|_{(\Phi,A)} 
&=
\mathrm{d}^{\Phi^*\nabla}\bigl( L(\Phi, A) \bigr),
\eas
also recall Eq.~\eqref{SliceOfBIiiigManifold}.

\hspace{0.3cm} Similarly, one shows for the pullback $\mathrm{ev}^!\omega$ of forms $\omega \in \Omega^k(N; V)$ that
\bas
\mleft. \mleft(\mathrm{ev}^! \omega\mright)\mright|_{(p,\Phi,A)}
\mleft( X_1, \dotsc, X_k \mright)
&=
\mleft. \mleft(\Phi^! \omega\mright)\mright|_{p}
\mleft( X_1, \dotsc, X_k \mright)
\eas
for all $X_1, \dotsc, X_k \in \mathfrak{X}(M)$. Hence, also the $\mathrm{ev}$-pullback of forms restricts to a $\Phi$-pullback of forms when fixing $(\Phi,A)$ and just evaluating at vector fields along $M$.
\end{remarks}

Therefore we define pullback functionals as in the following definition.

\begin{definitions}{Pullbacks as functionals}{PullbacksAsFunctionals}
Let $M, N$ be smooth manifolds, $E \to N$ a Lie algebroid, and $V \to N$ a vector bundle. For all $\omega \in \Gamma\mleft( V \mright)$ we define its \textbf{pullback functional $\gls{0*omega}$} as an element of $\mathcal{F}^0_E(M; {}^*V)$ by
\ba
{}^*v
&\coloneqq
\mathrm{ev}^*v.
\ea

For a vector bundle connection $\nabla$ on $V$ we define the \textbf{pullback connection $\gls{0*nabla}$ (to functionals)} by
\ba
{}^*\nabla
&\coloneqq
\mleft.\mleft(\mathrm{ev}^*\nabla\mright)\mright|_{\mathrm{T}M}.
\ea
Its induced exterior covariant derivative $\mathrm{d}^{{}^*\nabla}$ we view as an exterior covariant derivative on the space of functionals, especially
\ba
\mathrm{d}^{{}^*\nabla}:
\mathcal{F}^k_E(M; {}^*V)
&\to
\mathcal{F}^{k+1}_E(M; {}^*V)
\ea
for all $k \in \mathbb{N}_0$.

For all $\omega \in \Omega^k(N;V)$  ($k \in \mathbb{N}_0$) we define similarly its \textbf{form-pullback functional $\gls{0!omega}$} as an element of $\mathcal{F}_E^k(M; {}^*V)$ by
\ba
{}^!\omega
&\coloneqq
\mleft.\mleft(\mathrm{ev}^!\omega\mright)\mright|_{\bigwedge^k \mathrm{T}M}.
\ea
\end{definitions}

\begin{remarks}{}{}
Observe that
\ba
\mleft.({}^*v)(\Phi,A)\mright|_p
&\coloneqq
(\mathrm{ev}^*v)|_{(p, \Phi, A)}
=
\Phi^*v|_p
\ea
for all $(p, \Phi, A) \in M \times \mathfrak{M}_E$. Especially, $({}^*v)(\Phi, A) = \Phi^*v$, similarly to what we already pointed out for ${}^!w$ and ${}^*\nabla$ in Remark \ref{rem:NotionsOnFunctionals}. By construction, and as argued in Rem.~\ref{rem:NotionsOnFunctionals}, we also get
\ba
\mleft( \mathrm{d}^{{}^*\nabla} L \mright)(\Phi,A)
&=
\mathrm{d}^{\Phi^*\nabla}\bigl( L(\Phi,A) \bigr)
\ea
for all $L \in \mathcal{F}^k_E(M; {}^*V)$ ($k \in \mathbb{N}_0$) and $(\Phi, A) \in \mathfrak{M}_E(M;N)$. 

We can also locally write, using a frame $\mleft( e_a \mright)_a$ of $V$,
\ba
L
&=
L^a \otimes {}^*e_a,
\ea
using that $\mathrm{ev}$-pullbacks generate $\Gamma(\mathrm{ev}^*V)$,
where $L^a \in \mathcal{F}^k_E(M) = \Omega^{k,0}(M \times \mathfrak{M}_E)$ (restriction on open neighbourhood omitted).
\newline\newline
The first calculation of Remark \ref{rem:NotionsOnFunctionals} also shows that we have
\bas
\mathrm{D}
&=
\mathrm{Dev}|_{\mathrm{T}M}
\eas
as functionals, where we view $\mathrm{Dev}|_{\mathrm{T}M}$ as an element of $\mathcal{F}^1_E(M; {}^*\mathrm{T}N)$ given by Eq.~\eqref{DGleichDev}. This implies that we can apply Eq.~\eqref{EqPullBackFormelFuerVerschiedeneDefinitionen}, that is,
\bas
{}^!\omega
&=
\mleft.\mleft(\mathrm{ev}^!\omega\mright)\mright|_{\bigwedge^k \mathrm{T}M}
\stackrel{\eqref{EqPullBackFormelFuerVerschiedeneDefinitionen}}{=}
\frac{1}{k!}~
\mleft(\mathrm{ev}^*\omega\mright)\mleft(\mathrm{Dev}|_{\mathrm{T}M} \stackrel{\wedge}{,} \dotsc \stackrel{\wedge}{,} \mathrm{Dev}|_{\mathrm{T}M} \mright)
=
\frac{1}{k!}~
\mleft({}^*\omega\mright)\mleft(\mathrm{D} \stackrel{\wedge}{,} \dotsc \stackrel{\wedge}{,} \mathrm{D} \mright)
\eas
for all $\omega \in \Omega^k(N;V)$ ($k \in \mathbb{N}_0$). We are going to use this very often by just giving reference to Eq.~\eqref{EqPullBackFormelFuerVerschiedeneDefinitionen}.
\end{remarks}

\begin{examples}{Anchor as functional}{AnchorAsFunctional}
Recall Ex.~\ref{ex:ProjectionOntoGaugeBosonies}; the anchor gives also rise to a functional, especially needed for the minimal coupling. $({}^*\rho)(\varpi_2)$ is a functional in $\mathcal{F}^1_E(M; {}^*\mathrm{T}N)$, that is
\bas
\bigl(({}^*\rho)(\varpi_2)\bigr)(\Phi, A)
&=
(\Phi^*\rho)(A)
\eas
for all $(\Phi, A) \in \mathfrak{M}_E(M;N)$.
\end{examples}

We have now the setup to finally define the physical quantities.

\section{Physical Quantities}\label{NewPhysicQuants}
Let us first start with the definition of the field strength. The following definitions essentially are motivated by \cite{CurvedYMH}, however, we completely reformulated it with the previously-introduced notation in order to allow coordinate-free versions, also "free" with respect to $(\Phi, A) \in \mathfrak{M}_E$.

\begin{definitions}{Field of gauge bosons and their field strength, \newline \cite[especially Eq.~(11); $\Phi$ is denoted as $X$ there]{CurvedYMH}}{EichbosonenUndFeldstaerke}
Let $M, N$ be smooth manifolds, and $E \to N$ a Lie algebroid equipped with a connection $\nabla$ on $E$. We define the \textbf{field strength $\gls{F}$} as an element of $\mathcal{F}_E^2(M; {}^*E)$ by
\ba
F
&\coloneqq
\mathrm{d}^{{}^*\nabla}\varpi_2
	- \frac{1}{2} ({}^*t_{\nabla_\rho})\mleft( \varpi_2 \stackrel{\wedge}{,} \varpi_2 \mright),
\ea
that is
\ba\label{DefOfCovariantizedFieldStrengthF}
F(\Phi, A)
\coloneqq
\mathrm{d}^{\Phi^*\nabla} A
	- \frac{1}{2} \mleft( \Phi^* t_{\nabla_\rho} \mright)\mleft( A \stackrel{\wedge}{,} A \mright)
\ea
for all $\Phi \in C^\infty(M;N)$ and $A \in \Omega^1(M; \Phi^*E)$.
\end{definitions}

\begin{remark}\label{RemarkUeberDefinitionVonNormalerFeldUndA}
\leavevmode\newline
\indent $\bullet$ Recall Def.~\ref{def:CanonicalBasicConnection} and Prop.~\ref{prop:SnablamitREnabla} which imply $t_{\nabla_\rho} = -t_{\nabla^{\mathrm{bas}}}$, where $\nabla^{\mathrm{bas}}$ is the basic connection, such that
\bas
F
=
\mathrm{d}^{{}^*\nabla} \varpi_2
	+ \frac{1}{2} \mleft( {}^* t_{\nabla^{\mathrm{bas}}} \mright)\mleft( \varpi_2 \stackrel{\wedge}{,} \varpi_2 \mright).
\eas
We are going to use this often later.

$\bullet$ Let us recall the definition of the standard setting, recall Def.~\ref{def:ClassicFieldStrength}, and recall the bookkeeping trick before Prop.~\ref{prop:ClassicFunctionDerivativesAlongPsiEpsilon}, which we denoted by $\iota$: We then normally have $A \in \Omega^1(M; \mathfrak{g}), \Phi \in C^\infty(M;W)$ for a given Lie algebra $\mathfrak{g}$ and $W$ a vector space, then the field strength is normally defined as
\ba\label{StandardFDef}
F^{\mathrm{clas}}(\Phi, A)
\equiv
F^{\mathrm{clas}}(A)
&=
\mathrm{d}A^a \otimes e_a
	+ \frac{1}{2} \mleft[ A \stackrel{\wedge}{,} A \mright]_{\mathfrak{g}}
\ea
for some given basis $\mleft( e_a \mright)_a$ of $\mathfrak{g}$. $\mathfrak{g}$ is viewed as "trivial bundle" over $M$, $M \times \mathfrak{g}$, and $\mleft( e_a \mright)_a$ is a constant frame.

Now, let us instead restrict Eq.~\eqref{DefOfCovariantizedFieldStrengthF} to an action Lie algebroid $E = N \times \mathfrak{g}$ equipped with $\nabla$ as the canonical flat connection and $\mleft( e_a \mright)_a$ a global frame of constant sections, especially $\nabla e_a = 0$. Then $\mleft( \Phi^*e_a \mright)_a$ trivializes $\Phi^*E$ such that $\Phi^*E \cong M \times \mathfrak{g}$, $\mleft( \Phi^*e_a \mright)_a$ describes a constant frame, especially $(\Phi^* \nabla) (\Phi^*e_a) = \Phi^! (\nabla e_a) = 0$,
and all $\Phi^*E$-valued objects can be viewed as $\mathfrak{g}$-valued.
In that case, write $A = A^a \otimes \Phi^*e_a$, and observe that
\bas
- \frac{1}{2} \mleft( \Phi^* t_{\nabla_\rho} \mright)\mleft( A \stackrel{\wedge}{,} A \mright)
&=
- \frac{1}{2} \underbrace{\mleft( \Phi^* t_{\nabla_\rho} \mright)\mleft( \Phi^*e_a, \Phi^*e_b \mright)}_{\mathclap{= \Phi^*\mleft( t_{\nabla_\rho}(e_a, e_b) \mright)}}~
	A^a \wedge A^b
=
\frac{1}{2} \Phi^*\underbrace{\mleft( \mleft[ e_a, e_b \mright]_E \mright)}_{\mathclap{=\mleft[ e_a, e_b \mright]_{\mathfrak{g}} = \text{const.}}}~
	A^a \wedge A^b
=
\frac{1}{2} \mleft[ A \stackrel{\wedge}{,} A \mright]_{\mathfrak{g}}
\eas
and
\bas
\mathrm{d}^{\Phi^*\nabla} A
&=
\mathrm{d}A^a \otimes \Phi^*e_a
	- A^a \otimes \Phi^!(\nabla e_a)
=
\mathrm{d}A^a \otimes \Phi^*e_a
\eas
for all $A \in \Omega^1(M; \Phi^*E)$. Hence, we get
\bas
F
&=
\iota\mleft( F^{\mathrm{clas}} \mright).
\eas
%\footnote{Strictly spoken, the right hand side is $\iota\mleft( F^{\mathrm{clas}} \mleft( \iota^{-1}(A) \mright) \mright)$ due to that $A$ has values in the pullback bundle and is, thus, the field of gauge bosons of the classical theory with applied bookkeeping trick. But the bookkeeping trick is just that, bookkeeping, trivially $\iota(A) = A$. It just emphasizes the relationship }
%Hence, Eq.~\eqref{DefOfCovariantizedFieldStrengthF} restricts to the standard definition when restricted to the standard setting, comparing it with Eq.~\eqref{StandardFDef}.  That we have $\Phi^*e_a$ instead of $e_a$ as in the standard definition is just a distinction in the notation, both describe of course a global frame of constant sections, and that is all which is needed here. Due to the constancy of $e_a$, one could even argue to write $\Phi^*e_a = e_a$, \textit{i.e.}~one can omit the notation of the pull-back in the standard setting to emphasize that it is not affected by variations of $\Phi$ as it happens under a gauge transformation using the standard definition of gauge transformations in the standard setting, that is, one will use a canonical flat connection for the gauge transformations as argued in the last bullet point of Rem.~\ref{RemLeibnizeRegelaufProdukteWeshalbEConnectionNichtWichtigIst}. A similar argument holds for all other types of constant frames in the following, even when we do not explicitly mention it anymore. However, when one does not want to use canonical flat connections, then the pull-back should not be omitted; this will be important later.
\end{remark}

As we have seen in the definition of the action Lie algebroid, the anchor $\rho$ replaces the notion of Lie algebra actions and representations such that we now use the anchor to define the minimal coupling of $A$ to $\Phi$.

\begin{definitions}{Minimal coupling, \cite[Eq.~(3), $\Phi$ is denoted as $X$ there]{CurvedYMH}}{MinimalCoupling}
Let $M, N$ be smooth manifolds and $E \to N$ a Lie algebroid. Then we define the \textbf{minimal coupling $\mathfrak{D}$} as an element of $\mathcal{F}_E^1(M; {}^*\mathrm{T}N)$ by
\ba\label{MinimalCouplingInKurz}
\mathfrak{D}
&\coloneqq
\mathrm{D}
	- ({}^*\rho)(\varpi_2).
\ea

We also write
\ba
\gls{DAPhi}
&\coloneqq
\mathfrak{D}(\Phi, A)
=
\mathrm{D}\Phi
	- \mleft( \Phi^*\rho\mright)(A)
\ea
for all $\Phi \in C^\infty(M;N)$ and $A \in \Omega^1(M; \Phi^*E)$, and we say that \textbf{$\Phi$ is minimally coupled to $A$}.
\end{definitions}

\begin{remark}
\leavevmode\newline
Restricting this to the standard situation gives back the standard definition: Assume $N = W$ where $W$ is a vector space, $E = W \times \mathfrak{g}$ an action Lie algebroid over $W$,
whose action is induced by a Lie algebra representation $\psi: \mathfrak{g} \to \mathrm{End}(W)$. Then the minimal coupling is
\bas
\mleft.\mathfrak{D}^A \Phi\mright|_p
&=
\mleft.\mathrm{d}_p\Phi^\alpha \otimes \Phi^*\partial_\alpha\mright|_p
	+ \psi\bigl(A_p(Y)\bigr)\bigl(\Phi(p)\bigr)
\eas
for all $(p, \Phi, A) \in M \times \mathfrak{M}_E(M;W)$ and $Y \in \mathrm{T}_pM$,
where we use some global coordinates $\mleft(\partial_\alpha\mright)_\alpha$ of $W$ and Prop.~\ref{prop:LieRepAndLieAct}. Now we make use of the canonical identification of $W$'s tangent spaces with $W$ itself, especially, $v_\alpha = \partial_\alpha$ for some basis $\mleft( v_\alpha \mright)_\alpha$ on $W$. Then the first summand is clearly $\mathrm{d}\Phi^\alpha \otimes \Phi^*\partial_\alpha = \iota(\mathrm{d}\Phi)$. Hence, also here we arrive at the classical definition (under the bookkeeping trick), recall Def.~\ref{def:ClassicMinimalCoupling}.
\end{remark}

Finally we turn to the Lagrangian. 

\begin{definitions}{Yang-Mills-Higgs Lagrangian, \newline \cite[Eq.~(2) and (16); but a different field strength there which we will introduce later]{CurvedYMH}}{CurvedYMHLagrangian}
Let $M$ be a spacetime with a spacetime metric $\eta$, $N$ a smooth manifold, $E \to N$ a Lie algebroid, $\nabla$ a connection on $E$, and let $\kappa$ and $g$ be fibre metrics on $E$ and $\mathrm{T}N$, respectively. Also let $V \in C^\infty(N)$, which we call the \textbf{potential of the Higgs field}. Then we define the \textbf{Yang-Mills-Higgs Lagrangian $\gls{LYMH}$} as an element of $\mathcal{F}_E^{\mathrm{dim}(M)}(M)$ by
\ba
\mathfrak{L}_{\mathrm{YMH}}
&\coloneqq
- \frac{1}{2} \mleft( {}^*\kappa \mright)\mleft(F \stackrel{\wedge}{,} *F\mright)
	+ \mleft( {}^*g \mright)\mleft(\mathfrak{D} \stackrel{\wedge}{,} *\mathfrak{D} \mright)
	- *({}^*V),
\ea
that is
\ba
\mathfrak{L}_{\mathrm{YMH}}(\Phi, A)
&\coloneqq
- \frac{1}{2} \mleft( \Phi^*\kappa \mright)\mleft(F(\Phi, A) \stackrel{\wedge}{,} *F(\Phi, A)\mright)
	+ \mleft( \Phi^*g \mright)\mleft(\mathfrak{D}^A \Phi \stackrel{\wedge}{,} *\mathfrak{D}^A\Phi\mright)
	- *(V \circ \Phi)
\ea
for all $(\Phi, A) \in \mathfrak{M}_E(M;N)$, where $*$ is the Hodge star operator with respect to $\eta$.
\end{definitions}

A short summary:

\begin{corollaries}{Standard theory as action Lie algebroid, as motivated in \cite{CurvedYMH}}{StandardTheory}
Let $M$ be a spacetime with a spacetime metric $\eta$, $N = W$ be a vector space, equipped with a Riemannian metric $g$ on $\mathrm{T}W \cong W \times W$ canonically induced by a scalar product on $W$, and $E = N \times \mathfrak{g}$ an action Lie algebroid for a Lie algebra $\mathfrak{g}$, equipped with its canonical flat connection $\nabla$ and a fibre metric $\kappa$ which constantly extends a scalar product on $\mathfrak{g}$. The $\mathfrak{g}$-action $\gamma$ is induced by a Lie algebra representation $\psi: \mathfrak{g} \to \mathrm{End}(W)$, and we have a potential $V \in C^\infty(W)$.

Then Def.~\ref{def:EichbosonenUndFeldstaerke}, \ref{def:MinimalCoupling} and \ref{def:CurvedYMHLagrangian} are the same as for the standard formulation of gauge theory as introduced in Chapter \ref{ClassicGaugeTheory}.
\end{corollaries}

\begin{proof}[Proof of Cor.~\ref{cor:StandardTheory}]
\leavevmode\newline
By construction; also recall the remarks of Def.~\ref{def:EichbosonenUndFeldstaerke} and \ref{def:MinimalCoupling}. For $\kappa$ take a constant frame $\mleft( e_a \mright)_a$ of $E$ such that $\mleft( \Phi^*e_a \mright)_a$ trivializes $\Phi^*E \cong M \times \mathfrak{g}$ for all $\Phi \in C^\infty(M;N)$ and $\mleft( \Phi^*e_a \mright)_a$ is also a constant frame, and denote the scalar product on $\mathfrak{g}$ by $\widetilde{\kappa}$. Then observe
\bas
(\Phi^*\kappa)(\Phi^*e_a, \Phi^*e_b)
&=
\Phi^*\bigl( \kappa(e_a, e_b) \bigr)
=
\Phi^* \underbrace{(\widetilde{\kappa}(e_a, e_b))}_{\mathclap{= \text{const.}}}
=
\widetilde{\kappa}(e_a, e_b),
\eas
hence, ${}^*\kappa = \iota(\kappa) = \kappa$ a constant extension of $\widetilde{\kappa}$; similarly for $g$. Thence, we arrive at the standard definition of the Lagrangian, using the remarks of Def.~\ref{def:EichbosonenUndFeldstaerke} and \ref{def:MinimalCoupling},
\bas
\mathfrak{L}_{\mathrm{YMH}}(\Phi, A)
&=
- \frac{1}{2} \widetilde{\kappa} \mleft(F(\Phi, A) \stackrel{\wedge}{,} *F(\Phi, A)\mright)
	+ \widetilde{g} \mleft(\mathfrak{D}^A \Phi \stackrel{\wedge}{,} *\mathfrak{D}^A\Phi\mright)
	- *(V \circ \Phi),
\eas
where $\widetilde{g}$ is the scalar product on $W$; recall Def.~\ref{def:ClassicYMHLagrangian}.
\end{proof}

Now let us finally turn to the infinitesimal gauge transformation.

\section{Infinitesimal gauge transformations}\label{InfinitesimalGaugeTransformation}

\subsection{Infinitesimal gauge transformation of the Higgs field}

We will now do precisely the same, but more general, as in Section \ref{NewInfGaugeTrafoTrafos}. Infinitesimal gauge transformations of a functional $L \in \mathcal{F}^k(M; {}^*V)$ ($k \in \mathbb{N}_0$ and $V \to N$ a vector bundle) are derivatives along certain directions in $\mathfrak{M}_E(M; N)$, while the components of these directions as vector field will be identified with the infinitesimal gauge transformations of the corresponding fields, $\Phi$ and $A$. We want that these transformations satisfy the Leibniz rule, and we want to study the commutator of such two transformations. In order to do that easily, we require that such a derivative keeps a functional vertical, \textit{i.e.}~$\delta L \in \mathcal{F}^k(M; {}^*V)$, where $\delta$ denotes such a transformation, and for this we will use connections, especially ones induced by a Lie algebroid connection on $V$ itself. We will do that by using pull-backs, especially using Cor.~\ref{cor:VeryGeneralPullbackConnection}. That is, since functionals are forms on $M \times \mathfrak{M}_E$, we want to make the pullback along $\mathrm{ev}$, while avoiding the issue of lifting the evaluation map to a suitable vector bundle morphism by restricting to certain vector fields on $\mathfrak{M}_E$ satisfying the condition given in Cor.~\ref{cor:VeryGeneralPullbackConnection}; we will see that this will precisely give the formula of the infinitesimal gauge transformation of the Higgs field.

The arguments are precisely the same as in the discussion before Def.~\ref{def:ClassicGaugeTrafoOfHiggs}. Hence, we start now with a similar definition, but, as we also mentioned in the discussion of Def.~\ref{def:ClassicGaugeTrafoOfHiggs}, the Lie algebroid used for the mentioned Lie algebroid connection on $V$ does not need to be the same Lie algebroid used in the definition of $\mathfrak{M}_E(M;N)$. This is why there is now a second Lie algebroid $B$ over $N$, equipped with a Lie algebroid connection ${}^B\nabla$ on $V$; but when we turn to the infinitesimal gauge transformation of quantities like the minimal coupling, it is useful to have $E=B$, which we are then going to assume. However, one may want to do a similar construction using a typical vector bundle connection on $V$ which implies $B=\mathrm{T}N$; in order to allow those type of constructions we keep it that general for the basic definitions. Also recall Prop.~\ref{prop:TangentSpaceOfSpaceOfFields}.

\begin{definitions}{Vector fields along Lie algebroid paths}{VectorFieldAlongEPaths}
Let $M, N$ be two smooth manifolds and $\mleft(E, \rho_E, \mleft[ \cdot,\cdot \mright]_E \mright)$, $\mleft(B, \rho_B, \mleft[ \cdot,\cdot \mright]_B \mright)$ two Lie algebroids over $N$. For $(\Phi, A) \in \mathfrak{M}_E(M; N)$ we define $\mathrm{T}^B_{(\Phi,A)}\mathfrak{M}_E(M; N)$ as a subspace of $\mathrm{T}_{(\Phi,A)}\mathfrak{M}_E(M; N)$ by
\ba
\mathrm{T}^B_{(\Phi,A)}\mathfrak{M}_E(M; N)
&\coloneqq
\left\{ (\mathcal{v}, \mathcal{a}) \in \mathrm{T}_{(\Phi,A)}\bigl(\mathfrak{M}_E(M; N)\bigr)
~\middle|~
\exists \epsilon \in \Gamma(\Phi^*B):~
\mathcal{v} = - (\Phi^*\rho_B)(\epsilon)
\right\}.
\ea
The set of sections with values in these subspaces, called the set of \textbf{vector fields along $B$-paths}, is denoted by $\gls{XBM}$.
\end{definitions}

\begin{remark}\label{NotASubalgebraXB}
\leavevmode\newline
As images of the pullback of the anchor, it is clear that $\mathrm{T}^B_{(\Phi,A)}\bigl(\mathfrak{M}_E(M; N)\bigr)$ and $\mathfrak{X}^B\bigl(\mathfrak{M}_E(M; N)\bigr)$ are subspaces of $\mathrm{T}_{(\Phi,A)}\bigl(\mathfrak{M}_E(M; N)\bigr)$ and $\mathfrak{X}\bigl(\mathfrak{M}_E(M; N)\bigr)$, respectively.

For all $\Psi \in \mathfrak{X}^B(\mathfrak{M})$ there is by definition then an $\varepsilon \in \mathcal{F}^0_E(M; {}^*B)$ such that 
\ba\label{GaugeTrafoVektor}
\Psi
&=
\mleft( -({}^*\rho_B )(\varepsilon), \mathfrak{a} \mright)
\ea
where $({}^*\rho_B)(\varepsilon)$ is an element of $\mathcal{F}^0_E(M; {}^*\mathrm{T}N)$ given by $\mathfrak{M}_E(M; N) \ni (\Phi, A) \mapsto (\Phi^*\rho_B)(\varepsilon(\Phi, A))$, and $\mathfrak{a}$ is a map defined on $\mathfrak{M}_E(M; N)$ such that $\Psi|_{(\Phi,A)}$ is a tangent vector for all $(\Phi, A) \in \mathfrak{M}_E(M; N)$ as in Prop.~\ref{prop:TangentSpaceOfSpaceOfFields}. We will study $\mathfrak{a}$ in more detail later, but now it will not be important. We will write $\Psi \eqqcolon \Psi_\varepsilon$ to emphasize the relationship with an $\varepsilon \in \mathcal{F}^0_E(M; {}^*B)$. As in Remark \ref{PsiEpsilonDieErste}, for a given $\varepsilon$ there can be several $\Psi_\varepsilon$ as long as we do not fix $\mathfrak{a}$. Moreover, since $\varepsilon \in \mathcal{F}^0_E(M; {}^*B)$ we cannot expect in general that $\mathfrak{X}^B\bigl(\mathfrak{M}_E(M; N)\bigr)$ is a subalgebra of $\mathfrak{X}\bigl(\mathfrak{M}_E(M; N)\bigr)$. One may be able to show that if just allowing $\varepsilon = {}^*b$ ($b \in \Gamma(B)$), but since those more general $\varepsilon$ can have  very general dependencies on $(\Phi,A) \in \mathfrak{M}_E(M;N)$ one cannot expect a sub-algebraic behaviour at this point. We will come back to this after we will have defined the infinitesimal gauge transformation for the field of gauge bosons.
\end{remark}

By construction, the flows of those vector fields carry the structure of Lie algebroid paths which will allow us to do pullbacks of connections along these flows in order to define certain connections on functionals.

\begin{corollaries}{Flows of $\mathfrak{X}^B\bigl(\mathfrak{M}_E(M; N)\bigr)$}{ReasonWhyVectorFieldAlongAnBPath}
Let $M, N$ be two smooth manifolds and $\mleft(E, \rho_E, \mleft[ \cdot,\cdot \mright]_E \mright)$, $\mleft(B, \rho_B, \mleft[ \cdot,\cdot \mright]_B \mright)$ two Lie algebroids over $N$. For a $\Psi \in \mathfrak{X}^B(\mathfrak{M}_E(M; N))$ we denote its flow by $\gamma = (\Phi, A): I \to \mathfrak{M}_E(M; N)$, $t \mapsto \gamma(t) = (\Phi_t, A_t) \in \mathfrak{M}_E(M; N)$ through a fixed point $(\Phi_0, A_0) \in \mathfrak{M}_E(M; N)$ at $t=0$, where $I$ is an open interval of $\mathbb{R}$ containing 0, and we write $\Psi|_{\gamma(t)} = \mleft( - (\Phi_t^*\rho_B)(\epsilon_t), \mathcal{a}_t \mright) \in \mathrm{T}^B_{(\Phi_t, A_t)}\mathfrak{M}_E(M; N)$, where $\epsilon_t \in \Gamma(\Phi_t^*B)$ and $\mathcal{a}_t \in \Omega^1\mleft(M; \epsilon_t^*\mathrm{T}E\mright)$ (recall Prop.~\ref{prop:TangentSpaceOfSpaceOfFields}).

Then $-\epsilon(p) \coloneqq \mleft[ t\mapsto -\epsilon_t|_p\mright]$, viewed as a curve $I \to B$, is a $B$-path with base path $\Phi(p) \coloneqq \mleft[ t \mapsto \Phi_t(p) \mright]$ for all $p \in M$.
\end{corollaries}

\begin{proof}
\leavevmode\newline
For $p \in M$ fixed, it is clear by definition that the base path of $-\epsilon(p)$ is given by $\Phi(p)$ since $\epsilon_t|_p \in B_{\Phi_t(p)}$ for all $t \in I$, where $B_{\Phi_t(p)}$ is the fibre of $B$ at $\Phi_t(p)$. By definition of flows we have
\bas
\mleft.\frac{\mathrm{d}}{\mathrm{d}t}\mright|_t \gamma
&=
\Psi|_{\gamma(t)}
\eas
for all $t \in I$, and, so,
\bas
\mleft. \bigl((\Phi(p))^*\rho_B\bigr)\bigl(-\epsilon(p)\bigr)\mright|_t
&=
-\mleft. (\Phi_t^*\rho_B)(\epsilon_t)\mright|_p
= 
\mleft.\frac{\mathrm{d}}{\mathrm{d}t}\mright|_t \bigl(\Phi(p)\bigr),
\eas
which proves the claim.
\end{proof}

As in Section \ref{NewInfGaugeTrafoTrafos}, the first component of these vector fields also define the infinitesimal gauge transformation of the Higgs field.

\begin{definitions}{Infinitesimal gauge transformation of $\Phi$}{VariationenOfAundPhi}
Let $M, N$ be two smooth manifolds, $\mleft(E, \rho_E, \mleft[ \cdot,\cdot \mright]_E \mright)$, $\mleft(B, \rho_B, \mleft[ \cdot,\cdot \mright]_B \mright)$ two Lie algebroids over $N$, and $\varepsilon \in \mathcal{F}^0_E(M; {}^*B)$. For a $(\Phi, A) \in \mathfrak{M}_E(M; N)$ we define the \textbf{infinitesimal gauge transformation $\delta^B_{\varepsilon(\Phi, A)} \Phi$ of $\Phi$ along $\varepsilon(\Phi, A)$} as an element of $\Gamma (\Phi^*\mathrm{T}N)$ by
\ba\label{EqVariationOfHiggsField}
%\mleft( \gls{1delta0epsilon} \mathds{1}_{C^\infty(M;N)} \mright)(\Phi, A)
%&\coloneqq
\delta^B_{\varepsilon(\Phi, A)} \Phi
&\coloneqq
\bigl( -\mleft({}^*\rho_B\mright)(\varepsilon) \bigr) (\Phi, A)
=
- \mleft( \Phi^* \rho_B \mright)\bigl(\varepsilon(\Phi, A)\bigr),
\ea
shortly denoted as $\delta^B_\varepsilon \Phi \coloneqq - \mleft({}^*\rho_B\mright)(\varepsilon) \in \mathcal{F}^0_E(M; {}^*\mathrm{T}N)$.

In the case of $E=B$ we just write $\delta_\varepsilon \Phi \coloneqq - ({}^*\rho)(\varepsilon)$.
\end{definitions}

\begin{remark}\label{RemUeberVariationVonHiggs}
\leavevmode\newline
\indent $\bullet$ 
%For Eq.~\eqref{EqVariationOfHiggsField} recall Def.~\ref{def:VariationsInVariationalCalc}, \textit{i.e.}
%\bas
%\mleft(\delta_{\widetilde{\Phi}_\varepsilon} \Phi\mright)_p
%&=
%\mleft.\frac{\mathrm{d}}{\mathrm{d}t}\mright|_{t=0}\mleft[ t \mapsto \widetilde{\Phi}_{\varepsilon, t}(p) \mright]
%\stackrel{\text{Def.~\ref{def:EPaths}}}{=}
%\rho_{\Phi(p)}\mleft( \Psi_{\varepsilon, t=0}(p) \mright)
%\stackrel{\text{Eq.~\ref{EPfadWertBeiTGleich0}}}{=}
%- \mleft.\mleft( \Phi^* \rho \mright)(\varepsilon)\mright|_p
%\eas
%for all $p \in M$, where we used the abbreviated notation. 
Eq.~\eqref{EqVariationOfHiggsField} is also a generalization of a similar equation for a gauge transformation given in \cite[paragraph before Equation (10); we have a different sign in $\varepsilon$]{CurvedYMH}.
%
%$\bullet$ In the standard setting the infinitesimal gauge transformations are parametrised by sections $\epsilon$ of $M \times \mathfrak{g}$ for a given Lie algebra $\mathfrak{g}$. When $E = N \times \mathfrak{g}$ is an action Lie algebroid, we then take any constant global frame $\mleft( e_a \mright)_a$ of $E$, such that we have a trivialisation of $\Phi^*E \cong M \times \mathfrak{g}$ by $\mleft( \Phi^*e_a \mright)_a$. Thus, $\epsilon = \epsilon^a ~ \Phi^*e_a \in \Gamma(\Phi^*E)$, and with that notation there is a natural (frame-dependent) interpretation of $\epsilon$ as an element of $\mathcal{F}^0_E(M;{}^*E)$ given by $\epsilon = \epsilon^a ~ \Phi^*e_a \mapsto \varepsilon \coloneqq \epsilon^a ~ {}^*e_a \in \mathcal{F}^0_E(M;{}^*E)$. That identification is surely not surjective, so, hereby we see that $\mathcal{F}^0_E(M;{}^*E)$ is a richer space than $C^\infty(M;\mathfrak{g})$, and, thus, we are going to achieve a gauge invariance with respect to a bigger space, even when we restrict ourselves to the standard setting. 

$\bullet$ Finally let us observe why Eq.~\eqref{EqVariationOfHiggsField} recovers the standard formula of the infinitesimal gauge transformation of $\Phi$, Def.~\ref{def:ClassicTrafos}. As usual, use the setting as in Cor.~\ref{cor:StandardTheory}, \textit{i.e.}~let $W$ be a vector space and $N = W$ such that $\Phi \in C^\infty(M;W)$, and $E = N \times \mathfrak{g}$ an action Lie algebroid for a Lie algebra $\mathfrak{g}$ whose Lie algebra action $\gamma$ is induced by a Lie algebra representation $\psi: \mathfrak{g} \to \mathrm{End}(W)$. Also $E=B$. Then we can simply use Prop.~\ref{prop:LieRepAndLieAct}, using $\epsilon \coloneqq \varepsilon(\Phi, A)$, to get
\bas
\mleft(\delta_\varepsilon \Phi\mright)(p)
&=
- \epsilon^a(p) ~ \rho_{\Phi(p)}(e_a)
=
- \epsilon^a(p) ~ \gamma(e_a)_{\Phi(p)}
=
\epsilon^a(p)  ~ \psi(e_a)\bigl(\Phi(p)\bigr)
=
\psi\mleft(\epsilon_p\mright)\bigl(\Phi(p)\bigr)
\eas
for all $p \in M$ and $\epsilon \in \Gamma(\Phi^*E)$ viewed as an element of $C^\infty(M; \mathfrak{g})$, where $\mleft( e_a \mright)_a$ is a frame of constant sections. This is precisely the standard formula.
\end{remark}

There is a relationship similar to Cor.~\ref{cor:VeryGeneralPullbackConnection}, which summarizes the whole motivation of our construction; also recall Remark \ref{rem:CommutingDiagramOfPullbacks}.

\begin{corollaries}{Infinitesimal gauge transformation as condition for allowing pullbacks}{CoolesCommutingDiagramForHiggsTrafosStuff}
Let $M, N$ be two smooth manifolds and $\mleft(E, \rho_E, \mleft[ \cdot,\cdot \mright]_E \mright)$, $\mleft(B, \rho_B, \mleft[ \cdot,\cdot \mright]_B \mright)$ two Lie algebroids over $N$, and $\varepsilon \in \mathcal{F}^0_E(M; {}^*B)$. Then $\Psi \in \mathfrak{X}\bigl(\mathfrak{M}_E(M;N)\bigr)$ is an element of $\mathfrak{X}^B\bigl(\mathfrak{M}_E(M; N)\bigr)$ if and only if there is an $\varepsilon \in \mathcal{F}^0_E(M; {}^*B)$ such that the following diagram commutes
\begin{center}
	\begin{tikzcd}
		M \times \mathfrak{M}_E(M;N) \arrow{r}{-\varepsilon} \arrow{d}{(0, \Psi)}	& B \arrow{d}{\rho_B} 
		\\
		\mathrm{T}\bigl( M \times \mathfrak{M}_E(M;N) \bigr) \arrow{r}{\mathrm{Dev}} & \mathrm{T}N
	\end{tikzcd}
\end{center}
that is
\ba
\mathrm{Dev} \circ (0, \Psi)
&=
-\rho_B \circ \varepsilon,
\ea
where $(0, \Psi) \in \mathfrak{X}(M) \times \mathfrak{X}\bigl(\mathfrak{M}_E(M; N)\bigr)$ is the canonical embedding of $\Psi$ as a vector field on $M \times \mathfrak{M}_E(M; N)$.
\end{corollaries}

\begin{proof}
\leavevmode\newline
That is by construction. Let $\gamma = (\Phi, A): I \to \mathfrak{M}_E(M;N), t \mapsto \gamma(t)= (\Phi_t, A_t)$ ($I \subset \mathbb{R}$ an open interval containing 0) be the flow of $\Psi$ through $(\Phi_0, A_0) \in \mathfrak{M}_E(M;N)$ at $t=0$, as \textit{e.g.}~in Cor.~\ref{cor:ReasonWhyVectorFieldAlongAnBPath}. Then the local flow of $(0, \Psi)$ through $(p, \Phi_0, A_0) \in M \times \mathfrak{M}_E(M;N)$ is given by $(p, \Phi, A)$. Thus,
\bas
\mathrm{D}_{(p, \Phi_0, A_0)}\mathrm{ev}(0,\Psi)
&=
\mleft. \frac{\mathrm{d}}{\mathrm{d}t} \mright|_{t=0}\mleft(
	\mathrm{ev}(p, \Phi, A)
\mright)
=
\mleft. \frac{\mathrm{d}}{\mathrm{d}t} \mright|_{t=0}\mleft[
	t \mapsto \Phi_t(p)
\mright]
=
\mleft.\mleft(\mleft.\Psi^{(\Phi)}\mright|_{(\Phi_0, A_0)}\mright)\mright|_p
\in \mathrm{T}_{\Phi_0(p)}N,
\eas
where $\Psi^{(\Phi)}$ is the first component of $\Psi$, for this also recall Prop.~\ref{prop:TangentSpaceOfSpaceOfFields}. The commutation of the diagram is then equivalent to say that there is an $\varepsilon \in \mathcal{F}^0_E(M; {}^*B)$
\bas
\Psi^{(\Phi)}
&=
- ({}^*\rho_B)(\varepsilon),
\eas
which is precisely the definition for $\mathfrak{X}^B\bigl(\mathfrak{M}_E(M;N)\bigr)$ of Def.~\ref{def:VectorFieldAlongEPaths}.
\end{proof}

That immediately leads to:

\begin{propositions}{Parametrised variations of functionals}{VariationVonSkalarZeugsEasyPeasy}
Let $M, N$ be two smooth manifolds, $\mleft(E, \rho_E, \mleft[ \cdot,\cdot \mright]_E \mright)$, $\mleft(B, \rho_B, \mleft[ \cdot,\cdot \mright]_B \mright)$ two Lie algebroids over $N$, $V \to N$ a vector bundle, ${}^B\nabla$ a $B$-connection on $V$, and $\Psi_\varepsilon \in\mathfrak{X}^B(\mathfrak{M}_E(M; N))$ for $\varepsilon \in \mathcal{F}^0_E(M; {}^*B)$. Then there is a unique $\mathbb{R}$-linear map $\gls{1delta0Psiepsilon}: \mathcal{F}_E^\bullet(M;{}^*V) \to \mathcal{F}_E^\bullet(M;{}^*V)$ with
\ba\label{PullBackVariation}
\delta_{\Psi_\varepsilon} \mleft( {}^* v \mright)
&=
- {}^*\mleft({}^B\nabla_{\varepsilon} v \mright),
%\\
%\delta_{\Psi_\varepsilon} f
%&=
%\mathcal{L}_
\\
\iota_Y \delta_{\Psi_\varepsilon}
&=
\delta_{\Psi_\varepsilon} \iota_Y
\label{VertauschungMitVerjuengungVonEichtrafo}\\
\delta_{\Psi_\varepsilon}(f \wedge L)
&=
\mathcal{L}_{\Psi_\varepsilon}(f) \wedge L
	+ f \wedge \delta_{\Psi_\varepsilon} (L), \label{LeibnizForGauging}
\ea
for all $Y \in \mathfrak{X}(M)$, $v \in \Gamma(V)$, $L \in \mathcal{F}_E^k(M; {}^*V)$, and $f \in \mathcal{F}^m_E(M)$ ($k, m \in \mathbb{N}_0$), where $\mathcal{F}_E^\bullet(M;{}^*V) \coloneqq \bigoplus_{l\in \mathbb{N}_0} \mathcal{F}^l_E(M; {}^*V)$ while $\delta_{\Psi_\varepsilon}$ keeps a given degree invariant.
\end{propositions}

\begin{remark}
\leavevmode\newline
Since the notation of $\delta_{\Psi_\varepsilon}$ does not emphasize the used connection, we will often roughly write: \textbf{For the functional space $\mathcal{F}^\bullet_E(M;{}^*V)$ let $\delta_{\Psi_\varepsilon}$ be the unique operator of Prop.~\ref{prop:VariationVonSkalarZeugsEasyPeasy}, using ${}^B\nabla$ as a $B$-connection on $V$}, where $\bullet$ denotes an arbitrary degree.
\end{remark}

\begin{proof}[Proof of Prop.~\ref{prop:VariationVonSkalarZeugsEasyPeasy}]
\leavevmode\newline
That is a trivial consequence of Cor.~\ref{cor:CoolesCommutingDiagramForHiggsTrafosStuff} and Cor.~\ref{cor:VeryGeneralPullbackConnection}, that is, we have a unique $\mathbb{R}$-linear operator $\delta_{\Psi_\varepsilon}: \mathcal{F}^0_E(M; {}^*V) \to \mathcal{F}^0_E(M; {}^*V)$ such that
\bas
\delta_{\Psi_\varepsilon}(h s)
&=
\mathcal{L}_{\Psi_\varepsilon}(h) ~ s
	+ h ~ \delta_{\Psi_\varepsilon} s,
\\
\delta_{\Psi_\varepsilon}\underbrace{({}^*v)}_{\mathclap{ = \mathrm{ev}^*v }}
&=
-{}^*\mleft( {}^B\nabla_\varepsilon v \mright)
\eas
for all $s \in \Gamma(\mathrm{ev}^*V) = \mathcal{F}^0_E(M; {}^*V), h \in C^\infty(M \times \mathfrak{M}_E)$, and $v \in \Gamma(V)$. Eq.~\eqref{VertauschungMitVerjuengungVonEichtrafo} and linearity uniquely extends this operator to $\mathcal{F}^\bullet_E(M; {}^*V)$, that is,
\ba\label{DefOfGaugeTrafoWithBookkeep}
\mleft(\delta_{\Psi_\varepsilon} L\mright)(Y_1, \dotsc, Y_k)
&\coloneqq
\delta_{\Psi_\varepsilon}\bigl( L(Y_1, \dotsc, Y_k)\bigr)
\ea
for all $L \in \mathcal{F}^k_E(M; {}^*V)$ and $Y_1, \dotsc, Y_k \in \mathfrak{X}(M)$; similar to Def.~\ref{def:InfinitesimalGaugeTrafoClassicAsConnection} this is well-defined (recall also the remark after Def.~\ref{def:InfinitesimalGaugeTrafoClassicAsConnection}). This is not in violation with the desired Leibniz rule because $\Psi_\varepsilon$ are vector fields on $\mathfrak{M}_E(M;N)$ while $Y_1, \dotsc, Y_k$ are vector fields on $M$, thence, $[\Psi_\varepsilon, Y_i] = 0$ ($i \in \{1, \dotsc, k\}$) in $M \times \mathfrak{M}_E(M;N)$; also recall Eq.~\eqref{SliceOfBIiiigManifold}, the non-trivial information of $L$ is just stored in along $\mathrm{T}M$, then everything follows by the fact that $\mathcal{L}_{\Psi_\varepsilon}$ preserves the factors in $\mathrm{T}M \times \mathrm{T}\mathfrak{M}_E$ and that $\mathfrak{X}(M\times\mathfrak{M}_E)$ is generated by vector fields on $M$ and $\mathfrak{M}_E$ such that one can conclude that $\delta_\varepsilon L$ contains its non-trivial information just along $M$. The Leibniz rule in Eq.~\eqref{LeibnizForGauging} then just follows by this and the Leibniz rule inherited by Cor.~\ref{cor:VeryGeneralPullbackConnection}. In case this is unclear, it follows a more precise explanation why $\delta_\varepsilon L \in \Omega^{k,0}\bigl(M \times \mathfrak{M}_{E}; K\bigr)$ is implied by construction: Usually one would define
\ba\label{OriginalFormula}
\mleft(\delta_\varepsilon L \mright)\mleft( X_1, \dotsc, X_k \mright)
&\coloneqq
\delta_{\Psi_\varepsilon}\bigl(
	\iota(L)\mleft(X_1, \dotsc, X_k\mright)
\bigr)
	- \sum_{i=1}^k L\mleft( X_1, \dotsc, \mathcal{L}_{\Psi_\varepsilon} X_i, \dotsc, X_k \mright)
\ea
for all $X_1, \dotsc, X_k \in \mathfrak{X}\bigl( M \times \mathfrak{M}_{E} \bigr)$. As usual, such a definition leads to $C^\infty\bigl( M \times \mathfrak{M}_{E} \bigr)$-multilinearity, and vector fields of $M \times \mathfrak{M}_{E}$ are generated by vector fields of $M$ and $\mathfrak{M}_{E}$, such that we can restrict ourselves to vector fields of $M$ and $\mathfrak{M}_{E}$. If for example $X_1 = \Psi \in \mathfrak{X}\mleft(\mathfrak{M}_{E} \mright) \subset \mathfrak{X}\bigl( M \times \mathfrak{M}_{E} \bigr)$, then
\bas
\iota(L)\mleft( \Psi, X_2, \dotsc, X_k \mright)
&=
0
\eas
and
\bas
\sum_{i=1}^k L\mleft( X_1, \dotsc, \mathcal{L}_{\Psi_\varepsilon} X_i, \dotsc, X_k \mright)
&=
L( \mathcal{L}_{\Psi_\varepsilon} \Psi, X_2 \dotsc, X_k )
	+ \sum_{i=2}^k \underbrace{L\mleft( \Psi, X_2, \dotsc, \mathcal{L}_{\Psi_\varepsilon} X_i, \dotsc, X_k\mright)}_{=0}
\\
&=
L\bigl( \underbrace{[\Psi_\varepsilon, \Psi]}_{ \mathclap{\in ~ \mathfrak{X}\mleft( \mathfrak{M}_{E} \mright)} }, X_2 \dotsc, X_k \bigr)
\\
&=
0,
\eas
using $L \in \Omega^{k,0}\bigl(M \times \mathfrak{M}_{E}; K\bigr)$ and $\Psi_\varepsilon \in \mathfrak{X}\bigl( \mathfrak{M}_{E} \bigr)$. On the other hand if $X_1 = Y \in \mathfrak{X}(M)$, then 
\bas
L( \mathcal{L}_{\Psi_\varepsilon} Y, X_2 \dotsc, X_k )
&=
L\bigl( \underbrace{[\Psi_\varepsilon, Y]}_{ \mathclap{= 0} }, X_2 \dotsc, X_k \bigr)
=
0.
\eas
Using these relations, we can conclude that the non-trivial information of Def.~\eqref{OriginalFormula} is encoded completely on $\mathfrak{X}(M)$ as a $C^\infty(M)$-module,\footnote{Observe that $\mathcal{L}_{\Psi_\varepsilon}(f) = 0$ for all $f \in C^\infty(M)$.} as given in Def.~\eqref{DefOfGaugeTrafoWithBookkeep}. Therefore one can use Def.~\eqref{DefOfGaugeTrafoWithBookkeep} instead and canonically/trivially extend this definition to the "full" form.

Alternatively, use the flows given by Cor.~\ref{cor:ReasonWhyVectorFieldAlongAnBPath} and prove it in the same manner as in Prop.~\ref{prop:ClassicFunctionDerivativesAlongPsiEpsilon} (in combination with Def.~\ref{def:InfinitesimalGaugeTrafoClassicAsConnection}).
\end{proof}

\begin{remark}\label{RemLeibnizeRegelaufProdukteWeshalbEConnectionNichtWichtigIst}
\leavevmode\newline
\indent $\bullet$ Given by Remark \ref{JustLieDerivativeForGeneralPullbackAndlineBundle}, for $V = N \times \mathbb{R}$ we always take the canonical flat $B$-connection, \textit{i.e.}~the canonical flat vector bundle connection $\nabla^0 = \mathrm{d}$ and then ${}^B\nabla \coloneqq \nabla^0_{\rho_B}$ such that
\bas
\delta_{\Psi_\varepsilon}
&=
\mathcal{L}_{\Psi_\varepsilon}.
\eas
Thus, 
\ba
\delta_{\Psi_\varepsilon} \mathrm{d}
&=
\mathcal{L}_{\Psi_\varepsilon} \mathrm{d}
=
\mathrm{d} \mathcal{L}_{\Psi_\varepsilon}
=
\mathrm{d} \delta_{\Psi_\varepsilon}, \label{eqVariationVertauschtMitDifferential}
\ea
since coordinates on $\mathfrak{M}_E(M; N)$ and $M$ are independent; recall the end of Remark \ref{rem:Bigrading} for this. The Leibniz rule for $\delta_{\Psi_\varepsilon}$ can be then rewritten to
\ba
\delta_{\Psi_\varepsilon}(f \wedge L)
&=
\delta_{\Psi_\varepsilon}(f) \wedge L
	+ f \wedge \delta_{\Psi_\varepsilon} (L).
\ea

$\bullet$ For dual bundles $V^*$ we canonically take the dual connection to ${}^B\nabla$ in order to have Leibniz rules as usual. That also means the following (still keeping the same notation): Let $L \in \mathcal{F}^k_E(M; {}^*V)$ and $T \in \mathcal{F}^0_E(M; {}^*(V^*))$, then in a frame $\mleft( e_a \mright)_a$ of $V$ and $\mleft( f^a \mright)_a$ of $V^*$, $f^b(e_a) = \delta^b_a$, we locally write $L = L^a \otimes {}^*e_a$ and $T = T_b \cdot {}^*f^b$, where $L^a \in \mathcal{F}^k_E(M)$ and $T_b \in \mathcal{F}^0_E(M)$. Then with these conventions, including the previous bullet point,
\ba
\delta_{\Psi_\varepsilon} (T(L))
&=
\delta_{\Psi_\varepsilon} \underbrace{\mleft(
	T_a L^a
\mright)}_{\in \mathcal{F}^k_E(M)}
=
\mathcal{L}_{\Psi_\varepsilon} \mleft(
	T_a L^a
\mright)
=
\mathcal{L}_{\Psi_\varepsilon} (T_a) ~ L^a
	+ T_a ~ \mathcal{L}_{\Psi_\varepsilon}(L^a),
\ea
hence, one achieves an independence of the chosen ${}^B\nabla$. This emphasizes what we expect, that we can freely choose the chosen connections for the variations of the tensors involved in contractions, only the variations of their components matter in such situations; this is important for the gauge invariance of the Lagrangian later. As we have discussed at the end of Section \ref{NewInfGaugeTrafoTrafos}, we are going to take the basic connection to define $\delta_{\Psi_\varepsilon}$ for quantities like the field strength, which will not be related to the canonical flat connection when imposing the classical theory; also recall Thm.~\ref{thm:RecoverOfClassicInfgGaugeTrafo}. That is possible because the infinitesimal gauge transformation of the Lagrangian stays untouched by this, it is always just the Lie derivative. The connections only get important in explicit calculations when applying the Leibniz rule as in
\bas
\delta_{\Psi_\varepsilon} (T(L))
&=
\mleft(\delta_{\Psi_\varepsilon} T\mright)(L)
	+ T\mleft(\delta_{\Psi_\varepsilon} L\mright),
\eas
but the result will of course not change. Henceforth, the essential work is in defining $\Psi_\varepsilon$; we did not yet define the infinitesimal gauge transformation of $A$.
%
%$\bullet$ Let us shortly motivate what one does in the standard setting. In the standard setting $V$ will be a trivial vector bundle over $N$. Especially looking at Eq.~\eqref{DefVonUnsererCoolenEichung}, we want that the used $\mathrm{D}/\mathrm{d}t$ is the typical $\mathrm{d}/\mathrm{d}t$, that is, we then have $B = \mathrm{T}N$ and ${}^B\nabla$ is the canonical flat connection of $V$. Fix a global parallel (=constant) frame of $V$, $\mleft(e_a\mright)_a$, \textit{i.e.}~$\nabla e_a = 0$ such that 
%\bas
%(\delta_{\Psi_\varepsilon} {}^*e_a)(\Phi, A)
%&= 
%(\Phi^*\nabla)_{\varepsilon(\Phi, A)} (\Phi^*e_a) 
%= 
%\Phi^!\mleft(\nabla_{\varepsilon(\Phi, A)} e_a\mright) 
%= 0
%\eas
%for all $(\Phi, A ) \in \mathfrak{M}_E(M; N)$. We then get for all $L = L^a \otimes e_a \in \mathcal{F}^k_E(M; {}^*V)$ ($k \in \mathbb{N}_0$)
%\bas
%\delta_{\Psi_\varepsilon} L
%&=
%\delta_{\Psi_\varepsilon} L^a \otimes {}^*e_a,
%\eas
%thus, it is equivalent to a variation in each component with respect to a global parallel frame of the underlying connection (the pullback of $\nabla$), which is precisely what one normally does in the standard framework of gauge theory; see more about this later when we actually introduced more about the physical setting.
\end{remark}

This recovers the classical idea of infinitesimal gauge transformation, \textit{i.e.}~it is a Lie derivative of components with respect to flat connections; also recall Thm.~\ref{thm:RecoverOfClassicInfgGaugeTrafo}.

\begin{theorems}{Parametrised variations in the flat case}{NewFormulaRecoversOldGaugeTrafoYay}
Let $M, N$ be two smooth manifolds, $\mleft(E, \rho_E, \mleft[ \cdot,\cdot \mright]_E \mright)$, $\mleft(B, \rho_B, \mleft[ \cdot,\cdot \mright]_B \mright)$ two Lie algebroids over $N$, and $V \to N$ a trivial vector bundle. Also let $\nabla$ be the canonical flat connection of $V$, $\Psi_\varepsilon \in \mathfrak{X}^B\bigl( \mathfrak{M}_E(M;N) \bigr)$ for an $\varepsilon \in \mathcal{F}^0_E(M; {}^*B)$ and for $\mathcal{F}^\bullet_E(M; {}^*V)$ let $\delta_{\Psi_\varepsilon}$ be the unique operator of Prop.~\ref{prop:VariationVonSkalarZeugsEasyPeasy}, using ${}^B\nabla \coloneqq \nabla_{\rho_B}$ as a $B$-connection on $V$.

Then we have
\ba
\delta_{\Psi_\varepsilon} L
&=
\mleft(\mathcal{L}_{\Psi_\varepsilon}L^a\mright) \otimes {}^*e_a
\ea
for all $L \in \mathcal{F}^\bullet_E(M; {}^*V)$, where $\mleft( e_a \mright)_a$ is a global constant frame of $V$.
\end{theorems}

\begin{proof}
\leavevmode\newline
That is basically the same proof as in Thm.~\ref{thm:RecoverOfClassicInfgGaugeTrafo}. Take a global constant frame $\mleft( e_a \mright)_a$ of $V$, then
\bas
\nabla e_a &= 0,
\eas
and therefore
\bas
(\Phi^*\nabla)(\Phi^*e_a)
&=
\Phi^!(\nabla e_a)
=
0
\eas
for all $\Phi \in C^\infty(M;N)$. Hence, $({}^*\nabla)({}^*e_a) = {}^!(\nabla e_a) = 0$, such that, using the Leibniz rule,
\bas
\delta_{\Psi_\varepsilon} L
&=
\mleft(\mathcal{L}_{\Psi_\varepsilon}L^a\mright) \otimes {}^*e_a.
\eas
\end{proof}

As argued before, we can write $\Psi_\varepsilon = \mleft( -({}^*\rho_B )(\varepsilon), \mathfrak{a} \mright)$ (Eq.~\eqref{GaugeTrafoVektor}) and we want to identify its first and second component as the gauge transformation of $\Phi$ and $A$, respectively. Right now $\mathfrak{a}$ is just fixed by Prop.~\ref{prop:TangentSpaceOfSpaceOfFields} such that it is very arbitrary; as in the standard setting of gauge theory, we want that it is parametrised, which will be by $\varepsilon$, too.

\subsection{Infinitesimal gauge transformation of the field of gauge bosons}
Recall Prop.~\ref{prop:VerticalBundleOfFracM} and its discussion, the tangent vector along the "$A$-direction" is only in the same space as $A$ if the first component is zero, which is $\delta_\varepsilon \Phi$ because we want to think of $\delta_\varepsilon A$ as the second component of $\Psi_\varepsilon$. We cannot expect this to be zero in general, not even in the standard setting because a Lie algebra representation will not act trivially on $\Phi$, as we already discussed after Prop.~\ref{prop:VerticalBundleOfFracM}. However, as in the standard formulation, we want to formulate the gauge transformation of $A$ in such a way that it is somewhat in the same space; we will achieve this by fixing a connection on $E$ as we already did for functionals when defining $\delta_{\Psi_\varepsilon}$. Since $A$ has values in $\Phi^*E$, its image is also now affected by the gauge transformation of $\Phi$, this is why we can do something similar as for functionals; also recall Remark \ref{rem:BosonsAsFunctionalies}. 

One may argue that an involved horizontal projection in the definition for $\delta_\varepsilon A$ may lead to lost information about that object, especially important when one may want to integrate this theory, while we will not need the "full formula" for $\delta_\varepsilon A$ for the infinitesimal gauge transformation of the Lagrangian as we already argued earlier. However, since $A$ has values in $\Phi^*E$, one expects that $\delta_\varepsilon A$ encodes partially what $\delta_\varepsilon \Phi$ already encodes. Prop.~\ref{prop:TangentSpaceOfSpaceOfFields} shows us that $\delta_\varepsilon A$ is still somewhat vertical, because it is a form with values in the vector bundle $\mathrm{T}E \to \mathrm{T}N$, whose linear structure is essentially given by the vertical (prolonged) structure; $\delta_\varepsilon A$ is just shifted "horizontally" by $\delta_\varepsilon \Phi$ due to Eq.~\eqref{HorizontalCompOfDeltaA} and Prop.~\ref{prop:VerticalBundleOfFracM}. Henceforth, our idea is to shape the horizontal projection in such a way that we only "loose" the information we already know by $\delta_\varepsilon \Phi$; making use of Prop.~\ref{prop:TangentSpaceOfSpaceOfFields}.

Let us make it precise: Let us first look at a local trivialization of the Lie algebroid $E \stackrel{\pi}{\to} N$ is trivial. That is let us have base coordinates $\mleft( x^i \mright)_i$ of $N$, lifted to $E$ by $\pi^*x^i$, but we will omit all the given pullbacks in the notation now in the following rough discussion for simplicity; also let $\mleft( y^j \mright)_j$ be fibre coordinates. By Prop.~\ref{prop:TangentSpaceOfSpaceOfFields}, $\delta_\varepsilon A$ should be, for a given $(\Phi, A) \in \mathfrak{M}_E$, a form on $M$ with values in $\mathrm{T}E$ (along some function; but again, we omit the pullbacks and point evaluations for simplicity now). Hence, we expect
\bas
\delta_\varepsilon A
&=
\mleft(\delta_\varepsilon A\mright)^i ~ \frac{\partial}{\partial x^i}
	+ \mleft(\delta_\varepsilon A\mright)^j ~ \frac{\partial}{\partial y^j},
\eas
and $\delta_\varepsilon A$ is the second component of $\Psi_\varepsilon = (\delta_\varepsilon \Phi, \delta_\varepsilon A)$, which we used to define $\delta_{\Psi_\varepsilon}$. Again by Prop.~\ref{prop:TangentSpaceOfSpaceOfFields}, also recall Remark \ref{RemarkAboutThatWeStillHaveLinearStructureinDeltaA}, we know that
\bas
\delta_\varepsilon \Phi
&=
\mathrm{D}\pi \bigl((\delta_\varepsilon A)(Y)\bigr)
=
\mleft(\delta_\varepsilon A\mright)^i(Y) ~ \frac{\partial}{\partial x^i}
\eas
for all $Y \in \mathfrak{X}(M)$, where we used that $\partial/\partial y^j$ are vertical vector fields. Given that trivialization, $\partial/\partial x^i$ defines a canonical horizontal distribution. Hence, using that distribution for a horizontal projection, one could define the infinitesimal gauge transformation of $A$ in that trivialization just with $\mleft(\delta_\varepsilon A\mright)^j ~ \frac{\partial}{\partial y^j}$ which can be identified with a form with values in $E$ since $\partial/\partial y^j$ are vertical. While the components we "loose" because of the horizontal projection is something already encoded by $\delta_\varepsilon \Phi$, such that those are easy to reconstruct if one needs the "full formula" of $\delta_\varepsilon A$.

Globally that means we want to define $\delta_\varepsilon A$ as a form with values in $E$ using a Lie algebroid connection on $E$ as we did in Prop.~\ref{prop:VariationVonSkalarZeugsEasyPeasy} in such a way that $\Psi_\varepsilon$ is uniquely given. In order to do that we need to view $A$ as a functional, which is just $\varpi_2$ of Ex.~\ref{ex:ProjectionOntoGaugeBosonies}. So, we impose a formula for $\delta_\varepsilon \varpi_2$ in such a way that it uniquely defines $\Psi_\varepsilon$, and that we can derive the infinitesimal gauge invariance of the Lagrangian as usual.

But how does one fix the infinitesimal gauge transformation of $A$ normally when integrability is not used? One of the arguments in the standard formulation is given by looking at the transformation of the minimal coupling; we will do the same. Let us recall what that argument was: Again, let $N =W$ be a vector space, and $E = N \times \mathfrak{g}$ an action Lie algebroid associated to a Lie algebra $\mathfrak{g}$ whose Lie algebra action is induced by a Lie algebra representation $\psi: \mathfrak{g} \to \mathrm{End}(W)$. Then, for an $\epsilon \in C^\infty(M; \mathfrak{g})$, we have the infinitesimal gauge transformation $\delta_\epsilon \Phi = \psi(\epsilon)(\Phi)$ for all $\Phi \in C^\infty(M;W)$. The minimal coupling is then defined by $\mathfrak{D}^A \Phi = \mathrm{D}\Phi + \psi(A)(\Phi)$, where $A \in \Omega^1(M; \mathfrak{g})$; recall Def.~\ref{def:ClassicMinimalCoupling}. The (infinitesimal) gauge transformation of $A$ is then chosen in such a way that it is an element of $\Omega^1(M; \mathfrak{g})$, and such that one gets for the infinitesimal gauge transformation of the minimal coupling
\ba\label{StandardArgumenFuerDieMinimaleKopplungImBabyFall}
\mleft(\delta_\epsilon \mathfrak{D} \mright)(\Phi, A)
= 
\psi(\epsilon) \mleft( \mathfrak{D}^A \Phi \mright)
\ea
among the category of gauge theories, where $\delta_\epsilon$ denotes again the classical formulation of the infinitesimal gauge transformation as introduced in Chapter \ref{ClassicGaugeTheory}.

In order to provide a similar argument and since the minimal coupling $\mathfrak{D}$ is an element of $\mathcal{F}^1_E(M; {}^*\mathrm{T}N)$, we need to fix a connection on $\mathrm{T}N$ in order to use Prop.~\ref{prop:VariationVonSkalarZeugsEasyPeasy}. We want to use the basic connection, for this recall that for a given connection $\nabla$ on a Lie algebroid $E\to N$ we have the canonical basic connection $\nabla^{\mathrm{bas}}$, Def.~\ref{def:CanonicalBasicConnection},
\bas
\nabla^{\mathrm{bas}}_\mu \nu
&=
\mleft[ \mu, \nu \mright]_E + \nabla_{\rho(\nu)} \mu, \\
\nabla^{\mathrm{bas}}_\mu X
&=
\mleft[ \rho(\mu), X \mright] + \rho \mleft( \nabla_{X} \mu \mright)
\eas
for all $\mu, \nu \in \Gamma(E)$ and $X\in \mathfrak{X}(N)$. The reason why we want to use the basic connection is the following corollary about the recovery of Eq.~\eqref{StandardArgumenFuerDieMinimaleKopplungImBabyFall}.

\begin{corollaries}{Gauge transformation of the minimal coupling in the standard framework}{EichtrafovonDAPHIinClassicIstBabyEinfach}
Let $N=W$ be a vector space, $E = N \times \mathfrak{g}$ be an action Lie algebroid of a Lie algebra $\mathfrak{g}$ whose action is induced by a Lie algebra representation $\psi: \mathfrak{g} \to \mathrm{End}(W)$, $E$ is also equipped with its canonical flat connection $\nabla$. Also let $\Psi_\varepsilon \in \mathfrak{X}^E(\mathfrak{M}_E(M; N))$ for an $\varepsilon \in \mathcal{F}^0_E(M; {}^*E)$ and for the functional space $\mathcal{F}^\bullet_E(M; {}^*\mathrm{T}N)$ let $\delta_{\Psi_\varepsilon}$ be the unique operator of Prop.~\ref{prop:VariationVonSkalarZeugsEasyPeasy}, using $\nabla^{\mathrm{bas}}$ as $E$-connection on $\mathrm{T}N$. Then we have
\ba\label{EichtrafovonDAPHIinClassicIstBabyEinfachDieAequivalenz}
\bigl(\delta_{\Psi_\varepsilon} \mathfrak{D}\bigr)(\Phi, A)
&=
0
&\Leftrightarrow&&
\bigl(\delta_{\Psi_\varepsilon} \mathfrak{D}^\alpha \bigr)(\Phi, A)
&=
\mleft( \psi\bigl(\varepsilon(\Phi, A)\bigr) \mleft( \mathfrak{D}^A \Phi \mright) \mright)^\alpha
\ea
for all $(\Phi, A) \in \mathfrak{M}_E(M; N)$ and $\alpha \in \{1, \dotsc, \mathrm{dim}(W)\}$,
where the components are with respect to global coordinate vector fields $\mleft( \partial_\alpha \mright)_\alpha$ on $W$, and where we used the canonical trivializations $\mathrm{T}W \cong W\times W$ and $\Phi^*\mathrm{T}W \cong M \times W$ such that $\mathfrak{D}^A \Phi$ can be viewed as an element of $\Omega^1(M; W)$.
\end{corollaries}

\begin{proof}
\leavevmode\newline
Let $\mleft( e_a \mright)_a$ be a global and constant frame of $E$ and $\partial_\alpha$ coordinate vector fields on $N$, then we can write $\mathfrak{D} = \mathfrak{D}^\alpha \otimes {}^*\partial_\alpha$, and, thus, by the Leibniz rule and with $\epsilon \coloneqq \varepsilon(\Phi, A)$
\ba
\bigl(\delta_{\Psi_\varepsilon} \mathfrak{D}^\alpha\bigr)(\Phi, A)
	- \bigl( \underbrace{\mleft( \delta_{\Psi_\varepsilon} \mathfrak{D}\mright)}
	_{ \mathclap{ = \delta_{\Psi_\varepsilon} \mleft(\mathfrak{D}^\alpha\mright) \otimes {}^*\partial_\alpha
		+ \mathfrak{D}^\alpha \otimes \delta_{\Psi_\varepsilon} \mleft({}^*\partial_\alpha\mright) } }
	(\Phi, A) \bigr)^\alpha
&=
	- \biggl( \mleft( \mathfrak{D}^A \Phi \mright)^\beta \otimes \underbrace{\mleft(\delta_{\Psi_\varepsilon} \mleft( {}^* \partial_\beta \mright)\mright) (\Phi, A)}_{\mathclap{\stackrel{\text{Prop.~\ref{prop:VariationVonSkalarZeugsEasyPeasy}}}{=} - \Phi^* \mleft( \nabla^{\mathrm{bas}}_\epsilon \partial_\beta\mright)}} \biggr)^\alpha
\nonumber \\
&=
\epsilon^a ~ \Phi^*\mleft( 
	- \partial_\beta\rho_a^\alpha
	+ \rho^\alpha\mleft( \nabla_{\partial_\beta} e_a \mright) 
\mright) ~ \mleft( \mathfrak{D}^A \Phi \mright)^\beta \label{CompsVonDMinimalAlsErstes}
\ea
for all $\alpha$.
Let us write $\partial_\alpha = \partial/\partial w^\alpha$ for some coordinates $\mleft( w^\alpha \mright)_\alpha$ on $W$. Then by Prop.~\ref{prop:LieRepAndLieAct},
\ba\label{eqAbleitungVomAnkerGibtRepraesentierung}
- \partial_\beta\bigl[ w \mapsto \rho_a^\alpha(w) \bigr]
&=
- \partial_\beta\bigl[ w \mapsto \gamma_a^\alpha(w) \bigr]
=
\partial_\beta\mleft[ w \mapsto \bigl(\psi(e_a)(w)\bigr)^\alpha \mright]
=
\bigl( \psi(e_a) \bigr)^\alpha_\beta
\ea
for $w \in W$, because the differential is then just the differential of a matrix vector-product $W \ni w \mapsto \psi(e_a)(w)$. Since $\nabla$ is the canonical flat connection, constant sections are parallel, thus, we get in total
\bas
\mleft(\delta_{\Psi_\varepsilon} \mathfrak{D}^\alpha\mright)(\Phi, A)
	- \bigl( \mleft( \delta_{\Psi_\varepsilon} \mathfrak{D}\mright) (\Phi, A) \bigr)^\alpha
&=
\epsilon^a ~ \Phi^*\underbrace{\bigl( \psi(e_a) \bigr)^\alpha_\beta}_{\mathclap{\text{const.}}} ~ \mleft( \mathfrak{D}^A \Phi \mright)^\beta
=
\mleft( \psi(\epsilon) \mleft( \mathfrak{D}^A \Phi \mright) \mright)^\alpha
\eas
for all $\alpha$, having $\epsilon \in C^\infty(M; \mathfrak{g})$ and $\mathfrak{D}^A \Phi \in \Omega^1(M;W)$. That shows that we have 
\bas
\mleft(\delta_{\Psi_\varepsilon} \mathfrak{D}^\alpha\mright)(\Phi, A)
&=
\mleft( \psi(\epsilon) \mleft( \mathfrak{D}^A \Phi \mright) \mright)^\alpha
\eas
if and only if 
\bas
\delta_{\Psi_\varepsilon} \mathfrak{D}
&=
0.
\eas
\end{proof}

The right equation in the Equivalence \eqref{EichtrafovonDAPHIinClassicIstBabyEinfachDieAequivalenz} describes precisely the components of the expected infinitesimal gauge transformation of the minimal coupling in the standard formulation of gauge theory, and it is no coincidence that this is equivalent to $\delta_{\Psi_\varepsilon} \mathfrak{D} = 0$ when using the basic connection. 
%For the gauge invariance of the Lagrangian we just need to know the infinitesimal gauge transformation of the components due to the contractions in the definition of the Lagrangian, although the total formula looks different than the classical formula when restricting to the standard formulation.

\begin{lemmata}{Metric compatibilities and their imposed symmetries for gauge theory, \cite{CurvedYMH}}{MetricCompsAdInvUndLieAlgebraRepSymm}
Let $N=W$ be a vector space, $E = N \times \mathfrak{g}$ be an action Lie algebroid of a Lie algebra $\mathfrak{g}$ whose action is induced by a Lie algebra representation $\psi: \mathfrak{g} \to \mathrm{End}(W)$, $E$ is also equipped with its canonical flat connection $\nabla$. Also let $\kappa$ be a fibre metric on $E$ which is a constantly extended scalar product $\widetilde{\kappa}$ of $\mathfrak{g}$; similarly, let $g$ be a fibre metric which is a constant extension of a scalar product $\widetilde{g}$ of $W$.

Then we have
\ba
\nabla^{\mathrm{bas}} \kappa = 0
&\Leftrightarrow
\text{$\widetilde{\kappa}$ is $\mathrm{ad}$-invariant},
\\
\nabla^{\mathrm{bas}} g = 0
&\Leftrightarrow
\text{$\widetilde{g}$ is $\psi$-invariant},
\ea
and $\nabla^{\mathrm{bas}}$ on $E$ and $\mathrm{T}N$ are the adjoint and $\psi$ representation, respectively, when restricted on constant sections, \textit{i.e.}
\ba
\nabla^{\mathrm{bas}}_\mu \nu &= \mleft[ \mu, \nu \mright]_{\mathfrak{g}},
\\
\nabla^{\mathrm{bas}}_\mu Y
&=
\psi(\mu)(Y)
\ea
for all constant $\mu, \nu \in \Gamma(E)$ and constant $Y \in \mathrm{T}N \cong W \times W$.
\end{lemmata}

\begin{remark}
\leavevmode\newline
Here we see that the basic connection $\nabla^{\mathrm{bas}}$ replaces the canonical representations arising in the standard formulation of gauge theory. Moreover, we will later see that we need $R_\nabla^{\mathrm{bas}}=0$ to formulate the gauge theory, that implies that $\nabla^{\mathrm{bas}}$ is flat (both), recall Prop.~\ref{prop:SnablamitREnabla}, such that it makes sense to think about it as a representation in the context of this thesis.
\end{remark}

\begin{proof}
\leavevmode\newline
Let $\mleft( e_a \mright)_a$ be a frame of constant sections. Then $\kappa\mleft(e_a,e_b\mright) = \text{const.}$, and hence
\bas
0
&=
\mathcal{L}_{e_a} \bigl( \kappa(e_b, e_c) \bigr).
\eas
We also have
\bas
\mleft[ e_a, e_b \mright]_{\mathfrak{g}}
&=
\mleft[ e_a, e_b \mright]_{E}
	+ \nabla_{\rho(e_b)} e_a
=
\nabla^{\mathrm{bas}}_{e_a} e_b, 
\eas
because $\nabla$ is the canonical flat connection.
Therefore
\bas
&&
\widetilde{\kappa} &\text{ is $\mathrm{ad}$-invariant}
\\
&\Leftrightarrow&
0
&=
\widetilde{\kappa}\mleft( \mleft[ e_a, e_b \mright]_{\mathfrak{g}}, e_c \mright)
	+ \widetilde{\kappa} \mleft( e_b, \mleft[ e_a, e_c \mright]_{\mathfrak{g}} \mright)
\\
&\Leftrightarrow&
\mathcal{L}_{e_a} \mleft( \kappa(e_b, e_c) \mright)
&=
\kappa\mleft( \mleft[ e_a, e_b \mright]_{\mathfrak{g}}, e_c \mright)
	+ \kappa \mleft( e_b, \mleft[ e_a, e_c \mright]_{\mathfrak{g}} \mright)
\\
&\Leftrightarrow&
\mathcal{L}_{e_a} \mleft( \kappa(e_b, e_c) \mright)
&=
\kappa\mleft( \nabla^{\mathrm{bas}}_{e_a} e_b, e_c \mright)
	+ \kappa \mleft( e_b, \nabla^{\mathrm{bas}}_{e_a} e_c \mright)
\\
&\Leftrightarrow&
\nabla^{\mathrm{bas}}\kappa 
&=
0.
\eas
For $g$ recall Eq.~\eqref{eqAbleitungVomAnkerGibtRepraesentierung}, \textit{i.e.}
\bas
- \partial_\beta\rho_a^\alpha
&=
\bigl( \psi(e_a) \bigr)^\alpha_\beta,
\eas
where we use coordinate vector fields $\mleft( \partial_\alpha \mright)_\alpha$ on $N$ which also describes a constant frame for $\mathrm{T}W \cong W \times W$, and hence also, as before,
\bas
\bigl( \psi(e_a) \bigr)^\alpha_\beta
&=
\mleft[ \rho(e_a), \partial_\beta \mright]
=
\mleft[ \rho(e_a), \partial_\beta \mright]
	+ \rho \mleft( \nabla_{\partial_\beta} e_a \mright)
=
\nabla^{\mathrm{bas}}_{e_a} \partial_\beta,
\eas
and
\bas
0
&=
\mathcal{L}_{e_a} \bigl( g(\partial_\alpha, \partial_\beta) \bigr)
\eas
Thus,
\bas
&&
\widetilde{g} &\text{ is $\psi$-invariant}
\\
&\Leftrightarrow&
0
&=
\widetilde{g}\bigl( \psi(e_a)(\partial_\alpha), \partial_\beta \bigr)
	+ \widetilde{g} \bigl( \partial_\alpha, \psi(e_a) (\partial_\beta) \bigr)
\\
&\Leftrightarrow&
\mathcal{L}_{e_a} \bigl( g(\partial_\alpha, \partial_\beta) \bigr)
&=
g\bigl( \psi(e_a)(\partial_\alpha), \partial_\beta \bigr)
	+ g \bigl( \partial_\alpha, \psi(e_a) \mleft(\partial_\beta\mright) \bigr)
\\
&\Leftrightarrow&
\mathcal{L}_{e_a} \bigl( g(\partial_\alpha, \partial_\beta) \bigr)
&=
g \mleft( \nabla^{\mathrm{bas}}_{e_a} \partial_\beta, \partial_\beta \mright)
	+ g \mleft( \partial_\alpha, \nabla^{\mathrm{bas}}_{e_a} \partial_\beta \mright)
\\
&\Leftrightarrow&
\nabla^{\mathrm{bas}} g  
&=
0.
\eas
\end{proof}

Hence, when using the basic connection, we want that $\delta_{\Psi_\varepsilon} \mathfrak{D} = 0$ such that we can recover the classical formula in sense of Cor.~\ref{cor:EichtrafovonDAPHIinClassicIstBabyEinfach}. To study this and later results we need several auxiliary results, recall also Ex.~\ref{ex:ProjectionOntoGaugeBosonies}, \ref{ex:DAsFunctional} and \ref{ex:AnchorAsFunctional}.

\begin{lemmata}{Several identities related to variations with the basic connection}{VariationsIdentities}
Let $M, N$ be two smooth manifolds, $E \to N$ a Lie algebroid over $N$, $\nabla$ a connection on $E$, and $\Psi_\varepsilon \in \mathfrak{X}^E(\mathfrak{M}_E(M; N))$ for an $\varepsilon \in \mathcal{F}^0_E(M; {}^*E)$. For both functional spaces, $\mathcal{F}^\bullet_E(M; {}^*E)$ and $\mathcal{F}^\bullet_E(M; {}^*\mathrm{T}N)$, let $\delta_{\Psi_\varepsilon}$ be the unique operator of Prop.~\ref{prop:VariationVonSkalarZeugsEasyPeasy}, using $\nabla^{\mathrm{bas}}$ as $E$-connection on $E$ and $\mathrm{T}N$, respectively. Then
\ba
%\delta_\varepsilon \mleft[ \phi \mapsto \phi^*L \mright][\Phi]
%&=
%- \Phi^*\mleft( \nabla^{\mathrm{bas}}_\varepsilon L \mright), \label{PullBackVariation}
%\\
%\delta_\varepsilon \mleft[\phi \mapsto \phi^* \mu \mright][\Phi]
%&=
%- \Phi^* \mleft( {}^E\nabla_\varepsilon \mu \mright), \label{PullBackVariationE}
%\\
%\delta_\varepsilon \mleft[\phi \mapsto \phi^* Y \mright][\Phi]
%&=
%- \Phi^* \mleft( \nabla^{\mathrm{bas}}_\varepsilon Y \mright), \label{PullBackVariationTN}
%\\
%\delta_\varepsilon \mleft[\phi \mapsto \phi^* Y \mright][\Phi]
%&=
%- \Phi^* \mleft( \nabla^{\mathrm{bas}}_\varepsilon Y \mright), \label{PullBackVariationESternchen}
%\\
\delta_{\Psi_\varepsilon} \mathrm{D}
&=
- \mleft( {}^*\rho \mright) \bigl( {}^*\nabla \varepsilon \bigr), \label{DPhiVariation}
\\
\delta_{\Psi_\varepsilon} \mleft({}^*\rho\mright)
&=
0, \label{eqPhiRhoDieGeileSauIstnichtVariiert}
\\
\delta_{\Psi_\varepsilon} \bigl( ({}^*\rho)(\varpi_2) \bigr)
&=
\mleft( {}^* \rho \mright) \bigl( \delta_{\Psi_\varepsilon} \varpi_2 \bigr),
\label{eqRhoAVariation}
\\
\delta_{\Psi_\varepsilon} \mleft( {}^!\mleft(\nabla \mu \mright) \mright)
&=
- \biggl(
	{}^!\mleft(\nabla^{\mathrm{bas}}_\varepsilon \nabla \mu \mright)
	+ {}^*\mleft( \nabla_{({}^*\rho)\mleft( ({}^*\nabla) \varepsilon \mright)} \mu \mright)
\biggr) \label{EqVariationVonFormenBrrrr}
\ea
for all $\mu \in \Gamma(E)$, where we view $\nabla \mu$ as an element of $\Omega^1(N; E)$.
%\bas
%\delta_{\Psi_\varepsilon} \mleft( {}^!\mleft(\nabla \mu \mright) \mright)(\Phi, A)
%&=
%- \biggl(
	%\Phi^!\mleft(\nabla^{\mathrm{bas}}_{\varepsilon(\Phi, A)} \mleft( \nabla \mu \mright)\mright)
	%+ \Phi^*\mleft( \nabla_{(\Phi^*\rho)\mleft( (\Phi^*\nabla) \mleft(\varepsilon(\Phi, A)\mright) \mright)} \mu \mright)
%\biggr).
%\eas
%${}^E\nabla$ is either $\nabla_\rho$ for $j=1$ and $\nabla^{\mathrm{bas}}$ for $j=2$.
\end{lemmata}

\begin{remark}
\leavevmode\newline
We already introduced the notation for Eq.~\eqref{EqVariationVonFormenBrrrr} (also recall Remark \ref{RemarkNotationvonPullbackConnection}), but let us shortly write down what it is for each $(\Phi, A) \in \mathfrak{M}_E(M; N)$,
\bas
\mleft(\delta_{\Psi_\varepsilon} \mleft( {}^!\mleft(\nabla \mu \mright) \mright)\mright)(\Phi, A)
&=
- \biggl(
	\Phi^!\mleft(\nabla^{\mathrm{bas}}_{\epsilon} \mleft( \nabla \mu \mright)\mright)
	+ \Phi^*\mleft( \nabla_{(\Phi^*\rho)\mleft( (\Phi^*\nabla) \epsilon \mright)} \mu \mright)
\biggr)
\eas
where $\epsilon \coloneqq \varepsilon(\Phi, A)$. When $\varepsilon = {}^*\nu$ for a $\nu \in \Gamma(E)$, then $(\Phi^*\nabla ) (\Phi^*\nu) \stackrel{\text{Eq.~\eqref{EqGeilePullBackCommuteFormel}}}{=} \Phi^!(\nabla \nu)$, so, $({}^*\nabla) ({}^*\nu) = {}^!(\nabla \nu)$. Thus, we can then write
\ba\label{EqVariationVonFormenBrrrrVereinfacht}
\delta_{\Psi_{{}^*\nu}} \mleft( {}^!\mleft(\nabla \mu \mright) \mright)
&=
- {}^!\mleft(
	\nabla^{\mathrm{bas}}_\nu \nabla \mu
	+ \nabla_{\rho(\nabla \nu)} \mu
\mright).
\ea
\end{remark}

\begin{proof}[Proof for Lemma \ref{lem:VariationsIdentities}]
\leavevmode\newline
In the following $\mleft( e_a \mright)_a$ denotes a local frame of $E$, and $\partial_\alpha$ are local coordinate vector fields on $N$, and $(\Phi, A) \in \mathfrak{M}_E(M; N)$. Regarding $\varepsilon \in \mathcal{F}^0_E(M; {}^*E)$ we also write $\epsilon \coloneqq \varepsilon(\Phi, A)$.

$\bullet$ For Eq.~\eqref{DPhiVariation} we write locally
\bas
\mathrm{D} \Phi
&=
\mathrm{d} \Phi^\alpha \otimes \Phi^* \partial_\alpha,
\eas
where we view $(\Phi,A) \mapsto \Phi^\alpha$ as an element of $\mathcal{F}^0_E(M)$ (on an open subset of $M$), such that by $\delta_\varepsilon \Phi = - ({}^*\rho) (\varepsilon)$, and by using $\mathrm{d} \delta_{\Psi_\varepsilon} = \delta_{\Psi_\varepsilon} \mathrm{d}$ and $\delta_{\Psi_\varepsilon} = \mathcal{L}_{\Psi_\varepsilon}$ on $\mathcal{F}^0_E(M)$ (recall the discussion around Eq.~\eqref{eqVariationVertauschtMitDifferential}),
\bas
\mleft(\delta_\varepsilon \mathrm{d} \mleft[ (\Phi, A) \mapsto \Phi^\alpha \mright]\mright)(\Phi, A)
&=
\mleft(\mathrm{d} \mathcal{L}_{\Psi_\varepsilon} \mleft[ (\Phi, A) \mapsto \Phi^\alpha \mright]\mright)(\Phi, A)
=
- \mathrm{d} \mleft( \mleft( \rho^\alpha_a \circ \Phi \mright) ~ \epsilon^a \mright)
\eas
then by Eq.~\eqref{PullBackVariation} and the Leibniz rule of $\delta_{\Psi_\varepsilon}$
\bas
\mleft(\delta_{\Psi_\varepsilon} \mathrm{D} \mright)(\Phi, A)
&=
- \mathrm{d} \mleft( \mleft( \rho^\alpha_a \circ \Phi \mright) ~ \epsilon^a \mright) \otimes \Phi^* \partial_\alpha
	- \mathrm{d} \Phi^\alpha \otimes \epsilon^a~ \Phi^* \mleft( \nabla^{\mathrm{bas}}_{e_a}\partial_\alpha \mright) \\
&=
- \Bigl( \underbrace{\mathrm{d} \mleft( \rho^\alpha_a \circ \Phi \mright)}
	_{\mathclap{= ~ \mleft(\partial_\beta \rho^\alpha_a \circ \Phi \mright) ~ \mathrm{d} \Phi^\beta}}
 ~ \epsilon^a
	+ \mleft( \rho^\alpha_a \circ \Phi \mright) ~ \mathrm{d}\epsilon^a \Bigr) \otimes \Phi^* \partial_\alpha \\
&\hspace{1cm}
	- \mathrm{d} \Phi^\alpha \otimes \epsilon^a~ \Phi^* \mleft( 
	- \partial_\alpha \rho^\beta_a ~ \partial_\beta
	+ \rho \mleft( \nabla_{\partial_\alpha} e_a \mright)
	  \mright) \\
&=
- \mleft( \rho^\alpha_a \circ \Phi \mright) ~ \mathrm{d}\epsilon^a \otimes \Phi^* \partial_\alpha
	- \mathrm{d} \Phi^\beta \otimes \epsilon^b~ \mleft( \rho^\alpha_a \circ \Phi \mright) ~ \mleft(\omega_{b\beta}^a \circ \Phi \mright)
	~\Phi^*\partial_\alpha \\
&=
- \mleft( \rho^\alpha_a \circ \Phi \mright) \mleft(
\mathrm{d}\epsilon^a
	+ \epsilon^b~ \mleft(\omega_{b\beta}^a \circ \Phi \mright) ~\mathrm{d} \Phi^\beta
\mright) \otimes \Phi^*\partial_\alpha \\
&=
- \mleft( \Phi^*\rho \mright)\bigl( \mleft( \Phi^* \nabla\mright) \epsilon \bigr).
\eas

$\bullet$ By Eq.~\ref{PullBackVariation},
\bas
\delta_{\Psi_\varepsilon} \mleft( {}^*\rho \mright)
&=
- {}^*\mleft( \nabla^{\mathrm{bas}}_\varepsilon \rho \mright),
\eas
and by $\rho \circ \nabla^{\mathrm{bas}} = \nabla^{\mathrm{bas}} \circ \rho$ we get
\bas
\mleft( \nabla^{\mathrm{bas}} \rho \mright)(\mu)
&=
\nabla^{\mathrm{bas}}\mleft( \rho(\mu) \mright)
	- \rho\mleft( \nabla^{\mathrm{bas}}\mu \mright)
=
0
\eas
for all $\mu \in \Gamma(E)$. Hence,
\bas
\delta_{\Psi_\varepsilon} \mleft( {}^*\rho \mright)
&=
0.
\eas

$\bullet$ By the Leibniz rule and the previous result we also have
\bas
\delta_{\Psi_\varepsilon} \bigl( ({}^*\rho)(\varpi_2) \bigr)
&=
\mleft( {}^* \rho \mright) \bigl( \delta_{\Psi_\varepsilon} \varpi_2 \bigr).
%\delta_{\Psi_\varepsilon} \bigl[ (\phi, w) \mapsto \mleft( \phi^* \rho \mright)(w) \bigr](\Phi, A)
%&=
%\mleft(\delta_\varepsilon \mleft[ \phi \mapsto \phi^*\rho \mright][\Phi] \mright)(A)
	%+ \mleft(\Phi^*\rho\mright)\mleft( \delta_\varepsilon A \mright)
%\\
%&\stackrel{\mathclap{\text{Eq.~\eqref{eqPhiRhoDieGeileSauIstnichtVariiert}}}}{=}~~~~
%\mleft(\Phi^*\rho\mright)\mleft( \delta_\varepsilon A \mright)
%\\
%&\stackrel{\mathclap{\text{Def.~\ref{def:GaugeTrafoOfA}}}}{=}~~~~
%- \mleft(\Phi^*\rho\mright)\bigl( \mleft( \Phi^*\nabla \mright) \varepsilon \bigr).
\eas

$\bullet$ We view terms like $\nabla \mu$ as elements of $\Omega^1(N; E)$ for all $\mu \in \Gamma(E)$, $\mathfrak{X}(N) \ni Y \mapsto (\nabla \mu)(X) = \nabla_X \mu$, and therefore we can use the Leibniz rule on ${}^!(\nabla \mu) = \bigl({}^*(\nabla \mu)\bigr)(\mathrm{D}) = {}^*\mleft(\nabla_{\mathrm{D}} \mu\mright)$, \textit{i.e.}~due to
\bas
\Phi^!(\nabla \mu) &= \bigl(\Phi^*(\nabla \mu)\bigr)(\mathrm{D}\Phi)
\eas
we can view ${}^!(\nabla \mu)$ as a contraction of the functionals ${}^*(\nabla \mu)$ and $\mathrm{D}$. Hence,
\bas
\delta_{\Psi_\varepsilon} \mleft( {}^!\mleft(\nabla \mu \mright) \mright)
&=
\bigl( \delta_{\Psi_\varepsilon} ({}^*(\nabla \mu))\bigr)(\mathrm{D})
	+ {}^*\mleft(\nabla_{\delta_{\Psi_\varepsilon} \mathrm{D}} \mu\mright)
\\
&\stackrel{\mathclap{\text{Eq.~\eqref{PullBackVariation}}}}{=}~~~~
-\mleft({}^*\mleft(\nabla^{\mathrm{bas}}_\varepsilon \nabla \mu \mright)\mright)(\mathrm{D})
	+ {}^*\mleft(\nabla_{\delta_{\Psi_\varepsilon} \mathrm{D}} \mu\mright)
\\
&\stackrel{\mathclap{\text{Eq.~\eqref{DPhiVariation}}}}{=}~~~~
- \biggl(
	{}^!\mleft(\nabla^{\mathrm{bas}}_\varepsilon \nabla \mu \mright)
	+ {}^*\mleft( \nabla_{({}^*\rho)\mleft( ({}^*\nabla) \varepsilon \mright)} \mu \mright)
\biggr).
\eas
\end{proof}

Let us now fix the gauge transformation of $A$ using these results. Recall that we write $\Psi = \Psi_\varepsilon$ for a $\Psi \in \mathfrak{X}^E(\mathfrak{M}_E(M; N))$, where $\varepsilon \in \mathcal{F}^0_E(M;{}^*E)$ such that we can write (recall Eq.~\eqref{GaugeTrafoVektor})
\bas
\Psi_\varepsilon
&=
\mleft( -({}^*\rho_B )(\varepsilon), \mathfrak{a} \mright)
\eas
where $\mathfrak{a}$ is a map on $\mathfrak{M}_E(M; N)$ such that $\Psi|_{(\Phi,A)}$ is a tangent vector for all $(\Phi, A) \in \mathfrak{M}_E(M; N)$, \textit{i.e.}~satisfying the diagram of Prop.~\ref{prop:TangentSpaceOfSpaceOfFields} for all $(\Phi, A)$. For a given $\varepsilon$ such a $\Psi_\varepsilon$ is in general not unique. Recall that for a local frame $\mleft( e_a \mright)_a$ of $E$ and local coordinate functions $\mleft(\partial_\alpha\mright)_\alpha$ on $N$ we have
\bas
\mleft[ e_b, e_c \mright]_E
&=
C^a_{bc} e_a, &
\nabla e_b
&=
\omega^a_b \otimes e_a, &
\nabla_{\partial_\alpha} e_b
&=
\omega^a_{b\alpha} ~ e_a.
\eas

\begin{propositions}{Gauge transformation of the field of gauge bosons}{VariationOfA}
Let $M, N$ be two smooth manifolds, $E \to N$ a Lie algebroid over $N$, $\nabla$ a connection on $E$, $\varepsilon \in \mathcal{F}^0_E(M; {}^*E)$, and for the functional space $\mathcal{F}^\bullet_E(M; {}^*E)$ let $\delta_{\Psi_\varepsilon}$ be the unique operator of Prop.~\ref{prop:VariationVonSkalarZeugsEasyPeasy}, using $\nabla^{\mathrm{bas}}$ as $E$-connection on $E$ and any $\Psi_\varepsilon \in \mathfrak{X}^E\bigl( \mathfrak{M}_E(M;N) \bigr)$. Then there is a unique $\gls{1YPsiEpsilon} \in \mathfrak{X}^E\bigl(\mathfrak{M}_E(M; N)\bigr)$ such that
\ba\label{EichtrafoVonANochmal}
\delta_{\Psi_\varepsilon} \varpi_2
&=
- ({}^*\nabla) \varepsilon.
\ea
Locally with respect to a given frame $\mleft( e_a \mright)_a$
\ba
\mleft(\delta_{\Psi_\varepsilon} \varpi_2^a\mright)(\Phi, A)
&=
\mleft( C^a_{bc} \circ \Phi \mright) ~\epsilon^b A^c
	+ \mleft(\omega^a_{b\alpha} \circ \Phi \mright) ~ \mleft( \rho^\alpha_c \circ \Phi \mright)~\epsilon^b A^c
	- \mathrm{d}\epsilon^a - \epsilon^b ~ \Phi^!\mleft(\omega^a_{b} \mright)
\nonumber \\ \label{eqGaugeTrafoOfAacomps}
&=
\mleft( \epsilon^b A^c \otimes \Phi^*\mleft( \nabla^{\mathrm{bas}}_{e_b} e_c\mright)
	- \mleft(\Phi^*\nabla\mright)\epsilon \mright)^a
\ea
for all $(\Phi, A) \in \mathfrak{M}_E(M; N)$, where $\epsilon \coloneqq \varepsilon(\Phi, A)$.

Moreover, if we also have $\alpha, \beta \in \mathbb{R}$ and $\vartheta \in \mathcal{F}^0_E(M; {}^*E)$, then
\ba\label{LinearityOfPsiEpsilon}
\Psi_{\alpha \varepsilon + \beta \vartheta}
=
\alpha \Psi_\varepsilon + \beta \Psi_\vartheta,
\ea
where the vector fields are the ones uniquely given by Eq.~\eqref{EichtrafoVonANochmal}.
%\ba
%\mleft(\delta_{\Psi_\varepsilon} \mathrm{D}\mright)(\Phi, A)
%&=
%- \mleft( \Phi^* \rho \mright) \bigl( \mleft( \Phi^*\nabla \mright) \varepsilon \bigr), \label{DPhiVariation}
%\\
%\bigl(\delta_\varepsilon \mleft( \mathrm{D} - \mathfrak{D} \mright)\bigr)(\Phi, A)
%&=
%\mleft( \Phi^* \rho \mright) \bigl( \delta_{\Psi_\varepsilon} \varpi_2 \bigr)
%\label{eqRhoAVariation}
%\ea
\end{propositions}

\begin{proof}[Proof of Prop.~\ref{prop:VariationOfA}]
\leavevmode\newline
Since it is about a vector field on $\mathfrak{M}_E(M; N)$, we will classify $\Psi_\varepsilon$ by its flow, using Cor.~\ref{cor:ReasonWhyVectorFieldAlongAnBPath}: We denote its flow through a fixed point $(\Phi_0, A_0) \in \mathfrak{M}_E(M; N)$ by $\gamma: I \to \mathfrak{M}_E(M; N)$, $t \mapsto \gamma(t) \eqqcolon (\Phi_t, A_t) \in \mathfrak{M}_E(M; N)$, where $I$ is an open interval of $\mathbb{R}$ containing 0, and we write $\Psi|_{\gamma(t)} = \mleft( - (\Phi_t^*\rho)(\epsilon_t), \mathcal{a}_t \mright) \in \mathrm{T}^E_{(\Phi_t, A_t)}\mathfrak{M}_E(M; N)$, where $\epsilon_t \coloneqq \varepsilon(\Phi_t, A_t)\in \Gamma(\Phi^*_tE)$, and $\mathcal{a}_t$ is a morphism $\mathrm{T}M \to \mathrm{T}E$ satisfying the diagram in Prop.~\ref{prop:TangentSpaceOfSpaceOfFields}. So, we have a curve $\gamma$ with 
\bas
\gamma(0) &= (\Phi_0, A_0), \\
\frac{\mathrm{d}}{\mathrm{d}t} \gamma
&=
\Psi|_{\gamma(t)}
=
\mleft( - (\Phi_t^*\rho)(\epsilon_t), \mathcal{a}_t \mright).
\eas
$(\Phi_0, A_0)$ and $- (\Phi_t^*\rho)(\epsilon_t)$ are fixed, and we show that Eq.~\eqref{EichtrafoVonANochmal} will fix $\mathcal{a}_t$. Without loss of generality let us assume that everything is small and local enough such that we have frames and coordinates, like a frame $\mleft(e_a\mright)_a$ of $E$.\footnote{One could even fix a point $p \in M$ because we just need an interval for $t$ for $\mathrm{d}/\mathrm{d}t$.}
%, because then $t \mapsto \Phi_t(p)$ is a curve in $N$ over which we want to choose a trivialisation of $E$ for example; we are going to omit the notation of $p$ in the following. 
Making use of Prop.~\ref{prop:VariationVonSkalarZeugsEasyPeasy}, we get
\bas
\mleft(\delta_{\Psi_\varepsilon} \varpi_2 \mright) (\Phi_t, A_t)
&=
%\underbrace{
\mleft.\mathcal{L}_{\Psi_\varepsilon} \mleft(\varpi^a_2\mright)\mright|_{(\Phi_t, A_t)} \otimes \Phi^*_t e_a
%}_{\mathclap{= \mleft.\frac{\mathrm{d}}{\mathrm{d}t}\mright|_{t=0} \mleft(\varpi^a_2\circ\gamma\mright) }} \otimes ~ \Phi^*_0 e_a
	- A_t^a \otimes \Phi^*_t\mleft( \nabla^{\mathrm{bas}}_{\epsilon_t} e_a \mright).
%\\
%&=
%\mleft.\frac{\mathrm{d}}{\mathrm{d}t}\mright|_{t=0} \mleft[ t \mapsto A^a_t \mright] \otimes \Phi^*_0 e_a
	%- A_0^a \otimes \Phi^*_0\mleft( \nabla^{\mathrm{bas}}_\epsilon e_a \mright)
\eas
Let us first assume Eq.~\eqref{EichtrafoVonANochmal} does hold. Then
\bas
&\mleft.\mathcal{L}_{\Psi_\varepsilon} \mleft(\varpi^a_2\mright)\mright|_{(\Phi_t, A_t)} \otimes \Phi^*_t e_a
\\
&=
\epsilon_t^b A_t^c \otimes \Phi^*_t\mleft( \nabla^{\mathrm{bas}}_{e_b} e_c \mright)
	- \mleft(\Phi_t^* \nabla\mright) \epsilon_t
\\
&=
\mleft(
\mleft( C^a_{bc} \circ \Phi_t \mright) ~\epsilon_t^b A_t^c
	+ \mleft(\omega^a_{b\alpha} \circ \Phi_t \mright) ~ \mleft( \rho^\alpha_c \circ \Phi_t \mright)~\epsilon_t^b A_t^c
	- \mathrm{d}\epsilon_t^a - \epsilon_t^b ~ \Phi_t^!\mleft(\omega^a_{b} \mright)
\mright) \otimes \Phi^*_t e_a
\eas
which proves Eq.~\eqref{eqGaugeTrafoOfAacomps} (insert $t=0$). By the definition of $\gamma$ and the Lie derivative we also get
\bas
\mleft.\mathcal{L}_{\Psi_\varepsilon} \mleft(\varpi^a_2\mright)\mright|_{(\Phi_t, A_t)} 
&=
\frac{\mathrm{d}}{\mathrm{d}t} \mleft(\varpi^a_2\circ\gamma\mright)
=
\frac{\mathrm{d}}{\mathrm{d}t} \mleft[ t \mapsto A^a_t \mright],
\eas
and, thus,
\ba\label{DiffEqFuerAComp}
\frac{\mathrm{d}}{\mathrm{d}t} \mleft[ t \mapsto A^a_t \mright]
&=
\mleft( C^a_{bc} \circ \Phi_t \mright) ~\epsilon^b_t A_t^c
	+ \mleft(\omega^a_{b\alpha} \circ \Phi_t \mright) ~ \mleft( \rho^\alpha_c \circ \Phi_t \mright)~\epsilon^b_t A_0^c
	- \mathrm{d}\epsilon^a - \epsilon^b_t ~ \Phi_t^!\mleft(\omega^a_{b} \mright).
\ea
So, Eq.~\eqref{EichtrafoVonANochmal} is equivalent to a set of coupled differential equations: We have a curve $\gamma(t) = (\Phi_t, A_t)$, with $\Phi_{t=0} = \Phi_0$ and 
\bas
\frac{\mathrm{d}}{\mathrm{d}t} [t \mapsto \Phi_t]
&=
- (\Phi_t^*\rho) (\epsilon_t),
\eas
and $A_{t=0} = A_0$, while
\bas
\mathcal{a}_t
&=
\frac{\mathrm{d}}{\mathrm{d}t} \mleft[ t \mapsto A_t \mright]
=
\frac{\mathrm{d}}{\mathrm{d}t} \mleft[ t \mapsto A_t^a \otimes \Phi^*_t e_a \mright].
%\\
%&=
%\mleft.\frac{\mathrm{d}}{\mathrm{d}t}\mright|_{t=0} \mleft[ t \mapsto A^a_t \mright] \otimes \Phi^*_0 e_a
	%+ A_0^a \otimes \bigl(\Phi^*_0\mleft(\mathrm{D}e_a\mright)\bigr)\mleft( \mleft.\frac{\mathrm{d}}{\mathrm{d}t}\mright|_{t=0} \mleft[t\mapsto \Phi_t\mright] \mright)
%\\
%&=
%\mleft.\mathcal{L}_{\Psi_\varepsilon} \mleft(\varpi^a_2\mright)\mright|_{(\Phi_0, A_0)} \otimes \Phi^*_0 e_a
	%- A_0^a \otimes \bigl(\Phi^*_0\mleft(\mathrm{D}e_a\mright)\bigr)\bigl( (\Phi^*_0\rho)(\epsilon_0)\bigr)
\eas
$t \mapsto \Phi_t$ and $t \mapsto A^a_t$ are uniquely given by this system and the differential equation \eqref{DiffEqFuerAComp}, and, so, $t \mapsto A_t = A^a_t \otimes \Phi^*_t e_a$ is uniquely given, too. Hence, $\mathcal{a}_t$ is unique, and, thus, $\Psi_\varepsilon$. Alternatively, the differential equations for $\mathrm{d}/\mathrm{d}t ~ \Phi$ and $\mathrm{d}/\mathrm{d}t ~ A^a$ are the action of the vector field $\Psi_\varepsilon$ on the coordinates of $\mathfrak{M}_E$, and therefore defining $\Psi_\varepsilon$.

The linearity of $\psi_\varepsilon$ in $\varepsilon$ over $\mathbb{R}$ simply follows by the linearity given in the differential equations above: Define $\Theta \coloneqq \alpha \Psi_\varepsilon + \beta \Psi_\vartheta$ for $\alpha, \beta \in \mathbb{R}$ and $\vartheta \in \mathcal{F}^0_E(M; {}^*E)$, where $\Psi_\varepsilon$ and $\Psi_\vartheta$ are the unique vector fields as given above, \textit{i.e.}~$\delta_{\Psi_\varepsilon} \varpi_2 = - ({}^*\nabla) \varepsilon$ and $\delta_{\Psi_\vartheta} \varpi_2 = - ({}^*\nabla) \vartheta$, respectively. Observe that $\Theta \in \mathfrak{X}^E\bigl(\mathfrak{M}_E(M; N)\bigr)$, where the component along the "$\Phi$-direction" is by definition given by
\bas
-\alpha ~ ({}^*\rho)(\varepsilon) - \beta ~ ({}^*\rho)(\vartheta)
&=
- ({}^*\rho)(\alpha \varepsilon + \beta \vartheta),
\eas
then, using the linearity of Eq.~\eqref{DiffEqFuerAComp} in $\varepsilon$,
\bas
\delta_\Theta \varpi_2
&=
\mathcal{L}_\Theta (\varpi_2^a) \otimes {}^*e_a
	- \varpi_2^a \otimes {}^*\mleft(\nabla^{\mathrm{bas}}_{\alpha \varepsilon + \beta \vartheta} e_a\mright)
\\
&=
\mleft( \alpha \mathcal{L}_{\Psi_{\varepsilon}}  + \beta \mathcal{L}_{\Psi_\vartheta} \mright) (\varpi_2^a) \otimes {}^*e_a
	- \varpi_2^a \otimes {}^*\mleft(\nabla^{\mathrm{bas}}_{\alpha \varepsilon + \beta \vartheta} e_a\mright)
	\\
&\stackrel{\mathclap{ \text{ Eq.~\eqref{DiffEqFuerAComp}} }}{=}~~~
\mathcal{L}_{\Psi_{\alpha \varepsilon + \beta \vartheta}} (\varpi_2^a) \otimes {}^*e_a
	- \varpi_2^a \otimes {}^*\mleft(\nabla^{\mathrm{bas}}_{\alpha \varepsilon + \beta \vartheta} e_a\mright)
\\
&=
\delta_{\Psi_{\alpha \varepsilon + \beta \vartheta}} \varpi_2.
\eas
By the shown uniqueness of vector fields like $\Psi_{\alpha \varepsilon + \beta \vartheta}$, we get
\bas
\Theta
&=
\Psi_{\alpha \varepsilon + \beta \vartheta}.
\eas
%viewing $\Phi^*_t e_a$ canonically as a section along $\Phi_t$, hence, $\Phi^*_t e_a = e_a \circ \Phi_t: M \to E$, such that for all $p \in M$
%\bas
%\mleft.\frac{\mathrm{d}}{\mathrm{d}t}\mright|_{t=0} \mleft[ t \mapsto \mleft(\Phi^*_t(p)\mright) e_a \mright]
%&=
%\mathrm{D}_{\Phi_0(p)}e_a \mleft( \mleft.\frac{\mathrm{d}}{\mathrm{d}t}\mright|_{t=0} \mleft[t\mapsto \Phi_t(p)\mright] \mright)
%=
%\mleft.\bigl(\Phi^*_0\mleft(\mathrm{D}e_a\mright)\bigr)\mleft( \mleft.\frac{\mathrm{d}}{\mathrm{d}t}\mright|_{t=0} \mleft[t\mapsto \Phi_t\mright] \mright)\mright|_p
%\eas
%and using $\mathrm{D} e_a \in \Omega^1\mleft(N; e_a^*\mathrm{T}E \mright)$ (as we did for $\mathrm{D}\Phi$ before)\footnote{Rigorously, this form is just defined on an open subset of $N$.} such that $\Phi^*_0\mleft(\mathrm{D}e_a\mright) \in \Gamma\mleft( \Phi^*_0(\mathrm{T}^*N) \otimes (\Phi^*_0e_a)^*\mathrm{T}E \mright)$
\end{proof}

\begin{remark}\label{RemDifferentVersionsOfGaugeTrafos}
\leavevmode\newline
Eq.~\eqref{eqGaugeTrafoOfAacomps} is also \textit{e.g.}~defined in \cite[Eq.~(10); opposite sign of $\varepsilon$]{CurvedYMH}, but in this reference it was not known how a coordinate-free version can look like. This equation recovers the standard formula of the infinitesimal gauge transformation of $A$.
%As argued in the last statement of Remark \ref{RemLeibnizeRegelaufProdukteWeshalbEConnectionNichtWichtigIst} one does not necessarily need to know that, for the variation of the Lagrangian we only need to know $\delta_\varepsilon A^a$. 
In order to see why this restricts to the standard formula, let us look again at the standard setting: When $E= N \times \mathfrak{g}$ is an action Lie algebroid with Lie algebra $\mathfrak{g}$, equipped with its canonical flat connection $\nabla$, then we get the classical formula of gauge transformation by using a constant frame $\mleft( e_a \mright)_a$ for $E$, \textit{i.e.}
\bas
\mleft(\delta_{\Psi_\varepsilon} \varpi_2^a\mright)(\Phi, A)
&=
\Phi^*C^a_{bc} ~ \epsilon^b A^c - \mathrm{d}\epsilon^a 
=
\mleft(
	\mleft[ \epsilon \stackrel{\wedge}{,} A \mright]_{\mathfrak{g}}
	- \mathrm{d}^{\Phi^*\nabla} \epsilon 
\mright)^a
\eas
for all $(\Phi, A) \in \mathfrak{M}_E(M; N)$,
because $\omega^a_b = 0$ and $\Phi^*C^a_{bc} = C^a_{bc} = \text{const.}$, the structure constants of $\mathfrak{g}$. We can understand $\epsilon$ as an element of $C^\infty(M; \mathfrak{g})$ as usual in the standard setting. That is precisely the typical formula of the classical setting, because $\Phi^*\nabla$ is the canonical flat connection of $\Phi^*E \cong M \times \mathfrak{g}$. Moreover, we get in that situation
\bas
%\Phi^*\mleft(
	%\nabla_{\rho(\varepsilon)} e_a 
%\mright)
%&= 
%0, 
%\\
\Phi^*\mleft(
	\nabla^{\mathrm{bas}}_{\epsilon} e_a 
\mright)
&=
\epsilon^b ~ \Phi^*\mleft(\mleft[ e_b, e_a \mright]_E\mright),
\eas
which is the main reason why the transformations of the components recover the classical formula although the total formula, Eq.~\eqref{EichtrafoVonANochmal}, just carries the differential (as we saw in the proof). As already discussed, only the transformation of the components need the "correct form" when it is about the gauge invariance of the Lagrangian.
%
%Hence, this proposition provides two ways how to formulate the infinitesimal gauge transformation of $A$ given in \cite[Eq.~(10)]{CurvedYMH} in a coordinate-free way, that is given by Eq.~\eqref{EqVariationsBestimmungFuerAAaaaaah} and \eqref{EqVariationsBestimmungFuerADoeoeoeo}.
\end{remark}

Using such a $\Psi_\varepsilon$ results into an infinitesimal gauge transformation of the minimal coupling as in Cor.~\ref{cor:EichtrafovonDAPHIinClassicIstBabyEinfach}.

\begin{propositions}{Infinitesimal gauge transformation of the minimal Coupling}{InfinitesimalGaugeTrafoOfMinimalCoupleSmiley}
Let $M, N$ be two smooth manifolds, $E \to N$ a Lie algebroid over $N$, $\nabla$ a connection on $E$, and $\varepsilon \in \mathcal{F}^0_E(M; {}^*E)$ together with the unique $\Psi_\varepsilon \in \mathfrak{X}^E(\mathfrak{M}_E(M; N))$ as given in Prop.~\ref{prop:VariationOfA}. For both functional spaces, $\mathcal{F}^\bullet_E(M; {}^*E)$ and $\mathcal{F}^\bullet_E(M; {}^*\mathrm{T}N)$, let $\delta_{\Psi_\varepsilon}$ be the unique operator of Prop.~\ref{prop:VariationVonSkalarZeugsEasyPeasy}, using $\nabla^{\mathrm{bas}}$ as $E$-connection on $E$ and $\mathrm{T}N$, respectively.

Then we have
\ba
\delta_{\Psi_\varepsilon} \mathfrak{D}
&=
0.
\ea
\end{propositions}

\begin{remark}
\leavevmode\newline
We already have derived the variation of the components of $\mathfrak{D}$, for this recall the general calculation for Eq.~\eqref{CompsVonDMinimalAlsErstes}:
Let $\mleft( e_a \mright)_a$ be a local frame of $E$ and $\partial_\alpha$ coordinate vector fields on $N$, then we can write $\mathfrak{D} = \mathfrak{D}^\alpha \otimes {}^*\partial^\alpha$, and, thus, with $\epsilon \coloneqq \varepsilon(\Phi, A)$, 
\ba
\bigl(\delta_{\Psi_\varepsilon} \mathfrak{D}^\alpha\bigr)(\Phi, A)
&=
\epsilon^a ~ \Phi^*\mleft( 
	- \partial_\beta\rho_a^\alpha
	+ \rho^\alpha\mleft( \nabla_{\partial_\beta} e_a \mright) 
\mright) ~ \mleft( \mathfrak{D}^A \Phi \mright)^\beta.
\ea
That is precisely the same formula as given in \cite[Eq.~(12), different sign for $\epsilon$ there]{CurvedYMH}, but there only the formula for the components was known.
\end{remark}

\begin{proof}[Proof of Prop.~\ref{prop:InfinitesimalGaugeTrafoOfMinimalCoupleSmiley}]
\leavevmode\newline
This quickly follows by Lemma \ref{lem:VariationsIdentities}, especially Eq.~\eqref{DPhiVariation} and \eqref{eqRhoAVariation},
\bas
\delta_{\Psi_\varepsilon} \mathfrak{D}
&=
\delta_{\Psi_\varepsilon} \bigl( \mathrm{D} - ({}^*\rho)(\varpi_2) \bigr)
=
-({}^*\rho)({}^*\nabla\varepsilon)
	- \mleft( {}^* \rho \mright) \bigl( \delta_{\Psi_\varepsilon} \varpi_2 \bigr)
\stackrel{\text{Prop.~\ref{prop:VariationOfA}}}{=}
0.
\eas
\end{proof}

\begin{remark}
\leavevmode\newline
Following the proof of Prop.~\ref{prop:InfinitesimalGaugeTrafoOfMinimalCoupleSmiley} and using the uniqueness of Prop.~\ref{prop:VariationOfA} one could argue that $\Psi_\varepsilon$ is the unique element of $\mathfrak{X}^E(\mathfrak{M}_E(M; N))$ with $\delta_{\Psi_\varepsilon} \mathfrak{D} = 0$ for a given $\varepsilon$ in the category of Lie algebroids, because this must then \textit{e.g.}~hold for the tangent bundle $E = \mathrm{T}N$ as Lie algebroid, too, whose anchor is the identity.
\end{remark}

By this result and Cor.~\ref{cor:EichtrafovonDAPHIinClassicIstBabyEinfach} we define the following.

\begin{definitions}{Infinitesimal gauge transformation of gauge bosons}{GaugeTrafoOfA}
Let $M, N$ be two smooth manifolds, $E \to N$ a Lie algebroid over $N$, $\nabla$ a connection on $E$, and $\varepsilon \in \mathcal{F}^0_E(M; {}^*E)$ together with the unique $\Psi_\varepsilon \in \mathfrak{X}^E\bigl(\mathfrak{M}_E(M; N)\bigr)$ as given in Prop.~\ref{prop:VariationOfA}. For the functional space $\mathcal{F}^\bullet_E(M; {}^*E)$ let $\delta_{\Psi_\varepsilon}$ be the unique operator of Prop.~\ref{prop:VariationVonSkalarZeugsEasyPeasy}, using $\nabla^{\mathrm{bas}}$ as $E$-connection on $E$.

For a $(\Phi,A) \in \mathfrak{M}_E(M; N)$ we define the \textbf{infinitesimal gauge transformation $\delta_{\varepsilon(\Phi, A)} A$ of $A$} as an element of $\Omega^1(M; \Phi^*E)$ by
\ba
\delta_{\varepsilon(\Phi, A)} A
&\coloneqq
\mleft( \delta_{\Psi_\varepsilon} \varpi_2 \mright)(\Phi, A)
=
- (\Phi^*\nabla) \bigl( \varepsilon(\Phi, A) \bigr),
\ea
shortly denoted as $\delta_\varepsilon A \coloneqq \delta_{\Psi_\varepsilon} \varpi_2 = -({}^*\nabla) \varepsilon$. Given a local frame $\mleft( e_a \mright)_a$ of $E$, we also similarly define $\delta_\varepsilon A^a \coloneqq \delta \varpi_2^a$.
\end{definitions}

\begin{remark}\label{WhyNablaBasPartOne}
\leavevmode\newline
As discussed in Remark \ref{RemDifferentVersionsOfGaugeTrafos} we have seen that $\delta_\varepsilon A^a$ (using a frame $\mleft( e_a \mright)_a$ of $E$) recovers the classical formula of the infinitesimal gauge transformation. However, the total formula, $\delta_\varepsilon A$, does not recover it which is no problem due to that the variation of the Lagrangian just depends on the variation of the components; for this also recall that arising differentials of $A$ commute with $\delta_\varepsilon$, Eq.~\eqref{eqVariationVertauschtMitDifferential}, which is needed for the variation of the field strength. Later we will see this explicitly when showing the gauge invariance of the Lagrangian. 

Alternatively, one could use $\nabla_\rho$ as $E$-connection on $E$ instead of $\nabla^{\mathrm{bas}}$ for the definition of $\delta_{\Psi_\varepsilon}$; especially because of results like Thm.~\ref{thm:NewFormulaRecoversOldGaugeTrafoYay} and Thm.~\ref{thm:RecoverOfClassicInfgGaugeTrafo}, which imply that one recovers classical formulas when $\nabla$ is additionally flat.\footnote{A flat connection is locally canonically flat with respect to the trivialization given by a parallel frame; later we will also see that then $E$ is locally an action algebroid and $\nabla$ its canonical flat connection, if $\nabla$ is flat and has vanishing basic curvature.} When using $\nabla_\rho$, the same $\Psi_\varepsilon$ leads to
\ba
\delta_{\Psi_\varepsilon} \varpi_2
&=
-({}^*t_{\nabla_\rho})(\varepsilon, \varpi_2) - ({}^*\nabla)\varepsilon,
\ea
where $t_{\nabla_\rho}$ is the torsion of $\nabla_\rho$. As we have seen before, $\nabla$ will be the canonical flat connection in the standard setting such that then $\delta_{\Psi_\varepsilon} A^a = \mleft(\delta_{\Psi_\varepsilon} A \mright)^a$ by flatness and Thm.~\ref{thm:NewFormulaRecoversOldGaugeTrafoYay}. With similar calculations as before one also shows that the variation of the components, $\delta_{\Psi_\varepsilon} \varpi_2^a$, recovers the classical formula of the infinitesimal gauge transformation of the field of gauge bosons, thus, $\delta_{\Psi_\varepsilon} \varpi_2$ would restrict to the classical formula in the standard setting, too. Hence, $\nabla_\rho$ would look like the canonical choice, not $\nabla^{\mathrm{bas}}$. But we will later see that $\nabla_\rho$ is in general not flat, while $\nabla^{\mathrm{bas}}$ will be flat, such that only for the latter the infinitesimal gauge transformations in form of the operator $\delta_{\Psi_\varepsilon}$ will give rise to a Lie algebra in general. Moreover, we are not going to fix any separate connection on $\mathrm{T}N$ which would be identified with a canonical flat connection in the standard situation, such that the only canonical connection there is the basic connection; using the basic connections also for $E$-valued tensors is then in alignment to $\mathrm{T}N$-valued tensors.
\end{remark}

Hence, we finally arrived at defining the infinitesimal gauge transformation of functionals.

\begin{definitions}{Infinitesimal gauge transformation of functionals}{TotalInfGaugeTrafoYayy}
Let $M, N$ be two smooth manifolds, $E \to N$ a Lie algebroid over $N$, $V \to N$ a vector bundle, $\nabla$ a connection on $E$, ${}^E\nabla$ an $E$-connection on $V$, and $\varepsilon \in \mathcal{F}^0_E(M; {}^*E)$ together with the unique $\Psi_\varepsilon \in \mathfrak{X}^E(\mathfrak{M}_E(M; N))$ as given uniquely in Prop.~\ref{prop:VariationOfA}. For the functional space $\mathcal{F}^\bullet_E(M; {}^*V)$ let $\delta_{\Psi_\varepsilon}$ be the unique operator as in Prop.~\ref{prop:VariationVonSkalarZeugsEasyPeasy}, using ${}^E\nabla$ as $E$-connection on $V$.

Then we define the \textbf{infinitesimal gauge transformation $\gls{1delta0epsilon} L$ of $L \in \mathcal{F}^\bullet_E(M; {}^*V)$} as an element of $\mathcal{F}^\bullet_E(M; {}^*V)$ by
\ba
\delta_\varepsilon L
&\coloneqq
\delta_{\Psi_\varepsilon} L.
\ea

For $V=E$ or $V= \mathrm{T}N$ we take ${}^E\nabla = \nabla^{\mathrm{bas}}$ on $E$ and $\mathrm{T}N$, respectively; for all further tensor spaces constructed of $E$ and $\mathrm{T}N$, like their duals, we take the canonical extensions of the basic connection.
\end{definitions}

\begin{remark}
\leavevmode\newline
In the following we will have just one connection $\nabla$ on $E$ and ${}^E\nabla$ on $V$ given. Without mentioning it further, we always use these connections for the definition of $\delta_\varepsilon$ because it should be clear by context.
\end{remark}

We can quickly list two properties about $\delta_\varepsilon$.

\begin{corollaries}{Linearity in $\varepsilon$}{DeltaEpsilonIstLinearInEpsilon}
Let us assume the same as for Def.~\ref{def:TotalInfGaugeTrafoYayy}. Then
\ba
\delta_{\alpha \varepsilon + \beta \vartheta}
&=
\alpha \delta_\varepsilon
	+ \beta \delta_\vartheta
\ea
for all $\alpha, \beta \in \mathbb{R}$ and $\varepsilon, \vartheta \in \mathcal{F}^0_E(M; {}^*E)$.
\end{corollaries}

\begin{proof}
\leavevmode\newline
Let $k \in \mathbb{N}_0$, $L \in \mathcal{F}^k_E(M; {}^*V)$ and $\mleft( e_a \mright)_a$ a local frame of $V$. Then, using Eq.~\eqref{LinearityOfPsiEpsilon} and the Leibniz rule,
\bas
\delta_{\alpha \varepsilon + \beta \vartheta} L
&=
\underbrace{\mathcal{L}_{\Psi_{\alpha \varepsilon + \beta \vartheta}} L^a}
_{\mathclap{ \stackrel{Eq.~\eqref{LinearityOfPsiEpsilon}}{=} \mathcal{L}_{\alpha \Psi_\varepsilon + \beta \Psi_\vartheta} } }
 \otimes ~ {}^*e_a
	- L^a \otimes {}^*\mleft( {}^E\nabla_{\alpha \varepsilon + \beta \vartheta} e_a \mright)
\\
&=
\alpha ~ \mleft(
	\mathcal{L}_{\Psi_\varepsilon} L^a \otimes {}^*e_a
	- L^a \otimes {}^*\mleft( {}^E\nabla_\varepsilon e_a \mright)
\mright)
	+ \beta ~ \mleft(
	\mathcal{L}_{\Psi_\vartheta} L^a \otimes {}^*e_a
	- L^a \otimes {}^*\mleft( {}^E\nabla_\vartheta e_a \mright)
\mright)
\\
&=
\mleft(\alpha \delta_\varepsilon
	+ \beta \delta_\vartheta\mright) L,
\eas
where vector fields like $\Psi_\varepsilon$ are given by Def.~\ref{def:TotalInfGaugeTrafoYayy}.
\end{proof}

\begin{corollaries}{Independence of $\nabla$}{WennVonAUnabhaengigDannAuchVonNabla}
Let us assume the same as for Def.~\ref{def:TotalInfGaugeTrafoYayy}, and let $L \in \mathcal{F}^k_E(M; {}^*V)$ ($k \in \mathbb{N}_0$) be independent of $A$, \textit{i.e.}~$L(\Phi,A) = L(\Phi, A^\prime)$ for all $(\Phi,A), (\Phi, A^\prime) \in \mathfrak{M}_E(M; N)$.

Then the definition of $\delta_\varepsilon L$ is independent of $\nabla$.\footnote{But not of ${}^E\nabla$, so, if ${}^E\nabla = \nabla^{\mathrm{bas}}$, then there is still the dependency on $\nabla$ in the role of ${}^E\nabla$.}
\end{corollaries}

\begin{remark}
\leavevmode\newline
The independence mentioned in Remark \ref{RemLeibnizeRegelaufProdukteWeshalbEConnectionNichtWichtigIst} is about ${}^E\nabla$, not $\nabla$. Eq.~\eqref{eqGaugeTrafoOfAacomps} shows clearly that $\nabla$ contributes to $\delta_\varepsilon$ in general, that is, the definition of $\Psi_\varepsilon$ is certainly dependent on $\nabla$, where $\Psi_\varepsilon$ is given by Def.~\ref{def:TotalInfGaugeTrafoYayy}.
\end{remark}

\begin{proof}
\leavevmode\newline
Let $\mleft( e_a \mright)_a$ be a local frame of $V$, and write $L = L^a \otimes {}^*e_a$, then, using that $\delta_\varepsilon = \mathcal{L}_{\Psi_\varepsilon}$ on $\mathcal{F}^k_E(M)$ (recall Remark \ref{RemLeibnizeRegelaufProdukteWeshalbEConnectionNichtWichtigIst}, and $\Psi_\varepsilon$ is given by Def.~\ref{def:TotalInfGaugeTrafoYayy}),
\bas
\delta_\varepsilon L
&=
\mathcal{L}_{\Psi_\varepsilon} L^a \otimes {}^*e_a
	- L^a \otimes {}^*\mleft({}^E\nabla_\varepsilon e_a \mright).
\eas
The second summand is already independent of $\nabla$, so, let us look at the first summand. Recall that $\Psi_\varepsilon$ contains two components, the first is the differentiation along the "$\Phi$-direction", given by $-({}^*\rho)(\varepsilon)$, and the second for the "$A$-direction", fixed by Prop.~\ref{prop:VariationOfA} using $\nabla$. Due to the independence of $L$ with respect to $A$ we can conclude that $L^a$ must be independent of $A$ since ${}^* e_a$ is already independent of $A$, thus,
\bas
\mathcal{L}_{\Psi} L^a
&=
\mathcal{L}_{\Psi^\prime} L^a
\eas
for all $\Psi, \Psi^\prime \in \mathfrak{X}(\mathfrak{M}_E(M; N))$ whose first component, the derivative along "$\Phi$"-coordinates, coincide. Hence, regardless which connection $\nabla$ we choose to fix the second component of $\Psi_\varepsilon$ the definition of $\delta_\varepsilon L$ will be unaffected by this choice.
\end{proof}

\subsection{Curvature of gauge transformations}\label{CurvatureOfGaugePart1}

We want to calculate
\bas
\delta_\vartheta \delta_\varepsilon - \delta_\varepsilon \delta_\vartheta
\eas
for all $\varepsilon, \vartheta \in \mathcal{F}^0_E(M; {}^*E)$, and we want a behaviour similar to representations. For $\Phi \in C^\infty(M;N)$, $\Phi^*E$ is in general not a Lie algebroid, see \cite[\S 3.2ff.]{meinrenkensplitting} or \cite[\S 7.4; page 42ff.]{meinrenkenlie} about conditions on $\Phi$ which imply a natural Lie algebroid structure on $\Phi^*E$. Therefore we cannot expect to have a Lie bracket on sections of $\Phi^*E$. The essential problem is that we do not have an anchor on $\Phi^*E\to M$ in general such that one cannot try to construct first a bracket on pullbacks of sections and then to canonically extend such a bracket (similar to previous constructions), and this problem extends to $\mathcal{F}^0_E(M; {}^*E)$. But there is a better object measuring a "bracket-like" behaviour on this functional space; we will see at the end that this will be actually a Lie bracket.

\begin{definitions}{Pre-bracket on $\mathcal{F}^0_E(M; {}^*E)$}{PrebracketonPullbackLiealgebroid}
Let $M, N$ be smooth manifolds, $E \to N$ a Lie algebroid, and $\nabla$ a connection on $E$.

Then we define the \textbf{pre-bracket $\gls{1Delta}: \mathcal{F}^0_E(M; {}^*E) \times \mathcal{F}^0_E(M; {}^*E) \to \mathcal{F}^0_E(M; {}^*E)$} by
\ba
\Delta(\vartheta, \varepsilon)
&\coloneqq
\delta_\varepsilon \vartheta - \delta_\vartheta \varepsilon - \bigl( {}^*t_{\nabla^{\mathrm{bas}}} \bigr)\mleft( \vartheta, \varepsilon \mright)
%\mleft.\frac{\mathrm{D}_{\mleft(\Psi_{\varepsilon}(p), {}^E\nabla\mright)}}{\mathrm{d}t}\mright|_{t=0}
		%\biggl[ t \mapsto \vartheta_p\mleft[ \widetilde{\Phi}_{\varepsilon, t} \mright] \mleft[ \widetilde{A}_{\varepsilon, t} \mright] \biggr] \nonumber \\
%%%%%%%%%%%%%%%%%%%%%%%%%%%
%&\hspace{1cm}
	%- \mleft.\frac{\mathrm{D}_{\mleft(\Psi_{\vartheta}(p), {}^E\nabla\mright)}}{\mathrm{d}t}\mright|_{t=0}
	%\biggl[ t \mapsto \varepsilon_p\mleft[ \widetilde{\Phi}_{\vartheta, t} \mright] \mleft[ \widetilde{A}_{\vartheta, t} \mright] \biggr] \nonumber \\
%&\hspace{1cm}
	%- \mleft. \bigl( \Phi^*t_{{}^E\nabla} \bigr)\mleft( \vartheta, \varepsilon \mright)\mright|_p
\ea
for all $\varepsilon, \vartheta \in \mathcal{F}^0_E(M; {}^*E)$.
%When we want to emphasize which $E$-connection was used in the definition, then we write $\Delta^{(1)}$ when using $\nabla_\rho$ and $\Delta$ in the case of $\nabla^{\mathrm{bas}}$, thence,
%\bas
%\Delta(\vartheta, \varepsilon)
%&=
%\delta_\varepsilon \vartheta - \delta_\vartheta \varepsilon - \bigl( \Phi^*t_{{}^E\nabla} \bigr)\mleft( \vartheta, \varepsilon \mright).
%\eas
\end{definitions}

\begin{remark}\label{IdeaOfPrebracket}
\leavevmode\newline
Given an $E$-connection ${}^E\nabla$ on $E$, Lie brackets can be expressed as
\bas
\mleft[ \mu, \nu \mright]_E
&=
{}^E\nabla_\mu \nu
	- {}^E\nabla_\nu \mu
	- t_{{}^E\nabla}(\mu, \nu)
\eas
for all $\mu, \nu \in \Gamma(E)$. Recall that $\delta$ is strongly related to a certain pullback of $\nabla^{\mathrm{bas}}$; then the idea of the pre-bracket is to use the right-hand side as a definition. Since we know under which conditions and how to make pullbacks of $E$-connections and tensors, we circumvent the problem of defining a Lie bracket and anchor on a pullback bundle.
\end{remark}

Let us study this bracket.

\begin{propositions}{Properties of the pre-bracket}{PropertiesOfThePreBracket}
Let $M, N$ be smooth manifolds, $E \to N$ a Lie algebroid, and $\nabla$ a connection on $E$.

Then we have
\ba\label{DeltaIstZumGlueckANtisymm}
\Delta &\textit{ is antisymmetric}, \\
\Delta &\textit{ is $\mathbb{R}$-bilinear}, \\
%\Delta(\vartheta, f \varepsilon)_p
%&=
%f(p) ~ \Delta(\vartheta, \varepsilon)_p
	%+ \mleft.\frac{\mathrm{d}}{\mathrm{d}t} \mright|_{t=0} \biggl[ t \mapsto f \mleft[ \widetilde{\Phi}_{\vartheta, t} \mright] \mleft[ \widetilde{A}_{\vartheta, t} \mright](p) \biggr] ~ \varepsilon, \label{eqLeibnizRuleofPreBracket} \\
\Delta\mleft( {}^*\mu, {}^*\nu \mright)
&=
{}^*\bigl( \mleft[ \mu, \nu \mright]_E \bigr) \label{EqLieKlammerAufPullBackSections}
\ea
for all $\varepsilon, \vartheta \in \mathcal{F}^0_E(M; {}^*E)$, $f \in \mathcal{F}^0_E(M)$, $\mu, \nu \in \Gamma(E)$, and, when expressing everything with respect to a pull-back of a local frame $\mleft( e_a \mright)_a$ of $E$, we get
\ba\label{EqDeltaInFrameKoord}
\Delta\mleft( \vartheta, \varepsilon \mright)
&=
\delta_{\varepsilon} \vartheta^a ~ {}^*e_a
	- \delta_{\vartheta} \varepsilon^a ~ {}^*e_a
	+ \vartheta^a \varepsilon^b ~  {}^*\bigl( 
	\mleft[ e_a, e_b \mright]_E
	\bigr)
\ea
for all $\vartheta, \varepsilon \in \mathcal{F}^0_E(M; {}^*E)$.

Moreover, $\Delta(\vartheta, \varepsilon)$ is independent of the chosen connection $\nabla$ when both, $\varepsilon$ and $\vartheta$, are independent of $A$, that is, $\varepsilon(\Phi, A) = \varepsilon (\Phi, A^\prime)$ for all $(\Phi, A), (\Phi, A^\prime) \in \mathfrak{M}_E(M;N)$; similar for $\vartheta$.
\end{propositions}
%
%\begin{remark}
%\leavevmode\newline
%It is well-defined to speak about sections of $\Gamma(\Phi^*E)$ independent of $A$ as sections whose components are locally independent of $A$, because a change of the frame $\mleft(e_a\mright)_a$, $f_a = M_a^b e_b$ ($M_a^b$ an invertible matrix), would just contribute to $\Phi$-dependencies due to
%\bas
%\Phi^*\mleft( f_a \mright)
%&=
%\Phi^*\mleft( M_a^b e_b \mright)
%=
%\mleft( M_a^b \circ \Phi \mright) ~ \Phi^*e_b.
%\eas
%\end{remark}

\begin{remark}\label{ClassicalCommutatorRemark}
\leavevmode\newline
Eq.~\eqref{EqLieKlammerAufPullBackSections} and \eqref{EqDeltaInFrameKoord} emphasize that we have a suitable candidate in $\Delta$ as bracket. The latter actually proves that $\Delta$ is independent of the choice about whether or not one uses the basic connection to define $\delta_\varepsilon$ because the infinitesimal gauge transformation of scalar-valued functionals is just a Lie derivative, see also Remark \ref{RemarkBracketIsVeryIndependent}. Similar to how one can express $\mleft[ \cdot, \cdot \mright]_E$ using Lie algebroid connections as in Remark \ref{IdeaOfPrebracket}, but $\mleft[ \cdot, \cdot \mright]_E$ is of course independent of any choice of Lie algebroid connection.

Let $E = N \times \mathfrak{g}$ be an action Lie algebroid, the usual relationship in classical gauge theory is for $\varepsilon, \vartheta \in C^\infty(M;\mathfrak{g})$ that
\bas
\mleft[ \delta^{\mathrm{clas}}_\varepsilon, \delta^{\mathrm{clas}}_\vartheta \mright]A
&=
- \delta^{\mathrm{clas}}_{\mleft[\varepsilon, \vartheta\mright]_{\mathfrak{g}}}A,
\eas
where $\delta^{\mathrm{clas}}_\varepsilon$ is given by Def.~\ref{def:ClassFunctionalGaugeTrafoBlag}, and the negative sign on the right hand side is due to our choice of sign with respect to $\varepsilon$, which we prove later in full generality. As we discussed, we apply the "bookkeeping trick" to formulate infinitesimal gauge transformations, also recall Def.~\ref{def:InfinitesimalGaugeTrafoClassicAsConnection} and Thm.~\ref{thm:RecoverOfClassicInfgGaugeTrafo}. That is, for a constant frame $\mleft( e_a \mright)_a$ of $E$, we have the "bookkeeping trick" $\iota(\varepsilon)$ given by
\bas
\iota(\varepsilon)
&=
\varepsilon^a ~ {}^*e_a,
\eas
hence, the bookeeping trick is essentially a frame-dependent embedding of the functionals given in the classical gauge theory into $\mathcal{F}^\bullet_E$. $\varepsilon^a$ are in this case only functions depending on $M$, but not on $\mathfrak{M}_E(M;N)$, especially, $\delta^{\mathrm{clas}}_\vartheta \varepsilon^a = 0$. By Eq.~\eqref{EqDeltaInFrameKoord} we then have
\bas
\Delta\bigl( \iota(\vartheta), \iota(\varepsilon) \bigr)
&=
\vartheta^a \varepsilon^b ~ {}^*\bigl( 
	\mleft[ e_a, e_b \mright]_{\mathfrak{g}}
	\bigr)
=
\iota\mleft(\mleft[ \vartheta, \varepsilon \mright]_{\mathfrak{g}}\mright),
\eas
which is precisely what we want and expect of a generalized bracket.
\end{remark}

\begin{proof}[Proof of Prop.~\ref{prop:PropertiesOfThePreBracket}]
\leavevmode\newline
%The Leibniz rule \eqref{eqLeibnizRuleofPreBracket} is trivial and follows directly by the definition of $\mathrm{D}/\mathrm{d}t$, Prop.~\ref{prop:DerivationAlongEPath}.
The antisymmetry is clear, and the bilinearity follows by the linearity of $\delta_\varepsilon$ for all $\varepsilon \in \mathcal{F}^0_E(M; {}^*E)$, recall Cor.~\ref{cor:DeltaEpsilonIstLinearInEpsilon}. We have
\bas
\bigl( {}^*t_{\nabla^{\mathrm{bas}}} \bigr)\mleft( {}^*\mu, {}^*\nu \mright)
&=
{}^*\mleft( \bigl( t_{\nabla^{\mathrm{bas}}} \bigr)\mleft( \mu, \nu \mright) \mright)
=
{}^*\mleft( 
	\nabla^{\mathrm{bas}}_\mu \nu 
	- \nabla^{\mathrm{bas}}_\nu \mu 
	- \mleft[ \mu, \nu \mright]_E 
\mright)
\eas
for all $\mu, \nu \in \Gamma(E)$, and
\bas
\delta_{{}^*\nu} \mleft( {}^*\mu \mright)
&=
- {}^*\mleft( \nabla^{\mathrm{bas}}_\nu \mu \mright),
\eas
therefore
\bas
\Delta\mleft( {}^*\mu, {}^*\nu \mright)
&=
{}^*\mleft( \nabla^{\mathrm{bas}}_\mu \nu \mright)
	- {}^*\mleft( \nabla^{\mathrm{bas}}_\nu \mu \mright)
	- {}^*\mleft( 
	\nabla^{\mathrm{bas}}_\mu \nu 
	- \nabla^{\mathrm{bas}}_\nu \mu 
	- \mleft[ \mu, \nu \mright]_E 
\mright)
=
{}^*\bigl( \mleft[ \mu, \nu \mright]_E \bigr),
\eas
which proves Eq.~\eqref{EqLieKlammerAufPullBackSections}.
%We define the following notation
%\ba\label{DefKurzeNotationFuerDenGanzenScheiss}
%\mleft.\frac{\mathrm{d}}{\mathrm{d}t} \mright|_{t=0} \biggl[ t \mapsto f \mleft[ \widetilde{\Phi}_{-\vartheta, t} \mright] \mleft[ \widetilde{A}_{-\vartheta, t} \mright](p) \biggr]
%&\eqqcolon
%\dot{f}_{(\vartheta_p)}(0)
%\ea
%for all $f \in C^\infty(M)$, $\vartheta \in \Gamma(\Phi^*E)$ and $p \in M$. 
For $\varepsilon, \vartheta \in \mathcal{F}^0_E(M;{}^*E)$ we have, with respect to a frame $\mleft( e_a \mright)_a$ of $E$,
\bas
%\mleft.\frac{\mathrm{D}_{\mleft(\Psi_{-\vartheta}(p), {}^E\nabla\mright)}}{\mathrm{d}t}\mright|_{t=0}
	%\biggl[ t \mapsto \varepsilon_p\mleft[ \widetilde{\Phi}_{-\vartheta, t} \mright] \mleft[ \widetilde{A}_{-\vartheta, t} \mright] \biggr]
\delta_\vartheta \varepsilon
&=
\delta_{\vartheta} \varepsilon^a ~ {}^*e_a
	- \varepsilon^a \vartheta^b ~ {}^*\mleft(\nabla^{\mathrm{bas}}_{e_b} e_a\mright),
\eas
and so
\bas
\Delta(\vartheta, \varepsilon)
&=
\delta_{\varepsilon} \vartheta^a ~ {}^*e_a
	- \vartheta^a \varepsilon^b ~ {}^*\mleft(\nabla^{\mathrm{bas}}_{e_b} e_a\mright)
%\\
%&\hspace{1cm}
	- \delta_{\vartheta} \varepsilon^a ~ {}^*e_a
	+ \varepsilon^a \vartheta^b ~ {}^*\mleft(\nabla^{\mathrm{bas}}_{e_b} e_a\mright)
\\
&\hspace{1cm}
	- \vartheta^a \varepsilon^b ~ {}^*\mleft( 
	\nabla^{\mathrm{bas}}_{e_a} e_b 
	- \nabla^{\mathrm{bas}}_{e_b} e_a 
	- \mleft[ e_a, e_b \mright]_E
	\mright) \\
&=
\delta_{\varepsilon} \vartheta^a ~ {}^*e_a
	- \delta_{\vartheta} \varepsilon^a ~ {}^*e_a
	+ \vartheta^a \varepsilon^b ~ 
	{}^*\bigl( 
		\mleft[ e_a, e_b \mright]_E
	\bigr).
\eas
This expression for $\Delta(\vartheta, \varepsilon)$ shows that its value is independent of the chosen $\nabla$, when the functionals $\varepsilon = \varepsilon^a \otimes {}^*e_a$ and $\vartheta= \vartheta^a \otimes {}^*e_a$ are independent of $A$, since then also their components with respect to $\mleft({}^*e_a\mright)_a$ are independent of $A$ because ${}^*e_a$ is already independent of $A$. Then apply Cor.~\ref{cor:WennVonAUnabhaengigDannAuchVonNabla}.
\end{proof}

\begin{corollaries}{$\Delta$ a Lie bracket on the pull-backs of $\Gamma(E)$}{DeltaIstEineLieklammerAufPullbACkSections}
Let $M, N$ be smooth manifolds, $E \to N$ a Lie algebroid, and $\nabla$ a connection on $E$.

Then the restriction of $\Delta$ on pullback functionals is a Lie bracket.
\end{corollaries}

\begin{proof}
\leavevmode\newline
The antisymmetry, the bilinearity over $\mathbb{R}$ and the closedness follow by Prop.~\ref{prop:PropertiesOfThePreBracket}, the same also for the Jacobi identity by observing
\bas
\Delta \mleft( {}^*\mu, \Delta \mleft( {}^*\nu, {}^*\eta \mright) \mright)
&\stackrel{\text{Eq.~\eqref{EqLieKlammerAufPullBackSections}}}{=}
\Delta \mleft( {}^*\mu, {}^*\mleft(\mleft[ \nu,\eta \mright]_E\mright) \mright)
\stackrel{\text{Eq.~\eqref{EqLieKlammerAufPullBackSections}}}{=}
{}^* \mleft( \mleft[ \mu, \mleft[ \nu, \eta \mright]_E \mright]_E\mright)
\eas
for all $\mu, \nu, \eta \in \Gamma(E)$. Hence, the Jacobiator of the restriction of $\Delta$ on pullback functionals is given by the pullback of the Jacobiator of $\mleft[ \cdot, \cdot \mright]_E$, the latter is of course zero.
\end{proof}

We will see that $\Delta$ is actually always a Lie bracket, but for proving this we do not want to show the Jacobi identity directly, due to how we constructed it we rather are going to use the equivalence with Bianchi identities of curvatures; recall the proof of Thm.~\ref{thm:1stBianchi}. Hence, let us define the curvature we are interested into.

\begin{definitions}{Curvature of infinitesimal gauge transformations}{ErsteKruemmungsFormelFuerEichtrafos}
Let $M, N$ be smooth manifolds, $E \to N$ a Lie algebroid, $V \to N$ a vector bundle, $\nabla$ a connection on $E$, and ${}^E\nabla$ an $E$-connection on $V$.

Then we define the \textbf{curvature $\gls{Rdelta}$} by
\ba
\mathcal{F}^0_E(M;{}^*E) \times \mathcal{F}^0_E(M;{}^*E) \times \mathcal{F}^k_E(M;{}^*V) &\to \mathcal{F}^k_E(M;{}^*V)
\nonumber \\
(\vartheta, \varepsilon, L) &\mapsto R_{\delta}(\vartheta, \varepsilon)L
\coloneqq
	\delta_\vartheta \delta_\varepsilon L 
	- \delta_\varepsilon \delta_\vartheta L 
	+ \delta_{\Delta(\vartheta, \varepsilon)} L
\ea
for all $\vartheta, \varepsilon \in \mathcal{F}^0_E(M; {}^*E)$ and $L \in \mathcal{F}^k_E(M;{}^*V)$ ($k \in \mathbb{N}_0$ arbitrary).

In alignment to Def.~\ref{def:GaugeTrafoOfA} we denote $R_\delta(\cdot, \cdot) A \coloneqq R_\delta (\cdot, \cdot) \varpi_2$, and $R_\delta(\cdot, \cdot)A^a \coloneqq R_\delta(\cdot, \cdot)\varpi_2^a$ with respect to a frame $\mleft(e_a\mright)_a$ of $E$.
\end{definitions}

\begin{remark}
\leavevmode\newline
The sign in front of the third term depends on which sign one takes in the definition of $\delta_\varepsilon$. Changing the sign $\varepsilon$ in the definitions of the gauge tranformations would lead to a minus sign in front of the third summand.
\end{remark}

Using a frame of $E$ we can apply the Leibniz rule.

\begin{corollaries}{Relationships between curvatures}{RelationShipsOfCurvatures}
Let $M, N$ be smooth manifolds, $E \to N$ a Lie algebroid, $V \to N$ a vector bundle, $\nabla$ a connection on $E$, and ${}^E\nabla$ an $E$-connection on $V$. Then locally
\ba
R_\delta(\cdot, \cdot)L
&=
R_\delta(\cdot, \cdot)L^a \otimes {}^*e_a
	+ L^a \otimes {}^*\bigl( R_{{}^E\nabla}(\cdot, \cdot)e_a \bigr) 
\ea
for all $L \in \mathcal{F}^k_E(M; {}^*V)$ ($k \in \mathbb{N}_0$), where $\mleft( e_a \mright)_a$ is a local frame of $E$ and viewing $R_{{}^E\nabla}(\cdot, \cdot)e_a$ as an element of $\Omega^2(E;E)$.
\end{corollaries}

\begin{proof}
\leavevmode\newline
Let us first study terms like $R_{\delta}(\vartheta, \varepsilon)\mleft({}^*h\mright)$ for $\varepsilon, \vartheta \in \mathcal{F}^0_E(M;{}^*E)$ and $h \in \Gamma(V)$, using a local frame $\mleft( e_a \mright)_a$ of $E$,
\bas
\delta_\vartheta \delta_\varepsilon ({}^*h)
&=
- \delta_\vartheta \mleft(
	\varepsilon^a ~ {}^*\mleft({}^E\nabla_{e_a} h\mright)
\mright)
=
- \delta_\vartheta \varepsilon^a ~ {}^*\mleft({}^E\nabla_{e_a} h\mright)
	+ \varepsilon^a \vartheta^b ~ {}^*\mleft({}^E\nabla_{e_b} {}^E\nabla_{e_a} h\mright),
\eas
and
\bas
\delta_{\Delta(\vartheta, \varepsilon)} ({}^*h)
~~~~&\stackrel{\mathclap{\text{Eq.~\eqref{EqDeltaInFrameKoord}}}}{=}~~~~
- \mleft(
	\delta_{\varepsilon} \vartheta^a
	- \delta_{\vartheta} \varepsilon^a
	+ \vartheta^b ~ \varepsilon^c ~  \mleft({}^*\bigl( 
	\mleft[ e_b, e_c \mright]_E
	\bigr)\mright)^a
\mright) ~ {}^*\mleft({}^E\nabla_{e_a} h\mright)
\\
&=
\delta_\vartheta \varepsilon^a ~ {}^*\mleft({}^E\nabla_{e_a} h\mright)
	- \delta_\varepsilon \vartheta^a ~ {}^*\mleft({}^E\nabla_{e_a} h\mright)
	- \varepsilon^a \vartheta^b ~ {}^*\mleft( {}^E\nabla_{\mleft[ e_b, e_a \mright]_E} h \mright),
\eas
in total
\bas
R_{\delta}(\vartheta, \varepsilon)\mleft({}^*h\mright)
&=
\varepsilon^a \vartheta^b ~ {}^* \underbrace{\mleft(
	{}^E\nabla_{e_b} {}^E\nabla_{e_a} h
	- {}^E\nabla_{e_a} {}^E\nabla_{e_b} h
	- {}^E\nabla_{\mleft[ e_a, e_b \mright]_E} h
\mright)}_{R_{{}^E\nabla}(e_b,e_a)h}
=
\mleft(
	{}^*\bigl( R_{{}^E\nabla}(\cdot,\cdot)h \bigr)
\mright)(\vartheta, \varepsilon).
\eas
Therefore we arrive at
\bas
R_\delta(\vartheta, \varepsilon)\mleft(L^a \otimes {}^*e_a\mright)
&=
\delta_\vartheta \delta_\varepsilon L^a \otimes {}^*e_a
	+ \delta_\varepsilon L^a \otimes \delta_\vartheta\mleft({}^*e_a\mright)
	+ \delta_\vartheta L^a \otimes \delta_\varepsilon\mleft({}^*e_a\mright)
	+ L^a \otimes \delta_\vartheta \delta_\varepsilon \mleft({}^*e_a\mright)
\\
&\hspace{1cm}
	- (\vartheta \leftrightarrow \varepsilon)
\\
&\hspace{1cm}
	+ \delta_{\Delta(\vartheta, \varepsilon)} L^a \otimes {}^*e_a
	+ L^a \otimes \delta_{\Delta(\vartheta, \varepsilon)} {}^*e_a
\\
&=
R_\delta(\vartheta, \varepsilon)L^a \otimes {}^*e_a
	+ L^a \otimes R_\delta(\vartheta, \varepsilon) ({}^*e_a)
\\
&=
R_\delta(\cdot, \cdot)L^a \otimes {}^*e_a
	+ L^a \otimes \mleft({}^*\bigl( R_{{}^E\nabla}(\cdot, \cdot)e_a \bigr) \mright)(\vartheta, \varepsilon)
\eas
for all $L \in \mathcal{F}^k_E(M; {}^*V)$.
\end{proof}

Keep in mind that $R_\delta$ is not a typical curvature, for example $\delta_\varepsilon$ is not $C^\infty$-linear with respect to $\varepsilon$, such that it is not immediately clear whether this curvature is a tensor in all arguments, so, we need to prove this if we want to simplify calculations. We are first focusing on $R_\delta(\cdot, \cdot) A$. 

\begin{propositions}{$R_{\delta}$ is a tensor}{WirHabenEinenTensorBeiderTrafoKruemmung}
Let $M, N$ be smooth manifolds, $E \to N$ a Lie algebroid, and $\nabla$ a connection on $E$.

Then $R_{\delta}(\cdot, \cdot)A$ is an anti-symmetric tensor, \textit{i.e.}~anti-symmetric and $\mathcal{F}^0_E(M)$-bilinear, and we have
\ba\label{SplittingVonDerEichtrafo}
%R_{\delta}({}^*\mu, {}^*\nu)A
%&=
%\bigl(R_{\delta}({}^*\mu, {}^*\nu)A^a\bigr)
	%\otimes {}^*e_a
	%+ \mleft({}^*\bigl( R_{\nabla^{\mathrm{bas}}}(\mu, \nu)\bigr) \mright) A 
%\ea
%for all $\mu, \nu \in \Gamma(E)$, where $\mleft( e_a \mright)_a$ is a local frame of $E$, and
%%such that we write $A^a \otimes {}^*e_a$, and 
%we view $R_{\nabla^{\mathrm{bas}}}(\mu,\nu)$ as an element of $\Omega^1(E;E)$. If we introduce the notation $R_{\nabla^{\mathrm{bas}}}(\mu, \nu, X) \coloneqq R_{\nabla^{\mathrm{bas}}}(\mu, \nu) X$ for all $\mu, \nu \in \Gamma(E)$ and $X \in \mathfrak{X}(N)$, then we have a more general formula given by 
%\ba\label{SplittingVonDerEichtrafoGeneral}
R_{\delta}(\varepsilon, \vartheta)A
&=
R_{\delta}(\varepsilon, \vartheta)A^a
	\otimes {}^*e_a
	+ \mleft({}^* R_{\nabla^{\mathrm{bas}}} \mright)( \varepsilon, \vartheta) A
\ea
for all $\varepsilon, \vartheta \in \mathcal{F}^0_E(M; {}^*E)$.
\end{propositions}

\begin{proof}
\leavevmode\newline
%For simplicity we will omit to add $(\Phi, A)$ at the gauge transformations.
$\bullet$ The antisymmetry is clear by Prop.~\ref{prop:PropertiesOfThePreBracket}. Fix a local frame $\mleft( e_a \mright)_a$ of $E$, then we have
\bas
\delta_\vartheta \delta_{f\varepsilon}A
~~~~&\stackrel{\mathclap{\text{Def.~\ref{def:GaugeTrafoOfA}}}}{=}~~~~
- \delta_\vartheta \bigl( \mleft({}^*\nabla \mright) (f \varepsilon) \bigr) \\
&=
- \delta_\vartheta \bigl( \mathrm{d}f \otimes \varepsilon 
+ f ~ \mleft({}^*\nabla \mright) \varepsilon \bigr) \\
&=
- \delta_\vartheta \mathrm{d} f \otimes \varepsilon
	- \mathrm{d} f \otimes \delta_\vartheta \varepsilon
%- \delta_\vartheta \bigl( \mathrm{d}f ~ \varepsilon^a \bigr) \otimes {}^*e_a 
	%+ \mathrm{d}f ~ \varepsilon^a  \otimes {}^*\mleft(\nabla^{\mathrm{bas}}_\vartheta e_a \mright)
	- \delta_\vartheta f ~ \mleft({}^*\nabla \mright) \varepsilon 
	- f \delta_\vartheta \bigl(\mleft({}^*\nabla \mright) \varepsilon \bigr)
\\
&=
- \delta_\vartheta \mathrm{d} f \otimes \varepsilon
	- \mathrm{d} f \otimes \delta_\vartheta \varepsilon^a ~ {}^*e_a
	+ \mathrm{d} f \otimes \varepsilon^a \vartheta^b ~ {}^*\mleft(\nabla^{\mathrm{bas}}_{e_b} e_a \mright)
	- \delta_\vartheta f ~ \mleft({}^*\nabla \mright) \varepsilon 
	+ f \delta_\vartheta \delta_\varepsilon A
\eas
for all $\vartheta, \varepsilon \in \mathcal{F}^0_E(M; {}^*E)$ and $f \in \mathcal{F}^0_E(M)$, and
\bas
-\delta_{f\varepsilon} \delta_\vartheta A
&=
\delta_{f\varepsilon} \bigl( \mleft( {}^*\nabla \mright) \vartheta \bigr) \\
&=
\delta_{f\varepsilon} \mleft( \mathrm{d}\vartheta^a \otimes {}^*e_a + \vartheta^b ~ {}^! \mleft( \nabla e_b \mright) \mright) \\
&\stackrel{\mathclap{\text{Eq.~\eqref{EqVariationVonFormenBrrrr}}}}{=}~~~~
\delta_{f\varepsilon} \mathrm{d}\vartheta^a \otimes {}^*e_a 
	- \mathrm{d}\vartheta^a \otimes f \varepsilon^b~ {}^*\mleft( \nabla^{\mathrm{bas}}_{e_b} e_a \mright) \\
&\hspace{1cm}
	+ \delta_{f\varepsilon} \vartheta^b ~ {}^!\mleft( \nabla e_b \mright) 
	- f \vartheta^b ~ {}^!\mleft( \nabla^{\mathrm{bas}}_{\varepsilon} (\nabla e_b) \mright)
	- \vartheta^b ~\underbrace{{}^*\mleft( \nabla_{({}^*\rho)\mleft( ({}^*\nabla) (f \varepsilon) \mright)} e_b \mright)}
	_{\mathclap{= \mathrm{d}f \otimes {}^*\mleft( \nabla_{\mleft({}^*\rho\mright)(\varepsilon)} e_b \mright)
		+ f \cdot \underbrace{(\dotsc)}_{\mathclap{\text{indep. of }f}}
		%{}^*\mleft( \nabla_{({}^*\rho)\mleft( ({}^*\nabla) \varepsilon \mright)} e_b \mright)
		}}
\\
&=
\delta_{f\varepsilon} \mathrm{d}\vartheta^a \otimes {}^*e_a 
	+ \delta_{f\varepsilon} \vartheta^b ~ {}^! \mleft( \nabla e_b \mright) 
	- \vartheta^b \varepsilon^a\mathrm{d}f \otimes {}^*\mleft( \nabla_{\rho(e_a)} e_b \mright)
	+ f \cdot \underbrace{(\dotsc)}_{\mathclap{\text{independent of }f}}.
\eas
Since we want to check the tensorial property, we can ignore the terms proportional to $f$; 
we also have
\bas
\delta_{\Delta(\vartheta, f \varepsilon)} A
&=
\mleft( {}^*\nabla \mright)  \mleft( \Delta(f \varepsilon, \vartheta) \mright) \\
%&=
%\mleft( {}^*\nabla \mright) 
%\bigl(
		%\delta_\vartheta (f \varepsilon)
		%- \delta_{f \varepsilon} \vartheta
		%- \bigl( {}^* t_{\nabla^{\mathrm{bas}}} \bigr)(\vartheta, f\varepsilon)
	%%\delta_{f\varepsilon} \vartheta^a(\Phi, A) ~ {}^*e_a
	%%- \delta_{\vartheta} \mleft(f \varepsilon^a\mright)(\Phi, A) ~ {}^*e_a
	%%+ \vartheta^b ~ f \varepsilon^a ~ {}^*\bigl( 
	%%\mleft[ e_b, e_a \mright]_E
	%%\bigr)
%\bigr) \\
%&=
%\mleft( {}^*\nabla \mright) 
%\mleft(
		%\delta_\vartheta f ~ \varepsilon
		%+ f \delta_\vartheta \varepsilon^a ~ {}^*e_a
		%\vphantom{{}^*\mleft( \nabla^{\mathrm{bas}}_{e_b} e_a \mright)} 
		%- \delta_{f \varepsilon} \vartheta^b ~ {}^*e_b \mright.
%\\
%&\hspace{2cm}\mleft.
		%+ f \varepsilon^a  \vartheta^b ~ {}^*\mleft( \nabla^{\mathrm{bas}}_{e_b} e_a \mright)
		%- f \varepsilon^a \vartheta^b ~ {}^*\mleft( \nabla^{\mathrm{bas}}_{e_a} e_b \mright)
		%- f \varepsilon^a \vartheta^b ~ \bigl( {}^* t_{\nabla^{\mathrm{bas}}} \bigr)({}^*e_a, {}^*e_b)
%\mright)
%\\
&\stackrel{\mathclap{\text{Eq.~\eqref{EqDeltaInFrameKoord}}}}{=}~~~~
\mleft( {}^*\nabla \mright)
\mleft(
		\delta_\vartheta f ~ \varepsilon
		+ f \delta_\vartheta \varepsilon^a ~ {}^*e_a 
		- \delta_{f \varepsilon} \vartheta^b ~ {}^*e_b
		+ f \varepsilon^a \vartheta^b~ {}^*\mleft( \mleft[ e_a, e_b \mright]_E \mright)
\mright)
\\
&\stackrel{\mathclap{\text{Eq.~\eqref{eqVariationVertauschtMitDifferential}}}}{=}~~~~
\delta_\vartheta \mathrm{d} f \otimes \varepsilon
	+ \delta_\vartheta f ~ ({}^*\nabla)\varepsilon
	+ \mathrm{d} f \otimes \delta_\vartheta \varepsilon^a ~ {}^*e_a 
	- \delta_{f \varepsilon}\mathrm{d} \vartheta^b \otimes {}^*e_b
	- \delta_{f \varepsilon} \vartheta^b ~ {}^!\mleft( \nabla e_b \mright)
\\
&\hspace{1cm}~~~~
	+ \varepsilon^a \vartheta^b~\mathrm{d}f \otimes {}^*\mleft( \mleft[ e_a, e_b \mright]_E \mright)
	+ f \cdot \underbrace{(\dotsc)}_{\mathclap{\text{independent of }f}}.
\eas
Hence, we get in total
\bas
R_{\delta}(\vartheta, f \varepsilon)A
&=
\varepsilon^a \vartheta^b \mathrm{d} f \otimes {}^*\underbrace{\mleft( 
	\nabla^{\mathrm{bas}}_{e_b} e_a
	- \nabla_{\rho(e_a)} e_b
	+ \mleft[ e_a, e_b \mright]_E
\mright)}_{= \nabla^{\mathrm{bas}}_{e_b} e_a - \nabla^{\mathrm{bas}}_{e_b} e_a = 0}
	+ f \cdot \underbrace{(\dotsc)}_{\mathclap{\text{independent of }f}}
\\
&=
f \cdot \underbrace{(\dotsc)}_{\mathclap{\text{independent of }f}}
\eas
for all $\vartheta, \varepsilon \in \mathcal{F}^0_E(M; {}^*E)$ and $f \in \mathcal{F}^0_E(M)$. Using the antisymmetry proves that $R_{\delta}(\cdot, \cdot)A$ is a tensor because the shown equation also holds for $f \equiv 1$ such that the remaining terms in the $f$-independent bracket are precisely giving rise to $R_{\delta}(\vartheta, \varepsilon)A$.

$\bullet$ Eq.~\eqref{SplittingVonDerEichtrafo} just follows by Cor.~\ref{cor:RelationShipsOfCurvatures}.
%Now we write $A = A^a \otimes {}^*e_a$, then
%\bas
%\delta_{\Delta(\varepsilon, \vartheta)} A
%&=
%\delta_{\Delta(\varepsilon, \vartheta)} A^a \otimes {}^*e_a
	%- \varepsilon^a \vartheta^b A^a \otimes {}^*\mleft( \nabla^{\mathrm{bas}}_{\mleft[ e_a, e_b \mright]_E} e_a \mright),
%\eas
%and
%\bas
%\delta_{{}^*\mu} \delta_{{}^*\nu} A
%&=
%\delta_{{}^*\mu} \mleft(  
	%\delta_{{}^*\nu} A^a \otimes {}^*e_a
	%- A^a \otimes {}^*\mleft( \nabla^{\mathrm{bas}}_\nu e_a \mright)
%\mright)
%\\
%&=
%\delta_{{}^*\mu} \delta_{{}^*\nu} A^a \otimes {}^*e_a
	%- \delta_{{}^*\nu} A^a \otimes {}^*\mleft( \nabla^{\mathrm{bas}}_\mu e_a \mright)
%\\
%&\hspace{1cm}
	%- \delta_{{}^*\mu} A^a \otimes {}^*\mleft( \nabla^{\mathrm{bas}}_\nu e_a \mright)
	%+ A^a \otimes {}^*\mleft( \nabla^{\mathrm{bas}}_\mu \nabla^{\mathrm{bas}}_\nu e_a \mright).
%\eas
%In total the mixed terms cancel, and we get
%\bas
%R_{\delta}({}^*\mu, {}^*\nu)A
%&=
%\mleft(
%\delta_{{}^*\mu} \delta_{{}^*\nu} A^a
	%- \delta_{{}^*\nu}\delta_{{}^*\mu} A^a
	%+ \delta_{{}^*\mleft( \mleft[ \mu, \nu \mright]_E \mright)} A^a
	%\mright) \otimes {}^*e_a 
	%+ A^a \otimes {}^*\bigl( R_{\nabla^{\mathrm{bas}}}(\mu, \nu) e_a \bigr).
	%%_{= \mleft( {}^*\mleft( R_{\nabla^{\mathrm{bas}}}(\mu, \nu) \mright) \mright) A}.
%\eas
%In the same fashion one shows Eq.~\eqref{SplittingVonDerEichtrafoGeneral}, making use of expressions like $\varepsilon = \varepsilon^a \otimes {}^*e_a$ to derive the second summand.
\end{proof}

Due to the tensorial behaviour, we can study $R_{\delta}(\cdot, \cdot)A$ just with respect to pullback functionals, such that the notations and calculations can be simplified.

\begin{theorems}{Curvature of the infinitesimal gauge transformation measured by the basic curvature}{CurvatureOfBasicStuffIsEquivalentForGaugeTrafoCurvature}
Let $M, N$ be smooth manifolds, $E \to N$ a Lie algebroid, and $\nabla$ a connection on $E$. Then
\ba
R_{\delta}({}^*\mu, {}^*\nu)A
&=
- {}^!\mleft( R^{\mathrm{bas}}_\nabla(\mu, \nu) \mright)
\ea
for all $\mu, \nu \in \Gamma(E)$, viewing $R^{\mathrm{bas}}_\nabla(\mu, \nu)$ as an element of $\Omega^1(N;E)$.
\end{theorems}

\begin{remark}
\leavevmode\newline
\indent $\bullet$ One can then derive with Eq.~\eqref{EqPullBackFormelFuerVerschiedeneDefinitionen} that
\bas
{}^!\mleft( R^{\mathrm{bas}}_\nabla(\mu, \nu) \mright)
&=
\mleft({}^*\mleft( R^{\mathrm{bas}}_\nabla(\mu, \nu) \mright)\mright) \mathrm{D}
=
\mleft({}^* R^{\mathrm{bas}}_\nabla\mright) ({}^*\mu, {}^*\nu) \mathrm{D},
\eas
viewing $\mathrm{D}$ as an element of $\mathcal{F}^1_E(M; {}^*\mathrm{T}N)$; recall Ex.~\ref{ex:DAsFunctional}.
Using that $R_\delta(\cdot, \cdot) A$ is tensorial and that pullbacks are generators as usual, we get
\bas
R_{\delta}(\varepsilon, \vartheta)A
&=
- \mleft({}^* R^{\mathrm{bas}}_\nabla\mright) (\varepsilon, \vartheta) \mathrm{D}
\eas
for all $\varepsilon, \vartheta \in \mathcal{F}^0_E(M; {}^*E)$.

$\bullet$ One could also view this theorem as a physical interpretation of the basic curvature.
\end{remark}

\begin{proof}[Proof of Thm.~\ref{thm:CurvatureOfBasicStuffIsEquivalentForGaugeTrafoCurvature}]
\leavevmode\newline
We have
\bas
\delta_{{}^*\mu} \mleft( \delta_{{}^*\nu} A \mright)
&=
- \delta_{{}^*\mu} \mleft( {}^!\mleft( \nabla \nu \mright) \mright)
\\
&\stackrel{\mathclap{\text{Eq.~\eqref{EqVariationVonFormenBrrrrVereinfacht}}}}{=}~~~~
	%\mleft({}^*\mleft(\nabla^{\mathrm{bas}}_\mu \mleft( \nabla \nu \mright)\mright)\mright)(\mathrm{D}\Phi)
	%+ {}^*\mleft( \nabla_{\rho\mleft(\nabla_{\mathrm{D}\Phi} \mu \mright)} \nu \mright)
%\\
%&=
{}^! \mleft(
	\nabla^{\mathrm{bas}}_\mu \mleft( \nabla \nu \mright)
	+ \nabla_{\rho\mleft(\nabla \mu \mright)} \nu
\mright),
\eas
and
\bas
\mleft(\nabla^{\mathrm{bas}}_\mu \mleft( \nabla \nu \mright)
	+ \nabla_{\rho\mleft(\nabla \mu \mright)} \nu\mright)(Y)
&=
\nabla^{\mathrm{bas}}_\mu \nabla_Y \nu
	- \nabla_{\nabla^{\mathrm{bas}}_\mu Y} \nu
	+ \nabla_{\rho\mleft(\nabla_Y \mu \mright)} \nu
\\
&=
\mleft[ \mu, \nabla_Y \nu \mright]_E
	+ \nabla_{\rho\mleft( \nabla_Y \nu \mright)} \mu
	- \nabla_{\mleft[ \rho(\mu), Y \mright]} \nu
\eas
for all $Y \in \mathfrak{X}(M)$. In total we would then look at the pull-back of the following form, also using Eq.~\eqref{EqLieKlammerAufPullBackSections},
\bas
&\mleft(
	\nabla^{\mathrm{bas}}_\mu \mleft( \nabla \nu \mright)
	+ \nabla_{\rho\mleft(\nabla \mu \mright)} \nu
	- \nabla^{\mathrm{bas}}_\nu \mleft( \nabla \mu \mright)
	- \nabla_{\rho\mleft(\nabla \nu \mright)} \mu
	- \nabla \mleft( \mleft[ \mu, \nu \mright]_E \mright)
\mright)(Y)
\\
&=
\mleft[ \mu, \nabla_Y \nu \mright]_E
	+ \nabla_{\rho\mleft( \nabla_Y \nu \mright)} \mu
	- \nabla_{\mleft[ \rho(\mu), Y \mright]} \nu
	-	\mleft[ \nu, \nabla_Y \mu \mright]_E
	- \nabla_{\rho\mleft( \nabla_Y \mu \mright)} \nu
	+ \nabla_{\mleft[ \rho(\nu), Y \mright]} \mu
	- \nabla_Y \mleft( \mleft[ \mu, \nu \mright]_E \mright)
\\
&=
- \mleft(
	\nabla_Y\mleft(\mleft[\mu, \nu\mright]_E\mright) 
	- \mleft[ \nabla_Y \mu, \nu \mright]_E 
	- \mleft[ \mu, \nabla_Y \nu \mright]_E 
	- \nabla_{\nabla^{\mathrm{bas}}_\nu Y} \mu 
	+ \nabla_{\nabla^{\mathrm{bas}}_\mu Y} \nu
\mright)
\\
&\stackrel{\mathclap{\text{Def.~\ref{def:basiccurvature}}}}{=}~~~~
- R^{\mathrm{bas}}_\nabla(\mu, \nu)Y.
\eas
Therefore we arrive at
\bas
R_{\delta}({}^*\mu, {}^*\nu)A
&=
- {}^!\mleft( R^{\mathrm{bas}}_\nabla(\mu, \nu)Y \mright).
\eas
\end{proof}

We get immediately the following statement.

\begin{corollaries}{Flat infinitesimal gauge transformation}{FlatnessVonEichtrafos}
Let $M, N$ be smooth manifolds, $E \to N$ a Lie algebroid, and $\nabla$ a connection on $E$ with $R^{\mathrm{bas}}_\nabla=0$. Then
\ba\label{DieKruemmungIstNullVonDenEichtrafosGeeeeil}
R_{\delta}(\cdot, \cdot)A
&=
0.
\ea
%while
%\ba
%R_{\delta^{(1)}}\mleft( {}^*\mu, {}^*\nu \mright)A
%&=
%\mleft( {}^*\bigl( R_{\nabla_{\rho}}(\mu, \nu)  \bigr) \mright) A
%\ea
%for all $\mu, \nu \in \Gamma(E)$.
With respect to a frame $\mleft( e_a \mright)_a$ of $E$ we then also have
\ba\label{CoordFuerEichKruemmungsRegel}
R_\delta (\cdot, \cdot) A^a
&=
0
\ea
for all $\mu, \nu \in \Gamma(E)$.
\end{corollaries}

\begin{remark}\label{RemarkUeberNablaRhoCurvatureForGauegTrafo}
\leavevmode\newline
\indent $\bullet$ This discussion, especially Cor.~\ref{cor:FlatnessVonEichtrafos} and Thm.~\ref{thm:CurvatureOfBasicStuffIsEquivalentForGaugeTrafoCurvature}, are generalizations of statements in \cite[especially Prop.~8 and Thm.~1]{EichtrafoKruemmungUrspruenglich} and \cite[especially Eq.~9, 10 and 11; there the $S$ denotes the basic curvature]{mayerlieAuchEichtrafoStuff}.\footnote{The sign of $\varepsilon$ in the gauge transformations there is the opposite of our sign.} In both of these works a coordinate-free formulation of $\delta_\varepsilon A$ was not known, just $\delta_\varepsilon A^a$. It was known that $\delta_\varepsilon A^a$ is dependent on coordinates, but not how it can be written/defined such that it is again an element of $\Omega^1(M; \Phi^*E)$. \cite{EichtrafoKruemmungUrspruenglich} tries to formulate infinitesimal gauge transformations in a covariant way with a completely different approach by assuming a weaker form of equality, but only for a special situation and only for $\varepsilon$ as an element of $\Phi^*(\Gamma(E))$ (\textit{i.e.}~they only looked at pullback functionals, when we express that in our language). \cite{mayerlieAuchEichtrafoStuff} looks at the set $\Gamma(\Phi^*E)$ for $\varepsilon$ but assumes that $\varepsilon^a$ is independent of $\Phi$ and $A$ which is clearly a coordinate-dependent description, because a change of the pull-back frame would introduce a $\Phi$-dependency of the components $\varepsilon^a$ (in our words, they choose a coordinate-dependent embedding of $\Gamma(\Phi^*E)$ as functionals). In one way or the other, both works arrive at Eq.~\eqref{CoordFuerEichKruemmungsRegel}, but only evaluated at pullback functionals, that is, $R_\delta({}^*\mu, {}^*\nu)A^a=0$ for all $\mu, \nu \in \Gamma(E)$.

What we provide is a coordinate-independent and -free definition of such infinitesimal gauge transformations. Moreover, we have generalized Eq.~\eqref{CoordFuerEichKruemmungsRegel} in form of Eq.~\eqref{DieKruemmungIstNullVonDenEichtrafosGeeeeil}, in sense of not only assuming pullback functionals by defining the pre-bracket $\Delta$.

$\bullet$ Recall Remark \ref{WhyNablaBasPartOne}: One could also take $\nabla_\rho$ to define $\delta_\varepsilon$. It has the advantage that then $\delta_\varepsilon A$ directly restricts to the standard formula when restricting ourselves to the classical setting. When defining and calculating $R_\delta$ in a similar manner,
 %one probably still gets Eq.~\eqref{CoordFuerEichKruemmungsRegel} because that is for scalar-valued functionals $A^a$ whose infinitesimal gauge transformation is independent of whether one uses $\nabla_\rho$ or $\nabla^{\mathrm{bas}}$ on $E$-valued functionals. But 
we also get Eq.~\eqref{SplittingVonDerEichtrafo} where the curvature-term will be replaced with the curvature of $\nabla_\rho$ due to Cor.~\ref{cor:RelationShipsOfCurvatures}. Therefore one needs to impose at least flatness of $\nabla_\rho$ in order to get a similar result like Eq.~\eqref{DieKruemmungIstNullVonDenEichtrafosGeeeeil}; actually, one can check that one still needs a vanishing basic curvature, too.
%, but the condition about the vanishing basic curvature in Cor.~\ref{cor:FlatnessVonEichtrafos} would stay additionally. That is, for Eq.~\eqref{CoordFuerEichKruemmungsRegel} we would need a vanishing basic curvature \textbf{and} a flat $\nabla_\rho$. 
But we will later see that the basic connection will be in general flat, while $\nabla_\rho$ will not; especially we will see that the basic curvature will always vanish for the presented gauge theory. Thence, another reason for our choice to use the basic connection for the definition of $\delta$.
\end{remark}

\begin{proof}[Proof of Cor.~\ref{cor:FlatnessVonEichtrafos}]
\leavevmode\newline
That is a trivial consequence of Thm.~\ref{thm:CurvatureOfBasicStuffIsEquivalentForGaugeTrafoCurvature} and Prop.~\ref{prop:WirHabenEinenTensorBeiderTrafoKruemmung}, using $R_\nabla^{\mathrm{bas}}=0$ (and that then the basic connection is flat by Prop.~\ref{prop:SnablamitREnabla}) and that $R_\delta(\cdot, \cdot)A$ is $\mathcal{F}^0_E(M)$-bilinear such that one just needs to look at pullback functionals.
\end{proof}

These results motivate even further why we use the basic connection to define the infinitesimal gauge transformation. Moreover, $R^{\mathrm{bas}}_\nabla=0$ is also a condition which we will need for gauge invariance; see later. That we have this condition in the standard formulation of gauge theory is also emphasized in the following theorem:

\begin{theorems}{Relation of the basic curvature and action Lie algebroids, \newline \cite[discussion around Eq.~(9)]{CurvedYMH}, \cite[Prop.~2.12]{basicconn}, and \cite[\S 2.5, Theorem A]{blaomTangentBundleAsLieGroup}}{ActionLieALgebroid}
Let $E \to N$ be a Lie algebroid. Then $E$ is locally an action Lie algebroid if and only if it admits locally a flat connection $\nabla$ with $R_\nabla^{\mathrm{bas}} = 0$. If there is such a local isomorphism, then it can be chosen in such a way that $\nabla$ describes the canonical flat connection.
\end{theorems}

\begin{remark}\label{remSimplyConnectedEqualsGlobal}
\leavevmode\newline
As clarification of the last sentence, under that isomorphism we have (locally) $E = N \times \mathfrak{g}$ for some Lie algebra $\mathfrak{g}$, and a basis of $\mathfrak{g}$, that is, a constant frame of $E$, will be parallel with respect to $\nabla$. Especially, the canonical flat connection of every action Lie algebroid has a vanishing basic curvature. Furthermore, over a simply connected base the isomorphism is global as we will see in the proof (because one can then construct a global parallel frame for $\nabla$; see the proof).
\end{remark}

\begin{proof}
\leavevmode\newline
This basically follows by Eq.~\eqref{eq:compcondfast}, \textit{i.e.}
\bas
R_\nabla^{\mathrm{bas}}(\mu, \nu)Y
&= \mleft( \nabla_Y t_{\nabla^{\mathrm{bas}}} \mright)(\mu, \nu) 
- R_\nabla(\rho(\mu), Y) \nu + R_\nabla(\rho(\nu), Y) \mu
\eas
for all $\mu,\nu \in \Gamma(E)$ and $Y \in \mathfrak{X}(N)$.

\underline{"$\Rightarrow$"}: Assume $E|_U \cong U \times \mathfrak{g}$ is an action Lie algebroid for some open subset $U$ of $N$ for some Lie algebra $\mathfrak{g}$. Over $U$ take the canonical flat connection $\nabla$, and let $\mleft( e_a \mright)_a$ be a frame of constant sections on $U$. Then by Eq.~\eqref{eq:compcondfast}
\bas
R_\nabla^{\mathrm{bas}}(e_a, e_b)
&=
\mleft( \nabla t_{\nabla^{\mathrm{bas}}} \mright)(e_a, e_b) 
=
\nabla \bigl( \underbrace{t_{\nabla^{\mathrm{bas}}}(e_a, e_b)}_{\mathclap{= \mleft[ e_a, e_b \mright]_E}} \bigr)
=
\mathrm{d}C_{ab}^c \otimes e_c,
=
0
\eas
where $C_{ab}^c$ are the structure constants of $\mathfrak{g}$.

\underline{"$\Leftarrow$"}: Assume we have a flat connection $\nabla$ over some open subset $U$ with $R_\nabla^{\mathrm{bas}} = 0$. W.l.o.g. assume there is a parallel frame $\mleft( e_a \mright)_a$ for $\nabla$ on $U$ (otherwise restrict $U$ to a smaller subset). Then again by Eq.~\eqref{eq:compcondfast}
\bas
0
&=
\nabla \bigl( t_{\nabla^\mathrm{bas}}(e_a, e_b) \bigr)
=
\mathrm{d}C_{ab}^c \otimes e_c,
\eas
thus, the structure functions related to the parallel frame are constant. Therefore the parallel frame spans the same Lie algebra $\mathfrak{g}$ at each fibre, so, $E|_U \cong U \times \mathfrak{g}$ as vector bundles. Identifying elements of $\mathfrak{g}$ with constant sections, the anchor $\rho$ defines clearly an action for $\mathfrak{g}$ on $N$, and $\mleft[ \cdot,\cdot \mright]_E$ clearly restrict to $\mleft[ \cdot, \cdot \mright]_{\mathfrak{g}}$ on constant sections. The Lie algebroid is thence of the action type by the uniqueness given in Prop.~\ref{prop:ActionLieoidsAreOids}.
\end{proof}
%
%So, we only choose the basic connection when defining the variations/gauge transformations on $\Phi^*E$-valued objects, in order to know that the commutator of two transformations is again a transformation, because, as we will see, the basic curvature will always vanish. In that way we then have $R_{\delta}(\cdot, \cdot)A= 0$, especially Eq.~\eqref{CoordFuerEichKruemmungsRegel} suggest that we have a sort of representation (in sense of an anti-homomorphism due to our sign convention which could be easily reversed). We will later discuss whether $R_{\delta}(\cdot, \cdot)$ is also zero when acting on other type of tensors (see Thm.~\ref{thm:EichtrafoIstIWkrlichKrassFlach}). But for this we need to introduce more about the physical setting.

We now want to generalize Cor.~\ref{cor:FlatnessVonEichtrafos} by using Cor.~\ref{cor:RelationShipsOfCurvatures}, especially we need to understand the behaviour for scalar-valued functionals. For such functionals the infinitesimal gauge transformation is nothing else than the Lie derivative of some vector field in $\mathfrak{M}_E$, which we denoted by $\Psi_\varepsilon$.
Recall Remark \ref{NotASubalgebraXB}, we do in general not expect that $\Psi_\varepsilon \in \mathfrak{X}^E\bigl( \mathfrak{M}_E(M;N) \bigr)$ builds a subalgebra; however, since we restricted the set of those vector fields by defining $\delta_\varepsilon A$ in Prop.~\ref{prop:VariationOfA}, there may be hope for the structure of a subalgebra; this will be discussed now. 

\begin{theorems}{Bracket of gauge transformations a gauge transformation}{VektorfelderSindZumGlueckGeschlossen}
Let $M, N$ be smooth manifolds, $E \to N$ a Lie algebroid, $\nabla$ a connection on $E$ with $R^{\mathrm{bas}}_\nabla=0$. Furthermore let $\Psi_\varepsilon$ and $\Psi_\vartheta$ for $\varepsilon, \vartheta \in \mathcal{F}^0_E(M; {}^*E)$ be the unique elements of $\mathfrak{X}^E\bigl( \mathfrak{M}_E(M;N) \bigr)$ as given by Prop.~\ref{prop:VariationOfA}.\footnote{Recall that those $\Psi_\varepsilon$ are the vector fields describing the infinitesimal gauge transformation; see Def.~\ref{def:TotalInfGaugeTrafoYayy}.}

Then
\ba
\mleft[ \Psi_\varepsilon, \Psi_\vartheta \mright]
&=
- \Psi_{\Delta(\varepsilon, \vartheta)}
\ea
for all $\varepsilon, \vartheta \in \mathcal{F}^0_E(M; {}^*E)$, where $\Psi_{\Delta(\varepsilon, \vartheta)}$ is also the unique element of $\mathfrak{X}^E\bigl( \mathfrak{M}_E(M;N) \bigr)$ as given by Prop.~\ref{prop:VariationOfA}.
\end{theorems}

\begin{proof}
\leavevmode\newline
First recall that we have by Remark \ref{RemLeibnizeRegelaufProdukteWeshalbEConnectionNichtWichtigIst}
\bas
\delta_\varepsilon \omega
&=
\mathcal{L}_{\Psi_\varepsilon} \omega
\eas
for all $\omega \in \mathcal{F}^\bullet_E(M)$ and $\varepsilon \in \mathcal{F}^0_E(M; {}^*E)$. Therefore we want to use Cor.~\ref{cor:FlatnessVonEichtrafos}. As vector fields of $\mathfrak{M}_E(M;N)$, the action of $\mathcal{L}_{\Psi_\varepsilon}$ is uniquely given by its action on coordinates of $\mathfrak{M}_E(M;N)$, and these are essentially given by the components of the fields $(\Phi, A) \in \mathfrak{M}_E(M;N)$: Let $\mleft( x^i \mright)_i$ be local coordinate functions on $N$ and let $\mleft(e_a\mright)_a$ be a local frame of $E$, then coordinates of $\mathfrak{M}_E(M;N)$ are given by the functionals ${}^*\mleft(x^i\mright)$ and $\varpi_2^a$ because of
\bas
\mleft.{}^*\mleft(x^i\mright)\mright|_{(\Phi,A)}
&=
\Phi^i,
\\
\varpi_2^a(\Phi,A)
&=
A^a
\eas
for all $(\Phi, A) \in \mathfrak{M}_E(M;N)$. Recall the first calculation in the proof of Cor.~\ref{cor:RelationShipsOfCurvatures}, we get similarly
\bas
R_\delta(\varepsilon,\vartheta)\mleft( {}^*\mleft(x^i\mright) \mright) 
&= 
\varepsilon^a \vartheta^b ~ {}^*\underbrace{\mleft(
	\mathcal{L}_{\rho(e_a)} \mathcal{L}_{\rho(e_b)} x^i
	- \mathcal{L}_{\rho(e_b)} \mathcal{L}_{\rho(e_a)} x^i
	- \mathcal{L}_{\rho\mleft(\mleft[ e_a, e_b \mright]_E\mright)} x^i
\mright)}_{= \mleft( \mathcal{L}_{\mleft[ \rho(e_a), \rho(e_b) \mright]} - \mathcal{L}_{\rho\mleft(\mleft[ e_a, e_b \mright]_E\mright)} \mright) x^i = 0}
=
0
\eas
for all $\varepsilon, \vartheta \in \mathcal{F}^0_E(M; {}^*E)$, using that $\rho$ is a homomorphisma and Remark \ref{RemLeibnizeRegelaufProdukteWeshalbEConnectionNichtWichtigIst} such that $\delta_\varepsilon \mleft( {}^*\mleft( x^i \mright) \mright) = - \varepsilon^a ~ {}^*\mleft( \mathcal{L}_{\rho(e_a)} x^i \mright)$. By Cor.~\ref{cor:FlatnessVonEichtrafos} we also get
\bas
R_\delta (\varepsilon, \vartheta) \varpi_2^a
&=
0.
\eas
By $\delta_\varepsilon = \mathcal{L}_{\Psi_\varepsilon}$ on scalar-valued functionals we therefore get
\bas
\mleft(\mleft[ \mathcal{L}_{\Psi_\varepsilon}, \mathcal{L}_{\Psi_\vartheta} \mright]
+ \mathcal{L}_{\Psi_{\Delta(\varepsilon, \vartheta)}}\mright)f
&=
0
\eas
for all $f \in C^\infty\bigl(\mathfrak{M}_E(M;N)\bigr)$,
which finishes the proof.
\end{proof}

\begin{remarks}{Curvature of $\delta$ on $\Phi$}{WasIstMitDemHiggsFeldBeiDerDeltaKruemmung}
Keeping the same situation and notation as in the previous proof, observe that we have
\bas
\delta_{{}^*\nu} \delta_{{}^*\mu} \Phi
&=
- \delta_{{}^*\nu} \bigl( {}^*(\rho(\mu)) \bigr)
=
{}^*\mleft( \nabla^{\mathrm{bas}}_\nu \bigl( \rho(\mu) \bigr) \mright)
=
{}^*\mleft( \rho\mleft( \nabla^{\mathrm{bas}}_\nu \mu \mright) \mright)
\eas
for all $\mu, \nu \in \Gamma(E)$, hence,\footnote{Recall Eq.~\eqref{EqLieKlammerAufPullBackSections}.}
\bas
\delta_{{}^*\nu} \delta_{{}^*\mu} \Phi
	- \delta_{{}^*\mu} \delta_{{}^*\nu} \Phi
	+ \delta_{{}^*\mleft( \mleft[ \nu, \mu \mright]_E \mright)} \Phi
&=
{}^*\mleft( \rho\mleft( 
	\nabla^{\mathrm{bas}}_\nu \mu 
	- \nabla^{\mathrm{bas}}_\mu \nu 
	- \mleft[ \nu, \mu \mright]_E
\mright) \mright)
=
{}^*\Bigl( \rho\bigl( 
	t_{\nabla^{\mathrm{bas}}}(\nu, \mu)
\bigr) \Bigr).
\eas
Therefore, if we want that this is zero, too, we would need that the torsion of the basic connection has values in the kernel of the anchor which is in general not the case. However, it is no harm that we do not have a zero value in general here. That is due to the fact that on one hand $\Phi$ just contributes via pull-backs, as we will also see in the following sections; on the other hand $\Phi$ is not vector-bundle valued and hence will not arise in any other form than as the map for the pullbacks in any Lagrangian or physical quantity. Even in the classical case, recall Prop.~\ref{prop:LieRepAndLieAct}, a Lie algebra representation acting on $\Phi$ is just the evaluation of its induced action at $\Phi$.
\newline\newline
However, as we have seen in the proof, we got $R_\delta(\cdot,\cdot)\mleft( {}^*\mleft(x^i\mright) \mright) = 0$, and $\mleft.{}^*\mleft(x^i\mright)\mright|_{(\Phi,A)} = \Phi^i$ for all $(\Phi, A) \in \mathfrak{M}_E(M;N)$. That is, for the components of the Higgs field we have the desired behaviour, which is all we need.
\end{remarks}

Finally, we can generalize Cor.~\ref{cor:FlatnessVonEichtrafos}.

\begin{theorems}{Curvature of $\delta$ on arbitrary functionals}{AllgemEineGeileFormelFuerDieEichKruemmung}
Let $M, N$ be smooth manifolds, $E \to N$ a Lie algebroid, $\nabla$ a connection on $E$ with $R^{\mathrm{bas}}_\nabla=0$. Furthermore let $V\to N$ be a vector bundle, equipped with an $E$-connection ${}^E\nabla$ on $V$. Then
\ba
R_\delta(\varepsilon, \vartheta) L
&=
\mleft({}^*R_{{}^E\nabla} \mright)(\varepsilon, \vartheta)L
\ea
for all $L \in \mathcal{F}_E^k(M; {}^*V)$ ($k \in \mathbb{N}_0$) and $\varepsilon, \vartheta \in \mathcal{F}^0_E(M; {}^*E)$. In short, $R_\delta = {}^*R_{{}^E\nabla}$.
\end{theorems}

\begin{remark}
\leavevmode\newline
This also shows that $R_\delta$ is a tensor.
Moreover, as expected, for flat ${}^E\nabla$ we would get
\ba
R_\delta(\varepsilon, \vartheta) L
&=
0.
\ea
\end{remark}

\begin{proof}[Proof of Thm.~\ref{thm:AllgemEineGeileFormelFuerDieEichKruemmung}]
\leavevmode\newline
We want to use Cor.~\ref{cor:RelationShipsOfCurvatures}, so, for a given frame $\mleft( e_a \mright)_a$ we have
\bas
R_\delta(\varepsilon, \vartheta)L
&=
R_\delta(\varepsilon, \vartheta)L^a \otimes {}^*e_a
	+ \mleft({}^*R_{{}^E\nabla}\mright)(\varepsilon, \vartheta)L
\eas
for all $L \in \mathcal{F}^k_E(M; {}^*V)$ ($k \in \mathbb{N}_0$) and $\varepsilon, \vartheta \in \mathcal{F}^0_E(M; {}^*E)$. Hence, we just need to show that $R_\delta(\varepsilon, \vartheta)L^a = 0$. Again by Remark \ref{RemLeibnizeRegelaufProdukteWeshalbEConnectionNichtWichtigIst} we have $\delta_\varepsilon = \mathcal{L}_{\Psi_\varepsilon}$ on scalar-valued functionals, where $\Psi_\varepsilon$ still denotes vector fields as uniquely given by Prop.~\ref{prop:VariationOfA}. $\Psi_\varepsilon$ are elements of $\mathfrak{X}\bigl( \mathfrak{M}_E(M;N) \bigr)$, hence, 
\bas
(\underbrace{\delta_\varepsilon L^a}_{\mathclap{= \mathcal{L}_{\Psi_\varepsilon}L^a }})_p(Y_1, \dotsc, Y_k)
&=
\mathcal{L}_{\Psi_\varepsilon} \mleft(L^a_p(Y_1, \dotsc, Y_k)\mright)
\eas
for all $p \in M$ and $Y_1, \dotsc, Y_k \in \mathrm{T}_pM$. We know that $L^a\in\mathcal{F}^k_E(M)$, and therefore $L^a_p(Y_1, \dotsc, Y_k) \in C^\infty\bigl(\mathfrak{M}_E(M;N)\bigr)$, so, we just need to use Thm.~\ref{thm:VektorfelderSindZumGlueckGeschlossen} to get
\bas
\mleft(R_\delta(\varepsilon, \vartheta) L^a \mright)_p(Y_1, \dotsc, Y_k)
&=
\mleft(
	\mleft(\mleft[ \mathcal{L}_{\Psi_\varepsilon}, \mathcal{L}_{\Psi_\vartheta} \mright]
	+ \mathcal{L}_{\Psi_{\Delta(\varepsilon, \vartheta)}}\mright) L^a
\mright)_p(Y_1, \dotsc, Y_k)
\\
&=
\mleft(\mleft[ \mathcal{L}_{\Psi_\varepsilon}, \mathcal{L}_{\Psi_\vartheta} \mright]
	+ \mathcal{L}_{\Psi_{\Delta(\varepsilon, \vartheta)}}\mright)
\mleft(L^a_p(Y_1, \dotsc, Y_k)\mright)
\\
&\stackrel{\mathclap{ \text{Thm.~\ref{thm:VektorfelderSindZumGlueckGeschlossen}} }}{=}\quad~~
0,
\eas
which concludes the proof.
\end{proof} 

Let us conclude this section by showing that this finally implies that $\Delta$ is a Lie bracket.

\begin{theorems}{Pre-bracket a Lie bracket}{PreKlammerEineSuperLieKlammer}
Let $M, N$ be smooth manifolds, $E \to N$ a Lie algebroid, $\nabla$ a connection on $E$ with $R^{\mathrm{bas}}_\nabla=0$. Then $\Delta$ is a Lie bracket.
\end{theorems}

\begin{proof}
\leavevmode\newline
By Prop.~\ref{prop:PropertiesOfThePreBracket} we already know antisymmetry and $\mathbb{R}$-bilinearity. 
%and by Remark \ref{rem:PolynomeSindZumGlueckGeschlossenNachGaugeTrafo} we know that
%\bas
%\Delta(\vartheta, \varepsilon)
%&=
%\delta_\varepsilon \vartheta - \delta_\vartheta \varepsilon - \bigl( {}^*t_{\nabla^{\mathrm{bas}}} \bigr)\mleft( \vartheta, \varepsilon \mright)
%\in \mathcal{H}^0_E(M;{}^*E)
%\eas
%for all $\varepsilon, \vartheta \in \mathcal{H}_E^0(M; {}^*E)$, also using that clearly $\bigl( {}^*t_{\nabla^{\mathrm{bas}}} \bigr)\mleft( \vartheta, \varepsilon \mright) \in \mathcal{H}^0(M; {}^*E)$.
Thus, only the Jacobi identity is left to show, and the calculation is very similar to the calculation of the first Bianchi identity in Thm.~\ref{thm:1stBianchi},
\bas
\Delta\mleft( \eta, \Delta\mleft( \vartheta, \varepsilon \mright) \mright)
&=
\Delta\mleft( \eta, \delta_\varepsilon \vartheta - \delta_\vartheta \varepsilon - \bigl( {}^*t_{\nabla^{\mathrm{bas}}} \bigr)\mleft( \vartheta, \varepsilon \mright) \mright)
\\
&=
\underbrace{\delta_{\delta_\varepsilon \vartheta} \eta
	- \delta_{\delta_\vartheta \varepsilon} \eta
	- \delta_{\bigl( {}^*t_{\nabla^{\mathrm{bas}}} \bigr)\mleft( \vartheta, \varepsilon \mright)} \eta}
	_{\delta_{\Delta(\vartheta, \varepsilon)} \eta}
\\
&\hspace{1cm}
	- \delta_\eta \delta_\varepsilon \vartheta
	+ \delta_\eta \delta_\vartheta \varepsilon
	+ \bigl( {}^*t_{\nabla^{\mathrm{bas}}} \bigr)\mleft( \eta, \mleft( \bigl( {}^*t_{\nabla^{\mathrm{bas}}} \bigr)\mleft( \vartheta, \varepsilon \mright) \mright) \mright)
\\
&\hspace{1cm}
	+ \delta_\eta \mleft( \bigl( {}^*t_{\nabla^{\mathrm{bas}}} \bigr)\mleft( \vartheta, \varepsilon \mright) \mright)
	- \bigl( {}^*t_{\nabla^{\mathrm{bas}}} \bigr)\mleft( \eta, \delta_\varepsilon \vartheta \mright)
	+ \bigl( {}^*t_{\nabla^{\mathrm{bas}}} \bigr)\mleft( \eta, \delta_\vartheta \varepsilon \mright)
\\
&=
\delta_\eta \delta_\vartheta \varepsilon
	- \delta_\eta \delta_\varepsilon \vartheta
	+ \delta_{\Delta(\vartheta, \varepsilon)} \eta
\\
&\hspace{1cm}
	+ \delta_\eta \mleft( \bigl( {}^*t_{\nabla^{\mathrm{bas}}} \bigr)\mleft( \vartheta, \varepsilon \mright) \mright)
	- \bigl( {}^*t_{\nabla^{\mathrm{bas}}} \bigr)\mleft( \eta, \delta_\varepsilon \vartheta \mright)
	+ \underbrace{\bigl( {}^*t_{\nabla^{\mathrm{bas}}} \bigr)\mleft( \eta, \delta_\vartheta \varepsilon \mright)}_{\mathclap{= - \bigl( {}^*t_{\nabla^{\mathrm{bas}}} \bigr)\mleft( \delta_\vartheta \varepsilon, \eta \mright)}}
\\
&\hspace{1cm}
	+ \bigl( {}^*t_{\nabla^{\mathrm{bas}}} \bigr)\mleft( \eta, \mleft( \bigl( {}^*t_{\nabla^{\mathrm{bas}}} \bigr)\mleft( \vartheta, \varepsilon \mright) \mright) \mright)
\eas
for all $\varepsilon, \vartheta, \eta \in \mathcal{F}_E^0(M; {}^*E)$. Taking the cyclic sum, we collect the terms as in the proof of Thm.~\ref{thm:1stBianchi}, and hence we get, using that $\nabla^{\mathrm{bas}}$ is used for the definition of $\delta$ on $E$-valued functionals,
\bas
&\Delta\mleft( \eta, \Delta\mleft( \vartheta, \varepsilon \mright) \mright)
	+ \Delta\mleft( \vartheta, \Delta\mleft( \varepsilon, \eta \mright) \mright)
	+ \Delta\mleft( \varepsilon, \Delta\mleft( \eta, \vartheta \mright) \mright)
\\
&=
\underbrace{R_{\delta}( \eta, \vartheta ) \varepsilon
	+ R_{\delta}( \varepsilon,\eta ) \vartheta
	+ R_{\delta}( \vartheta, \varepsilon) \eta}_{\stackrel{\text{Thm.~\ref{thm:AllgemEineGeileFormelFuerDieEichKruemmung}}}{=}0}
\\
&\hspace{1cm}
	+ \bigl( {}^*t_{\nabla^{\mathrm{bas}}} \bigr)\mleft( \eta, \bigl( {}^*t_{\nabla^{\mathrm{bas}}} \bigr)\mleft( \vartheta, \varepsilon \mright) \mright) 
	+ \bigl( {}^*t_{\nabla^{\mathrm{bas}}} \bigr)\mleft( \varepsilon, \bigl( {}^*t_{\nabla^{\mathrm{bas}}} \bigr)\mleft( \eta, \vartheta \mright) \mright) 
\\
&\hspace{1cm}
	+ \bigl( {}^*t_{\nabla^{\mathrm{bas}}} \bigr)\mleft( \vartheta, \bigl( {}^*t_{\nabla^{\mathrm{bas}}} \bigr)\mleft( \varepsilon, \eta \mright) \mright) 
\\
&\hspace{1cm}
	+ \underbrace{\mleft( \delta_\eta \mleft({}^*t_{\nabla^{\mathrm{bas}}} \mright) \mright)}_{= - {}^*\mleft( \nabla^{\mathrm{bas}}_\eta t_{\nabla^{\mathrm{bas}}} \mright)} (\vartheta, \varepsilon)
	+ \mleft( \delta_\varepsilon \mleft({}^*t_{\nabla^{\mathrm{bas}}} \mright) \mright) (\eta, \vartheta)
	+ \mleft( \delta_\vartheta \mleft({}^*t_{\nabla^{\mathrm{bas}}} \mright) \mright) (\varepsilon, \eta)
\\
&=
- \vartheta^a \varepsilon^b \eta^c ~ {}^*\biggl(
		t_{\nabla^{\mathrm{bas}}} \mleft( t_{\nabla^{\mathrm{bas}}}(e_a, e_b), e_c \mright)
		+ t_{\nabla^{\mathrm{bas}}} \mleft( t_{\nabla^{\mathrm{bas}}}(e_b, e_c), e_a \mright)
		+ t_{\nabla^{\mathrm{bas}}} \mleft( t_{\nabla^{\mathrm{bas}}}(e_c, e_a), e_b \mright)
\\
&\hspace{1cm}\hphantom{- \vartheta^a \varepsilon^b \eta^c ~ {}^*\biggl(}
	+ \mleft( \nabla^{\mathrm{bas}}_{e_c} t_{\nabla^{\mathrm{bas}}} \mright) (e_a, e_b)
	+ \mleft( \nabla^{\mathrm{bas}}_{e_a} t_{\nabla^{\mathrm{bas}}} \mright) (e_b, e_c)
	+ \mleft( \nabla^{\mathrm{bas}}_{e_b} t_{\nabla^{\mathrm{bas}}} \mright) (e_c, e_a)
\biggr)
\\
&\stackrel{\mathclap{\text{Thm.~\ref{thm:1stBianchi}}}}{=}~~~~~
0
\eas
for all $\varepsilon, \vartheta, \eta \in \mathcal{F}^0_E(M; {}^*E)$, where $\mleft( e_a \mright)_a$ is a local frame of $E$, and we also used that $\nabla^{\mathrm{bas}}$ is flat by Prop.~\ref{prop:SnablamitREnabla}; the flatness was applied when we used Thm.~\ref{thm:1stBianchi} and Thm.~\ref{thm:AllgemEineGeileFormelFuerDieEichKruemmung}.\footnote{But flatness is not actually needed here; see also the following remark.} Thence, the Jacobi identity follows.
\end{proof}

\begin{remark}\label{RemarkBracketIsVeryIndependent}
\leavevmode\newline
The proof is essentially based on the first Bianchi identity of curvatures. Hence, taking any other $E$-connection $\nabla^\prime$ on $E$ one could define the bracket $\Delta$ by using the torsion of $\nabla^\prime$ instead of $\nabla^{\mathrm{bas}}$, and then also define the $\delta$ operator with respect to $\nabla^\prime$ on $E$-valued form. By Thm.~\ref{thm:AllgemEineGeileFormelFuerDieEichKruemmung} we could not expect $R_\delta=0$ in general, but $\Delta$ should be nevertheless a Lie bracket due to the fact that the first Bianchi identity always holds and that Thm.~\ref{thm:AllgemEineGeileFormelFuerDieEichKruemmung} provides the needed curvature terms for the Bianchi identity. Furthermore, already Eq.~\eqref{EqDeltaInFrameKoord} points out that the definition of $\Delta$ is independent of the choice of $\nabla^\prime$ because $\delta_\varepsilon$ is just a Lie derivative on scalar-valued functionals, so that it is clear that it is always the same Lie bracket. By the very last statement of Prop.~\ref{prop:PropertiesOfThePreBracket}, we achieve a Lie bracket completely independent of connections, if the parameters are just functionals depending on the Higgs field $\Phi$. Recall Remark \ref{ClassicalCommutatorRemark} (the part about the bookkeeping trick of the parameters) and the first bullet point of Remark \ref{RemarkUeberNablaRhoCurvatureForGauegTrafo} about that it is in general unavoidable to assume that the parameters depend on $\Phi$.
\end{remark}

\section{Infinitesimal gauge invariance}\label{InfInvariance}

Let us now calculate the infinitesimal gauge transformations needed for the Lagrangian.

\begin{propositions}{Infinitesimal gauge transformations of the field strength}{GaugeTrafosOfFieldStrengthAndMinimalCoupling}
Let $M, N$ be smooth manifolds, $E \to N$ a Lie algebroid, and $\nabla$ a connection on $E$. Then we have
\ba
\delta_\varepsilon F
&=
- \mleft(
	\frac{1}{2} ~ \mleft(	{}^* R_{\nabla} \mright)\mleft( \mathfrak{D} \stackrel{\wedge}{,} \mathfrak{D} \mright) \varepsilon
	+ \mleft({}^* R_\nabla^{\mathrm{bas}} \mright) \mleft(\varepsilon \stackrel{\wedge}{,} \varpi_2  \stackrel{\wedge}{,} \mathrm{D} \mright)
\mright) \label{EqVariationForF}
\ea
for all $\varepsilon \in \mathcal{F}^0_E(M; {}^*E)$, where we write $\Gamma(E) \times \Gamma(E) \times \mathfrak{X}(N) \ni (\mu, \nu, Y) \mapsto R_\nabla^{\mathrm{bas}}(\mu, \nu, Y) \coloneqq R_\nabla^{\mathrm{bas}}(\mu, \nu) Y$.
\end{propositions}

\begin{proof}
\leavevmode\newline
%For simplicity we will omit to denote $(\Phi, A)$. 
Let $\mleft( e_a \mright)_a$ be a frame of $E$, then by Eq.~\eqref{eqGaugeTrafoOfAacomps}
\bas
\mathrm{d} \delta_\varepsilon \varpi_2^a \otimes {}^*e_a
&=
\mathrm{d}\mleft( \varepsilon^b \varpi_2^c \otimes {}^*\mleft( \nabla^{\mathrm{bas}}_{e_b} e_c\mright)
	- \mleft({}^*\nabla\mright)\varepsilon \mright)^a \otimes {}^*e_a
\\
&=
\mathrm{d}\varepsilon^b \wedge \varpi_2^c \otimes {}^*\mleft( \nabla^{\mathrm{bas}}_{e_b} e_c\mright)
	+ \varepsilon^b ~ \mathrm{d}\varpi_2^c \otimes {}^*\mleft( \nabla^{\mathrm{bas}}_{e_b} e_c\mright)
\\
&\hspace{1cm}
	- \varepsilon^b \varpi_2^c \wedge \mathrm{d}\mleft( {}^*\mleft( \nabla^{\mathrm{bas}}_{e_b} e_c\mright) \mright)^a \otimes {}^*e_a
	- \mathrm{d} \bigl( \mleft({}^*\nabla\mright)\varepsilon \bigr)^a \otimes {}^*e_a
\eas
also recall Eq.~\eqref{eqVariationVertauschtMitDifferential}, and \eqref{EqVariationVonFormenBrrrr} (and also the calculation for Eq.~\eqref{EqVariationVonFormenBrrrrVereinfacht}),
then, using the previous calculation,
\bas
\delta_\varepsilon \mleft(\mathrm{d}^{{}^*\nabla} \varpi_2 \mright)
&=
\delta_\varepsilon \mleft(
	\mathrm{d} \varpi_2^a \otimes {}^*e_a 
	- \varpi_2^b \wedge {}^!(\nabla e_b)
\mright)
\\
&=
\mathrm{d} \delta_\varepsilon \varpi_2^a \otimes {}^*e_a 
	- \mathrm{d} \varpi_2^a \otimes {}^*\mleft( \nabla^{\mathrm{bas}}_\varepsilon e_a \mright)
	- \delta_\varepsilon \varpi_2^b \wedge {}^!(\nabla e_b)
\\
&\hspace{1cm}
	+ \varpi_2^b \wedge \biggl(
	\underbrace{\mleft({}^*\mleft(\nabla^{\mathrm{bas}}_\varepsilon \mleft( \nabla e_b \mright)\mright)\mright)(\mathrm{D})
	+ {}^*\mleft( \nabla_{({}^*\rho)\mleft( ({}^*\nabla) \varepsilon \mright)} e_b \mright)}
	_{\mathclap{= \varepsilon^a  ~ {}^! \mleft( \nabla^{\mathrm{bas}}_{e_a} \mleft( \nabla e_b \mright) \mright)
		+ \varepsilon^a ~ {}^!\mleft( \nabla_{  \rho\mleft( \nabla e_a \mright)} e_b \mright)
		+ \mathrm{d}\varepsilon^a \otimes {}^*\mleft( \nabla_{  \rho\mleft( e_a \mright) }e_b \mright)}}
\biggr)
\\
&=
\mathrm{d}\varepsilon^a \wedge \varpi_2^b \otimes {}^*\underbrace{\mleft( 
	\nabla^{\mathrm{bas}}_{e_a} e_b
	- \nabla_{  \rho\mleft( e_a \mright) }e_b
\mright)}
_{\mathclap{= t_{\nabla^{\mathrm{bas}}}(e_a, e_b)}}
\\
&\hspace{1cm}
	- \varepsilon^a \varpi_2^c \wedge \mathrm{d}\mleft( {}^*\mleft( \nabla^{\mathrm{bas}}_{e_a} e_c\mright) \mright)^b \otimes {}^*e_b
	- \mathrm{d} \bigl( \mleft({}^*\nabla\mright)\varepsilon \bigr)^b \otimes {}^*e_b
\\
&\hspace{1cm}
	- \mleft( \varepsilon^a \varpi_2^c \otimes {}^*\mleft( \nabla^{\mathrm{bas}}_{e_a} e_c\mright)
	- \mleft({}^*\nabla\mright)\varepsilon \mright)^b \wedge {}^!(\nabla e_b)
\\
&\hspace{1cm}
	+ \varepsilon^a \varpi_2^b \wedge {}^! \biggl(
	  \nabla^{\mathrm{bas}}_{e_a} \mleft( \nabla e_b \mright)
		+ \nabla_{  \rho\mleft( \nabla e_a \mright)} e_b
\biggr)
\\
&=
\mathrm{d}\varepsilon^a \wedge \varpi_2^b \otimes {}^* \mleft(t_{\nabla^{\mathrm{bas}}}(e_a, e_b) \mright)
	- \varepsilon^a \varpi_2^c \wedge \underbrace{\mathrm{d}^{{}^*\nabla} \mleft( {}^*\mleft( \nabla^{\mathrm{bas}}_{e_a} e_c \mright) \mright)}
	_{\mathclap{\stackrel{\text{Eq.~\eqref{EqGeilePullBackCommuteFormel}}}{=} {}^!\mleft( \nabla \mleft( \nabla^{\mathrm{bas}}_{e_a} e_c \mright) \mright)}}
	- \underbrace{\mleft( \mathrm{d}^{{}^*\nabla} \mright)^2 \varepsilon}
	_{\mathclap{= R_{{}^*\nabla}(\cdot, \cdot) \varepsilon}}
\\
&\hspace{1cm}
	+ \varepsilon^a  \varpi_2^b \wedge {}^!  \biggl(
	\nabla^{\mathrm{bas}}_{e_a} \mleft( \nabla e_b \mright)
		+ \nabla_{  \rho\mleft( \nabla e_a \mright)} e_b
\biggr)
\\
&=
\mathrm{d}\varepsilon^a \wedge \varpi_2^b ~ {}^* \mleft(t_{\nabla^{\mathrm{bas}}}(e_a, e_b) \mright)
	- \varepsilon^a ~ {}^!\bigl( R_{\nabla}(\cdot, \cdot) e_a \bigr)
\\
&\hspace{1cm}
	+ \varepsilon^a  \varpi_2^b \wedge {}^!  \biggl(
	\underbrace{
	\nabla^{\mathrm{bas}}_{e_a} \mleft( \nabla e_b \mright)
		- \nabla \mleft( \nabla^{\mathrm{bas}}_{e_a} e_b \mright)
		+ \nabla_{  \rho\mleft( \nabla e_a \mright)} e_b
		}
		_{\mathclap{\mathfrak{X}(N) \ni Y \mapsto \mleft[ e_a, \nabla_Y e_b \mright]_E
	+ \nabla_{\rho\mleft( \nabla_Y e_b \mright)} e_a
	- \nabla_{\mleft[ \rho(e_a), Y \mright]} e_b
	- \nabla_Y\mleft( \mleft[ e_a, e_b \mright]_E \mright)
	- \nabla_Y \nabla_{\rho(e_b)} e_a}}
	\biggr),
%\\
%&=
%\mathrm{d}^{{}^*\nabla} \mleft( \delta_\varepsilon \varpi_2 \mright)
	%- \mathrm{d} \varpi_2^a \otimes {}^*\mleft( \nabla^{\mathrm{bas}}_\varepsilon e_a \mright)
%\\
%&\hspace{1cm}
	%- \varpi_2^b \otimes \biggl(
	%\mleft({}^*\mleft(\nabla^{\mathrm{bas}}_\varepsilon \mleft( \nabla e_b \mright)\mright)\mright)(\mathrm{D})
	%+ {}^*\mleft( \nabla_{({}^*\rho)\mleft( ({}^*\nabla) \varepsilon \mright)} e_b \mright)
%\biggr)
\eas
using the second calculation in the proof of Thm.~\ref{thm:CurvatureOfBasicStuffIsEquivalentForGaugeTrafoCurvature}. Moreover,
\bas
\mleft( \nabla^{\mathrm{bas}}_\eta t_{\nabla^{\mathrm{bas}}} \mright) (\mu, \nu)
&\stackrel{\text{Thm.~\ref{thm:modBianchithm}}}{=}
R_{\nabla_\rho}(\mu, \nu) \eta
%&=
%\nabla^{\mathrm{bas}}_\eta \bigl( t_{\nabla^{\mathrm{bas}}} (\mu, \nu) \bigr)
	%- \underbrace{t_{\nabla^{\mathrm{bas}}}}_{\mathclap{= - t_{\nabla_\rho}}} \mleft(\nabla^{\mathrm{bas}}_\eta \mu,\nu\mright)
	%- t_{\nabla^{\mathrm{bas}}} \mleft( \mu, \nabla^{\mathrm{bas}}_\eta \nu\mright)
%\\
%&=
%\bigl[\eta, t_{\nabla^{\mathrm{bas}}} (\mu, \nu) \bigr]_E
	%+ \nabla_{\rho \bigl( t_{\nabla^{\mathrm{bas}}} (\mu, \nu) \bigr)} \eta
%\\
%&\hspace{1cm}
	%+ \nabla_{\rho\mleft( \nabla^{\mathrm{bas}}_\eta \mu \mright)} \nu
	%- \nabla_{\rho(\nu)} \nabla^{\mathrm{bas}}_\eta \mu
	%- \mleft[ \nabla^{\mathrm{bas}}_\eta \mu, \nu \mright]_E
%\\
%&\hspace{1cm}
	%+ \nabla_{\rho(\mu)} \nabla^{\mathrm{bas}}_\eta \nu
	%- \nabla_{\rho\mleft( \nabla^{\mathrm{bas}}_\eta \nu \mright)} \mu
	%- \mleft[ \mu, \nabla^{\mathrm{bas}}_\eta \nu \mright]_E
\eas
for all $\mu, \nu, \eta \in \Gamma(E)$, such that, also using Eq.~\eqref{PullBackVariation},
\bas
\delta_\varepsilon \mleft( \frac{1}{2} \mleft( {}^* t_{\nabla^{\mathrm{bas}}} \mright)\mleft( \varpi_2 \stackrel{\wedge}{,} \varpi_2 \mright) \mright)
&=
-\frac{1}{2} \biggl(
	\mleft( {}^* \mleft( \nabla^{\mathrm{bas}}_\varepsilon t_{\nabla^{\mathrm{bas}}} \mright) \mright) \mleft( \varpi_2 \stackrel{\wedge}{,} \varpi_2 \mright)
\\
&\hspace{1cm}\hphantom{-\frac{1}{2} \biggl(}
	+ \mleft( {}^* t_{\nabla^{\mathrm{bas}}} \mright)\bigl( \mleft( {}^*\nabla\mright) \varepsilon \stackrel{\wedge}{,} \varpi_2 \bigr)
	+ \underbrace{\mleft( {}^* t_{\nabla^{\mathrm{bas}}} \mright)\bigl( \varpi_2 \stackrel{\wedge}{,} \mleft( {}^*\nabla\mright) \varepsilon \bigr)}
	_{\mathclap{= \mleft( {}^* t_{\nabla^{\mathrm{bas}}} \mright)\bigl( \mleft( {}^*\nabla\mright) \varepsilon \stackrel{\wedge}{,} \varpi_2 \bigr)}}
\biggr)
\\
&=
-\frac{\varepsilon^a}{2} ~
	\bigl( {}^* \mleft( R_{\nabla_\rho}(\cdot, \cdot) e_a \mright) \bigr) \mleft( \varpi_2 \stackrel{\wedge}{,} \varpi_2 \mright)
\\
&\hspace{1cm}
	- \mathrm{d}\varepsilon^a \wedge \varpi_2^b ~ {}^*\bigl( t_{\nabla^{\mathrm{bas}}} \mleft( e_a, e_b \mright)\bigr)
	+ \varepsilon^a ~ \varpi_2^b \wedge {}^! \bigl( t_{\nabla^{\mathrm{bas}}} \mleft( \nabla e_a, e_b \mright) \bigr)
\eas
where we used that the torsion is anti-symmetric such that by Prop.~\ref{prop:GradedExtensionPlusAntiSymm}
\ba\label{IchMussDasDringendVerallgemeinern}
\mleft( {}^* t_{\nabla^{\mathrm{bas}}} \mright)\bigl( \varpi_2 \stackrel{\wedge}{,} \mleft( {}^*\nabla\mright) \varepsilon \bigr) = \mleft( {}^* t_{\nabla^{\mathrm{bas}}} \mright)\bigl( \mleft( {}^*\nabla\mright) \varepsilon \stackrel{\wedge}{,} \varpi_2 \bigr),
\ea
because both arguments are 1-forms. We also have
\bas
&\mleft[ e_a, \nabla_Y e_b \mright]_E
	+ \nabla_{\rho\mleft( \nabla_Y e_b \mright)} e_a
	- \nabla_{\mleft[ \rho(e_a), Y \mright]} e_b
	- \nabla_Y\mleft( \mleft[ e_a, e_b \mright]_E \mright)
	- \nabla_Y \nabla_{\rho(e_b)} e_a
	+ t_{\nabla^{\mathrm{bas}}} \mleft( \nabla_Y e_a, e_b \mright)
\\
&=
\mleft[ e_a, \nabla_Y e_b \mright]_E
	+ \nabla_{\rho\mleft( \nabla_Y e_b \mright)} e_a
	- \nabla_{\mleft[ \rho(e_a), Y \mright]} e_b
	- \nabla_Y\mleft( \mleft[ e_a, e_b \mright]_E \mright)
	- \nabla_Y \nabla_{\rho(e_b)} e_a
\\
&\hspace{1cm}
	+ \mleft[ \nabla_Y e_a, e_b \mright]_E
	- \nabla_{\rho\mleft( \nabla_Y e_a \mright)} e_b
	+ \nabla_{\rho(e_b)} \nabla_Y e_a
\\
&=
- \nabla_Y\mleft( \mleft[ e_a, e_b \mright]_E \mright)
	+ \mleft[ e_a, \nabla_Y e_b \mright]_E
	+ \mleft[ \nabla_Y e_a, e_b \mright]_E
	+ \nabla_{\nabla^{\mathrm{bas}}_{e_b} Y} e_a
	- \nabla_{\nabla^{\mathrm{bas}}_{e_a} Y} e_b
\\
&\hspace{1cm}
	+ \nabla_{\rho(e_b)} \nabla_Y e_a
	- \nabla_Y \nabla_{\rho(e_b)} e_a
	- \nabla_{\mleft[ \rho(e_b), Y \mright]} e_a
\\
&\stackrel{\mathclap{\text{Def.~\ref{def:basiccurvature}}}}{=}~~~~
- R_\nabla^{\mathrm{bas}}(e_a, e_b) Y
	+ R_\nabla \mleft( \rho(e_b), Y \mright) e_a
\eas
for all $Y\in \mathfrak{X}(N)$, and we are going to view $Y \mapsto - R_\nabla^{\mathrm{bas}}(e_a, e_b) Y + R_\nabla \mleft( \rho(e_b), Y \mright) e_a$ as an element of $\Omega^1(N; E)$ (locally). Hence, altogether
\bas
\delta_\varepsilon F
~~~~&\stackrel{\mathclap{\text{Def.~\ref{def:EichbosonenUndFeldstaerke}}}}{=}~~~~
- \varepsilon^a ~ {}^!\bigl( R_{\nabla}(\cdot, \cdot) e_a \bigr)
	-\frac{\varepsilon^a}{2} ~
	\mleft( {}^* \mleft( R_{\nabla_\rho}(\cdot, \cdot) e_a \mright) \mright) \mleft( \varpi_2 \stackrel{\wedge}{,} \varpi_2 \mright)
\\
&\hspace{1cm}~~~~
	+ \varepsilon^a \varpi_2^b \wedge {}^! \mleft( R_\nabla(\rho(e_b), \cdot) e_a - R_\nabla^{\mathrm{bas}}(e_a, e_b) \mright)
\\
&\stackrel{\mathclap{\text{Eq.~\eqref{EqPullBackFormelFuerVerschiedeneDefinitionen}}}}{=}~~~~
-\frac{1}{2} ~ \Bigl(
 \mleft(	{}^* R_{\nabla} \mright)\mleft(\mathrm{D} \stackrel{\wedge}{,} \mathrm{D}\mright) \varepsilon
	+ \underbrace{\mleft( {}^* R_{\nabla_\rho} \mright)\mleft( \varpi_2 \stackrel{\wedge}{,} \varpi_2 \mright) \varepsilon}
	_{\mathclap{= \mleft( {}^* R_{\nabla} \mright)\mleft( \mleft({}^*\rho\mright)(\varpi_2) ~\stackrel{\wedge}{,} ~\mleft({}^*\rho\mright)(\varpi_2) \mright) \varepsilon}}
	\Bigr)
\\
&\hspace{1cm}~~~~
	+ \underbrace{\mleft( {}^* R_\nabla \mright) \bigl(\mleft({}^*\rho\mright)(\varpi_2) \stackrel{\wedge}{,} \mathrm{D}\bigr) \varepsilon}
	_{\mathclap{= \frac{1}{2}\mleft( 
		\mleft( {}^* R_\nabla \mright) \bigl(\mleft({}^*\rho\mright)(\varpi_2) ~\stackrel{\wedge}{,}~ \mathrm{D}\bigr) \varepsilon
		+ \mleft( {}^* R_\nabla \mright) \bigl(\mathrm{D} ~\stackrel{\wedge}{,}~ \mleft({}^*\rho\mright)(\varpi_2) \bigr) \varepsilon
	\mright) }}
	- \mleft({}^* R_\nabla^{\mathrm{bas}} \mright) \mleft(\varepsilon \stackrel{\wedge}{,} \varpi_2  \stackrel{\wedge}{,} \mathrm{D}\mright)
\\
&\stackrel{\mathclap{\text{Def.~\ref{def:MinimalCoupling}}}}{=}~~~~
-\frac{1}{2} \Bigl(
	 \mleft(	{}^* R_{\nabla} \mright)\mleft( \mathfrak{D} \stackrel{\wedge}{,} \mathrm{D}\mright) \varepsilon
	- \mleft( {}^* R_{\nabla} \mright)\mleft( \mathfrak{D} \stackrel{\wedge}{,} \mleft({}^*\rho\mright)(\varpi_2) \mright) \varepsilon
\Bigr)
\\
&\hspace{1cm}~~~~
	- \mleft({}^* R_\nabla^{\mathrm{bas}} \mright) \mleft(\varepsilon \stackrel{\wedge}{,} \varpi_2  \stackrel{\wedge}{,} \mathrm{D}\mright)
\\
&=
- \mleft(
	\frac{1}{2} ~ \mleft(	{}^* R_{\nabla} \mright)\mleft( \mathfrak{D} \stackrel{\wedge}{,} \mathfrak{D} \mright) \varepsilon
	+ \mleft({}^* R_\nabla^{\mathrm{bas}} \mright) \mleft(\varepsilon \stackrel{\wedge}{,} \varpi_2  \stackrel{\wedge}{,} \mathrm{D}\mright)
\mright),
\eas
where we introduced the notation $\Gamma(E) \times \Gamma(E) \times \mathfrak{X}(N) \ni (\mu, \nu, Y) \mapsto R_\nabla^{\mathrm{bas}}(\mu, \nu, Y) = R_\nabla^{\mathrm{bas}}(\mu, \nu) Y$ in order to emphasize the anti-symmetrization when applying the graded extension on $R_\nabla^{\mathrm{bas}}$, and we used the same argument on $\mleft( {}^* R_\nabla \mright) \bigl(\mleft({}^*\rho\mright)(\varpi_2) \stackrel{\wedge}{,} \mathrm{D}\bigr) \varepsilon$ as in Eq.~\eqref{IchMussDasDringendVerallgemeinern}.
\end{proof}

\begin{remark}\label{RemVergleicheVonVariationenVonFUndDAPhi}
\leavevmode\newline
These formulas look different when comparing it with the standard formulas, but that is again related to that we use the basic connection for the variations instead. As introduced, we should look at the variation of the components to see how the variation affects the variation of the Lagrangian.

$\bullet$ In order to define gauge invariance the idea is as in \cite{CurvedYMH}, $\delta_\varepsilon F^a$ should be proportional to $F$ which is not the case here for both terms. Explicitly we need that $\delta_\varepsilon F = 0$; in that case we would have for the components (with respect to a frame $\mleft( e_a \mright)_a$ of $E$)
\ba
\delta_\varepsilon F^a
&=
\underbrace{\mleft(\delta_\varepsilon F\mright)^a}_{= 0}
	- F^b ~ \bigl( \delta_\varepsilon ({}^* e_b) \bigr)^a
=
\mleft({}^*\mleft( \nabla^{\mathrm{bas}}_\varepsilon e_b \mright) \mright)^a  F^b
=
\varepsilon^c ~ \mleft({}^*\mleft( \mleft[ e_c, e_b \mright]_E + \nabla_{\rho(e_b)} e_c \mright) \mright)^a F^b \label{eqVariationVonFKomps}
\ea
such that the variation of the components is proportional to themselves and we can then formulate the symmetry on scalar products as usual as a symmetry under (infinitesimal) "rotations", see also the next theorem.

In the proof we saw that we can also write
\bas
\delta_\varepsilon F
&=
-\frac{1}{2} ~ \Bigl(
 \mleft(	{}^* R_{\nabla} \mright)\mleft(\mathrm{D} \stackrel{\wedge}{,} \mathrm{D} \mright) \varepsilon
	+ \mleft( {}^* R_{\nabla_\rho} \mright)\mleft( \varpi_2 \stackrel{\wedge}{,} \varpi_2 \mright) \varepsilon
	\Bigr)
\\
&\hspace{1cm}~~~~
	+ \mleft( {}^* R_\nabla \mright) \bigl(\mleft({}^*\rho\mright)(\varpi_2) \stackrel{\wedge}{,} \mathrm{D} \bigr) \varepsilon
	- \mleft({}^* R_\nabla^{\mathrm{bas}} \mright) \mleft(\varepsilon \stackrel{\wedge}{,} \varpi_2  \stackrel{\wedge}{,} \mathrm{D} \mright).
\eas
Since $\Phi$ and $A$ are regarded as the fields with respect to which the theory gets varied and $M$, $N$ \textit{etc.}~are completely arbitrary up to this point, so, thinking about the whole category of possible manifolds, $\mathrm{D}$ and $\varpi_2$ can be viewed as (in general) independent functionals while $\varepsilon$ is very arbitrary. Thus, in order to get $\delta_\varepsilon F = 0$ we need $R_\nabla = 0$ and $R_\nabla^{\mathrm{bas}} = 0$ in general. $R_\nabla^{\mathrm{bas}} = 0$ sounds reasonable as we discussed in the previous section, recall the discussion around Cor.~\ref{cor:FlatnessVonEichtrafos}, but the condition that $\nabla$ is flat is not a good condition because this will lead to that we have locally the standard formulation of gauge theory which is not the aim of this new formulation. The problems with flatness we are going to discuss later, instead let us discuss why this formula recovers the standard formula when using again action Lie algebroids with canonical flat connections. 

$\bullet$ As usual we use again Cor.~\ref{cor:StandardTheory}, for this assume that $E = N \times \mathfrak{g}$ is an action Lie algebroid for some Lie algebra $\mathfrak{g}$, equipped with the canonical flat connection $\nabla$; as in the proof of Thm.~\ref{thm:ActionLieALgebroid} the canonical flat connection satisfies $R_\nabla^{\mathrm{bas}}= 0$. Thus, we have then $\delta_\varepsilon F = 0$, and by the previous calculation
\bas
\delta_\varepsilon F^a
&=
\varepsilon^c ~ \underbrace{\Phi^*\mleft( \mleft[ e_c, e_b \mright]_{\mathfrak{g}} \mright)^a}_{\mathclap{\text{const.}}} ~ F^b
=
\mleft( \mleft[ \varepsilon, F \mright]_{\mathfrak{g}} \mright)^a
\eas
for $\mleft( e_a \mright)_a$ a constant frame. This is again precisely the expected formula, recall Prop.~\ref{prop:ClassicGaugeTrafoOfFieldStrengthAndMinimalCoupling}, and this is also shown and argued in \cite[see the second paragraph after Eq.~(11), keep in mind that the different sign for $\varepsilon$]{CurvedYMH}, where also the general formula with the curvature got stated, but again only for the components without knowing the full tensor. 
%
%Finally, observe in the general situation that when we would rewrite the formula using $\delta^{(1)}$ instead (recall the discussion around Def.~\ref{def:TotalInfinGaugeTransform}) such that $\delta_\varepsilon^{(1)} F^a = \delta_\varepsilon F^a$, then we get, using the previous calculations,
%\bas
%\delta_\varepsilon^{(1)}F
%&=
%\delta^{(1)}_\varepsilon F^a \otimes \Phi^*e_a
	%+ F^b \otimes \delta^{(1)}_\varepsilon \mleft( \Phi^*e_b \mright)
%\\
%&=
%\varepsilon^c F^b \otimes \Phi^*\mleft( \mleft[ e_c, e_b \mright]_E + \nabla_{\rho(e_b)} e_c \mright)
	%- \varepsilon^c F^b \otimes \Phi^*\mleft( \nabla_{\rho(e_c)} e_b \mright)
%\\
%&\hspace{1cm}
	%- \mleft(
	%\frac{1}{2} ~ \mleft(	\Phi^* R_{\nabla} \mright)\mleft( \mathfrak{D}^A \Phi \stackrel{\wedge}{,} \mathfrak{D}^A \Phi \mright) \varepsilon
	%+ \mleft(\Phi^* R_\nabla^{\mathrm{bas}} \mright) \mleft(\varepsilon \stackrel{\wedge}{,} A  \stackrel{\wedge}{,} \mathrm{D}\Phi\mright)
%\mright)
%\\
%&=
%\mleft( \Phi^* t_{\nabla_\rho} \mright)\mleft( \varepsilon, F \mright) 
	%- \mleft(
	%\frac{1}{2} ~ \mleft(	\Phi^* R_{\nabla} \mright)\mleft( \mathfrak{D}^A \Phi \stackrel{\wedge}{,} \mathfrak{D}^A \Phi \mright) \varepsilon
	%+ \mleft(\Phi^* R_\nabla^{\mathrm{bas}} \mright) \mleft(\varepsilon \stackrel{\wedge}{,} A  \stackrel{\wedge}{,} \mathrm{D}\Phi\mright)
%\mright),
%\eas
%where we used Eq.~\eqref{eqVariationVonFKomps} but now with non-zero $\delta_\varepsilon F$.
\end{remark}

Using this and Remark \ref{RemVergleicheVonVariationenVonFUndDAPhi} we can finally formulate what we need to have a gauge-invariant Lagrangian; for this we need to calculate $\delta_\varepsilon \mathfrak{L}_{\mathrm{YMH}}$ (Def.~\ref{def:CurvedYMHLagrangian}).

\begin{theorems}{The gauge invariance of the Lagrangian, \newline \cite[especially the discussion around Eq.~(16)]{CurvedYMH}}{GaugeInvariantStandardLagrangian}
Let $M$ be a spacetime with a spacetime metric $\eta$, $N$ a smooth manifold, $E \to N$ a Lie algebroid, $\nabla$ a connection on $E$, $\kappa$ and $g$ fibre metrics on $E$ and $\mathrm{T}N$, respectively. Also let $V \in C^\infty(N)$ and assume that the following \textbf{compatibility conditions} hold:
\ba
	R_\nabla &= 0, \\
	R_\nabla^{\mathrm{bas}} &= 0, \\
	\nabla^{\mathrm{bas}} \kappa &= 0, \\
	\nabla^{\mathrm{bas}} g &= 0, \\
	{}^*\mleft(\mathcal{L}_{({}^*\rho)(\varepsilon)} V\mright) &= 0 \label{PotentialCompatibility}
\ea
for all $\varepsilon \in \mathcal{F}^0_E(M; {}^*E)$. Then we have
\ba
\delta_\varepsilon \mathfrak{L}_{\mathrm{YMH}}
&=
0
\ea
for all $\varepsilon \in \mathcal{F}^0_E(M; {}^*E)$.
\end{theorems}

\begin{remark}\label{RemarkUeberPotentialCompatibility}
\leavevmode\newline
Since Lie derivatives describe the canonical flat connection on smooth functions (canonical flatness with respecto to the trivial line bundle over $N$, the notation of Eq.~\eqref{PotentialCompatibility} is the same as introduced in Remark \ref{RemarkNotationvonPullbackConnection} and as in other similar terms, \textit{i.e.}
\bas
\mleft.\mleft({}^*\mleft(\mathcal{L}_{({}^*\rho)(\varepsilon)} V\mright)\mright)(\Phi,A)\mright|_p
&=
\mleft.\Phi^*\mleft(\mathcal{L}_{(\Phi^*\rho)(\epsilon)} V\mright)\mright|_p
=
\mathcal{L}_{\mleft(\rho_{\Phi(p)}\mright)\mleft(\epsilon_p\mright)} V
\eas
for all $(p,\Phi,A) \in M \times \mathfrak{M}_E(M;N)$, where $\epsilon \coloneqq \varepsilon(\Phi,A) \in \Gamma(\Phi^*E)$. It is clear that Eq.~\eqref{PotentialCompatibility} generalizes Eq.~\eqref{ClassicPotential} if $E$ is an action Lie algebroid.
\end{remark}

\begin{proof}
\leavevmode\newline
Observe that $* ({}^*V ) = {}^*V ~ \mathrm{dvol}_\eta$, where $\mathrm{dvol}_\eta$ is the canonical volume form of $\eta$ and the sign might differ depending on the definition of the Hodge star operator. Using that, we only need to look at the variation of ${}^*V$ because $\mathrm{dvol}_\eta$ is clearly not affected by $\delta$, hence,
\bas
\delta_\varepsilon ({}^*V)
&=
- {}^*\mleft(\mathcal{L}_{({}^*\rho)(\varepsilon)} V\mright)
=
0
\eas
for all $\varepsilon \in \mathcal{F}^0_E(M; {}^*E)$, where we used the last condition. Up to a sign we also have\footnote{As also defined in \cite[\S 7.2, Definition 7.2.4; page 408]{hamilton}.}
\bas
\omega \wedge *\psi
&=
\langle \omega, \psi \rangle ~ \mathrm{dvol}_\eta
\eas
for all $\omega, \psi \in \Omega^k(M)$ $(k \in \mathbb{N}_0)$, where $\langle \cdot, \cdot \rangle$ is the standard scalar product defined on $\Omega^k(M)$ using $\eta$, \textit{i.e.}
\bas
\langle \omega, \psi \rangle
&=
\frac{1}{k!} ~ \omega_{\alpha_1, \dotsc, \alpha_k} \psi^{\alpha_1, \dotsc, \alpha_k}
\eas
where we express the forms with respect to coordinate vector fields $\mleft(\partial_\alpha\mright)_\alpha$ on $M$ and raising an index is done by using $\eta$; especially, $\delta_\varepsilon$ satisfies the Leibniz rule on $\langle \cdot, \cdot \rangle$ because $\delta_\varepsilon \eta = 0$. Hence, similar to before,
\bas
\delta_\varepsilon\mleft(\omega \wedge *\psi\mright)
&=
\delta_\varepsilon \bigl( \langle \omega, \psi \rangle ~ \mathrm{dvol}_\eta \bigr)
=
\bigl( 
	\langle \delta_\varepsilon \omega, \psi \rangle
	+ \langle \omega, \delta_\varepsilon \psi \rangle
 \bigr) ~ \mathrm{dvol}_\eta
=
\delta_\varepsilon \omega \wedge *\psi
	+ \omega \wedge *\mleft(\delta_\varepsilon\psi\mright)
\eas
for all $\varepsilon \in \mathcal{F}^0_E(M; {}^*E)$. This clearly extends to Def.~\ref{def:GradingOfProducts} by the Leibniz rule (\textit{e.g.}~this is immediate by the coordinate expression of graded extensions), in the sense of
\bas
\delta_\varepsilon \bigl(
\mleft( {}^*\kappa \mright)\mleft(F \stackrel{\wedge}{,} *F\mright)
\bigr)
&=
\mleft( \delta_\varepsilon \mleft({}^*\kappa \mright) \mright)\mleft(F \stackrel{\wedge}{,} *F\mright)
	+ \mleft( {}^*\kappa \mright)\mleft(\delta_\varepsilon F \stackrel{\wedge}{,} *F\mright)
	+ \mleft( {}^*\kappa \mright)\bigl(F \stackrel{\wedge}{,} *(\delta_\varepsilon F)\bigr)
\eas
for all $\varepsilon \in \mathcal{F}^0_E(M; {}^*E)$, similarly for other all terms of that form.
Observe that we have $\delta_\varepsilon F = 0$ additionally to $\delta_\varepsilon \mathfrak{D} = 0$ by Prop.~\ref{prop:InfinitesimalGaugeTrafoOfMinimalCoupleSmiley} and \ref{prop:GaugeTrafosOfFieldStrengthAndMinimalCoupling} and due to $R_\nabla = 0$ and $R_\nabla^{\mathrm{bas}} = 0$.
So, we get in total, using the result of the variation of the potential $V$,
\bas
\delta_\varepsilon \mathfrak{L}_{\mathrm{YMH}}
&=
\delta_\varepsilon \mleft(
	- \frac{1}{2} \mleft( {}^*\kappa \mright)\mleft(F \stackrel{\wedge}{,} *F\mright)
	+ \mleft( {}^*g \mright)\mleft(\mathfrak{D} \stackrel{\wedge}{,} *\mathfrak{D} \mright)
	- *({}^*V)
\mright)
\\
&=
- \frac{1}{2} \bigl( \delta_\varepsilon \mleft( {}^*\kappa \mright)\bigr)\mleft(F \stackrel{\wedge}{,} *F\mright)
	+ \bigl( \delta_\varepsilon \mleft( {}^*g \mright) \bigr)\mleft(\mathfrak{D} \stackrel{\wedge}{,} *\mathfrak{D} \mright)
\\
&\stackrel{\mathclap{\text{Eq.~\eqref{PullBackVariation}}}}{=}~~~~
\frac{1}{2} \biggl( {}^*\mleft( \nabla^{\mathrm{bas}}_\varepsilon \kappa \mright)\biggr)\mleft(F \stackrel{\wedge}{,} *F\mright)
	- \biggl( {}^*\mleft( \nabla^{\mathrm{bas}}_\varepsilon g \mright) \biggr)\mleft(\mathfrak{D}  \stackrel{\wedge}{,} *\mathfrak{D} \mright)
\\
&=
0
\eas
for all $\varepsilon \in \mathcal{F}^0_E(M; {}^*E)$, using the metric compatibilities in the assumed conditions.
\end{proof}

Lie algebroids equipped with a connection with vanishing basic curvature are also called \textbf{Cartan algebroids} as \textit{e.g.}~defined in \cite[\S 2.3]{blaomTangentBundleAsLieGroup}; hence, this special type of Lie algebroid seems to be the relevant one for gauge theories, as we already have noticed in the discussion about gauge transformations. Let us collect all the results we got along the way in relation to the standard formulation.

\begin{theorems}{Standard formulation of gauge theory is recovered, \cite{CurvedYMH}}{StandardEichtheorieStecktInDenBedingung}
Assume that $N=W$ is a vector space, $E= N \times \mathfrak{g}$ an action Lie algebroid for a Lie algebra $\mathfrak{g}$ whose Lie algebra action $\gamma$ is induced by a Lie algebra representation $\psi: \mathfrak{g} \to \mathrm{End}(W)$, and assume that $\nabla$ is the canonical flat connection of $E$. Moreover, let $\kappa$ be a fibre metric of $E$ which is a canonical extension of an $\mathrm{ad}$-invariant scalar product of $\mathfrak{g}$, similarly $g$ is a metric on $\mathrm{T}W \cong W \times W$ constantly extending an $\psi$-invariant scalar product of $W$. Finally, let $V \in C^\infty(N)$ such that it satisfies Eq.~\eqref{PotentialCompatibility}.

Then the compatibility conditions of Thm.~\ref{thm:GaugeInvariantStandardLagrangian} are satisfied, and we recover the standard theory: The Lagrangian $\mathfrak{L}_{\mathrm{YMH}}$ is as in the standard formulation and gauge-invariant, as does the field strength $F$, the minimal coupling $\mathfrak{D}$, the field of gauge bosons $A$, the field $\Phi$, and its variation $\delta_\varepsilon \Phi$; with respect to a constant frame $\mleft( e_a \mright)_a$ of $E$ and a constant frame $\mleft(\partial_\alpha\mright)_\alpha$ of $\mathrm{T}W$, $\delta_\varepsilon A^a$ coincide with the components of the variation of $A$ of the standard formulation, as does $\delta_\varepsilon F^a$ and $\delta_\varepsilon \mleft( \mathfrak{D} \mright)^\alpha$.
\end{theorems}

\begin{remark}
\leavevmode\newline
As discussed in subsection \ref{InfinitesimalGaugeTransformation}, the infinitesimal gauge transformation of the Lagrangian is just $\delta_\varepsilon \mathfrak{L}_{\mathrm{YMH}} = \mathcal{L}_{\Psi_\varepsilon} \mathfrak{L}_{\mathrm{YMH}}$. Thence, the definition of $\Psi_\varepsilon$ is of importance for the gauge invariance of the Lagrangian, that is, how $\Phi$ and how the components of $A$ transform; recall Prop.~\ref{prop:VariationOfA}. Given that unique $\Psi_\varepsilon$ of Prop.~\ref{prop:VariationOfA} (for a fixed $\nabla$) one can take any other connection on $E$ to formulate $\delta_\varepsilon A$ and $\delta_\varepsilon$ in general, one will always get the gauge invariance of the Lagrangian, and the components of $A$ \textit{etc.}~will also transform the same. Hence, the statement about the transformations of the components could also be formulated as that $\Psi_\varepsilon$ reduces to the same vector field on the space of fields as in the classical situation.

However, as already mentioned before, the definition of $\Psi_\varepsilon$ depends on $\nabla$; but \textbf{given a $\Psi_\varepsilon$} the choice of connections for the definition of $\delta_\varepsilon$ does not affect the gauge invariance of the Lagrangian.

When we would use $\nabla_\rho$ to define the gauge transformations of $E$-valued functionals, then many of the total formulas would also restrict to standard formulas due to the flatness of $\nabla$ in the standard situation, not just their components, recall Thm.~\ref{thm:NewFormulaRecoversOldGaugeTrafoYay}. That is especially due to that $\nabla_\rho$ will be a canonical flat connection, while the basic connection is flat but it may not have a parallel frame due to the kernel of the anchor. If we would use $\nabla_\rho$, we would loose the flatness of the gauge transformations as discussed in Cor.~\ref{cor:FlatnessVonEichtrafos} whenever $\nabla_\rho$ is not flat anymore. However, we have now seen that $\nabla$ needs to be flat for the gauge invariance of the Lagrangian such that this does seemingly not matter; but we will see later that there is the possibility to allow non-flat $\nabla$.
\end{remark}

\begin{proof}[Proof of Thm.~\ref{thm:StandardEichtheorieStecktInDenBedingung}]
\leavevmode\newline
First recall Thm.~\ref{thm:ActionLieALgebroid}, especially, the canonical flat connection satisfies $R_\nabla^{\mathrm{bas}}=0$; the metric compatibilities follow by Lemma \ref{lem:MetricCompsAdInvUndLieAlgebraRepSymm}, hence, all compatibility conditions of Thm.~\ref{thm:GaugeInvariantStandardLagrangian} are satisfied. That the formulas restrict to the standard ones we have discussed in Cor.~\ref{cor:StandardTheory} and \ref{cor:EichtrafovonDAPHIinClassicIstBabyEinfach}, and Remarks \ref{RemUeberVariationVonHiggs}, \ref{RemDifferentVersionsOfGaugeTrafos}, and \ref{RemVergleicheVonVariationenVonFUndDAPhi}.
\end{proof}

But due to the compatibility condition about the flatness we arrive locally now at an action Lie algebroid, regardless of the specific choice of $E$; and as we have seen multiple times, action Lie algebroids recovers the classical theory.

\begin{corollaries}{Gauge invariance implies standard theory, \newline \cite[the discussion around Eq.~(9)ff.]{CurvedYMH}}{ManBrauchZetaWahrscheinlich}
Let us have the same conditions as in Thm.~\ref{thm:GaugeInvariantStandardLagrangian}. Then $E$ is locally isomorphic to an action Lie algebroid $N \times \mathfrak{g}$ such that $\nabla$ is its canonical flat connection and $N =W$ is a vector space, also, $\delta_\varepsilon A^a$ are then of the form as in the standard formulation of gauge theory with respect to a constant frame $\mleft( e_a \mright)_a$, as does $\delta_\varepsilon F^a$.
\end{corollaries}

\begin{remark}
\leavevmode\newline
Using Thm.~\ref{thm:StandardEichtheorieStecktInDenBedingung} one can also derive the other classical formulas depending on the conditions about the structure, like a given Lie algebra representation. But those are just technicalities, the important part is to have an action Lie algebroid and its canonical flat connection.
\end{remark}

\begin{proof}[Proof of Cor.~\ref{cor:ManBrauchZetaWahrscheinlich}]
\leavevmode\newline
By Thm.~\ref{thm:ActionLieALgebroid} we immediately know that $E\cong N \times \mathfrak{g}$ is an action Lie algebroid for a Lie algebra $\mathfrak{g}$ with Lie algebra action $\gamma: \mathfrak{g} \to \mathfrak{X}(N)$ on some open neighbourhood around each point, in such a way that $\nabla$ is its canonical flat connection. Restricting the neighbourhood even further results into $N=W$ for some vector space $W$. The remaining proof is exactly as in Thm.~\ref{thm:StandardEichtheorieStecktInDenBedingung}.
\end{proof}

Hence, we arrive locally always at the standard situation; at least at something very similar to it. 
%(up to whether the fibre metrics\footnote{Not speaking about the spacetime metric of course.} are extensions of scalar products, the potential is a polynomial and whether there is a Lie algebra representation behind the action and so on, but those are rather technical than important details). 
The Lie algebra action might not come from a Lie algebra representation and the metrics might look exotic, but these are just technicalities which are not important for us, especially when one recalls that the aim of this theory is that gauge theory is covariantized in order to easily replace $\nabla$ with non-flat connections. However, there is a possibility in allowing non-flat connections, and for this we need to change the field strength to compensate the curvature term in Prop.~\ref{prop:GaugeTrafosOfFieldStrengthAndMinimalCoupling} which is mainly the reason behind the compatibility condition about flatness, as also argued as an ansatz in \cite[second paragraph after Equation (11)]{CurvedYMH}. We want to motivate this change by a field redefinition instead, a transformation which keeps the Lagrangian invariant after a modification, but breaking the condition about flatness.

Before we do this let us shortly summarize an aspect of the classical theory which is now obvious due to this formulation.

\begin{corollaries}{Abelian Lie algebras and zero torsion}{AbelianIffNablaBasIsLeviCivita}
Let $E = N \times \mathfrak{g}$ be an action Lie algebroid over $N$ for a Lie algebra $\mathfrak{g}$, equipped with the canonical flat connection $\nabla$. Then
\ba
t_{\nabla^{\mathrm{bas}}}=0
&\Leftrightarrow
\mathfrak{g} \text{ is abelian}.
\ea
\end{corollaries}

\begin{remark}
\leavevmode\newline\label{remELEVICITAOfBasnbala}
Given a fixed fibre metric $\kappa$ such that $\nabla^{\mathrm{bas}} \kappa = 0$, as in one of the compatibility conditions, we would therefore know that $\nabla^{\mathrm{bas}}$ is an $E$-Levi-Civita connection if and only if $\mathfrak{g}$ is abelian.\footnote{See \textit{e.g.}~\cite[\S 2.5]{ELeviCivita} for a definition of such Levi-Civita connections. However, it is precisely defined as usual.}
\end{remark}

\begin{proof}
\leavevmode\newline
We only need to check under which conditions the tensor of the torsion of the basic connection is zero for constant sections $\mu, \nu$ since these generate all sections, especially we have $\nabla \mu = \nabla \nu = 0$ and $\mleft[ \mu, \nu \mright]_E = \mleft[ \mu,\nu \mright]_{\mathfrak{g}}$:
\bas
&&
0
&=
\underbrace{t_{\nabla^{\mathrm{bas}}}( \mu, \nu)}_{\mathclap{= - t_{\nabla_{\rho}}( \mu, \nu)}}
\\
&\Leftrightarrow&
0
&=
t_{\nabla_{\rho}}( \mu, \nu)
\\
&\Leftrightarrow&
0
&=
\mleft[ \mu, \nu \mright]_{\mathfrak{g}}.
\eas
\end{proof}
\section{Field redefinition}\label{FieldRedefSection}

We want to study a certain transformation which keeps the action invariant; for this recall first Sylvester's determinant theorem (\cite[Appendix B; page 271]{DeterminantenTheorem}), also called Weinstein-Aronszajn identity, which says
\ba\label{SylvestersDeterminante}
\mathrm{det}\mleft( \mathds{1}_n + C B \mright)
&=
\mathrm{det}\mleft( \mathds{1}_m + B C \mright),
\ea
where $n, m \in \mathbb{N}$, $\mathds{1}_n$ and $\mathds{1}_m$ are the identity matrices on $\mathbb{R}^n$ and $\mathbb{R}^m$, respectively, and $C \in \mathbb{R}^{n \times m}$ and $B \in \mathbb{R}^{m \times n}$. 

Abstractly spoken, the typical idea of field redefinitions is the same as for covariantizing physical theories and definitions. One applies a non-constant change of coordinates in such a way that one leaves the "inertial frame" as in classical mechanics, resulting to that one gets extra terms in several formulas like contributions coming from "inertial forces"; but one still has the same physics, because the Lagrangian is actually invariant under that change of coordinates. Usually one reformulates the same theory naturally supporting those extra terms, leading to a theory naturally invariant under the observed changes of coordinates in all definitions, which is often referred to as covariantization by physicists. Up to this point it is just something aesthetic one could say, however, the next step is then to study whether the mentioned extra terms always vanish in some coordinate system. Think \textit{e.g.}~of connection 1-forms of connections and one started with a theory with an underlying flat connection such that the initial coordinate system was also the parallel frame where the 1-forms are zero, and the connection 1-forms then arise as those extra terms in other coordinate systems. Studying whether those connection 1-forms always can vanish in some coordinate system, means, whether or not non-trivial curvatures are possible.

In our case the "coordinates" we speak of is the structural data, especially $A$, a coordinate of $\mathfrak{M}_E$, but also for example $\nabla$, and, so, the extra terms are going to be in the compatibility condition about the curvature of $\nabla$. To keep the same physics, that is, the Lagrangian stays invariant, we need to correct especially the field strength since the field strength is of course directly affected by non-trivial changes of $A$. Since the previously-discussed flatness of $\nabla$ is given by the infinitesimal gauge transformation of the field strength, there is the hope that whatever we need to add to "correct" the field strength will also lead to a gauge invariant theory allowing non-flat connections. As a next step it is then natural to rewrite gauge theory allowing those extra terms, leading to a theory naturally invariant under the chosen change of "coordinates" (as in coordinate-independence), while the classical theory is just the same theory, written with respect to "coordinates" where those extra terms are zero. Finally, one may want to discuss what happens when these extra terms actually never vanish, even after such changes of "coordinates". So, precisely the same as in the previous paragraph, just happening with a different type of "coordinate", which is why we are not going to say covariantization but field redefinition.

Let us start defining that field redefinition.

\begin{definitions}{Field redefinition}{FieldRedefinition}
Let $M, N$ be smooth manifolds, $E \to N$ a Lie algebroid, $\nabla$ a connection on $E$, and $\kappa$ and $g$ fibre metrics on $E$ and $\mathrm{T}N$, respectively. Also let $\gls{1lambda} \in \Omega^1(N; E)$ such that $\gls{1Lambda} \coloneqq \mathds{1}_E - \lambda \circ \rho$ is an element of $\sAut(E)$. We then define the \textbf{field redefinition} by 
\ba\label{EqFieldRedefFuerA}
\gls{1pivarwidetildelambda}
&\coloneqq 
\mleft( {}^* \Lambda \mright) (\varpi_2)+ {}^! \lambda,
\\
\gls{0nabla0widetildelambda}
&\coloneqq
\nabla
	+ \mleft( \Lambda \circ \mathrm{d}^{\nabla^{\mathrm{bas}}} \circ \Lambda^{-1} \mright) \lambda,
\label{FieldTrafoOfNabla} \\
\gls{1kappawidetildelambda}
&\coloneqq
\kappa \circ \mleft( \Lambda^{-1}, \Lambda^{-1} \mright),
\label{FieldTrafoOfKappa} \\
\gls{gwidetildelambda}
&\coloneqq
g \circ \mleft( \widehat{\Lambda}^{-1}, \widehat{\Lambda}^{-1} \mright), \label{FieldTrafoOfG}
\ea
where $\gls{1Lambdatilde} \coloneqq \mathds{1}_{\mathrm{T}N} - \rho \circ \lambda$.
\end{definitions}

\begin{remark}
\leavevmode\newline
\indent $\bullet$ $\widehat{\Lambda}$ and $\Lambda$ are already endomorphisms by definition, and, so, by Eq.~\eqref{SylvestersDeterminante} we know that $\widehat{\Lambda} \in \sAut(\mathrm{T}N)$ if and only if $\Lambda \in \sAut(E)$. Also recall that we view elements of $\Omega^1(N;E)$ also as elements of $\Omega^{1,0}(N,E;E)$, Def.~\ref{def:ExteriorCovariantDerivatives}, therefore $\mleft( \Lambda \circ \mathrm{d}^{\nabla^{\mathrm{bas}}} \circ \Lambda^{-1} \mright) \lambda \in \Omega^{1,1}(N,E;E) \cong \Omega^1(N; \mathrm{End}(E))$. 

$\bullet$ 
We can rewrite $\widetilde{\varpi_2}^\lambda$ to
\ba
\widetilde{\varpi_2}^\lambda
&= 
\mleft( {}^* \Lambda \mright) (\varpi_2)+ {}^! \lambda
\stackrel{\text{ Eq.~\eqref{EqPullBackFormelFuerVerschiedeneDefinitionen} }}{=}
\varpi_2
	- \bigl(\underbrace{{}^*(\lambda \circ \rho)}_{\mathclap{ = ({}^*\lambda) \circ ({}^*\rho) }}\bigr)(\varpi_2)
	+ ({}^*\lambda)(\mathrm{D})
= 
\varpi_2 + \mleft( {}^* \lambda \mright) \mleft( \mathfrak{D} \mright).
\ea
With respect to points $(\Phi, A) \in \mathfrak{M}_E(M;N)$ this implies
\ba\label{EqAlternativeFormelFuerFieldredefofA}
\mleft(\widetilde{\varpi_2}^\lambda\mright)(\Phi,A)
&=
\gls{a0widetildelambda}
= 
\mleft( \Phi^* \Lambda \mright) (A)+ \Phi^! \lambda
= 
A + \mleft( \Phi^* \lambda \mright) \mleft( \mathfrak{D}^A \Phi \mright).
\ea
Viewing $A$ and $\varpi_2$ as coordinates on $\mathfrak{M}_E(M;N)$, the idea of the field redefinition is a change of coordinates, consisting of a translation and a rotation with $\Lambda$ which is basically a first order approximation of the typical rotation given by an exponential. The other formulas of the field redefinition are taken in such a way to keep all compatibility conditions in Thm.~\ref{thm:GaugeInvariantStandardLagrangian} but the one about the curvature of $\nabla$. We will see this in the following.

$\bullet$ If we additionally have $R_\nabla^{\mathrm{bas}} = 0$, then we have
\bas
\mleft( \mathrm{d}^{\nabla^{\mathrm{bas}}} \mright)^2
&=
0
\eas
by Prop.~\ref{prop:SnablamitREnabla}, thus, also
\bas
\mleft( \Lambda \circ \mathrm{d}^{\nabla^{\mathrm{bas}}} \circ \Lambda^{-1} \mright)^2
&=
\Lambda \circ \mleft( \mathrm{d}^{\nabla^{\mathrm{bas}}} \mright)^2 \circ \Lambda^{-1} 
=
0,
\eas
hence, we add then an exact term to $\nabla$.

$\bullet$ Eq.~\eqref{EqFieldRedefFuerA} was suggested by one of my supervisors, Thomas Strobl, and the first task of my PhD was to calculate all the remaining formulas and properties needed for the following discussions. In \cite[the example at the very end, right before the conclusion]{CurvedYMH} some transformation was discussed which is a special and simplified situation of the field redefinition. Thomas Strobl got this special example of the field redefinition after a private dialogue with Edward Witten.
\end{remark}

\begin{remarks}{An important note about notation}{TrickyLeibnizRuleForConnections}
Due to $\lambda \in \Omega^{1,0}(N,E;E)$ one may want to write
\bas
\mleft( \Lambda \circ \mathrm{d}^{\nabla^{\mathrm{bas}}} \circ \Lambda^{-1} \mright) \lambda
&=
\mleft( \Lambda \circ \nabla^{\mathrm{bas}} \circ \Lambda^{-1} \mright) \lambda
=
\mathrm{d}^{\Lambda \circ \nabla^{\mathrm{bas}} \circ \Lambda^{-1}} \lambda,
\eas
but the first equality is \textbf{not} correct with our notation! Keep in mind that we have two degrees in form of the spaces $\Omega^{p,q}(N,E;E)$ ($p,q \in \mathbb{N}_0$), so, there are Leibniz rules involved on the $p$-degree if $p \neq 0$, here $p = 1$. That is, for $Y \in \mathfrak{X}(N)$ and $\nu \in \Gamma(E)$, compare
\bas
\mleft(\mleft( \Lambda \circ \mathrm{d}^{\nabla^{\mathrm{bas}}} \circ \Lambda^{-1} \mright) \lambda\mright)(Y, \nu)
&=
\Lambda\mleft( 
	\mleft(\nabla_\nu^{\mathrm{bas}}\mleft( \Lambda^{-1}\circ \lambda \mright)\mright)(Y)
\mright)
\\
&=
\Lambda\mleft(
	\nabla_\nu^{\mathrm{bas}}\mleft( \mleft(\Lambda^{-1} \circ \lambda\mright)(Y) \mright)
\mright)
	- \lambda \mleft( \nabla_\nu^{\mathrm{bas}} Y \mright)
\\
&=
\mleft( \Lambda \circ \nabla^{\mathrm{bas}}_\nu \circ \Lambda^{-1} \mright)\bigl(\lambda(Y)\bigr)
	- \lambda\mleft( \nabla^{\mathrm{bas}}_\nu Y \mright)
\eas
with
\bas
\mleft(\mathrm{d}^{\Lambda \circ \nabla^{\mathrm{bas}} \circ \Lambda^{-1}} \lambda\mright)(Y, \nu)
&=
\mleft( \Lambda \circ \nabla^{\mathrm{bas}}_\nu \circ \Lambda^{-1} \mright)\bigl(\lambda(Y)\bigr)
	- \lambda\mleft( \mleft(\Lambda \circ \nabla^{\mathrm{bas}}_\nu \circ \Lambda^{-1}\mright)Y \mright).
\eas
Hence, due to the Leibniz rules, a composition of maps with connections is not the same as usual compositions of maps, here with a differential. With $\Lambda \circ \nabla^{\mathrm{bas}} \circ \Lambda^{-1}$ we mean the whole object as a connection, so, acting on $\lambda$, extending $\Lambda \circ \nabla^{\mathrm{bas}} \circ \Lambda^{-1}$ as an $E$-connection to $\Omega^1(N;E)$. While each component in $\Lambda \circ \mathrm{d}^{\nabla^{\mathrm{bas}}} \circ \Lambda^{-1}$ acts separately on forms like $\lambda$, and $\nabla^{\mathrm{bas}}$ is extended as $E$-connection to $\Omega^1(N;E)$ (without the conjugation). Therefore one needs to be very careful about how to use conjugations like $\Lambda \circ \dotsc \circ \Lambda^{-1}$ and how to put square brackets, especially when connections are involved. Thus, also
\ba\label{OneofmanyformulasForTildeNabla}
\mleft(\mleft( \Lambda \circ \mathrm{d}^{\nabla^{\mathrm{bas}}} \circ \Lambda^{-1} \mright) \lambda\mright)(\cdot, \nu)
&=
\Lambda\mleft(
	\nabla^{\mathrm{bas}}_\nu\mleft(
		\Lambda^{-1} \circ \lambda
	\mright)
\mright)
\neq
\mleft( \Lambda \circ \nabla^{\mathrm{bas}}_\nu \circ \Lambda^{-1} \mright)\lambda.
\ea
If one always wants to write $\mathrm{d}^{\nabla^{\mathrm{bas}}} = \nabla^{\mathrm{bas}}$ for elements of $\Omega^{p,0}(N,E;E)$ as at the beginning of this remark, then one needs to introduce a notation for extensions as of $\nabla^{\mathrm{bas}}$ to $\Omega^1(N;E)$ in order to avoid precisely the confusion of notation discussed here.
\end{remarks}

We have actually the following corollary relating both notations/notions.

\begin{corollaries}{Conjugation of differentials}{ConjugationOfDifferentialsAreShitty}
Let $N$ be smooth manifolds, $E \to N$ a Lie algebroid, and $\nabla$ a connection on $E$. Also let $\lambda \in \Omega^1(N; E)$ such that $\Lambda = \mathds{1}_E - \lambda \circ \rho$ is an element of $\sAut(E)$. Then
\ba\label{eqKrassWieDasBeiETileNablaAussieht}
&\left(\mathrm{d}^{ \Lambda \circ \nabla^{\mathrm{bas}} \circ \Lambda^{-1}} \omega\right)\left( X_1, \dotsc, X_p, \nu_0, \dotsc, \nu_q \right) \nonumber \\
&= 
\Biggl( \mleft( \Lambda \circ \mathrm{d}^{\nabla^{\mathrm{bas}}} \circ \Lambda^{-1} \mright)
\mleft( \omega \circ \mleft( \vphantom{\widehat{\Lambda}, \dotsc, \widehat{\Lambda}} \smash{\underbrace{\widehat{\Lambda}, \dotsc, \widehat{\Lambda}}_{\mathclap{p \text{ times}}}}, \vphantom{\mathds{1}_{E}, \dotsc, \mathds{1}_E} \smash{\underbrace{\mathds{1}_{E}, \dotsc, \mathds{1}_E}_{q \text{ times}}} \mright) \mright)
 \Biggr)
\mleft( \widehat{\Lambda}^{-1}(X_1), \dotsc, \widehat{\Lambda}^{-1}(X_p), \nu_0, \dotsc, \nu_q \mright)
\ea
for all $\omega \in \Omega^{p,q}(N,E;E)$ ($p, q \in \mathbb{N}_0$), $X_1, \dotsc, X_p \in \mathfrak{X}(N)$ and $\nu_0, \dotsc, \nu_q \in \Gamma(E)$. Equivalently,
\ba
&\mathrm{d}^{ \Lambda \circ \nabla^{\mathrm{bas}} \circ \Lambda^{-1}} \biggl(
	\Lambda \circ \omega \circ \Bigl( \underbrace{\widehat{\Lambda}^{-1}, \dotsc, \widehat{\Lambda}^{-1}}_{p \text{ times}}, \underbrace{\mathds{1}_{E}, \dotsc, \mathds{1}_E}_{q \text{ times}} \Bigr)
\biggr)
\nonumber\\
&=
\Lambda \circ \mleft(\mathrm{d}^{\nabla^{\mathrm{bas}}} \omega\mright) \circ \Bigl( \underbrace{\widehat{\Lambda}^{-1}, \dotsc, \widehat{\Lambda}^{-1}}_{p \text{ times}}, \underbrace{\mathds{1}_{E}, \dotsc, \mathds{1}_E}_{q+1 \text{ times}} \Bigr).
\ea
\end{corollaries}

\begin{remark}
\leavevmode\newline
The second formulation emphasizes that it is roughly about a commutation relation between the conjugation with $\Lambda$ and the differential with the basic connection.
\end{remark}

\begin{proof}[Proof of Cor.~\ref{cor:ConjugationOfDifferentialsAreShitty}]
\leavevmode\newline
That is a straightforward calculation, writing $ {}^E\widetilde{\nabla} \coloneqq \Lambda \circ \nabla^{\mathrm{bas}} \circ \Lambda^{-1}$,
\bas
&\left(\mathrm{d}^{{}^E\widetilde{\nabla}} \omega\right)\left( X_1, \dots, X_p, \nu_0, \dots, \nu_q \right) \nonumber \\
&= \sum_{i=0}^q (-1)^i \biggl( {}^E\widetilde{\nabla}_{\nu_i} \mleft( \omega\mleft( \mleft(\widehat{\Lambda} \circ \widehat{\Lambda}^{-1}\mright) (X_1), \dots, \mleft(\widehat{\Lambda} \circ \widehat{\Lambda}^{-1}\mright) (X_p), \nu_0, \dots, \widehat{\nu}_i, \dots \nu_q\mright) \mright) \nonumber \\
&\qquad\qquad\qquad - \sum_{j=1}^p \quad \underbrace{\omega}_{\mathclap{\Lambda \circ \Lambda^{-1} \circ \omega}}\mleft( \mleft(\widehat{\Lambda} \circ \widehat{\Lambda}^{-1}\mright) (X_1), \dots, {}^E\widetilde{\nabla}_{\nu_i} X_j, \dots, \mleft(\widehat{\Lambda} \circ \widehat{\Lambda}^{-1}\mright) (X_p), \nu_0, \dots, \widehat{\nu}_i, \dots, \nu_q \mright) \biggr) \nonumber \\
&\quad + \sum_{0 \leq i < j \leq q} (-1)^{i+j} \underbrace{\omega}_{\mathclap{= \Lambda \circ \Lambda^{-1} \circ \omega}}\mleft( \mleft(\widehat{\Lambda} \circ \widehat{\Lambda}^{-1}\mright) (X_1), \dots, \mleft(\widehat{\Lambda} \circ \widehat{\Lambda}^{-1}\mright) (X_p), [\nu_i, \nu_j]_E, \nu_0, \dots, \widehat{\nu}_i, \dots, \widehat{\nu}_j, \dots, \nu_q \mright) \nonumber \\
&= 
\Bigg( \mleft( \Lambda \circ \mathrm{d}^{\nabla^{\mathrm{bas}}} \circ \Lambda^{-1} \mright)
\mleft( \omega \circ \mleft( \vphantom{\widehat{\Lambda}, \dots, \widehat{\Lambda}} \smash{\underbrace{\widehat{\Lambda}, \dots, \widehat{\Lambda}}_{\mathclap{p \text{ times}}}}, \vphantom{\mathds{1}_{E}, \dots, \mathds{1}_E} \smash{\underbrace{\mathds{1}_{E}, \dots, \mathds{1}_E}_{q \text{ times}}} \mright) \mright)
 \Bigg)
\mleft( \widehat{\Lambda}^{-1}(X_1), \dots, \widehat{\Lambda}^{-1}(X_p), \nu_0, \dots, \nu_q \mright)
\eas
for all $\omega \in \Omega^{p,q}(N,E;E)$ ($p, q \in \mathbb{N}_0$), $X_1, \dotsc, X_p \in \mathfrak{X}(N)$ and $\nu_0, \dotsc, \nu_q \in \Gamma(E)$. The second equation is of course just that formula applied to 
\bas
\Lambda \circ \omega \circ \Bigl( \underbrace{\widehat{\Lambda}^{-1}, \dotsc, \widehat{\Lambda}^{-1}}_{p \text{ times}}, \underbrace{\mathds{1}_{E}, \dotsc, \mathds{1}_E}_{q \text{ times}} \Bigr).
\eas
\end{proof}

Before we can study and discuss this field redefinition let us list several useful properties.

\begin{propositions}{Properties of $\Lambda$ and $\widehat{\Lambda}$}{PropsOfBigLambdas}
Let $N$ be a smooth manifold, $E \to N$ a Lie algebroid, $\nabla$ a connection on $E$, and $\kappa$ and $g$ fibre metrics on $E$ and $\mathrm{T}N$, respectively. Also let $\lambda \in \Omega^1(N; E)$ such that $\Lambda = \mathds{1}_E - \lambda \circ \rho$ is an element of $\sAut(E)$. Then we have
\ba
\Lambda^{-1}
&=
\sum_{k=0}^l \mleft( \lambda \circ \rho \mright)^k
	+ \Lambda^{-1} \circ \mleft( \lambda \circ \rho \mright)^{l+1},
&
\widehat{\Lambda}^{-1}
&=
\sum_{k=0}^l \mleft( \rho \circ \lambda \mright)^k
	+ \widehat{\Lambda}^{-1} \circ \mleft( \rho \circ \lambda \mright)^{l+1},
\\
\label{basicconnectionTrafoRefield}
\mleft(\widetilde{\nabla}^{\lambda}\mright)^{\mathrm{bas}}
&=
\Lambda \circ \nabla^{\mathrm{bas}} \circ \Lambda^{-1},
&
\mleft(\widetilde{\nabla}^{\lambda}\mright)^{\mathrm{bas}}
&=
\widehat{\Lambda} \circ \nabla^{\mathrm{bas}} \circ \widehat{\Lambda}^{-1},
\\
\rho \circ \Lambda
&=
\widehat{\Lambda} \circ \rho,
&
\Lambda \circ \lambda
&=
\lambda \circ \widehat{\Lambda},
\label{EqCommutationWithLambda}\\
\rho \circ \Lambda^{-1}
&=
\widehat{\Lambda}^{-1} \circ \rho,
&
\Lambda^{-1} \circ \lambda
&=
\lambda \circ \widehat{\Lambda}^{-1}
\label{EqCommutationWithLambdaInverse}
\ea
for all $l \in \mathbb{N}_0$, where we mean the basic connection on $E$ on the left and the one on $\mathrm{T}N$ on the right in the second line. Moreover, we have several identities for the redefinition of the connection
\ba\label{AndereFormelFuerNablaTrafoBesserFuerDasRechnen}
\widetilde{\nabla}^\lambda
&=
\nabla^\prime
	- \mleft( \mathrm{d}^{\nabla^\prime} \lambda \mright) \circ \mleft( \mathds{1}_{\mathrm{T}N}, \rho \mright)
	+ \Lambda \circ t_{\nabla_\rho} \circ \mleft( \Lambda^{-1} \circ \lambda, \mathds{1}_E \mright),
\ea
where $\nabla^\prime \coloneqq \Lambda\circ\nabla\circ\Lambda^{-1}$, and
\ba\label{dievielBessereFormuelFuersRechnenFragezeichen}
\widetilde{\nabla}^\lambda_Y \mu
&=
\Lambda \mleft( \nabla_{\widehat{\Lambda}^{-1}(Y)} \mu
- \mleft[ \mleft( \Lambda^{-1} \circ \lambda \mright)(Y), \mu \mright]_E \mright)
+ \lambda \big([Y, \rho(\mu)] \big)
\ea
for all $\mu\in \Gamma(E)$ and $Y \in \mathfrak{X}(N)$, finally also
\ba\label{KuerzesteFormelForRedefOfNabla}
\widetilde{\nabla}^\lambda_{\widehat{\Lambda}}
&=
\nabla_{\widehat{\Lambda}}
	+ \mathrm{d}^{\nabla^{\mathrm{bas}}}\lambda.
\ea
\end{propositions}

\begin{remark}
\leavevmode\newline
We especially need the formulas of the inverse for $l=0$, \textit{i.e.}
\bas
\Lambda^{-1}
&=
\mathds{1}_E
	+ \Lambda^{-1} \circ \lambda \circ \rho,
\\
\widehat{\Lambda}^{-1}
&=
\mathds{1}_{\mathrm{T}N}
	+ \widehat{\Lambda}^{-1} \circ \lambda \circ \rho.
\eas
\end{remark}

\begin{proof}
\leavevmode\newline
\indent $\bullet$ The Eq.~\eqref{EqCommutationWithLambda} simply follow by definition, and inverting these with respect to $\Lambda$ and $\widehat{\Lambda}$ gives Eq.~\eqref{EqCommutationWithLambdaInverse}.
Using these, we also have
\bas
\Lambda \circ 
\mleft( 
\sum_{k=0}^l \mleft( \lambda \circ \rho \mright)^k
	+ \Lambda^{-1} \circ \mleft( \lambda \circ \rho \mright)^{l+1} 
\mright)
&=
\sum_{k=0}^l \underbrace{\mleft( \mathds{1}_E - \lambda \circ \rho \mright) \circ \mleft( \lambda \circ \rho \mright)^k}
_{= \mleft( \lambda \circ \rho \mright)^k - \mleft( \lambda \circ \rho \mright)^{k+1}}
	+ ~\mleft( \lambda \circ \rho \mright)^{l+1} 
\\
&\stackrel{\mathclap{\text{telescoping sum}}}{=}\qquad~
\mleft( \lambda \circ \rho \mright)^0
	- \mleft( \lambda \circ \rho \mright)^{l+1}
	+ \mleft( \lambda \circ \rho \mright)^{l+1}
\\
&=
\mathds{1}_E,
\eas
which proves the claim. In the same manner one shows the formula for $\widehat{\Lambda}^{-1}$.

$\bullet$ We have
\bas
\mleft(\mleft( \Lambda \circ \mathrm{d}^{\nabla^{\mathrm{bas}}} \circ \Lambda^{-1} \mright) \lambda \mright)(Y, \mu)
&=
\Lambda\biggl(
\nabla^{\mathrm{bas}}_\mu \mleft( \mleft(\Lambda^{-1} \circ \lambda\mright) (Y) \mright)
	- \mleft(\Lambda^{-1} \circ \lambda\mright)\mleft(\nabla^{\mathrm{bas}}_\mu Y\mright)
\biggr)
\\
&=
\Lambda \biggl(
	- \mleft[ \mleft( \Lambda^{-1} \circ \lambda \mright) (Y), \mu \mright]_E
	+ \nabla_{\widehat{\Lambda}^{-1} \circ \rho \circ \lambda(Y)} \mu
\biggr)
	+ \lambda\mleft( \mleft[ Y, \rho(\mu) \mright] \mright) 
\\
&\hspace{1cm}
\underbrace{- \lambda \circ \rho\mleft( \nabla_Y \mu \mright)
	+ \nabla_Y \mu}_{\Lambda \mleft( \nabla_Y \mu \mright)} - \nabla_Y \mu 
\\
&=
\Lambda \biggl( 
	\nabla_{\widehat{\Lambda}^{-1}(Y)} \mu
	- \mleft[ \mleft(\Lambda^{-1} \circ \lambda\mright)(Y), \mu \mright]_E
\biggr)
	+ \lambda\mleft( \mleft[ Y, \rho(\mu) \mright] \mright)
	- \nabla_Y \mu,
\eas
which proves Eq.~\eqref{dievielBessereFormuelFuersRechnenFragezeichen} by using Def.~\eqref{FieldTrafoOfNabla}.
Let $\nabla^\prime \coloneqq \Lambda \circ \nabla \circ \Lambda^{-1}$, then by Prop.~\ref{prop:PropsOfBigLambdas}
\bas
&\nabla^{\prime}_Y\mu
- \mleft( \mathrm{d}^{\nabla^{\prime}}\lambda \mright)(Y, \rho(\mu))
+ \Lambda \Big( t_{\nabla_\rho}\mleft( \Lambda^{-1} (\lambda(Y)), \mu \mright) \Big) 
\\
&=
\underbrace{\nabla^{\prime}_Y\mu
- \nabla^{\prime}_Y \big( (\lambda \circ \rho)(\mu) \big)}_{= \nabla^\prime_Y \mleft( \Lambda(\mu) \mright)}
+ \nabla^{\prime}_{\rho(\mu)} \big( \lambda(Y) \big)
+ \lambda \big([Y, \rho(\mu)] \big) 
\\
&\hspace{1cm}
	+ \Lambda \bigg( - \mleft[ \mleft( \Lambda^{-1} \circ \lambda \mright)(Y), \mu \mright]_E
	+ \nabla_{\mleft( \rho \circ \Lambda^{-1} \circ \lambda \mright)(Y)} \mu
	- \nabla_{\rho(\mu)} \Big( \mleft( \Lambda^{-1} \circ \lambda \mright) (Y) \Big) \bigg) 
\\
&=
\Lambda \mleft( \nabla_{\widehat{\Lambda}^{-1}(Y)} \mu
- \mleft[ \mleft( \Lambda^{-1} \circ \lambda \mright)(Y), \mu \mright]_E \mright)
+ \lambda \big([Y, \rho(\mu)] \big),
\eas
comparing it with the previous formula, we arrive at
\bas
\widetilde{\nabla}^\lambda
&=
\nabla^\prime
	- \mleft( \mathrm{d}^{\nabla^\prime} \lambda \mright) \circ \mleft( \mathds{1}_{\mathrm{T}N}, \rho \mright)
	+ \Lambda \circ t_{\nabla_\rho} \circ \mleft( \Lambda^{-1} \circ \lambda, \mathds{1}_E \mright)
%\\
%&=
%\mleft[
%\mathfrak{X}(N) \times \Gamma(E) \ni (Y, \mu) \mapsto
%\Lambda \mleft( \nabla_{\widehat{\Lambda}^{-1}(Y)} \mu
%- \mleft[ \mleft( \Lambda^{-1} \circ \lambda \mright)(Y), \mu \mright]_E \mright)
%+ \lambda \big([Y, \rho(\mu)] \big)
%\mright]
.
\eas
For $I \coloneqq \mleft( \Lambda \circ \mathrm{d}^{\nabla^{\mathrm{bas}}} \circ \Lambda^{-1} \mright) \lambda \in \Omega^1(N; \mathrm{End}(E)) \cong \Omega^{1,1}(N,E;E)$ we also have
\bas
I(Y, \nu)
&=
\mleft( \Lambda \circ \nabla^{\mathrm{bas}}_\nu \circ \Lambda^{-1} \circ \lambda - \lambda \circ \nabla^{\mathrm{bas}}_\nu \mright)(Y)
\stackrel{\text{Eq.~\eqref{FieldTrafoOfNabla}}}{=}
\widetilde{\nabla}^\lambda_\nu Y
	- \nabla_\nu Y
\eas
for all $\nu \in \Gamma(E)$ and $Y \in \mathfrak{X}(N)$; especially with $\rho \circ \nabla^{\mathrm{bas}} = \nabla^{\mathrm{bas}} \circ \rho$ we get
\bas
I\mleft(\widehat{\Lambda}(Y), \nu\mright)
&=
\mleft( 
	\Lambda \circ \nabla^{\mathrm{bas}}_\nu \circ \lambda 
	- \lambda \circ \nabla^{\mathrm{bas}}_\nu \circ \widehat{\Lambda} 
\mright)(Y)
\\
&=
\mleft( 
	\nabla^{\mathrm{bas}}_\nu \circ \lambda 
	- \lambda \circ \nabla^{\mathrm{bas}}_\nu
	- \lambda \circ \rho \circ \nabla^{\mathrm{bas}}_\nu \circ \lambda 
	+ \lambda \circ \nabla^{\mathrm{bas}}_\nu \circ \rho \circ\lambda
\mright)(Y)
\\
&=
\mleft( 
	\nabla^{\mathrm{bas}}_\nu \circ \lambda 
	- \lambda \circ \nabla^{\mathrm{bas}}_\nu
\mright)(Y)
\\
&=
\mleft(\mathrm{d}^{\nabla^{\mathrm{bas}}} \lambda \mright)(Y, \nu),
\eas
which proves the last equation. Alternatively, use Cor.~\ref{cor:ConjugationOfDifferentialsAreShitty}.

$\bullet$ Finally, using the things just shown,
\bas
\mleft(\widetilde{\nabla}^{\lambda}\mright)^{\mathrm{bas}}_\mu \nu
&=
\mleft[ \mu, \nu \mright]_E
	+ \widetilde{\nabla}^{\lambda}_{\rho(\nu)} \mu
\\
&=
\mleft[ \mu, \nu \mright]_E
	+ \Lambda \mleft( \nabla_{\mleft(\widehat{\Lambda}^{-1}\circ\rho\mright)(\nu)} \mu 
	- \mleft[ \mleft( \Lambda^{-1} \circ \lambda \circ \rho \mright)(\nu), \mu \mright]_E \mright)
	+ \lambda \big([\rho(\nu), \rho(\mu)] \big)
\\
&=
\underbrace{\mleft[ \mu, \nu \mright]_E
	+ \mleft[ \mu, \mleft( \Lambda^{-1} \circ \lambda \circ \rho \mright)(\nu) \mright]_E}
	_{= \mleft[ \mu, \Lambda^{-1}(\nu) \mright]_E}
	+ \Lambda \mleft( \nabla_{\mleft(\rho \circ \Lambda^{-1}\mright)(\nu)} \mu \mright)
\\
&\hspace{1cm}
	+ \underbrace{(\lambda\circ\rho)\mleft( \mleft[ \mleft( \Lambda^{-1} \circ \lambda \circ \rho \mright)(\nu), \mu \mright]_E \mright)
	+ (\lambda\circ\rho) \bigl([\nu, \mu]_E \bigr)}
	_{= (\lambda \circ \rho)\mleft( \mleft[ \Lambda^{-1}(\nu), \mu \mright]_E \mright)}
\\
&=
\Lambda\mleft(\mleft[ \mu, \Lambda^{-1}(\nu) \mright]_E\mright)
	+ \Lambda \mleft( \nabla_{\mleft(\rho \circ \Lambda^{-1}\mright)(\nu)} \mu \mright)
\\
&=
\Lambda\biggl(
	\nabla^{\mathrm{bas}}_\mu \mleft( \Lambda^{-1} (\nu) \mright)
\biggr)
\eas
for all $\mu,\nu \in \Gamma(E)$. Similarly,
\bas
\mleft(\widetilde{\nabla}^{\lambda}\mright)^{\mathrm{bas}}_\mu Y
&=
[\rho(\mu), Y]
	+ \rho \mleft( \widetilde{\nabla}^\lambda_Y \mu \mright)
\\
&=
[\rho(\mu), Y]
	+ \rho\biggl( 
		\Lambda \mleft( \nabla_{\widehat{\Lambda}^{-1}(Y)} \mu
		- \mleft[ \mleft( \Lambda^{-1} \circ \lambda \mright)(Y), \mu \mright]_E \mright)
		+ \lambda \big([Y, \rho(\mu)] \big)
	\biggr)
\\
&=
\underbrace{[\rho(\mu), Y]
	+ \mleft[ \rho(\mu), \mleft( \widehat{\Lambda}^{-1} \circ\rho \circ \lambda \mright)(Y) \mright]}
	_{= \mleft[ \rho(\mu), \widehat{\Lambda}^{-1}(Y) \mright]}
	+ (\rho \circ \Lambda )\mleft( \nabla_{\widehat{\Lambda}^{-1}(Y)} \mu \mright)
\\
&\hspace{1cm}
	+ \underbrace{(\rho \circ \lambda )\mleft( \mleft[ \mleft( \widehat{\Lambda}^{-1} \circ \rho \circ \lambda \mright)(Y) + Y, \rho(\mu) \mright] \mright)}
	_{- (\rho\circ\lambda) \mleft(\mleft[ \rho(\mu), \widehat{\Lambda}^{-1}(Y) \mright]\mright)}
\\
&=
\widehat{\Lambda} \biggl(
	\nabla^{\mathrm{bas}}_\mu \mleft( \widehat{\Lambda}^{-1} (Y) \mright)
\biggr)
\eas
for all $\mu \in \Gamma(E)$ and $Y\in \mathfrak{X}(N)$.
\end{proof}

We will use these small results all the time, and we will not necessarily mention each equation each time when we use it. Using the formulas of the inverse, we can show the following.

\begin{lemmata}{Invertible field redefinition}{FieldRedefinitionIsInvertible}
Let $M, N$ be smooth manifolds, $E \to N$ a Lie algebroid, $\nabla$ a connection on $E$, and $\kappa$ and $g$ fibre metrics on $E$ and $\mathrm{T}N$, respectively. Also let $\lambda \in \Omega^1(N; E)$ such that $\Lambda = \mathds{1}_E - \lambda \circ \rho$ is an element of $\sAut(E)$. Then
\ba
\widehat{\varpi_2}^{-\lambda}
&=
\varpi_2,
\\
\widehat{\nabla}^{-\lambda}
&=
\nabla,
\\
\widehat{\kappa}^{-\lambda}
&=
\kappa,
\\
\widehat{g}^{-\lambda}
&=
g,
\ea
where we denote 
\bas
\widehat{\varpi_2}^{-\lambda}
&\coloneqq
\widetilde{ \widetilde{\varpi_2}^\lambda }^{- \Lambda^{-1} \circ \lambda}
\eas
and so on.
\end{lemmata}

\begin{remark}
\leavevmode\newline
All following formulas implied by the field redefinition, like a field redefinition of the basic connection, are defined by taking their typical definition and replacing the terms with the field redefinitions given in Def.~\ref{def:FieldRedefinition}. That will imply similar inversion formulas for those terms.
\end{remark}

\begin{proof}
\leavevmode\newline
First observe that, using Prop.~\ref{prop:PropsOfBigLambdas},
\bas
\mathfrak{\Lambda}
&\coloneqq
\mathds{1}_E
	- \mleft( - \Lambda^{-1} \circ \lambda \mright) \circ \rho
=
\mathds{1}_E
	+ \Lambda^{-1} \circ \lambda \circ \rho
=
\Lambda^{-1},
\\
\widehat{\mathfrak{\Lambda}}
&\coloneqq
\mathds{1}_{\mathrm{T}N}
	- \rho \circ \mleft( - \Lambda^{-1} \circ \lambda \mright) 
=
\mathds{1}_{\mathrm{T}N}
	+ \widehat{\Lambda}^{-1} \circ \lambda \circ \rho
=
\widehat{\Lambda}^{-1}.
\eas
Those are invertible, thus, we can apply the field redefinition using $-\Lambda^{-1} \circ \lambda$. Using these formulas, we get trivially,
\bas
\widehat{\kappa}^{-\lambda}
&=
\mleft(\kappa \circ \mleft( \Lambda^{-1}, \Lambda^{-1} \mright)\mright) \circ 
\mleft( \mathfrak{\Lambda}^{-1}, \mathfrak{\Lambda}^{-1} \mright)
=
\kappa,
\eas
similarly for $g$. Moreover,
\bas
\widehat{\varpi_2}^{-\lambda}
&=
\mleft( {}^* \mathfrak{\Lambda} \mright) (\widetilde{\varpi_2}^\lambda)
	- {}^! \mleft( \Lambda^{-1} \circ \lambda \mright)
\\
&=
\mleft( {}^* \mathfrak{\Lambda} \mright) \mleft(\mleft( {}^* \Lambda \mright) (\varpi_2)+ {}^! \lambda\mright)
	- {}^! \mleft( \Lambda^{-1} \circ \lambda \mright)
\\
&=
\varpi_2
	+ {}^! \mleft( \Lambda^{-1} \circ \lambda \mright)
	- {}^! \mleft( \Lambda^{-1} \circ \lambda \mright)
\\
&=
\varpi_2,
\eas
and
\bas
\widehat{\nabla}^{-\lambda}
&=
\widetilde{\nabla}^\lambda
	- \mleft( \mathfrak{\Lambda} \circ \mathrm{d}^{\mleft(\widetilde{\nabla}^\lambda\mright)^{\mathrm{bas}}} \circ \mathfrak{\Lambda}^{-1} \mright) \mleft( \Lambda^{-1} \circ \lambda\mright)
\\
&\stackrel{\mathclap{ \text{Cor.~\ref{cor:ConjugationOfDifferentialsAreShitty}} }}{=}\quad~
\nabla
	+ \mleft( \Lambda \circ \mathrm{d}^{\nabla^{\mathrm{bas}}} \circ \Lambda^{-1} \mright) \lambda
	- \mleft(\mathrm{d}^{\nabla^{\mathrm{bas}}} \lambda \mright) \circ \mleft( \widehat{\Lambda}^{-1}, \mathds{1}_E \mright)
\\
&\stackrel{\mathclap{ \text{Eq.~\eqref{KuerzesteFormelForRedefOfNabla}} }}{=}\quad~~
\nabla
	+ \mleft( \Lambda \circ \mathrm{d}^{\nabla^{\mathrm{bas}}} \circ \Lambda^{-1} \mright) \lambda
	- \mleft( \Lambda \circ \mathrm{d}^{\nabla^{\mathrm{bas}}} \circ \Lambda^{-1} \mright) \lambda
\\
&=
0,
\eas
viewing $\mathrm{d}^{\nabla^{\mathrm{bas}}} \lambda$ as an element of $\Omega^{1,1}(N,E;E)$.
\end{proof}

\section{Redefined gauge theory}\label{NastyCalculationsForTheseFieldRedefsBaeaeaeae}

We now want to calculate what the field redefinition changes, especially with respect to the field strength.

\begin{theorems}{Field redefinition of the field strength}{FieldRedefofstandardFieldStrengthF}
Let $M, N$ be smooth manifolds, $E \to N$ a Lie algebroid, and $\nabla$ a connection on $E$. Also let $\lambda \in \Omega^1(N; E)$ such that $\Lambda = \mathds{1}_E - \lambda \circ \rho$ is an element of $\sAut(E)$. Then we have
\ba
\widetilde{\mathfrak{D}}^\lambda
&=
\mleft( {}^* \widehat{\Lambda} \mright)\mleft(\mathfrak{D}\mright),
\\
\widetilde{F}^\lambda
&=
\mleft( {}^* \Lambda \mright) \mleft(
	F
	- \frac{1}{2} \mleft({}^* \xi \mright) \mleft( \mathfrak{D} \stackrel{\wedge}{,} \mathfrak{D} \mright)
\mright),\label{FieldRedefOfClassicF}
\ea
where
\ba
\widetilde{\mathfrak{D}}^\lambda
&\coloneqq
\mathrm{D}
	- ({}^*\rho)\mleft(\widetilde{\varpi_2}^\lambda\mright),
\\
\widetilde{F}^\lambda
&\coloneqq
\mathrm{d}^{{}^*\widetilde{\nabla}^\lambda} \widetilde{\varpi_2}^\lambda
	- \frac{1}{2} \mleft( {}^* t_{\widetilde{\nabla}^\lambda_\rho} \mright)\mleft( \widetilde{\varpi_2}^\lambda \stackrel{\wedge}{,} \widetilde{\varpi_2}^\lambda \mright),
\\
\xi
&\coloneqq
\Lambda^{-1} \circ \widehat{\zeta}^\lambda \circ \mleft( \widehat{\Lambda}, \widehat{\Lambda} \mright)
\ea
and $\gls{1fZetaTilHat}$ is an element of $\Omega^2(N; E)$ defined by
\ba\label{FormulaForZetaTildeWithZetaEqualzero}
\mleft(-\widehat{\zeta}^\lambda \circ \mleft( \widehat{\Lambda}, \widehat{\Lambda} \mright)\mright) (X,Y)
&\coloneqq
\mleft(\mathrm{d}^{\widetilde{\nabla}^\lambda} \lambda
	- t_{\widetilde{\nabla}^\lambda_\rho} \circ (\lambda, \lambda)\mright)(X,Y)
\nonumber\\
&=
\mleft( \mathrm{d}^\nabla \lambda \mright)(X,Y)
%}_{\mathclap{= \nabla_X \bigl( \lambda(Y) \bigr) - \nabla_Y \bigl( \lambda(X) \bigr) - \lambda\bigl( [X,Y] \bigr)}}
	+ \lambda\Bigl(
		\nabla^{\mathrm{bas}}_{\lambda(X)} Y
		- \nabla^{\mathrm{bas}}_{\lambda(Y)} X
		%}_{= [(\rho\circ\lambda)(Y), X] + \rho\mleft( \nabla_X ( \lambda(Y) ) \mright)}
	\Bigr)
	- \mleft[ \lambda(X), \lambda(Y) \mright]_E
\ea
for all $X, Y \in \mathfrak{X}(N)$.
\end{theorems}

\begin{remark}\label{OtherNotationForZetaTransform}
\leavevmode\newline
When we define the \textbf{formal torsion}\footnote{It is formal because $\nabla^{\mathrm{bas}}_\lambda$ is not a connection due to the fact that $\rho \circ \lambda \neq \mathds{1}_{\mathrm{T}N}$, otherwise $\widehat{\Lambda} = 0$ and, so, $\Lambda$ would not be invertible by Sylvester's determinant theorem. Therefore the Leibniz rule is not as usual. That is, $\nabla^{\mathrm{bas}}_\lambda: \mathrm{T}N \to \mathfrak{D}(E)$ is in general not anchor-preserving.} $t_{\nabla^{\mathrm{bas}}_\lambda}$ of $\nabla^{\mathrm{bas}}_\lambda$, $\mathfrak{X}(N) \times \Gamma(E) \ni (Y, \nu) \mapsto \nabla^{\mathrm{bas}}_{\lambda(Y)} \nu$, as an element of $\Omega^2(N; \mathrm{T}N)$ by
\ba
t_{\nabla^{\mathrm{bas}}_\lambda}(X,Y)
&\coloneqq
\nabla^{\mathrm{bas}}_{\lambda(X)} Y
	- \nabla^{\mathrm{bas}}_{\lambda(Y)} X
	- [X,Y]
\ea
for all $X, Y \in \mathfrak{X}(N)$, then recall Def.~\ref{def:GeneralDefOfCurvMorphisms} for
\bas
R_\lambda(X,Y)
=
\bigl[ \lambda(X), \lambda(Y) \bigr]
	- \lambda \bigl( [X,Y] \bigr),
\eas
hence, we can write
\bas
\mleft(
	\lambda\mleft(t_{\nabla^{\mathrm{bas}}_\lambda}\mright)
	- R_\lambda
\mright)(X,Y)
&=
\lambda\Bigl(
		\nabla^{\mathrm{bas}}_{\lambda(X)} Y
		- \nabla^{\mathrm{bas}}_{\lambda(Y)} X
	\Bigr)
	- \mleft[ \lambda(X), \lambda(Y) \mright]_E,
\eas
in total arriving to
\ba\label{TollsteFormelFuerZetaTrafoFragezeichen}
-\widehat{\zeta}^\lambda \circ \mleft( \widehat{\Lambda}, \widehat{\Lambda} \mright)
&=
\mathrm{d}^\nabla \lambda
	+ \lambda\mleft(t_{\nabla^{\mathrm{bas}}_\lambda}\mright)
	- R_\lambda.
\ea
Observe the (very rough) similarity with the Maurer-Cartan equation; especially for Lie algebra bundles, that is, zero anchor, this will look like a covariantized Maurer-Cartan equation with inhomogeneity. We will see this later.
\end{remark}

\begin{proof}[Proof of Thm.~\ref{thm:FieldRedefofstandardFieldStrengthF}]
\leavevmode\newline
In the following let $(\Phi,A) \in \mathfrak{M}_E(M;N)$.

$\bullet$ The field redefinition of the minimal coupling directly follows by Def.~\eqref{EqFieldRedefFuerA}, so,
\bas
\mathfrak{D}^{\widetilde{A}^\lambda} \Phi
&=
\underbrace{\mathrm{D}\Phi}_{\mathclap{= (\Phi^* \mathds{1}_{\mathrm{T}N}) (\mathrm{D}\Phi)}}
	- (\Phi^*\rho)\bigl( 
		\mleft( \Phi^* \Lambda \mright) (A)
		+ (\Phi^* \lambda)(\mathrm{D}\Phi) 
	\bigr)
\\
&=
\mleft( \Phi^* \widehat{\Lambda} \mright) (\mathrm{D}\Phi)
	- \mleft(\Phi^* \mleft(\widehat{\Lambda} \circ \rho\mright)\mright) (A)
\\
&=
\mleft( \Phi^* \widehat{\Lambda} \mright)\mleft(\mathfrak{D}^{A} \Phi\mright).
\eas

$\bullet$ With respect to a local frame $\mleft( e_a \mright)_a$ of $E$ and viewing terms like $\widetilde{\nabla}^\lambda - \nabla$ as an element of $\Omega^1(N; \mathrm{End}(E))$,
\bas
\mathrm{d}^{\Phi^*\widetilde{\nabla}^\lambda} \bigl( \mleft( \Phi^* \Lambda \mright) (A) \bigr)
&=
\mathrm{d}^{\Phi^*\mleft( \nabla
	+ \widetilde{\nabla}^\lambda - \nabla
	\mright)}
	\bigl(
		(\Phi^*\Lambda)(A)
	\bigr)
\\
%&\stackrel{\mathclap{\text{Eq.~\eqref{eqDifferentialSplit}}}}{=}~~~
%\mathrm{d}^{\Phi^*\nabla}
%\mleft(
		%A^a \otimes \Phi^*\bigl( \Lambda(e_a) \bigr)
%\mright)
	%+ \Phi^!\Biggl( \biggl(\mleft( \Lambda \circ \mathrm{d}^{\nabla^{\mathrm{bas}}} \circ \Lambda^{-1} \mright) \lambda \biggr) \bigl( \Lambda(e_a) \bigr) \Biggr) \wedge  A^a 
%\\
%&=
%\mathrm{d}^{\Phi^*\nabla}
%\mleft(
		%A^a \otimes \Phi^*\bigl( \Lambda(e_a) \bigr)
%\mright)
	%+ \Phi! 
	%\mleft( \mleft( \widetilde{\nabla}^\lambda  - \nabla \mright) (\Lambda(e_a)) \mright)
	%\wedge A^a
%\\
&\stackrel{\mathclap{\text{Eq.~\eqref{eqDifferentialSplit}}}}{=}~~~
\mathrm{d}^{\Phi^*\nabla} \mleft( A^a \otimes \Phi^*\bigl( \Lambda(e_a) \bigr) \mright)
	+ \Phi^!\mleft( \widetilde{\nabla}^\lambda - \nabla \mright) \wedge \mleft( A^a \otimes \Phi^*\bigl( \Lambda(e_a) \bigr)\mright)
\\
&=
\mathrm{d}A^a \otimes \Phi^*\bigl( \Lambda(e_a) \bigr)
	- A^a \wedge \Phi^!\underbrace{\Bigl( \nabla \bigl(\Lambda(e_a)\bigr) \Bigr)}
	_{\mathclap{= (\nabla \Lambda)(e_a) + \Lambda(\nabla e_a)}}
\\
&\hspace{1cm}
	- A^a \wedge \Phi! \mleft(  \widetilde{\nabla}^\lambda (\Lambda(e_a))  - (\nabla \Lambda)(e_a) - \Lambda(\nabla e_a)) \mright)
\\
&=
\mathrm{d}A^a \otimes \Phi^*\bigl( \Lambda(e_a) \bigr)
	- A^a \wedge \underbrace{\Phi^! \mleft( \Lambda(\nabla e_a) \mright)}
		_{\mathclap{= \mleft( \Phi^* \Lambda \mright) \mleft( \Phi^! (\nabla e_a) \mright)}}
	+ \Phi! \mleft(  \widetilde{\nabla}^\lambda (\Lambda(e_a))  - \Lambda(\nabla e_a)) \mright) \wedge A^a
\\
&=
\mleft( \Phi^* \Lambda \mright)\mleft( \mathrm{d}^{\Phi^*\nabla} A \mright)
	+ \mleft( \Phi^! \mleft( \widetilde{\nabla}^\lambda \circ \Lambda - \Lambda \circ \nabla \mright) \mright) (A)
\\
&\stackrel{\mathclap{\text{Eq.~\eqref{EqPullBackFormelFuerVerschiedeneDefinitionen}}}}{=}~~~
\mleft( \Phi^* \Lambda \mright)\mleft( \mathrm{d}^{\Phi^*\nabla} A \mright)
	+ \mleft( \Phi^* \mleft( \widetilde{\nabla}^\lambda \circ \Lambda - \Lambda \circ \nabla \mright) \mright) \mleft( \mathrm{D}\Phi\stackrel{\wedge}{,} A \mright),
\eas
and
\bas
\mathrm{d}^{\Phi^*\widetilde{\nabla}^\lambda} \mleft( \Phi^!\lambda \mright)
&\stackrel{\text{Eq.~\eqref{EqGeilePullBackCommuteFormel}}}{=}
\Phi^!\mleft( \mathrm{d}^{\widetilde{\nabla}^\lambda} \lambda \mright)
\stackrel{\text{Eq.~\eqref{EqPullBackFormelFuerVerschiedeneDefinitionen}}}{=}
\frac{1}{2} \mleft( \Phi^* \mleft( \mathrm{d}^{\widetilde{\nabla}^\lambda} \lambda \mright) \mright) \mleft( \mathrm{D}\Phi \stackrel{\wedge}{,} \mathrm{D} \Phi \mright),
\eas
also
%\bas
 %t_{\widetilde{\nabla}^\lambda_\rho} \mleft( \mu, \nu \mright)
	%- t_{\nabla_\rho}(\mu, \nu)
%~~~~&\stackrel{\mathclap{\text{Def.~\eqref{FieldTrafoOfNabla}}}}{=}~~~~
%\mleft(\mleft( \Lambda \circ \mathrm{d}^{\nabla^{\mathrm{bas}}} \circ \Lambda^{-1} \mright) \lambda\mright) \mleft( \rho(\mu), \nu \mright)
	%+ \mleft(\mleft( \Lambda \circ \mathrm{d}^{\nabla^{\mathrm{bas}}} \circ \Lambda^{-1} \mright) \lambda\mright) \mleft( \rho(\nu), \mu \mright)
%\eas
\bas
\frac{1}{2} \mleft( \Phi^* t_{\widetilde{\nabla}^\lambda_\rho} \mright) \mleft( \widetilde{A}^\lambda \stackrel{\wedge}{,} \widetilde{A}^\lambda \mright)
~~~~~&\stackrel{\mathclap{\text{Prop.~\ref{prop:GradedExtensionPlusAntiSymm}}}}{=}~~~~~
\frac{1}{2} \Biggl(
	\mleft( \Phi^* t_{\widetilde{\nabla}^\lambda_\rho} \mright) \bigl( (\Phi^*\Lambda) (A) \stackrel{\wedge}{,} (\Phi^* \Lambda) (A) \bigr)
	+  \mleft( \Phi^* t_{\widetilde{\nabla}^\lambda_\rho} \mright) \mleft( \Phi^!\lambda \stackrel{\wedge}{,} \Phi^!\lambda \mright) 
\Biggr)
\\
&\hspace{1cm}~~~~~
	+ \mleft( \Phi^* t_{\widetilde{\nabla}^\lambda_\rho} \mright) \mleft( \Phi^!\lambda \stackrel{\wedge}{,} (\Phi^* \Lambda) (A) \mright)
\\
&\stackrel{\mathclap{\text{Eq.~\eqref{EqPullBackFormelFuerVerschiedeneDefinitionen}}}}{=}~~~
\frac{1}{2} \Biggl(\Phi^* \mleft( t_{\widetilde{\nabla}^\lambda_\rho} \circ (\Lambda, \Lambda) \mright) \Biggr) \mleft( A \stackrel{\wedge}{,} A \mright)
	+ \Biggl( \Phi^* \mleft( t_{\widetilde{\nabla}^\lambda_\rho} \circ \mleft( \lambda, \Lambda \mright) \mright) \Biggr) (\mathrm{D}\Phi \stackrel{\wedge}{,} A)
\\
&\hspace{1cm}~~~
	+ \frac{1}{2} \Biggl( \Phi^*\mleft( t_{\widetilde{\nabla}^\lambda_\rho} \circ (\lambda, \lambda) \mright) \Biggr) \mleft( \mathrm{D}\Phi \stackrel{\wedge}{,} \mathrm{D}\Phi \mright).
\eas
So, in total we get, adding the missing term of the torsion in the definition of the field strength,
\bas
\widetilde{F}^\lambda(\Phi, A)
&=
\mleft( \Phi^* \Lambda \mright)\mleft( \mathrm{d}^{\Phi^*\nabla} A \mright)
	- \frac{1}{2} (\Phi^* \Lambda) \mleft( \mleft( \Phi^* t_{\nabla_\rho} \mright) \mleft( A \stackrel{\wedge}{,} A \mright) \mright)
\\
&\hspace{1cm}
	+ \frac{1}{2} (\Phi^* \Lambda) \mleft( \mleft( \Phi^* t_{\nabla_\rho} \mright) \mleft( A \stackrel{\wedge}{,} A \mright) \mright)
	+ \mleft( \Phi^* \mleft( \widetilde{\nabla}^\lambda \circ \Lambda - \Lambda \circ \nabla \mright) \mright) \mleft( \mathrm{D}\Phi\stackrel{\wedge}{,} A \mright)
\\
&\hspace{1cm}
	+ \frac{1}{2} \mleft( \Phi^* \mleft( \mathrm{d}^{\widetilde{\nabla}^\lambda} \lambda \mright) \mright) \mleft( \mathrm{D}\Phi \stackrel{\wedge}{,} \mathrm{D} \Phi \mright)
	- \frac{1}{2} \Biggl(\Phi^* \mleft( t_{\widetilde{\nabla}^\lambda_\rho} \circ (\Lambda, \Lambda) \mright) \Biggr) \mleft( A \stackrel{\wedge}{,} A \mright)
\\
&\hspace{1cm}
	- \Biggl( \Phi^* \mleft( t_{\widetilde{\nabla}^\lambda_\rho} \circ \mleft( \lambda, \Lambda \mright) \mright) \Biggr) (\mathrm{D}\Phi \stackrel{\wedge}{,} A)
	- \frac{1}{2} \Biggl( \Phi^*\mleft( t_{\widetilde{\nabla}^\lambda_\rho} \circ (\lambda, \lambda) \mright) \Biggr) \mleft( \mathrm{D}\Phi \stackrel{\wedge}{,} \mathrm{D}\Phi \mright)
\\
%%%%%%%%%%%%%%%%%%%%%%%%%%%%%%%%%%%%%%%%%%%%%%%%%%%%%%
&=
(\Phi^* \Lambda) (F)
\\
&\hspace{1cm}
+ \Biggl(\Phi^* \mleft(
	\widetilde{\nabla}^\lambda \circ \Lambda - \Lambda \circ \nabla
	- t_{\widetilde{\nabla}^\lambda_\rho} \circ \mleft( \lambda, \Lambda \mright)
\mright) \Biggr) \mleft( \mathrm{D}\Phi \stackrel{\wedge}{,} A \mright)
\\
&\hspace{1cm}
+ \frac{1}{2}\Biggl( \Phi^* \mleft(
	\Lambda \circ t_{\nabla_\rho}
	- t_{\widetilde{\nabla}^\lambda_\rho} \circ (\Lambda, \Lambda)
\mright) \Biggr) \mleft( A \stackrel{\wedge}{,} A \mright)
\\
&\hspace{1cm}
+ \frac{1}{2}\Biggl( \Phi^* \mleft(
	\mathrm{d}^{\widetilde{\nabla}^\lambda} \lambda
	- t_{\widetilde{\nabla}^\lambda_\rho} \circ (\lambda, \lambda)
\mright) \Biggr) \mleft( \mathrm{D}\Phi \stackrel{\wedge}{,} \mathrm{D} \Phi \mright).
\eas
Now we need to insert the definition of $\widetilde{\nabla}^\lambda$,
\bas
\mleft( \mathrm{d}^{\widetilde{\nabla}^\lambda} \lambda \mright)(X, Y)
&=
\mleft( \mathrm{d}^{
		\nabla
	+ \mleft( \Lambda \circ \mathrm{d}^{\nabla^{\mathrm{bas}}} \circ \Lambda^{-1} \mright) \lambda} \lambda \mright)(X, Y)
\\
&\stackrel{\mathclap{\text{Eq.~\eqref{eqDifferentialSplit}}}}{=}~~~
\mleft( \mathrm{d}^\nabla \lambda \mright)(X, Y)
	+ \Lambda\biggl(
\nabla^{\mathrm{bas}}_{\lambda(Y)} \mleft( \mleft(\Lambda^{-1} \circ \lambda\mright) (X) \mright)
	- \mleft(\Lambda^{-1} \circ \lambda\mright)\mleft(\nabla^{\mathrm{bas}}_{\lambda(Y)} X\mright)
\biggr)
\\
&\hspace{1cm}
	- \Lambda\biggl(
\nabla^{\mathrm{bas}}_{\lambda(X)} \mleft( \mleft(\Lambda^{-1} \circ \lambda\mright) (Y) \mright)
	- \mleft(\Lambda^{-1} \circ \lambda\mright)\mleft(\nabla^{\mathrm{bas}}_{\lambda(X)} Y\mright)
\biggr)
\\
&=
\mleft( \mathrm{d}^\nabla \lambda \mright)(X,Y)
	+ \lambda\mleft(
		\nabla^{\mathrm{bas}}_{\lambda(X)} Y
		- \nabla^{\mathrm{bas}}_{\lambda(Y)} X
	\mright)
\\
&\hspace{1cm}
	+ \Lambda\biggl(
		\nabla^{\mathrm{bas}}_{\lambda(Y)} \mleft( \mleft(\Lambda^{-1} \circ \lambda\mright) (X) \mright)
		- \nabla^{\mathrm{bas}}_{\lambda(X)} \mleft( \mleft(\Lambda^{-1} \circ \lambda\mright) (Y) \mright)
	\biggr)
\\
&=
\Lambda\mleft( \nabla_X \bigl( \lambda(Y) \bigr) \mright)
	- \Lambda\mleft( \nabla_Y \bigl( \lambda(X) \bigr) \mright)
	+ \lambda\mleft(
		\mleft[ \widehat{\Lambda}(Y), X \mright]
		+ \mleft[ (\rho\circ\lambda)(X), Y \mright]
	\mright)
\\
&\hspace{1cm}
	+ \Lambda\biggl(
		\nabla^{\mathrm{bas}}_{\lambda(Y)} \mleft( \mleft(\Lambda^{-1} \circ \lambda\mright) (X) \mright)
		- \nabla^{\mathrm{bas}}_{\lambda(X)} \mleft( \mleft(\Lambda^{-1} \circ \lambda\mright) (Y) \mright)
	\biggr)
\eas
for all $X, Y \in \mathfrak{X}(N)$, and, by using the results about the field redefinition of the basic connection,
\bas
- t_{\widetilde{\nabla}^\lambda_\rho} (\lambda(X), \lambda(Y))
&=
t_{\mleft(\widetilde{\nabla}^\lambda\mright)^{\mathrm{bas}}} (\lambda(X), \lambda(Y))
\\
&=
\Lambda\biggl( \nabla^{\mathrm{bas}}_{\lambda(X)} \mleft( \mleft(\Lambda^{-1} \circ \lambda\mright) (Y) \mright) \biggr)
	- \Lambda\biggl( \nabla^{\mathrm{bas}}_{\lambda(Y)} \mleft( \mleft(\Lambda^{-1} \circ \lambda\mright) (X) \mright) \biggr)
	-\mleft[ \lambda(X),\lambda(Y) \mright]_E.
\eas
Then
\bas
\mleft(-\widehat{\zeta}^\lambda \circ \mleft( \widehat{\Lambda}, \widehat{\Lambda} \mright)\mright) (X,Y)
&\coloneqq
\mleft(\mathrm{d}^{\widetilde{\nabla}^\lambda} \lambda
	- t_{\widetilde{\nabla}^\lambda_\rho} \circ (\lambda, \lambda)\mright)(X,Y)
\\
&=
\mleft( \mathrm{d}^\nabla \lambda \mright)(X,Y)
%}_{\mathclap{= \nabla_X \bigl( \lambda(Y) \bigr) - \nabla_Y \bigl( \lambda(X) \bigr) - \lambda\bigl( [X,Y] \bigr)}}
	+ \lambda\Bigl(
		\nabla^{\mathrm{bas}}_{\lambda(X)} Y
		- \nabla^{\mathrm{bas}}_{\lambda(Y)} X
		%}_{= [(\rho\circ\lambda)(Y), X] + \rho\mleft( \nabla_X ( \lambda(Y) ) \mright)}
	\Bigr)
	- \mleft[ \lambda(X), \lambda(Y) \mright]_E
%\\
%&=
%\Lambda\mleft( \nabla_X \bigl( \lambda(Y) \bigr) \mright)
	%- \Lambda\mleft( \nabla_Y \bigl( \lambda(X) \bigr) \mright)
%\\
%&\hspace{1cm}
	%+ \lambda\mleft(
		%\mleft[ \widehat{\Lambda}(Y), X \mright]
		%+ \mleft[ (\rho\circ\lambda)(X), Y \mright]
	%\mright)
	%- \mleft[ \lambda(X), \lambda(Y) \mright]_E,
\eas
and, using $\rho \circ \nabla^{\mathrm{bas}} = \nabla^{\mathrm{bas}} \circ \rho$ and $t_{\nabla_\rho} = t_{\nabla^{\mathrm{bas}}}$,
\bas
\mleft(\Lambda \circ t_{\nabla_\rho}
	- t_{\widetilde{\nabla}^\lambda_\rho} \circ (\Lambda, \Lambda)\mright)(\mu, \nu)
&=
t_{\mleft(\widetilde{\nabla}^\lambda\mright)^{\mathrm{bas}}} (\Lambda(\mu), \Lambda(\nu))
	- \mleft(\Lambda \circ t_{\nabla^{\mathrm{bas}}}\mright)(\mu,\nu)
\\
&=
\Lambda\mleft( 
	\nabla^{\mathrm{bas}}_{\Lambda(\mu)} \nu
	- \nabla^{\mathrm{bas}}_{\Lambda(\nu)} \mu 
	- \nabla^{\mathrm{bas}}_\mu \nu + \nabla^{\mathrm{bas}}_\nu \mu
\mright)
\\
&\hspace{1cm}
	-\mleft[ \Lambda(\mu),\Lambda(\nu) \mright]_E
	+ \Lambda\mleft( \mleft[ \mu, \nu \mright]_E \mright)
\\
&=
\Lambda
\mleft(
	\nabla^{\mathrm{bas}}_{(\lambda \circ \rho)(\nu)} \mu
	- \nabla^{\mathrm{bas}}_{(\lambda \circ \rho)(\mu)} \nu
\mright)
	-\mleft[ (\lambda \circ \rho)(\mu),(\lambda \circ \rho) (\nu) \mright]_E
\\
&\hspace{1cm}
	- \mleft[ \mu, \nu \mright]_E
	+ \mleft[ (\lambda \circ \rho)(\mu), \nu \mright]_E
	+ \mleft[ \mu, (\lambda \circ \rho)(\nu) \mright]_E
\\
&\hspace{1cm}
	+ \mleft[ \mu, \nu \mright]_E
	- (\lambda\circ\rho)\mleft( \mleft[ \mu, \nu \mright]_E \mright)
\\
&=
\lambda\mleft(
	\nabla^{\mathrm{bas}}_{(\lambda \circ \rho)(\mu)} \bigl( \rho(\nu) \bigr)
	- \nabla^{\mathrm{bas}}_{(\lambda \circ \rho)(\nu)} \bigl( \rho(\mu) \bigr)
\mright)
	-\mleft[ (\lambda \circ \rho)(\mu),(\lambda \circ \rho) (\nu) \mright]_E
\\
&\hspace{1cm}
	+ \mleft[ (\lambda\circ\rho)(\nu), \mu \mright]_E
	+ \nabla_{\rho(\mu)} \mleft( (\lambda\circ\rho)(\nu) \mright)
\\
&\hspace{1cm}
	- \mleft[ (\lambda\circ\rho)(\mu), \nu \mright]_E
	- \nabla_{\rho(\nu)} \mleft( (\lambda\circ\rho)(\mu) \mright)
\\
&\hspace{1cm}
	+ \mleft[ (\lambda \circ \rho)(\mu), \nu \mright]_E
	+ \mleft[ \mu, (\lambda \circ \rho)(\nu) \mright]_E
	- \lambda\mleft( \mleft[ \rho(\mu), \rho(\nu) \mright] \mright)
\\
&=
\mleft( \mathrm{d}^\nabla \lambda \mright)(\rho(\mu),\rho(\nu))
	+ \lambda\mleft(
	\nabla^{\mathrm{bas}}_{(\lambda \circ \rho)(\mu)} \bigl( \rho(\nu) \bigr)
	- \nabla^{\mathrm{bas}}_{(\lambda \circ \rho)(\nu)} \bigl( \rho(\mu) \bigr)
\mright)
\\
&\hspace{1cm}
	-\mleft[ (\lambda \circ \rho)(\mu),(\lambda \circ \rho) (\nu) \mright]_E
\\
&=
\mleft(-\widehat{\zeta}^\lambda \circ \mleft( \widehat{\Lambda}, \widehat{\Lambda} \mright)\mright) (\rho(\mu), \rho(\nu))
\\
&=
\mleft(-\widehat{\zeta}^\lambda \circ \mleft( \widehat{\Lambda}\circ \rho, \widehat{\Lambda}\circ\rho \mright)\mright) (\mu, \nu)
\eas
for all $\mu, \nu \in \Gamma(E)$. In a similar very straightforward fashion,
\bas
&\mleft(
\widetilde{\nabla}^\lambda \circ \Lambda - \Lambda \circ \nabla
	- t_{\widetilde{\nabla}^\lambda_\rho} \circ \mleft( \lambda, \Lambda \mright)
\mright) (Y, \mu)
\\
&=
\biggl(
	\nabla \circ \Lambda - \Lambda \circ \nabla 
	+ t_{\mleft(\widetilde{\nabla}^\lambda\mright)^{\mathrm{bas}}} (\lambda, \Lambda)
%\\
%&\hspace{1cm}\hphantom{\biggl(}
	+ \mleft(\mleft( \Lambda \circ \mathrm{d}^{\nabla^{\mathrm{bas}}} \circ \Lambda^{-1} \mright) \lambda\mright) \circ (\mathds{1}_{\mathrm{T}N}, \Lambda)
\biggr) (Y, \mu)
\\
&=
\nabla_Y \bigl( \Lambda(\mu) \bigr)
	- \Lambda (\nabla_Y \mu)
%\\
%&\hspace{1cm}
	+ \Lambda\mleft( \nabla^{\mathrm{bas}}_{\lambda(Y)} \mu \mright)
	- \mleft(\Lambda \circ \nabla^{\mathrm{bas}}_{\Lambda(\mu)} \circ \Lambda^{-1}\mright) \bigl( \lambda(Y) \bigr)
	-\mleft[ \lambda(Y),\Lambda(\mu) \mright]_E
\\
&\hspace{1cm}
	+ \mleft(\Lambda \circ \nabla^{\mathrm{bas}}_{\Lambda(\mu)} \circ \Lambda^{-1} \mright) \bigl( \lambda(Y)\bigr) 
	- \lambda\mleft(\nabla^{\mathrm{bas}}_{\Lambda(\mu)} Y \mright)
\\
&= \dotsc
\\
&=
	\nabla_Y \Bigl( \bigl(\lambda \circ(-\rho)\bigr)(\mu) \Bigr)
	- \nabla_{-\rho(\mu)} \bigl( \lambda(Y) \bigr)
	- \lambda\bigl( \mleft[ Y, - \rho(\mu) \mright] \bigr)
\\
&\hspace{1cm}
	- \mleft[ \lambda(Y), \bigl(\lambda \circ (-\rho)\bigr)(\mu) \mright]_E
	+ \lambda 
	\mleft(
		\nabla^{\mathrm{bas}}_{\lambda(Y)}\bigl(- \rho(\mu)\bigr) 
		- \nabla^{\mathrm{bas}}_{\mleft(\lambda \circ (-\rho) \mright)(\mu)} Y
	\mright)
\\
&=
\mleft( \mathrm{d}^\nabla \lambda \mright) \bigl( Y, - \rho(\mu) \bigr)
	+ \lambda 
	\mleft(
		\nabla^{\mathrm{bas}}_{\lambda(Y)}\bigl(- \rho(\mu)\bigr) 
		- \nabla^{\mathrm{bas}}_{\mleft(\lambda \circ (-\rho) \mright)(\mu)} Y
	\mright)
	- \mleft[ \lambda(Y), \bigl(\lambda \circ (-\rho)\bigr)(\mu) \mright]_E
\\
&=
\mleft(-\widehat{\zeta}^\lambda \circ \mleft( \widehat{\Lambda}, \widehat{\Lambda} \mright)\mright) \bigl(Y, -\rho(\mu) \bigr)
\\
&=
\mleft(-\widehat{\zeta}^\lambda \circ \mleft( \widehat{\Lambda}, \widehat{\Lambda} \circ (-\rho) \mright)\mright) \bigl(Y, \mu \bigr)
\eas
for all $\mu \in \Gamma(E)$ and $Y \in \mathfrak{X}(N)$. Finally, we can therefore conclude, by using that $-\widehat{\zeta}^\lambda \circ \mleft( \widehat{\Lambda}, \widehat{\Lambda} \mright)$ is clearly an antisymmetric tensor by definition,
\bas
\widetilde{F}^\lambda(\Phi, A)
&=
(\Phi^* \Lambda) (F)
+ \underbrace{\Biggl(\Phi^* \biggl(
	-\widehat{\zeta}^\lambda \circ \mleft( \widehat{\Lambda}, \widehat{\Lambda} \circ (-\rho) \mright)
\biggr) \Biggr) \mleft( \mathrm{D}\Phi \stackrel{\wedge}{,} A \mright)}
_{\mathclap{\stackrel{\text{Prop.~\ref{prop:GradedExtensionPlusAntiSymm}}}{=}  \frac{1}{2}\mleft(  
	\mleft(\Phi^* \mleft( -\widehat{\zeta}^\lambda \circ \mleft( \widehat{\Lambda}, \widehat{\Lambda} \mright) \mright)\mright)\mleft(\mathrm{D}\Phi \stackrel{\wedge}{,} -(\Phi^*\rho)(A) \mright)
	+ \mleft(\Phi^* \mleft( -\widehat{\zeta}^\lambda \circ \mleft( \widehat{\Lambda}, \widehat{\Lambda} \mright) \mright)\mright)\mleft( -(\Phi^*\rho)(A) \stackrel{\wedge}{,} \mathrm{D}\Phi \mright)
	\mright)}}
\\
&\hspace{1cm}
+ \frac{1}{2}\underbrace{\Biggl( \Phi^* \biggl(
	-\widehat{\zeta}^\lambda \circ \mleft( \widehat{\Lambda}\circ \rho, \widehat{\Lambda}\circ\rho \mright)
\biggr) \Biggr) \mleft( A \stackrel{\wedge}{,} A \mright)}
_{= \mleft( \Phi^* \mleft(
	-\widehat{\zeta}^\lambda \circ \mleft( \widehat{\Lambda}, \widehat{\Lambda} \mright)
\mright) \mright) \mleft( -(\Phi^*\rho)(A) \stackrel{\wedge}{,} - (\Phi^*\rho)(A) \mright)}
+ \frac{1}{2}\Biggl( \Phi^* \biggl(
	-\widehat{\zeta}^\lambda \circ \mleft( \widehat{\Lambda}, \widehat{\Lambda} \mright)
\biggr) \Biggr) \mleft( \mathrm{D}\Phi \stackrel{\wedge}{,} \mathrm{D} \Phi \mright)
\\
&=
(\Phi^*\Lambda)(F)
	+ \frac{1}{2} \Biggl(
		\Phi^* \biggl(
			-\widehat{\zeta}^\lambda \circ \mleft( \widehat{\Lambda}, \widehat{\Lambda} \mright)
		\biggr)
	\Biggr) \mleft( \mathfrak{D}^A\Phi \stackrel{\wedge}{,} \mathfrak{D}^A \Phi \mright)
\\
&=
\mleft( \Phi^* \Lambda \mright) \mleft(
	F
	- \frac{1}{2} \Biggl(\Phi^*\biggl(\Lambda^{-1} \circ \widehat{\zeta}^\lambda \circ \mleft( \widehat{\Lambda}, \widehat{\Lambda} \mright)\biggr)\Biggr) \mleft( \mathfrak{D}^A\Phi \stackrel{\wedge}{,} \mathfrak{D}^A \Phi \mright)
\mright)
\\
&=
\mleft( \Phi^* \Lambda \mright) \mleft(
	F
	- \frac{1}{2} \mleft(\Phi^*\xi\mright) \mleft( \mathfrak{D}^A\Phi \stackrel{\wedge}{,} \mathfrak{D}^A \Phi \mright)
\mright).
\eas
\end{proof}

Let us now look at the compatibility conditions of Thm.~\ref{thm:GaugeInvariantStandardLagrangian} and how they change under the field redefinition. For this we need the following auxiliary results.

\begin{propositions}{Change of (basic) curvature under a change of the connection}{ChangeofCurvaturesUnderCHangesOfConnections}
Let $E\to N$ be a Lie algebroid, equipped with a vector bundle connection $\nabla$. For any other connection $\nabla^\prime$ write $\nabla^\prime = \nabla + I$ where $I \in \Omega^1(N; \mathrm{End}(E))$. Then we have
\ba
R^{\mathrm{bas}}_{\nabla^\prime}
&=
R^{\mathrm{bas}}_\nabla
	- \mathrm{d}^{\nabla^{\mathrm{bas}}} I
	- I \wedge (\rho \circ I).
\ea
For the curvatures of the connections we get
\ba
R_{\nabla^\prime}
&=
R_\nabla
	+ \mathrm{d}^\nabla I
	+ I \wedge I.
\ea
\end{propositions}

\begin{remark}\label{Wedgies}
\leavevmode\newline
$I \wedge (\rho \circ I)$ is similarly defined to Def.~\eqref{DefVonWedgedemitEnd} although $\rho \circ I$ has values in $\mathrm{T}N$, the first factor $I$ simply acts on the $\mathrm{T}N$ part then, \textit{i.e.}~$I \wedge (\rho \circ I)$ is an element of $\Omega^{1,2}(N, E;E)$ defined by
\bas
\bigl(I \wedge (\rho \circ I)\bigr)(Y, \mu, \nu)
&=
I \bigl( (\rho\circ I)\bigl(Y, \nu \bigr), \mu\bigr)
	- I\bigl((\rho\circ I)\bigl(Y, \mu \bigr), \nu \bigr)
\eas
for all $\mu, \nu \in \Gamma(E)$ and $Y \in \mathfrak{X}(N)$.

$I \wedge I \in \Omega^2(N; \mathrm{End}(E))$ makes direct use of Def.~\eqref{DefVonWedgedemitEnd}, but the second factor is directly contracted with a section of $E$, that is
\bas
\mleft( I \wedge I \mright)(X, Y, \nu )
&=
I\bigl( X, I(Y, \nu) \bigr)
	- I\bigl( Y, I(X, \nu) \bigr)
\eas
for all $\nu \in \Gamma(E)$ and $X, Y \in \mathfrak{X}(N)$. Using the definition of derivations $\mathcal{D}(V)$ of vector bundles $V$ one could also write
\bas
(I \wedge I)(X,Y, \cdot)
&=
\mleft[ I(X, \cdot), I(Y, \cdot) \mright]_{\mathcal{D}(E)}
\eas
for all $X, Y \in \mathfrak{X}(N)$.
\end{remark}

\begin{proof}[Proof of Prop.~\ref{prop:ChangeofCurvaturesUnderCHangesOfConnections}]
\leavevmode\newline
We have
\bas
\mleft(\nabla^\prime\mright)^{\mathrm{bas}}_\nu Y
&=
\mleft[ \rho(\nu), Y \mright]
	+ \rho\mleft( \nabla^\prime_Y \nu \mright)
=
\nabla^{\mathrm{bas}}_\nu Y
	+ \rho\bigl( I(Y, \nu) \bigr),
%\\
%\mleft(\nabla^\prime\mright)^{\mathrm{bas}}_\nu \mu
%&=
%\mleft[ \nu, \mu \mright]_E
	%+ \nabla^\prime_{\rho(\mu)} \nu
%=
%\nabla^{\mathrm{bas}}_\nu \mu
	%+ I\bigl( \rho(\mu), \nu \bigr)
\eas
for all $\mu, \nu \in \Gamma(E)$ and $Y \in \mathfrak{X}(N)$. Using these identities we get
\bas
R_{\nabla^\prime}^\mathrm{bas}(\mu, \nu) Y
&=
\nabla^\prime_Y\mleft(\mleft[\mu, \nu\mright]_E\mright) 
	- \mleft[ \nabla^\prime_Y \mu, \nu \mright]_E 
	- \mleft[ \mu, \nabla^\prime_Y \nu \mright]_E 
	- \nabla^\prime_{\mleft(\nabla^\prime\mright)^{\mathrm{bas}}_\nu Y} \mu 
	+ \nabla^\prime_{\mleft(\nabla^\prime\mright)^{\mathrm{bas}}_\mu Y} \nu
\\
&=
\underbrace{\nabla_Y\mleft(\mleft[\mu, \nu\mright]_E\mright) 
	- \mleft[ \nabla_Y \mu, \nu \mright]_E 
	- \mleft[ \mu, \nabla_Y \nu \mright]_E
	- \nabla_{\nabla^{\mathrm{bas}}_\nu Y} \mu 
	+ \nabla_{\nabla^{\mathrm{bas}}_\mu Y} \nu}
	_{= R_\nabla^{\mathrm{bas}}(\mu, \nu)Y}
\\
&\hspace{1cm}
	- \mleft[ I(Y, \mu), \nu \mright]_E 
	- \mleft[ \mu, I(Y, \nu) \mright]_E
	+ I\mleft(Y, \mleft[\mu, \nu\mright]_E\mright)
\\
&\hspace{1cm}
	- \nabla_{\mleft(\rho \circ I \mright)(Y, \nu)} \mu
	+ \nabla_{\mleft(\rho \circ I \mright)(Y, \mu)} \nu
\\
&\hspace{1cm}
	- I \mleft( \nabla_\nu^{\mathrm{bas}} Y, \mu\mright)
	+ I\mleft(\nabla^{\mathrm{bas}}_\mu Y, \nu \mright)
	- I \Bigl( (\rho\circ I)\bigl(Y, \nu \bigr), \mu\Bigr)
	+ I\Bigl((\rho\circ I)\bigl(Y, \mu \bigr), \nu \Bigr)
\\
&=
R_\nabla^{\mathrm{bas}}(\mu, \nu)Y
\\
&\hspace{1cm}
	+ \nabla^{\mathrm{bas}}_\nu \mleft(
		I(Y, \mu)
	\mright)
	- I \mleft( \nabla_\nu^{\mathrm{bas}} Y, \mu\mright)
\\
&\hspace{1cm}
	- \nabla^{\mathrm{bas}}_\mu \mleft(
		I(Y, \nu)
	\mright)
	+ I\mleft(\nabla^{\mathrm{bas}}_\mu Y, \nu \mright)
\\
&\hspace{1cm}
	+ I\mleft(Y, \mleft[\mu, \nu\mright]_E\mright)
	- I \Bigl( (\rho\circ I)\bigl(Y, \nu \bigr), \mu\Bigr)
	+ I\Bigl((\rho\circ I)\bigl(Y, \mu \bigr), \nu \Bigr)
\\
&=
\mleft(
R^{\mathrm{bas}}_\nabla
	- \mathrm{d}^{\nabla^{\mathrm{bas}}} I
	- I \wedge (\rho \circ I)
\mright) (Y, \mu, \nu)
\eas
for all $\mu, \nu \in \Gamma(E)$ and $Y \in \mathfrak{X}(N)$.
 For the curvatures we get
\bas
R_{\nabla^\prime}(\cdot, \cdot) \nu
&=
\mathrm{d}^{\nabla^\prime} \mleft( \nabla^\prime \nu \mright)
%\\
%&=
%\mathrm{d}^{\nabla^\prime} (\nabla\nu)
	%+ \mathrm{d}^{\nabla^\prime} \bigl(I(\cdot, \nu) \bigr)
\\
&\stackrel{\mathclap{\text{Eq.~\eqref{eqDifferentialSplit}}}}{=}~~~
\mathrm{d}^{\nabla} \mleft( \nabla^\prime \nu \mright) + I \wedge \nabla^\prime \nu
\\
&=
	R_\nabla(\cdot, \cdot)\nu
	+ \underbrace{\mathrm{d}^\nabla \bigl( I(\cdot, \nu) \bigr)}
	_{\mathclap{\stackrel{\text{Eq.~\eqref{TypischerSplitdesDifferentialsaufdasWedgeProdukt}}}{=} \mleft( \mathrm{d}^\nabla I \mright)(\nu)- I \wedge \nabla \nu}}
	+ ~I \wedge \nabla \nu
	+ I \wedge I(\cdot, \nu)
\\
&=
\mleft(
R_\nabla
	+ \mathrm{d}^\nabla I
	+ I \wedge I
\mright) (\nu)
\eas
for all $\nu \in \Gamma(E)$, where we used that $T \wedge \nu = T(\nu) \in \Omega^\bullet(N; E)$ for all $T \in \Omega^\bullet(N;\mathrm{End}(E))$.
%By Eq.~\eqref{eqDifferentialSplit} we can write $\mathrm{d}^{\nabla^\prime} I - I \wedge I = \mathrm{d}^\nabla I + I \wedge I$. 
%\bas
%\mleft(
	%\mathrm{d}^\nabla \bigl( I(\cdot, \nu) \bigr)
	%+ I \wedge \nabla \nu
%\mright)(X,Y)
%&=
%\nabla_X \bigl( I(Y, \nu) \bigr)
	%- \nabla_Y \bigl( I(X, \nu) \bigr)
	%- I\bigl( [X,Y], \nu \bigr)
%\\
%&\hspace{1cm}	
	%+ I\mleft( X, \nabla_Y \nu \mright)
	%- I\mleft( Y, \nabla_X \nu \mright)
%\eas
%for all $X, Y \in \mathfrak{X}(N)$ and $\nu \in \Gamma(E)$.
\end{proof}

Let us first look at the compatibility conditions besides the curvature of $\nabla$; we want that these are preserved with the field redefinition.

\begin{theorems}{Field redefinition of the compatibility conditions except curvature}{FieldRedefDerEinfacherenCompatibilities}
Let $N$ be smooth manifolds, $E \to N$ a Lie algebroid, $\nabla$ a connection on $E$, and $\kappa$ and $g$ fibre metrics on $E$ and $\mathrm{T}N$, respectively. Assume that the compatibility conditions of Thm.~\ref{thm:GaugeInvariantStandardLagrangian} are satisfied, but $\nabla$ is allowed to be non-flat. Also let $\lambda \in \Omega^1(N; E)$ such that $\Lambda = \mathds{1}_E - \lambda \circ \rho$ is an element of $\sAut(E)$. Then we have
\ba
\mleft(\widetilde{\nabla}^{\lambda}\mright)^{\mathrm{bas}} \widetilde{\kappa}^\lambda
&=
0,
\\
\mleft(\widetilde{\nabla}^{\lambda}\mright)^{\mathrm{bas}} \widetilde{g}^\lambda
&=
0,
\\
R_{\widetilde{\nabla}^\lambda}^\mathrm{bas}
&=
0.
%\\
%R_{\widetilde{\nabla}^\lambda}
%&=
%-\mathrm{d}^{\mleft(\widetilde{\nabla}^\lambda\mright)^{\mathrm{bas}}} \widehat{\zeta}^\lambda,
\ea
%where $\widehat{\zeta}^\lambda$ is the element of $\Omega^2(N;E)$ as defined in Thm.~\ref{thm:FieldRedefofstandardFieldStrengthF}.
\end{theorems}

\begin{proof}
\leavevmode\newline
For the compatibilities with the metrics use Eq.~\eqref{FieldTrafoOfKappa}, \eqref{FieldTrafoOfG} and \eqref{basicconnectionTrafoRefield}, so,
\bas
&\mleft(\mleft(\widetilde{\nabla}^{\lambda}\mright)^{\mathrm{bas}} \widetilde{g}^\lambda\mright)
\mleft( \widehat{\Lambda}(X), \widehat{\Lambda}(Y) \mright)
\\
&=
\mathrm{d}\mleft(  
	\widetilde{g}^\lambda \mleft( \widehat{\Lambda}(X), \widehat{\Lambda}(Y) \mright)
\mright)
	- \widetilde{g}^\lambda \mleft( \mleft(\widetilde{\nabla}^{\lambda}\mright)^{\mathrm{bas}} \mleft( \widehat{\Lambda}(X) \mright), \widehat{\Lambda}(Y) \mright)
	- \widetilde{g}^\lambda \mleft( \widehat{\Lambda}(X), \mleft(\widetilde{\nabla}^{\lambda}\mright)^{\mathrm{bas}} \mleft( \widehat{\Lambda}(Y) \mright) \mright)
\\
&=
\mathrm{d}\mleft(  
	g\mleft( X, Y \mright)
\mright)
	- g \mleft( \nabla^{\mathrm{bas}} X, Y \mright)
	- g \mleft( X, \nabla^{\mathrm{bas}} Y \mright)
\\
&=
\mleft(\nabla^{\mathrm{bas}} g\mright)(X, Y)
\\
&=
0,
\eas
for all $X, Y \in \mathfrak{X}(N)$,
similarly for $\kappa$. For $I \coloneqq \mleft( \Lambda \circ \mathrm{d}^{\nabla^{\mathrm{bas}}} \circ \Lambda^{-1} \mright) \lambda \in \Omega^1(N; \mathrm{End}(E)) \cong \Omega^{1,1}(N,E;E)$ we also have
\bas
I(Y, \nu)
%&=
%\mleft( \Lambda \circ \nabla^{\mathrm{bas}}_\nu \circ \Lambda^{-1} \circ \lambda - \lambda \circ \nabla^{\mathrm{bas}}_\nu \mright)(Y)
&\stackrel{\text{Eq.~\eqref{FieldTrafoOfNabla}}}{=}
\widetilde{\nabla}^\lambda_\nu Y
	- \nabla_\nu Y
\eas
for all $\nu \in \Gamma(E)$ and $Y \in \mathfrak{X}(N)$, and
%and by Eq.~\eqref{KuerzesteFormelForRedefOfNabla}
%\bas
%I\mleft(\widehat{\Lambda}(Y), \nu\mright)
%&=
%\mleft(\mathrm{d}^{\nabla^{\mathrm{bas}}} \lambda \mright)(Y, \nu).
%\eas
%We also have
\bas
%\mleft(\widetilde{\nabla}^\lambda\mright)^{\mathrm{bas}}_\nu Y
%&=
%\mleft[ \rho(\nu), Y \mright]
	%+ \rho\mleft( \widetilde{\nabla}^\lambda_Y \nu \mright)
%=
%\nabla^{\mathrm{bas}}_\nu Y
	%+ \rho\bigl( I(Y, \nu) \bigr),
%\\
\underbrace{\nabla^{\mathrm{bas}}_\nu \mu}
_{\mathclap{ = \mleft[ \nu, \mu \mright]_E + \nabla_{\rho(\mu)} \nu }}
	+ I\bigl( \rho(\mu), \nu \bigr)
&=
\mleft[ \nu, \mu \mright]_E
	+ \widetilde{\nabla}^\lambda_{\rho(\mu)} \nu
=
\mleft(\widetilde{\nabla}^\lambda\mright)^{\mathrm{bas}}_\nu \mu
\stackrel{\text{Eq.~\eqref{basicconnectionTrafoRefield}}}{=}
\mleft( \Lambda \circ \nabla^{\mathrm{bas}}_\nu \circ \Lambda^{-1}\mright) \mu
\eas
for all $\mu, \nu \in \Gamma(E)$.
Using these identities and $R_\nabla^\mathrm{bas} = 0$, we can show
\bas
\mleft(
\mathrm{d}^{\nabla^{\mathrm{bas}}} I
	+ I \wedge (\rho \circ I) 
\mright)(Y, \mu,\nu)
&=
\nabla^{\mathrm{bas}}_\mu \bigl( I(Y, \nu) \bigr)
	- I \mleft( \nabla^{\mathrm{bas}}_\mu Y, \nu \mright)
\\
&\hspace{1cm}
	- \nabla^{\mathrm{bas}}_\nu \bigl( I(Y, \mu) \bigr)
	+ I \mleft( \nabla^{\mathrm{bas}}_\nu Y, \mu \mright)
\\
&\hspace{1cm}
	- I \mleft( Y, \mleft[ \mu, \nu \mright]_E \mright)
	+ I\mleft( (\rho \circ I)(Y, \nu), \mu \mright)
	- I\mleft( (\rho \circ I)(Y, \mu), \nu \mright)
\\
&=
\mleft( \Lambda \circ \nabla^{\mathrm{bas}}_\mu \circ \Lambda^{-1}\mright) \bigl( I(Y, \nu) \bigr)
	- I \mleft( \nabla^{\mathrm{bas}}_\mu Y, \nu \mright)
\\
&\hspace{1cm}
	- \mleft( \Lambda \circ \nabla^{\mathrm{bas}}_\nu \circ \Lambda^{-1}\mright) \bigl( I(Y, \mu) \bigr)
	+ I \mleft( \nabla^{\mathrm{bas}}_\nu Y, \mu \mright)
\\
&\hspace{1cm}
	- I \mleft( Y, \mleft[ \mu, \nu \mright]_E \mright)
\\
&=
\mleft(\mleft( \Lambda \circ \mathrm{d}^{\nabla^{\mathrm{bas}}} \circ \Lambda^{-1} \mright) I \mright) (Y, \nu ,\mu)
\\
&=
\mleft(\mleft( \Lambda \circ \mathrm{d}^{\nabla^{\mathrm{bas}}} \circ \Lambda^{-1} \mright)^2 \lambda \mright) (Y, \nu ,\mu)
\\
&=
\Biggl(\biggl( \Lambda \circ \underbrace{\mleft(\mathrm{d}^{\nabla^{\mathrm{bas}}}\mright)^2}
_{\mathclap{\stackrel{\text{Prop.~\ref{prop:SnablamitREnabla}}}{=}~ 0}} 
\circ ~\Lambda^{-1} \biggr) \lambda \Biggr) (Y, \nu ,\mu)
\\
&=
0.
\eas
for all $\mu, \nu \in \Gamma(E)$ and $Y \in \mathfrak{X}(N)$.
Using this and $R_\nabla^\mathrm{bas} = 0$, we get
\bas
R_{\widetilde{\nabla}^\lambda}^\mathrm{bas}
&\stackrel{\text{Prop.~\ref{prop:ChangeofCurvaturesUnderCHangesOfConnections}}}{=}
R_\nabla^{\mathrm{bas}}
	- \mathrm{d}^{\nabla^{\mathrm{bas}}} I
	- I \wedge (\rho \circ I)
=
0.
\eas
\end{proof}

Let us now look at what happens with the curvature of $\nabla$. 

\begin{theorems}{Flatness breaking}{BrokenFlatness}
Let $N$ be smooth manifolds, $E \to N$ a Lie algebroid, and $\nabla$ a connection on $E$ with vanishing basic curvature. Also let $\lambda \in \Omega^1(N; E)$ such that $\Lambda = \mathds{1}_E - \lambda \circ \rho$ is an element of $\sAut(E)$. Then
\ba
R_{\widetilde{\nabla}^\lambda}
&=
\Lambda \circ R_\nabla \circ \mleft( \widehat{\Lambda}^{-1}, \widehat{\Lambda}^{-1} \mright)
	- \mathrm{d}^{\mleft(\widetilde{\nabla}^\lambda\mright)^{\mathrm{bas}}} \widehat{\zeta}^\lambda,
\ea
where $\widehat{\zeta}^\lambda$ is defined as in Thm.~\ref{thm:FieldRedefofstandardFieldStrengthF} and viewing the curvatures as elements of $\Omega^2(N; \mathrm{End}(E))$.
\end{theorems}

\begin{proof}[Sketch of the proof]
\leavevmode\newline
\indent $\bullet$ The proof of this theorem is extremely tedious and long, but very straightforward. Essentially, just insert all the formulas of the field redefinition on both sides, then compare both sides, making use of the vanishing of the basic curvature. However, you may want to use certain tricks to make the calculation less tedious (but it is still extremely tedious with tricks). Hence, we show the first steps until one "just" needs to insert all definitions.

First let us observe that we can rewrite $\mathrm{d}^{\mleft(\widetilde{\nabla}^\lambda\mright)^{\mathrm{bas}}} \widehat{\zeta}^\lambda$ using Cor.~\ref{cor:ConjugationOfDifferentialsAreShitty}, also recall Remark \ref{OtherNotationForZetaTransform},
\bas
- \mleft(\mathrm{d}^{\mleft(\widetilde{\nabla}^\lambda\mright)^{\mathrm{bas}}} \widehat{\zeta}^\lambda\mright)\mleft( \widehat{\Lambda}(X), \widehat{\Lambda}(Y), \nu \mright)
&=
- \Biggl( \mleft( \Lambda \circ \mathrm{d}^{\nabla^{\mathrm{bas}}} \circ \Lambda^{-1} \mright)
\mleft( \widehat{\zeta}^\lambda \circ \mleft( \widehat{\Lambda}, \widehat{\Lambda} \mright) \mright)
 \Biggr) (X, Y, \nu)
\\
&=
\Biggl( \mleft( \Lambda \circ \mathrm{d}^{\nabla^{\mathrm{bas}}} \circ \Lambda^{-1} \mright)
	\mleft(\mathrm{d}^\nabla \lambda
	+ \lambda\mleft(t_{\nabla^{\mathrm{bas}}_\lambda}\mright)
	- R_\lambda\mright)
 \Biggr) (X, Y, \nu)
\eas
for all $X, Y \in \mathfrak{X}(N)$ and $\nu \in \Gamma(E)$, where $-\widehat{\zeta}^\lambda \circ \mleft( \widehat{\Lambda}, \widehat{\Lambda} \mright)$ is given by Eq.~\eqref{FormulaForZetaTildeWithZetaEqualzero}, also recall Eq.~\eqref{TollsteFormelFuerZetaTrafoFragezeichen}. We also have
\bas
&\mleft(\mleft( \Lambda \circ \mathrm{d}^{\nabla^{\mathrm{bas}}} \circ \Lambda^{-1} \mright)\mleft(
	\lambda\mleft( t_{\nabla^{\mathrm{bas}}_\lambda} \mright)
	- R_\lambda
\mright)\mright) \mleft(X, Y, \nu\mright)
\\
&=
\mleft(\Lambda \circ \nabla^{\mathrm{bas}}_\nu \circ \Lambda^{-1}\mright)\mleft(
	\lambda\Bigl(
		\nabla^{\mathrm{bas}}_{\lambda(X)} Y
		- \nabla^{\mathrm{bas}}_{\lambda(Y)} X
	\Bigr)
	- \mleft[ \lambda(X), \lambda(Y) \mright]_E
\mright)
\\
&\hspace{1cm}
	- \lambda\mleft(
		\nabla^{\mathrm{bas}}_{\lambda\mleft(\nabla^{\mathrm{bas}}_\nu X\mright)} Y
		- \nabla^{\mathrm{bas}}_{\lambda(Y)} \nabla^{\mathrm{bas}}_\nu X
		%}_{= [(\rho\circ\lambda)(Y), X] + \rho\mleft( \nabla_X ( \lambda(Y) ) \mright)}
	\mright)
	+ \mleft[ \lambda\mleft(\nabla^{\mathrm{bas}}_\nu X\mright), \lambda(Y) \mright]_E
\\
&\hspace{1cm}
	- \lambda\mleft(
		\nabla^{\mathrm{bas}}_{\lambda(X)} \nabla^{\mathrm{bas}}_\nu Y
		- \nabla^{\mathrm{bas}}_{\lambda\mleft(\nabla^{\mathrm{bas}}_\nu Y\mright)} X
		%}_{= [(\rho\circ\lambda)(Y), X] + \rho\mleft( \nabla_X ( \lambda(Y) ) \mright)}
	\mright)
	+ \mleft[ \lambda(X), \lambda\mleft(\nabla^{\mathrm{bas}}_\nu Y\mright) \mright]_E.
\eas
%
%For the curvature $R_{\widetilde{\nabla}^\lambda}$ we instead calculate $R_{\widetilde{\nabla}^\lambda}\mleft( \widehat{\Lambda}(X), \widehat{\Lambda}(Y) \mright) \nu$, because then we can then use Eq.~\eqref{KuerzesteFormelForRedefOfNabla}

Now let us start to calculate the left hand side given by $R_{\widetilde{\nabla}^\lambda}$, using the second equation in Prop.~\ref{prop:ChangeofCurvaturesUnderCHangesOfConnections}, especially we need to calculate
\bas
\mathrm{d}^\nabla\mleft( \mleft(\Lambda \circ \mathrm{d}^{\nabla^{\mathrm{bas}}} \circ \Lambda^{-1} \mright) \lambda \mright),
\eas
and for this we want to use Cor.~\ref{cor:commutationS=0}. Using the commutator of operators, we see
\bas
\mleft[ \mathrm{d}^\nabla, \Lambda \circ \mathrm{d}^{\nabla^{\mathrm{bas}}} \circ \Lambda^{-1} \mright]
&=
\mleft[ \mathrm{d}^\nabla, \Lambda \mright] \circ \mathrm{d}^{\nabla^{\mathrm{bas}}} \circ \Lambda^{-1}
	+ \Lambda \circ \mleft[ \mathrm{d}^\nabla, \mathrm{d}^{\nabla^{\mathrm{bas}}} \mright] \circ \Lambda^{-1}
	+ \Lambda \circ \mathrm{d}^{\nabla^{\mathrm{bas}}} \circ \mleft[ \mathrm{d}^\nabla, \Lambda^{-1} \mright],
\eas
with that we can write
\bas
\mathrm{d}^\nabla\mleft( \mleft(\Lambda \circ \mathrm{d}^{\nabla^{\mathrm{bas}}} \circ \Lambda^{-1} \mright) \lambda \mright)
&=
\mleft[ \mathrm{d}^\nabla, \Lambda \circ \mathrm{d}^{\nabla^{\mathrm{bas}}} \circ \Lambda^{-1} \mright](\lambda)
	+ \mleft( \Lambda \circ \mathrm{d}^{\nabla^{\mathrm{bas}}} \circ \Lambda^{-1} \mright)
	\mleft(\mathrm{d}^\nabla \lambda\mright).
\eas
One needs to calculate the first summand, the summand in the middle in the formula of $\mleft[ \mathrm{d}^\nabla, \Lambda \circ \mathrm{d}^{\nabla^{\mathrm{bas}}} \circ \Lambda^{-1} \mright]$ is given by Cor.~\ref{cor:commutationS=0} due to the vanishing basic curvature of $\nabla$, so,
\bas
\mleft[ \mathrm{d}^\nabla, \mathrm{d}^{\nabla^{\mathrm{bas}}} \mright]\mleft( \Lambda^{-1} \circ \lambda \mright)(X,Y, \nu)
&=
R_\nabla\mleft(X, \mleft(\rho \circ \Lambda^{-1} \circ \lambda\mright)(Y)\mright)\nu
	- R_\nabla\mleft(Y, \mleft(\rho \circ \Lambda^{-1} \circ \lambda\mright)(X)\mright)\nu
\\
&\hspace{1cm}
	- \mleft(\Lambda^{-1} \circ \lambda \circ \rho \mright)\bigl( R_\nabla(X,Y)\nu \bigr)
\eas
for all $X, Y \in \mathfrak{X}(N)$ and $\nu \in \Gamma(E)$,
and
\bas
\mleft[ \mathrm{d}^\nabla, \Lambda \mright]
&=
\mleft[ \mathrm{d}^\nabla, \mathds{1}_E - \lambda \circ \rho \mright]
=
- \mleft[ \mathrm{d}^\nabla, \lambda \circ \rho \mright],
\eas
and for the last summand in the second equation of Prop.~\ref{prop:ChangeofCurvaturesUnderCHangesOfConnections} we have, also recall Remark \ref{Wedgies} and Eq.~\eqref{KuerzesteFormelForRedefOfNabla},
\bas
\mleft[ I\mleft(\widehat{\Lambda}(X), \cdot\mright), I\mleft(\widehat{\Lambda}(Y), \cdot\mright) \mright]_{\mathcal{D}(E)}(\nu)
&=
\nabla^{\mathrm{bas}}_{\nabla^{\mathrm{bas}}_{\nu}(\lambda(Y)) - \lambda\mleft( \nabla^{\mathrm{bas}}_\nu Y \mright)}\bigl(\lambda(X)\bigr)
%\\
%&\hspace{1cm}
	-\lambda\mleft(
		\nabla^{\mathrm{bas}}_{\nabla^{\mathrm{bas}}_{\nu}(\lambda(Y)) - \lambda\mleft( \nabla^{\mathrm{bas}}_\nu Y \mright)} X
	\mright)
\\
&\hspace{1cm}
	- (Y \leftrightarrow X \text{ of all previous lines}).
\eas
Now the purely tedious but straightforward part comes. Insert $X, Y, \nu$ everywhere\footnote{In general use $\widehat{\Lambda}(X)$ instead of $X$, similar for $Y$, as we did at the beginning and at the end, then it will be easier to compare the terms since a lot of $\Lambda$ will get canceled.} and the definition of the basic connection on both sides of the desired equation; although you may already recognize some similar terms of the calculation of the right hand side at the beginning, for those terms one does not need to insert the definition of the basic connection. Also make heavily use of Prop.~\ref{prop:PropsOfBigLambdas}, and also directly use the vanishing of the basic curvature on the right hand side (which implies flatness of the basic connection). We already got three curvature terms, and there is one additional by Prop.~\ref{prop:ChangeofCurvaturesUnderCHangesOfConnections}; there is actually one missing, but that term will be produced by the other remaining terms, for example by some of the form "$\nabla^{\mathrm{bas}}_{\nabla^{\mathrm{bas}}}$".

$\bullet$ As a proof of concept, you can also look at \cite[proof of Theorem 3.6, the first equation for the transformed curvature there]{My1stpaper} where I have calculated this for Lie algebra bundles; the structure of the calculation there is, abstractly-spoken, the same, but extremely shorter and less tedious due to a vanishing anchor. However, we will actually not need this theorem for the gauge invariance of the transformed Lagrangian as we are going to see, and we will argue later why the gauge invariance of the Lagrangian in general proves this theorem, too, avoiding the tedious calculation.
\end{proof}

%and observe that for all $f \in \Omega^1(E;E)$ that
%\bas
%\mleft[ \mathrm{d}^\nabla, f \mright]
%&=
%\mathrm{d}^\nabla f \wedge,
%\eas
%that is for all $\omega \in \Omega^{1, q}(N,E;E)$ ($q \in \mathbb{N}_0$)
%\bas
%\mleft(\mathrm{d}^\nabla f \wedge \omega \mright)(X, Y, \nu_1, \dotsc, \nu_q)
%&=
%\mleft(\mathrm{d}^{\nabla}f \mright) \bigl(X, \omega(Y, \nu_1, \dotsc, \nu_q)\bigr)
	%- \mleft(\mathrm{d}^{\nabla}f \mright) \bigl(Y, \omega(X, \nu_1, \dotsc, \nu_q)\bigr)
%\\
%&=
%\nabla_X\bigl( (f \circ \omega)(Y, \nu_1, \dotsc, \nu_q) \bigr)
	%- f\Bigl( \nabla_X \bigl( \omega(Y, \nu_1, \dotsc, \nu_q) \bigr) \Bigr)
%\eas
%for all $X, Y \in \mathfrak{X}(N)$ and $\nu_1, \dotsc, \nu_q \in \Gamma(E)$.

%\bas
%\widetilde{\nabla}^\lambda_{\widehat{\Lambda}(X)} \widetilde{\nabla}^\lambda_{\widehat{\Lambda}(X)} \nu
%&=
%\nabla_{\widehat{\Lambda}(X)} \nabla_{\widehat{\Lambda}(Y)} \nu
	%+ \nabla_{\widehat{\Lambda}(X)} \mleft( \mleft(\mathrm{d}^{\nabla^{\mathrm{bas}}} \lambda \mright)\mleft(Y, \nu\mright) \mright)
%\\
%&\hspace{1cm}
	%+ \mleft(\mathrm{d}^{\nabla^{\mathrm{bas}}} \lambda \mright)
		%\mleft( X , \nabla_{\widehat{\Lambda}(Y)} \nu + \mleft(\mathrm{d}^{\nabla^{\mathrm{bas}}} \lambda \mright)\mleft(Y, \nu\mright) \mright)
%\eas
Therefore we see that the curvature is not necessarily flat after a field redefinition. We have seen that the other remaining compatibility conditions are still satisfied, but what about infinitesimal gauge invariance when flatness is gone? Eq.~\eqref{FieldRedefOfClassicF} shows us that we get an offset in the field strength, which one may want to correct for preserving gauge invariance and the Lagrangian itself, and Thm.~\ref{thm:BrokenFlatness} motivates that the derivative of this offset using a basic connection has something to do with the curvature of $\nabla$ such that there is hope that the offset compensates the curvature, leading to a gauge invariant theory with a non-flat connection! Let us prove this.

\begin{theorems}{Infinitesimal gauge transformation after field redefinition}{FieldRedefOfGaugeTrafo}
Let $M, N$ be smooth manifolds, $E \to N$ a Lie algebroid, and $\nabla$ a connection on $E$. Also let $\lambda \in \Omega^1(N; E)$ such that $\Lambda = \mathds{1}_E - \lambda \circ \rho$ is an element of $\sAut(E)$. Then
\ba
\widetilde{\delta}_\varepsilon^\lambda
&=
{}^*\Lambda \circ \delta_\varepsilon \circ {}^*\mleft(\Lambda^{-1}\mright)
\ea
on $E$ and
\ba
\widetilde{\delta}_\varepsilon^\lambda
&=
{}^*\widehat{\Lambda} \circ \delta_\varepsilon \circ {}^*\mleft(\widehat{\Lambda}^{-1}\mright)
\ea
on $\mathrm{T}N$
for all $\varepsilon \in \mathcal{F}^0_E(M; {}^*E)$, where $\widetilde{\delta}_\varepsilon^\lambda$ is similarly defined to $\delta_\varepsilon$ but using $\widetilde{\nabla}^\lambda$ instead of $\nabla$ and $\widetilde{\varpi_2}^\lambda$ instead of $\varpi_2$ in Def.~\ref{def:TotalInfGaugeTrafoYayy}.\footnote{$\varpi_2$ was needed for fixing the vector fields like $\Psi_\varepsilon \in \mathfrak{X}^E\bigl( \mathfrak{M}_E(M;N) \bigr)$ by Prop.~\ref{prop:VariationOfA}.} Moreover, on scalar-valued functionals we have
\ba
\widetilde{\delta}_\varepsilon^\lambda
&=
\mathcal{L}_{\Psi_\varepsilon}
=
\delta_\varepsilon,
\ea
where $\Psi_\varepsilon \in \mathfrak{X}^E(\mathfrak{M}_E(M; N))$ is the vector field behind the definition of $\delta_\varepsilon$, recall Def.~\ref{def:TotalInfGaugeTrafoYayy}.
\end{theorems}

\begin{remark}
\leavevmode\newline
Observe how $\Psi_\varepsilon$ is unaffected by the field redefinition although $\varpi_2$ and $\nabla$ transform by the field redefinition, both of which were essential in the construction of infinitesimal gauge transformations.
\end{remark}

\begin{proof}[Proof of Thm.~\ref{thm:FieldRedefOfGaugeTrafo}]
\leavevmode\newline
We will prove this by using the uniqueness behind the construction of operators like $\delta_\varepsilon$, especially recall Prop.~\ref{prop:VariationVonSkalarZeugsEasyPeasy} and \ref{prop:VariationOfA}. We write 
\bas
\delta_\varepsilon^\prime
&\coloneqq
{}^*\Lambda \circ \delta_\varepsilon \circ {}^*\mleft(\Lambda^{-1}\mright)
\eas
and first observe that
\bas
\delta_\varepsilon^\prime ({}^*\nu)
&=
{}^*\Lambda \biggl(
	\delta_\varepsilon\mleft( {}^*\mleft( \Lambda^{-1}(\nu) \mright) \mright)
\biggr)
=
- {}^*\Lambda \mleft(
	{}^*\mleft( \nabla^{\mathrm{bas}}_\varepsilon \mleft( \Lambda^{-1} (\nu) \mright) \mright)
\mright)
\stackrel{ \text{Eq.~\eqref{basicconnectionTrafoRefield}} }{=}
- {}^*\mleft(
	\mleft(\widetilde{\nabla}^{\lambda}\mright)^{\mathrm{bas}}_\varepsilon \nu
\mright)
\eas
for all $\nu \in \Gamma(E)$. Hence, it shares this property with $\widetilde{\delta}_\varepsilon^\lambda$, $\delta_\varepsilon^\prime$ is also clearly $\mathbb{R}$-linear and satisfies Eq.~\eqref{VertauschungMitVerjuengungVonEichtrafo}. In order to use the uniqueness of Prop.~\ref{prop:VariationVonSkalarZeugsEasyPeasy} we need to check the Leibniz rule \eqref{LeibnizForGauging}. $\delta_\varepsilon^\prime$ certainly satisfies the Leibniz rule by
\bas
\delta_\varepsilon^\prime (f ~ L)
&=
{}^*\Lambda\mleft(
	\delta_\varepsilon \mleft(
		f ~ \mleft({}^*\Lambda^{-1}\mright)(L)
	\mright)
\mright)
\\
&=
{}^*\Lambda\mleft(
	f ~ \delta_\varepsilon \biggl(
		\mleft({}^*\mleft(\Lambda^{-1}\mright)\mright)(L)
	\biggr)
	+ \mathcal{L}_{\Psi_\varepsilon}(f) ~ \mleft({}^*\Lambda^{-1}\mright)(L)
\mright)
\\
&=
f ~ \delta_\varepsilon^\prime L
	+ \mathcal{L}_{\Psi_\varepsilon} (f) ~ L
\eas
for all $L\in \mathcal{F}^\bullet_E(M; {}^*E)$ and $f \in C^\infty\bigl( M \times \mathfrak{M}_E(M;N) \bigr)$. Therefore $\delta_\varepsilon^\prime$ is of the type of operator as in Prop.~\ref{prop:VariationVonSkalarZeugsEasyPeasy}, it even uses precisely the same vector field $\Psi_\varepsilon$. So, we only need to check whether $\Psi_\varepsilon$ is the same vector field as the one behind the definition of $\widetilde{\delta}^\lambda_\varepsilon$.

For this let us use the uniqueness given in the Prop.~\ref{prop:VariationOfA}, there it was about the uniqueness of vector fields like $\Psi_\varepsilon \in \mathfrak{X}^E(\mathfrak{M}_E(M; N))$ behind the Leibniz rule. The component along the direction of the Higgs field is of course always $({}^*\rho)(\varepsilon)$ by definition. Hence, we only need to check the second component fixed by Eq.~\eqref{EichtrafoVonANochmal}. So, using Prop.~\ref{prop:VariationOfA} for $\delta_\varepsilon$,
\bas
\delta_\varepsilon^\prime \widetilde{\varpi_2}^\lambda\quad~
&\stackrel{\mathclap{ \text{Def.~\eqref{EqFieldRedefFuerA}} }}{=}\quad~
\delta_\varepsilon^\prime \mleft(
	\mleft( {}^* \Lambda \mright) (\varpi_2)+ {}^! \lambda
\mright)
\\
&\stackrel{\mathclap{ \text{Eq.~\eqref{EqPullBackFormelFuerVerschiedeneDefinitionen}} }}{=}\quad
\mleft( {}^*\Lambda \circ \delta_\varepsilon \circ {}^*\mleft(\Lambda^{-1}\mright) \mright) \bigl(
	\mleft( {}^* \Lambda \mright) (\varpi_2)+ ({}^* \lambda)(\mathrm{D})
\bigr)
\\
&=
{}^*\Lambda\biggl(
	\delta_\varepsilon \varpi_2
	+ \delta_\varepsilon \biggl( \mleft( {}^* \mleft( \Lambda^{-1} \circ \lambda \mright) \mright) (\mathrm{D}) \biggr)
\biggr)
\\
&\stackrel{\mathclap{ \text{Eq.~\eqref{DPhiVariation}} }}{=}\quad~
{}^*\Lambda\Biggl(
	- ({}^*\nabla) \varepsilon
	- \mleft( {}^* \biggl( \nabla^{\mathrm{bas}}_\varepsilon \mleft(\Lambda^{-1} \circ \lambda \mright) \biggr) \mright) (\mathrm{D})
	- \mleft( {}^* \mleft( \Lambda^{-1} \circ \lambda \mright) \mright) \bigl( ({}^*\rho)\bigl(({}^*\nabla) \varepsilon\bigr) \bigr)
\Biggr)
\\
&=
\underbrace{- \mleft({}^*\Lambda\mright)\bigl(({}^*\nabla)\varepsilon\bigr)
	- \bigl( {}^*(\lambda \circ \rho) \bigr)\bigl( ({}^*\nabla) \varepsilon \bigr)}
	_{= - ({}^*\nabla) \varepsilon}
	- \underbrace{\mleft( {}^*\biggl( \Lambda\mleft( \nabla^{\mathrm{bas}}_\varepsilon \mleft(\Lambda^{-1} \circ \lambda \mright) \mright)\biggr)\mright) (\mathrm{D})}
	_{\mathclap{ \stackrel{\text{Eq.~\eqref{EqPullBackFormelFuerVerschiedeneDefinitionen} }}{=} {}^! \mleft(\Lambda\mleft( \nabla^{\mathrm{bas}}_\varepsilon \mleft(\Lambda^{-1} \circ \lambda \mright) \mright) \mright) }}
\\
&\stackrel{\mathclap{ \text{Eq.~\eqref{OneofmanyformulasForTildeNabla}} }}{=}\quad~
- \mleft({}^*\mleft(
	\widetilde{\nabla}^\lambda 
\mright)\mright) \varepsilon
\eas
using that ${}^*\mleft( \nabla^\prime \mright) = {}^*\nabla + {}^!I$ for all other connections $\nabla^\prime = \nabla + I$, where $I \in \Omega^1(N; \mathrm{End}(E))$; this just follows by the definition of pullbacks of vector bundle connections. Hence, the vector field behind $\widetilde{\delta}^\lambda_\varepsilon$ is precisely the one of $\delta^\prime_\varepsilon$, that is, $\Psi_\varepsilon$, using the uniqueness of Prop.~\ref{prop:VariationOfA}.

Finally, we have shown everything what we need to use the uniqueness of Prop.~\ref{prop:VariationVonSkalarZeugsEasyPeasy}, hence,
\bas
\widetilde{\delta}^\lambda_\varepsilon
&=
\delta_\varepsilon^\prime.
\eas
Similarly one shows this for the one on $\mathrm{T}N$, and that $\widetilde{\delta}_\varepsilon^\lambda = \mathcal{L}_{\Psi_\varepsilon}$ on scalar-valued functionals we have already shown by observing that $\Psi_\varepsilon$ is behind the definition of $\widetilde{\delta}_\varepsilon^\lambda$; also recall Remark \ref{RemLeibnizeRegelaufProdukteWeshalbEConnectionNichtWichtigIst}.
\end{proof}

That leads to the following important statement.

\begin{theorems}{Still a gauge theory after field redefinition}{WeHaveGladlyStillAGaugeTheoryAfterTheFieldRedefinition}
Let $M$ be a spacetime with a spacetime metric $\eta$, $N$ a smooth manifold, $E \to N$ a Lie algebroid, $\nabla$ a connection on $E$, $\kappa$ and $g$ fibre metrics on $E$ and $\mathrm{T}N$, respectively. Also let $V \in C^\infty(N)$, assume that the compatibility conditions of Thm.~\ref{thm:GaugeInvariantStandardLagrangian} hold, and let $\lambda \in \Omega^1(N; E)$ such that $\Lambda = \mathds{1}_E - \lambda \circ \rho$ is an element of $\sAut(E)$. Then we have
\ba
	R_{\widetilde{\nabla}^\lambda}
&=
	- \mathrm{d}^{\mleft(\widetilde{\nabla}^\lambda\mright)^{\mathrm{bas}}} \widehat{\zeta}^\lambda, \\
	R_{\widetilde{\nabla}^\lambda}^{\mathrm{bas}} &= 0, \\
	\mleft(\widetilde{\nabla}^\lambda \mright)^{\mathrm{bas}} \widetilde{\kappa}^\lambda 
	&= 0, \\
	\mleft(\widetilde{\nabla}^\lambda \mright)^{\mathrm{bas}} \widetilde{g}^\lambda  
	&= 0, \\
	{}^*\mleft(\mathcal{L}_{({}^*\rho)(\varepsilon)} V\mright) &= 0
\ea
for all $\varepsilon \in \mathcal{F}^0_E(M; {}^*E)$. Then we have
\ba
\widetilde{\mathfrak{L}}^\lambda_{\mathrm{YMH}}
&=
\mathfrak{L}_{\mathrm{YMH}},
\ea
and
\ba
\widetilde{\delta}^\lambda_\varepsilon \widetilde{\mathfrak{L}}^\lambda_{\mathrm{YMH}}
&=
0
\ea
for all $\varepsilon \in \mathcal{F}^0_E(M; {}^*E)$, where
\ba
\widetilde{\mathfrak{L}}^\lambda_{\mathrm{YMH}}
&\coloneqq
- \frac{1}{2} \biggl( {}^*\mleft(\widetilde{\kappa}^\lambda\mright) \biggr)\mleft(\widetilde{G}^\lambda
\stackrel{\wedge}{,} *\mleft(\widetilde{G}^\lambda
 \mright)\mright)
	+ \biggl( {}^*\mleft(\widetilde{g}^\lambda\mright) \biggr)\mleft(\widetilde{\mathfrak{D}}^\lambda \stackrel{\wedge}{,} *\mleft( \widetilde{\mathfrak{D}}^\lambda \mright) \mright)
	- *({}^*V),
\ea
with
\ba\label{MaybeANewFieldStrength}
\widetilde{G}^\lambda
&\coloneqq
\widetilde{F}^\lambda 
	+ \frac{1}{2} \biggl({}^* \mleft(\widehat{\zeta}^\lambda\mright) \biggr) \mleft( \widetilde{\mathfrak{D}}^\lambda  \stackrel{\wedge}{,} \widetilde{\mathfrak{D}}^\lambda\mright)
\ea
and $\widetilde{F}^\lambda$, $\widehat{\zeta}^\lambda$ and $\widetilde{\mathfrak{D}}^\lambda$ are defined in Thm.~\ref{thm:FieldRedefofstandardFieldStrengthF}.
\end{theorems}

\begin{remark}
\leavevmode\newline
Recall our discussion about Cor.~\ref{cor:FlatnessVonEichtrafos}, where we mentioned that the vanishing basic curvature is essential.
\end{remark}

\begin{proof}[Proof of Thm.~\ref{thm:WeHaveGladlyStillAGaugeTheoryAfterTheFieldRedefinition}]
\leavevmode\newline
The first four equations we have proven by Thm.~\ref{thm:FieldRedefDerEinfacherenCompatibilities} and \ref{thm:BrokenFlatness}, for the first equation recall that the first compatibility condition in Thm.~\ref{thm:GaugeInvariantStandardLagrangian} imposes that $\nabla$ is flat, and the fifth equation is just the same compatibility condition as of Thm.~\ref{thm:GaugeInvariantStandardLagrangian}.

Using Thm.~\ref{thm:FieldRedefofstandardFieldStrengthF},
 %and \ref{thm:FieldRedefOfGaugeTrafo},
%\bas
%\widetilde{\delta}_\varepsilon^\lambda \widetilde{\mathfrak{D}}^\lambda
%&=
%\mleft({}^*\widehat{\Lambda} \circ \delta_\varepsilon \circ {}^*\mleft(\widehat{\Lambda}^{-1}\mright) \mright)
	%\mleft( \mleft( {}^* \widehat{\Lambda} \mright)\mleft(\mathfrak{D}\mright) \mright)
%=
%\mleft({}^*\widehat{\Lambda} \mright)\mleft(\delta_\varepsilon \mathfrak{D} \mright)
%\stackrel{\text{Prop.~\ref{prop:InfinitesimalGaugeTrafoOfMinimalCoupleSmiley}} }{=}
%0
%\eas
%and, additionally using the compatibility conditions, Prop.~\ref{prop:GaugeTrafosOfFieldStrengthAndMinimalCoupling} and Thm.~\ref{thm:BrokenFlatness},
%\bas
%\widetilde{\delta}_\varepsilon^\lambda \widetilde{F}^\lambda
%&=
%\mleft({}^*\Lambda \circ \delta_\varepsilon \circ {}^*\mleft(\Lambda^{-1}\mright) \mright)
%\mleft(
		%\mleft( {}^* \Lambda \mright) \mleft(
		%F
		%- \frac{1}{2} \mleft({}^* \xi \mright) \mleft( \mathfrak{D} \stackrel{\wedge}{,} \mathfrak{D} \mright)
	%\mright)
%\mright)
%\\
%&=
%\mleft( {}^*\Lambda \mright) \mleft(  
	%\frac{1}{2} \mleft({}^* \mleft( \nabla^{\mathrm{bas}}_\varepsilon \xi \mright) \mright) \mleft( \mathfrak{D} \stackrel{\wedge}{,} \mathfrak{D} \mright)
%\mright)
%\\
%&\stackrel{\mathclap{ \text{Cor.~\ref{cor:ConjugationOfDifferentialsAreShitty}} }}{=}\quad~
%\frac{1}{2} \mleft({}^* \mleft( \mleft(\widetilde{\nabla}^\lambda\mright)^{\mathrm{bas}}_\varepsilon \widehat{\zeta}^\lambda \mright) \mright) \mleft( \mathfrak{D} \stackrel{\wedge}{,} \mathfrak{D} \mright)
%\eas
%and observe
\ba\label{FieldRedefOfGWithZeroZeta}
\widetilde{G}^\lambda
&=
\widetilde{F}^\lambda 
	+ \frac{1}{2} \mleft({}^* \widehat{\zeta}^\lambda \mright) \mleft( \widetilde{\mathfrak{D}}^\lambda  \stackrel{\wedge}{,} \widetilde{\mathfrak{D}}^\lambda\mright)
=
\mleft( {}^* \Lambda \mright) \mleft(
		F
		- \frac{1}{2} \mleft({}^* \xi \mright) \mleft( \mathfrak{D} \stackrel{\wedge}{,} \mathfrak{D} 	\mright)
\mright)
	+ \frac{1}{2} \mleft({}^* \widehat{\zeta}^\lambda \mright) \mleft( \widetilde{\mathfrak{D}}^\lambda  \stackrel{\wedge}{,} \widetilde{\mathfrak{D}}^\lambda\mright)
=
\mleft( {}^* \Lambda \mright) (F),
\ea
where $\xi = \Lambda^{-1} \circ \widehat{\zeta}^\lambda \circ \mleft( \widehat{\Lambda}, \widehat{\Lambda} \mright)$. 
%Therefore, as for the minimal coupling, additionally using the compatibility conditions and Prop.~\ref{prop:GaugeTrafosOfFieldStrengthAndMinimalCoupling},
%\bas
%\widetilde{\delta}_\varepsilon^\lambda \widetilde{G}^\lambda
%=
%\mleft( {}^*\Lambda \mright)(\delta_\varepsilon F)
%=
%0.
%\eas
Thence, we immediately have by Def.~\ref{def:FieldRedefinition} and Thm.~\ref{thm:FieldRedefofstandardFieldStrengthF}
\bas
\widetilde{\mathfrak{L}}^\lambda_{\mathrm{YMH}}
&=
\mathfrak{L}_{\mathrm{YMH}},
\eas
and finally, by Thm.~\ref{thm:FieldRedefOfGaugeTrafo}, 
\bas
\widetilde{\delta}^\lambda_\varepsilon
&=
\delta_\varepsilon,
\eas
such that by Thm.~\ref{thm:GaugeInvariantStandardLagrangian}
\bas
\widetilde{\delta}^\lambda_\varepsilon \widetilde{\mathfrak{L}}^\lambda_{\mathrm{YMH}}
&=
\delta_\varepsilon \mathfrak{L}_{\mathrm{YMH}}
=
0.
\eas
\end{proof}

That theorem is a good starting point of formulating a new version of gauge theory allowing non-flat connections, especially because the physics stay the same due to the invariance of the Lagrangian under the field redefinition. Indeed, using theorems like Thm.~\ref{thm:FieldRedefOfGaugeTrafo} and \ref{thm:BrokenFlatness} we could have shown the gauge invariance of the adjusted and transformed Lagrangian similarly to Thm.~\ref{thm:GaugeInvariantStandardLagrangian}.

Let us now redefine gauge theory, using these results.

\section{Curved Yang-Mills-Higgs gauge theory}\label{SectionAboutCYMHGTs}

Let us first redefine the field strength adding the correction term in Eq.~\eqref{MaybeANewFieldStrength}.

\begin{definitions}{New field strength, \cite[Equation (14)]{CurvedYMH}}{FinallyIAmAtTheNewFieldStrength}
Let $M, N$ be smooth manifolds, $E \to N$ a Lie algebroid equipped with a connection $\nabla$ on $E$, and $\gls{1fZeta}\in \Omega^2(N;E)$, the \textbf{primitive of $\nabla$}. We define the \textbf{(generalized) field strength $\gls{G}$} as an element of $\mathcal{F}_E^2(M; {}^*E)$ by
\ba
G
&\coloneqq
F
	+ \frac{1}{2} ({}^*\zeta)\mleft( \mathfrak{D} \stackrel{\wedge}{,} \mathfrak{D} \mright).
\ea
\end{definitions}

Let us quickly state its infinitesimal gauge transformation.

\begin{corollaries}{Infinitesimal gauge transformation of the new field strength}{NewGaugeTrafoOfFieldStrengthG}
Let $M, N$ be smooth manifolds, $E \to N$ a Lie algebroid equipped with a connection $\nabla$ on $E$, and $\zeta \in \Omega^2(N;E)$. Then
\ba
\delta_\varepsilon G
&=
- \Biggl(
	\frac{1}{2} ~\biggl( 
		\mleft(	{}^* R_{\nabla} \mright)\mleft( \mathfrak{D} \stackrel{\wedge}{,} \mathfrak{D} \mright) \varepsilon
		+ \mleft({}^* \mleft( \nabla^{\mathrm{bas}}_\varepsilon\zeta \mright) \mright)\mleft( \mathfrak{D} \stackrel{\wedge}{,} \mathfrak{D} \mright)
	\biggr)
	+ \mleft({}^* R_\nabla^{\mathrm{bas}} \mright) \mleft(\varepsilon \stackrel{\wedge}{,} \varpi_2  \stackrel{\wedge}{,} \mathrm{D} \mright)
\Biggr)
\ea
for all $\varepsilon \in \mathcal{F}^0_E(M; {}^*E)$.
\end{corollaries}

\begin{remark}
\leavevmode\newline
That is a generalized version of \cite[Equation (15)]{CurvedYMH}.
\end{remark}

\begin{proof}
\leavevmode\newline
Observe, using Prop.~\ref{prop:InfinitesimalGaugeTrafoOfMinimalCoupleSmiley} and \ref{prop:VariationVonSkalarZeugsEasyPeasy},
\bas
\delta_\varepsilon\bigl(
	({}^*\zeta)\mleft( \mathfrak{D} \stackrel{\wedge}{,} \mathfrak{D} \mright)
\bigr)
&=
- \mleft({}^*\mleft( \nabla^{\mathrm{bas}}_\varepsilon \zeta\mright) \mright)\mleft( \mathfrak{D} \stackrel{\wedge}{,} \mathfrak{D} \mright),
\eas
such that the statement follows by Prop.~\ref{prop:GaugeTrafosOfFieldStrengthAndMinimalCoupling}.
\end{proof}

Now towards the Lagrangian.

\begin{definitions}{Curved Yang-Mills-Higgs Lagrangian, \newline \cite[Eq.~(2) and (16)]{CurvedYMH}}{NowReallyTheFinalLagrangian}
Let $M$ be a spacetime with a spacetime metric $\eta$, $N$ a smooth manifold, $E \to N$ a Lie algebroid, $\nabla$ a connection on $E$, $\zeta \in \Omega^2(N;E)$, and let $\kappa$ and $g$ be fibre metrics on $E$ and $\mathrm{T}N$, respectively. Also let $V \in C^\infty(N)$, which we still call the \textbf{potential of the Higgs field}. Then we define the \textbf{curved Yang-Mills-Higgs Lagrangian $\gls{LZYMH}$} as an element of $\mathcal{F}_E^{\mathrm{dim}(M)}(M)$ by
\ba
\mathfrak{L}_{\mathrm{CYMH}}
&\coloneqq
- \frac{1}{2} \mleft( {}^*\kappa \mright)\mleft(G \stackrel{\wedge}{,} *G\mright)
	+ \mleft( {}^*g \mright)\mleft(\mathfrak{D} \stackrel{\wedge}{,} *\mathfrak{D} \mright)
	- *({}^*V),
\ea
where $*$ is the Hodge star operator with respect to $\eta$.
\end{definitions}

The gauge invariance is immediate by the previous results.

\begin{theorems}{Infinitesimal gauge invariance of the curved Yang-Mills-Higgs Lagrangian}{FinallyTheGaugeInvarianceWeWant}
Let $M$ be a spacetime with a spacetime metric $\eta$, $N$ a smooth manifold, $E \to N$ a Lie algebroid, $\nabla$ a connection on $E$, $\zeta \in \Omega^2(N;E)$, $\kappa$ and $g$ fibre metrics on $E$ and $\mathrm{T}N$, respectively. Also let $V \in C^\infty(N)$ and assume that the following \textbf{compatibility conditions} hold:
\ba
	R_\nabla &= - \mathrm{d}^{\nabla^{\mathrm{bas}}} \zeta,\label{EqMyFormulationOfZetaCondition} \\
	R_\nabla^{\mathrm{bas}} &= 0, \label{VanishingBasicCurvComp} \\
	\nabla^{\mathrm{bas}} \kappa &= 0, \\
	\nabla^{\mathrm{bas}} g &= 0, \\
	{}^*\mleft(\mathcal{L}_{({}^*\rho)(\varepsilon)} V\mright) &= 0
\ea
for all $\varepsilon \in \mathcal{F}^0_E(M; {}^*E)$. Then we have
\ba
\delta_\varepsilon \mathfrak{L}_{\mathrm{CYMH}}
&=
0
\ea
for all $\varepsilon \in \mathcal{F}^0_E(M; {}^*E)$.
\end{theorems}

\begin{remarks}{}{CYMH}
We call a setup like this a \textbf{curved Yang-Mills-Higgs gauge theory}, short as \gls{CYMH}, or also \textbf{CYMH GT} for emphasizing the part with gauge theory.

We speak of that we have found a CYMH GT structure, if we were able to define $\nabla$, $\kappa$ and $g$ for $E \to N$ satisfying the first four compatibility conditions. The spacetime and the potential are not our focus and thoroughly discussed elsewhere, so, we always assume that these exist in a suitable way.
\end{remarks}

\begin{remark}
\leavevmode\newline
This is basically the essential statement of \cite[especially the discussion around Equation (16)]{CurvedYMH}, but Eq.~\eqref{EqMyFormulationOfZetaCondition} has there a different form, see \cite[Equation (13)]{CurvedYMH}. We have reformulated that equation, and this equation and the other compatibility conditions naturally arise if using the basic connection in the definition of the infinitesimal gauge transformation.

Eq.~\eqref{EqMyFormulationOfZetaCondition} means
\bas
R_\nabla(\cdot, \cdot)\nu
&=
- \nabla^{\mathrm{bas}}_\nu \zeta
\eas
for all $\nu \in \Gamma(E)$.
\end{remark}

\begin{proof}[Proof of Thm.~\ref{thm:FinallyTheGaugeInvarianceWeWant}]
\leavevmode\newline
By Eq.~\eqref{EqMyFormulationOfZetaCondition}, the vanishing of the basic curvature and Cor.~\ref{cor:NewGaugeTrafoOfFieldStrengthG} we immediately get
\bas
\delta_\varepsilon G
&=
0
\eas
for all $\varepsilon \in \mathcal{F}^0_E(M; {}^*E)$. Therefore the remaining part of the proof is precisely as in Thm.~\ref{thm:GaugeInvariantStandardLagrangian}.
\end{proof}

Finally, we now arrived at a covariantized formulation of gauge theory allowing non-flat $\nabla$. We can still apply Thm.~\ref{thm:ActionLieALgebroid}, so, a flat connection locally still applies the structure of an action Lie algebroid such that one may argue that flatness already implies a classical theory. However, $\zeta$ is not necessarily zero, it is then just constant with respect to the basic connection by compatibility condition \eqref{EqMyFormulationOfZetaCondition}; we will actually see some examples for this later. Hence, one cannot expect that the field strength looks as in the classical formulation if $\nabla$ is flat, and, so, we can only apply Thm.~\ref{thm:StandardEichtheorieStecktInDenBedingung} if both, $R_\nabla$ and $\zeta$ vanish. This motivates the following definitions.

\begin{definitions}{Classical gauge theory}{ClassicalGT}
Let us assume the same structure as in Thm.~\ref{thm:FinallyTheGaugeInvarianceWeWant}.
Then we say that we have a \textbf{pre-classical gauge theory}, if $\nabla$ is flat.

If we have additionally $\zeta = 0$, then we say that we have a \textbf{classical gauge theory}.
\end{definitions}

\begin{remark}
\leavevmode\newline
If we have a classical CYMH GT, then also a pre-classical one by compatibility condition \ref{EqMyFormulationOfZetaCondition}.
\end{remark}

However, we motivated $\zeta$ by the field redefinition; there might be of course a field redefinition making $\nabla$ flat and/or $\zeta$ zero. This is what we mainly study in the remaining part of this thesis. We have seen that we needed to add the part with $\zeta$ to the classical field strength $F$ after the field redefinition in order to get the same Lagrangian. That can be seen as that the "actual field redefinition" of $F$ was not just given by the field redefinition of $\varpi_2$ and $\nabla$; or, in other words, that means we need a field redefinition of $\zeta$, too, while $\zeta$ was zero in Thm.~\ref{thm:WeHaveGladlyStillAGaugeTheoryAfterTheFieldRedefinition} and $\widehat{\zeta}^\lambda$ was the field redefinition of $\zeta \equiv 0$.

\begin{definitions}{Field redefinition of the primitive}{FieldRedefinitionOfThePrimitive}
Let $E \to N$ a Lie algebroid over a smooth manifold $N$, $\nabla$ a connection on $E$, $\zeta \in \Omega^2(N;E)$, and $\lambda \in \Omega^1(N;E)$ such that $\Lambda = \mathds{1} - \lambda \circ \rho \in \sAut(E)$. Then we define the \textbf{field redefinition $\gls{1fZetaTilde}$ of $\zeta$} by
\ba
\widetilde{\zeta}^\lambda
&\coloneqq
\Lambda \circ \zeta \circ \mleft( \widehat{\Lambda}^{-1}, \widehat{\Lambda}^{-1} \mright)
	+ \widehat{\zeta}^\lambda,
\ea
where $\widehat{\zeta}^\lambda$ is given as in Thm.~\ref{thm:FieldRedefofstandardFieldStrengthF}, that is,
\bas
\widetilde{\zeta}^\lambda\mleft(\widehat{\Lambda}(X),\widehat{\Lambda}(Y)\mright)
&=
\Lambda\bigl(
	\zeta\mleft( X, Y \mright)
\bigr)
	- \mleft(\mathrm{d}^{\widetilde{\nabla}^\lambda} \lambda\mright)(X,Y)
	+ t_{\widetilde{\nabla}^\lambda_\rho}(\lambda(X), \lambda(Y))
\\
&=
\Lambda\bigl(
	\zeta\mleft( X, Y \mright)
\bigr)
	- \mleft( \mathrm{d}^\nabla \lambda \mright)(X,Y)
%}_{\mathclap{= \nabla_X \bigl( \lambda(Y) \bigr) - \nabla_Y \bigl( \lambda(X) \bigr) - \lambda\bigl( [X,Y] \bigr)}}
	- \lambda\Bigl(
		\nabla^{\mathrm{bas}}_{\lambda(X)} Y
		- \nabla^{\mathrm{bas}}_{\lambda(Y)} X
		%}_{= [(\rho\circ\lambda)(Y), X] + \rho\mleft( \nabla_X ( \lambda(Y) ) \mright)}
	\Bigr)
	+ \mleft[ \lambda(X), \lambda(Y) \mright]_E
\eas
for all $X, Y \in \mathfrak{X}(N)$.
\end{definitions}

\begin{remarks}{Field redefinition of CYMH GTs}{FieldRedefOfFullCYMHGT}
The field redefinition is therefore given by using Def.~\ref{def:FieldRedefinition} and \ref{def:FieldRedefinitionOfThePrimitive} altogether, so, when we speak of the field redefinition of anything else besides the quantities in these definitions, then it is just canonically given; for example the field redefinition of something depending on $\zeta$ is then the same definition but replacing $\zeta$ with $\widetilde{\zeta}^\lambda$; similarly for dependencies on $\nabla$, $\varpi_2$ and the metrics $\kappa$ on $E$ and $g$ on $\mathrm{T}N$ as we already did before. We call this procedure the \textbf{field redefinition of a CYMH GT} on a given spacetime $M$, a smooth manifold $N$ and Lie algebroid $E \to N$. We are going to show that the Lagrangian stays invariant under the field redefinition and that this describes an equivalence relation of CYMH GTs on given $M, N$ and $E$.
\end{remarks}

For the invariance of the Lagrangian we do not need to prove everything again, we just need to check the field redefinition of the field strength $G$ and whether compatibility condition \eqref{EqMyFormulationOfZetaCondition} stays form-invariant.

\begin{lemmata}{Field redefinition of the new field strength and compatibility condition}{FinallyNiceTrafoOfEverything}
Let $M, N$ be smooth manifolds, $E \to N$ a Lie algebroid, $\nabla$ a connection on $E$, $\zeta \in \Omega^2(N;E)$, and $\lambda \in \Omega^1(N;E)$ such that $\Lambda = \mathds{1} - \lambda \circ \rho \in \sAut(E)$. Then we have
\ba
\widetilde{G}^\lambda
&=
\mleft( {}^*\Lambda \mright)(G),
\ea
where
\ba
\widetilde{G}^\lambda
&\coloneqq
\widetilde{F}^\lambda
	+ \frac{1}{2} \biggl({}^*\mleft(\widetilde{\zeta}^\lambda\mright)\biggr)\mleft( \widetilde{\mathfrak{D}}^\lambda \stackrel{\wedge}{,} \widetilde{\mathfrak{D}}^\lambda \mright),
\ea
for which $\widetilde{F}^\lambda$ and $\widetilde{\mathfrak{D}}^\lambda$ are given by Thm.~\ref{thm:FieldRedefofstandardFieldStrengthF}.

If the basic curvature of $\nabla$ vanishes additionally and satisfies $R_\nabla = - \mathrm{d}^{\nabla^{\mathrm{bas}}} \zeta$, then we have
\ba
R_{\widetilde{\nabla}^\lambda}
&= 
- \mathrm{d}^{\mleft(\widetilde{\nabla}^\lambda\mright)^{\mathrm{bas}}} \widetilde{\zeta}^\lambda.
\ea
\end{lemmata}

\begin{proof}
\leavevmode\newline
Those results are an immediate consequence of our calculations in the previous section, that is,
\bas
\widetilde{G}^\lambda
&=
\underbrace{\widetilde{F}^\lambda
	+ \frac{1}{2} \biggl({}^*\mleft(\widehat{\zeta}^\lambda\mright)\biggr)\mleft( \widetilde{\mathfrak{D}}^\lambda \stackrel{\wedge}{,} \widetilde{\mathfrak{D}}^\lambda \mright)}
	_{\stackrel{ \text{Eq.~\eqref{FieldRedefOfGWithZeroZeta}} }{=} ({}^*\Lambda)(F)}
	+ \frac{1}{2} \biggl({}^*\mleft( \Lambda \circ \zeta \circ \mleft( \widehat{\Lambda}^{-1}, \widehat{\Lambda}^{-1} \mright) \mright)\biggr)\mleft( \widetilde{\mathfrak{D}}^\lambda \stackrel{\wedge}{,} \widetilde{\mathfrak{D}}^\lambda \mright)
\\
&\stackrel{\mathclap{ \text{Thm.~\ref{thm:FieldRedefofstandardFieldStrengthF}} }}{=}\quad~
({}^*\Lambda)(F)
	+ \frac{1}{2} \bigl(
		{}^*\mleft( \Lambda \circ \zeta \mright)
	\bigr)\mleft( \mathfrak{D} \stackrel{\wedge}{,} \mathfrak{D} \mright)
\\
&=
({}^*\Lambda)\mleft(
	F
	+ \frac{1}{2} \mleft( {}^*\zeta \mright)\mleft( \mathfrak{D} \stackrel{\wedge}{,} \mathfrak{D} \mright)
\mright)
\\
&=
({}^*\Lambda)(G),
\eas
and, by Thm.~\ref{thm:BrokenFlatness} (for which we need the vanishing of the basic curvature), Prop.~\ref{prop:PropsOfBigLambdas} and compatibility condition \eqref{EqMyFormulationOfZetaCondition},
\bas
R_{\widetilde{\nabla}^\lambda}
&=
\Lambda \circ R_\nabla \circ \mleft( \widehat{\Lambda}^{-1}, \widehat{\Lambda}^{-1} \mright)
	- \mathrm{d}^{\mleft(\widetilde{\nabla}^\lambda\mright)^{\mathrm{bas}}} \widehat{\zeta}^\lambda
\\
&=
- \Lambda \circ \mathrm{d}^{\nabla^{\mathrm{bas}}} \zeta \circ \mleft( \widehat{\Lambda}^{-1}, \widehat{\Lambda}^{-1} \mright)
	- \mathrm{d}^{\mleft(\widetilde{\nabla}^\lambda\mright)^{\mathrm{bas}}} \widehat{\zeta}^\lambda
\\
&\stackrel{\mathclap{ \text{Cor.~\ref{cor:ConjugationOfDifferentialsAreShitty}} }}{=}\quad~
- \mathrm{d}^{\mleft( \widetilde{\nabla}^\lambda \mright)^{\mathrm{bas}}} \mleft(
	\Lambda \circ \zeta \circ \mleft( \widehat{\Lambda}^{-1}, \widehat{\Lambda}^{-1} \mright)
\mright)
	- \mathrm{d}^{\mleft(\widetilde{\nabla}^\lambda\mright)^{\mathrm{bas}}} \widehat{\zeta}^\lambda
\\
&=
- \mathrm{d}^{\mleft( \widetilde{\nabla}^\lambda \mright)^{\mathrm{bas}}} \mleft(
	\widetilde{\zeta}^\lambda
\mright).
\eas
\end{proof}

Hence, we immediately get:

\begin{theorems}{Gauge theory invariant under the field redefinition}{InvarianceUnderTheFieldRedefinition}
Let $M$ be a spacetime with a spacetime metric $\eta$, $N$ a smooth manifold, $E \to N$ a Lie algebroid, $\nabla$ a connection on $E$, $\zeta \in \Omega^2(N;E)$, $\kappa$ and $g$ fibre metrics on $E$ and $\mathrm{T}N$, respectively. Also let $V \in C^\infty(N)$, assume that the compatibility conditions of Thm.~\ref{thm:FinallyTheGaugeInvarianceWeWant} hold, and let $\lambda \in \Omega^1(N; E)$ such that $\Lambda = \mathds{1}_E - \lambda \circ \rho$ is an element of $\sAut(E)$. Then we have
\ba
	R_{\widetilde{\nabla}^\lambda}
&=
	- \mathrm{d}^{\mleft(\widetilde{\nabla}^\lambda\mright)^{\mathrm{bas}}} \widetilde{\zeta}^\lambda, \\
	R_{\widetilde{\nabla}^\lambda}^{\mathrm{bas}} &= 0, \\
	\mleft(\widetilde{\nabla}^\lambda \mright)^{\mathrm{bas}} \widetilde{\kappa}^\lambda 
	&= 0, \\
	\mleft(\widetilde{\nabla}^\lambda \mright)^{\mathrm{bas}} \widetilde{g}^\lambda  
	&= 0, \\
	{}^*\mleft(\mathcal{L}_{({}^*\rho)(\varepsilon)} V\mright) &= 0
\ea
for all $\varepsilon \in \mathcal{F}^0_E(M; {}^*E)$. Then we have
\ba
\widetilde{\mathfrak{L}}^\lambda_{\mathrm{CYMH}}
&=
\mathfrak{L}_{\mathrm{CYMH}},
\ea
and
\ba
\widetilde{\delta}^\lambda_\varepsilon \widetilde{\mathfrak{L}}^\lambda_{\mathrm{CYMH}}
&=
0
\ea
for all $\varepsilon \in \mathcal{F}^0_E(M; {}^*E)$, where
\ba
\widetilde{\mathfrak{L}}^\lambda_{\mathrm{CYMH}}
&\coloneqq
- \frac{1}{2} \biggl( {}^*\mleft(\widetilde{\kappa}^\lambda\mright) \biggr)\mleft(\widetilde{G}^\lambda
\stackrel{\wedge}{,} *\mleft(\widetilde{G}^\lambda
 \mright)\mright)
	+ \biggl( {}^*\mleft(\widetilde{g}^\lambda\mright) \biggr)\mleft(\widetilde{\mathfrak{D}}^\lambda \stackrel{\wedge}{,} *\mleft( \widetilde{\mathfrak{D}}^\lambda \mright) \mright)
	- *({}^*V),
\ea
and where $\widetilde{G}^\lambda$ is given as in Lemma \ref{lem:FinallyNiceTrafoOfEverything}, $\widetilde{\mathfrak{D}}^\lambda$ is defined as in Thm.~\ref{thm:FieldRedefofstandardFieldStrengthF} and $\widetilde{\delta}^\lambda_\varepsilon$ as in Thm~\ref{thm:FieldRedefOfGaugeTrafo}.
\end{theorems}

\begin{remark}
\leavevmode\newline
It is important to note for future proofs that the field redefinition already preserves the vanishing of the basic curvature if $\nabla$ has vanishing basic curvature, so, this is independent to whether or not the other compatibility conditions are satisfied. Similar for the metric compatibilities. However, for the invariance of compatibility condition \eqref{EqMyFormulationOfZetaCondition} one not only needs the condition itself but also additionally the vanishing of the basic curvature as stated in Lemma \ref{lem:FinallyNiceTrafoOfEverything}. We sometimes make use of this information when speaking about compatibility conditions in the context of the field redefinition. However, we will not necessarily mention it again; recall the previous calculations and proofs.
\end{remark}

\begin{proof}
\leavevmode\newline
This is precisely the same proof as in Thm.~\ref{thm:WeHaveGladlyStillAGaugeTheoryAfterTheFieldRedefinition}, using Lemma \ref{lem:FinallyNiceTrafoOfEverything} and $\widetilde{\zeta}^\lambda$ instead of just $\widehat{\zeta}^\lambda$.
\end{proof}

\begin{remarks}{Avoidance of the calculation in the proof of Thm.~\ref{thm:BrokenFlatness}}{HaesslicherBeweisUnwichtig}
As we have seen in the proofs for Thm.~\ref{thm:InvarianceUnderTheFieldRedefinition} and \ref{thm:WeHaveGladlyStillAGaugeTheoryAfterTheFieldRedefinition} we only needed Thm.~\ref{thm:BrokenFlatness} for the proof about the relationship of $R_\nabla$ with $\zeta$ after the field redefinition, everything else follows independent of Thm.~\ref{thm:BrokenFlatness}, especially the other compatibility conditions and the gauge invariance of the Lagrangian. Hence, one may want to argue, given the gauge invariance of the Lagrangian and the other compatibility conditions after the field redefinition, that the gauge transformation of the transformed field strength has to vanish, using similar calculations. By Cor.~\ref{cor:NewGaugeTrafoOfFieldStrengthG} one may then be able to argue in general that the compatibility condition of $\zeta$ has to be preserved by the field redefinition. However, for this one needs to discuss certain edge cases and that the contraction with $\kappa$ can be ignored (to avoid an argument about orthogonality). If one is able to argue like this, then one can avoid the tedious calculation behind the proof of Thm.~\ref{thm:BrokenFlatness}.
\end{remarks}

Therefore the field redefinition is now a transformation of the curved Yang-Mills-Higgs (infinitesimal) gauge theory which keeps the Lagrangian invariant. Furthermore, the field redefinition is an equivalence of CYMH GTs, which we now prove. We start with something similar to Lemma \ref{lem:FieldRedefinitionIsInvertible} but for the primitive.

\begin{lemmata}{Invertible behaviour of the field redefinition of the primitive}{InverseOfZetaLambda}
Let $E \to N$ a Lie algebroid over a smooth manifold $N$, $\nabla$ a connection on $E$, $\zeta \in \Omega^2(N;E)$, and $\lambda \in \Omega^1(N;E)$ such that $\Lambda = \mathds{1}_E - \lambda \circ \rho \in \sAut(E)$. Then
\ba
\overline{\zeta}^{-\lambda}
&=
\zeta,
\ea
where
\bas
\overline{\zeta}^{-\lambda}
&\coloneqq
\widetilde{ \widetilde{\zeta}^\lambda }^{- \Lambda^{-1} \circ \lambda}.
\eas
\end{lemmata}

\begin{proof}
\leavevmode\newline
That is similar to the proof of Lemma \ref{lem:FieldRedefinitionIsInvertible}, hence, let us summarize what we have derived there,
\bas
\mathfrak{\Lambda}
&\coloneqq
\mathds{1}_E
	- \mleft( - \Lambda^{-1} \circ \lambda \mright) \circ \rho
=
\Lambda^{-1},
\\
\widehat{\mathfrak{\Lambda}}
&\coloneqq
\mathds{1}_{\mathrm{T}N}
	- \rho \circ \mleft( - \Lambda^{-1} \circ \lambda \mright)
=
\widehat{\Lambda}^{-1},
\eas
those are invertible, thus, we can apply the field redefinition using $-\Lambda^{-1} \circ \lambda$. Then by Def.~\ref{def:FieldRedefinitionOfThePrimitive}, especially also recall Def.~\eqref{FormulaForZetaTildeWithZetaEqualzero},
\bas
\overline{\zeta}^{-\lambda}
&=
\mathfrak{\Lambda} \circ \widetilde{\zeta}^\lambda \circ \mleft( \widehat{\mathfrak{\Lambda}}^{-1}, \widehat{\mathfrak{\Lambda}}^{-1} \mright)
	+ \widehat{\widetilde{\zeta}^\lambda}^{-\Lambda^{-1} \circ \lambda},
\eas
where, recalling Eq.~\eqref{AndereFormelFuerNablaTrafoBesserFuerDasRechnen},
\bas
\mathfrak{\Lambda} \circ \widetilde{\zeta}^\lambda \circ \mleft( \widehat{\mathfrak{\Lambda}}^{-1}, \widehat{\mathfrak{\Lambda}}^{-1} \mright)
&=
\zeta
	+ \Lambda^{-1} \circ \widehat{\zeta}^\lambda \circ \mleft( \widehat{\Lambda}, \widehat{\Lambda} \mright)
\\
&=
\zeta
	- \Lambda^{-1} \circ \mleft(
		\mathrm{d}^{\widetilde{\nabla}^\lambda} \lambda
		- t_{\widetilde{\nabla}^\lambda_\rho} \circ (\lambda, \lambda)
		\mright)
\\
&\stackrel{\mathclap{ \eqref{eqDifferentialSplit} }}{=}~ 
\zeta
	- \Lambda^{-1} \circ \Bigl(
		\underbrace{\mathrm{d}^{\Lambda\circ\nabla\circ \Lambda^{-1}} \lambda}
		_{\mathclap{ = \mleft( \Lambda \circ \mathrm{d}^\nabla \circ \Lambda^{-1} \mright) \lambda }}
		+	D \wedge \lambda
		- t_{\widetilde{\nabla}^\lambda_\rho} \circ (\lambda, \lambda)
	\Bigr)
\\
&=
\zeta
	- \mathrm{d}^\nabla \mleft( \Lambda^{-1} \circ \lambda \mright)
\\
&\hspace{1cm}
	+ \mathrm{d}^\nabla \mleft( \Lambda^{-1} \circ \lambda \mright) \circ \mleft( \mathds{1}_{\mathrm{T}N}, \rho \circ \lambda \mright)
	+ \mathrm{d}^\nabla \mleft( \Lambda^{-1} \circ \lambda \mright) \circ \mleft( \rho \circ \lambda, \mathds{1}_{\mathrm{T}N} \mright)
\\
&\hspace{1cm}
	- t_{\nabla_\rho} \circ \mleft( \Lambda^{-1} \circ \lambda, \lambda \mright)
	- t_{\nabla_\rho} \circ \mleft( \lambda, \Lambda^{-1} \circ \lambda \mright)
	+ \Lambda^{-1} \circ t_{\widetilde{\nabla}^\lambda_\rho} \circ (\lambda, \lambda)
\eas
viewing $D \coloneqq - \mleft( \mathrm{d}^{\Lambda\circ\nabla\circ \Lambda^{-1}} \lambda \mright) \circ \mleft( \mathds{1}_{\mathrm{T}N}, \rho \mright) + \Lambda \circ t_{\nabla_\rho} \circ \mleft( \Lambda^{-1} \circ \lambda, \mathds{1}_E \mright)$ as an element of $\Omega^1(N;\mathrm{End}(E))$,
and, using Prop.~\ref{prop:PropsOfBigLambdas},
\bas
&\Bigl(- t_{\nabla_\rho} \circ \mleft( \Lambda^{-1} \circ \lambda, \lambda \mright)
	- t_{\nabla_\rho} \circ \mleft( \lambda, \Lambda^{-1} \circ \lambda \mright)
	+ \Lambda^{-1} \circ \underbrace{t_{\widetilde{\nabla}^\lambda_\rho} \circ (\lambda, \lambda)}
	_{\mathclap{ = - t_{\mleft(\widetilde{\nabla}^\lambda\mright)^{\mathrm{bas}}} \circ (\lambda, \lambda) }}
\Bigr)(X, Y)
\\
&\hspace{1cm}=
- \nabla_{(\rho \circ \lambda)(X)} \mleft( \mleft(\Lambda^{-1} \circ \lambda\mright)(Y) \mright)
	+ \nabla_{\mleft(\rho \circ \Lambda^{-1} \circ \lambda\mright)(Y)} \bigl( \lambda(X) \bigr)
	+ \mleft[ \lambda(X), \mleft(\Lambda^{-1} \circ \lambda\mright)(Y) \mright]_E
\\
&\hspace{2cm}
	+ \nabla_{(\rho \circ \lambda)(Y)} \mleft( \mleft(\Lambda^{-1} \circ \lambda\mright)(X) \mright)
	- \nabla_{\mleft(\rho \circ \Lambda^{-1} \circ \lambda\mright)(X)} \bigl( \lambda(Y) \bigr)
	+ \mleft[ \mleft(\Lambda^{-1} \circ \lambda\mright)(X), \lambda(Y) \mright]_E
\\
&\hspace{2cm}
	- \nabla^{\mathrm{bas}}_{\lambda(X)} \mleft( \mleft( \Lambda^{-1} \circ \lambda \mright) (Y) \mright)
	+ \nabla^{\mathrm{bas}}_{\lambda(Y)} \mleft( \mleft( \Lambda^{-1} \circ \lambda \mright) (Y) \mright)
	+ \Lambda^{-1} \mleft( \mleft[ \lambda(X), \lambda(Y) \mright]_E \mright)
\\
&\hspace{1cm}=
- \nabla_{(\rho \circ \lambda)(X)} \mleft( \mleft(\Lambda^{-1} \circ \lambda\mright)(Y) \mright)
	+ \nabla_{\mleft(\rho \circ \Lambda^{-1} \circ \lambda\mright)(Y)} \bigl( \lambda(X) \bigr)
	+ \mleft[ \lambda(X), \mleft(\Lambda^{-1} \circ \lambda\mright)(Y) \mright]_E
\\
&\hspace{2cm}
	+ \nabla_{(\rho \circ \lambda)(Y)} \mleft( \mleft(\Lambda^{-1} \circ \lambda\mright)(X) \mright)
	- \nabla_{\mleft(\rho \circ \Lambda^{-1} \circ \lambda\mright)(X)} \bigl( \lambda(Y) \bigr)
	+ \mleft[ \mleft(\Lambda^{-1} \circ \lambda\mright)(X), \lambda(Y) \mright]_E
\\
&\hspace{2cm}
	- \mleft[ \lambda(X), \mleft( \Lambda^{-1} \circ \lambda \mright) (Y) \mright]_E
	- \nabla_{\mleft(\rho \circ \Lambda^{-1} \circ \lambda \mright) (Y)} \bigl( \lambda(X) \bigr)
\\
&\hspace{2cm}
	+ \mleft[ \lambda(Y), \mleft( \Lambda^{-1} \circ \lambda \mright) (X) \mright]_E
	+ \nabla_{\mleft(\rho \circ \Lambda^{-1} \circ \lambda \mright) (X)} \bigl( \lambda(Y) \bigr)
\\
&\hspace{2cm}
	+ \Lambda^{-1} \mleft( \mleft[ \lambda(X), \lambda(Y) \mright]_E \mright)
\\
&\hspace{1cm}=
- \nabla_{(\rho \circ \lambda)(X)} \mleft( \mleft(\Lambda^{-1} \circ \lambda\mright)(Y) \mright)
	+ \nabla_{(\rho \circ \lambda)(Y)} \mleft( \mleft(\Lambda^{-1} \circ \lambda\mright)(X) \mright)
	+ \Lambda^{-1} \mleft( \mleft[ \lambda(X), \lambda(Y) \mright]_E \mright)
\eas
for all $X, Y \in \mathfrak{X}(N)$,
and, using additionally Lemma \ref{lem:FieldRedefinitionIsInvertible},
\bas
\widehat{\widetilde{\zeta}^\lambda}^{-\Lambda^{-1} \circ \lambda}
&\coloneqq
\mleft(\mathrm{d}^{\widehat{\nabla}^{-\lambda}} \mleft( \Lambda^{-1}\circ \lambda\mright)
	+ t_{\widehat{\nabla}^{-\lambda}_\rho} \circ (\Lambda^{-1}\circ \lambda, \Lambda^{-1}\circ \lambda)\mright)
	\circ \mleft(\widehat{\mathfrak{\Lambda}}^{-1}, \widehat{\mathfrak{\Lambda}}^{-1}\mright)
\\
&=
\mleft(\mathrm{d}^{\nabla} \mleft( \Lambda^{-1}\circ \lambda\mright)
	+ t_{\nabla_\rho} \circ \mleft(\Lambda^{-1}\circ \lambda, \Lambda^{-1}\circ \lambda\mright)\mright)
	\circ \mleft(\widehat{\Lambda}, \widehat{\Lambda}\mright)
\\
&=
\mathrm{d}^{\nabla} \mleft( \Lambda^{-1}\circ \lambda\mright) \circ \mleft(\widehat{\Lambda}, \widehat{\Lambda}\mright)
	+ t_{\nabla_\rho} \circ \mleft(\lambda, \lambda\mright)
\eas
Therefore altogether, using $\widehat{\Lambda} = \mathds{1}_{\mathrm{T}N} - \rho \circ \lambda$ and again Prop.~\ref{prop:PropsOfBigLambdas},
\bas
\overline{\zeta}^{-\lambda}(X, Y)
&=
\zeta(X, Y)
	+ \mathrm{d}^{\nabla} \mleft( \Lambda^{-1}\circ \lambda\mright) \bigl((\rho \circ \lambda)(X), (\rho\circ\lambda)(Y)\bigr)
	+ t_{\nabla_\rho} \bigl(\lambda(X), \lambda(Y)\bigr)
\\
&\hspace{1cm}
	- \nabla_{(\rho \circ \lambda)(X)} \mleft( \mleft(\Lambda^{-1} \circ \lambda\mright)(Y) \mright)
	+ \nabla_{(\rho \circ \lambda)(Y)} \mleft( \mleft(\Lambda^{-1} \circ \lambda\mright)(X) \mright)
	+ \Lambda^{-1} \mleft( \mleft[ \lambda(X), \lambda(Y) \mright]_E \mright)
\\
&=
\zeta(X, Y)
	+ \mathrm{d}^{\nabla} \mleft( \Lambda^{-1}\circ \lambda\mright) \bigl((\rho \circ \lambda)(X), (\rho\circ\lambda)(Y)\bigr)
\\
&\hspace{1cm}
	+ \nabla_{(\rho \circ \lambda)(X)}\bigl( \lambda(Y) \bigr)
	- \nabla_{(\rho \circ \lambda)(Y)}\bigl( \lambda(X) \bigr)
	- \mleft[ \lambda(X), \lambda(Y) \mright]_E
\\
&\hspace{1cm}
	- \nabla_{(\rho \circ \lambda)(X)} \mleft( \mleft(\Lambda^{-1} \circ \lambda\mright)(Y) \mright)
	+ \nabla_{(\rho \circ \lambda)(Y)} \mleft( \mleft(\Lambda^{-1} \circ \lambda\mright)(X) \mright)
	+ \Lambda^{-1} \mleft( \mleft[ \lambda(X), \lambda(Y) \mright]_E \mright)
\\
&=
\zeta(X, Y)
	+ \mathrm{d}^{\nabla} \mleft( \Lambda^{-1}\circ \lambda\mright) \bigl((\rho \circ \lambda)(X), (\rho\circ\lambda)(Y)\bigr)
\\
&\hspace{1cm}
	- \nabla_{(\rho \circ \lambda)(X)} \mleft( \mleft(\Lambda^{-1} \circ \lambda \circ \rho \circ \lambda\mright)(Y) \mright)
	+ \nabla_{(\rho \circ \lambda)(Y)} \mleft( \mleft(\Lambda^{-1} \circ \lambda \circ \rho \circ \lambda\mright)(X) \mright)
\\
&\hspace{1cm}
	+ \underbrace{\mleft(\Lambda^{-1} \circ \lambda \circ \rho \mright) \mleft( \mleft[ \lambda(X), \lambda(Y) \mright]_E \mright)}
	_{\mathclap{ = \mleft(\Lambda^{-1} \circ \lambda\mright) \mleft( \mleft[ (\rho\circ\lambda)(X), (\rho\circ\lambda)(Y) \mright]_E \mright) }}
\\
&=
\zeta(X, Y)
	+ \mathrm{d}^{\nabla} \mleft( \Lambda^{-1}\circ \lambda\mright) \bigl((\rho \circ \lambda)(X), (\rho\circ\lambda)(Y)\bigr)
\\
&\hspace{1cm}
	- \mathrm{d}^{\nabla} \mleft( \Lambda^{-1}\circ \lambda\mright) \bigl((\rho \circ \lambda)(X), (\rho\circ\lambda)(Y)\bigr)
\\
&=
\zeta(X, Y).
\eas
\end{proof}

The field redefinition, Def.~\ref{def:FieldRedefinition} and \ref{def:FieldRedefinitionOfThePrimitive}, is also transitive.

\begin{lemmata}{Transitivity of the field redefinition}{TransFieldRedef}
Let $M, N$ be smooth manifolds, $E \to N$ a Lie algebroid, $\nabla$ a connection on $E$, $\zeta \in \Omega^2(N;E)$, $\kappa$ and $g$ fibre metrics on $E$ and $\mathrm{T}N$, respectively. Moreover, let $\lambda, \lambda^\prime \in \Omega^1(N;E)$ such that $\Lambda = \mathds{1}_E - \lambda \circ \rho, \Lambda^\prime \coloneqq \mathds{1}_E - \lambda^\prime \circ \rho \in \sAut(E)$.

Then the field redefinition with $\lambda^\prime$ composed with the field redefinition of $\lambda$ is equivalent to a field redefinition with $\lambda + \lambda^\prime - \lambda^\prime \circ \rho \circ \lambda$.
\end{lemmata}

\begin{remark}
\leavevmode\newline
With this one can also quickly show Lemma \ref{lem:FieldRedefinitionIsInvertible} and \ref{lem:InverseOfZetaLambda} by defining $\lambda^\prime \coloneqq - \Lambda^{-1} \circ \lambda$ such that
\bas
\lambda + \lambda^\prime - \lambda^\prime \circ \rho \circ \lambda
&=
\lambda \underbrace{- \Lambda^{-1} \circ \lambda + \Lambda^{-1} \circ \lambda \circ \rho \circ \lambda}
_{= - \Lambda^{-1} \circ \Lambda \circ \lambda}
=
0,
\eas
which gives trivial transformations.
\end{remark}

\begin{proof}[Proof of Lemma \ref{lem:TransFieldRedef}]
\leavevmode\newline
First observe that
\bas
\Lambda^\prime \circ \Lambda
&=
\mleft( \mathds{1}_E - \lambda^\prime \circ \rho \mright)
\circ \mleft( \mathds{1}_E - \lambda \circ \rho \mright)
\\
&=
\mathds{1}_E 
	- \lambda \circ \rho 
	- \lambda^\prime \circ \rho 
	+ \lambda^\prime \circ \rho \circ \lambda \circ \rho
\\
&=
\mathds{1}_E 
	- \mleft( \lambda 
		+ \lambda^\prime 
		- \lambda^\prime \circ \rho \circ \lambda \mright) \circ \rho
\\
&\eqqcolon
\mathfrak{\Lambda}
\eas
so, $\lambda + \lambda^\prime - \lambda^\prime \circ \rho \circ \lambda$ is a valid element of $\Omega^1(N;E)$ with which one can apply the field redefinition due to the fact that $\Lambda^\prime \circ \Lambda \in \sAut(E)$, thence, also $\mathfrak{\Lambda}\in \sAut(E)$; we also define and calculate similarly
\bas
\widehat{\mathfrak{\Lambda}} 
&\coloneqq 
\widehat{\Lambda}^\prime \circ \widehat{\Lambda}
= 
\mathds{1}_{\mathrm{T}N} 
	- \rho \circ (\lambda + \lambda^\prime - \lambda^\prime \circ \rho \circ \lambda)
\eas
which is an element of $\sAut(\mathrm{T}N)$ (similarly to why $\widehat{\Lambda}$ is), where we denote $\widehat{\Lambda}^\prime \coloneqq \mathds{1}_{\mathrm{T}N} - \rho \circ \lambda^\prime$.
By Remark \ref{rem:FieldRedefOfFullCYMHGT} we only need to check the basic field redefinition of Def.~\ref{def:FieldRedefinition} and \ref{def:FieldRedefinitionOfThePrimitive}, so,
\bas
\widetilde{\widetilde{\varpi_2}^\lambda}^{\lambda^\prime}
&=
\mleft( {}^* \Lambda^\prime \mright)\Bigl(\mleft( {}^* \Lambda \mright) (\varpi_2)+ \underbrace{{}^! \lambda}_{\mathclap{ \stackrel{\eqref{EqPullBackFormelFuerVerschiedeneDefinitionen}}{=} ~ ({}^*\lambda)(\mathrm{D}) }}\Bigr)
	+ {}^!\lambda^\prime
\\
&=
\mleft( {}^*\Lambda^\prime \circ {}^* \Lambda \mright) (\varpi_2)
	+ \underbrace{\bigl({}^*\mleft(\Lambda^\prime \circ \lambda\mright)\bigr)(\mathrm{D})}
	_{= {}^!\mleft(\Lambda \circ \lambda\mright)}
	+ {}^!\lambda^\prime
\\
&=
({}^*\mathfrak{\Lambda})(\varpi_2)
	+ {}^!\mleft( \lambda + \lambda^\prime - \lambda^\prime \circ \rho \circ \lambda \mright).
\eas
For the metrics we immediately have
\bas
\widetilde{\widetilde{\kappa}^\lambda}^{\lambda^\prime}
&=
\kappa \circ \mleft( \Lambda^{-1}, \Lambda^{-1} \mright) \circ \mleft( (\Lambda^\prime)^{-1}, (\Lambda^\prime)^{-1} \mright)
=
\kappa \circ \mleft( \mathfrak{\Lambda}^{-1}, \mathfrak{\Lambda}^{-1} \mright),
\eas
similarly for $g$. Recall again Prop.~\ref{prop:PropsOfBigLambdas} and Cor.~\ref{cor:ConjugationOfDifferentialsAreShitty}, then
\bas
\widetilde{\widetilde{\nabla}^\lambda}^{\lambda^\prime}
&=
\widetilde{\nabla}^\lambda
	+ \mleft( \Lambda^\prime \circ \mathrm{d}^{\mleft(\widetilde{\nabla}^\lambda\mright)^{\mathrm{bas}}} \circ (\Lambda^\prime)^{-1} \mright) \lambda^\prime
\\
&=
\nabla
	+ \underbrace{\mleft( \Lambda \circ \mathrm{d}^{\nabla^{\mathrm{bas}}} \circ \Lambda^{-1} \mright) \lambda}
	_{\mathclap{ = \mleft( \Lambda^\prime \mright)^{-1} \circ \mleft(\mleft( \mathfrak{\Lambda} \circ \mathrm{d}^{\nabla^{\mathrm{bas}}} \circ \mathfrak{\Lambda}^{-1} \mright)\mleft( \Lambda^\prime \circ \lambda \mright)\mright) }}
	+ \Lambda^\prime \circ \Lambda \circ \mleft(
		\mleft(\mathrm{d}^{\nabla^{\mathrm{bas}}} \circ \Lambda^{-1} \circ (\Lambda^\prime)^{-1}\mright)
		\mleft(
			\lambda^\prime \circ \widehat{\Lambda}
		\mright)
	\mright) \circ \mleft( \widehat{\Lambda}^{-1}, \mathds{1}_E \mright)
\\
&=
\nabla
	+ \mleft( \mathfrak{\Lambda} \circ \mathrm{d}^{\nabla^{\mathrm{bas}}} \circ \mathfrak{\Lambda}^{-1} \mright)\underbrace{\mleft( \lambda + \lambda^\prime \circ \widehat{\Lambda} \mright)}
	_{= \lambda + \lambda^\prime - \lambda^\prime \circ \rho \circ \lambda}
\\
&\hspace{1cm}
	- \mleft( \mathfrak{\Lambda} \circ \mathrm{d}^{\nabla^{\mathrm{bas}}} \circ \mathfrak{\Lambda}^{-1} \mright)\mleft( \lambda^\prime \circ \rho \circ \lambda \mright)
\\
&\hspace{1cm}
	+ \mleft( \Lambda^\prime \mright)^{-1} \circ \lambda^\prime \circ \rho \circ \mleft(\mleft( \mathfrak{\Lambda} \circ \mathrm{d}^{\nabla^{\mathrm{bas}}} \circ \mathfrak{\Lambda}^{-1} \mright)\mleft( \Lambda^\prime \circ \lambda \mright)\mright)
\\
&\hspace{1cm}
	+ \mleft(\mleft( \mathfrak{\Lambda} \circ \mathrm{d}^{\nabla^{\mathrm{bas}}} \circ \mathfrak{\Lambda}^{-1} \mright)\mleft( \lambda^\prime \circ \widehat{\Lambda} \mright) \mright) \circ \mleft( \widehat{\Lambda}^{-1} \circ \rho \circ \lambda, \mathds{1}_E \mright)
\\
&=
\widetilde{\nabla}^{\lambda + \lambda^\prime - \lambda^\prime \circ \rho \circ \lambda}
\\
&\hspace{1cm}
	- \mleft( \mathfrak{\Lambda} \circ \mathrm{d}^{\nabla^{\mathrm{bas}}} \mright)\mleft( \mathfrak{\Lambda}^{-1} \circ \lambda^\prime \circ \rho \circ \lambda \mright)
\\
&\hspace{1cm}
	+ \lambda^\prime \circ \rho \circ \Lambda \circ \mathrm{d}^{\nabla^{\mathrm{bas}}} \mleft( \Lambda^{-1} \circ \lambda \mright)
\\
&\hspace{1cm}
	+ \mleft(\mleft( \mathfrak{\Lambda} \circ \mathrm{d}^{\nabla^{\mathrm{bas}}} \circ \mathfrak{\Lambda}^{-1} \mright)\mleft( \lambda^\prime \circ \widehat{\Lambda} \mright) \mright) \circ \mleft( \widehat{\Lambda}^{-1} \circ \rho \circ \lambda, \mathds{1}_E \mright)
\\
&=
\widetilde{\nabla}^{\lambda + \lambda^\prime - \lambda^\prime \circ \rho \circ \lambda}
\\
&\hspace{1cm}
	- \mathfrak{\Lambda} \circ \nabla^{\mathrm{bas}} \circ \mathfrak{\Lambda}^{-1} \circ \lambda^\prime \circ \rho \circ \lambda 
	+ \lambda^\prime \circ \rho \circ \lambda \circ \nabla^{\mathrm{bas}}
\\
&\hspace{1cm}
	+ \lambda^\prime \circ \rho \circ \Lambda \circ \nabla^{\mathrm{bas}} \circ \Lambda^{-1} \circ \lambda
	- \lambda^\prime \circ \rho \circ \Lambda \circ \Lambda^{-1} \circ \lambda \circ \nabla^{\mathrm{bas}}
\\
&\hspace{1cm}
	+ \mathfrak{\Lambda} \circ \nabla^{\mathrm{bas}} \circ \mathfrak{\Lambda}^{-1} \circ \lambda^\prime \circ \widehat{\Lambda} \circ \widehat{\Lambda}^{-1} \circ \rho \circ \lambda
	- \mathfrak{\Lambda} \circ \mathfrak{\Lambda}^{-1} \circ \lambda^\prime \circ \underbrace{\widehat{\Lambda} \circ \nabla^{\mathrm{bas}} \circ \widehat{\Lambda}^{-1} \circ \rho}
	_{\mathclap{ \stackrel{\text{Cor.~\ref{cor:ENablaMitRhoVertauschung}}}{=} \rho \circ \Lambda \circ \nabla^{\mathrm{bas}} \circ \Lambda^{-1} }}
	\circ \lambda
\\
&=
\widetilde{\nabla}^{\lambda + \lambda^\prime - \lambda^\prime \circ \rho \circ \lambda},
\eas
rewriting definitions like $\mathrm{d}^{\nabla^{\mathrm{bas}}} \lambda = \nabla^{\mathrm{bas}} \circ \lambda - \lambda \circ \nabla^{\mathrm{bas}}$, where the basic connection in the first summand is the one on $E$ and the one on $\mathrm{T}N$ in the second summand, \textit{i.e.}
\bas
\mleft(\mathrm{d}^{\nabla^{\mathrm{bas}}} \lambda\mright)(Y, \nu)
&= 
\nabla^{\mathrm{bas}}_\nu \bigl(\lambda(Y)\bigr) 
	- \lambda \mleft(\nabla^{\mathrm{bas}}_\nu Y \mright)
\eas
for all $\nu \in \Gamma(E)$ and $Y \in \mathfrak{X}(N)$. Finally let us look at the field redefinitions of $\zeta$, the calculation is very similar to the proof of Lemma \ref{lem:InverseOfZetaLambda}; the calculation is purely straightforward, just compare the definitions of $\widetilde{\widetilde{\zeta}^\lambda}^{\lambda^\prime}$ with $\widetilde{\zeta}^{\lambda + \lambda^\prime - \lambda^\prime \circ \rho \circ \lambda}$. However, it is very tedious and long, hence, we will omit the calculation; we are going to motivate it differently, using the field redefinition of the field strength provided in Lemma \ref{lem:FinallyNiceTrafoOfEverything}. That is,
\bas
\widetilde{G}^{\lambda + \lambda^\prime - \lambda^\prime \circ \rho \circ \lambda}
&=
\widetilde{F}^{\lambda + \lambda^\prime - \lambda^\prime \circ \rho \circ \lambda}
	+ \frac{1}{2} \biggl({}^*\mleft(\widetilde{\zeta}^{\lambda + \lambda^\prime - \lambda^\prime \circ \rho \circ \lambda}\mright)\biggr)\mleft( \widetilde{\mathfrak{D}}^{\lambda + \lambda^\prime - \lambda^\prime \circ \rho \circ \lambda} \stackrel{\wedge}{,} \widetilde{\mathfrak{D}}^{\lambda + \lambda^\prime - \lambda^\prime \circ \rho \circ \lambda} \mright),
\eas
but also Lemma \ref{lem:FinallyNiceTrafoOfEverything}
\bas
\widetilde{G}^{\lambda + \lambda^\prime - \lambda^\prime \circ \rho \circ \lambda}
&=
({}^*\mathfrak{\Lambda})(G)
=
\mleft( \mleft( {}^*\Lambda^\prime \mright) \circ ({}^*\Lambda) \mright) (G)
=
\mleft( {}^*\Lambda^\prime \mright)\mleft( \widetilde{G}^\lambda \mright)
=
\widetilde{\widetilde{G}^\lambda}^{\lambda^\prime}.
\eas
By the previous results we immediately get
\bas
\widetilde{F}^{\lambda+ \lambda^\prime - \lambda^\prime \circ \rho \circ \lambda}
&=
\widetilde{\widetilde{F}^\lambda}^{\lambda^\prime},
\eas
because $F$ is independent of $\zeta$. Similarly as for $G$ we get by Thm.~\ref{thm:FieldRedefofstandardFieldStrengthF}
\bas
\widetilde{\mathfrak{D}}^{\lambda + \lambda^\prime - \lambda^\prime \circ \rho \circ \lambda}
&=
\widetilde{\widetilde{\mathfrak{D}}^\lambda}^{\lambda^\prime}.
\eas
Then simply compare both sides in $\widetilde{G}^{\lambda + \lambda^\prime - \lambda^\prime \circ \rho \circ \lambda} = \widetilde{\widetilde{G}^\lambda}^{\lambda^\prime}$ to get
\bas
\mleft({}^*\mleft(\widetilde{\zeta}^{\lambda + \lambda^\prime - \lambda^\prime \circ \rho \circ \lambda} - \widetilde{\widetilde{\zeta}^\lambda}^{\lambda^\prime}\mright)\mright)\mleft( \widetilde{\mathfrak{D}}^{\lambda + \lambda^\prime - \lambda^\prime \circ \rho \circ \lambda} \stackrel{\wedge}{,} \widetilde{\mathfrak{D}}^{\lambda + \lambda^\prime - \lambda^\prime \circ \rho \circ \lambda} \mright)
&=
0.
\eas
Since $\mathrm{Dev}$ and $\mathfrak{D}$ are in general non-zero, and by $\widetilde{\mathfrak{D}} = \mathrm{D} - ({}^*\rho)(\varpi_2)$ (so, the minimal coupling stays non-zero if it was initially non-zero), one can conclude
\bas
\widetilde{\zeta}^{\lambda + \lambda^\prime - \lambda^\prime \circ \rho \circ \lambda} 
&=
\widetilde{\widetilde{\zeta}^\lambda}^{\lambda^\prime},
\eas
however, there are edge cases where this argument fails: $M$ could be a point for example, but it is clear that the field redefinition of $\zeta$ is independent of the choice of $M$ such that one can quickly circumvent this problem. Another edge case is $N$ as a point, but then $\zeta \equiv 0$ such that everything is trivially concluded.
\end{proof}

\begin{remarks}{Field redefinition as equivalence of CYMH GTs}{FieldredefAsEquivalence}
This finally shows that the field redefinition is an equivalence of CYMH GTs (for fixed $M, N$ and $E$). Reflexivity simply follows due to that $\lambda \equiv 0$ is a valid parameter for the field redefinition, symmetry by Lemma \ref{lem:FieldRedefinitionIsInvertible} and \ref{lem:InverseOfZetaLambda}, and transitivity by Lemma \ref{lem:TransFieldRedef}. Furthermore, by Thm.~\ref{thm:InvarianceUnderTheFieldRedefinition}, the physics stay the same after a field redefinition, which is why one may speak of a \emph{physical} equivalence.
\end{remarks}

As we already argued, starting with a non-flat $\nabla$ and/or a non-zero $\zeta$, it is now natural to ask whether or not there is a field redefinition making $\nabla$ flat and/or $\zeta$ zero, equivalently, whether or not there is an equivalence class with pre-classical and/or classical representative, respectively. We will do this in the next chapter, but let us first state some basic properties of a CYMH GT.

\section{Properties of CYMH GT}\label{PropertiesOFNewTOlleGTs}

\begin{theorems}{Curvature closed under basic connections, by Alexei Kotov}{CurvatureClosed}
Let $E \to N$ be a Lie algebroid over a smooth manifold $N$, and $\nabla$ be a connection on $E$ with vanishing basic curvature. Then
\ba
\mathrm{d}^{\nabla^{\mathrm{bas}}} R_\nabla &= 0.
\ea
\end{theorems}

\begin{remark}
\leavevmode\newline
Alexei Kotov has found this identity, too, with a different approach; this was communicated in a private communication but there is a paper planned about that by Alexei Kotov and Thomas Strobl, planned for 2021.
\end{remark}

\begin{proof}[Proof of Thm.~\ref{thm:CurvatureClosed}]
\leavevmode\newline
We know how the connection acts on the Lie bracket of $E$ due to the vanishing of the basic curvature, hence, let us look at how the curvature acts on the Lie bracket, also using the Jacobi identity of $[ \cdot, \cdot]$,
\bas
R_\nabla(Y, Z) \mleft(\mleft[ \mu, \nu\mright]_E\mright)
&=~
\stackrel{\text{Use } R_\nabla^{\mathrm{bas}} = 0}{\dotsc}
\\
&=
\mleft[ \nabla_Y \nabla_Z \mu, \nu \mright]_E
	+ \mleft[ \nabla_Z \mu, \nabla_Y \nu \mright]_E
	+ \nabla_{\nabla^{\mathrm{bas}}_\nu Y} \nabla_Z \mu
	- \nabla_{\nabla^{\mathrm{bas}}_{\nabla_Z \mu} Y} \nu
	+ \mleft[ \nabla_Y \mu, \nabla_Z \nu \mright]_E \\
	&\quad+ \mleft[ \mu, \nabla_Y \nabla_Z \nu \mright]_E
	+ \nabla_{\nabla^{\mathrm{bas}}_{\nabla_Z \nu} Y} \mu
	- \nabla_{\nabla^{\mathrm{bas}}_\mu Y} \nabla_Z \nu
	+ \nabla_Y \nabla_{\nabla^{\mathrm{bas}}_\nu Z} \mu
	- \nabla_Y \nabla_{\nabla^{\mathrm{bas}}_\mu Z} \nu \\
	&\quad- \Big( Y \leftrightarrow Z \text{ of previous two lines} \Big) \\
	&\quad- \mleft[ \nabla_{[Y, Z]} \mu, \nu \mright]_E
	- \mleft[ \mu, \nabla_{[Y, Z]}\nu \mright]_E
	- \nabla_{\nabla^{\mathrm{bas}}_\nu \mleft( [Y, Z] \mright)} \mu
	+ \nabla_{\nabla^{\mathrm{bas}}_\mu \mleft( [Y, Z] \mright)} \nu \\
&=
R_\nabla\mleft( \nabla^{\mathrm{bas}}_\nu Y, Z \mright)\mu
	+ R_\nabla\mleft( Y, \nabla^{\mathrm{bas}}_\nu Z \mright)\mu
	- R_\nabla\mleft( \nabla^{\mathrm{bas}}_\mu Y, Z \mright)\nu
	- R_\nabla\mleft( Y, \nabla^{\mathrm{bas}}_\mu Z \mright)\nu \\
	&\quad+ \underbrace{\nabla_{\mleft[ \nabla^{\mathrm{bas}}_\nu Y, Z \mright]} \mu}_{= \nabla_{\mleft[ \mleft[ \rho(\nu), Y \mright] + \rho \mleft( \nabla_Y \nu \mright), Z \mright]} \mu}
	+ \nabla_{\mleft[ Y, \nabla^{\mathrm{bas}}_\nu Z \mright]} \mu
	- \nabla_{\mleft[ \nabla^{\mathrm{bas}}_\mu Y, Z \mright]} \nu
	- \nabla_{\mleft[ Y, \nabla^{\mathrm{bas}}_\mu Z \mright]} \nu \\
	&\quad+ \nabla_{\mleft[ \rho \mleft( \nabla_Z \nu \mright), Y \mright] + \rho \mleft( \nabla_Y \nabla_Z \nu\mright)} \mu
	- \nabla_{\mleft[ \rho \mleft( \nabla_Z \mu \mright), Y \mright] + \rho \mleft( \nabla_Y \nabla_Z \mu\mright)} \nu
	- \Big( Y \leftrightarrow Z \Big) \\
	&\quad+ \nabla_{\mleft[ \rho(\mu), [Y, Z] \mright] + \rho \mleft(\nabla_{[Y, Z]}\mu\mright)} \nu
	- \nabla_{\mleft[ \rho(\nu), [Y, Z] \mright] + \rho \mleft(\nabla_{[Y, Z]}\nu\mright)} \mu \\
	&\quad+ \mleft[ \mu, R_\nabla(Y, Z) \nu \mright]_E
	- \mleft[ \nu, R_\nabla(Y, Z) \mu \mright]_E \\
&=
R_\nabla\mleft( \nabla^{\mathrm{bas}}_\nu Y, Z \mright)\mu
	+ R_\nabla\mleft( Y, \nabla^{\mathrm{bas}}_\nu Z \mright)\mu
	- R_\nabla\mleft( \nabla^{\mathrm{bas}}_\mu Y, Z \mright)\nu
	- R_\nabla\mleft( Y, \nabla^{\mathrm{bas}}_\mu Z \mright)\nu \\
	&\quad+ \nabla^{\mathrm{bas}}_\mu \mleft( R_\nabla(Y, Z) \nu \mright)
	- \nabla^{\mathrm{bas}}_\nu \mleft( R_\nabla(Y, Z) \mu \mright) \\
&=
\mleft( \mathrm{d}^{\nabla^{\mathrm{bas}}} R_\nabla \mright)(Y, Z, \mu, \nu)
	+ R_\nabla(Y, Z) \mleft(\mleft[ \mu, \nu\mright]_E\mright) \\
\Leftrightarrow\qquad
0
&= 
\mleft( \mathrm{d}^{\nabla^{\mathrm{bas}}} R_\nabla \mright)(Y, Z, \mu, \nu)
\eas
for all $Y, Z \in \mathfrak{X}(N)$ and $\nu, \mu \in \Gamma(E)$
\end{proof}

So, we know that the basic connection is flat when the basic curvature vanishes, recall Prop.~\ref{prop:SnablamitREnabla}, and that the curvature $R_\nabla$ is closed with respect to the differential induced by the basic connection. The compatibility condition \ref{EqMyFormulationOfZetaCondition} then imposes that the curvature even needs to be exact in order to formulate a gauge theory.

We know that curvatures satisfy a Bianchi identity, let us therefore check what this implies about $\zeta$.

\begin{theorems}{Bianchi identity for the primitives of the connection}{BianchiIdentityForZeta}
Let $E \to N$ be a Lie algebroid over a smooth manifold $N$, and $\nabla$ a connection on $E$ with vanishing basic curvature and for whose curvature there is a $\zeta \in \Omega^2(N;E)$ such that $R_\nabla = - \mathrm{d}^{\nabla^{\mathrm{bas}}} \zeta$. Then
\ba
0
&=
\mleft( \nabla^{\mathrm{bas}}_{\nu_0} \mleft( \mathrm{d}^\nabla \zeta \mright) \mright)(Y_0, Y_1, Y_2)
	- \mleft( \nabla^{\mathrm{bas}}_{\nu_0} \bigl( \zeta \circ \mleft( \mathds{1}_{\mathrm{T}N}, \rho \circ \zeta \mright) \bigr) \mright)(Y_0, Y_1, Y_2)
\nonumber \\
&\hspace{1cm}
	- \mleft( \nabla^{\mathrm{bas}}_{\nu_0} \bigl( \zeta \circ \mleft( \mathds{1}_{\mathrm{T}N}, \rho \circ \zeta \mright) \bigr) \mright)(Y_1, Y_2, Y_0)
	- \mleft( \nabla^{\mathrm{bas}}_{\nu_0} \bigl( \zeta \circ \mleft( \mathds{1}_{\mathrm{T}N}, \rho \circ \zeta \mright) \bigr) \mright)(Y_2, Y_0, Y_1)
\ea
for all $Y_0, Y_1, Y_2 \in \mathfrak{X}(N)$ and $\nu_0 \in \Gamma(E)$, where 
\bas
\bigl(\zeta \circ \mleft( \mathds{1}_{\mathrm{T}N}, \rho \circ \zeta \mright) \bigr) (Y_0, Y_1, Y_2) = \zeta\bigl( Y_0, (\rho \circ \zeta)(Y_1, Y_2) \bigr).
\eas
\end{theorems}

\begin{proof}
\leavevmode\newline
$R_\nabla$ satisfies the Bianchi identity, \textit{i.e.}~
\bas
\mathrm{d}^\nabla R_\nabla
&=
0,
\eas
where we view the curvature as an element of $\Omega^2(N; \mathrm{End}(E))$.
Then use Cor.~\ref{cor:commutationS=0} to get
\bas
0
&=
\mleft( -\mathrm{d}^\nabla R_\nabla \mright) \mleft(Y_0, Y_1, Y_2, \nu_0\mright)
\\
&=
\mleft(\mathrm{d}^\nabla \mathrm{d}^{\nabla^{\mathrm{bas}}} \zeta \mright) \mleft(Y_0, Y_1, Y_2, \nu_0\mright)
\\
&=
\mleft( \mathrm{d}^{\nabla^{\mathrm{bas}}} \mathrm{d}^\nabla \zeta \mright) \mleft(Y_0, Y_1, Y_2, \nu_0\mright)
\\
&\hspace{1cm}
	+ R_\nabla \bigl( Y_0, \mleft(\rho \circ \zeta\mright)(Y_1, Y_2) \bigr) \nu_0
	- R_\nabla \bigl( Y_1, \mleft(\rho \circ \zeta\mright)(Y_0, Y_2) \bigr) \nu_0
	+ R_\nabla \bigl( Y_2, \mleft(\rho \circ \zeta\mright)(Y_0, Y_1) \bigr) \nu_0
\\
&\hspace{1cm}
	- \zeta \bigl( \mleft(\rho \circ R_\nabla\mright)(Y_0, Y_1)\nu_0, Y_2 \bigr)
	+ \zeta \bigl( \mleft(\rho \circ R_\nabla\mright)(Y_0, Y_2)\nu_0, Y_1 \bigr)
	- \zeta \bigl( \mleft(\rho \circ R_\nabla\mright)(Y_1, Y_2)\nu_0, Y_0 \bigr)
\\
&=	
\mleft( \mathrm{d}^{\nabla^{\mathrm{bas}}} \mathrm{d}^\nabla \zeta \mright) \mleft(Y_0, Y_1, Y_2, \nu_0\mright)
\\
&\hspace{1cm}
	- \mleft(\nabla^{\mathrm{bas}}_{\nu_0} \zeta\mright) \bigl( Y_0, \mleft(\rho \circ \zeta\mright)(Y_1, Y_2) \bigr)
	+ \mleft(\nabla^{\mathrm{bas}}_{\nu_0} \zeta\mright) \bigl( Y_1, \mleft(\rho \circ \zeta\mright)(Y_0, Y_2) \bigr) 
\\
&\hspace{1cm}
	- \mleft(\nabla^{\mathrm{bas}}_{\nu_0} \zeta\mright) \bigl( Y_2, \mleft(\rho \circ \zeta\mright)(Y_0, Y_1) \bigr) 
\\
&\hspace{1cm}
	+ \zeta \mleft( \mleft(\nabla^{\mathrm{bas}}_{\nu_0} ( \rho \circ \zeta ) \mright)(Y_0, Y_1), Y_2 \mright)
	- \zeta \mleft( \mleft(\nabla^{\mathrm{bas}}_{\nu_0} ( \rho \circ \zeta )\mright)(Y_0, Y_2), Y_1 \mright)
\\
&\hspace{1cm}
	+ \zeta \mleft( \mleft(\nabla^{\mathrm{bas}}_{\nu_0} ( \rho \circ \zeta)\mright)(Y_1, Y_2), Y_0 \mright)
\eas
for all $Y_0, Y_1, Y_2 \in \mathfrak{X}(N)$ and $\nu_0 \in \Gamma(E)$,
using that $\zeta \in \Omega^{2,0}(N, E;E) \cong \Omega^2(N;E)$, $R_\nabla = - \mathrm{d}^{\nabla^{\mathrm{bas}}} \zeta$ and $\rho \circ \nabla^{\mathrm{bas}} = \nabla^{\mathrm{bas}} \circ \rho$ such that
\bas
\mleft( \rho \circ \nabla^{\mathrm{bas}}_{\nu_0} \zeta \mright) (Y_0,Y_1)
&=
\rho\mleft(\mleft(\nabla^{\mathrm{bas}}_{\nu_0} \zeta \mright)(Y_0, Y_1)\mright)
\\
&=
\rho\mleft(
	\nabla^{\mathrm{bas}}_{\nu_0} \bigl(\zeta (Y_0, Y_1) \bigr)
	- \zeta \mleft( \nabla^{\mathrm{bas}}_{\nu_0} Y_0, Y_1 \mright)
	- \zeta \mleft( Y_0, \nabla^{\mathrm{bas}}_{\nu_0} Y_1 \mright)
\mright)
\\
&=
	\nabla^{\mathrm{bas}}_{\nu_0} \bigl( (\rho \circ \zeta) (Y_0, Y_1) \bigr)
	- (\rho \circ \zeta) \mleft( \nabla^{\mathrm{bas}}_{\nu_0} Y_0, Y_1 \mright)
	- (\rho \circ \zeta) \mleft( Y_0, \nabla^{\mathrm{bas}}_{\nu_0} Y_1 \mright)
\\
&=
\mleft(\nabla^{\mathrm{bas}}_{\nu_0} ( \rho \circ \zeta ) \mright)(Y_0, Y_1).
\eas
We can also write
\bas
\mleft(\nabla^{\mathrm{bas}}_{\nu_0} \zeta\mright) \bigl( Y_0, \mleft(\rho \circ \zeta\mright)(Y_1, Y_2) \bigr)
&=
\nabla^{\mathrm{bas}}_{\nu_0} \Bigl(
	\zeta \bigl( Y_0, \mleft(\rho \circ \zeta\mright)(Y_1, Y_2) \bigr)
\Bigr)
\\
&\hspace{1cm}
	- \zeta \mleft( \nabla^{\mathrm{bas}}_{\nu_0} Y_0, \mleft(\rho \circ \zeta\mright)(Y_1, Y_2) \mright)
	- \zeta \mleft( Y_0, \nabla^{\mathrm{bas}}_{\nu_0} \bigl( \mleft(\rho \circ \zeta\mright)(Y_1, Y_2) \bigr) \mright),
\eas
and (again)
\bas
\mleft(\nabla^{\mathrm{bas}}_{\nu_0}  (\rho \circ \zeta)\mright)(Y_0, Y_1)
&=
\nabla^{\mathrm{bas}}_{\nu_0} \bigl(
	(\rho \circ \zeta) \mleft( Y_0, Y_1 \mright)
\bigr)
	- (\rho \circ \zeta) \mleft( \nabla^{\mathrm{bas}}_{\nu_0} Y_0, Y_1 \mright)
	- (\rho \circ \zeta) \mleft( Y_0, \nabla^{\mathrm{bas}}_{\nu_0} Y_1 \mright),
\eas
such that in total
\bas
0
&=
\mleft( \mathrm{d}^{\nabla^{\mathrm{bas}}} \mathrm{d}^\nabla \zeta \mright) \mleft(Y_0, Y_1, Y_2, \nu_0\mright)
\\
&\hspace{1cm}
	- \nabla^{\mathrm{bas}}_{\nu_0} \Bigl(
	\zeta \bigl( Y_0, \mleft(\rho \circ \zeta\mright)(Y_1, Y_2) \bigr)
\Bigr)
	+ \zeta \mleft( \nabla^{\mathrm{bas}}_{\nu_0} Y_0, \mleft(\rho \circ \zeta\mright)(Y_1, Y_2) \mright)
	+ \zeta \mleft( Y_0, \nabla^{\mathrm{bas}}_{\nu_0} \bigl( \mleft(\rho \circ \zeta\mright)(Y_1, Y_2) \bigr) \mright)
\\
&\hspace{1cm}
	+ \nabla^{\mathrm{bas}}_{\nu_0} \Bigl(
	\zeta \bigl( Y_1, \mleft(\rho \circ \zeta\mright)(Y_0, Y_2) \bigr)
\Bigr)
	- \zeta \mleft( \nabla^{\mathrm{bas}}_{\nu_0} Y_1, \mleft(\rho \circ \zeta\mright)(Y_0, Y_2) \mright)
	- \zeta \mleft( Y_1, \nabla^{\mathrm{bas}}_{\nu_0} \bigl( \mleft(\rho \circ \zeta\mright)(Y_0, Y_2) \bigr) \mright)
\\
&\hspace{1cm}
	- \nabla^{\mathrm{bas}}_{\nu_0} \Bigl(
	\zeta \bigl( Y_2, \mleft(\rho \circ \zeta\mright)(Y_0, Y_1) \bigr)
\Bigr)
	+ \zeta \mleft( \nabla^{\mathrm{bas}}_{\nu_0} Y_2, \mleft(\rho \circ \zeta\mright)(Y_0, Y_1) \mright)
	+ \zeta \mleft( Y_2, \nabla^{\mathrm{bas}}_{\nu_0} \bigl( \mleft(\rho \circ \zeta\mright)(Y_0, Y_1) \bigr) \mright)
\\
&\hspace{1cm}
	- \zeta \mleft( Y_2,
		\nabla^{\mathrm{bas}}_{\nu_0} \bigl(
	(\rho \circ \zeta) \mleft( Y_0, Y_1 \mright)
\bigr)
	- (\rho \circ \zeta) \mleft( \nabla^{\mathrm{bas}}_{\nu_0} Y_0, Y_1 \mright)
	- (\rho \circ \zeta) \mleft( Y_0, \nabla^{\mathrm{bas}}_{\nu_0} Y_1 \mright)
	\mright)
\\
&\hspace{1cm}
	+ \zeta \mleft( Y_1,
	\nabla^{\mathrm{bas}}_{\nu_0} \bigl(
	(\rho \circ \zeta) \mleft( Y_0, Y_2 \mright)
\bigr)
	- (\rho \circ \zeta) \mleft( \nabla^{\mathrm{bas}}_{\nu_0} Y_0, Y_2 \mright)
	- (\rho \circ \zeta) \mleft( Y_0, \nabla^{\mathrm{bas}}_{\nu_0} Y_2 \mright)
	\mright)
\\
&\hspace{1cm}
	- \zeta \mleft( Y_0,
	\nabla^{\mathrm{bas}}_{\nu_0} \bigl(
	(\rho \circ \zeta) \mleft( Y_1, Y_2 \mright)
\bigr)
	- (\rho \circ \zeta) \mleft( \nabla^{\mathrm{bas}}_{\nu_0} Y_1, Y_2 \mright)
	- (\rho \circ \zeta) \mleft( Y_1, \nabla^{\mathrm{bas}}_{\nu_0} Y_2 \mright)
	\mright)
\\
%%%%%%%%%%%%%%%%%%%%%%%%%%%%%%%%%%%%%%%%%%%%%%%%%%%%%%%%%%%%%%%%%%%%%%%%%%%%%%%%%
&=
\underbrace{\mleft( \mathrm{d}^{\nabla^{\mathrm{bas}}} \mathrm{d}^\nabla \zeta \mright) \mleft(Y_0, Y_1, Y_2, \nu_0\mright)}
_{\mathclap{ = \mleft( \nabla^{\mathrm{bas}}_{\nu_0} \mleft( \mathrm{d}^\nabla \zeta \mright) \mright)(Y_0, Y_1, Y_2) }}
\\
&\hspace{1cm}
	- \nabla^{\mathrm{bas}}_{\nu_0} \Bigl(
	\zeta \bigl( Y_0, \mleft(\rho \circ \zeta\mright)(Y_1, Y_2) \bigr)
\Bigr)
	+ \zeta \mleft( \nabla^{\mathrm{bas}}_{\nu_0} Y_0, \mleft(\rho \circ \zeta\mright)(Y_1, Y_2) \mright)
\\
&\hspace{1cm}
	+ \zeta \mleft( Y_0, 
	(\rho \circ \zeta) \mleft( \nabla^{\mathrm{bas}}_{\nu_0} Y_1, Y_2 \mright)
	+ (\rho \circ \zeta) \mleft( Y_1, \nabla^{\mathrm{bas}}_{\nu_0} Y_2 \mright)
	\mright)
\\
&\hspace{1cm}
	- \nabla^{\mathrm{bas}}_{\nu_0} \Bigl(
	\zeta \bigl( Y_1, \mleft(\rho \circ \zeta\mright)(Y_2, Y_0) \bigr)
\Bigr)
	+ \zeta \mleft( \nabla^{\mathrm{bas}}_{\nu_0} Y_1, \mleft(\rho \circ \zeta\mright)(Y_2, Y_0) \mright)
\\
&\hspace{1cm}
	+ \zeta \mleft( Y_1,
	(\rho \circ \zeta) \mleft( \nabla^{\mathrm{bas}}_{\nu_0} Y_2, Y_0 \mright)
	+ (\rho \circ \zeta) \mleft( Y_2, \nabla^{\mathrm{bas}}_{\nu_0} Y_0 \mright)
	\mright)
\\
&\hspace{1cm}
	- \nabla^{\mathrm{bas}}_{\nu_0} \Bigl(
	\zeta \bigl( Y_2, \mleft(\rho \circ \zeta\mright)(Y_0, Y_1) \bigr)
\Bigr)
	+ \zeta \mleft( \nabla^{\mathrm{bas}}_{\nu_0} Y_2, \mleft(\rho \circ \zeta\mright)(Y_0, Y_1) \mright)
\\
&\hspace{1cm}
	+ \zeta \mleft( Y_2,
	(\rho \circ \zeta) \mleft( \nabla^{\mathrm{bas}}_{\nu_0} Y_0, Y_1 \mright)
	+ (\rho \circ \zeta) \mleft( Y_0, \nabla^{\mathrm{bas}}_{\nu_0} Y_1 \mright)
	\mright)
%%%%%%%%%%%%%%%%%%%%%%%%%%%%%%%%%%%%%%%%%%%%%%%%%%%%%%%%%%%%%%%%%%%%%%%%%%%%%%%%
\\
&=
\mleft( \nabla^{\mathrm{bas}}_{\nu_0} \mleft( \mathrm{d}^\nabla \zeta \mright) \mright)(Y_0, Y_1, Y_2)
	- \mleft( \nabla^{\mathrm{bas}}_{\nu_0} \bigl( \zeta \circ \mleft( \mathds{1}_{\mathrm{T}N}, \rho \circ \zeta \mright) \bigr) \mright)(Y_0, Y_1, Y_2)
\\
&\hspace{1cm}
	- \mleft( \nabla^{\mathrm{bas}}_{\nu_0} \bigl( \zeta \circ \mleft( \mathds{1}_{\mathrm{T}N}, \rho \circ \zeta \mright) \bigr) \mright)(Y_1, Y_2, Y_0)
	- \mleft( \nabla^{\mathrm{bas}}_{\nu_0} \bigl( \zeta \circ \mleft( \mathds{1}_{\mathrm{T}N}, \rho \circ \zeta \mright) \bigr) \mright)(Y_2, Y_0, Y_1).
\eas
\end{proof}

Recall Thm.~\ref{thm:modBianchithm} for the following statement.

\begin{theorems}{Primitives of the connection along the foliation of the anchor}{BAlongL}
Let $E \to N$ be a Lie algebroid over a smooth manifold $N$, and $\nabla$ a connection on $E$ with vanishing basic curvature. Then all $\zeta \in \Omega^2(N;E)$ satisfying
\ba
\zeta \circ (\rho, \rho)
&= - t_{\nabla^{\mathrm{bas}}} + H,\label{eq:BrhoaufOrbit}
\ea
where $H \in \Omega^2(E;E)$ with $\nabla^{\mathrm{bas}} H = 0$, also satisfy
\ba
R_\nabla \circ (\rho, \rho)
&=
-\mleft(\mathrm{d}^{\nabla^{\mathrm{bas}}} \zeta \mright) \circ (\rho, \rho, \mathds{1}_E),
\ea
that is,
\bas
R_\nabla \bigl( \rho(\mu), \rho(\nu) \bigr) \eta
&=
-\mleft(\mathrm{d}^{\nabla^{\mathrm{bas}}} \zeta \mright) \bigl(\rho(\mu), \rho(\nu), \eta\bigr)
\eas
for all $\mu, \nu, \eta \in \Gamma(E)$.
\end{theorems}

\begin{proof}[Proof of Thm.~\ref{thm:BAlongL}]
\leavevmode\newline
That is a trivial consequence of Cor.~\ref{cor:LemmaCurvatureOfDualConnections} and Lemma \ref{lem:commutationanchordifferential}, that is,
\bas
R_\nabla \circ (\rho, \rho)
&=
R_{\nabla_\rho}
=
\nabla^{\mathrm{bas}} t_{\nabla^{\mathrm{bas}}}
\stackrel{\nabla^{\mathrm{bas}} H = 0}{=}
- \nabla^{\mathrm{bas}} \bigl( \zeta \circ (\rho, \rho) \bigr).
\stackrel{ \text{Lem.~\ref{lem:commutationanchordifferential}} }{=} 
\left(- \mathrm{d}^{\nabla^{\mathrm{bas}}} \zeta\right) \circ (\rho, \rho).
\eas
\end{proof}

Therefore one can view the negative of the torsion of the basic connection as a canonical choice for $\zeta$ along the foliation of the anchor. 
In case we decide to take $\zeta \in \Omega^2(N;E)$ such that $\zeta \circ (\rho, \rho) = - t_{\nabla^{\mathrm{bas}}}$, we get:

\begin{corollaries}{Certain classical CYMH GTs implying an abelian structure}{ClassicalTheoriesAreAbelianWithCanonicalChoices}
Let us have the same setup and notation as in Thm.~\ref{thm:FinallyTheGaugeInvarianceWeWant}, \textit{i.e.}~let us assume a CYMH GT. Moreover, assume we have $\zeta \circ (\rho, \rho) = - t_{\nabla^{\mathrm{bas}}}$ and that $N$ is simply connected.

If this CYMH GT is classical, then it is isomorphic to an abelian action Lie algebroid such that $\nabla$ is its canonical flat connection.

In case of tangent bundles, $E = \mathrm{T}N$, this statement is an equivalence, that is, this CYMH GT is classical if and only if it is isomorphic to an abelian action Lie algebroid such that $\nabla$ is its canonical flat connection.
\end{corollaries}

\begin{remark}
\leavevmode\newline
In general one could study whether it is possible to have a connection with vanishing basic curvature on a Lie algebroid which is locally never an action Lie algebroid; in that case the connection could not be flat by Thm.~\ref{thm:ActionLieALgebroid}. However, this is a difficult task; this statement may simplify that, one could just look at abelian action Lie algebroids. With that particular choice for $\zeta$ one would have then a non-classical gauge theory, in case one has a Lie algebroid which is not isomorphic to an abelian action Lie algebroid.
\end{remark}

\begin{proof}[Proof of Cor.~\ref{cor:ClassicalTheoriesAreAbelianWithCanonicalChoices}]
\leavevmode\newline
Classical means that $\nabla$ is flat, and, thus, we have a global isomorphism to an action Lie algebroid $N \times \mathfrak{g}$ for a Lie algebra $\mathfrak{g}$, using that $N$ is simply connected and Thm.~\ref{thm:ActionLieALgebroid}; also recall Remark \ref{remSimplyConnectedEqualsGlobal}. $\nabla$ is then its canonical flat connection.

Classical also implies that $\zeta \equiv 0$, hence, the torsion of $\nabla^{\mathrm{bas}}$ vanishes.\footnote{By the metric compatibility with $\kappa$, $\nabla^{\mathrm{bas}}$ is an $E$-Levi-Civita connection, as we also discussed in Rem.~\ref{remELEVICITAOfBasnbala}.} By Cor.~\ref{cor:AbelianIffNablaBasIsLeviCivita}, $\mathfrak{g}$ is abelian.

If we have $E = \mathrm{T}N$, then just use the equivalence in Cor.~\ref{cor:AbelianIffNablaBasIsLeviCivita}, so, assuming that $E$ is isomorphic to an abelian action Lie algebroid and $\nabla$ is its canonical flat connection, implies that the basic connection has no torsion; since the anchor is now bijective we have $\zeta \equiv 0$.
\end{proof}

Along the transversal directions it will be a bit more difficult as we will see in the next chapter. However, as a first approach one can look at the following proposition, which is based on the assumption that one has partially a parallel frame of the basic connection along the foliation, also using Thm.~\ref{thm:BAlongL}; recall Section \ref{DirectProdsOfLieAlgoids}, and also recall that BLA means bundle of Lie algebras. The setup of the following proposition is basically for Lie algebroids restricted on a suitable neighbourhood of regular points.

\begin{propositions}{Local mixed terms of the primitive of the connection}{MixedTermsOfB}
Let $N$ be a parallelizable smooth manifold, $K \to S$ a BLA over a smooth manifold $S$, and $E = \mathrm{T}N \times K \to N \times S$ as direct product of Lie algebroids, equipped with a connection $\nabla$ with a vanishing basic curvature. Furthermore, assume that there is a global trivialisation $\mleft( f_i \mright)_i$ of $\mathrm{T}N$ such that $\nabla^{\mathrm{bas}} f_i = 0$ (on $E$) for all $i$, and assume that we have a $\zeta \in \Omega^2(N;E)$ with $\zeta \circ (\rho, \rho) = - t_{\nabla^{\mathrm{bas}}}$.

If $\zeta$ additionally satisfies $\zeta(Y, f_i) = \nabla_{Y} f_i$ for all $Y \in \mathfrak{X}(S) \subset \mathfrak{X}(N \times S)$, then
\ba\label{MixeDZetaTermEquation}
R_\nabla\bigl( Y, \rho(\mu) \bigr) \nu
&=
- \mleft( \mathrm{d}^{\nabla^{\mathrm{bas}}} \zeta \mright) \bigl(Y, \rho(\mu), \nu \bigr)
\ea
for all $\mu, \nu \in \Gamma(E)$ and $Y \in \mathfrak{X}(S)$.
\end{propositions}

\begin{remark}
\leavevmode\newline
With $\mathfrak{X}(S) \subset \mathfrak{X}(N \times S)$ we emphasize that we view vector fields of a factor of the base, here $S$, as vector fields on $N \times S$ with values in $S$ and constant along $N$, \textit{i.e.}~the canonical embedding. That is important to keep in mind if one sees notations like $\mathfrak{X}(S)$ in this context.

A word on why we wrote "$\nabla^{\mathrm{bas}} f_i = 0$ (on $E$)". One needs to be careful here, with the basic connection we always mean two connections. However, we have for example $\rho(f_i) = f_i$ such that both versions of the basic connection can act on $f_i$, and as long as $K$ has not zero rank we can not expect that both connections give the same, that is, let $\nu \in \Gamma(K)$, then, on $E$,
\bas
\nabla^{\mathrm{bas}}_\nu f_i
&=
\mleft[ \nu, f_i \mright]_E
	+ \nabla_{f_i} \nu,
\eas
and, on $\mathrm{T}N$,
\bas
\nabla^{\mathrm{bas}}_\nu f_i
&=
\rho(\nabla_{f_i} \nu),
\eas
which is clearly different, even if $\mleft[ \nu, f_i \mright]_E = 0$. However, our imposed condition is about that $f_i$ as an element of $\Gamma(E)$ should be parallel to the basic connection, then we use the usual commutation with the anchor to get
\bas
0
&=
\rho\mleft( \nabla^{\mathrm{bas}} f_i \mright)
=
\nabla^{\mathrm{bas}} \bigl( \rho(f_i) \bigr),
\eas
where we did not write $\rho(f_i)$ as $f_i$ to emphasize that $f_i$ is viewed as an element of $\mathfrak{X}(N)$ on the right hand side. Hence, $\nabla^{\mathrm{bas}} f_i = 0$ in sense of $\mathrm{T}N$ is implied here. In the proof we sometimes write $\rho(f_i)$ for similar reasons of accentuation.
\end{remark}

\begin{proof}[Proof of Prop.~\ref{prop:MixedTermsOfB}]
\leavevmode\newline
We prove Eq.~\eqref{MixeDZetaTermEquation} locally using frames due to its tensorial nature. Let $\mleft( f_a \mright)_a$ be a local frame of $E$, which is given by the frame $\mleft(f_i \mright)_i$ of $\mathrm{T}N$ and by a frame $\mleft(f_\alpha \mright)_\alpha$ of $K$, both frames are canonically embedded into $E$; that is, $f_i$ are constant along $S$, and $f_\alpha$ along $N$. 
Other Latin indices still denote the frame of $\mathrm{T}N$, and other Greek ones the part of $K$, and we clearly have $\rho(f_i) = f_i, \rho(f_\alpha) = 0$; especially, $f_i$ also span the image of the anchor. 
Then
\bas
&&
\nabla^{\mathrm{bas}}_{f_i} Y
&=
\underbrace{\mleft[ f_i, Y \mright]}_{=0}
	+ ~\rho\mleft( \nabla_Y f_i \mright)
=
\rho\mleft( \nabla_Y f_i \mright),
\\
&&
\nabla^{\mathrm{bas}}_{f_\alpha} Y
&=
[ \underbrace{\rho(f_\alpha)}_{=0}, Y ]
	+ \rho\mleft( \nabla_Y f_\alpha \mright)
=
\rho\mleft( \nabla_Y f_\alpha \mright),
\\
&\Rightarrow&
\nabla^{\mathrm{bas}}_{f_a} Y
&=
\rho\mleft( \nabla_Y f_a \mright)
\eas
for all $Y \in \mathfrak{X}(S)$.
%\bas
%&&
%\rho\underbrace{\mleft(\nabla^{\mathrm{bas}}_{f_j} f_i \mright)}_{=0}
%&=
%0,
%\\
%&&
%\rho\mleft(\nabla^{\mathrm{bas}}_{f_\alpha} f_i \mright)
%&=
%\rho\bigl( 
	%\underbrace{\mleft[ f_\alpha, f_i \mright]_E}_{=0}
	%+ \nabla_{f_i} f_\alpha
%\bigr)
%=
%\rho\bigl( \nabla_{f_i} f_\alpha \bigr)
%\\
%&\Rightarrow&
%\eas
By the vanishing of the basic curvature we get
\bas
\nabla_Y\mleft( \mleft[ f_a, f_b \mright]_E \mright)
&=
\mleft[ \nabla_Y f_a, f_b \mright]_E
	+ \mleft[ f_a, \nabla_Y f_b \mright]_E
	+ \nabla_{\nabla^{\mathrm{bas}}_{f_b} Y} f_a
	- \nabla_{\nabla^{\mathrm{bas}}_{f_a} Y} f_b
\\
&=
\mleft[ \nabla_Y f_a, f_b \mright]_E
	+ \mleft[ f_a, \nabla_Y f_b \mright]_E
	+ \nabla_{\rho(\nabla_Y f_b)} f_a
	- \nabla_{\rho(\nabla_Y f_a)} f_b,
\eas
such that, additionally using $t_{\nabla^{\mathrm{bas}}} \stackrel{\text{Cor.~\ref{cor:TorsionOfDualTorsions}}}{=} - t_{\nabla_\rho}$ and the assumptions about $\zeta$,
\bas
\mleft(- \nabla^{\mathrm{bas}}_{f_a} \zeta \mright)\bigl(Y, \rho(f_i)\bigr)
&=
- \nabla^{\mathrm{bas}}_{f_a} \underbrace{\mleft(\zeta\bigl(Y, \rho(f_i)\bigr)\mright)}_{\mathclap{= \nabla_Y f_i}}
	+ \underbrace{\zeta\mleft( \nabla^{\mathrm{bas}}_{f_a} Y, \rho(f_i) \mright)}
		_{= \zeta\mleft( \rho(\nabla_Y f_a), \rho(f_i) \mright)}
	+ \zeta\mleft( Y, \vphantom{\nabla^{\mathrm{bas}}_{f_a} \bigl( \rho(f_i) \bigr)} \smash{\underbrace{\nabla^{\mathrm{bas}}_{f_a} \bigl( \rho(f_i) \bigr)}_{\mathclap{= \rho\mleft( \nabla^{\mathrm{bas}}_{f_a} f_i \mright) = 0 }}} \mright) 
\\
&=
- \mleft[ f_a, \nabla_Y f_i \mright]_E
	- \nabla_{\rho\mleft(\nabla_Y f_i \mright)} f_a
\\
&\hspace{1cm}
	+ \nabla_{\rho(\nabla_Y f_a)} f_i
	- \nabla_{f_i} \nabla_Y f_a
	- \mleft[ \nabla_Y f_a, f_i \mright]
\\
&=
\nabla_Y \underbrace{\mleft( \mleft[ f_i, f_a \mright]_E \mright)}
_{\mathclap{ = - \nabla^{\mathrm{bas}}_{f_a} f_i + \nabla_{f_i} f_a }}
	- \nabla_{f_i} \nabla_Y f_a
\\
&=
\nabla_Y \nabla_{f_i} f_a - \nabla_{f_i} \nabla_Y f_a
\\
&\stackrel{\mathclap{ [Y, f_i] = 0 }}{=}\quad 
R_\nabla(Y, f_i) f_a
\\
&=
R_\nabla\bigl(Y, \rho(f_i)\bigr) f_a.
\eas
\end{proof}
\chapter{Obstruction for CYMH GT}\label{ObstructionStuff}

Let us finally turn to the question whether or not there is always a field redefinition making $\nabla$ flat or $\zeta$ zero.
As we know by the splitting theorem of Lie algebroids, Thm.~\ref{thm:DirectSplitting}, around regular points every Lie algebroid is the sum of a tangent bundle and a bundle of Lie algebras (BLAs). The discussion about general Lie algebroids is very difficult, hence, let us first focus on both factors separately.

\section{Lie algebra bundles}\label{ObstrLAB}

We only want to discuss Lie algebra bundles (LABs) actually, not BLAs in general. That is motivated by the following theorem.

\begin{theorems}{BLA $\stackrel{?}{=}$ LAB, \newline \cite[Theorem 6.4.5, see also the last note at the beginning of \S 6.4; page 238f.]{mackenzieGeneralTheory} \newline \cite[Proposition 2.13]{basicconn}}{BLALAB}
Let $K \to N$ be a bundle of Lie algebras (BLA) over a connected manifold $N$ whose field of Lie brackets is denoted by $\mleft[ \cdot, \cdot \mright]_K$. Then $K$ is an LAB if and only if it admits a vector bundle connection $\nabla$ with vanishing basic curvature, that is
\bas
\nabla_Y \mleft( \mleft[ \mu, \nu \mright]_K \mright)
&=
\mleft[ \nabla_Y \mu, \nu \mright]_K
	+ \mleft[ \mu, \nabla_Y \nu \mright]_K
\eas
for all $\mu, \nu \in \Gamma(K)$ and $Y \in \mathfrak{X}(N)$.
\end{theorems}

\begin{remark}
\leavevmode\newline
Even if the Lie algebras of the fibres of a BLA are not isomorphic as Lie algebras recall that each BLA is a vector bundle, hence, the rank is constant.
\end{remark}

\begin{proof}[Sketch of the proof]
\leavevmode\newline
For "$\Rightarrow$", that is, $K$ is assumed to be an LAB, just take locally the canonical flat connection related to a local trivialization $K|_U \cong U \times \mathfrak{g}$, where $U$ is an open subset of $N$ and $\mathfrak{g}$ the Lie algebra describing $K$ as LAB; recall Def.~\ref{def:LAB}. Such a connection has trivially a vanishing basic curvature, \textit{e.g.}~use that the basic curvature is a tensor and test the vanishing against a frame of constant sections. Then use a partition of unity subordinate to a covering of such trivializations in order to get a globally defined connection with vanishing basic curvature.

The essential idea for the other direction is to observe that in the case of BLAs (zero anchor) we have
\bas
t_{\nabla^{\mathrm{bas}}}
&\stackrel{ \text{Cor.~\ref{cor:TorsionOfDualTorsions}} }{=}
- t_{\nabla_\rho}
=
\mleft[ \cdot, \cdot \mright]_K
\eas
for all vector bundle connections $\nabla$ on $K$. In case of a vanishing basic curvature we get by Eq.~\eqref{eq:compcondfast}
\bas
\nabla\mleft(\mleft[ \cdot, \cdot \mright]_K\mright)
&=
0,
\eas
\textit{i.e.}~the field of Lie brackets is parallel with respect to all $\nabla$ with vanishing basic curvature. In \cite[\S 6.4; page 236ff.]{mackenzieGeneralTheory} it is then shown that $\mleft[ \cdot, \cdot \mright]_K$ is deformable under the conjugation of vector space isomorphisms between two fibres of $K$, that is, the bracket of $\mu,\nu \in E_{p_2}$ at $p_2 \in N$ can be calculated by the value of the bracket at another base point $p_1 \in N$ using a conjugation of the bracket;\footnote{$p_1, p_2$ need to be connected by a path which is why one assumes a connected base manifold.} given an vector space isomorphism $\xi: E_{p_1} \to E_{p_2}$ the mentioned conjugation is given by $\xi\mleft( \mleft[ \xi^{-1}(\mu), \xi^{-1}(\nu) \mright]_K \mright)$. That implies that $\xi$ must be a Lie algebra isomorphism, and, extending this, $K$ is an LAB. This argument can be proven with arguments of the holonomy theory of connections, especially one uses that the values of a parallel section at two points connected by a curve are related by the parallel transport along that curve, or, in other words, the value at one point is the value at the other point conjugated by the parallel transport.

Alternatively (but very similar), one argues as in \cite[Proposition 2.13]{basicconn}; that is, as we have seen, $\nabla_X$ is a linear vector field on $K$ as a derivation on a vector bundle (recall Section \ref{DerivationsOnvector}, especially Thm.~\ref{thm:DerivationsSindEigentlichLineareVektorfelderKrass}). One can argue that linear vector fields are infinitesimal automorphisms of a vector bundle.\footnote{See also the beginning of \cite{meinrenkensplitting}.} Since the vanishing of the basic curvature is just the infinitesimal version of a Lie algebra homomorphism, the connection encodes the infinitesimal information of a Lie algebra isomorphism, therefore one can show that parallel transports by $\nabla$ are then Lie algebra isomorphisms with which one can construct a suitable LAB trivialization of $K$.
\end{proof}

So, this theorem implies that a vanishing basic curvature means that a bundle of Lie algebras is an LAB (over a connected base manifold). So, in our context bundle of Lie algebras are not so important, which is why we just want to focus on LABs.

\subsection{CYMH GT for LABs}\label{SumamryForLABSituation}

Let us now start to look at the situation of LABs; recall Def.~\ref{def:LAB}. Let us summarize the important previous results about CYMH GTs restricted onto LABs. The following section about LABs is also discussed in my paper \cite{My1stpaper}, slightly differently written. Also observe that for a zero anchor the basic connection $\nabla^{\mathrm{bas}}$ on $\mathrm{T}N$ is just zero, making the compatibility condition on the metric $g$ on $\mathrm{T}N$ trivial, and on $E$ it is the adjoint representation. This and the zero anchor in general simplifies all the involved equations:

\begin{situations}{CYMH GT for Lie algebra bundles}{CYMHGTForLABsToDoList}
Let $\mathfrak{g}$ be a real finite-dimensional Lie algebra with Lie bracket $\mleft[ \cdot, \cdot \mright]_{\mathfrak{g}}$. With
%\newline
\begin{center}
	\begin{tikzcd}
		\mathfrak{g} \arrow{r}	& \mleft(K, \mleft[ \cdot, \cdot \mright]_K\mright) \arrow{d} \\
			& N
	\end{tikzcd}
\end{center}
we denote LAB over a smooth manifold $N$ with Lie algebra structure inherited by $\mathfrak{g}$, with its field $\mleft[ \cdot, \cdot \mright]_K \in \Gamma\mleft(\bigwedge^2 K^* \otimes K \mright)$ of Lie brackets which restricts on the Lie bracket $\mleft[ \cdot, \cdot \mright]_{\mathfrak{g}}$ on each fibre. The gauge theory we look at is then now with respect to $E = K$.
%
%With respect to some trivialization over an open neighbourhood $U$, consider a fixed basis $\mleft( e_a \mright)_a$ of $\mathfrak{g}$, and extend that to constant sections of $U \times \mathfrak{g}$ which we still denote by $\mleft( e_a \mright)_a$. We denote the structure constants then by
%\bas
%\mleft[ e_a, e_b \mright]_K
%&\equiv
%\mleft[ e_a, e_b \mright]_{\mathfrak{g}}
%=
%C^c_{ab} e_c,
%\eas
%using Einstein's sum convention.

In the classical setting that would be a gauge theory where the gauge bosons are not paired to another fields via the minimal coupling because LABs are action Lie algebroids with zero action.

Let $(M, \eta)$ be a spacetime $M$ with its spacetime metric $\eta$, and $\Phi: M \to N$ a smooth map, representing the Higgs field. $\Phi^*K$ has also the structure of an LAB with a field of Lie brackets denoted by $\mleft[ \cdot, \cdot \mright]_{\Phi^*K} \in \Gamma\mleft(\bigwedge^2 \Phi^*\mleft(K^*\mright) \otimes \Phi^*K \mright)$, which restricts to $\mleft[ \cdot, \cdot \mright]_{\mathfrak{g}}$ on each fibre, too. This bracket is given by
\bas
\mleft[ \cdot, \cdot \mright]_{\Phi^*K}
&=
\Phi^*\mleft(\mleft[ \cdot, \cdot \mright]_{K}\mright).
\eas
%\bas
%\mleft[ \Phi^*\mu, \Phi^*\nu \mright]_{\Phi^*K}
%=
%\Phi^*\mleft(\mleft[ \mu, \nu \mright]_K\mright)
%\eas
%for all $\mu, \nu \in \Gamma(K)$ and $q \in M$, similar to \eqref{DefPullBackOfSuperTensors}. 
%Hence, we arrive at:
%\begin{center}
	%\begin{tikzcd}
		  %\mleft(\Phi^*K, \mleft[ \cdot, \cdot \mright]_{\Phi^*K}\mright) \arrow{d} & \mleft(K, \mleft[ \cdot, \cdot \mright]_K\mright) \arrow{d} \\
			%(M, \eta) \arrow{r}{\Phi} & N
	%\end{tikzcd}
%\end{center}
Let us also fix a vector bundle connection $\nabla$ on $K$ for which there is a $\zeta \in \Omega^2(N; K)$ such that
\ba\label{CondSGleichNullLAB}
\nabla_Y\mleft( \mleft[ \mu, \nu \mright]_K \mright)
&=
\mleft[ \nabla_Y \mu, \nu \mright]_K
	+ \mleft[ \mu, \nabla_Y \nu \mright]_K, \\
R_\nabla(Y, Z) \mu
&=
\mleft[ \zeta(Y, Z), \mu \mright]_K \label{CondKruemmungmitBLAB}
\ea
for all $Y, Z \in \mathfrak{X}(N)$ and $\mu, \nu \in \Gamma(K)$.

The \textbf{field of gauge bosons} (for a given Higgs field) will be represented by
\bas
A &\in \Omega^1(M; \Phi^*K).
\eas
The \textbf{field strength} $G$ is then defined as an element of $\mathcal{F}^2_K(M; {}^*K)$ by
\ba\label{defNewFieldStrengthG}
G(\Phi, A)
&\coloneqq
\mathrm{d}^{\Phi^*\nabla}A
	+ \frac{1}{2} \mleft[ A \stackrel{\wedge}{,} A \mright]_{\Phi^*K}
	+ \frac{1}{2} \mleft( \Phi^*\zeta \mright)\mleft( \mathrm{D}\Phi \stackrel{\wedge}{,} \mathrm{D}\Phi \mright) \nonumber \\
&=
\mathrm{d}^{\Phi^*\nabla}A
	+ \frac{1}{2} \mleft[ A \stackrel{\wedge}{,} A \mright]_{\Phi^*K}
	+ \Phi^!\zeta.
\ea
%, $\mleft[ A \stackrel{\wedge}{,} A \mright]_{\Phi^*K}$ is defined as an element of $\Omega^2(M; \Phi^*K)$ by
%\ba
%\mleft[ A \stackrel{\wedge}{,} A \mright]_{\Phi^*K}(\Phi, \Psi)
%&\coloneqq
%2 ~ \mleft[ A(\Phi) , A(\Psi) \mright]_{\Phi^*K}
%\ea
%for all vector fields $\Phi, \Psi \in \mathfrak{X}(M)$; similarly one defines $\mleft( \Phi^*\zeta \mright)\mleft( \mathrm{D}\Phi \stackrel{\wedge}{,} \mathrm{D}\Phi \mright)$ as an element of $\Omega^2(M; \Phi^*K)$ by
%\ba
%\mleft( \Phi^*\zeta \mright)\mleft( \mathrm{D}\Phi \stackrel{\wedge}{,} \mathrm{D}\Phi \mright)(\Phi, \Psi)(q)
%&\coloneqq
%2 ~ \zeta_{\Phi(q)}\Big( \mathrm{D}_qX(\Phi_q), \mathrm{D}_qX(\Psi_q) \Big)
%\ea
%for all $\Phi, \Psi \in \mathfrak{X}(M)$ and $q \in M$.

The curved Yang-Mills-Higgs Lagrangian is then defined as a top-degree-form $\mathcal{L}_{\mathrm{CYMH}} \in \mathcal{F}_K^{\mathrm{dim}(M)}(M)$ given by 
\ba\label{defLagrangianForLABs}
\mathcal{L}_{\mathrm{CYMH}}(\Phi, A)
&\coloneqq
- \frac{1}{2} \mleft( \Phi^*\kappa \mright)(G \stackrel{\wedge}{,} *G)
	+ \mleft( \Phi^*g \mright)(\mathrm{D}\Phi \stackrel{\wedge}{,} *\mathrm{D}\Phi)
	+ *(V \circ \Phi),
\ea
where $*$ is the Hodge star operator w.r.t. to $\eta$, $V \in C^\infty(N)$ is the \textbf{potential} for $\Phi$, $g$ is a Riemannian metric on $N$ and $\kappa$ a fibre metric on $K$.

We only allow Lie algebras $\mathfrak{g}$ admitting an \textbf{$\mathrm{ad}$-invariant scalar product} to which $\kappa$ shall restrict to on each fibre. Doing so, we achieve infinitesimal gauge invariance for $\mathcal{L}_{\mathrm{CYMH}}$.
\end{situations}
\newpage
\begin{remark}
\leavevmode\newline
\indent $\bullet$ In the following we want to test whether a given connection $\nabla$ satisfies the compatibility conditions \eqref{CondSGleichNullLAB} and \eqref{CondKruemmungmitBLAB}. Especially about the latter we say that a connection $\nabla$ satisfies compatibility condition \eqref{CondKruemmungmitBLAB} if there is a $\zeta \in \Omega^2(N; K)$ such that this condition is satisfied. So, we are not going to study this condition with respect to a fixed $\zeta$. Moreover, for simplicity for LABs we only mean \eqref{CondSGleichNullLAB} and \eqref{CondKruemmungmitBLAB} with compatibility conditions because the compatibility conditions on the metrics are either trivial or well-understood.

$\bullet$ Recall Remark \ref{RemarkUeberNablaRhoCurvatureForGauegTrafo}; if we would use $\nabla_\rho$ in general to define the infinitesimal gauge transformation for $K$-valued forms, then we can only expect $R_\delta(\cdot, \cdot) A = 0$ if the basic curvature vanishes and $\nabla_\rho$ is flat; the latter is now trivially satisfied, while the former is one of the compatibility conditions. If doing so, the essential gauge transformations have again the very familiar form,
\ba\label{EqInfGaugeTrafoLABs}
\delta_{\varepsilon(\Phi, A)} A
&=
\mleft(\delta_{\varepsilon} \varpi_2\mright)(\Phi, A)
=
\mleft[ \varepsilon(\Phi, A), A \mright]_{\Phi^*K}
		- \mathrm{d}^{\Phi^*\nabla}\bigl( \varepsilon(\Phi, A) \bigr), \\
\delta_{\varepsilon(\Phi, A)} \Phi &= 0
\ea
for all $\varepsilon \in \mathcal{F}^0_K(M; {}^*K)$ and $(\Phi, A) \in \mathfrak{M}_K(M;N)$.
As usual, the infinitesimal gauge transformation $\delta_\varepsilon G$ of $G$ is then given by (recall Thm.~\ref{thm:RecoverOfClassicInfgGaugeTrafo} and \ref{thm:NewFormulaRecoversOldGaugeTrafoYay})
\ba
(\delta_\varepsilon G)(\Phi, A)
=
\mleft.\frac{\mathrm{d}}{\mathrm{d}t}\mright|_{t=0} \mleft[ t \mapsto
 G\mleft( \Phi, A + t \cdot \delta_{\varepsilon(\Phi, A)} A \mright)
\mright]
\ea
for $t \in \mathbb{R}$. Because of the compatibility conditions \eqref{CondSGleichNullLAB} and \eqref{CondKruemmungmitBLAB} we can derive that $\delta_\varepsilon G$ has the following form
\ba
(\delta_\varepsilon G)(\Phi, A)
&=
\mleft[ \varepsilon(\Phi,A), G(\Phi, A) \mright]_{\Phi^*K}.
\ea
However, we will not need those since we have discussed the gauge transformations thoroughly before, which is why we do not prove this; but it is easy to check as an exercise.
\end{remark}

That is the situation regarding gauge theory and its formalism on Lie algebra bundles. The field redefinition defined earlier has the following simplified form. Recall its properties shown earlier.

\begin{fieldredefinitions}{In the situation of LABs}{FieldRedefForLABs}
Let $\lambda \in \Omega^1(N; K)$, then the field redefinition in the case of LABs leads to the following formulas
\ba
\widetilde{A}^\lambda
&=
A
	+ \mleft( \Phi^*\lambda \mright)(\mathrm{D}\Phi)
=
A
	+ \Phi^! \lambda, \\
\widetilde{\zeta}^\lambda
&=
\zeta
	- \mathrm{d}^\nabla \lambda
	+ \frac{1}{2} \mleft[ \lambda \stackrel{\wedge}{,} \lambda \mright]_K, \label{EqZetaTrafoForLAB}
\ea
and
\ba\label{EqWennFlachDannExaktOderHaltInner}
\widetilde{\nabla}_Y^\lambda \mu
&=
\nabla_Y \mu
	- \mleft[ \lambda(Y) , \mu \mright]_K
\ea
for all $Y \in \mathfrak{X}(N)$ and $\mu \in \Gamma(K)$. The metrics $\kappa$ and $g$ stay the same.
\end{fieldredefinitions}

\begin{remark}
\leavevmode\newline
For Eq.~\eqref{EqWennFlachDannExaktOderHaltInner} we can write 
\ba
\widetilde{\nabla}^\lambda
&=
\nabla
- \mathrm{ad} \circ \lambda,
\ea
where $\mathrm{ad} \circ \lambda \in \Omega^1(N; \mathrm{End}(K))$, $\mleft(\mathrm{ad} \circ \lambda \mright)(Y)(\mu) \coloneqq \mleft[ \lambda(Y), \mu \mright]_K$ for all $Y \in \mathfrak{X}(N)$ and $\mu \in \Gamma(K)$. This implies that 
\bas
(\mathrm{ad} \circ \lambda)(\mu)
&=
\mleft[ \lambda, \mu \mright]_K
=
\mleft[ \lambda \stackrel{\wedge}{,} \mu \mright]_K.
\eas
Similarly, we get $\mathrm{ad} \circ \omega \in \Omega^l(N; \mathrm{End}(K))$.
\end{remark}
\subsection{Relation of vector bundle connections in gauge theories with certain Lie derivation laws} \label{ConnectionIsALieDerivation}

Starting with a CYMH GT using LABs, there is the natural question whether or not one arrives at a (pre-)classical gauge theory by using the field redefinition \ref{fieldredef:FieldRedefForLABs}. We now especially need what we have discussed in Section \ref{SectionOfLABStuff}, most importantly Ex.~\ref{ex:BigCoolDiagramOfMackenzieAboutLABsStuff} which was about the following commuting diagram of Lie algebroid morphisms:
\be\label{theFullDiagramForLABStuff}
	\begin{tikzcd}
		Z(K) \arrow[hook]{d} \arrow[equal]{r} & Z(K) \arrow[hook]{d} \\
		K \arrow{d}{\mathrm{ad}} \arrow[equal]{r} & K \arrow{d} \\
		\mathrm{Der}(K) \arrow[two heads]{d}{\sharp^+} \arrow[hook]{r} & \mathcal{D}_{\mathrm{Der}}(K) \arrow[two heads]{d}{\sharp} \arrow[two heads]{r}{a} & \mathrm{T}N \arrow[equal]{d} \\
		\mathrm{Out}(K) \arrow[hook]{r} & \mathrm{Out}\mleft(\mathcal{D}_{\mathrm{Der}}(K)\mright) \arrow[two heads]{r}{\overline{a}} & \mathrm{T}N
	\end{tikzcd}
\ee
where $K \to N$ is an LAB over a smooth manifold $N$, $Z(K)$ its centre, $\mathcal{D}_{\mathrm{Der}}(K)$ derivations of $K$ which are also Lie bracket derivations, $\mathrm{Der}(K)$ are the same but as endomorphisms, so, the kernel of $a$; and the $\mathrm{Out}$ denotes the quotient over the adjoint of $K$, $\mathrm{ad}(K)$.

In order to understand CYMH GT using LABs, it is important to understand what type of connection $\nabla$ we have due to the compatibility conditions \eqref{CondSGleichNullLAB} and \eqref{CondKruemmungmitBLAB}. We understand vector bundle connections as an anchor-preserving (and base-preserving) vector bundle morphism $\mathrm{T}N \to \mathcal{D}(K)$. For all $Y \in \mathfrak{X}(N)$, compatibility condition \eqref{CondSGleichNullLAB} implies that $\nabla_Y$ is a derivation of the Lie bracket $\mleft[ \cdot, \cdot \mright]_K$ and so of $\mleft[ \cdot, \cdot \mright]_{\mathfrak{g}}$ on each fibre. Thence, the vector bundle morphism $\nabla$ has values in $\mathcal{D}_{\mathrm{Der}}(K)$.

$\mathcal{D}_{\mathrm{Der}}(K)$ is also a Lie subalgebroid of $\mathcal{D}(K)$ as discussed earlier. So, by compatibility condition \eqref{CondSGleichNullLAB}, we arrive at that $\nabla$ has to be what we will call a Lie derivation law:

\begin{definitions}{Lie derivation law, \newline \cite[\S 7.2, special form of Definition 7.2.9, page 275.]{mackenzieGeneralTheory}}{LieConnection}
Let $K \to N$ be an LAB. A \textbf{Lie derivation law} for $\mathrm{T}N$ with coefficients in $K$ is an anchor- and base-preserving vector bundle morphism $\nabla: \mathrm{T}N \to \mathcal{D}_{\mathrm{Der}}(K)$, that is, a connection $\nabla$ on $K$ in the usual sense such that
\ba
\nabla_Y\mleft( \mleft[ \mu, \nu \mright]_K \mright)
&=
\mleft[ \nabla_Y \mu, \nu \mright]_K
	+ \mleft[ \mu, \nabla_Y \nu \mright]_K
\ea
for all $Y \in \mathfrak{X}(N)$ and $\mu, \nu \in \Gamma(K)$.
\end{definitions}

\begin{remark}
\leavevmode\newline
By Thm.~\ref{thm:BLALAB} such a connection always exists for LABs.

In \cite[\S 5.2, second part of Example 5.2.12; page 188f.]{mackenzieGeneralTheory} such a connection is also called Lie connection; Lie derivation laws are actually a bit more general defined, using general Lie algebroids in place of $\mathrm{T}N$. However, we will not need this generalization, but all the references in the following are actually about more general connections; in order to make it easier for the reader who looks up those references, we decided to still use the term Lie derivation law instead to avoid confusion.
\end{remark}
%
%
%
%\begin{proof}
%\leavevmode\newline
%As discussed earlier a connection has a 1:1 correspondence with an anchor-preserving vector bundle morphism $\mathrm{T}N \to \mathcal{D}(V)$, thence, we only need to check the preservation of the Lie brackets in the flat case. We have for the curvature $R_\nabla$
%\bas
%R_\nabla(Y, Z)
%&=
%\mleft[ \nabla_Y, \nabla_Z \mright]_{\mathcal{D}(V)}
	%- \nabla_{[Y, Z]}
%\eas
%such that we have the following set of equivalent statements
%\bas
%&&
%\nabla
%&\text{ is flat} \\
%&\Leftrightarrow&
%\nabla_{[Y, Z]}
%&=
%\mleft[ \nabla_Y, \nabla_Z \mright]_{\mathcal{D}(V)} \\
%&\stackrel{\mathclap{\nabla \text{ morphism of anchored vector bundles}}}{\Leftrightarrow}&
%\nabla
%&\text{ is a base-preserving morphism of Lie algebroids.}
%\eas
%\end{proof}

Now about understanding the compatibility condition \eqref{CondKruemmungmitBLAB}: In the context of the field redefinition, if it would be possible to make $\nabla$ flat by a field redefinition, then there would be a parallel frame $\mleft( e_a \mright)_a$ locally for $\widetilde{\nabla}^\lambda$ such that by Eq. \eqref{EqWennFlachDannExaktOderHaltInner}
\bas
\nabla_Y e_a
&=
\mleft[ \lambda(Y), e_a \mright]_K
\eas
for all $Y \in \mathfrak{X}(N)$. That is, with respect to that frame, the Lie bracket derivation $\nabla_Y$ looks like an adjoint of $\lambda(Y)$, an inner Lie bracket derivation. Thence, it makes sense to look at the previously discussed Lie algebroid of outer derivations \textit{etc.}, which is why we emphasize again to recall the discussion around diagram \eqref{theFullDiagramForLABStuff} in Section \ref{SectionOfLABStuff}.

With diagram \eqref{theFullDiagramForLABStuff} we can now also study compatibility condition \eqref{CondKruemmungmitBLAB}. The curvature $R_\nabla$ of a Lie connection $\nabla:\mathrm{T}N \to \mathcal{D}_{\mathrm{Der}}(K)$ is clearly an element of $\Omega^2\mleft(N; \mathcal{D}_{\mathrm{Der}}(K)\mright)$ since
\bas
R_\nabla(Y, Z)
&=
\underbrace{\mleft[ \nabla_Y, \nabla_Z \mright]_{\mathcal{D}_{\mathrm{Der}}(K)}}_{\in ~ \Gamma(\mathcal{D}_{\mathrm{Der}}(K))}
	- \underbrace{\nabla_{[Y, Z]}}_{\mathclap{\in ~ \Gamma(\mathcal{D}_{\mathrm{Der}}(K))}}
	\in \Gamma(\mathcal{D}_{\mathrm{Der}}(K))
\eas
for all $Y, Z \in \mathfrak{X}(N)$. Compatibility condition \eqref{CondKruemmungmitBLAB} is then equivalent to
\ba
\sharp\mleft( R_\nabla(Y, Z) \mright) = 0
\ea
for all $Y, Z \in \mathfrak{X}(N)$. We will show that this implies that $\nabla$ is a Lie derivation law covering what is called a pairing of $\mathrm{T}N$ with $K$. For that we need to define what a pairing is.\footnote{Mackenzie called the following construction a \textbf{coupling} and not \textbf{pairing}. I renamed it to avoid confusion with couplings in a physical context. Thanks for this suggestion, Alessandra Frabetti.}

\begin{definitions}{Pairing of $\mathrm{T}N$, \cite[\S 7.2, Definitions 7.2.2; page 272]{mackenzieGeneralTheory}}{pairingsOfTNWithK}
A \textbf{pairing} of $\mathrm{T}N$ is a pair of an LAB $K \to N$ together with a (base-preserving) morphism of Lie algebroids $\Xi: \mathrm{T}N \to \mathrm{Out}(\mathcal{D}_{\mathrm{Der}}(K))$. We also say that $\mathrm{T}N$ and $K$ are \textbf{paired by $\Xi$}.
\end{definitions}

Now we can define a special type of connection.

\begin{definitions}{Lie derivation law covering $\Xi$, \newline\cite[\S 7.2, see discussion after Definition 7.2.2; page 272]{mackenzieGeneralTheory}}{LieDerivationLawOverApairingXi}
Let $K \to N$ be an LAB and $\nabla: \mathrm{T}N \to \mathcal{D}_{\mathrm{Der}}(K)$ a Lie derivation law. Assume that $\mathrm{T}N$ and $K$ are paired by a (base-preserving) Lie algebroid morphism $\Xi: \mathrm{T}N \to \mathrm{Out}(\mathcal{D}_{\mathrm{Der}}(K))$.
Then we say that $\nabla$ is a \textbf{Lie derivation law covering $\Xi$} if
\ba
\sharp \circ \nabla
&=
\Xi.
\ea
\end{definitions}

\begin{remark}
\leavevmode\newline
So, while a Lie derivation law is not necessarily a morphism of Lie algebroids, $\sharp \circ \nabla$ is of that type when $\nabla$ covers a pairing.
\end{remark}

This type of connection is exactly the type we need for gauge theory on LABs.

\begin{theorems}{(C)YMH GT only allows Lie derivation laws covering $\Xi$}{GaugeTheoryNeedsLieDerivLawsCoveringApairing}
Let $K \to N$ be an LAB. Then a map $\nabla: \mathrm{T}N \to \mathcal{D}_{\mathrm{Der}}(K)$ is a Lie derivation law covering some (base-preserving) Lie algebroid morphism $\Xi: \mathrm{T}N \to \mathrm{Out}(\mathcal{D}_{\mathrm{Der}}(K))$ if and only if it is a connection on $K$ satisfying the compatibility conditions \eqref{CondSGleichNullLAB} and \eqref{CondKruemmungmitBLAB}, \textit{i.e.}
\bas
\nabla_Y\mleft( \mleft[ \mu, \nu \mright]_K \mright)
&=
\mleft[ \nabla_Y \mu, \nu \mright]_K
	+ \mleft[ \mu, \nabla_Y \nu \mright]_K, \\
\sharp\mleft(R_\nabla(Y, Z)\mright)
&=
0
\eas
for all $Y, Z \in \mathfrak{X}(N)$ and $\mu, \nu \in \Gamma(K)$.
\end{theorems}

\begin{remark}\label{remExistenceOfLieDerivationLawsCoveringApairing}
\leavevmode\newline
So, we have seen that compatibility condition \eqref{CondSGleichNullLAB} implies that $\nabla$ has to be a Lie derivation law, and compatibility condition \eqref{CondKruemmungmitBLAB} then implies that it covers a pairing of $\mathrm{T}N$ and $K$.

As argued in \cite[\S 7.2, discussion after Definition 7.2.2, replace the $A$ there with $\mathrm{T}N$; page 272]{mackenzieGeneralTheory}, for a given $\Xi$ there is always a Lie derivation law covering it. As a sketch, that follows by the construction and definition of $\sharp$ given by Prop.~\ref{prop:QuotientsOfTransitiveLAOids}, \textit{i.e.}~it is a surjective submersion, such that the existence of a map $\nabla: \mathrm{T}N \to \mathcal{D}_{\mathrm{Der}}(K)$ with $\sharp \circ \nabla = \Xi$ follows, $\nabla$ is a vector bundle morphism, since $\sharp$ and $\Xi$ are; finally, we have by diagram \eqref{theFullDiagramForLABStuff} $\overline{a} \circ \sharp = a$ and $\Xi$ is anchor-preserving, so, $\overline{a}\circ\Xi= \mathds{1}_{\mathrm{T}N}$, such that we can apply $\overline{a}$ on both side of $\sharp\circ\nabla= \Xi$ to get 
\bas
a\circ\nabla&= \mathds{1}_{\mathrm{T}N}.
\eas
Therefore $\nabla$ is also anchor-preserving and, thus, a vector bundle connection.
\end{remark}

\begin{proof}
\leavevmode\newline
We already have seen that a connection $\nabla$ satisfying compatibility condition \eqref{CondSGleichNullLAB} has a 1:1 correspondence to an anchor-preserving vector bundle morphism $\nabla: \mathrm{T}N \to \mathcal{D}_{\mathrm{Der}}(K)$, \textit{i.e.} a Lie derivation law. So, we only have to care about compatibility condition \eqref{CondKruemmungmitBLAB}.

"$\Leftarrow$": So, let us have a Lie derivation law with additionally $\sharp\mleft(R_\nabla(Y, Z)\mright) = 0$ for all $Y, Z \in \mathfrak{X}(N)$. Define $\Xi \coloneqq \sharp \circ \nabla$, and recall that $\sharp: \mathcal{D}_{\mathrm{Der}}(K) \to \mathrm{Out}(\mathcal{D}_{\mathrm{Der}}(K))$ is a Lie algebroid morphism such that $\Xi$ is an anchor-preserving vector bundle morphism by definition, using that $\nabla$ is a Lie derivation law,
\bas
\overline{a} \circ \Xi
&=
\overline{a} \circ \sharp \circ \nabla
=
a \circ \nabla
= \mathds{1}_{\mathrm{T}N}.
\eas
Using that $\sharp$ is a homormorphism of Lie brackets, and by $\sharp\mleft(R_\nabla(Y, Z)\mright) = 0$ for all $Y, Z \in \mathfrak{X}(N)$, we also get
\bas
\Xi([Y, Z])
&=
\sharp \mleft( \nabla_{[Y, Z]} \mright)
\\
&=
\sharp \mleft( \mleft[ \nabla_Y, \nabla_Z \mright]_{\mathcal{D}_{\mathrm{Der}}(K)} \mright) 
\\
&=
\mleft[ \sharp\mleft(\nabla_Y\mright), \sharp\mleft(\nabla_Z\mright) \mright]_{\mathrm{Out}(\mathcal{D}_{\mathrm{Der}}(K))}
\\
&=
\mleft[ \Xi(Y), \Xi(Z) \mright]_{\mathrm{Out}(\mathcal{D}_{\mathrm{Der}}(K))},
\eas
\textit{i.e.} $\Xi$ is a Lie algebroid morphism (base-preserving), and it is covered by $\nabla$ due to its definition.

"$\Rightarrow$": This part of the proof is as in \cite[\S 7.2, discussion after Definition 7.2.2; page 272]{mackenzieGeneralTheory} and similar to the previous calculation. Let $\nabla$ be a Lie derivation law covering some Lie algebroid morphism $\Xi$, especially, $\sharp \circ \nabla = \Xi$. That implies
\bas
\sharp\mleft( R_\nabla(Y, Z) \mright)
&=
\sharp\mleft( \mleft[ \nabla_Y, \nabla_Z \mright]_{\mathcal{D}_{\mathrm{Der}}(K)} - \nabla_{[Y, Z]} \mright) \\
&=
\mleft[ \sharp(\nabla_Y), \sharp(\nabla_Z) \mright]_{\mathrm{Out}(\mathcal{D}_{\mathrm{Der}}(K))} - \sharp\mleft(\nabla_{[Y, Z]}\mright) \\
&=
\mleft[ \Xi(Y), \Xi(Z) \mright]_{\mathrm{Out}(\mathcal{D}_{\mathrm{Der}}(K))} - \Xi([Y, Z]) \\
&=
0
\eas
for all $Y, Z \in \mathfrak{X}(N)$, using that both, $\sharp$ and $\Xi$, are homomorphisms of the corresponding Lie brackets. This finishes the proof.
\end{proof}

Given a Lie derivation law covering some $\Xi$, we get that $\nabla$ is an anchor-preserving vector bundle morphism and $\sharp \circ \nabla = \Xi$ is a Lie algebroid morphism. When we want that $\nabla$ is not flat, in the hope of finding a new gauge theory (recall Cor. \ref{cor:ManBrauchZetaWahrscheinlich}), we do not want that $\nabla$ itself is a Lie algebroid morphism by Cor. \ref{cor:FlatConnectionsAreLieAlgebroidMorphisms}, while $\sharp$ is a Lie algebroid morphism and $\Xi = \sharp \circ \nabla$, too. That looks like a tightrope walk. But there are a lot of non-flat Lie derivation laws covering some $\Xi$, we may see some in the following parts, so, constructing non-flat connections for a gauge theory is not impossible. But the field redefinition \ref{fieldredef:FieldRedefForLABs} may still lead to a flat connection while keeping the same physics, \textit{i.e.} the Lagrangian stays the same.

%When one could interpret $\sharp \circ \nabla$ as typical vector bundle connection on $K$ (which it isn't in general, we just assume that for now), then Cor. \ref{cor:FlatConnectionsAreLieAlgebroidMorphisms} would imply that $\sharp \circ \nabla$ is a flat connection such that locally a parallel frame of $\sharp \circ \nabla$ would exist. That would suggest that there is locally a frame $\mleft( e_a \mright)_a$ such that
%\bas
%\nabla_Y e_a
%&=
%\mleft[ \lambda(Y), e_a \mright]_K
%\eas
%for some $\lambda \in \Omega^1(N; K)$. By the field redefinition Eq. \eqref{EqWennFlachDannExaktOderHaltInner} that would lead to a flat connection $\widetilde{\nabla}^\lambda$, and so we arrive at a description using a flat connection.
%
%$\sharp \circ \nabla$ is in general not a connection of $K$, so this was not really a proof, just a gedankenexperiment to motivate possible results. But indeed, we are going to show that there is locally always a field redefinition in the sense of \ref{fieldredef:FieldRedefForLABs} such that $\widetilde{\nabla}^\lambda$ is flat. But globally that might be different.

To study this we now need to construct an invariant for the field redefinition.
%We will show that this is exactly the so-called \textit{obstruction class} as Mackenzie constructed in \cite[\S 7.2; page 271ff.]{mackenzieGeneralTheory}. 
Observe the following, using the notation as introduced in \eqref{theFullDiagramForLABStuff}.

\begin{propositions}{Field redefinition preserves the pairing}{FieldRedefPreservespairing}
Let $(K, \Xi)$ be a pairing of $\mathrm{T}N$, $\nabla$ be a Lie derivation law covering $\Xi$ and $\zeta \in \Omega^2(N;K)$ satisfying compatibility condition \eqref{CondKruemmungmitBLAB} with respect to $\nabla$.

Then the field redefinition \ref{fieldredef:FieldRedefForLABs} preserves the pairing, \textit{i.e.}~$\widetilde{\nabla}^\lambda$ is also a Lie derivation law covering $\Xi$ for all $\lambda \in \Omega^1(N;K)$. Moreover, for every other Lie derivation law $\nabla^\prime$ covering $\Xi$ there is a $\lambda \in \Omega^1(N; K)$ such that
\bas
\nabla^\prime &= \widetilde{\nabla}^\lambda
\eas
and for its curvature
\bas
R_{\nabla^\prime} &= \mathrm{ad} \circ \widetilde{\zeta}^\lambda.
\eas
\end{propositions}

\begin{remark}
\leavevmode\newline
These are exactly the same formulas as in \cite[\S 7.2, Proposition 7.2.7, identifying Mackenzie's 1-form $l$ with $- \lambda$, also keep in mind that Mackenzie defines curvatures with an opposite sign; page 274]{mackenzieGeneralTheory}. In this reference Mackenzie studies the form given by the difference of two Lie derivation laws covering the same pairing and arrives exactly at our formulas of the field redefinition which we have derived from a more general context of gauge theory on Lie algebroids.

In this work the context is given by field redefinitions of a gauge theory, while Mackenzie studies these connections in the context of extending Lie algebroids by Lie algebra bundles (over the same base) such that their Whitney sum admits a Lie algebroid structure. Hence, in the following we will see that Mackenzie's study about extensions has a 1:1 correspondence to the question whether one can find a field redefinition such that $\widetilde{\nabla}^\lambda$ is flat.
\end{remark}

\begin{proof}[Proof of Prop.~\ref{prop:FieldRedefPreservespairing}]
\leavevmode\newline
By Thm.~\ref{thm:InvarianceUnderTheFieldRedefinition} we know that the field redefinition preserves the compatibility conditions \eqref{CondSGleichNullLAB} and \eqref{CondKruemmungmitBLAB}, \textit{i.e.}
\bas
\widetilde{\nabla}^\lambda_Y\mleft( \mleft[ \mu, \nu \mright]_K \mright)
&=
\mleft[ \widetilde{\nabla}^\lambda_Y \mu, \nu \mright]_K
	+ \mleft[ \mu, \widetilde{\nabla}^\lambda_Y \nu \mright]_K, \\
R_{\widetilde{\nabla}^\lambda}(Y, Z) \mu
&=
\mleft[ \widetilde{\zeta}^\lambda(Y, Z), \mu \mright]_K,
\eas
that implies by Thm.~\ref{thm:GaugeTheoryNeedsLieDerivLawsCoveringApairing} that $\widetilde{\nabla}^\lambda$ is a Lie derivation law covering $\widetilde{\Xi}^\lambda \coloneqq \sharp \circ \widetilde{\nabla}^\lambda$. Moreover, using the notation \eqref{theFullDiagramForLABStuff},
\bas
\sharp \circ \widetilde{\nabla}^\lambda
&=
\sharp \circ \mleft( \nabla - \mathrm{ad} \circ \lambda \mright)
=
\sharp \circ \nabla
=
\Xi
\eas
for all $\lambda \in \Omega^1(N; K)$, using $\sharp \circ \mathrm{ad} = 0$. This shows that $\widetilde{\nabla}^\lambda$ covers $\Xi$.

Now let $\nabla^\prime$ be another Lie derivation law covering $\Xi$, then clearly
\bas
a|_{\mathcal{D}_{\mathrm{Der}}(K)}(\nabla^\prime_Y - \nabla_Y)
&= Y- Y = 0
\eas
for all $Y \in \mathfrak{X}(N)$, such that $\nabla^\prime - \nabla \in \Omega^1(N; \mathrm{Der}(K))$ by \eqref{theFullDiagramForLABStuff}, and
\bas
0
&=
\Xi - \Xi
=
\sharp \circ \nabla^\prime
	- \sharp \circ \nabla
=
\sharp \circ \underbrace{\mleft( \nabla^\prime - \nabla \mright)}_{\mathclap{\in ~ \Omega^1(N; \mathrm{Der}(K))}}
=
\sharp^+ \circ \mleft( \nabla^\prime - \nabla \mright).
\eas
Again by \eqref{theFullDiagramForLABStuff}, there is a $\mu(Y) \in \Gamma(K)$ such that $\nabla^\prime_Y - \nabla_Y = \mathrm{ad}(\mu(Y))$ for all $Y \in \mathfrak{X}(N)$, and due to the $C^\infty$-linearity w.r.t. $Y$ we get $\nabla^\prime - \nabla = \mathrm{ad} \circ \mu$ for a $\mu \in \Omega^1(N; K)$. By field redefinition \ref{fieldredef:FieldRedefForLABs} we can take $\lambda = - \mu$ to get $\nabla^\prime = \widetilde{\nabla}^\lambda$.

Since $\nabla$ satisfies compatibility condition \eqref{CondKruemmungmitBLAB} by Thm.~\ref{thm:GaugeTheoryNeedsLieDerivLawsCoveringApairing} and since this condition is preserved by a field redefinition, the last statement follows, $R_{\nabla^\prime}(Y, Z) = \mathrm{ad}\mleft( \widetilde{\zeta}^\lambda(Y, Z) \mright)$ for all $Y, Z \in \mathfrak{X}(N)$.
\end{proof}

%This result is basically equivalent to the result given in \cite[\S 7.2, Proposition 7.2.7; page 274]{mackenzieGeneralTheory}, but coming from a different context:

%\begin{proposition}[Difference of two Lie derivation laws covering the same pairing] \cite[\S 7.2, Proposition 7.2.7, identifying Mackenzie's 1-form $l$ with $- \lambda$, also beware that Mackenzie defines curvatures with an opposite sign; page 274]{mackenzieGeneralTheory}
%\leavevmode\newline
%%Let $K \to N$ be an LAB, and let $\nabla$ and $\nabla^\prime$ be two Lie derivation laws covering the same (base-preserving) Lie algebroid morphism $\Xi: \mathrm{T}N \to \mathrm{Out}(\mathcal{D}_{\mathrm{Der}}(K))$.
%Let $(K, \Xi)$ be a pairing of $\mathrm{T}N$, and let $\nabla$ and $\nabla^\prime$ be two Lie derivation laws covering $\Xi$.
%
%Then there exists a $\lambda \in \Omega^1(N; K)$ such that
%\ba
%\nabla^\prime
%&=
%\nabla - \mathrm{ad} \circ \lambda,
%\ea
%and for their curvatures we get
%\ba
%R_{\nabla^\prime}
%&=
%R_\nabla
	%+ \mathrm{ad}\mleft( - \mathrm{d}^\nabla \lambda + \frac{1}{2} \mleft[ \lambda \stackrel{\wedge}{,} \lambda \mright]_K \mright).
%\ea
%\end{proposition}

Locally we can say the following.

\begin{corollaries}{Local existence of a flat Lie derivation law covering a pairing}{CorLocalerFlacherZusammenhangFuerIrgendeineKopplung}
Let $K$ be an LAB. Then locally there is always a flat Lie derivation law covering some (base-preserving) Lie algebroid morphism $\Xi: \mathrm{T}N \to \mathrm{Out}(\mathcal{D}_{\mathrm{Der}}(K))$.
\end{corollaries}

\begin{remark}
\leavevmode\newline
So, locally, by using Prop.~\ref{prop:FieldRedefPreservespairing}, the question whether or not one can transform to a flat connection with the field redefinition breaks down to the question if there is a flat connection covering the same pairing.
\end{remark}

\begin{proof}
\leavevmode\newline
Locally there is a trivialization $K \cong U \times \mathfrak{g}$ as LABs on some open subset $U \subset N$. Then define $\nabla$ as the canonical flat connection, and by Thm.~\ref{thm:ActionLieALgebroid} we know that it has vanishing basic curvature, so, it satisfies compatibility condition \eqref{CondSGleichNullLAB}; compatibility condition \eqref{CondKruemmungmitBLAB} is trivially satisfied by the flatness.

By Thm.~\ref{thm:GaugeTheoryNeedsLieDerivLawsCoveringApairing} the statement follows.
\end{proof}

\subsection{Obstruction for non-pre-classical gauge theories}\label{MackenzieZeugsUndExistenzvonPreclassical}

Using the previous subsection, let us now look at whether or not we can make the connection flat by a field redefinition. For such questions it is useful to have an invariant; actually, $\mathrm{d}^\nabla \zeta$ is invariant under the field redefinition.

\begin{propositions}{$\mathrm{d}^\nabla \zeta$ an invariant of the field redefinition, \newline \cite[\S 7.2, Proposition 7.2.11, last statement, there $\zeta$ is denoted by $\Lambda$ and $\mathrm{d}^\nabla \zeta$ by $f(\nabla, \Lambda)$; page 276]{mackenzieGeneralTheory}}{InvarianteFuerFieldRedefImFallLAB}
Let $(K, \Xi)$ be a pairing of $\mathrm{T}N$ and $\nabla$ be a Lie derivation law covering $\Xi$. Also let $\zeta$ be any element of  $\Omega^2(N; K)$ that satisfies compatibility condition \eqref{CondKruemmungmitBLAB} with respect to $\nabla$.

Then $\mathrm{d}^\nabla \zeta$ is invariant under the field redefinition \ref{fieldredef:FieldRedefForLABs}, \textit{i.e.}
\ba
\mathrm{d}^{\widetilde{\nabla}^\lambda} \widetilde{\zeta}^\lambda
&=
\mathrm{d}^\nabla \zeta.
\ea
\end{propositions}

\begin{proof}
\leavevmode\newline
Recall that in general curvatures satisfy
\bas
\mleft( \mathrm{d}^\nabla \mright)^2 \omega = R_\nabla \wedge \omega
\eas
for all $\omega \in \Omega^l(N;K)$, viewing $R_\nabla$ as an element of $\Omega^2(N; \mathrm{End}(K))$. Then we have
\bas
\mleft( \mathrm{d}^\nabla \mright)^2 \lambda
&=
R_\nabla \wedge \lambda
\stackrel{\text{Eq.~\eqref{CondKruemmungmitBLAB}}}{=}
(\mathrm{ad} \circ \zeta) \wedge \lambda
\stackrel{\text{Eq.~\eqref{wedgeproduktmitadLambdaergibtLieklammer}}}{=}
\mleft[ \zeta \stackrel{\wedge}{,} \lambda \mright]_K, \\
\mathrm{d}^\nabla \mleft( \mleft[ \lambda \stackrel{\wedge}{,} \lambda \mright]_K \mright)
~~~~&\stackrel{\mathclap{\text{Eq.~\eqref{eqDerivationOfDifferentialOnBracketonK}}}}{=}~~~~
\mleft[ \mathrm{d}^\nabla \lambda \stackrel{\wedge}{,} \lambda \mright]_K
	- \mleft[ \lambda \stackrel{\wedge}{,} \mathrm{d}^\nabla \lambda \mright]_K
\stackrel{\text{Eq.~\eqref{VertauschungsregelForKKlammerAufFormen}}}{=}
2 ~ \mleft[ \mathrm{d}^\nabla \lambda \stackrel{\wedge}{,} \lambda \mright]_K, \\
( \mathrm{ad} \circ \lambda ) \wedge \widetilde{\zeta}^\lambda
~~~~&\stackrel{\mathclap{\text{Eq.~\eqref{wedgeproduktmitadLambdaergibtLieklammer}}}}{=}~~~~
\mleft[ \lambda \stackrel{\wedge}{,} \widetilde{\zeta}^\lambda \mright]_K
\stackrel{\text{Eq.~\eqref{VertauschungsregelForKKlammerAufFormen}}}{=}
- \mleft[ \widetilde{\zeta}^\lambda \stackrel{\wedge}{,} \lambda \mright]_K
\stackrel{\text{Eq.~\eqref{EqZetaTrafoForLAB},~\eqref{JacobiIdentityForFormBracket}}}{=}
- \mleft[ \zeta \stackrel{\wedge}{,} \lambda \mright]_K
	+ \mleft[ \mathrm{d}^\nabla \lambda \stackrel{\wedge}{,} \lambda \mright]_K,
\eas
and, by combining everything, we arrive at
\bas
\mathrm{d}^{\widetilde{\nabla}^\lambda} \widetilde{\zeta}^\lambda
&=
\mathrm{d}^{\nabla - \mathrm{ad} \circ \lambda} \mleft( \widetilde{\zeta}^\lambda \mright)
\stackrel{\text{Eq.~\eqref{eqDifferentialSplit},~\eqref{EqZetaTrafoForLAB}}}{=}
\mathrm{d}^\nabla \mleft( \zeta
	- \mathrm{d}^\nabla \lambda
	+ \frac{1}{2} \mleft[ \lambda \stackrel{\wedge}{,} \lambda \mright]_K \mright)
	- \mleft( \mathrm{ad} \circ \lambda \mright) \wedge \widetilde{\zeta}^\lambda
=
\mathrm{d}^\nabla \zeta
\eas
for all $\lambda \in \Omega^1(N;K)$.
\end{proof}

Therefore let us study $\mathrm{d}^\nabla \zeta$. Earlier we have shown what the (second) Bianchi identity for $R_\nabla$, $\mathrm{d}^\nabla R_\nabla = 0$, implies for $\zeta$ under using the compatibility condition \eqref{CondKruemmungmitBLAB}; recall Thm.~\ref{thm:BianchiIdentityForZeta}. Let us state what this means in the situation of LABs.

\begin{propositions}{Bianchi identity for $\zeta$}{BianchiIdentityForZeta}
Let $(K, \Xi)$ be a pairing of $\mathrm{T}N$ and $\nabla$ be a Lie derivation law covering $\Xi$. Also let $\zeta$ be any element of  $\Omega^2(N; K)$ that satisfies compatibility condition \eqref{CondKruemmungmitBLAB} with respect to $\nabla$.

Then we have 
\bas
\mathrm{d}^\nabla \zeta &\in \Omega^3(N; Z(K)),
\eas
\textit{i.e.}~$\mathrm{d}^\nabla \zeta$ has always values in the centre of $K$.
\end{propositions}

\begin{remark}
\leavevmode\newline
This is equivalent to \cite[\S 7.2, Lemma 7.2.4, $\zeta$ is denoted as $\Lambda$ there; page 273]{mackenzieGeneralTheory}. Mackenzie shows it by direct calculation in that special situation, while we derive it from the previous, more general result.
\end{remark}

\begin{proof}
\leavevmode\newline
By Thm.~\ref{thm:BianchiIdentityForZeta}, which clearly reduces to the following in the case of LABs (insert $\rho = 0$)
\bas
\mleft[ \mathrm{d}^\nabla \zeta(Y_1, Y_2, Y_3), \mu \mright]_K
&= 0
\eas
for all $Y_1, Y_2, Y_3 \in \mathfrak{X}(N)$, and $\mu \in \Gamma(K)$. That proves the claim.
\end{proof}

In fact, $\mathrm{d}^\nabla$ is a differential on centre-valued forms.

\begin{theorems}{Differential on centre-valued forms, \newline \cite[\S 7.2, Definition 7.2.3 and the discussion directly before; page 273]{mackenzieGeneralTheory}}{DifferentialAufZentrumsDinge}
Let $(K, \Xi)$ be a pairing. Then every Lie derivation law $\nabla$ covering $\Xi$ restricts to a flat connection $\nabla^{Z(K)}$ on $Z(K)$.

Moreover, $\Xi$ induces a differential $\mathrm{d}^\Xi: \Omega^\bullet(N; Z(K)) \to \Omega^{\bullet+1}(N; Z(K))$ by choosing $\mathrm{d}^\Xi \coloneqq \mathrm{d}^{\nabla^{Z(K)}} = \mleft.\mathrm{d}^\nabla\mright|_{\Omega^\bullet(N; Z(K))}$ for any Lie derivation law $\nabla$ covering $\Xi$. $\mathrm{d}^\Xi$ is independent of the choice of $\nabla$.

We call this differential \textbf{central representation of $\Xi$}.
\end{theorems}

\begin{remark}
\leavevmode\newline
Recall the second paragraph of Remark \ref{remExistenceOfLieDerivationLawsCoveringApairing}, \textit{i.e.}~there is a Lie derivation Law $\nabla: \mathrm{T}N \to \mathcal{D}_{\mathrm{Der}}(K)$ covering $\Xi$. Hence, $\mathrm{d}^\Xi$ always exists for a given $\Xi$.
\end{remark}

\begin{proof}[Proof of Thm.~\ref{thm:DifferentialAufZentrumsDinge}]
\leavevmode\newline
By Thm.~\ref{thm:GaugeTheoryNeedsLieDerivLawsCoveringApairing} $\nabla$ satisfies compatibility conditions
\bas
\nabla_Y\mleft( \mleft[ \mu, \nu \mright]_K \mright)
&=
\mleft[ \nabla_Y \mu, \nu \mright]_K
	+ \mleft[ \mu, \nabla_Y \nu \mright]_K, \\
R_\nabla(Y, Z)
&=
\mathrm{ad}(\zeta(Y, Z))
\eas
for all $Y, Z \in \mathfrak{X}(N)$, $\mu, \nu \in \Gamma(K)$ and for some $\zeta \in \Omega^2(N; K)$. Let $\mu \in \Gamma(Z(K))$, then the first compatibility condition implies
\bas
0 &= \mleft[ \nabla_Y \mu, \nu \mright]_K
\eas
for all $Y \in \mathfrak{X}(N)$, $\nu \in \Gamma(K)$ and $\mu \in \Gamma(Z(K))$. That implies that $\nabla_Y \mu \in \Gamma(Z(K))$ such that $\nabla$ is also a connection on $\Gamma(Z(K))$, which we now denote by $\nabla^{Z(K)}$. Restricting the second compatibility condition onto $Z(K)$ then immediately implies
\bas
R_{\nabla^{Z(K)}} &= 0,
\eas
\textit{i.e.}~$\nabla^{Z(K)}$ is flat, and therefore, by the definition of the exterior covariant derivative,
\bas
\mathrm{d}^\Xi &\coloneqq \mleft.\mathrm{d}^\nabla\mright|_{\Omega^\bullet(N; Z(K))} = \mathrm{d}^{\nabla^{Z(K)}}
\eas
is a differential. Now take any other Lie derivation law $\nabla^\prime$ covering $\Xi$. By Prop. \ref{prop:FieldRedefPreservespairing}, there is a $\lambda \in \Omega^1(N; K)$ such that
\bas
\nabla^\prime
&=
\nabla - \mathrm{ad} \circ \lambda,
\eas
\textit{i.e.}
\bas
\nabla^\prime_Y \mu
&=
\nabla_Y \mu
\eas
for all $Y \in \mathfrak{X}(N)$ and $\mu \in \Gamma(Z(K))$. Hence, $\mathrm{d}^\Xi$ is independent of the choice of $\nabla$.
\end{proof}

One can now check that $\mathrm{d}^\nabla \zeta$ is closed under $\mathrm{d}^\Xi$. Be aware of that for non-flat Lie derivation laws $\nabla$ covering $\Xi$ this is not an obviously trivial question; due to compatibility condition \eqref{CondKruemmungmitBLAB}, $\zeta$ is not centre-valued in general such that $\mathrm{d}^\nabla \zeta$ cannot be written as $\mathrm{d}^\Xi \zeta$.

\begin{lemmata}{Closedness of $\mathrm{d}^\nabla \zeta$ under the central representation, \newline \cite[\S 7.2, Lemma 7.2.5, $\mathrm{d}^\nabla \zeta$ is denoted by $f$ and $\mathrm{d}^\Xi$ as $d$, and without written proof there; page 274]{mackenzieGeneralTheory}}{DNablaZetaIsClosedUnderDXi}
Let $(K, \Xi)$ be a pairing of $\mathrm{T}N$ and $\nabla$ be a Lie derivation law covering $\Xi$. Also let $\zeta$ be any element of  $\Omega^2(N; K)$ that satisfies compatibility condition \eqref{CondKruemmungmitBLAB} with respect to $\nabla$.

Then 
\ba
\mathrm{d}^\Xi \mathrm{d}^\nabla \zeta
&=
0
\ea
\textit{i.e.}~$\mathrm{d}^\nabla \zeta \in \Omega^3(N; Z(K))$ is closed under $\mathrm{d}^\Xi$.
\end{lemmata}

\begin{proof}
\leavevmode\newline
%This follows by Eq.~\eqref{DNablaZetaIsClosed}.
We have
\bas
\mleft( \mathrm{d}^\nabla \mright)^2 \zeta
&=
R_\nabla \wedge \zeta
\stackrel{\text{Eq.~\eqref{CondKruemmungmitBLAB}}}{=}
\mleft( \mathrm{ad} \circ \zeta \mright) \wedge \zeta
\stackrel{\text{Eq.~\eqref{wedgeproduktmitadLambdaergibtLieklammer}}}{=}
\mleft[ \zeta \stackrel{\wedge}{,} \zeta \mright]_K,
\eas
but also, using that $\zeta \in \Omega^2(N;K)$,
\bas
\mleft[ \zeta \stackrel{\wedge}{,} \zeta \mright]_K
\stackrel{\text{Eq.~\eqref{VertauschungsregelForKKlammerAufFormen}}}{=}
- \mleft[ \zeta \stackrel{\wedge}{,} \zeta \mright]_K,
\eas
such that $\mleft( \mathrm{d}^\nabla \mright)^2 \zeta = - \mleft( \mathrm{d}^\nabla \mright)^2 \zeta$. Hence, the last statement follows.
\end{proof}

We need to know how $\mathrm{d}^\nabla \zeta$ changes by varying $\zeta$.

\begin{lemmata}{Varying $\zeta$ in $\mathrm{d}^\nabla \zeta$, \newline \cite[\S 7.2, Lemma 7.2.6, Mackenzie denotes $\zeta$ by $\Lambda$, $\mathrm{d}^\nabla \zeta$ by $f$ and $\mathrm{d}^\Xi$ by $d$; page 274]{mackenzieGeneralTheory}}{ZetaKannGutGeaendertWerden}
Let $(K, \Xi)$ be a pairing of $\mathrm{T}N$ and $\nabla$ be a Lie derivation law covering $\Xi$. Also let $\zeta$ and $\zeta^\prime$ be two elements of  $\Omega^2(N; K)$ which satisfy compatibility condition \eqref{CondKruemmungmitBLAB} with respect to $\nabla$.

Then
\ba
\zeta^\prime - \zeta \in \Omega^2(N; Z(K)).
\ea
Especially, $\mathrm{d}^\nabla\zeta^\prime - \mathrm{d}^\nabla\zeta$ is $\mathrm{d}^\Xi$-exact. 
\end{lemmata}

\begin{proof}
\leavevmode\newline
This simply follows by the compatibility condition \eqref{CondKruemmungmitBLAB}, \textit{i.e.}
\bas
\mleft[ \zeta^\prime(Y, Z) - \zeta(Y, Z), \mu \mright]_K
&=
R_\nabla(Y, Z) \mu - R_\nabla(Y, Z) \mu
= 0
\eas
for all $Y, Z \in \mathfrak{X}(N)$ and $\mu \in \Gamma(K)$. Thence, $\xi \coloneqq \zeta^\prime - \zeta$ is an element of $\Omega^2(N; Z(K))$. By Thm. \ref{thm:DifferentialAufZentrumsDinge} we get
\bas
\mathrm{d}^\nabla\zeta^\prime - \mathrm{d}^\nabla\zeta
&=
\mathrm{d}^\nabla\underbrace{\mleft(\zeta^\prime - \zeta\mright)}_{\mathclap{\in \Omega^2(N; Z(K))}}
=
\mathrm{d}^\Xi\mleft(\zeta^\prime - \zeta\mright),
\eas
\textit{i.e.} $\mathrm{d}^\nabla\zeta^\prime - \mathrm{d}^\nabla\zeta$ is exact with respect to $\mathrm{d}^\Xi$ since $\zeta^\prime - \zeta$ has values in $Z(K)$.
\end{proof}

Since $\mathrm{d}^\nabla \zeta$ is invariant under the field redefinition, this finally shows that $\mathrm{d}^\nabla \zeta$ is a useful object to study in the context of the field redefinition. By Lemma \ref{lem:DNablaZetaIsClosedUnderDXi} this is a closed form, and it is clear that in the flat situation $\zeta$ has values in $Z(K)$ by compatibility condition \eqref{CondKruemmungmitBLAB}. By Thm.~\ref{thm:DifferentialAufZentrumsDinge} we would get $\mathrm{d}^\nabla \zeta = \mathrm{d}^\Xi \zeta$, \textit{i.e.}~$\mathrm{d}^\nabla \zeta$ would be then exact. Hence, it makes sense to study the cohomology class of $\mathrm{d}^\nabla \zeta$ with respect to $\mathrm{d}^\Xi$ if one is interested into whether or not the gauge theory can be transformed into a pre-classical\footnote{Recall Def.~\ref{def:ClassicalGT}.} gauge theory by the field redefinitions.

We denote the space of cohomology classes of $\mathrm{d}^\Xi$-closed elements of $\Omega^\bullet(N; Z(K))$ by
\ba
\mathcal{H}^\bullet\mleft(\mathrm{T}N, \mathrm{d}^\Xi, Z(K)\mright)
\ea
as in \cite[Theorem 7.2.12, replace $A$ with $\mathrm{T}N$ and $\rho^\Xi$ with $\mathrm{d}^\Xi$; page 277]{mackenzieGeneralTheory}, and the classes by $\mleft[ \cdot \mright]_\Xi$. Thus,
\bas
\mleft[ \mathrm{d}^\nabla \zeta \mright]_\Xi &\in \mathcal{H}^3\mleft(\mathrm{T}N, \mathrm{d}^\Xi, Z(K)\mright),
\eas
using that $\mathrm{d}^\nabla \zeta$ is $\mathrm{d}^\Xi$-closed by Lemma \ref{lem:DNablaZetaIsClosedUnderDXi}.

\begin{theorems}{Cohomology of $\mathrm{d}^\nabla \zeta$ an invariant, \newline \cite[\S 7.2, Theorem 7.2.12, Mackenzie denotes $\mathrm{d}^\Xi$ with $\rho^\Xi$, $\zeta$ with $\Lambda$, $\mathrm{d}^\nabla \zeta$ with $f(\nabla, \Lambda)$, and replace $A$ with $\mathrm{T}N$; page 277]{mackenzieGeneralTheory}}{ObstructionClassIstGeileInvariante}
Let $(K, \Xi)$ be a pairing of $\mathrm{T}N$ and $\nabla$ be a Lie derivation law covering $\Xi$. Also let $\zeta$ be any element of  $\Omega^2(N; K)$ that satisfies compatibility condition \eqref{CondKruemmungmitBLAB} with respect to $\nabla$.

Then $\mleft[ \mathrm{d}^\nabla \zeta \mright]_\Xi$ only depends on $\Xi$ and not on the particular choice of $\nabla$ and $\zeta$.
\end{theorems}

\begin{proof}
\leavevmode\newline
This follows by Lemma \ref{lem:ZetaKannGutGeaendertWerden} and Prop.~\ref{prop:InvarianteFuerFieldRedefImFallLAB}. The former shows that changing $\zeta$ with another element $\zeta^\prime$ of $\Omega^2(N; K)$ satisfying compatibility condition \eqref{CondKruemmungmitBLAB} results into
\bas
\mathrm{d}^\nabla \zeta^\prime
&=
\mathrm{d}^\nabla \zeta
	+ \underbrace{\mathrm{d}^\nabla \mleft( \zeta^\prime - \zeta \mright)}_{\mathrm{d}^\Xi\text{-exact}}
\in \mleft[ \mathrm{d}^\nabla \zeta \mright]_\Xi,
\eas
\textit{i.e.} $\mleft[\mathrm{d}^\nabla \zeta^\prime\mright]_\Xi = \mleft[\mathrm{d}^\nabla \zeta \mright]_\Xi$, and the latter shows 
\bas
\mleft[ \mathrm{d}^{\widetilde{\nabla}^\lambda} \widetilde{\zeta}^\lambda \mright]_\Xi
&=
\mleft[ \mathrm{d}^{\nabla} \zeta \mright]_\Xi.
\eas
Thence, by using Prop.~\ref{prop:FieldRedefPreservespairing}, \textit{i.e.}~one can reach every other Lie derivation law covering $\Xi$ by using the field redefinition \ref{fieldredef:FieldRedefForLABs}, one can freely change the Lie derivation law covering $\Xi$ by Prop.~\ref{prop:InvarianteFuerFieldRedefImFallLAB}, and by Lemma \ref{lem:ZetaKannGutGeaendertWerden} it does not matter which $\zeta$ is used.
\end{proof}

This clearly motivates the following definition of Mackenzie's obstruction class.

\begin{definitions}{The obstruction class of pairings, \newline \cite[\S 7.2, comment after Theorem 7.2.12; page 277]{mackenzieGeneralTheory}}{ObstructionClassOfXi}
Let $(K, \Xi)$ be a pairing of $\mathrm{T}N$, and let $\nabla$ be any Lie derivation law covering $\Xi$. Also let $\zeta$ be any element of  $\Omega^2(N; K)$ that satisfies compatibility condition \eqref{CondKruemmungmitBLAB} with respect to $\nabla$.

Then we define the \textbf{obstruction class of $\Xi$} by
\ba
\mathrm{Obs}(\Xi)
&\coloneqq
\mleft[ \mathrm{d}^\nabla \zeta \mright]_\Xi.
\ea
\end{definitions}

We immediately get a first result related to CYMH GT.

\begin{corollaries}{First approach of obstruction for CYMH GT on LABs}{FirstApproachOfLABConstruction}
Let $(K, \Xi)$ be a pairing of $\mathrm{T}N$, and let $\nabla$ be a fixed Lie derivation law covering $\Xi$.

Then we have
\bas
\exists \text{ a field redefinition as in \ref{fieldredef:FieldRedefForLABs}}: ~ \widetilde{\nabla}^\lambda \text{ is flat}
\quad&\Rightarrow\quad
\mathrm{Obs}(\Xi) = 0\in\mathcal{H}^3\mleft(\mathrm{T}N, \mathrm{d}^\Xi, Z(K)\mright).
\eas
Or, equivalently, if there is a flat Lie derivation law covering $\Xi$, then $\mathrm{Obs}(\Xi) = 0$.
\end{corollaries}
%
%\begin{remark}
%\leavevmode\newline
%This implies that for any given pairing $(K, \Xi)$ with $\mathrm{Obs}(\Xi) \neq 0$ and with fibre type $\mathfrak{g}$ admitting an $\mathrm{ad}$-invariant scalar product we can define a gauge theory which can not be transformed to a (pre-)classical gauge theory by using a field redefinition. For that, recall the second paragraph of Remark \ref{remExistenceOfLieDerivationLawsCoveringApairing} about the existence of $\nabla$ for a given $\Xi$.
%\end{remark}

\begin{proof}[Proof of Cor.~\ref{cor:FirstApproachOfLABConstruction}]
\leavevmode\newline
Let $\zeta$ be any element of $\Omega^2(N; K)$ that satisfies compatibility condition \eqref{CondKruemmungmitBLAB} with respect to $\nabla$. When there is a field redefinition such that $\widetilde{\nabla}^\lambda$ is flat then we can conclude that $\widetilde{\zeta}^\lambda$ has only values in $Z(K)$ by compatibility condition \eqref{CondKruemmungmitBLAB}. But then we arrive at
\bas
\mathrm{Obs}(\Xi)
&=
\mleft[ \mathrm{d}^\nabla \zeta \mright]_\Xi
\stackrel{\text{Prop. \ref{prop:InvarianteFuerFieldRedefImFallLAB}}}{=}
\mleft[ \mathrm{d}^{\widetilde{\nabla}^\lambda} \widetilde{\zeta}^\lambda \mright]_\Xi
\stackrel{\text{Thm. \ref{thm:DifferentialAufZentrumsDinge}}}{=}
\mleft[ \mathrm{d}^{\Xi} \widetilde{\zeta}^\lambda \mright]_\Xi
=
0.
\eas
The equivalence to the last statement simply follows by using Prop. \ref{prop:FieldRedefPreservespairing}.
\end{proof}

\subsection{Mackenzie's theory about extensions of tangent bundles}\label{MackenzieStuff}

We now want to study when the obstruction is zero and when it implies the existence of a flat Lie derivation law covering $\Xi$. To understand this, we need to understand why Mackenzie studied this obstruction class. Mackenzie was interested into whether or not a Lie algebroid can be extended by an LAB; we are going to state Mackenzie's statements in the special situation of having $\mathrm{T}N$ as the Lie algebroid. But the arguments and calculations do not really differ; in the context of gauge theory we just need to study $\mathrm{T}N$. Recall Def.~\ref{def:ExtensionOfTNByLABs} about extensions and transversals; there will be now another Lie algebroid $E$ besides the LAB $K$, and the anchor of $E$ we will denote by $\pi$ instead of $\rho$ to avoid confusion with $\rho = 0$ of $K$. This $E$ is not the same $E$ as in the context of CYMH GT; the Lie algebroid for CYMH GT in this section is $K$ as we have introduced it.

To a given transversal we are able to define a Lie derivation law covering some Lie algebroid morphism $\Xi: \mathrm{T}N \to \mathrm{Out}(\mathcal{D}_{\mathrm{Der}}(K))$.

\begin{propositions}{Lie derivation law of a transversal, \newline \cite[\S 7.3, Proposition 7.3.2 and Lemma 7.3.3, replace $A$ with $\mathrm{T}N$ and $A^\prime$ with $E$; page 278]{mackenzieGeneralTheory}}{TransversalAndItsLieDerivationLaw}
Let
\begin{center}
	\begin{tikzcd}
		K \arrow[hook]{r}{\iota} & E \arrow[two heads]{r}{\pi} & \mathrm{T}N.
	\end{tikzcd}
\end{center}
be an extension of $\mathrm{T}N$ by an LAB $K \to N$, and let $\chi$ be any transversal. Then a connection $\nabla^\chi$ on $K$, given by
\ba\label{DefTransversalConnection}
\iota\mleft( \nabla^\chi_Y \mu \mright)
&=
\mleft[ \chi(Y), \iota(\mu) \mright]_E
\ea
for all $Y \in \mathfrak{X}(N)$ and $\mu \in \Gamma(K)$, describes a Lie derivation law covering some Lie algebroid morphism $\Xi: \mathrm{T}N \to \mathrm{Out}(\mathcal{D}_{\mathrm{Der}}(K))$.
\end{propositions}

\begin{proof}
\leavevmode\newline
Let us discuss why Eq. \eqref{DefTransversalConnection} is well-defined and giving rise to a vector bundle morphism $\nabla^\chi:\mathrm{T}N \to \mathcal{D}(K)$. $\iota$ is an injective\footnote{This follows by the exactness of the given sequence.} Lie algebroid morphism and embedding such that we can identify $K$ and $\iota(K)$ as LABs; since the kernel of $\pi$ is given by the image of $\iota$ we know that any element $\xi \in \Gamma(E)$ with $\pi(\xi) = 0$ is also an element of $\Gamma(\iota(K))$ and has, thus, a 1:1 correspondence in $\Gamma(K)$ given by $\iota^{-1}(\xi)$. Due to that $\pi$ is a homomorphism of of Lie brackets and by $\pi \circ \iota = 0$, we have
\bas
\pi\mleft( \mleft[ \chi(Y), \iota(\mu) \mright]_E \mright)
&= 0
\eas
for all $Y \in \mathfrak{X}(N)$ and $\mu \in \Gamma(K)$. It follows that the right hand side of Eq. \eqref{DefTransversalConnection} defines an element of $\Gamma(K)$. Hence, it is valid to define $\nabla^\chi_Y$ as some map on $\Gamma(K)$ by using Eq. \eqref{DefTransversalConnection} for all $Y \in \mathfrak{X}(N)$. Additionally, for all $Y, Z \in \mathfrak{X}(N)$, $\mu, \nu \in \Gamma(K)$, $f, h \in C^\infty(N)$ and $\alpha, \beta \in \mathbb{R}$ we have
\bas
\iota\mleft( \nabla^\chi_{fY+hZ} \mu \mright)
&=
\mleft[ \chi(fY+hZ), \iota(\mu) \mright]_E
\\
&=
\mleft[ f \chi(Y)+h \chi(Z), \iota(\mu) \mright]_E 
\\
&\stackrel{\mathclap{\pi \circ \iota = 0}}{=}\quad
f ~ \mleft[ \chi(Y), \iota(\mu) \mright]_E
	+ h ~ \mleft[ \chi(Z), \iota(\mu) \mright]_E
\\
&=
\iota\mleft( f \nabla^\chi_{Y} \mu + h \nabla^\chi_{Z} \mu \mright),
\eas
also
\bas
\iota\bigl( \nabla^\chi_{Y} \mleft( \alpha \mu + \beta \nu \mright) \bigr)
&=
\mleft[ \chi(Y), \iota(\alpha \mu + \beta \nu) \mright]_E
=
\alpha \mleft[ \chi(Y), \iota(\mu) \mright]_E
	+ \beta \mleft[ \chi(Y), \iota(\nu) \mright]_E
=
\iota\mleft( \alpha \nabla^\chi_{Y} \mu + \beta \nabla^\chi_{Y} \nu \mright),
\eas
and
\bas
\iota\mleft( \nabla^\chi_{Y}(f \mu) \mright)
&=
\mleft[ \chi(Y), f \iota(\mu) \mright]_E
\stackrel{\pi \circ \chi = \mathds{1}_{\mathrm{T}N}}{=}
f~ \iota\mleft( \nabla^\chi_{Y} \mu \mright)
	+ \mathcal{L}_Y(f) ~ \iota(\mu)
=
\iota\mleft( f ~\nabla^\chi_{Y} \mu + \mathcal{L}_Y(f) ~ \mu \mright).
\eas
Moreover,
\bas
\iota \mleft( \nabla^\chi_Y \mleft( \mleft[ \mu, \nu \mright]_K \mright) \mright)
&=
[ \chi(Y), \underbrace{\iota(\mleft[ \mu, \nu \mright]_K)}_{=\mleft[ \iota(\mu), \iota(\nu) \mright]_E} ]_E
\\
&=
\mleft[ \mleft[ \chi(Y), \iota(\mu) \mright]_E, \iota(\nu) \mright]_E
	+ \mleft[ \iota(\mu), \mleft[ \chi(Y), \iota(\nu) \mright]_E \mright]_E 
\\
&=
\mleft[ \iota\mleft( \nabla^\chi_Y \mu \mright), \iota(\nu) \mright]_E
	+ \mleft[ \iota(\mu), \iota\mleft( \nabla^\chi_Y \nu \mright) \mright]_E
\\
&=
\iota\mleft(\mleft[ \nabla^\chi_Y \mu, \nu \mright]_K \mright)
	+ \iota\mleft(\mleft[ \mu, \nabla^\chi_Y \nu \mright]_K \mright) 
\\
&=
\iota\mleft(\mleft[ \nabla^\chi_Y \mu, \nu \mright]_K + \mleft[ \mu, \nabla^\chi_Y \nu \mright]_K \mright)
\eas
using the Jacobi identity for $\mleft[ \cdot, \cdot \mright]_E$. Thence, $\nabla^\chi$ is a Lie derivation law. By Thm. \ref{thm:GaugeTheoryNeedsLieDerivLawsCoveringApairing} we are left showing whether $\sharp \circ R_{\nabla^\chi} = 0$,
\ba
\iota\mleft( R_{\nabla^\chi}(Y, Z) \mu \mright)
&=
\mleft[ \chi(Y), \mleft[ \chi(Z), \iota(\mu) \mright]_E \mright]_E
	- \mleft[ \chi(Z), \mleft[ \chi(Y), \iota(\mu) \mright]_E \mright]_E
	- \mleft[ \chi([Y, Z]), \iota(\mu) \mright]_E \nonumber\\
&=
\mleft[ \mleft[ \chi(Y), \chi(Z) \mright]_E, \iota(\mu) \mright]_E
	- \mleft[ \chi([Y, Z]), \iota(\mu) \mright]_E \nonumber\\
&= [ \underbrace{\mleft[ \chi(Y), \chi(Z) \mright]_E - \chi([Y, Z])}_{= R_\chi (Y, Z)}, \iota(\mu) ]_E \nonumber\\
&=\label{EqKruemmungderTransversalen}
\mleft[R_\chi (Y, Z), \iota(\mu) \mright]_E,
\ea
using again the Jacobi identity for $\mleft[ \cdot, \cdot \mright]_E$ and that $\iota$ is a Lie algebroid morphism, where $R_\chi$ is the \textbf{curvature of $\chi$} as defined in Def.~\ref{def:GeneralDefOfCurvMorphisms}, which is a tensor by Lemma \ref{lem:KruemmungenSindTensorenMitAnkerErhaltung} and by the fact that $\chi$ is a transversal, that is, $\chi$ is anchor-preserving. Observe
\bas
\pi\mleft( R_\chi(Y, Z) \mright)
&=
\mleft[ (\pi\circ\chi)(Y), (\pi\circ\chi)(Z) \mright] - (\pi\circ\chi)([Y, Z])
\stackrel{\pi \circ \chi = \mathds{1}_{\mathrm{T}N}}{=}
0,
\eas
using that $\pi$ is a Lie algebroid morphism. Therefore $R_\chi(Y, Z) \in \iota(K)$ for all $Y, Z \in \mathfrak{X}(N)$, and, so, Eq. \eqref{EqKruemmungderTransversalen} implies
\ba
R_{\nabla^\chi}(Y, Z)
&=
\mleft(\mathrm{ad} \circ \iota^{-1}\mright)(R_\chi(Y, Z))
\ea
using that $\iota$ is an injective Lie algebroid morphism. By \eqref{theFullDiagramForLABStuff} we get $\sharp \circ R_{\nabla^\chi} = 0$, and the statement follows.
\end{proof}

Furthermore, the pairing covered by $\nabla^\chi$ is the same for all transversals $\chi$.

\begin{corollaries}{All transversals results into the same covered pairing, \newline \cite[\S 7.3, comment after Lemma 7.3.3, replace $A$ with $\mathrm{T}N$ and $A^\prime$ with $E$; page 278]{mackenzieGeneralTheory}}{TransversalsCoverTheSamepairing}
Let
\begin{center}
	\begin{tikzcd}
		K \arrow[hook]{r}{\iota} & E \arrow[two heads]{r}{\pi} & \mathrm{T}N.
	\end{tikzcd}
\end{center}
be an extension of $\mathrm{T}N$ by an LAB $K \to N$, and let $\chi$ and $\chi^\prime$ be two transversals.

Then
\bas
\sharp \circ \nabla^\chi
&=
\sharp \circ \nabla^{\chi^\prime}.
\eas
\end{corollaries}

\begin{proof}
\leavevmode\newline
Since $\chi$ and $\chi^\prime$ are transversals we get
\bas
\pi \circ \mleft( \chi(Y) - \chi^\prime(Y) \mright)
&=
Y -Y
= 0,
\eas
for all $Y \in \mathfrak{X}(N)$, such that, again by the exactness of the sequence, there is a $\mu(Y) \in \Gamma(K)$ with $\chi(Y) - \chi^\prime(Y) = \iota(\mu(Y))$. Due to the $C^\infty$-linearity of the transversals we even have a vector bundle morphism $\mu: \mathrm{T}N \to K$ such that
\bas
\chi - \chi^\prime
&=
\iota \circ \mu,
\eas
such that
\bas
\nabla^\chi_Y \nu
&\stackrel{\text{Eq. \eqref{DefTransversalConnection}}}{=}
\mleft[ \chi(Y), \iota(\nu) \mright]_E
=
\mleft[ \chi^\prime(Y), \iota(\nu) \mright]_E
	+ \underbrace{\mleft[ \iota(\mu(Y)), \iota(\nu) \mright]_E}_{= \iota\mleft( \mleft[ \mu(Y), \nu \mright]_K \mright)}
=
\iota\mleft( \nabla^{\chi^\prime}_Y \nu + \mleft[ \mu(Y), \nu \mright]_K \mright)
\eas
for all $Y \in \mathfrak{X}(N)$ and $\nu \in \Gamma(K)$. Therefore
\bas
\nabla^\chi
&=
\nabla^{\chi^\prime}
	+ \mathrm{ad}\circ\mu,
\eas
thus, by \eqref{theFullDiagramForLABStuff},
\bas
\sharp \circ \nabla^\chi
&=
\sharp \circ \nabla^{\chi^\prime}.
\eas
\end{proof}

This immediately leads to the following definition.

\begin{definitions}{Pairing induced by an extension, \newline \cite[\S7.3, Definition 7.3.4, replace $A$ with $\mathrm{T}N$ and $A^\prime$ with $E$; page 278]{mackenzieGeneralTheory}}{pairingsOfExtensions}
Let
\begin{center}
	\begin{tikzcd}
		K \arrow[hook]{r}{\iota} & E \arrow[two heads]{r}{\pi} & \mathrm{T}N.
	\end{tikzcd}
\end{center}
be an extension of $\mathrm{T}N$ by an LAB $K \to N$, and let $\chi$ be any transversal.

Then the pairing $\Xi_{\mathrm{ext}} \coloneqq \sharp \circ \nabla^\chi: \mathrm{T}N \to \mathrm{Out}\mleft( \mathcal{D}_{\mathrm{Der}}(K) \mright)$ is the \textbf{pairing of $\mathrm{T}N$ with $K$ induced by the extension}.
\end{definitions}

Finally we can state what Mackenzie has shown about the obstruction class.

\begin{theorems}{Obstruction of an extension, \newline \cite[\S 7.3, Proposition 7.3.6, page 279, Corollary 7.3.9 and the comment afterwards, page 281; replace $A$ with $\mathrm{T}N$ and $A^\prime$ with $E$]{mackenzieGeneralTheory}}{ObstructionOfExtensions}
Let $(K, \Xi)$ be a pairing of $\mathrm{T}N$.

Then there is an extension
\begin{center}
	\begin{tikzcd}
		K \arrow[hook]{r}{\iota} & E \arrow[two heads]{r}{\pi} & \mathrm{T}N
	\end{tikzcd}
\end{center}
of $\mathrm{T}N$ by $K$ such that $\Xi_{\mathrm{ext}} = \Xi$ if and only if $\mathrm{Obs}(\Xi) = 0 \in \mathcal{H}^3\mleft(\mathrm{T}N, \mathrm{d}^\Xi, Z(K)\mright)$. Moreover, given such an extension, then for all Lie derivation laws $\nabla$ covering $\Xi$ there is a transversal $\chi$ such that
\bas
\nabla
&=
\nabla^\chi.
\eas
\end{theorems}

\begin{proof}
\leavevmode\newline
We only give a sketch; for the full proof please see the reference. We especially need the part of the proof starting with a zero obstruction class. Given a zero obstruction class, fix a Lie derivation law $\nabla$ covering $\Xi$, and let $\zeta$ be any element of  $\Omega^2(N; K)$ that satisfies compatibility condition \eqref{CondKruemmungmitBLAB} with respect to $\nabla$. First, additionally following \cite[Proposition 7.2.13; page 277]{mackenzieGeneralTheory}, that is, $\mathrm{Obs}(\Xi) = 0$ implies that there is an $h \in \Omega^2(N; Z(K))$ with
\bas
\mathrm{d}^\nabla \zeta 
&=
\mathrm{d}^\Xi h
\stackrel{ \text{Thm.~\ref{thm:DifferentialAufZentrumsDinge}} }{=}
\mathrm{d}^\nabla h,
\eas 
then define $\zeta^\prime \coloneqq \zeta - h$ such that clearly $\mathrm{d}^\nabla \zeta^\prime = 0$. Observe,
\bas
R_\nabla
&\stackrel{ \eqref{CondKruemmungmitBLAB} }{=}
\mathrm{ad}\circ\zeta
=
\mathrm{ad} \circ \zeta^\prime.
\eas
Define 
\bas
E 
&\coloneqq
\mathrm{T}N \oplus K
\eas
be the vector bundle given as the Whitney sum of $K$ and $\mathrm{T}N$. The anchor is just the projection onto the first factor, and define the bracket by
\bas
\mleft[
	(Y, \nu), (Z, \mu)
\mright]_E
&\coloneqq
\mleft(
	[Y, Z], \mleft[ \nu, \mu \mright]_K + \nabla_Y \mu - \nabla_Z \nu - \zeta^\prime(Y, Z)
\mright)
\eas
for all $(Y, \nu), (Z, \mu) \in E$. It is trivial to check that the Leibniz rule is with respect to the chosen anchor, bilinearity and antisymmetry are also clear. Hence, one essentially needs to check the Jacobi identity: This is a straightforward calculation resulting into a big sum. All the terms will cancel each other by the Jacobi identity of $\mleft[ \cdot, \cdot \mright]_K$; and there will be terms where $\nabla$ will act on the Lie bracket and terms where adjoints act on $\nabla$ such that these cancel each other by using that $\nabla$ has values in $\mathcal{D}_{\mathrm{Der}}(K)$; moreover, one also gets clearly the curvature of $\nabla$ and adjoints of $\zeta^\prime$ which will cancel the curvature terms by $R_\nabla = \mathrm{ad} \circ \zeta^\prime$; finally, there are also terms where $\nabla$ acts on $\zeta^\prime$ and $\zeta^\prime$ is contracted in one factor with terms like $[Y, Z]$, and all these terms will result into $\mathrm{d}^\nabla \zeta^\prime$ which is zero by construction. Hence, Jacobi identity will be given and, thus, a Lie algebroid structure.

For the other direction, that is, now assume that we have an extension with $\Xi_{\mathrm{ext}} = \Xi$, one first shows that there is a transversal $\chi$ with $\nabla^\chi = \nabla$; this is as in the proof of \cite[Proposition 7.3.6; page 279]{mackenzieGeneralTheory}, and we also omit the notation of $\iota$ now again, assuming the standard inclusion, for simplicity in the notation. For any transversal $\chi^\prime$ we have $\sharp \circ \nabla = \sharp \circ \nabla^{\chi^\prime}$ due to $\Xi = \Xi_{\mathrm{ext}}$, that leads to that there is a field redefinition by Prop.~\ref{prop:FieldRedefPreservespairing} with $\lambda \in \Omega^1(N; K)$ such that
\bas
\nabla
&=
\nabla^{\chi^\prime} 
	+ \mathrm{ad} \circ \lambda
=
\mathrm{ad} \circ \mleft(\chi^\prime + \lambda\mright)
=
\nabla^\chi,
\eas
using the definition of connections like $\nabla^{\chi^\prime}$, where $\chi \coloneqq \chi^\prime + \lambda$ and $\mathrm{ad}$ is of course using the Lie bracket of $E$, possibly restricting onto the bracket of $K$. Recall Def.~\ref{def:GeneralDefOfCurvMorphisms}, by the calculation of Eq.~\eqref{EqKruemmungderTransversalen} we have
\bas
R_\nabla
&=
R_{\nabla^\chi}
=
\mathrm{ad} \circ R_\chi,
\eas
hence, $R_\chi$ is a possible primitive (which is how we actually called $\zeta$), satisfying compatibility condition \eqref{CondKruemmungmitBLAB} with respect to $\nabla$. We want to calculate $\mathrm{d}^{\nabla^\chi} R_\chi$ in order to study $\mathrm{Obs}(\Xi)$, so,
\bas
\mleft(\mathrm{d}^{\nabla^\chi} R_\chi\mright)(X, Y, Z)
&=
\underbrace{\nabla^\chi_X \bigl( R_\chi(Y, Z) \bigr)}
_{= \mleft[ \chi(X), R_\chi(Y, Z) \mright]_E}
	- \nabla^\chi_Y \bigl( R_\chi(X, Z) \bigr)
	+ \nabla^\chi_Z \bigl( R_\chi(X, Y) \bigr)
\\
&\hspace{1cm}
	- R_\chi([X, Y], Z)
	+ R_\chi([X, Z], Y)
	- R_\chi([Y, Z], X)
\\
&=
\sigma\Bigl(
	\mleft[ \chi(X), \mleft[ \chi(Y), \chi(Z) \mright]_E \mright]_E
	- \mleft[ \chi(X), \chi\bigl( [Y, Z] \bigr) \mright]_E
\\
&\hspace{1cm}\hphantom{\sigma \Bigl(}
	- \mleft[\chi\bigl([X, Y]\bigr), \chi(Z)\mright]_E
	+ \chi\bigl( [X, Y], Z \bigr)
\Bigr)
\\
&=
0
\eas
for all $X, Y, Z \in \mathfrak{X}(N)$, where $\sigma$ denotes the cyclic sum through $X, Y, Z$ and where we used the Jacobi identity of $\mleft[ \cdot, \cdot \mright]$ and $\mleft[ \cdot,\cdot \mright]_E$. Thus, trivially $\mathrm{Obs}(\Xi) = 0$.
\end{proof}

By Cor. \ref{cor:FirstApproachOfLABConstruction} we see that the question about whether there is a field redefinition in sense of \ref{fieldredef:FieldRedefForLABs} to arrive at a pre-classical gauge theory, \textit{i.e.}~when $\nabla$ is flat, is related to the existence of an extension of $\mathrm{T}N$ by $K$.

When we are just interested into local behaviours then we might assume that $N$ is contractible.

\begin{theorems}{Extensions over contractible manifolds, \newline \cite[\S 8.2, Theorem 8.2.1, replace $A$ with $E$, $L$ with $K$ and $TM$ with $\mathrm{T}N$; page 314ff.]{mackenzieGeneralTheory}}{ExtensionWennNContrahierbar}
Let
\begin{center}
	\begin{tikzcd}
		K \arrow[hook]{r}{\iota} & E \arrow[two heads]{r}{\pi} & \mathrm{T}N.
	\end{tikzcd}
\end{center}
be an extension of $\mathrm{T}N$ by an LAB $K$ over a contractible manifold $N$. Then there is a flat Lie derivation law covering $\Xi_{\mathrm{Ext}}$.\footnote{Mackenzie stated that $E$ admits a flat connection, with that they actually mean that it is a flat Lie derivation law covering $\Xi_{\mathrm{Ext}}$.}
\end{theorems}

\begin{proof}
\leavevmode\newline
The proof of this theorem is very long and needs a lot of preparation, therefore this would sadly exceed this work; thence, see the reference of this statement. The essential idea is that this is the generalization of the infinitesimal analogue about that a principal bundle admits a global section over a contractible base. Mackenzie's proof is about generalizing the proof of principal bundles where the base is contracted and homotopy classification of bundles is used. In order to do something similar, Mackenzie introduces a certain cohomology theory in \cite[\S 7; page 257ff.]{mackenzieGeneralTheory}; in parts we already introduced the basics for it.
\end{proof}

\subsection{Results}\label{LABResultsWooooo}

In total we derive therefore the following two statements, the first can be seen as a generalization of Cor.~\ref{cor:CorLocalerFlacherZusammenhangFuerIrgendeineKopplung}.

\begin{theorems}{Local existence of pre-classical gauge theory}{LokalLeiderImmerPreklassisch}
Let $(K, \Xi)$ be a pairing of $\mathrm{T}N$ over a contractible manifold $N$, and let $\nabla$ be a fixed Lie derivation law covering $\Xi$.

Then we have a field redefinition in sense of \ref{fieldredef:FieldRedefForLABs} making $\nabla$ flat, \textit{i.e.}~there is a $\lambda\in\Omega^1(N;K)$ such that $\widetilde{\nabla}^\lambda$ is flat.
\end{theorems}

\begin{proof}
\leavevmode\newline
We only need to show that $\mathrm{Obs}(\Xi) = \mleft[ \mathrm{d}^\nabla \zeta \mright]_\Xi = 0$, where $\zeta \in \Omega^2(N; K)$ such that compatibility condition \eqref{CondKruemmungmitBLAB} is satisfied. As given in Thm.~\ref{thm:DifferentialAufZentrumsDinge} the central representation $\mathrm{d}^\Xi$ of $\Xi$ is basically $\mathrm{d}^{\nabla^{Z(K)}}$ where $\nabla^{Z(K)}$ is $\nabla$ restricted on the subbundle $Z(K)$, and we have shown that $\nabla^{Z(K)}$ is flat by compatibility condition \eqref{CondKruemmungmitBLAB}. Due to the fact that $N$ is contractible, we have a global parallel frame $\mleft( e_a \mright)_a$ for $Z(K)$ with respect to $\nabla^{Z(K)}$.

By Prop.~\ref{prop:BianchiIdentityForZeta} we have $\mathrm{d}^\nabla \zeta \in \Omega^3(N; Z(K))$, thence, we can write $\mathrm{d}^\nabla \zeta = \omega^a \otimes e_a$ with $\omega^a \in \Omega^3(N)$. We arrive at
\bas
\mathrm{d}^\Xi \mathrm{d}^\nabla \zeta
&=
\mathrm{d}\omega^a \otimes e_a,
\eas
where $\mathrm{d}$ is the standard de-Rham differential. So, the differential breaks down to the standard differential in each component, especially closedness and exactness mean to be closed and exact in each component with respect to $\mleft( e_a \mright)_a$, respectively. By Lemma \ref{lem:DNablaZetaIsClosedUnderDXi} we have $\mathrm{d}^\Xi \mathrm{d}^\nabla \zeta=0$, thus, $\mathrm{d}\omega^a = 0$. Again due to that $N$ is contractible, we can conclude that closedness implies exactness by the Poincaré lemma. Thence, $\mathrm{Obs}(\Xi) = 0$.

By Thm.~\ref{thm:ObstructionOfExtensions} we have an extension
\begin{center}
	\begin{tikzcd}
		K \arrow[hook]{r}{\iota} & E \arrow[two heads]{r}{\pi} & \mathrm{T}N.
	\end{tikzcd}
\end{center}
such that $\Xi_{\mathrm{ext}} = \Xi$, and, hence, a flat Lie derivation law covering $\Xi$ by Thm.~\ref{thm:ExtensionWennNContrahierbar}. By Prop.~\ref{prop:FieldRedefPreservespairing} the existence of the field redefinition to a flat derivation law covering $\Xi$ follows.
\end{proof}

\begin{theorems}{Possible new and curved gauge theories on LABs}{NeueLABGTs}
Let $(K, \Xi)$ be a pairing of $\mathrm{T}N$ with $\mathrm{Obs}(\Xi) \neq 0$ and such that the fibre Lie algebra $\mathfrak{g}$ admits an $\mathrm{ad}$-invariant scalar product.

Then we can construct a CYMH GT for which there is no field redefinition with what it would become pre-classical.
\end{theorems}

\begin{proof}
\leavevmode\newline
Take any Lie derivation law $\nabla$ covering $\Xi$ (recall the second paragraph of Remark \ref{remExistenceOfLieDerivationLawsCoveringApairing} about the existence of $\nabla$ for a given $\Xi$). By Thm.~\ref{thm:GaugeTheoryNeedsLieDerivLawsCoveringApairing} this connection satisfies compatibility conditions \eqref{CondSGleichNullLAB} and \eqref{CondKruemmungmitBLAB}. Together with the existence of an $\mathrm{ad}$-invariant scalar product we have everything what we need to construct a CYMH GT in sense of \ref{sit:CYMHGTForLABsToDoList}.

Due to $\mathrm{Obs}(\Xi) \neq 0$ and Cor.~\ref{cor:FirstApproachOfLABConstruction} the statement follows.
\end{proof}

Hence, we have shown that $\mathrm{Obs}(\Xi)$ is not just an obstruction for extensions of $\mathrm{T}N$, it also leads to an obstruction for the question about whether or not a CYMH GT can be transformed to a pre-classical gauge theory by a field redefinition. However, Mackenzie also has shown that there are examples with zero obstruction class but without a flat Lie derivation law covering the pairing. Thus, there is in general only for contractible $N$ an equivalence of $\mathrm{Obs}(\Xi) = 0$ and the existence of flat Lie derivation laws covering a pairing.

\begin{examples}{The isotropy of a Hopf fibration, \newline \cite[Example 7.3.20; page 287]{mackenzieGeneralTheory}}{HopfBuendelEventuellSuperFragezeichen}
$\bullet$ Let $P$ be the Hopf fibration
\begin{center}
	\begin{tikzcd}
		\mathrm{SU}(2) \arrow{r}	& \mathds{S}^7 \arrow{d} \\
			& \mathds{S}^4
	\end{tikzcd}
\end{center}
Then for the adjoint bundle
\bas
K
&\coloneqq
P \times_{\mathrm{SU}(2)} \mathrm{su}(2)
\coloneqq 
\mleft( \mathds{S}^7 \times \mathrm{su}(2) \mright) \Big/ \mathrm{SU}(2)
\eas
we have the Atiyah sequence
\begin{center}
	\begin{tikzcd}
		K \arrow[hook]{r}{\iota} & E\coloneqq \mathrm{T}P \Big/ \mathrm{SU}(2) \arrow[two heads]{r}{\pi} & \mathrm{T}\mathds{S}^4.
	\end{tikzcd}
\end{center}
of $\mathrm{T}\mathds{S}^4$ by $K$. We can view this sequence as an extension.

Then $\mathrm{Obs}(\Xi_{\mathrm{Ext}}) = 0$ because of the fact that $K$ is semisimple, but there is no flat derivation law, especially no flat derivation law covering $\Xi_{\mathrm{Ext}}$.

$\bullet$ We are not going to prove this, because introducing Atiyah sequences \textit{etc.}~would certainly exceed this work, since we will not need these notions in the following again. Hence, see the reference for the proof; for the definition of Atiyah sequences see \cite[\S 3.1 and \S3.2; page 86ff.]{mackenzieGeneralTheory}. The main idea about the definition of Atiyah sequences however is to observe that the Lie group behind the definition of a principal bundle $P \stackrel{p}{\to} N$, $N$ a smooth manifold, also acts on $\mathrm{T}P$ by the differential of left- (or right-) multiplication. Due to how the Lie group acts on $P$ it is trivial to see that it also restricts to an action on the vertical bundle, which is isomorphic to $P \times \mathfrak{g}$ since its trivialization are the induced fundamental vector fields. $\mathrm{D}p$ projects $\mathrm{T}P$ onto $\mathrm{T}\mathds{S}^4$ and the vertical bundle is its kernel; one can show that this is preserved by the chosen quotients over the Lie group action. This leads to such short exact sequences, the Atiyah sequences.

$\bullet$ In case you do not know the construction of this Hopf bundle, see \textit{e.g.}~\cite[Example 4.2.14; page 214ff.]{hamilton}; the construction is basically that we view $\mathds{S}^7$ as unit octonions and $\mathrm{SU}(2) \cong \mathds{S}^3$ as unit quaternions, an action of $\mathds{S}^3$ on $\mathds{S}^7$ is then canonically given. Taking the quotient of $\mathds{S}^7$ over $\mathds{S}^3$ is precisely the quaternionic projective line which is isomorphic to $\mathds{S}^4$.

$\bullet$ As other Hopf fibrations, this Hopf fibration is not trivial. Hence, the idea of the proof is to show that a flat Lie derivation law covering $\Xi_{\mathrm{ext}}$ would imply a trivialization of this Hopf fibration. A sketch: First observe that the adjoint of $E$ of any section of $E$ induces an element of $\mathcal{D}_{\mathrm{Der}}(K)$ if restricted onto $K$; due to that $K$ is the kernel of $E$'s anchor, this even defines an $E$-connection on $K$. Since $\mathrm{su}(2)$ is semisimple this induces an isomorphism $E \to \mathcal{D}_{\mathrm{Der}}(K)$. Then one can argue that a flat Lie derivation law would induce a flat connection on the Hopf bundle; $\mathds{S}^4$ is simply connected such that this implies a trivialization of this Hopf bundle. Which would be clearly a contradiction.
\end{examples}

\begin{remarks}{Hopf bundle as an example for CYMH GT}{HopfBundleIstEventuellDasBeispiel}
The fibre of $K$ is given by $\mathrm{su}(2)$, and, thence, the existence of an $\mathrm{ad}$-invariant scalar product is given. Therefore this gives an example of a CYMH GT as in~\ref{sit:CYMHGTForLABsToDoList} by taking any fibre metric $\kappa$ on $K$ which restricts to an $\mathrm{ad}$-invariant scalar product on each fibre, and taking any Lie derivation law $\nabla$ covering $\Xi_{\mathrm{Ext}}$, and, so, the existence of a $\zeta \in \Omega^2(N;K)$ as in compatibility condition~\eqref{CondKruemmungmitBLAB} is given. By Prop.~\ref{prop:FieldRedefPreservespairing} this example shows that there is no field redefinition as in~\ref{fieldredef:FieldRedefForLABs} such that this gauge theory would become pre-classical.

In \cite{TwoQubits} is a relationship of two-qubit systems, as arising in quantum computational science, and precisely this Hopf fibration shown. This may or may not prove any physical significance of this example. At least it may give hints towards a further study related to this example.
\end{remarks}

\begin{remark}
\leavevmode\newline
Observe that a trivial semisimple LAB would not work: Fix any global frame $\mleft( e_a \mright)_a$ of the trivial LAB, then we would have $\nabla e_a = \mleft[ \lambda , e_a \mright]_K$ for a $\lambda \in \Omega^1(N; K)$ because all bracket derivations are inner derivations for semisimple Lie algebras; for this, simply view the connection 1-forms $\omega_a^b$, given by $\nabla e_a = \omega_a^b \otimes e_b$, as matrices acting on constant (w.r.t. $\mleft( e_a \mright)_a$) sections. Then $\widetilde{\nabla}^\lambda$ would be flat, and its parallel frame is \textit{e.g.}~given by $\mleft( e_a \mright)_a$. This argument just depends on the triviality of the LAB, regardless whether the base is contractible or not. The obstruction class is of course always trivial for semisimple LABs because their centre is zero.
\end{remark}

\subsection{Existence of non-vanishing primitives stable under the field redefinition} \label{NonclassicalStuff}

When one is interested into perturbation theory, especially just in a local theory, then Thm.~\ref{thm:LokalLeiderImmerPreklassisch} seems to show that locally one can not hope for new gauge theories, especially ones related to non-flat $\nabla$. However, we still have the two-form $\zeta$. We can transform every CYMH GT locally to pre-classical ones by Thm.~\ref{thm:LokalLeiderImmerPreklassisch}, but not always to classical ones as we are now going to see.

\begin{theorems}{Existence of LABs giving rise to non-classical gauge theories}{AbelschIstGeileNeueTheorie}
Let $K \to N$ be an LAB , $\nabla$ a connection satisfying compatibility conditions~\eqref{CondSGleichNullLAB} and~\eqref{CondKruemmungmitBLAB} with respect to a given $\zeta \in \Omega^2(N; K)$ such that $\mathrm{d}^\nabla \zeta \neq 0$.

Then there is no $\lambda\in\Omega^1(N;K)$ as in~\ref{fieldredef:FieldRedefForLABs} such that $\widetilde{\zeta}^\lambda = 0$.
\end{theorems}

\begin{proof}
\leavevmode\newline
We have a 2-form $\zeta \in \Omega^2(N; K)$ such that
\bas
\mathrm{d}^\nabla \zeta &\neq 0.
\eas
By Prop.~\ref{prop:InvarianteFuerFieldRedefImFallLAB} we have $\mathrm{d}^{\widetilde{\nabla}^\lambda} \widetilde{\zeta}^\lambda= \mathrm{d}^\nabla \zeta$ for all $\lambda \in \Omega^1(N;K)$. When there would be a field redefinition leading to a classical gauge theory, then $\widetilde{\zeta}^\lambda = 0$ but then also $\mathrm{d}^{\widetilde{\nabla}^\lambda} \widetilde{\zeta}^\lambda = 0$. Thence, by $\mathrm{d}^\nabla \zeta \neq 0$ the statement follows.
\end{proof}

Starting with a standard Yang-Mills gauge theory with an additional free physical field $\Phi$ with a Lagrangian similar to the Higgs field, we have a canonical construction when the centre of the Lie algebra is non-trivial.

\begin{corollaries}{Canonical construction of non-classical gauge theories}{CanonicalConstructionOfGaugeTheories}
Let $\mathfrak{g}$ be a Lie algebra with non-zero centre and admitting an $\mathrm{ad}$-invariant scalar product. Also let $(N, g)$ be any Riemannian manifold with at least three dimensions, and $K = N \times \mathfrak{g}$ be a trivial LAB over $N$, equipped with the canonical flat connection $\nabla$ and a metric $\kappa$ which restricts to an $\mathrm{ad}$-invariant scalar product on each fibre.

Then there is a $\zeta \in \Omega^2(N; Z(K))$ in sense of~\ref{sit:CYMHGTForLABsToDoList}, with $\mathrm{d}^\nabla \zeta \neq 0$, such that this set-up describes a non-classical CYMH GT with respect to an arbitrary spacetime $M$. Additionally, there is no $\lambda\in\Omega^1(N;K)$ as in~\ref{fieldredef:FieldRedefForLABs} such that $\widetilde{\zeta}^\lambda = 0$.
\end{corollaries}

\begin{proof}
\leavevmode\newline
By the assumptions we have everything we need to formulate a YMH GT for a given spacetime $M$, following~\ref{sit:CYMHGTForLABsToDoList}; by Thm.~\ref{thm:ActionLieALgebroid} compatibility condition~\eqref{CondSGleichNullLAB} follows. For compatibility condition~\eqref{CondKruemmungmitBLAB} just take any element of $\Omega^2(N; Z(K))$, denoted as $\zeta$, then this condition is trivially satisfied because $\nabla$ is flat and $\zeta$ only has values in the centre of $K$.

Since $N$ is three-dimensional and $Z(K)$ is non-zero, we can then conclude the existence of $\mathrm{d}^\nabla \zeta \neq 0$. For this recall that $\mathrm{d}^\nabla \zeta$ is still a centre-valued form by Prop.~\ref{prop:BianchiIdentityForZeta} and that $\mathrm{d}^\nabla$ is then just the differential $\mathrm{d}^\Xi$ for $\Xi \coloneqq \sharp \circ \nabla$ as in Thm.~\ref{thm:DifferentialAufZentrumsDinge}. Therefore we only need to take any non-$\mathrm{d}^\Xi$-closed centre-valued form $\zeta$, of which there are plenty. The non-existence of a $\lambda$ with $\widetilde{\zeta}^\lambda = 0$ then follows by Thm.~\ref{thm:AbelschIstGeileNeueTheorie}.
\end{proof}

\subsection{The Bianchi identity of the new field strength} \label{BianchiStuff}

We conclude this paper with an interpretation of $\mathrm{d}^\nabla \zeta$, and for this we need to calculate the Bianchi identity of the field strength. Hence, we need to understand how $\Phi^*\nabla$ behaves.

\begin{propositions}{Pull-Back of a Lie derivation law covering a pairing}{PullbackvonUnseremGaugeNabla}
Let $K \to N$ be an LAB, equipped with a connection $\nabla$ satisfying compatibility condition~\eqref{CondSGleichNullLAB}; also let $M$ be another smooth manifold and $\Phi: M \to N$ a smooth map. Then $\Phi^*\nabla$ also satisfies compatibility condition~\eqref{CondSGleichNullLAB} with respect to $\Phi^*K$.
\newline

When $\nabla$ satisfies compatibility condition~\eqref{CondKruemmungmitBLAB} with respect to a $\zeta \in \Omega^2(N; K)$, not necessarily assuming~\eqref{CondSGleichNullLAB}, then this extends to $\Phi^*K$, too, \textit{i.e.}
\ba\label{EqCompCondFuerPullbackCurvature}
R_{\Phi^*\nabla} = \mathrm{ad}^* \circ \Phi^!\zeta,
\ea
viewing the curvature as an element of $\Omega^2(M; \mathrm{End}(\Phi^*K))$ and $\mathrm{ad}^*$ denotes the adjoint of $\Phi^*K$.
\end{propositions}

\begin{remark}
\leavevmode\newline
By Thm.~\ref{thm:GaugeTheoryNeedsLieDerivLawsCoveringApairing}, we get that the pull-back of a Lie derivation law of $K$ covering the Lie algebroid morphism $\sharp \circ \nabla$ is a Lie derivation law of $\Phi^*K$ covering the Lie algebroid morphism $\sharp \circ \Phi^*\nabla$.
\end{remark}

\begin{proof}
\leavevmode\newline
\indent$\bullet$ We can show
\bas
\Phi^*\nabla \underbrace{\mleft( \mleft[ \Phi^*\mu, \Phi^*\nu \mright]_{\Phi^*K} \mright)}
_{=~ \Phi^*\mleft( \mleft[ \mu, \nu \mright]_K \mright)}
~~~&\stackrel{\mathclap{\text{Eq.~\eqref{eqShortNotationForPullbackConnections}}}}{=}~~~
\Phi^!\mleft( \nabla\mleft( \mleft[ \mu, \nu \mright]_K \mright) \mright)\\
&\stackrel{\mathclap{\text{Eq.~\eqref{CondSGleichNullLAB}}}}{=}~~
\Phi^!\mleft(  \mleft[ \nabla\mu, \nu \mright]_K + \mleft[ \mu, \nabla\nu \mright]_K \mright)\\
&\stackrel{\mathclap{\text{Eq.~\eqref{eqPullbackofLiebracketStuff}}}}{=}~~~
\mleft[ \Phi^!(\nabla\mu), \Phi^*\nu \mright]_{\Phi^*K} + \mleft[ \Phi^*\mu, \Phi^!(\nabla\nu) \mright]_{\Phi^*K} \\
&\stackrel{\mathclap{\text{Eq.~\eqref{eqShortNotationForPullbackConnections}}}}{=}~~~
\mleft[ (\Phi^*\nabla)(\Phi^*\mu), \Phi^*\nu \mright]_{\Phi^*K} + \mleft[ \Phi^*\mu, (\Phi^*\nabla)(\Phi^*\nu) \mright]_{\Phi^*K}
\eas
for all $\mu, \nu \in \Gamma(K)$. Since pull-backs of $\Gamma(K)$ generate $\Gamma(\Phi^*K)$ and since~\eqref{CondSGleichNullLAB} is a tensorial equation, we can derive that $\Phi^*\nabla$ also satisfies compatibility condition~\eqref{CondSGleichNullLAB} with respect to the LAB $\Phi^*K$. 

$\bullet$ Now let $\nabla$ satisfy compatibility condition~\eqref{CondKruemmungmitBLAB}, and recall that in general curvatures satisfy
\bas
R_\nabla(\cdot,\cdot)\nu
=
R_\nabla \nu &= \mleft( \mathrm{d}^\nabla \mright)^2 \nu \in \Omega^2(N; K)
\eas
for all $\nu \in \Gamma(K)$ (see also \cite[\S 5, third part of Exercise 5.15.12; page 316]{hamilton}). Then apply Eq.~\eqref{EqGeilePullBackCommuteFormel} to get
\bas
R_{\Phi^*\nabla} (\Phi^*\nu)
&=
\mleft( \mathrm{d}^{\Phi^*\nabla} \mright)^2 (\Phi^*\nu)
=
\Phi^!\mleft( \mleft(\mathrm{d}^\nabla\mright)^2 \nu \mright)
\stackrel{\text{Eq.~\eqref{CondKruemmungmitBLAB}}}{=}
\Phi^!\mleft( \mleft[ \zeta, \nu \mright]_K \mright)
\stackrel{\text{Eq.~\eqref{eqPullbackofLiebracketStuff}}}{=}
\mleft[ \Phi^!\zeta, \Phi^*\nu \mright]_{\Phi^*K},
\eas
such that $R_{\Phi^*\nabla} = \mathrm{ad}^* \circ \Phi^!\zeta$ follows, by using again that pull-backs of $\Gamma(K)$ generate $\Gamma(\Phi^*K)$.
\end{proof}

Using this we calculate the Bianchi identity for the field strength $G$. 
%In the following statement the ${}^*$ is still related to the pullback with the evaluation map, so, a pullback to $M \times \mathfrak{M}_K(M;N)$; recall Def.~\ref{def:EvaluationMap} and \ref{def:PullbacksAsFunctionals}.

\begin{theorems}{Bianchi identity of the field strength}{BianchiIdentityOfFieldStrength}
Let $M$ and $N$ be smooth manifolds, $K \to N$ an LAB, $\Phi \in C^\infty(M;N)$, and $\nabla$ a connection satisfying compatibility conditions~\eqref{CondSGleichNullLAB} and~\eqref{CondKruemmungmitBLAB} with respect to a given $\zeta \in \Omega^2(N; K)$.

Then
\ba
\mathrm{d}^{\Phi^*\nabla}\bigl(G(\Phi,A)\bigr) + \mleft[ A \stackrel{\wedge}{,} G(\Phi,A) \mright]_{\Phi^*K}
&=
\Phi^! \mleft( \mathrm{d}^\nabla \zeta \mright),
\ea
where
\bas
G(\Phi,A)
&=
\mathrm{d}^{\Phi^*\nabla}A
	+ \frac{1}{2} \mleft[ A \stackrel{\wedge}{,} A \mright]_{\Phi^*K}
	+ \Phi^!\zeta
\eas
was the field strength.
\end{theorems}

\begin{remark}
\leavevmode\newline
This clearly generalizes the standard Bianchi identity for field strengths as in Thm.~\ref{thm:ClassicBianchiIdenityOfFieldstrength}: Take a trivial LAB $K$ equipped with its canonical flat connection and $\zeta \equiv 0$. Then we arrive at the typical Bianchi identity. In general, we get $\mathrm{d}^{\Phi^*\nabla}G + \mleft[ A \stackrel{\wedge}{,} G \mright]_{\Phi^*K}=0$ if $\mathrm{d}^\nabla \zeta = 0$, which resembles strongly the standard Bianchi identity, but covariantized. Hence, we say that $G$ satisfies the \textbf{Bianchi identity} if and only if $\mathrm{d}^{\Phi^*\nabla}G + \mleft[ A \stackrel{\wedge}{,} G \mright]_{\Phi^*K}=0$.
\end{remark}

\begin{proof}
\leavevmode\newline
The calculation is similarly to the standard calculation of the standard formulation of the Bianchi identity as in \cite[\S 5, Theorem 5.14.2; page 311]{hamilton}, making use of compatibility condition~\eqref{CondSGleichNullLAB} needed for Eq.~\eqref{eqDerivationOfDifferentialOnBracketonK}. We have, viewing the curvature $R_{\Phi^*\nabla}$ as an element of $\Omega^2(M; \mathrm{End}(\Phi^*K))$,
\bas
&\mleft( \mathrm{d}^{\Phi^*\nabla} \mright)^2 A
=
R_{\Phi^*\nabla} \wedge A
\stackrel{\text{Eq.~\eqref{EqCompCondFuerPullbackCurvature}}}{=}
\mleft( \mathrm{ad}^* \circ \Phi^!\zeta \mright) \wedge A
\stackrel{\text{Eq.~\eqref{wedgeproduktmitadLambdaergibtLieklammer}}}{=}
\mleft[ \Phi^!\zeta \stackrel{\wedge}{,} A \mright]_{\Phi^*K}
\stackrel{\text{Eq.~\eqref{VertauschungsregelForKKlammerAufFormen}}}{=}
- \mleft[ A \stackrel{\wedge}{,} \Phi^!\zeta \mright]_{\Phi^*K}, \\
&\mathrm{d}^{\Phi^*\nabla}\mleft( \mleft[ A \stackrel{\wedge}{,} A \mright]_{\Phi^*K} \mright)
\stackrel{\text{Eq.~\eqref{eqDerivationOfDifferentialOnBracketonK}}}{=}
\mleft[ \mathrm{d}^{\Phi^*\nabla} A \stackrel{\wedge}{,} A \mright]_{\Phi^*K}
	- \mleft[ A \stackrel{\wedge}{,} \mathrm{d}^{\Phi^*\nabla} A \mright]_{\Phi^*K}
\stackrel{\text{Eq.~\eqref{VertauschungsregelForKKlammerAufFormen}}}{=}
- 2 ~ \mleft[ A \stackrel{\wedge}{,} \mathrm{d}^{\Phi^*\nabla} A \mright]_{\Phi^*K}, \\
&\mleft[ A \stackrel{\wedge}{,} \mleft[ A \stackrel{\wedge}{,} A \mright]_{\Phi^*K} \mright]_{\Phi^*K}
\stackrel{\text{Eq.~\eqref{JacobiIdentityForFormBracket}}}{=}
0, \\
&\mathrm{d}^{\Phi^*\nabla} \mleft( \Phi^!\zeta \mright)
\stackrel{\text{Eq.~\eqref{EqGeilePullBackCommuteFormel}}}{=}
\Phi^! \mleft( \mathrm{d}^\nabla \zeta \mright),
\eas
and, using all of these, we arrive at
\bas
\mathrm{d}^{\Phi^*\nabla}\bigl(G(\Phi,A) \bigr) + \mleft[ A \stackrel{\wedge}{,} G(\Phi,A) \mright]_{\Phi^*K}
~~~&\stackrel{\mathclap{\text{Def.~\eqref{defNewFieldStrengthG}}}}{=}~~~
\Phi^! \mleft( \mathrm{d}^\nabla \zeta \mright).
\eas
\end{proof}

Thence, $\mathrm{d}^\nabla \zeta$ measures the failure of the Bianchi identity of the field strength $G$. For example, applying Cor.~\ref{cor:CanonicalConstructionOfGaugeTheories} to the Yang-Mills gauge theory of electromagnetism, \textit{i.e.}~the Lie algebra is given by $\mathfrak{g} = \mathrm{u}(1)$, would result into a gauge theory where there is no (vector) potential of the field strength as usual, so, $G$ could not be written as $\mathrm{d}^\nabla \widehat{A}$ for some $\widehat{A} \in \Omega^1(N;\Phi^*K)$.\footnote{Recall that $\mathrm{d}^\nabla$ is a differential since $\nabla$ is flat in that situation.} This concludes our discussion about LABs in the context of CYMH GTs.

%\begin{theorems}{Existence of CYMH GT on Lie algebra bundles}{LABCYMHGT}
%l
%\end{theorems}
%\newpage
\section{Tangent bundles} \label{TangentBundles}

Let us look at the next extreme of possible Lie algebroids: The tangent bundles themselves.

\subsection{General situation}\label{GeneralSituForTangent}

Let us quickly summarize what we need for tangent bundles in the context of CYMHG GT.

\begin{situations}{Compatibility conditions for tangent bundles}{SituationForTangentBundles}
We now have $E = \mathrm{T}N$, and, thus, the Lie bracket is just the typical one for vector fields. The anchor is the identity on $\mathrm{T}N$, $\rho = \mathds{1}_{\mathrm{T}N}$. Therefore there is now a coupling between the fields of gauge bosons and the Higgs field; however, since tangent bundles are transitive Lie algebroids, there is no transversal structure, hence, no Higgs bosons, only Nambu-Goldstone bosons if assuming a classical structure.\footnote{Recall, that the components of the Higgs field along the orbits are the Nambu-Goldstone bosons which can often be "gauged away" by the unitary gauge, thus, not relevant for the Higgs effect; see \cite[\S 8; page 445ff.]{hamilton}.} Thus, also now we still have no real Higgs effect.

Both basic connections clearly now coincide, especially we have for a connection $\nabla$ on $E$,
\bas
\nabla^{\mathrm{bas}}_Y Z
&=
[Y, Z]
	+ \nabla_Z Y
\eas
for all $Y, Z \in \mathfrak{X}(N)$,
so, $\nabla^{\mathrm{bas}}$ is also a vector bundle connection and has a 1:1 correspondence with $\nabla$.
The compatibility condition \eqref{VanishingBasicCurvComp} reduces to
\ba
R_{\nabla^{\mathrm{bas}}}
&=
0
\ea
by Prop.~\ref{prop:SnablamitREnabla}, hence, $\nabla^{\mathrm{bas}}$ shall be a flat connection as compatibility condition.

The other compatibility conditions do not really change their form. However, we assume for simplicity that the fibre metric $\kappa$ on $E$ and Riemannian metric $g$ on $\mathrm{T}N$ coincide, such that the number of compatibility conditions is reduced by one; thus, we only have compatibility condition related to the metrics
\ba
\nabla^{\mathrm{bas}} g
&=
0.
\ea
Moreover, for a gauge invariance of the theory we need $\zeta \in \Omega^2(N;E)$ such that 
\ba
R_\nabla
&=
- \mathrm{d}^{\nabla^{\mathrm{bas}}}\zeta.
\ea
That a $\zeta$ exists in this situation we already know by Thm.~\ref{thm:BAlongL} and Cor.~\ref{cor:TorsionOfDualTorsions} that $\zeta= t_\nabla$ is a solution of this compatibility condition; this also implies that $\nabla$. Choosing that $\zeta$, what we do, means that we only have two compatibility conditions. Essentially we only need to construct a \textbf{flat metric connection $\nabla^{\mathrm{bas}}$}, and due to the 1:1 correspondence to $\nabla$ we have then everything needed for a CYMH GT as in Thm.~\ref{thm:FinallyTheGaugeInvarianceWeWant}, modulo the potential which is not important for the discussion since we always assume that a suitable potential is given.

Every other structure needed for a CYMHG GT still looks the same in its form. Hence, we will now not recall the field strength and the Lagrangian as we did for LABs.
\end{situations}

\begin{remark}
\leavevmode\newline
We used a lot of exterior covariant derivatives in the past, especially we had two degrees in forms like $\Omega^{p,q}(N,E;E)$ ($p,q \in \mathbb{N}_0$), hence, a degree with respect to both $\mathrm{T}N$ and $E$. Now both bundles coincide, but for the purpose of calculating with such forms it is still important to distinguish them. For example the combatibility condition about $\zeta \in \Omega^2(N;E) \cong \Omega^{2,0}(N,E;E)$ reads
\bas
\mleft(\mathrm{d}^{\nabla^{\mathrm{bas}}}\zeta\mright)(X, Y, Z)
&=
\nabla^{\mathrm{bas}}_Z\bigl( \zeta(X, Y) \bigr)
	- \zeta\mleft( \nabla^{\mathrm{bas}}_Z X, Y \mright)
	- \zeta\mleft( X, \nabla^{\mathrm{bas}}_Z Y \mright)
\eas
for all $X, Y, Z \in \mathfrak{X}(N)$, but "only $Z$ as a section of $E$". If we view all three arguments as sections of $E$, that is, $\zeta$ as an element of $\Omega^2(E;E) \cong \Omega^{0,2}(N,E;E)$, we would get instead that
\bas
\mleft(\mathrm{d}^{\nabla^{\mathrm{bas}}}\zeta\mright)(X, Y, Z)
&=
\nabla^{\mathrm{bas}}_X\bigl( \zeta(Y, Z) \bigr)
	- \nabla^{\mathrm{bas}}_Y\bigl( \zeta(X, Z) \bigr)
	+ \nabla^{\mathrm{bas}}_Z\bigl( \zeta(X, Y) \bigr)
\\
&\hspace{1cm}
	- \zeta\bigl( [X, Y], Z \bigr)
	+ \zeta\bigl( [X, Z], Y \bigr)
	- \zeta\bigl( [Y, Z], X \bigr),
\eas
which is clearly different. Hence, it is still important to distinguish between $\mathrm{T}N$ as the Lie algebroid $E$ and as tangent bundle $\mathrm{T}N$. However, in that case, for $\zeta \in \Omega^2(N;E)$ we know that
\bas
\mathrm{d}^{\nabla^{\mathrm{bas}}}\zeta
&=
\nabla^{\mathrm{bas}} \zeta,
\eas
and the right hand side would be in alignment with both interpretations of $\zeta$ as form.
\end{remark}

For the field redefinition there is not much to say additionally, besides that for $\lambda \in \Omega^1(N;E)$ we have $\Lambda = \mathds{1}_E - \lambda = \widehat{\Lambda}$. There are important results with respect to whether we have a (pre-)classical gauge theory.

\begin{corollaries}{Pre-classical theories have constant torsion}{TorsionConstancyAndFlatness}
Let $N$ be a smooth manifold, equipped with a connection $\nabla$ on $E \coloneqq \mathrm{T}N$ with vanishing basic curvature. Then there is a $\lambda \in \Omega^1(N;E)$ such that $\widetilde{\nabla}^\lambda$ is flat if and only if there is a $\lambda \in \Omega^1(N;E)$ such that $t_{\mleft(\widetilde{\nabla}^\lambda \mright)^{\mathrm{bas}}} = - t_{\widetilde{\nabla}^\lambda}$ is constant with respect to $\mleft(\widetilde{\nabla}^\lambda \mright)^{\mathrm{bas}}$, that is, 
\ba
\mleft(\widetilde{\nabla}^\lambda \mright)^{\mathrm{bas}} t_{\mleft(\widetilde{\nabla}^\lambda \mright)^{\mathrm{bas}}}
&= 
0.
\ea
\end{corollaries}

\begin{remark}
\leavevmode\newline
Recall Cor.~\ref{cor:TOrsionCanBeLieBracketIfFlat}; in the case of a flat $\nabla_\rho = \nabla$ (or its field redefinition) its torsion would be another Lie bracket on $E$, but tensorial.
\end{remark}

\begin{proof}[Proof of Cor.~\ref{cor:TorsionConstancyAndFlatness}]
\leavevmode\newline
That quickly follows by Cor.~\ref{cor:LemmaCurvatureOfDualConnections}, using the vanishing of the basic curvature which is here equivalent to that $\nabla^{\mathrm{bas}}$ is flat, \textit{i.e.}
\bas
R_\nabla
&=
\nabla^{\mathrm{bas}} t_{\nabla^{\mathrm{bas}}},
\eas
hence, $\nabla$ is flat if and only if $\nabla^{\mathrm{bas}} t_{\nabla^{\mathrm{bas}}} = 0$. By Thm.~\ref{thm:InvarianceUnderTheFieldRedefinition} and its remark afterwards the vanishing of the basic curvature is preserved, hence,
\bas
R_{\widetilde{\nabla}^\lambda}
&=
\mleft(\widetilde{\nabla}^\lambda \mright)^{\mathrm{bas}} t_{\mleft(\widetilde{\nabla}^\lambda \mright)^{\mathrm{bas}}}.
\eas
Hence, the statement follows immediately.
\end{proof}

Of special importance is the next theorem.

\begin{theorems}{Certain classical CYMH GTs are Lie groups, \newline \cite[\S 3.1 and the references therein]{blaomTangentBundleAsLieGroup} and \cite[Comment after Proposition 2.12]{basicconn}}{LieGroupIsomorphisms}
Let $N$ be a smooth compact and simply connected manifold, and assume we have a connection $\nabla$ on $E\coloneqq\mathrm{T}N$ such that $\nabla$ is flat and has vanishing basic curvature. Then $N$ is diffeomorphic to a Lie group.
\end{theorems}

\begin{proof}[Sketch of the proof for Thm.~\ref{thm:LieGroupIsomorphisms}]
\leavevmode\newline
We only give a sketch of the proof, see the references for all details. First of all, as we already discussed, the vanishing of the basic curvature and the fact that $N$ is simply connected imply there is an isomorphism to an action Lie algebroid $N \times \mathfrak{g}$, $\mathfrak{g}$ a Lie algebra, such that $\nabla$ is its canonical flat connection by Thm.~\ref{thm:ActionLieALgebroid}. Then define $\omega \in \Omega^1(N; \mathfrak{g})$ by the composition of the given isomorphism\footnote{We will use this isomorphism all the time in the following, without further extra notation.} $\mathrm{T}N \to N \times \mathfrak{g}$ and the projection onto the second factor $N \times \mathfrak{g} \to \mathfrak{g}$. $\omega_p: \mathrm{T}_pN \to \mathfrak{g}$ is then clearly an isomorphism of vector spaces for all $p \in N$; such forms are also equivalent to absolute parallelisms, a trivialization of the tangent bundle, because specifying such a form gives clearly a trivialization (also in the case if $\mathfrak{g}$ is just a vector space). 

The idea is that the parallel frames of $\nabla$ will be left-invariant vector fields of a Lie group. Let us denote the parallel frame of $\nabla$ by $\mleft( e_a \mright)_a$, which is also a constant frame of $N \times \mathfrak{g}$, making it obvious why that frame will be the left-invariant vector fields (their generators); it is global due to the fact that $N$ is simply connected. So, $\nabla e_a = 0$ and
let us study
\bas
\mleft(\mathrm{d} \omega\mright)(X,Y)
&=
\mleft(\mathrm{d}^\nabla \omega\mright)(X,Y)
=
\nabla_X\bigl( \omega(Y) \bigr)
	- \nabla_Y\bigl( \omega(X) \bigr)
	- \omega([X, Y])
\eas
for all $X, Y \in \mathfrak{X}(N)$. In coordinates, especially for the constant frame, we have by definition
\bas
\omega(\nu)
&=
\nu
\eas
for all constant $\nu \in \Gamma(N \times \mathfrak{g}) \cong \mathfrak{X}(N)$,
thus,
\bas
(\mathrm{d}\omega)(\mu, \nu) 
&= 
- \omega\underbrace{\mleft( \mleft[ \mu, \nu \mright]_{\mathfrak{g}} \mright)}_{\text{const.}}
=
-\mleft[ \mu, \nu \mright]_{\mathfrak{g}}
=
- \mleft[ \omega(\mu), \omega(\nu) \mright]_{\mathfrak{g}}
=
- \mleft(\frac{1}{2} \mleft[ \omega \stackrel{\wedge}{,} \omega \mright]_{\mathfrak{g}}\mright)(\mu, \nu)
\eas
for all constant $\mu, \nu \in \Gamma(N \times \mathfrak{g})$. Since this is a tensorial equation this holds for all sections/vector fields, so, the Maurer-Cartan equation is satisfied. Hence, $\omega$ will be the Maurer-Cartan form, infinitesimally decoding the Lie group structure related to the differential of the Left multiplication. The Maurer-Cartan equation is the integrability condition, that is, one can locally define an exponential, generating a Lie group structure locally.\footnote{The Maurer-Cartan equation as a "zero curvature condition" encodes basically the infinitesimal information about that there is a unique group element connecting two other group elements.} By compactness and conectedness one can do this globally leading to that $M$ is diffeomorphic to a Lie group integrating $\mathfrak{g}$.
\end{proof}

Especially looking at manifolds which are not Lie groups can help to find CYMH GTs on tangent bundle which are not pre-classical, also under the field redefinition.

\subsection{Local picture}\label{LocalTangentBundles}

Having Thm.~\ref{thm:LieGroupIsomorphisms} in mind, one expects that tangent bundles as CYMH GT are locally always a pre-classical CYMH GT.

\begin{theorems}{Tangent bundles are locally pre-classical as CYMH GT}{NoGoLocalTangentBundle}
Let $N = \mathbb{R}^n$ ($n \in \mathbb{N}_0$) be an Euclidean space as smooth manifold and $\nabla$ a connection on $E \coloneqq \mathrm{T}N$ with vanishing basic curvature. Then there is a $\lambda \in \Omega^1(N;E)$ such that $\widetilde{\nabla}^\lambda$ is flat.
\end{theorems}

\begin{proof}
\leavevmode\newline
That will essentially follow by Cor.~\ref{cor:TorsionConstancyAndFlatness}, we need to find a field redefinition such that 
\bas
\mleft(\widetilde{\nabla}^\lambda\mright)^{\mathrm{bas}} t_{\mleft(\widetilde{\nabla}^\lambda\mright)^{\mathrm{bas}}}
&=
0,
\eas
so, $\widetilde{\nabla}^\lambda$ is flat if and only if $t_{\mleft(\widetilde{\nabla}^\lambda\mright)^{\mathrm{bas}}}$ is constant w.r.t.~$\mleft(\widetilde{\nabla}^\lambda\mright)^{\mathrm{bas}}$. As we have discussed in \ref{sit:SituationForTangentBundles} we know that there is a parallel frame $(e_a)_a$ of $E$ for $\nabla^{\mathrm{bas}}$, globally defined since $N = \mathbb{R}^n$, especially simply connected. Then also
\bas
t_{\nabla^{\mathrm{bas}}}(e_a, e_b) 
&= 
-\left[e_a, e_b\right]_E
=
-C_{ab}^c ~ e_c,
\eas
where $C_{ab}^c$ are structure functions, and
\bas
\left(\nabla^{\mathrm{bas}} t_{\nabla^{\mathrm{bas}}}\right) (e_a, e_b)
&=
\nabla^{\mathrm{bas}} \left( t_{\nabla^{\mathrm{bas}}}(e_a, e_b)\right)
=
- \nabla^{\mathrm{bas}} \big( \left[ e_a, e_b \right]_E \big)
=
- \mathrm{d}(C^c_{ab}) \otimes e_c.
\eas
When the structure functions are already constants we're done, otherwise we will now use the transformation formulas in Def.~\ref{def:FieldRedefinition}. By Eq.~\eqref{basicconnectionTrafoRefield} it is clear that $\widetilde{e}_a \coloneqq \Lambda(e_a)$ defines a parallel frame for $\widetilde{\nabla}^{\mathrm{bas}}$ and, thus, similarly
\bas
\mleft(\mleft( \widetilde{\nabla}^\lambda\mright)^{\mathrm{bas}}
\widetilde{t}_{\mleft(\widetilde{\nabla}^\lambda\mright)^{\mathrm{bas}}}\mright)(\widetilde{e}_a, \widetilde{e}_b)
&=
- \mleft( \widetilde{\nabla}^\lambda\mright)^{\mathrm{bas}} \bigl( \mleft[\widetilde{e}_a, \widetilde{e}_b \mright]_E \bigr)
= - \mathrm{d} \mleft(\widetilde{C}_{ab}^c\mright) \otimes \widetilde{e}_c,
\eas
where $\widetilde{C}_{ab}^c$ are the structure functions related to $\mleft(\widetilde{e}_a\mright)_a$.
Thence, $\widetilde{\nabla}^\lambda$ is flat if and only if $\widetilde{C}_{ab}^c$ are constants.
%
%Therefore,
%\bas
%&&
%\exists \lambda \in \Omega^1(U;E): ~ 
 %R_{\widetilde{\nabla}} &=0 \\
%&\Leftrightarrow& \exists \lambda \in \Omega^1(U;E): ~
%0&=
%\widetilde{\nabla}^{\mathrm{bas}}\big( \left[ \Lambda(e_a), \Lambda(e_b) \right]_E \big) \\
%&\stackrel{\mathclap{\rho \text{ bijective}}}{\Leftrightarrow}& \exists \lambda \in \Omega^1(U;E): ~
%0&=
%\rho\left( \widetilde{\nabla}^{\mathrm{bas}}\big( \left[ \Lambda(e_a), \Lambda(e_b) \right]_E \big) \right) \\
%&&
%&\stackrel{\mathclap{\text{Def. } \ref{def:basicconn}}}{=}
%\quad~\widetilde{\nabla}^{\mathrm{bas}}\left(\rho\big( \left[ \Lambda(e_a), \Lambda(e_b) \right]_E \big)\right) \\
%&&
%&=
%\widetilde{\nabla}^{\mathrm{bas}}\big( \left[ \rho(\Lambda(e_a)), \rho(\Lambda(e_b)) \right]_E  \big) \\
%&&
%&=
%\widetilde{\nabla}^{\mathrm{bas}}\Big( \left[ \widehat{\Lambda}(\rho(e_a)), \widehat{\Lambda}(\rho(e_b)) \right]_E \Big).
%\eas
%$(\rho(e_a))_a$ is a linear independent system because $\rho$ is bijective, \textit{i.e.} it describes a local frame of $\mathrm{T}M$. 

$\Lambda \in \sAut(E)$ can be taken in such a way that $\left(\Lambda(e_a)\right)_a$ are global coordinate vector fields $\partial_i$, because then
\bas
\lambda
&=
\mathds{1}_{\mathrm{T}N} 
	- \Lambda
\eas
is a valid definition for $\lambda \in \Omega^1(N;E)$. Using such a $\lambda$ implies
\bas
\mleft[ \widetilde{e}_a, \widetilde{e}_b \mright]_E
&= 
0,
\eas
thus, $\widetilde{C}_{ab}^c = 0$. So, we have found a field redefinition to a flat connection by Cor.~\ref{cor:TorsionConstancyAndFlatness}.
\end{proof}

\subsection{Unit octonions}\label{UnitoctonionsasGT}

By Thm.~\ref{thm:LieGroupIsomorphisms}, we now show that there is an example for a CYMH GT by using a manifold which is not a Lie group; of course we study the canonical example of such a manifold, the seven dimensional sphere $\mathds{S}^7$. $\mathds{S}^7$ can be understood as the set of unit octonions. It would certainly exceed the purpose of this thesis to discuss those in full detail, hence, we only introduce and show parts of the basics needed for the proof such that one should be able to understand the motivation and structure behind the following definitions. See the following reference for a thorough discussion. We will follow \cite[\S 3.10, page 170ff.; Exercise 3.12.15, page 189f.; Example 4.5.10, page 229]{hamilton}, using the exceptional Lie group $G_2$ to define octonions.

In this subsection let $V \coloneqq \mathbb{R}^7$, and we denote its standard Euclidean scalar product by $\langle \cdot, \cdot \rangle$, its orthonormal base by $\left(e_j\right)_{j=1}^7$ and $\left(w^i\right)_{i=1}^7$ its dual basis, \textit{i.e.}~$w^i\mleft(e_j\mright) = \delta_j^i$, the Kronecker delta. We also define a shorter notation for products of $w^i$, for example
\bas
w^{ij} &\coloneqq w^i \wedge w^j,
\eas
similar with more than two factors.

\begin{definitions}{Multiplication form for octonions, \newline \cite[Definition 3.10.1; page 171]{hamilton}}{MultiplcationTableOfOctonions}
We define a 3-form $\phi \in \bigwedge^3 V^*$ by
\ba
\phi 
&\coloneqq
w^{123}
	+ w^1 \wedge \left( w^{45} + w^{67} \right)
	+ w^2 \wedge \left( w^{46} - w^{57} \right)
	- w^3 \wedge \left( w^{47} + w^{56} \right).
\ea
\end{definitions}
%
%\begin{remark}
%\leavevmode\newline
%One can use such a form to define the exceptional Lie group $G_2$, which we will not introduce. See \textit{e.g.}~\cite[Definition 3.10.3; page 171]{hamilton}.
%\end{remark}

This 3-form will essentially define the multiplication table for octonions; but before we do so, let us define $G_2$ for which we need a $\mathrm{GL}(7, \mathbb{R})$-action on $\bigwedge^k V^*$.

\begin{definitions}{$\mathrm{GL}(7, \mathbb{R})$-action on $\bigwedge^k V^*$, \newline \cite[comment before Definition 3.10.3]{hamilton}}{GL7action}
We define
\ba
(q \alpha)(v_1, \dotsc, v_k)
&\coloneqq
\alpha \mleft(
	q^{-1}v_1, \dotsc, q^{-1} v_k
\mright)
\ea
for all $\alpha \in \bigwedge^k V^*$ ($k \in \mathbb{N}_0$), $q \in \mathrm{GL}(7, \mathbb{R})$, and $v_1, \dotsc, v_k \in V$, where $q$ acts on $V$ as usual by the standard representation.
\end{definitions}

Using this notion, we can define $G_2$.

\begin{definitions}{Exceptional Lie group $G_2$, \cite[Definition 3.10.3; page 171]{hamilton}}{ExceptionalLieGroup}
We define the \textbf{exceptional Lie group $G_2$} as a subset of $\mathrm{GL}(7, \mathbb{R})$ by
\ba
G_2
&\coloneqq
\left\{
q \in \mathrm{GL}(7, \mathbb{R})
~\middle|~
q \phi
=
\phi
\right\}.
\ea
\end{definitions}

\begin{remark}
\leavevmode\newline
$G_2$ is clearly a subgroup of $\mathrm{GL}(7, \mathbb{R})$ as the isotropy of $\phi$. As argued in \cite{hamilton}, it is therefore also a closed embedded Lie subgroup; furthermore, in \cite[Corollary 3.10.7; page 173]{hamilton} it is also shown that $G_2$ is a compact embedded Lie subgroup of $\mathrm{SO}(7)$. That also implies that 
\ba
\langle qx, qy\rangle
&=
\langle x, y \rangle
\ea
for all $x, y \in V$ and $q \in G_2$. We will not prove this because because it is on one hand straighforward but a bit tedious to prove, and we assume that the exceptional Lie group $G_2$ is a known object for the reader.
\end{remark}

\begin{definitions}{\cite[Definition 3.10.8; page 175]{hamilton}}{PhiAsP}
Let us define a map $P: V \times V \to V$ by
\ba
\langle
P(x,y), z
\rangle
&\coloneqq
\phi(x, y,z)
\ea
for all $x,y,z \in V$.
\end{definitions}

By definition we get.

\begin{propositions}{Properties of $P$, \cite[Proposition 3.10.9]{hamilton}}{PIsNice}
The map $P$ is antisymmetric, bilinear and $G_2$-equivariant, that is
\ba
q \bigl( P(x, y) \bigr)
&=
P(qx, qy)
\ea
for all $q \in G_2$ and $x, y \in V$.
\end{propositions}

\begin{proof}
\leavevmode\newline
Antisymmetry and bilinearity follow immediately by definition. For the third property we use that $G_2 \subset \mathrm{SO}(7)$ and the definition of $G_2$, so,
\bas
\langle q \bigl( P(x, y) \bigr), z \rangle
&=
\mleft\langle P(x, y), q^{-1} z \mright\rangle
=
\phi\mleft(
	x, y, q^{-1}z
\mright)
=
\underbrace{(q\phi)}_{=\phi}\mleft(
	qx, qy, z
\mright)
=
\langle P(qx, qy), z \rangle
\eas
for all $x, y, z \in V$ and $q \in G_2$.
\end{proof}

We will also need some additional technical result for $P$.

\begin{lemmata}{Additonal properties of $P$, \newline \cite[first part of Exercise 3.12.16; page 190]{hamilton}}{PPFormula}
We have
\ba
P\bigl(x, P(x, y)\bigr)
&=
- \langle x, x \rangle y
	+ \langle x, y \rangle x
\ea
for all $x, y \in V$.
\end{lemmata}

\begin{proof}[Sketch of proof for Lemma \ref{lem:PPFormula}]
\leavevmode\newline
\indent $\bullet$ Let $x, y \in V$. Then there is a $q \in G_2$ such that
\bas
qx
&=
x_1 e_1,
&
qy
&=
y_1 e_1
	+ y_2 e_2
\eas
for some $x_1, y_1, y_2 \in \mathbb{R}^2$ (not necessarily the components of $x$ and $y$, which is why the indices are at lower position).
This is given in \cite[first part of Exercise 3.12.15; page 189]{hamilton}; we only give a sketch of this part of the proof actually, see the references for all the calculations. First assume that $x$ and $y$ are linear independent, then apply the Gram-Schmidt process to get orthonormal vectors
\bas
x^\prime
&\coloneqq
\frac{x}{||x||},
&
y^\prime
&\coloneqq
\frac{y - \langle x^\prime, y \rangle x^\prime}{\mleft|\mleft| y - \langle x^\prime, y \rangle x^\prime \mright|\mright|}.
\eas
Let 
\bas
V_2\mleft(\mathbb{R}^7\mright)
&\coloneqq
\left\{
(v_1, v_2)
~\middle|~
v_i \in \mathbb{R}^7, \langle v_i, v_j \rangle = \delta_{ij}
\right\}
\eas
where $i,j\in \{1,2\}$; this is known as a certain \textbf{Stiefel manifold}, see for example \cite[Example 3.9.1; page 168]{hamilton} for an introduction and discussion. We have $(x^\prime, y^\prime), (e_1, e_2) \in V_2\mleft( \mathbb{R}^7 \mright)$, and then there is an element $q \in G_2$ such that $qx^\prime = e_1$ and $qy^\prime = e_2$; this is given by \cite[Theorem 3.10.15; page 177]{hamilton}, where it is shown that $G_2$ acts transitively on $V_2\mleft(\mathbb{R}^7\mright)$ by $q \cdot (v_1, v_2) = (qv_1, q v_2)$ for all $q \in G_2$ and $(v_1, v_2) \in V_2\mleft(\mathbb{R}^7\mright)$. With that we can derive
\bas
qx
&=
q\bigl( ||x|| ~ x^\prime \bigr)
=
x_1 e_1,
\\
qy
&=
q \mleft(
	\langle x^\prime, y \rangle x^\prime
	+ \mleft|\mleft| y - \langle x^\prime, y \rangle x^\prime \mright|\mright| ~ y^\prime
\mright)
=
y_1 e_1 + y_2 e_2
\eas
where $x_1 \coloneqq ||x||, y_1 \coloneqq \langle x^\prime, y \rangle, y_2 \coloneqq \mleft|\mleft| y - \langle x^\prime, y \rangle x^\prime \mright|\mright|$. Hence, we have found the desired element $q \in G_2$; in case $x$ and $y$ are linear dependent and one element is unzero (it is a trivial task if both are zero), one extends the non-zero element first to a basis of a 2-dimensional subspace of $\mathbb{R}^7$ and applies then the same argument as in the previous situation.

$\bullet$ We now want to fix such a $q$ for a given pair $x$ and $y$; it allows us to simplify the calculation by reducing the involved dimensions, using the $G_2$-equivariance of $P$. So,
\bas
\langle P(x, P(x,y)), z \rangle
&=
\langle qP(x, P(x,y)), qz \rangle
\\
&=
\langle P(qx, qP(x,y)), qz \rangle
\\
&=
\langle P(qx, P(qx,qy)), qz \rangle
\\
&=
\mleft\langle \mleft( x_1 \mright)^2 y_2 ~ P(e_1, P(e_1,e_2)), qz \mright\rangle
\\
&=
\mleft\langle\mleft( x_1 \mright)^2 y_2 ~ P(e_1, e_3), qz \mright\rangle
\\
&=
\mleft\langle 
	- \mleft( x_1 \mright)^2 y_2 e_2 
	+ \mleft( x_1 \mright)^2 y_1 e_1 
	- \mleft( x_1 \mright)^2 y_1 e_1, 
	qz 
\mright\rangle
\\
&=
\Bigl\langle 
	- \underbrace{\mleft( x_1 \mright)^2 (y_1 e_1 + y_2 e_2 )}
		_{= \langle qx, qx \rangle qy}
	+ \underbrace{ x_1 y_1 ~ x_1 e_1}
		_{= \langle qx, qy \rangle qx}, 
	qz 
\Bigr\rangle
\\
&=
- \langle x, x \rangle \langle qy, qz\rangle
	+ \langle x, y \rangle \langle qx, qz \rangle
\\
&=
\mleft\langle
	- \langle x, x \rangle y + \langle x, y \rangle x, z
\mright\rangle
\eas
for all $x, y, z \in V$, using $G_2 \subset \mathrm{SO}(7)$, the antisymmetry of $P$, and the definition of $\phi$ to calculate that
\bas
\langle P(e_1, e_2), v \rangle
&=
\phi(e_1, e_2, v)
=
v^3
\eas
for all $v \in V$, such that $P(e_1, e_2) = e_3$, and similarly one derives $P(e_1, e_3) = -e_2$. Therefore
\bas
P(x, P(x,y))
&=
- \langle x, x \rangle y + \langle x, y \rangle x.
\eas
\end{proof}

Now let us define the octonions.

\begin{definitions}{Octonions, \cite[third part of Exercise 3.12.15; page 189f.]{hamilton}}{OctonionsDef}
We define the \textbf{octonions $\mathbb{O}$} by
\ba
\mathbb{O}
&\coloneqq
\mathbb{R}e_0 \oplus V
\cong 
\mathbb{R}^8,
\ea
where $\mathbb{R}e_0$ denotes $\mathbb{R}$ emphasizing that $e_0$ denotes a basis along that factor, and define an $\mathbb{R}$-bilinear multiplication $\cdot$ on $\mathbb{O}$ by
\ba
e_0 \cdot e_0 &\coloneqq e_0,
&e_0 \cdot x &\coloneqq x \cdot e_0 \coloneqq x,
& x \cdot y \coloneqq - \langle x, y \rangle e_0 + P(x,y),
\ea
for all $x, y \in V$. Furthermore, let $(\cdot,\cdot)$ be the scalar product on $\mathbb{O}$ sucht that $\mleft( e_a \mright)_{a=0}^7$ is its orthonormal basis.
\end{definitions}

\begin{remark}
\leavevmode\newline
As one trivially sees and pointed out in \cite[last part of Example 4.5.10; page 229]{hamilton}, one has 
\bas
e_j^2
&=
- e_0
\eas
for all $j \in \{1, \dotsc, 7\}$, using the antisymmetry of $P$.
\end{remark}

With the norm $||\cdot||$ induced by $(\cdot,\cdot)$ one can show that $\mathbb{O}$ is a normed division algebra, but $\cdot$ is not an associative multiplication, see \textit{e.g.}~\cite[third and sixth part of Exercise 3.12.15; page 189f.]{hamilton}. This especially means that
\bas
||z \cdot w||
&=
||z|| ~ ||w||
\eas
for all $z, w \in \mathbb{O}$, and by defining the \textbf{octonionic conjugation}
\bas
\overline{z}
&\coloneqq
x_0e_0
	- x
\eas
for $z = x^0e_0 + x$, where $x^0 \in \mathbb{R}$ and $x \in V$, one can show that
\bas
z\cdot \overline{z}
&=
\overline{z} \cdot z
=
||z||^2~ e_0,
\eas
such that every non-zero octonion has a multiplicative inverse. Especially, the multiplication is closed on the elements with norm 1, that is, for all $z,w \in \mathbb{O}\cong\mathbb{R}^8$ with $||z||= ||w||=1$ we have $||zw||=1$. $\mathds{S}^7$ can be then interpreted as those octonions with unit norm, the unit octonions, and henceforth it carries their non-associative algebra. It is a well-known fact that $\mathds{S}^7$ does not admit a Lie group structure, so, especially one cannot get rid of the non-associativity.

These properties are straightforward calculations and very well-known, hence, we are not proving these explicitly, see the mentioned reference for example. But the non-associativity can be quickly seen by (recall the end of the proof of Lemma \ref{lem:PPFormula} in order to see how to calculate values of $P$),
\bas
(e_1 \cdot e_2) \cdot e_4
&=
P(e_1, e_2) \cdot e_4
=
e_3 \cdot e_4
=
P(e_3, e_4)
=
- e_7
\eas
and
\bas
e_1 \cdot (e_2 \cdot e_4)
&=
e_1 \cdot P(e_2, e_4)
=
e_1 \cdot e_6
=
P(e_1, e_6)
=
e_7,
\eas
hence, $(e_1 \cdot e_2) \cdot e_4 \neq e_1 \cdot (e_2 \cdot e_4)$,
as also mentioned in \cite[sixth part of Exercise 3.12.15; page 190]{hamilton}.

$\mathbb{S}^7$ is a parallelizable manifold. To see this we also need the following.

\begin{propositions}{Compatibility of the multiplication in $\mathbb{O}$ with $(\cdot,\cdot)$, \newline \cite[motivated by Example 4.5.10; page 229]{hamilton}}{ImportantRelationOfScalarproductonO}
We have
\ba
\mleft(e_j z, w\mright)
&=
- \mleft(z, e_j w\mright)
\ea
for all $z, w \in \mathbb{O}$ and $j \in \{1, \dotsc, 7\}$.
\end{propositions}

\begin{proof}
\leavevmode\newline
For $z, w \in \mathbb{O}$ let us write $z= x^0 e_0 + x$ and $w = y^0e_0 + y$, where $x^0, y^0 \in \mathbb{R}$ and $x,y \in V$. Then, using $i,j \in \{1, \dotsc, 7\}$,
\bas
e_j z
&=
x^0e_j
	- \langle e_j, x \rangle e_0
	+ P(e_j, x)
=
x^0e_j
	- x^j e_0
	+ x^i ~ P(e_j, e_i),
\eas
then, using $k \in \{1, \dotsc, 7\}$,
\bas
\mleft(e_j z, w\mright)
&=
\mleft(
	x^0 e_j
	- x^j e_0
	+ x^i ~ P(e_j, e_i), 
	y^0 e_0 
	+ y^k e_k
\mright)
\\
&=
x^0 y^j
	- x^j y^0
	+ x^iy^k ~ \underbrace{\bigl( P(e_j, e_i), e_k \bigr)}_{\mathclap{ = \langle P(e_j, e_i), e_k \rangle }}
\\
&=
x^0 y^j
	- x^j y^0
	+ x^iy^k ~ \underbrace{\phi(e_j, e_i , e_k )}
		_{\mathclap{ = - \phi(e_j, e_k , e_i) = - \langle P(e_j, e_k), e_i \rangle }}
\\
&=
- \mleft(
	x^j y^0
	- x^0 y^j
	+ x^i y^k ~ \bigl( e_i, P(e_j, e_k) \bigr)
\mright)
\\
&=
- (z, e_j w).
\eas
\end{proof}

With that one can construct a trivialization of $\mathrm{T}\mathds{S}^7$.

\begin{theorems}{$\mathrm{T}\mathds{S}^7$ is trivial, \cite[last part of Example 4.5.10; page 229]{hamilton}}{OktonionenFuerParalellilitaet}
$\mathds{S}^7$ is a parallelizable manifold, and a possible trivialization is given by vector fields $Y_j \in \mathfrak{X}\mleft(\mathds{S}^7\mright)$ ($j \in \{1,\dotsc,7\}$), defined by
\ba
\mleft.Y_j\mright|_z
&\coloneqq 
e_j \cdot z
\ea
for all $z \in \mathds{S}^7$, which is also a orthonormal frame for $(\cdot, \cdot)$ (restricted to a scalar product for $\mathrm{T}\mathbb{S}^7$).
\end{theorems}

\begin{proof}
\leavevmode\newline
Observe
\bas
\mleft(Y_j|_z, z\mright)
&=
(e_j \cdot z, z)
\stackrel{\text{Prop.~\ref{prop:ImportantRelationOfScalarproductonO}}}{=}
- (z, e_j z)
=
- (e_j z, z)
=
- (Y_j|_z, z)
\eas
for all $z \in \mathds{S}^7$,
hence, $(Y_j|_z, z) = 0$, so, perpendicular to $z$, which is why one can view $Y_j \in \mathfrak{X}(\mathds{S}^7)$.
We also have, $k$ also an element of $\{1, \dotsc, 7\}$,
\bas
\mleft( Y_j, Y_k \mright)
&=
\mleft( e_j \cdot z, e_k \cdot z \mright)
\\
&\stackrel{\mathclap{ \text{Prop.~\ref{prop:ImportantRelationOfScalarproductonO}} }}{=}\quad~~
- \bigl( z, e_j \cdot (e_k \cdot z) \bigr)
\\
&=
- \mleft( 
	z, 
	e_j \cdot 
	\mleft(
		x^0 e_k
		- x^k e_0
		+ x^i ~ P(e_k, e_i)
	\mright) 
\mright)
\\
&=
- \mleft( 
	x^0 e_0 + x,
	- x^0 \delta_{jk} e_0
	+ x^0 P(e_j, e_k)
	- x^k e_j
	- x^i \langle e_j, P(e_k, e_i) \rangle ~ e_0
	+ x^i P(e_j, P(e_k, e_i))
\mright)
\\
&=
\mleft( x^0 \mright)^2 \delta_{jk}
	+ x^0x^i \langle e_j, P(e_k, e_i) \rangle
	- x^0 x^i \underbrace{\langle e_i, P(e_j, e_k) \rangle}
		_{\mathclap{ = \phi(e_j, e_k, e_i) = \phi(e_k, e_i, e_j) = \langle e_j, P(e_k, e_i) \rangle }}
	+ x^k x^j
	- \mleft( x, x^i P(e_j, P(e_k, e_i)) \mright)
\\
&=
\mleft( x^0 \mright)^2 \delta_{jk}
	+ x^k x^j
	- \mleft( x, x^i P(e_j, P(e_k, e_i)) \mright)
\eas
writing $z = x^0 e_0 + x$, where $x^0 \in \mathbb{R}$ and $x \in V$; also recall similar calculations of the previous proofs like at the beginning of the proof of Prop.~\ref{prop:ImportantRelationOfScalarproductonO}. Using Lemma \ref{lem:PPFormula},
\bas
\mleft( x, x^i P(e_j, P(e_k, e_i)) \mright)
&=
\langle
	x,
	P(e_j, P(e_k, x))
\rangle
\\
&=
\phi\bigl(e_j, P(e_k, x), x\bigr)
\\
&=
\phi\bigl( x, P(x, e_k), e_j \bigr)
\\
&=
\langle
	P(x, P(x, e_k)),
	e_j
\rangle
\\
&\stackrel{\mathclap{ \text{Lemma \ref{lem:PPFormula}} }}{=}\qquad
\langle
	- \langle x, x \rangle e_k
	+ \langle x, e_k \rangle x,
	e_j
\rangle
\\
&=
- \langle x, x \rangle \delta_{jk}
	+ x^k x^j,
\eas
and, so,
\bas
\mleft( Y_j, Y_k \mright)
&=
\mleft( \mleft(x^0\mright)^2 + \langle x, x \rangle \mright) \delta_{jk}
=
||z||^2 ~ \delta_{jk}
=
\delta_{jk},
\eas
using that $z$ is a unit octonion. Hence, $\mleft( Y_j \mright)_j$ is an orthonormal frame, globally defined, especially linear independent by the orthogonality. Thus, we have a global trivialization of $\mathrm{T}\mathbb{S}^7$.
\end{proof}

We can therefore finally prove that the unit octonions as $\mathds{S}^7$ give rise to a CYMH GT.

\begin{theorems}{Global example: Unit octonions}{UnitOctonionsAreExamples}
$\mathds{S}^7$ admits a CYMH GT as in Thm.~\ref{thm:FinallyTheGaugeInvarianceWeWant} such that the related connection $\nabla$ on $E \coloneqq \mathrm{T}\mathds{S}^7$ is not flat. Moreover, there is no field redefinition $\widetilde{\nabla}^\lambda$ of $\nabla$ such that $\widetilde{\nabla}^\lambda$ is flat, where $\lambda \in \Omega^1(N;E)$ such that $\Lambda = \mathds{1}_{\mathrm{T}\mathbb{S}^7} - \lambda \in \sAut(E)$.
\end{theorems}

\begin{remark}
\leavevmode\newline
The following constructions for this CYMHG GT structure is also very similar to the construction of a flat metric connection in \cite[\S 4]{flatmetricconn}, where a Clifford algebra is used instead.
\end{remark}

\begin{proof}[Proof of Thm.~\ref{thm:UnitOctonionsAreExamples}]
\leavevmode\newline 
Recall the situation as described in \ref{sit:SituationForTangentBundles}; we only need to construct a flat metric connection $\nabla^{\mathrm{bas}}$ on $\mathrm{T}\mathds{S}^7$, because we are going to assume that the metrics on $\mathrm{T}\mathds{S}^7$ as Lie algebroid and tangent bundle are the same. The connection $\nabla$ is then uniquely given by $\nabla^{\mathrm{bas}}$, and we will define the primitive of $\nabla$ by $\zeta \coloneqq t_\nabla$.

The construction follows by Thm.~\ref{thm:OktonionenFuerParalellilitaet}, so, let $\mleft( Y_j \mright)_j$ ($j \in \{1, \dotsc, 7\}$) be the global trivialization of $\mathrm{T}\mathds{S}^7$ defined by $\mathds{S}^7 \ni z \mapsto e_j \cdot z$ for all $j$. Then define $\nabla^{\mathrm{bas}}$ by
\bas
\nabla^{\mathrm{bas}} Y_j
&=
0,
\eas
uniquely extended to a connection of $\mathrm{T}\mathds{S}^7$, using that $Y_j$ is a global frame. Flatness is an immediate consequence, since $\mleft( Y_j \mright)_j$ is a parallel frame by definition.

Moreover, $\mleft( Y_j \mright)_j$ are an orthonormal frame of $(\cdot, \cdot)$; hence, for the CYMH GT we take $(\cdot, \cdot)$ restricted on $\mathrm{T}\mathds{S}^7$ as fibre metric. Then
\bas
\mleft( \nabla^{\mathrm{bas}} (\cdot, \cdot) \mright)(Y_j, Y_k)
&=
\mathrm{d}\bigl( \underbrace{(Y_j, Y_k)}_{= \delta_{jk}} \bigr)
	- \mleft( \nabla^{\mathrm{bas}} Y_j, Y_k \mright)
	- \mleft( Y_j, \nabla^{\mathrm{bas}} Y_k \mright)
=
0
\eas
for all $j,k$. Thus, we have now everything for a CYMH GT, especially, we have a $\nabla$ with vanishing basic curvature. Moreover, by Thm.~\ref{thm:LieGroupIsomorphisms} $\nabla$ cannot be flat, otherwise $\mathds{S}^7$ would admit a Lie group structure. Furthermore, by Thm.~\ref{thm:InvarianceUnderTheFieldRedefinition} the field redefinition preserves the vanishing of the basic curvature such that we can apply the same argument to $\widetilde{\nabla}^\lambda$, thence, $\widetilde{\nabla}^\lambda$ cannot be flat for all $\lambda \in \Omega^1(N;E)$.
\end{proof}

\begin{remarks}{Stability with respect to other transformations}{OktonionenSehrStabil}
As one can see by the proof, the base ingredient is Thm.~\ref{thm:LieGroupIsomorphisms}. Hence, one can probably apply the same statement to every transformation preserving the vanishing of the basic curvature.
\end{remarks}

Hence, we have a CYMH GT on $\mathds{S}^7$ which is not pre-classical (stable under the field redefinition). It was essential that $\mathds{S}^7$ cannot admit a Lie group structure, strongly related to the non-associativity. As we also have seen in Cor.~\ref{cor:TorsionConstancyAndFlatness} and \ref{cor:LemmaCurvatureOfDualConnections}, also recall the proof of the former, the flatness of $\nabla$ is equivalent to the constancy of the structure functions with respect to a parallel frame of $\nabla^{\mathrm{bas}}$. The parallel frame we took in the last proof was the trivialization $\mleft( Y_j \mright)_j$ ($j \in \{1, \dotsc, 7\}$) given in Thm.~\ref{thm:OktonionenFuerParalellilitaet}; summarising all of that, we can conclude that the non-associativity is directly related to the non-constancy of the structure functions for $\mleft( Y_j \mright)_j$. In \cite[Equation (4); an ArXiv preprint]{octonions} is a formula derived for precisely those structure functions, emphasizing this argument since the non-constant term there is directly related to the non-associativity.

This concludes our discussion of tangent bundles; let us now turn to general Lie algebroids. The octonions will not appear anymore, hence, the notation will not be used anymore and the following notation will resemble the previous notations again.
\section{General Lie algebroids}\label{GeneralObstrAoids}

\subsection{General situation}\label{GeneralGeneral}

Let us now go to more general Lie algebroids as also used in the discussion until and around Thm.~\ref{thm:FinallyTheGaugeInvarianceWeWant}.

The previously discussed constancy of the torsion and its relationship to flatness in the case of tangent bundles we also have partially for general Lie algebroids.

\begin{corollaries}{Pre-classical theories have constant torsion}{TorsionConstancyAndFlatnessGeneral}
Let $E\to N$ be a Lie algebroid over a smooth manifold $N$, equipped with a connection $\nabla$ on $E$ with vanishing basic curvature. Then there is a $\lambda \in \Omega^1(N;E)$ such that $\widetilde{\nabla}^\lambda_\rho$ is flat if and only if there is a $\lambda \in \Omega^1(N;E)$ such that $t_{\mleft(\widetilde{\nabla}^\lambda \mright)^{\mathrm{bas}}} = - t_{\widetilde{\nabla}^\lambda_\rho}$ is constant with respect to $\mleft(\widetilde{\nabla}^\lambda \mright)^{\mathrm{bas}}$, that is, 
\ba
\mleft(\widetilde{\nabla}^\lambda \mright)^{\mathrm{bas}} t_{\mleft(\widetilde{\nabla}^\lambda \mright)^{\mathrm{bas}}}
&= 
0.
\ea
\end{corollaries}

\begin{remark}
\leavevmode\newline
As for tangent bundles also recall here Cor.~\ref{cor:TOrsionCanBeLieBracketIfFlat}; in the case of a flat $\nabla_\rho$ (or its field redefinition) its torsion would be another Lie bracket on $E$, but tensorial. One could clearly generalize this statement by just imposing flatness of $\nabla^\mathrm{bas}$ on $E$.
\end{remark}

\begin{proof}[Proof of Cor.~\ref{cor:TorsionConstancyAndFlatnessGeneral}]
\leavevmode\newline
The proof is exactly as in Cor.~\ref{cor:TorsionConstancyAndFlatness}, the only exception is that Cor.~\ref{cor:LemmaCurvatureOfDualConnections} (in combination with Prop.~\ref{prop:SnablamitREnabla}) in general implies
\bas
R_{\nabla_\rho}
&=
\nabla^{\mathrm{bas}} t_{\nabla^{\mathrm{bas}}},
\eas
which is why we can extend Cor.~\ref{cor:TorsionConstancyAndFlatness} only to $\nabla_\rho$ in general.
\end{proof}

\subsection{Direct products of CYMH GTs}\label{GeneralSitDirectProducts}

As we know, Lie algebroids are the direct product of a tangent bundle and a bundle of Lie algebras around regular points, Thm.~\ref{thm:DirectSplitting}. Hence, there is hope to extend some of the previous results to direct products of Lie algebroids. Therefore let us first define the direct product of CYMH GTs, especially recall Remark \ref{rem:NotationAboutProductStructures}, Lemma \ref{lem:LemmaUniquenessOfDirectProductStructure} and Section \ref{DirectProdsOfLieAlgoids} in general. We will make use of the direct product of Lie algebroids without further explaining again how the anchor and bracket \textit{etc.}~are defined.

\begin{theorems}{Direct products of CYMH GTs is a CYMH GT}{DirectProductsOfCYMHGT}
Let $i \in \{1,2\}$ and $E_i \to N_i$ be Lie algebroids over smooth manifolds $N_i$, both equipped with a connection $\nabla^i$, a fibre metric $\kappa_i$ on $E_i$ and a Riemannian metric $g_i$ of $N_i$ such that the compatibility conditions are satisfied for each $i$, where we denote the primitives of $R_{\nabla^i}$ by $\zeta^i$. 

Then the direct product of Lie algebroids $E_1 \times E_2$ is a CYMH GT, equipped with $\nabla \coloneqq \nabla^1 \times \nabla^2$, $\kappa_1 \times \kappa_2$, and $g_1 \times g_2$, where the primitive of the curvature $R_{\nabla^1\times\nabla^2}$ is for example given by $\zeta^1\times\zeta^2$.
\end{theorems}

\begin{proof}
\leavevmode\newline
That is trivial to see by recalling Remark \ref{rem:NotationAboutProductStructures}, especially we have
\bas
\mleft( \nabla^1 \times \nabla^2 \mright)^{\mathrm{bas}}
&=
\mleft(\nabla^1\mright)^{\mathrm{bas}} \times \mleft(\nabla^2\mright)^{\mathrm{bas}},
\\
R_{\nabla^1 \times \nabla^2}^{\mathrm{bas}}
&=
R_{\nabla^1}^{\mathrm{bas}}
\times
R_{\nabla^2}^{\mathrm{bas}},
\\
R_{\nabla^1\times\nabla^2}
&=
R_{\nabla^1}
\times
R_{\nabla^2},
\\
\mathrm{d}^{\mleft( \nabla^1 \times \nabla^2 \mright)^{\mathrm{bas}}}
\mleft( \zeta^1\times\zeta^2 \mright)
&=
\mathrm{d}^{\mleft( \nabla^1 \mright)^{\mathrm{bas}}}
\mleft( \zeta^1\mright)
\times
\mathrm{d}^{\mleft( \nabla^2 \mright)^{\mathrm{bas}}}
\mleft( \zeta^2 \mright)
\eas
Hence, using the compatibility conditions on $E_i$,
\bas
R_{\nabla}
&=
- \nabla^{\mathrm{bas}} \mleft(\zeta^1\times\zeta^2\mright),
\\
R_\nabla^{\mathrm{bas}}
&=
0,
\eas
and
\bas
\nabla^{\mathrm{bas}}\mleft( \kappa^1 \times \kappa^2 \mright)
&=
\mleft(\mleft(\nabla^1\mright)^{\mathrm{bas}} \kappa^1\mright)
\times \mleft( \mleft(\nabla^2\mright)^{\mathrm{bas}} \kappa^2\mright)
=
0,
\eas
similarly for $g^1\times g^2$.
\end{proof}

\begin{definitions}{Direct product of CYMH GT}{DefinitionDerDirectProdCYMHGT}
Assume the same as in Thm.~\ref{thm:DirectProductsOfCYMHGT}. Then we call $E_1\times E_2$ with its natural CYMH GT structure defined there the \textbf{direct product of CYMH GTs}.
\end{definitions}

In the following statement we study a certain CYMH GT, as it is given around regular points, and we will not always denote all the structures; for example, we just denote the connections when we are not going to use the compatibilities with the metrics.

\begin{theorems}{Direct products of CYMHG GTs around regular points are flat}{DirectProductsSadlyAlwaysFlat}
Let $N \coloneqq \mathbb{R}^n$ ($n \in \mathbb{N}_0$) be a smooth manifold such that its tangent bundle admits a CYMH GT, whose connection satisfying the compatibility conditions we denote by $\nabla^{N}$, and let $K \to S$ be an LAB over a smooth contractible manifold $S$ which also admits a CYMH GT, equipped with a connection $\nabla^K$ satisfying the compatibility conditions.

Then there is a field redefinition with respect to the direct product of CYMH GTs, $E \coloneqq \mathrm{T}N \times K \to N \times S$, such that $\widetilde{\nabla}^\lambda$ is flat, where $\nabla \coloneqq \nabla^{N} \times \nabla^K$ and $\lambda \in \Omega^1(N;E)$ such that $\Lambda = \mathds{1}_E - \lambda \circ \rho \in \sAut(E)$.
\end{theorems}

\begin{proof}
\leavevmode\newline
%We have seen earlier that this is possible when just looking at $N$ or $K$, see Thm.~\ref{thm:LokalLeiderImmerPreklassisch} and \ref{thm:NoGoLocalTangentBundle}. Recall Eq.~\eqref{dievielBessereFormuelFuersRechnenFragezeichen}, that is
%\bas
%\widetilde{\nabla}^\lambda_Y \mu
%&=
%\Lambda \mleft( 
	%\nabla_{\widehat{\Lambda}^{-1}(Y)} \mu
	%- \mleft[ \mleft( \Lambda^{-1} \circ \lambda \mright)(Y), \mu \mright]_E 
%\mright)
%+ \lambda \bigl([Y, \rho(\mu)] \bigr)
%\eas
%for all $\mu \in \Gamma(E)$ and $Y \in \mathfrak{X}(N)$.
We need to check whether we can apply Thm.~\ref{thm:LokalLeiderImmerPreklassisch} and \ref{thm:NoGoLocalTangentBundle} separately. We will do so by studying the field redefinition only for $\nabla$ with respect to $\lambda$ of the form
\bas
\lambda
&=
\lambda^N \times \lambda^K
=
\mathrm{pr}_1^!\mleft( \lambda^N \mright)
	\oplus \mathrm{pr}_2^!\mleft( \lambda^K \mright),
\eas
where $\mathrm{pr}_i$ ($i \in \{1,2\}$) is the projection onto the $i$-th factor in $N \times S$, $\lambda^N \in \Omega^1(N; \mathrm{T}N)$, and $\lambda^K \in \Omega^1(S;K)$. Using such a $\lambda$ implies 
\bas
\Lambda
&=
\underbrace{\mathds{1}_{\mathrm{T}N \times K}}_{\mathclap{ = \mathds{1}_{\mathrm{T}N} \times \mathds{1}_{K} }}
	- \lambda \circ \underbrace{\rho_{\mathrm{T}N \times K}}_{= \rho_{\mathrm{T}N} \times \rho_{K} = \mathds{1}_{\mathrm{T}N} \times 0 }
=
\Lambda^{N} \times \Lambda^K,
\eas
where $\Lambda^{N} \coloneqq \mathds{1}_{\mathrm{T}N} - \lambda^{N}$ and $\Lambda^{K} \coloneqq \mathds{1}_{K}$. Therefore
\bas
\Lambda^{-1}
&=
\mleft(\Lambda^{N}\mright)^{-1} \times \mleft(\Lambda^K\mright)^{-1},
\eas
similarly for $\widehat{\Lambda}$. Again by Remark \ref{rem:NotationAboutProductStructures} we have
\bas
\nabla^{\mathrm{bas}}
&=
\mleft(\nabla^N\mright)^{\mathrm{bas}}
	\times \mleft(\nabla^K\mright)^{\mathrm{bas}},
\eas
and, so, the following completely splits as direct product
\bas
\mleft(\Lambda \circ \mathrm{d}^{\nabla^{\mathrm{bas}}} \circ \Lambda^{-1} \mright)\lambda
&=
\mleft( \mleft(\Lambda^N \circ \mathrm{d}^{\mleft(\nabla^N\mright)^{\mathrm{bas}}} \circ \mleft(\Lambda^N\mright)^{-1} \mright)\lambda^N  \mright)
	\times \mleft( \mleft(\Lambda^K \circ \mathrm{d}^{\mleft(\nabla^K\mright)^{\mathrm{bas}}} \circ \mleft(\Lambda^K\mright)^{-1} \mright)\lambda^K  \mright),
\eas
by Def.~\eqref{FieldTrafoOfNabla} we get, using $\nabla= \nabla^N \times \nabla^K$,
\bas
\widetilde{\nabla}^\lambda
&=
\mleft(\widetilde{\nabla}^N\mright)^{\lambda^N}
	\times \mleft(\widetilde{\nabla}^K\mright)^{\lambda^K}.
\eas
This means that we can calculate the field redefinition of the curvature as if we would just look at either $\mathrm{T}N$ or $K$ as in the previous sections because then the curvature splits, too, as usual. So, define $\lambda^N$ in such a way that $\mleft(\widetilde{\nabla}^N\mright)^{\lambda^{N}}$ is flat by using Thm.~\ref{thm:NoGoLocalTangentBundle}; in the same fashion choose $\lambda^K$ such that $\mleft(\widetilde{\nabla}^K\mright)^{\lambda^K}$ is flat using Thm.~\ref{thm:LokalLeiderImmerPreklassisch}.
\end{proof}

As one has seen in the proof, the idea is to take a $\lambda = \lambda^N \times \lambda^K$. It is natural to assume that we can extend and generalize previous statements which were just about the existence of a $\lambda$. However, statements about the stability of a CYMH GT under the field redefinition like Thm.~\ref{thm:AbelschIstGeileNeueTheorie} and \ref{thm:UnitOctonionsAreExamples}, or the construction of the obstruction class for LABs. The reason for this are the mixed terms in the formulas of the field redefinition if $\lambda \neq \lambda^N \times \lambda^K$ such that the connection of $K$ could contribute to the curvature of $\mathrm{T}N$, for example assume, using the same notation as in the previous statement and proof, $\lambda \in \Omega^1(N; K)$, so, a form along $N$ but having values in $K$. Then by Eq.~\eqref{dievielBessereFormuelFuersRechnenFragezeichen}, similar calculations as before and using that $\lambda$ has values in $K$,
\bas
\mleft(\widetilde{\nabla}^N\mright)^\lambda_{\partial_i} \partial_j
&=
\Lambda\mleft(
	\nabla_{\widehat{\Lambda}^{-1}(\partial_i)}^N \partial_j
	- \mleft[ \lambda(\partial_i), \partial_j \mright]_E
\mright)
\\
&=
\mleft( \mathds{1} - \lambda \mright)\mleft(
	\nabla_{\partial_i}^N \partial_j
\mright)
	- \mleft[ \lambda(\partial_i), \partial_j \mright]_E
\\
&=
\nabla_{\partial_i}^N \partial_j
	- \lambda\mleft(
	\nabla_{\partial_i}^N \partial_j
\mright)
	- \mleft[ \lambda(\partial_i), \partial_j \mright]_E,
\eas
observe that $\widehat{\Lambda} = \mathds{1}$ such that every $\lambda \in \Omega^1(N;K)$ is allowed by Sylvester's determinant theorem. The first summand has values in $\mathrm{T}N$ and the second and third in $K$. Hence, in general the formulas will not split anymore for general $\lambda$. However, I personally hope and assume the following conjecture.

\begin{conjectures}{Existence of a splitted field redefinition}{DoesMyFieldRedefSplit}
Let $N$ be a smooth manifold such that its tangent bundle admits a CYMH GT, and let $K \to S$ be an LAB over a smooth manifold $S$ which also admits a CYMH GT.

If there is a field redefinition such that the direct product of CYMH GTs,  $E \coloneqq \mathrm{T}N \times K \to N \times S$, is pre-classical or classical, then there is also a field redefinition with respect to a $\lambda$ of the form $\lambda^N \times \lambda^K$ such that the direct product of CYMH GTs is pre-classical or classical, respectively, where $\lambda^N \in \Omega^1(N; \mathrm{T}N)$ and $\lambda^K \in \Omega^1(S;K)$ are valid parameters for field redefinitions for each factor.
\end{conjectures}

If it is possible to show this, then the whole discussion about field redefinition towards pre-classical or classical structures would reduce to parameters of the form $\lambda = \lambda^N \times \lambda^K$, essentially, one could look at both factors separately in a direct product of CYMH GTs.

Due to the fact that the general situation is very difficult to study this is the final conclusion of CYMH GTs. What will follow are loose ideas and ansatzes, very loosely structured, for a possible following discussion and study after the thesis. Hence, the reader can ignore the following subsection if wanted.

\subsection{Loose ideas and ansatzes}\label{LastAnsatzes}

As a first ansatz one may want to assume a connection which can restrict to the isotropy of the anchor, in the hope to generalize the discussion about the LABs; especially recall the discussion about LABs in the context of CYMH GTs, we will strongly refer to that without much further notice.

\begin{lemmata}{Invariance of connection restricting on the isotropy}{lemmaIsotropyInvariant}
Let $E\to N$ be a Lie algebroid over a smooth manifold $N$, and $L$ a subbundle of $E$ with $\rho(\nu) = 0$ and $\mleft[ \nu, \mu \mright]_E \in \Gamma(L)$ for all $\nu \in \Gamma(L)$ and $\mu \in \Gamma(E)$, \textit{i.e.}~$\Gamma(L)$ is an ideal of $\Gamma(E)$, living in the kernel of $\rho$. Moreover, let $\nabla$ be a connection on $E$ and $L$ with $\nabla \bigl( \Gamma(L) \bigr) \subset \Gamma(L)$.

Then
\ba
\widetilde{\nabla}^\lambda \bigl( \Gamma(L) \bigr) \subset \Gamma(L).
\ea
\end{lemmata}

\begin{proof}
\leavevmode\newline
By Eq.~\eqref{dievielBessereFormuelFuersRechnenFragezeichen} we have
\bas
\widetilde{\nabla}^\lambda_Y \mu
&=
\Lambda \mleft( \nabla_{\widehat{\Lambda}^{-1}(Y)} \mu
- \mleft[ \mleft( \Lambda^{-1} \circ \lambda \mright)(Y), \mu \mright]_E \mright)
+ \lambda \big([Y, \rho(\mu)] \big)
\eas
for all $\mu\in \Gamma(E)$ and $Y \in \mathfrak{X}(N)$. The statement follows now for $\mu \in \Gamma(L)$ because of the assumptions and $\Lambda|_{\mathrm{Ker}(\rho)} = \mathds{1}_{\mathrm{Ker}(\rho)}$.
\end{proof}

Let us interpret this algebraically for the flat situation; recall Def.~\ref{def:IsotropyForLieAlgeoids} and its discussion.

\begin{propositions}{Algebraic meaning in the flat situation}{AlgebraicMeaningOfTheFirstInvariantIveFound}
Let $E= N \times \mathfrak{g}$ be an action Lie algebroid over a smooth manifold $N$ of a Lie algebra $\mathfrak{g}$, whose Lie algebra action is induced by a Lie group action of a Lie group $G$ on $N$, $G \times N \ni (g,p) \mapsto gp \in N$. Moreover, let $\nabla$ be the canonical flat connection for which we assume $\rho(\nabla \nu) = 0$ for all $\nu \in \Gamma(E)$ with $\rho(\nu)=0$.

Then $\mathrm{Ker}(\rho_p) = \mathrm{Ker}(\rho_q)$, and $\mathrm{Ker}(\rho_p)$ is an ideal of $\mathfrak{g}$, where $p, q \in N$ are arbitrary regular points of the same connected component of regular points.
\end{propositions}

\begin{remark}
\leavevmode\newline
%Thanks to my professor of my master thesis, Mark J.~D.~Hamilton, for discussing this proposition.
Recall Thm.~\ref{thm:ActionLieALgebroid}: Having a flat connection $\nabla$ with vanishing basic curvature implies that locally we have a similar situation as in this proposition, just with additional integrability of the underlying Lie algebra assumed here.

Since every action Lie algebroid can be integrated to a Lie groupoid and due to a generalization of $\mathrm{Ad}$ as in \cite[Section 3.7, especially Prop. 3.7.1 (iii); page 141ff.]{mackenzieGeneralTheory}, one might be able to proof that statement (locally) for any Lie algebroid with a flat CYMH-compatible connection $\nabla$.
\end{remark}

\begin{proof}[Proof of Prop.~\ref{prop:AlgebraicMeaningOfTheFirstInvariantIveFound}]
\leavevmode\newline
%As we already know, $E$ will be locally isomorphic to an action Lie algebroid, see Thm.~\ref{thm:ActionLieALgebroid}. Hence, w.l.o.g.~assume $E \cong N \times \mathfrak{g}$ as an action Lie algebroid for some Lie algebra $\mathfrak{g}$, and the anchor is induced. Hence, we only have to show that $K$ is an ideal of $\mathfrak{g}$.
By definition parallel sections of $\nabla$ are precisely constant sections, so, fix a basis $\mleft( e_a \mright)_a$ of $\mathfrak{g}$, constantly extended to $E$, such that $\nabla e_a = 0$. W.l.o.g.~assume that $N$ is connected and just consists of regular points (fix \textit{e.g.}~a connected component of regular points on $N$), hence, $K \coloneqq \mathrm{Ker}(\rho)$ has constant rank and describes a bundle of Lie algebras. Then due to $\nabla\bigl( \Gamma(K) \bigr) \subset \Gamma(K)$ by assumption, we know that $\nabla|_K$ is also flat which implies that a subset of the parallel sections (= constant sections) describes a frame of $K$. Thus, we can choose $\mleft(e_a\mright)_a$ in such a way that there is a subframe $\mleft( f_\alpha \mright)_\alpha$ (locally) spanning $K$.\footnote{Technical: A space of parallel sections are finite-dimensional subspaces of, here, $\Gamma(E)$, whose basis is \textit{e.g.}~the frame we choose here. Then one can just apply standard analysis of vector spaces, \textit{i.e.}~take any finite-dimensional basis of parallel sections of $K$, and then extend that basis to a basis of parallel sections of $E$.} Since $\mleft( f_\alpha \mright)_\alpha$ consists of constant sections, we can conclude that the isotropy subalgebra of $\mathfrak{g}$ is the same for all points of $N$, \textit{i.e.}~
\bas
K_p=
\mathfrak{g}_p &= \mathfrak{g}_{q} = K_q
\eas 
for all $p, q \in N$, where $K_p= \mathfrak{g}_p$ and $K_q=\mathfrak{g}_q$ is the isotropy algebra at $p$ and $q$, respectively.

Also recall Cor.~\ref{cor:IsotropyVonLieAlgMitAdjoint}, that is, also using the just shown equality $K_p = K_{gp}$ for all $p \in N$ and $g \in G$, we get
\bas
&&&
\mathrm{Ad}(g)(w) \in K_{p}
\\
&\Rightarrow&
&
\mathrm{Ad}(\exp(tv))(w)
\in K_p \\
&\stackrel{\mathclap{\mathfrak{g}_p \text{ closed subalgebra of } \mathfrak{g}}}{\Rightarrow}&
&
\mleft[ v, w \mright]_{\mathfrak{g}}
\in K_p,
\eas
for all $p \in N$, $g \in G$, $w \in \mathfrak{g}_p = K_p$, $t \in \mathbb{R}$, and $v \in \mathfrak{g}$.
Thus, $K_p$ is an ideal of $\mathfrak{g}$.
\end{proof}

\begin{remark}
\leavevmode\newline
For simplicity assume now that the rank of the anchor is constant. Also assume we have an action Lie algebroid, related to a Lie algebra $\mathfrak{g}$, with a non-flat connection $\nabla$ such that we have a CYMH gauge theory and $\nabla(\Gamma(K)) \subset \Gamma(K)$, where $K \coloneqq \mathrm{Ker}(\rho)$. Moreover, assume that the action behind the anchor can be integrated to a Lie group action. If the anchor has a non-trivial kernel (so, nonzero and not all of the Lie algebroid), then one may try the following argument: Assume there is a $\lambda \in \Omega^1(N;E)$ such that $\widetilde{\nabla}^\lambda$ is flat. By Lemma \ref{lem:lemmaIsotropyInvariant} we have $\widetilde{\nabla}^\lambda(\Gamma(K)) \subset \Gamma(K)$. Locally we still have an action Lie algebroid related to a Lie algebra $\mathfrak{g}^\prime$ by Thm.~\ref{thm:ActionLieALgebroid} such that $\widetilde{\nabla}^\lambda$ is the canonical flat connection. Then by Prop.~\ref{prop:AlgebraicMeaningOfTheFirstInvariantIveFound} we know that the kernel of $\rho_p$ at a regular point $p\in N$ is an ideal of the Lie algebra $\mathfrak{g}^\prime$ of the new action Lie algebroid; this ideal is nontrivial (not zero and not $\mathfrak{g}^\prime$) because the anchor's kernel is nontrivial. When we start \textit{e.g.}~with a simple Lie algebra $\mathfrak{g}$, we get clearly a contradicion if the new Lie algebra $\mathfrak{g}^\prime$ is still simple.

However, we cannot expect that $\mathfrak{g}^\prime$ is of a similar type as $\mathfrak{g}$ when the anchor is nonzero. For example take the two dimensional non-abelian Lie algebra $\mathfrak{g} \coloneqq \mathbb{R}^2 = \mathrm{span}\langle e_1, e_2 \rangle$, $[e_1, e_2]_{\mathfrak{g}} = e_2$, equipped with an action $\gamma$ on $N \coloneqq \mathbb{R}^2$ defined by
\bas
\gamma(e_1) &\coloneqq \partial_x, 
\\
\gamma(e_2) &\coloneqq 0,
\eas
where we denote the coordinates of $N$ by $x$ and $y$.
It is trivial to check that $\gamma$ is a Lie algebra action, hence, we have a corresponding action Lie algebroid $E = N \times \mathfrak{g}$ with anchor $\rho$ induced by $\gamma$ and Lie algebroid bracket $\mleft[ \cdot, \cdot \mright]_E$ induced by $\mleft[ \cdot, \cdot \mright]_{\mathfrak{g}}$. $e_1$ and $e_2$ are a global frame when viewed as constant sections.

Now we make a change of the frame: $\tilde{e}_1 \coloneqq e_1$, and $\tilde{e}_2 \coloneqq \e^{-x} e_2$. We still have $\rho(\tilde{e}_1) = \partial_x$ and $\rho(\tilde{e}_2) = 0$, but by the Leibniz rule we arrive at
\bas
\mleft[ \tilde{e}_1, \tilde{e}_2 \mright]_E
&=
\e^{-x} \underbrace{[e_1, e_2]_{\mathfrak{g}}}_{= e_2}
	- \e^{-x} e_2
=
0.
\eas
Therefore, the frame given by $\tilde{e}_1$ and $\tilde{e}_2$ gives rise to an isomorphism $E \cong N \times \mathfrak{g}^\prime$ as action Lie algebroid, where $\mathfrak{g}^\prime$ is the two-dimensional abelian Lie algebra. So, we could have also started with the abelian Lie algebra instead of the non-abelian one to define precisely the same action Lie algebroid, both equipped with an action inducing the same anchor. 

This ambiguous behaviour depends on the rank of the anchor. For a zero anchor, that is, for bundle of Lie algebras, like the BLA induced by the kernel of an anchor around regular points, that can certainly not happen. But recall the splitting theorem, Section \ref{SectionAboutSplitting}, one part of the Lie algebroid also comes from the tangent bundle of the leaves, and as we know, the structure functions of a tangent bundle can be very arbitrary. For example start with the coordinate vector fields, hence, zero structure functions (abelian). Then there is obviously a non-constant change of the frame such that the structure functions are not zero anymore because of the Leibniz rule in the bracket; for example choose a frame which is not a full set of coordinate vector fields.
\end{remark}

As in the case of LABs, having a connection restricting to the kernel (or an ideal of it) would imply that we have an LAB structure there due to the vanishing of the basic curvature; recall the the isotropy is a bundle of Lie algebras around regular points.

\begin{corollaries}{Lie derivation laws and vanishing basic curvature}{LieDerivationGleichVanishingBasicCurvature}
Let $E \to N$ be a Lie algebroid, where $N$ is a connected manifold just consisting of regular points, $L$ be a subbundle of Lie algebras of $K \coloneqq \mathrm{Ker}(\rho)$, and $\nabla$ a connection on $E$ with $\nabla \bigl( \Gamma(L) \bigr) \subset \Gamma(K)$. Then
\ba
\nabla^{\mathrm{bas}}_\nu Y
&=
0
\ea
for all $\nu \in \Gamma(L)$ and $Y \in \mathfrak{X}(N)$.

If we additionally have $\nabla \bigl( \Gamma(L) \bigr) \subset \Gamma(L)$, then the following are equivalent:
\begin{enumerate}
	\item $\nabla$ a Lie derivation law on $L$.
	\item The basic curvature of $\nabla$ restricted on $L$ is zero, \textit{i.e.}~
	\bas
	R_\nabla^{\mathrm{bas}}(\mu, \nu) Y
	&=
	0
	\eas
	for all $\mu, \nu \in \Gamma(L)$ and $Y \in \mathfrak{X}(N)$.
\end{enumerate}
\end{corollaries}

\begin{proof}
\leavevmode\newline
Those are trivial consequences of $\nabla \bigl( \Gamma(L) \bigr) \subset \Gamma(K)$, \textit{i.e.}~
\bas
\rho (\nabla \nu)
&=
0
\eas
for all $\nu \in \Gamma(L)$, hence,
\bas
\nabla^{\mathrm{bas}}_\nu Y
&=
[\underbrace{\rho(\nu)}_{=0}, Y]
	+ \underbrace{\rho\bigl( \nabla_Y \nu \bigr)}_{=0}
=
0
\eas
for all $\nu \in \Gamma(L)$ and $Y \in \mathfrak{X}(N)$. With additionally $\nabla \bigl( \Gamma(L) \bigr) \subset \Gamma(L)$ then also
\bas
R_\nabla^{\mathrm{bas}}(\mu, \nu)Y
&=
\nabla_Y\mleft(\mleft[\mu, \nu\mright]_E\mright) 
	- \mleft[ \smash{\underbrace{\nabla_Y \mu}_{\in \Gamma(L)}}, \nu \mright]_E 
	- \mleft[ \mu, \nabla_Y \nu \mright]_E 
	\underbrace{- \nabla_{\nabla^{\mathrm{bas}}_\nu Y} \mu 
	+ \nabla_{\nabla^{\mathrm{bas}}_\mu Y} \nu}
	_{= 0}
\\
&=
\nabla_Y\mleft(\mleft[\mu, \nu\mright]_L\mright) 
	- \mleft[ \nabla_Y \mu, \nu \mright]_L 
	- \mleft[ \mu, \nabla_Y \nu \mright]_L
\eas
for all $\mu, \nu \in \Gamma(L)$ and $Y \in \mathfrak{X}(N)$. Therefore, $\nabla$ has a vanishing basic curvature restricted on $L$ if and only if it is a Lie derivation law on $L$ (a Lie bracket derivation of $L$).
\end{proof}

Using Thm.~\ref{thm:BLALAB}, $L$ has to be an LAB in such a case; hence having such an $L$ and $\nabla$ there is hope to generalize our results with respect to LABs. In the study about LABs, the obstruction class was given by $\mathrm{d}^\nabla \zeta$ and we have argued that this is exact with respect to $\mathrm{d}^\Xi$ in the case of flatness, which was the differential for centre-valued forms induced by a pairing $\Xi$ of an LAB with a tangent bundle, induced by $\nabla$ which restricted to centre-valued forms by the vanishing of the basic curvature. The essential argument about the exactness of $\mathrm{d}^\nabla \zeta$ was the compatibility condition for $\zeta$, implying that $\zeta$ is centre-valued in the case of LABs and flatness, and another argument was that $\nabla$ restricts to such centre-valued sections. In general, flatness now implies closedness of $\zeta$ with respect to the basic connection. Therefore let us study whether $\nabla$ restricts to closed forms also in general.

\begin{corollaries}{$\nabla$ preserving $\nabla^{\mathrm{bas}}$-closedness}{NablaErhaeltBasicConnKonstanz}
Let $E \to N$ be a Lie algebroid over a smooth manifold $N$, and $\nabla$ a connection on $E$ with vanishing basic curvature. Then we have
\ba
\mathrm{d}^{\nabla^{\mathrm{bas}}} \mathrm{d}^\nabla \omega
&=
0
\ea
for all $\omega \in \Omega^q(E;E)$ ($q \in \mathbb{N}_0$) with $\mathrm{d}^{\nabla^{\mathrm{bas}}} \omega = 0$ and $\rho (\omega) = 0$.
\end{corollaries}

\begin{remark}
\leavevmode\newline
By Cor.~\ref{cor:commutationS=0} we immediately have
\ba
\mathrm{d}^{\nabla^{\mathrm{bas}}} \mathrm{d}^\nabla \omega
&=
\mathrm{d}^\nabla \mathrm{d}^{\nabla^{\mathrm{bas}}} \omega
\ea
for all $\omega \in \Omega^{p,q}(N,E;E)$ ($p,q \in \mathbb{N}_0$), when $\nabla$ is flat. Thus, 
\ba
\mathrm{d}^{\nabla^{\mathrm{bas}}} \mathrm{d}^\nabla \omega
&=
0
\ea
for all $\nabla^{\mathrm{bas}}$-closed $\omega \in \Omega^{p,q}(N,E;E)$ and flat $\nabla$ with vanishing basic curvature.
\end{remark}

\begin{proof}[Proof of Cor.~\ref{cor:NablaErhaeltBasicConnKonstanz}]
\leavevmode\newline
That is a trivial consquence of Cor.~\ref{cor:commutationS=0}, using $\Omega^q(E;E) \cong \Omega^{(p=0,q)}(N,E;E)$,
\bas
\mathrm{d}^{\nabla^{\mathrm{bas}}} \mathrm{d}^\nabla \omega
&\stackrel{\rho (\omega) = 0}{=}
\mathrm{d}^\nabla \underbrace{\mathrm{d}^{\nabla^{\mathrm{bas}}} \omega}_{=0}
=
0.
\eas
\end{proof}

 Hence, in general it is natural to assume that it is about exactness with respect to the basic connection, a replacement of the centre-valued forms in the study about LABs.
However, in order to define a differential on such parallel sections similar to $\mathrm{d}^\Xi$, we require flatness of $\nabla$ restricted to these sections, regardless whether $\nabla$ itself was flat; otherwise it is difficult to study non-flat $\nabla$ similar to the discussion for LABs. In the case of LABs this was trivially given by the compatibility condition between the curvature and $\zeta$, which immediately implied that $R_\nabla(\cdot, \cdot)\nu = 0$ for all centre-value sections $\nu$. But in general this would mean
\bas
0
&=
R_\nabla(\cdot, \cdot)\nu 
=
- \nabla^{\mathrm{bas}}_\nu \zeta
\eas
for all $\nu$. Hence, centre-valued sections, onto which $\nabla$ shall restrict, seems not only be about closed sections, but also about sections $\nu$ with $\nabla^{\mathrm{bas}}_\nu = 0$,\footnote{Recall the similarity to the condition in Lemma \ref{lem:ParallelFramesForEConnections}.} which makes sense, because the basic connection on $E$ is in the case of LABs an adjoint representation in both arguments, so, there is an ambiguity in how to generalize centre-valued sections in this context.

\begin{definitions}{The centre of basic connections}{CentreOfBasicConnections}
Let $E \to N$ be a Lie algebroid over a smooth manifold $N$, $V\to N$ a vector bundle, and ${}^E\nabla$ an $E$-connection on $V$. Then we define the \textbf{centre $\gls{ZENabla}$ of ${}^E\nabla$} by
\ba
Z\mleft( {}^E\nabla \mright)
&\coloneqq
\left\{
	\nu \in E ~ \middle| ~
	{}^E\nabla_\nu = 0
\right\}.
\ea
In the case of ${}^E\nabla = \nabla^{\mathrm{bas}}$ we mean both, $\nabla^{\mathrm{bas}}$ on $E$ and $\mathrm{T}N$, \textit{i.e.}~$\nabla^{\mathrm{bas}}_\nu = 0$ for both connections simultaneously when $\nu \in Z \mleft( \nabla^{\mathrm{bas}} \mright)$.
\end{definitions}

\begin{remark}
\leavevmode\newline
Since ${}^E\nabla_\nu$ is tensorial in $\nu$, we can restrict this definition to a point $p \in N$, giving rise to a definition of the centre at $p$, denoted by $Z_p \mleft( {}^E\nabla \mright)$; the tensorial behaviour clearly also implies that this is a vector space. Similarly, sections with values in $Z\mleft( {}^E\nabla \mright)$ are a vector space subset of $\Gamma(E)$ by definition, but it is not necessarily a module with constant rank as we are going to see.

Thus, for the following proofs about the structure of $Z \mleft( \nabla^{\mathrm{bas}} \mright)$ we will often use (local) sections $\nu \in \Gamma(E)$ with values in $Z \mleft( \nabla^{\mathrm{bas}} \mright)$, extending a certain element of $E$. That is mainly for convenience due to the fact how connections are normally denoted, and in order to use the definition of $\nabla^{\mathrm{bas}}$.
\end{remark}

Recall that the kernel of the anchor $\rho$ at a point $p \in N$ is a Lie algebra, whose Lie algebra is inherited by $\mleft[ \cdot, \cdot \mright]_E$, and that we denote centres of Lie algebras $\mathfrak{g}$ by $Z(\mathfrak{g})$ (similar for Lie algebra bundles). We denote the Lie bracket of $\mleft[ \cdot, \cdot \mright]_E$ on the kernel by $\mleft[ \cdot, \cdot \mright]_{\mathrm{Ker}(\rho)}$ (similar for the Lie algebra structure on each fibre or for any subalgebras). Around regular points of $E$ the kernel of the anchor is a bundle of Lie algebras as previously mentioned, and by Thm.~\ref{thm:BLALAB} it will be a Lie algebra bundle (LAB) when there is a Lie derivation law.

\begin{propositions}{Properties of the centre}{PropsofCentreOfBasicConnections}
Let $E \to N$ be a Lie algebroid over a smooth manifold $N$, $V\to N$ a vector bundle of at least rank 1, and ${}^E\nabla$ an $E$-connection on $V$. Then $Z_p \mleft( {}^E\nabla \mright)$ is a subset of $\mathrm{Ker}\mleft(\rho_p\mright)$ for all $p \in N$.

If we have a vector bundle connection $\nabla$ on $E$, then $Z_p \mleft( \nabla^{\mathrm{bas}} \mright)$ is an abelian subalgebra of $Z \mleft(\mathrm{Ker}\mleft(\rho_p\mright)\mright)$. Moreover, we have
\ba\label{NablaZenterIstImKernyippie}
\rho(\nabla \nu)
&=0
\ea
for all (local) sections $\nu$ of $E$ with values in $Z\mleft( \nabla^{\mathrm{bas}} \mright)$, that is $\nabla \nu$ is an element of the kernel of the anchor.
\end{propositions}

\begin{remark}
\leavevmode\newline
The dimension of the kernel of $\rho$ is in general not constant such that we cannot expect that $Z\mleft( {}^E\nabla \mright)$ gives rise to a module with constant rank; but even if we just look at neighbourhoods around regular points of $E$ we cannot expect a constant rank. For example take $E = \mathrm{T}N \times K \to N \times S$, where we mean the direct sum of Lie algebroids of $\mathrm{T}N \to N$ and $K \to S$, where $K \to S$ is a Lie algebra bundle (zero-anchor) over a manifold $S$. Then take a coordinate frame $\mleft( \partial_i \mright)_i$ of $\mathrm{T}N$ and $\mleft( f_\alpha \mright)_\alpha$ of $K$, both constantly extended to $E$ such that $\mleft[ \partial_i, f_\alpha \mright]_E = 0$ and the total collection is denoted by $\mleft( e_a \mright)_a$. 
%For simplicity assume that this is a global frame such that we can take its canonical flat connection $\nabla$, defined by $\nabla e_a = 0$.
Let us look at $Z \mleft( \nabla^{\mathrm{bas}} \mright) \ni \nu = \nu^\alpha f_\alpha$ (using Prop.~\ref{prop:PropsofCentreOfBasicConnections}, especially $\nu$ is an element of the kernel)
\bas
0
&=
\nabla^{\mathrm{bas}}_\nu \partial_i
=
\nu^\alpha ~ \nabla^{\mathrm{bas}}_{f_\alpha} \partial_i
=
\nu^\alpha ~ \underbrace{\nabla_{\partial_i} f_\alpha}_{\mathclap{\eqqcolon \omega^a_{\alpha i} ~ e_a}}
=
\nu^\alpha \omega^a_{\alpha i} ~ e_a,
\eas
where we viewed $\partial_i$ as an element of the tangent bundle as Lie algebroid, \textit{i.e.}~we took the definition of $\nabla^{\mathrm{bas}}$ on Lie algebroids (denoted by $E$ usually). Hence, this is then a purely algebraic equation and depends also on the kernel of $\omega^a_{i \alpha}$ such that a general statement about the rank of the centre is not possible without further information about $\nabla$.
\end{remark}

\begin{proof}[Proof of Prop.~\ref{prop:PropsofCentreOfBasicConnections}]
\leavevmode\newline
We have, using the definition of $E$-Lie derivatives,
\bas
0
&=
{}^E\nabla_\nu (f v)
=
\mathcal{L}_\nu (f) ~ v
	+ f ~ \underbrace{{}^E\nabla_\nu v}_{=0}
=
\mathcal{L}_\nu (f) ~ v
=
\mathcal{L}_{\rho(\nu)} (f) ~ v
\eas
for all $v \in \Gamma(V)$, $\nu \in Z \mleft( {}^E\nabla \mright)$ and $f \in C^\infty(N)$. Since $V$ has at least rank 1, we can conclude that $\rho(\nu) = 0$. Hence, $\nu_p \in \mathrm{Ker}(\rho_p)$ for all $p \in N$. 

Furthermore, in the case of $\nabla^{\mathrm{bas}}$ we get additionally
\bas
0
&=
\nabla^{\mathrm{bas}}_{\nu_p} \mu_p
=
\mleft[ \nu_p, \mu_p \mright]_E
=
\mleft[ \nu_p, \mu_p \mright]_{\mathrm{Ker}(\rho_p)}
\eas
for all $\mu_p \in \mathrm{Ker}(\rho_p)$ and $p \in N$, where we used that $\nabla^{\mathrm{bas}}_{\nu_p}$ is tensorial due to $\rho(\nu_p)=0$ such that $\nabla^{\mathrm{bas}}_{\nu_p}$ can be viewed as a tensor (similar for $\mleft[ \cdot, \cdot \mright]_E$), and that the basic connection on $E$ is just the Lie bracket when acting on the kernel of the anchor. Hence, $\nu_p \in Z_p \mleft( \mathrm{Ker}(\rho_p) \mright)$, and, since $Z_p \mleft( \mathrm{Ker}(\rho_p) \mright)$ is abelian, it immediately follows that $Z_p\mleft(\nabla^{\mathrm{bas}}\mright)$ is an abelian subalgebra.

Finally, let $\nu \in \Gamma(E)$ with values in $Z\mleft( \nabla^{\mathrm{bas}} \mright)$, then we have
\bas
0
&=
\nabla^{\mathrm{bas}}_\nu Y
=
[\rho(\nu), Y]
	+ \rho(\nabla_Y \nu)
\eas
for all $Y \in \mathfrak{X}(N)$. Previously we have shown that $\rho(\nu) = 0$, this implies $\rho(\nabla_Y \nu) = 0$, which finishes the proof.
\end{proof}

Around regular points we can say a bit more, recall Thm.~\ref{thm:DirectSplitting}.

\begin{lemmata}{Centre of the basic connection around regular points}{CentreOfBasicConnectionForRegularPointsPlusFlatness}
Let $N$ be a smooth manifold and $K \to S$ be a bundle of Lie algebras over a smooth manifold $S$ such that $Z(K)$ is a subbundle of abelian Lie algebras, that is $Z(K)$ has constant rank. Then define the Lie algebroid $E$ as the direct product of Lie algebroids, $E \coloneqq \mathrm{T}N \times K \to N \times S$, equipped with a connection $\nabla = \nabla^{\mathrm{T}N} \times \nabla^K$, where $\nabla^{\mathrm{T}N}$ and $\nabla^K$ are connections on $\mathrm{T}N$ and $K$, respectively.
Then
%\begin{enumerate}
	%\item 
\ba
Z\mleft( \nabla^{\mathrm{bas}} \mright)= Z \mleft( K \mright).
\ea
	%\item $\nabla \nu \in \Gamma(Z(K))$ for all $\nu \in \Gamma(Z(K))$.
%\end{enumerate}
%With $\Gamma(Z(K))$ we mean sections of $E$ with values in $Z(K)$.
\end{lemmata}

\begin{remark}
\leavevmode\newline
In that case, $Z\mleft( \nabla^{\mathrm{bas}} \mright)$ has constant rank and is independent of the choice of $\nabla$.
\end{remark}

\begin{proof}[Proof of Lemma \ref{lem:CentreOfBasicConnectionForRegularPointsPlusFlatness}]
\leavevmode\newline
By definition of $E$, there are coordinates $\mleft( \partial_i \mright)_i$ of $N$ and a frame of $E$ consisting of two parts, $\mleft( f_i \mright)_i$ locally spanning $\mathrm{T}N$ (as Lie algebroid) and $\mleft( f_\alpha \mright)_\alpha$ locally spanning $K$, both (locally) constantly extended along the base of the other factor in $E = \mathrm{T}N \times K$, such that
\bas
\rho(f_i) &= \partial_i, &
\rho(f_\alpha) &= 0, \\
\mleft[ f_i, f_j \mright]_E &= 0, &
\mleft[ f_i, f_\alpha \mright]_E &= 0.
\eas
Since $Z(K)$ is a subbundle of Lie subalgebras of $K$ we can assume that $\mleft( f_\alpha\mright)_\alpha$ contains a subframe $\mleft( f_{\mathcal{r}} \mright)_{\mathcal{r}}$ spanning $Z(K)$.
Then for all $\nu = \nu^{\mathcal{r}} f_{\mathcal{r}} \in \Gamma(E)$ ($\nu^\alpha \in C^\infty(N\times S)$) with values in $Z(K)$ we then have by definition,
\ba\label{GesplitteteFormelVonNablaBas2}
\nabla^{\mathrm{bas}}_\nu f_i
&=
\nu^{\mathcal{r}} ~ \nabla^{\mathrm{bas}}_{f_{\mathcal{r}}} f_i
=
\nu^{\mathcal{r}} ~ \mleft(
	\mleft[ f_{\mathcal{r}}, f_i \mright]
	+ \nabla_{\rho(f_i)} f_{\mathcal{r}}
\mright)
=
\nu^{\mathcal{r}} ~ \nabla_{\partial_i} f_{\mathcal{r}},&
\nabla^{\mathrm{bas}}_\nu f_\alpha
&=
\mleft[ \nu, f_\alpha \mright]_K
=
0.
\ea
Similar to before, $\nabla^{\mathrm{bas}}_\nu$ is a tensor due to $\rho(\nu)=0$ such that Eq.~\eqref{GesplitteteFormelVonNablaBas2} are fully encoding $\nabla^{\mathrm{bas}}_\nu$ on $E$. Therefore we are interested into whether $\nabla^{\mathrm{bas}}_\nu f_i$ is zero. By definition $\nabla_\rho$ is flat when restricted onto $Z(K)$, \textit{i.e.}~on $Z(K)$-valued sections of $K$ which are constantly extended along $N$, that is, we have
\bas
\nabla_{\rho} f_{\mathcal{r}} &= 0.
\eas
Then for all $\nu = \nu^\mathcal{r} f_\mathcal{r}$ ($\nu^\mathcal{r}$ can depend on $N$) we get by Eq.~\eqref{GesplitteteFormelVonNablaBas2}
\bas
\nabla^{\mathrm{bas}}_\nu f_i
&=
\nu^\mathcal{r}  ~ \nabla_{\partial_i} f_\mathcal{r}
=
0
\eas
for all $i$. 
%\underline{\textbf{"2. $\Rightarrow$ 1.":}} 
By definition we also have 
\bas
\nabla \nu &\in \Gamma(K)
\eas
for all sections $\nu$ with values in the centre of $K$.
 %since $\nabla$ restricts on $\Gamma(Z(K))$ and since $Z(K)$ is abelian, it follows that $\nabla$ is a Lie derivation law on $Z(K)$, thus, we can apply Thm.~\ref{thm:BLALAB}, that is, $Z(K)$ is an LAB. That $Z(K)$ is abelian, is clear. 
Therefore, by Cor.~\ref{cor:LieDerivationGleichVanishingBasicCurvature}, we know
\bas
\nabla^{\mathrm{bas}}_\nu Y &= 0
\eas
for all $\nu \in \Gamma(E)$ with values in $Z(K)$ and $Y \in \mathfrak{X}(N)$. 

Hence,
\bas
\nabla^{\mathrm{bas}}_\nu &= 0
\eas
for all section $\nu$ with values in $Z(K)$. So, $Z(K) \subset Z \mleft( \nabla^{\mathrm{bas}} \mright)$. Recall Prop.~\ref{prop:PropsofCentreOfBasicConnections} such that we already know that
\bas
Z \mleft( \nabla^{\mathrm{bas}} \mright)
&\subset
Z \mleft( K \mright),
\eas
hence, $Z \mleft( \nabla^{\mathrm{bas}} \mright) = Z(K)$.
%
%\underline{\textbf{"1. $\Rightarrow$ 2.":}} Now we assume that $Z \mleft( \nabla^{\mathrm{bas}} \mright) = Z(K)$, and that it is an LAB, that means that the rank is constant. Hence, we can now again assume that $\mleft( f_\alpha\mright)_\alpha$ contains a subframe $\mleft( f_{\mathcal{r}} \mright)_{\mathcal{r}}$ spanning $Z(K)$. By Eq.~\eqref{GesplitteteFormelVonNablaBas2} we get
%\bas
%0
%&=
%\nabla^{\mathrm{bas}}_\nu f_i
%=
%\nu^{\mathcal{r}} ~ \nabla_{\partial_i} f_{\mathcal{r}}
%\eas
%for all $\nu \in \Gamma(Z(K))$. Therefore also
%\bas
%\nabla_\rho \nu
%&=
%\mleft(\mathrm{d}\nu^{\mathcal{r}} \circ \rho \mright) ~ f_{\mathcal{r}}
%\in \Gamma(Z(K)).
%\eas
\end{proof}

As already motivated, we have then a flat curvature in the case of CYMH GT.

\begin{corollaries}{Zero curvature on the centre}{FlacheKruemmungBeiNablaBasZentrum}
Let $E \to N$ be a Lie algebroid over a smooth manifold $N$, and $\nabla$ a connection on $E$ such that $R_\nabla$ is exact with respect to $\mathrm{d}^{\nabla^{\mathrm{bas}}}$, \textit{i.e.}~there is a $\zeta \in \Omega^2(N;E)$ with $R_\nabla(\cdot, \cdot) \mu = - \nabla^{\mathrm{bas}}_\mu \zeta$ for all $\mu \in \Gamma(E)$.\footnote{Here $\mathrm{d}^{\nabla^{\mathrm{bas}}}$ is not necessarily a differential.} Then
\ba
R_\nabla(\cdot, \cdot) \nu
&=
0
\ea
for all $\nu \in Z \mleft( \nabla^{\mathrm{bas}} \mright)$.
\end{corollaries}

\begin{proof}
\leavevmode\newline
That is a simple consequence of the $\mathrm{d}^{\nabla^{\mathrm{bas}}}$-exactness and $\nabla_\nu^{\mathrm{bas}} = 0$ for all $\nu \in Z \mleft( \nabla^{\mathrm{bas}} \mright)$.
\end{proof}

The vanishing of the basic curvature also implies in the general situation that $\nabla$ preserves such centres, similar to LABs.

\begin{lemmata}{Stability of the kernel of the adjoint representation}{StableKernelOfAdjointRepresentation}
Let $E \to N$ be a Lie algebroid over a smooth manifold $N$, and $\nabla$ a connection on $E$ with vanishing basic curvature and such that $R_\nabla$ is exact with respect to $\mathrm{d}^{\nabla^{\mathrm{bas}}}$. Moreover, we require
\bas
\rho(\nabla \nu)
&=
0
\eas
for all $\nu \in \Gamma(E)$ with $\rho(\nu) = 0$.

Then
\ba
\nabla^{\mathrm{bas}}_{\nabla \nu}
&=
0
\ea
for all $\nu \in \Gamma(E)$ with $\nabla^{\mathrm{bas}}_\nu = 0$, where we mean with $\nabla^{\mathrm{bas}}$ both connections, on $E$ and on $\mathrm{T}N$.
\end{lemmata}

\begin{proof}
\leavevmode\newline
We have, using Cor.~\ref{cor:FlacheKruemmungBeiNablaBasZentrum} and the vanishing basic curvature,
\bas
\nabla^{\mathrm{bas}}_{\nabla_Y \nu} \mu
&=
\mleft[ \nabla_Y \nu, \mu \mright]_E
	+ \nabla_{\rho(\mu)} \nabla_Y \nu
\\
&=
\mleft[ \nabla_Y \nu, \mu \mright]_E
	+ \nabla_Y \underbrace{\nabla_{\rho(\mu)} \nu}_{= \mleft[ \mu, \nu \mright]_E}
	+ \nabla_{[\rho(\mu), Y]} \nu
\\
&=
\underbrace{\mleft[ \nabla_Y \nu, \mu \mright]_E
	+ \mleft[ \mu, \nabla_Y \nu \mright]_E}
		_{=0}
	+ \mleft[ \nabla_Y \mu, \nu \mright]_E
	+ \underbrace{\nabla_{\nabla^{\mathrm{bas}}_\nu Y} \mu}
		_{=0}
	\underbrace{- \nabla_{\nabla^{\mathrm{bas}}_\mu Y} \nu
	+ \nabla_{[\rho(\mu), Y]} \nu}
		_{= - \nabla_{\rho\mleft( \nabla_Y \mu \mright)} \nu}
\\
&=
- \nabla^{\mathrm{bas}}_{\nu} \nabla_Y \mu
\\
&=
0
\eas
for all $\mu, \nu \in \Gamma(E)$, where $\nabla^{\mathrm{bas}}_\nu = 0$, and $Y \in \mathfrak{X}(N)$. Hence, only the basic connection on $\mathrm{T}N$ is left. We know $\rho(\nabla \nu) = 0$ by Eq.~\eqref{NablaZenterIstImKernyippie}, hence, by the condition on $\nabla$ about kernel-valued sections we have 
\bas
\rho\mleft( \nabla_X \nabla_Y \nu \mright)
&=
0
\eas
for all $X \in \mathfrak{X}(N)$, and so
\bas
\nabla^{\mathrm{bas}}_{\nabla_Y \nu} X
&=
[ \underbrace{\rho(\nabla_Y \nu)}_{=0}, X ]
	+ \rho(\nabla_X \nabla_Y \nu).
\eas
This proves the claim.
\end{proof}

With Cor.~\ref{cor:NablaErhaeltBasicConnKonstanz}, Lemma \ref{lem:StableKernelOfAdjointRepresentation} and Cor.~\ref{cor:FlacheKruemmungBeiNablaBasZentrum} we may have everything for doing something similar as for LABs. However, another important result for LABs was that $\mathrm{d}^\nabla\zeta$ is centre-valued; this was given by the Bianchi identity \ref{thm:BianchiIdentityForZeta}. This identity does now not immediately imply that $\mathrm{d}^\nabla \zeta$ is closed with respect to the basic connection; and even if, for example because it has values in the isotropy, we would still need that $\mathrm{d}^\nabla \zeta$ has also values in the centre of the basic connection in order to use Cor.~\ref{cor:FlacheKruemmungBeiNablaBasZentrum} to define a cohomology class. This is not given, not even by the Bianchi identity.

Summarizing, the problem is that we cannot simply generalize the discussion about LABs. The Bianchi identity for $\zeta$ suggests that a possible differential for a cohomology is a differential induced by $\nabla$ restricted on $\nabla^{\mathrm{bas}}$-closed forms. But the compatibility condition on $R_\nabla$ and $\zeta$ only implies flatness on sections with values in the centre of the basic connection. Even if we are able to construct suitable $\zeta$, satisfying all of that for $\mathrm{d}^\nabla \zeta$, it is not given that this construction is "stable enough" under the field redefinition, which is important in order to show that $\mathrm{d}^\nabla \zeta$ is an invariant of the field redefinition.

Concluding, this means one needs in general a (completely?) different construction; maybe hoping for that Conjecture \ref{conj:DoesMyFieldRedefSplit} holds. Nevertheless, one may see that the general situation is highly more complicated.

\chapter{Future works}\label{ConclusionTheEnd}
One may take these results as a motivation to always assume that a CYMH GT is pre-classical. 
There is hope to generalize the construction of the obstruction class to every Lie algebroid by assuming that the isotropy of the Lie algebroid is stable under the chosen connection. As we have seen, this stability condition is invariant under the field redefinition, and it may allow to reduce the study "roughly" to a study of Lie algebra bundles because the isotropy is a Lie algebra bundle around \textbf{regular points} in our case, also recall Thm.~\ref{thm:BLALAB}. Of course, a Lie algebroid consists of more than an isotropy. To take care of the remaining structure one could "decouple" the Lie algebroid along the foliation and along a transversal submanifold using the splitting theorem. However, we also have seen that there are certain difficulties in that approach.

%Sadly, nothing of this is published yet, but I am going to write two or even three preprints soon. This task was way more difficult than expected such that I couldn't publish anything earlier yet. I attached a section of my thesis which will be roughly the same as my first preprint which I want to upload in the next weeks.

Future plans for research could be studying a possible generalized definition of the obstruction class, using the previously-mentioned idea or another ansatz; in general, there are still a lot of open questions regarding general Lie algebroids which need to be answered. The question about the (physical) significance of the tensor $\zeta$ is interesting, too. For this it is also necessary to quantize this theory.

One could also think about integrating this theory, probably using Lie groupoids instead of Lie groups. Often it is of advantage if underlying curvatures are flat when it is about integrability, which may mean that $\nabla$ needs to be flat for a suitable integration and that may be a further argument for assuming that the theory is already pre-classical. However, since we used the basic connection to define infinitesimal gauge transformations, which is always flat in our context, we may or may not have solved a certain problem in integrating CYMH GTs.

Another possible plan is to go back to the example of unit octonions. $\mathbb{S}^7$ is a Moufang loop and its corresponding tangent space at its neutral element is an algebra known as Malcev algebra. Hence, this example may show that a suitable new formulation of gauge theory may be in replacing Lie groups and Lie algebras with Moufang loops and Malcev algebras, respectively. In a private talk to Alessandra Frabetti I learned that one seemingly only needs the structure of Moufang loops for renormalizations such that it might be fruitful to develop a gauge theory using that notion.

\textbf{\emph{Thanks}} for reading and your support! Do not hesitate to ask me further questions. I wish you a nice and pleasant time.

\textbf{Acknowledgements about finances:} This work was produced within the scope of the NCCR SwissMAP which was funded by the Swiss National Science Foundation. I would like to thank the Swiss National Science Foundation for their financial support.

This thesis was also supported by the LABEX MILYON (ANR-10-LABX-0070) of Universit\'{e} de Lyon, within the program "Investissements d'Avenir" (ANR-11-IDEX- 0007) operated by the French National Research Agency (ANR).
%oder \include für anderen Seitenanfang?
%\setcounter{equation}{0}

%%%%%%%%%%%%%%%%%%%%%%%%%%%% Hier beginnt der Anhang %%%%%%%%%%%%%%%%%%%%%%%%%%%%

\newpage

%\listoftables % Tabellenverzeichnis

%\listoffigures %Abbildungsverzeichnis

\appendix
\setcounter{equation}{0}
\renewcommand{\theequation}{\Alph{chapter}.\arabic{equation}} %Reset first and then add section to number
%\section{Identities for the calculus given in (C)YMH GT}\label{CalculusIdentitiesNeeded}

%%%%%%%%%%%%%%%%%%%%%%%%%%%% Hier beginnt der Anhang %%%%%%%%%%%%%%%%%%%%%%%%%%%%

%\newpage

%\listoftables % Tabellenverzeichnis

%\listoffigures %Abbildungsverzeichnis

%\appendix
%\setcounter{equation}{0}
%\renewcommand{\theequation}{\Alph{section}.\arabic{equation}} %Reset first and then add section to number
\chapter{Certain useful identities}\label{CalculusIdentitiesNeeded}

\section{Lie algebra bundles}

In this appendix we prove and define very basic notions, which are often direct generalizations of typical relations known in gauge theory. It is recommended to read this part at the beginning of Chapter \ref{GeneralizedGTfas}, especially if one is interested into all the calculations. Recall the following wedge product\footnote{As also defined in \cite[\S 5, third part of Exercise 5.15.12; page 316]{hamilton}.} of forms with values in a vector bundle $E$ and values in its space of endomorphisms $\mathrm{End}(E)$,
\bas
\wedge: \Omega^k(N; \mathrm{End}(E)) \times \Omega^l(N; E)
&\mapsto
\Omega^{k+l}(N; E) \\
(T, \omega) &\mapsto T \wedge \omega
\eas
for all $k, l \in \mathbb{N}_0$, given by
\ba\label{DefVonWedgedemitEnd}
\mleft( T \wedge \omega \mright) \mleft( Y_1, \dotsc, Y_{k+l} \mright)
&\coloneqq
\frac{1}{k! l!} \sum_{\sigma \in S_{k+l}} \mathrm{sgn}(\sigma) ~
	T \mleft( Y_{\sigma(1)}, \dotsc, Y_{\sigma(k)} \mright)
		\mleft( \omega\mleft( Y_{\sigma(k+1)}, \dotsc, Y_{\sigma(k+l)} \mright) \mright),
\ea
where $S_{k+l}$ is the group of permutations $\{1, \dotsc, k+l\}$. This is then locally given by, with respect to a frame $\mleft( e_a \mright)_a$ of $E$,
\bas
T \wedge \omega &= T(e_a) \wedge w^a,
\eas
where $T$ acts as an endomorphism on $e_a$, \textit{i.e.}~$T(e_a) \in \Omega^k(N; E)$, and $\omega = \omega^a \otimes e_a$. Also recall that there is the canonical extension of $\nabla$ on $\mathrm{End}(E)$ by forcing the Leibniz rule. We still denote this connection by $\nabla$, too.

\begin{propositions}{Several useful identities}{SeveralIdentitiesFortheCalculusWithPullbackandBlah}
Let $M$ and $N$ be two smooth manifolds, $K \to N$ a vector bundle, $\Phi: M \to N$ a smooth map, $\nabla$ a connection on $K$, and $k,l, m \in \mathbb{N}_0$. Then we have
\ba\label{EqGeilePullBackCommuteFormel}
\mathrm{d}^{\Phi^*\nabla}\mleft( \Phi^!\omega \mright)
&=
\Phi^! \mleft( \mathrm{d}^\nabla \omega \mright), \\
\mathrm{d}^{\nabla+D} \omega
&=
\mathrm{d}^\nabla \omega + D \wedge \omega, \label{eqDifferentialSplit}, \\
\mathrm{d}^\nabla \mleft( T \wedge \omega \mright)
&=
\mathrm{d}^\nabla T \wedge \omega
	+ (-1)^m ~ T \wedge \mathrm{d}^\nabla \omega \label{TypischerSplitdesDifferentialsaufdasWedgeProdukt}
\ea
for all $\omega \in \Omega^l(N; K)$, $\psi \in \Omega^k(N;K)$, $D \in \Omega^1(N; \mathrm{End}(K))$, and $T \in \Omega^m(N; \mathrm{End}(K))$.
\newline

If $K$ is additionally an LAB, then we also have
\ba
\mleft( \mathrm{ad} \circ \omega \mright) \wedge \psi
&=
\mleft[ \omega \stackrel{\wedge}{,} \psi \mright]_K, \label{wedgeproduktmitadLambdaergibtLieklammer} \\
\Phi^!\mleft( \mleft[ \omega \stackrel{\wedge}{,} \psi \mright]_K \mright)
&=
\mleft[ \Phi^!\omega \stackrel{\wedge}{,} \Phi^!\psi \mright]_{\Phi^*K}, \label{eqPullbackofLiebracketStuff} \\
\mleft[ \omega \stackrel{\wedge}{,} \psi \mright]_K
&=
- (-1)^{lk} ~ \mleft[ \psi \stackrel{\wedge}{,} \omega \mright]_K, \label{VertauschungsregelForKKlammerAufFormen}\\
\mleft[ \omega \stackrel{\wedge}{,} \mleft[ \omega \stackrel{\wedge}{,} \omega \mright]_K \mright]_K
&=
0 \label{JacobiIdentityForFormBracket}, \\
\mathrm{ad}^* \circ \Phi^!\omega
&=
\Phi^!\mleft( \mathrm{ad} \circ \omega \mright) \label{EqCommutationRelation}
\ea
for all $\omega \in \Omega^l(N; K)$, $\psi \in \Omega^k(N;K)$, and smooth maps $\Phi: M \to N$, where we write $\mathrm{ad}^*$ for the adjoint representation with respect to $\mleft[ \cdot, \cdot \mright]_{\Phi^*K}$.
\end{propositions}

\begin{remarkohne}
\leavevmode\newline
Eq.~\eqref{VertauschungsregelForKKlammerAufFormen} and Eq.~\eqref{JacobiIdentityForFormBracket} are generalizations of similar expressions just using the Lie algebra bracket $\mleft[ \cdot, \cdot\mright]_{\mathfrak{g}}$ of a Lie algebra $\mathfrak{g}$, which basically is the formulation on trivial LABs, see \cite[\S 5, first and second statement of Exercise 5.15.14; page 316]{hamilton}. Eq.~\eqref{TypischerSplitdesDifferentialsaufdasWedgeProdukt} is of course the typical Leibniz rule of the exterior covariant derivative just extended to the wedge-product with $\mathrm{End}(K)$-valued forms, and Eq.~\eqref{EqGeilePullBackCommuteFormel} is a generalization of the well-known $\Phi^! \circ \mathrm{d} = \mathrm{d} \circ \Phi^!$, where $\mathrm{d}$ is the de-Rham differential (we omit to clarify on which manifold; this should be given by the context).
\end{remarkohne}

\begin{proof}
\leavevmode\newline
\indent $\bullet$ Recall that we have the following property of the pullback connection
\bas
\mleft( \Phi^*\nabla \mright)_Y \mleft( \Phi^* \mu \mright)
&=
\Phi^*\mleft( \nabla_{\mathrm{D}\Phi(Y)} \mu \mright)
\eas
for all $Y \in \mathfrak{X}(M)$, smooth maps $\Phi: M \to N$, connections $\nabla$, and $\mu \in \Gamma(K)$, shortly writing as\footnote{Recall that the pull-back of forms is denoted with an exclamation mark.}
\ba\label{eqShortNotationForPullbackConnections}
\mleft( \Phi^*\nabla \mright) \mleft( \Phi^* \mu \mright)
&=
\Phi^*\mleft( \nabla_{\mathrm{D}\Phi} \mu \mright)
=
\Phi^!(\nabla \mu),
\ea
viewing terms like $\nabla \mu$ as an element of $\Omega^1(N; K)$, $\mathfrak{X}(N) \ni \xi \mapsto \nabla_\xi \mu$, such that we can apply Eq.~\eqref{EqPullBackFormelFuerVerschiedeneDefinitionen}. That extends to exterior covariant derivatives by fixing a local frame $\mleft(e_a\mright)_a$ of $K$ (also used in the following), then we have $\omega^a \in \Omega^l(U)$ ($l \in \mathbb{N}_0$) such that locally
\bas
\omega
&=
\omega^a \otimes e_a
\eas
for all $\omega \in \Omega^l(N; K)$. The exterior covariant derivative generally (locally) writes
\bas
\mathrm{d}^{\Phi^*\nabla}w
&=
\mathrm{d}w^a \otimes \Phi^*e_a
	+ (-1)^l w^a \wedge \underbrace{\mleft( \Phi^*\nabla \mright)\mleft( \Phi^*e_a \mright)}_{\stackrel{\text{Eq.~\eqref{eqShortNotationForPullbackConnections}}}{=} \Phi^!\mleft( \nabla e_a \mright)}
=
\mathrm{d}w^a \otimes \Phi^*e_a
	+ (-1)^l w^a \wedge \Phi^!\mleft( \nabla e_a \mright)
\eas
for all $w \in \Omega^l(M; \Phi^*K)$,
and the pull-back of forms clearly splits over this tensor product by its definition, \textit{i.e.}
\bas
\Phi^!\omega
&=
\Phi^!\omega^a \otimes \Phi^* e_a,
\eas
and, so,
\bas
\mathrm{d}^{\Phi^*\nabla}\mleft( \Phi^!\omega \mright)
&=
\underbrace{\mathrm{d}\mleft( \Phi^! w^a \mright)}_{\mathclap{= \Phi^! \mleft(\mathrm{d} \omega^a\mright)}} \otimes~ \Phi^*e_a
	+ (-1)^l ~ \Phi^!w^a \wedge \Phi^!\mleft( \nabla e_a \mright) 
\nonumber \\
&=
\Phi^!\mleft( \mathrm{d}\omega^a \otimes e_a + (-1)^l ~ \omega^a \wedge \nabla e_a \mright) \nonumber \\
&=
\Phi^! \mleft( \mathrm{d}^\nabla \omega \mright). 
\eas

$\bullet$ Observe
\bas
\mathrm{d}^{\nabla+D} \omega
&=
\mathrm{d}\omega^a \otimes e_a + (-1)^l ~ \omega^a \wedge \mleft(\nabla + D\mright) e_a
= \mathrm{d}^\nabla \omega + D \wedge \omega
\eas
for all $\omega\in\Omega^l(N;K)$, $D \in \Omega^1(N;K)$, and connections $\nabla$ on $K$.

$\bullet$ Now let $T \in \Omega^m(N; \mathrm{End}(K))$ and $\mleft( L_a \mright)_a$ a frame of $\mathrm{End}(K)$, such that we can write $T= T^a \otimes L_a$, then
\bas
\mathrm{d}^\nabla(T \wedge \omega)
&=
\mathrm{d}^\nabla(T(e_a) \wedge \omega^a)
=
\mathrm{d}^\nabla(T(e_a)) \wedge \omega^a
	+ (-1)^m ~ T(e_a) \wedge \mathrm{d} \omega^a
\eas
for all $\omega \in \Omega^l(N; K)$, and 
\bas
&&
\mleft( \mathrm{d}^\nabla T \mright)(e_a)
&=
\mathrm{d}T^b \otimes L_b(e_a)
	+ (-1)^m ~ T^b \wedge \underbrace{(\nabla L_b)(e_a)}
	_{\mathclap{= ~ \nabla (L_b(e_a)) - L_b(\nabla e_a)}} \\
&&&=
\mathrm{d}^\nabla(T(e_a))
	- (-1)^m ~ T^b \wedge L_b(\nabla e_a) \\
&&&=
\mathrm{d}^\nabla(T(e_a))
	- (-1)^m ~ \underbrace{\mleft(T^b \otimes L_b\mleft( e_c \mright) \mright)}_{= ~ T(e_c)} \wedge ~ \mleft(\nabla e_a\mright)^c \\
&&&=
\mathrm{d}^\nabla(T(e_a))
	- (-1)^m ~ T \wedge \nabla e_a \\
&\Leftrightarrow&
\mathrm{d}^\nabla(T(e_a))
&=
\mleft( \mathrm{d}^\nabla T \mright)(e_a) + (-1)^m ~ T \wedge \nabla e_a.
\eas
Combining both equations, we arrive at
\bas
\mathrm{d}^\nabla(T \wedge \omega)
&=
\mathrm{d}^\nabla T \wedge \omega
	+ (-1)^m ~ T(e_a) \wedge \mleft( \mathrm{d}\omega^a + (-1)^l ~ w^b \wedge \mleft(\nabla e_b\mright)^a \mright) \\
&=
\mathrm{d}^\nabla T \wedge \omega
	+ (-1)^m ~ T \wedge \mathrm{d}^\nabla \omega.
\eas

In the following let $K$ also be an LAB.

$\bullet$ We also have
\bas
&(\underbrace{\mleft( \mathrm{ad} \circ \omega \mright)}_{\mathclap{\in ~ \Omega^l(N;~ \mathrm{End}(K))}} \wedge ~\psi)(Y_1, \dotsc, Y_{l+k}) \\
&\hspace{1cm}\stackrel{\mathclap{\text{Def.~\eqref{DefVonWedgedemitEnd}}}}{=}~~~
\frac{1}{k!l!}
\sum_{\sigma \in S_{k+l}}
\mathrm{sgn}(\sigma) ~
	\mleft[ \omega\mleft(Y_{\sigma(1)}, \dotsc, Y_{\sigma(l)}\mright), \psi\mleft(Y_{\sigma(l+1)}, \dotsc, Y_{\sigma(l+k)}\mright) \mright]_K \\
&\hspace{1cm}\stackrel{\mathclap{\text{Def.~\ref{def:GradingOfProducts}}}}{=}~~~
\mleft[ \omega \stackrel{\wedge}{,} \psi \mright]_K(Y_1, \dotsc, Y_{l+k})
\eas
for all $w\in\Omega^l(N;K)$, $\psi \in \Omega^k(N;K)$, and $Y_1, \dotsc, Y_{l+k} \in \mathfrak{X}(N)$, where $S_{k+l}$ is the group of permutations $\{1, \dotsc, k+l\}$.

$\bullet$ By definition of $\Phi^*K$ we have
\bas
\mleft[ \Phi^*\mu, \Phi^*\nu \mright]_{\Phi^*K}
&=
\Phi^*\mleft( \mleft[ \mu, \nu \mright]_{K} \mright)
\eas
for all smooth maps $\Phi: M \to N$ and $\mu, \nu \in \Gamma(K)$. Let $\mleft( e_a \mright)_a$ be again a fixed frame of $K$, $\omega =  \omega^a \otimes e_a \in \Omega^l(N;K)$ and $\psi = \psi^a \otimes e_a \in \Omega^k(N;K)$, then, again using Def.~\ref{def:GradingOfProducts},
\bas
\Phi^!\mleft( \mleft[ \omega \stackrel{\wedge}{,} \psi \mright]_K \mright)
&=
\Phi^!\mleft( \mleft[ e_a , e_b \mright]_K \otimes \omega^a \wedge \psi^b \mright)
=
\underbrace{\Phi^*\mleft(\mleft[ e_a , e_b \mright]_K\mright)}_{\mathclap{=~\mleft[ \Phi^*e_a, \Phi^*e_b \mright]_{\Phi^*K}}} \otimes \Phi^!\omega^a \wedge \Phi^!\psi^b
=
\mleft[ \Phi^!\omega \stackrel{\wedge}{,} \Phi^!\psi \mright]_{\Phi^*K}.
\eas

$\bullet$ The antisymmetry of the Lie bracket generalizes to
\bas
\mleft[ \omega \stackrel{\wedge}{,} \psi \mright]_K
&=
\underbrace{\mleft[ e_a, e_b \mright]_K}
_{=~ -\mleft[ e_b, e_a \mright]_K} 
\otimes \underbrace{\omega^a \wedge \psi^b}
_{=~(-1)^{lk} \psi^b \wedge \omega^a}
=
- (-1)^{lk} ~ \mleft[ \psi \stackrel{\wedge}{,} \omega \mright]_K
\eas
for all $\omega \in \Omega^l(N;K)$ and $\psi \in \Omega^k(N;K)$.

$\bullet$ Let $\mleft( e_a \mright)_a$ be still a local frame of $K$, then
\bas
&&
\mleft[ \omega \stackrel{\wedge}{,} \mleft[ \omega \stackrel{\wedge}{,} \omega \mright]_K \mright]_K
~~~&\stackrel{\mathclap{\text{Eq.~\eqref{VertauschungsregelForKKlammerAufFormen}}}}{=}~~~
- (-1)^{2l^2} ~
\mleft[ \mleft[ \omega \stackrel{\wedge}{,} \omega \mright]_K \stackrel{\wedge}{,} \omega \mright]_K \\
&&
&=
- \underbrace{\mleft[ \mleft[ e_a, e_b \mright]_K, e_c \mright]_K}
_{\stackrel{\text{Jacobi}}{=}~ \mleft[ e_a, \mleft[ e_b, e_c \mright]_K \mright]_K + \mleft[ e_b, \mleft[ e_c, e_a \mright]_K \mright]_K}
 \otimes ~\omega^a \wedge \omega^b \wedge \omega^c \\
&&
&=
- \mleft[ \omega \stackrel{\wedge}{,} \mleft[ \omega \stackrel{\wedge}{,} \omega \mright]_K \mright]_K
	- \mleft[ e_b, \mleft[ e_c, e_a \mright]_K \mright]_K \otimes 
	\underbrace{\omega^a \wedge \omega^b \wedge \omega^c}_{\mathclap{=~ (-1)^{2l^2} \omega^b \wedge \omega^c \wedge \omega^a}} \\
&&
&=
-2~ \mleft[ \omega \stackrel{\wedge}{,} \mleft[ \omega \stackrel{\wedge}{,} \omega \mright]_K \mright]_K \\
&\Leftrightarrow&
\mleft[ \omega \stackrel{\wedge}{,} \mleft[ \omega \stackrel{\wedge}{,} \omega \mright]_K \mright]_K
&= 0
\eas
for all $\omega \in \Omega^l(N;K)$.

$\bullet$ We also have
\bas
\mleft[ \Phi^!\omega, \Phi^*\mu \mright]_{\Phi^*K}
&\stackrel{\text{Eq.~\eqref{eqPullbackofLiebracketStuff}}}{=}
\Phi^!\mleft( \mleft[ \omega, \mu \mright]_K \mright)
=
\Phi^!\Big( (\mathrm{ad} \circ \omega)(\mu) \Big)
=
\underbrace{\mleft(\Phi^!\mleft( \mathrm{ad} \circ \omega \mright)\mright)}_{\mathclap{\in ~ \Omega^1(M; ~\mathrm{End}(\Phi^*K))}}(\Phi^*\mu)
\eas
for all $\mu \in \Gamma(K)$, $\omega \in \Omega^l(N;K)$, and smooth maps $\Phi: M \to N$, where we used $(\Phi^*T)(\Phi^*\mu) = \Phi^*(T(\mu))$ for all $T \in \Gamma(\mathrm{End}(K))$ for the last equality. Since sections of $\Phi^*K$ are generated by pullbacks of sections of $K$, we can conclude
\bas
\mathrm{ad}^* \circ \Phi^!\omega
&=
\Phi^!\mleft( \mathrm{ad} \circ \omega \mright).
\eas
\end{proof}

When we add the compatibility conditions~\eqref{CondSGleichNullLAB}, then we have a few more identities.

\begin{corollaries}{Identities related to Lie bracket derivations}{IdentitiesFuerBianchiZeugs}
Let $K \to N$ be an LAB, equipped with a connection $\nabla$ satisfying compatibility condition~\eqref{CondSGleichNullLAB}; also let $M$ be another smooth manifold and $\Phi: M \to N$ a smooth map. Then
\ba\label{eqDerivationOfDifferentialOnBracketonK}
\mathrm{d}^\nabla\bigl( \mleft[ \omega\stackrel{\wedge}{,} \psi \mright]_K \bigr)
&=
\mleft[ \mathrm{d}^\nabla \omega \stackrel{\wedge}{,} \psi \mright]_K
	+ (-1)^l~ \mleft[ \omega \stackrel{\wedge}{,} \mathrm{d}^\nabla \psi \mright]_K, \\
\mathrm{d}^\nabla \mleft( \mathrm{ad} \circ \omega \mright)
&=
\mathrm{ad} \circ \mathrm{d}^\nabla \omega \label{DifferentialvonNabalVertauschmitAd}
\ea
for all $\omega \in \Omega^l(N; K)$ and $\psi \in \Omega^k(N; K)$.
%\newline
%
%When $\nabla$ satisfies both compatibility conditions~\eqref{CondSGleichNullLAB} and~\eqref{CondKruemmungmitBLAB} with respect to a $\zeta \in \Omega^2(N; K)$, we then get
%\ba
%\mathrm{ad} \circ \mathrm{d}^\nabla \zeta
%&=
%0,
%\ea
%\textit{i.e.}~$\mathrm{d}^\nabla \zeta$ has only values in the centre of $K$. 
\end{corollaries}

\begin{remarkohne}
\leavevmode\newline
Eq.~\eqref{eqDerivationOfDifferentialOnBracketonK} is a direct generalization of \cite[\S 5, third statement of Exercise 5.15.14 where it is stated for $\mathfrak{g}$ (trivial LAB with canonical flat connection); page 316]{hamilton}.
\end{remarkohne}

\begin{proof}
\leavevmode\newline
\indent $\bullet$ Using compatibility condition~\eqref{CondSGleichNullLAB} and a local frame $\mleft( e_a \mright)_a$ of $K$,
\bas
\mathrm{d}^\nabla \mleft( \mleft[ \omega \stackrel{\wedge}{,} \psi \mright]_K \mright)
&=
\mathrm{d}^\nabla\mleft( \mleft[ e_a, e_b \mright]_K \otimes \omega^a \wedge \psi^b \mright) \\
&=
\underbrace{\nabla \mleft( \mleft[ e_a, e_b \mright]_K \mright)}
_{=~ \mleft[ \nabla e_a, e_b \mright]_K + \mleft[ e_a, \nabla e_b \mright]_K}
	\wedge ~ \omega^a \wedge \psi^b
	+ \mleft[ e_a, e_b \mright]_K \otimes \mathrm{d}\omega^a \wedge \psi^b \\
&\hspace{1cm}
	+ (-1)^l ~ \mleft[ e_a, e_b \mright]_K \otimes \omega^a \wedge \mathrm{d} \psi^b \\
&=
 \mleft[ e_a, e_b \mright]_K \otimes \mleft( \nabla e_c \mright)^a \wedge \omega^c \wedge \psi^b
	+ (-1)^l ~ \mleft[ e_a, e_b \mright]_K \otimes \omega^a \wedge \mleft( \nabla e_c \mright)^b \wedge \psi^c \\
&\hspace{1cm}
	+ \mleft[ e_a, e_b \mright]_K \otimes \mathrm{d} \omega^a \wedge \psi^b
	+ (-1)^l ~ \mleft[ e_a, e_b \mright]_K \otimes \omega^a \wedge \mathrm{d} \psi^b \\
&=
\mleft[ e_a, e_b \mright]_K \otimes \Big(
	\Big( \underbrace{\mleft( \nabla e_c \mright)^a \wedge \omega^c + \mathrm{d} \omega^a}
	_{=~ \mleft( \mathrm{d}^\nabla \omega \mright)^a} \Big) \wedge \psi^b
	+ (-1)^l ~ \omega^a \wedge \mleft( \mleft( \nabla e_c \mright)^b \wedge \psi^c + \mathrm{d} \psi^b \mright)
\Big) \\
&=
\mleft[ \mathrm{d}^\nabla \omega \stackrel{\wedge}{,} \psi \mright]_K
	+ (-1)^l~ \mleft[ \omega \stackrel{\wedge}{,} \mathrm{d}^\nabla \psi \mright]_K
\eas
for all $\omega \in \Omega^l(N;K)$ and $\psi \in \Omega^k(N;K)$.

$\bullet$ Then by Eq.~\eqref{TypischerSplitdesDifferentialsaufdasWedgeProdukt} and~\eqref{wedgeproduktmitadLambdaergibtLieklammer}, we get
\bas
\mathrm{d}^\nabla \mleft( \mleft[ \omega \stackrel{\wedge}{,} \psi \mright]_K \mright)
&=
\mathrm{d}^\nabla \mleft( (\mathrm{ad} \circ \omega) \wedge \psi \mright)
=
\mathrm{d}^\nabla \mleft( \mathrm{ad} \circ \omega \mright) \wedge \psi
	+ (-1)^l ~ (\mathrm{ad} \circ \omega) \wedge \mathrm{d}^\nabla \psi,
\eas
and we can rewrite Eq.~\eqref{eqDerivationOfDifferentialOnBracketonK}
\bas
\mathrm{d}^\nabla \mleft( \mleft[ \omega \stackrel{\wedge}{,} \psi \mright]_K \mright)
&=
\mleft( \mathrm{ad} \circ \mathrm{d}^\nabla \omega \mright) \wedge \psi
	+ (-1)^l ~ (\mathrm{ad} \circ \omega) \wedge \mathrm{d}^\nabla \psi.
\eas
Combining both, we have
\bas
\mathrm{d}^\nabla \mleft( \mathrm{ad} \circ \omega \mright) \wedge \psi
&=
\mleft( \mathrm{ad} \circ \mathrm{d}^\nabla \omega \mright) \wedge \psi
\eas
for all $\omega \in \Omega^l(N;K)$ and $\psi \in \Omega^k(N;K)$. By (locally) using the 0-forms $\psi = e_a$ for all $a$, this implies Eq.~\eqref{DifferentialvonNabalVertauschmitAd}.
\end{proof}

\newpage

\renewcommand\refname{List of References}
%\begin{thebibliography}{99}
%\bibitem[I]{Anl01} \url{http://adsabs.harvard.edu/abs/1971ApJ...170..319D}, Datum: 01.11.2014
%\end{thebibliography}

%\printbibliography 
\bibliography{Literatur/Literatur}
\bibliographystyle{unsrt}

\printnoidxglossary[type=symbols,style=long4col,title={List of Symbols}]
%\printunsrtglossary[type=symbols,style=alttreegroup,title={List of Symbols}]
%\newpage
%\pagestyle{empty}
%\cleardoublepage
%
%%\newpage\thispagestyle{empty}\hspace{1em}\newpage
%
%Name: \fullname \hfill Matrikelnummer: \matnr \vspace{2cm}
%
 %\begin{center}
%{\large \textbf{Ehrenwörtliche Erklärung}}
%\end{center}
%
%\vspace{1cm}
%
%\noindent
%Ich erkläre hiermit ehrenwörtlich, dass ich die vorliegende Arbeit
%selbstständig angefertigt habe; die aus fremden Quellen direkt oder indirekt 
%übernommenen Gedanken sind als solche kenntlich gemacht. Die Arbeit wurde 
%bisher keiner anderen Prüfungsbehörde vorgelegt und auch noch nicht 
%veröffentlicht.\\
%
%\noindent
%Ich bin mir bewusst, dass eine unwahre Erklärung rechtliche Folgen haben wird.
%
%\vspace{2cm}
%
%
%
%\noindent Genf, den \dotfill
%
%\hspace{10cm} {\footnotesize \fullname}

\end{document}